\documentclass[12pt]{article}
\pdfoutput=1
\usepackage[T1]{fontenc}
\usepackage{lmodern}

\usepackage{graphicx}
\usepackage{epsfig}
\usepackage{epstopdf}
\usepackage{url}
\usepackage[implicit=false]{hyperref}
\DeclareGraphicsExtensions{.pdf,.eps,.png,.jpg,.mps}
\usepackage{amsmath}
\usepackage{amsfonts}
\usepackage{color}

\usepackage{cite}
\usepackage{makeidx}
\makeindex
\usepackage{fancyhdr}
\renewcommand\indexname{\'Indice}

\def\lsim{\mathrel{\rlap{\lower3pt\hbox{\hskip0pt$\sim$}}
     \raise1pt\hbox{$<$}}}         
\def\gsim{\mathrel{\rlap{\lower4pt\hbox{\hskip1pt$\sim$}}
     \raise1pt\hbox{$>$}}}         

\begin{document}
\emergencystretch 3em
\begin{titlepage}
\lhead{
\vspace{-3.0in}
\textcolor{blue}{Esto es la versi{\'o}n completa del siguiente libro:}
Z. Kakushadze y J.A. Serur. {\em 151 Estrategias de Trading} (Spanish Edition, 2019), 398 pp; ISBN 978-1071261873; \url{https://www.amazon.com/dp/1071261878}.\\
\textcolor{blue}{Copyright \copyright{} 2018 Zura Kakushadze and Juan Andr\'{e}s Serur. All Rights Reserved.}\\
}
\cfoot{\thepage}
\centerline{\Large \bf 151 Estrategias de Trading}
\medskip

\centerline{
Zura Kakushadze$^\S$$^\dag$\footnote{\, Zura Kakushadze, Ph.D., es el Presidente y CEO de Quantigic$^\circledR$ Solutions LLC,
y Profesor Titular en la Free University of Tbilisi. Email: \href{mailto:zura@quantigic.com}{zura@quantigic.com}
}
y
Juan Andr\'{e}s Serur$^\sharp$\footnote{\, Juan Andr\'{e}s Serur, M.Fin., es Profesor Asistente en la Universidad del CEMA.
Email: \href{mailto:jaserur15@ucema.edu.ar}{jaserur15@ucema.edu.ar}}
}
\bigskip

\centerline{\em $^\S$ Quantigic$^\circledR$ Solutions LLC}
\centerline{\em 1127 High Ridge Road \#135, Stamford, CT 06905\,\,\footnote{\, DESCARGO: Esta direcci\'on es utilizada por el autor correspondiente con el \'unico prop\'osito de indicar su afiliaci\'on profesional como es habitual en las publicaciones. En particular, el contenido de este documento no pretende ser un consejo de inversi\'on, legal, fiscal o de otro tipo, y de ninguna manera representa una opini\'on de Quantigic$^\circledR$ Solutions LLC, el sitio web \url{www.quantigic.com} o cualquiera de sus otros afiliados.
}}
\centerline{\em $^\dag$ Free University of Tbilisi, Business School \& School of Physics}
\centerline{\em 240, David Agmashenebeli Alley, Tbilisi, 0159, Georgia}
\centerline{\em $^\sharp$ Universidad del CEMA}
\centerline{\em Av. C\'{o}rdoba 374, C1054AAP, Ciudad de Buenos Aires, Argentina}
\medskip
\centerline{(La versi\'on en ingl\'es: 17 de agosto de 2018; la versi\'on en espa\~{n}ol: 30 de mayo de 2019)}

\bigskip
\medskip
\centerline{\it ZK: A mi madre Mila y mis hijos Mirabelle y Maximilien}
\medskip
\centerline{\it JAS: A mis padres, Claudio y Andrea, y mi hermano Emiliano}
\bigskip

\begin{abstract}
{}{}Proporcionamos descripciones detalladas, que incluyen m\'as de 550 f\'ormulas matem\'aticas, para m\'as de 150 estrategias de trading para una gran cantidad de clases de activos (y estilos de trading). Esto incluye acciones, opciones, bonos (renta fija), futuros, ETFs, \'indices, commodities, divisas, bonos convertibles, activos estructurados, volatilidad (como clase de activos), bienes inmuebles, activos en distress, efectivo, criptomonedas, miscel\'aneos (como clima, energ\'ia, inflaci\'on), macro global, infraestructura y arbitraje impositivo. Algunas estrategias se basan en algoritmos de aprendizaje autom\'atico (como redes neuronales artificiales, Bayes, k vecinos m\'as cercanos). El libro tambi\'en incluye: c\'odigo para backtesting fuera de la muestra con notas explicativas; cerca de 2,000 referencias bibliogr\'aficas; m\'as de 900 t\'erminos que comprenden el glosario, acr\'onimos y definiciones matem\'aticas. La presentaci\'on pretende ser descriptiva y pedag\'ogica, y de particular inter\'es para los profesionales de las finanzas, traders, investigadores, acad\'emicos y estudiantes de escuelas de negocios y programas de finanzas.
\end{abstract}
\bigskip
\textcolor{blue}{Esto es la versi{\'o}n completa del siguiente libro:}
Z. Kakushadze y J.A. Serur. {\em 151 Estrategias de Trading} (Spanish Edition, 2019), 398 pp; ISBN 978-1071261873; \url{https://www.amazon.com/dp/1071261878}. \textcolor{blue}{Copyright \copyright{} 2018 Zura Kakushadze and Juan Andr\'es Serur. All Rights Reserved.}
(Traducci\'on del ingl\'es al espa\~{n}ol realizada por JAS, con revisi\'on realizada por ZK.)

\end{titlepage}

\newpage

\vspace*{\stretch{1}}

\noindent Translation from the English language edition:\\
151 Trading Strategies by Zura Kakushadze and Juan Andr\'es Serur\\
Copyright \copyright{} The Editor(s) (if applicable) and The Author(s) 2018. All Rights Reserved.

\vspace*{\stretch{1}}

\newpage

\tableofcontents

\newpage
\phantomsection
\addcontentsline{toc}{section}{Comentarios de {\em 151 Estrategias de Trading}}
\section*{Comentarios de {\em 151 Estrategias de Trading}}

\noindent
``Si usted desea trabajar como trader o analista cuantitativo en Wall Street, debe prometer y cumplir. Este libro \'unico es una introducci\'on completa a una amplia variedad de estrategias de trading probadas y testeadas. !`Recomiendo altamente una estrategia de trading n\'umero 152 llamada comprar este libro!''\\
\indent{--{\bf Peter Carr}, Profesor y Presidente del Departamento de Finanzas e Ingenier\'ia de Riesgos, NYU's Tandon School
of Engineering; e Ingeniero Financiero del A\~{n}o 2010, International Association for Quantitative
Finance \& Sungard}\\
\\
\noindent
``Este libro es una visita guiada enciclop\'edica de estrategias de inversiones ``cuantitativas'', desde las m\'as simples (como las de seguimiento de tendencias) a otras mucho m\'as ex\'oticas que utilizan sofisticados contratos derivados. No se hace ninguna afirmaci\'on sobre la rentabilidad de estas estrategias: uno sabe muy bien cu\'an importantes son los detalles de la implementaci\'on y los costos de transacci\'on. Sin embargo, ning\'un trader cuantitativo puede permitirse ignorar lo que existe, como fuente de inspiraci\'on o como punto de referencia para nuevas ideas.''\\
\indent --{\bf Jean-Philippe Bouchaud}, Presidente y Jefe Cient\'ifico, Capital Fund Management; Profesor, \'{E}cole
Normale Sup\'{e}rieure; Miembro, Academia de Ciencias de Francia; y Co-Director, CFM-Imperial Institute of Quantitative Finance\\
\\
\noindent
``Zura Kakushadze y Juan Andr\'{e}s Serur han creado una enciclopedia magistral de estrategias de trading cuantitativo. Los autores nos ofrecen un tratamiento riguroso pero accesible de los fundamentos matem\'aticos de estas estrategias. La cobertura es completa, comenzando con estrategias simples y bien conocidas, como las call cubiertas y luego avanzando naturalmente a estrategias que involucran criptomonedas. El material de apoyo, tal como un glosario detallado y una extensa lista de referencias har\'an de este libro una referencia esencial para los economistas financieros y profesionales de las inversiones.''\\
\indent --{\bf Hossein Kazemi}, Profesor Dotado de Michael \& Cheryl Philipp en Finanzas, Universidad de Massachusetts en Amherst; y Jefe Editor, {\em The Journal of Alternative Investments}\\
\\
\noindent
``El trading exitoso de instrumentos financieros es tanto una ciencia como un arte, de la misma forma que los esfuerzos de un chef reflejan tanto el arte gastron\'omico como los procesos qu\'imicos y t\'ermicos subyacentes de cocinar. En {\em 151 Estrategias de Trading} se proporciona a los traders financieros un compendio de recetas v\'alidas, que abarca la amplia gama de m\'etodos que se pueden aplicar en la pr\'actica de la inversi\'on moderna. La exposici\'on de ambas, las matem\'aticas y la intuici\'on de cada estrategia descrita, es clara y concisa. Los lectores apreciar\'an la inclusi\'on de un extenso c\'odigo de computadora para reducir el esfuerzo necesario para implementar cualquier c\'alculo requerido.''\\
\indent --{\bf Dan diBartolomeo}, Presidente, Northfield Information Services; y Editor, {\em Journal of Asset Management}\\
\\
\noindent
``Un verdadero tour de force--{\em 151 Estrategias de Trading} proporciona la revelaci\'on m\'as exhaustiva de las estrategias populares de los fondos de cobertura. Al revelar toda la salsa secreta de los fondos de cobertura, Kakushadze y Serur ahora han hecho todo como estrategias beta. !`Es momento de bajar las tarifas!''\\
\indent --{\bf Jim Kyung-Soo Liew}, Profesor Asistente de Finanzas, Carey Business School, Universidad de Johns Hopkins;
Miembro del Consejo Asesor, {\em The Journal of Portfolio Management}; y Co-Fundador, SoKat\\
\\
\noindent
``Este libro es una impresionante concentraci\'on de estrategias y f\'ormulas para expandir el conocimiento en finanzas cuantitativas; es una lectura obligada para cualquier persona que quiera mejorar dr\'asticamente su experiencia en la din\'amica de los mercados financieros.''\\
\indent --{\bf Daniele Bernardi}, CEO, DIAMAN Capital; y Presidente de la Junta, {\em INVESTORS' Magazine Italia}

\newpage
\phantomsection
\addcontentsline{toc}{section}{Biograf\'ias de los Autores}
\section*{Biograf\'ias de los Autores}

{\bf{Zura Kakushadze}} recibi\'o su Ph.D. en f\'isica te\'orica en la Universidad de Cornell, Estados Unidos, a los 23 a\~{n}os, fue becario postdoctoral en la Universidad de Harvard, Estados Unidos, y Profesor Asistente en el Instituto C.N. Yang de F\'isica Te\'orica en la Universidad de Stony Brook, Estados Unidos. Recibi\'o una beca de la Fundaci\'on Alfred P. Sloan en 2001. Despu\'es de expandirse a las finanzas cuantitativas, fue Director de RBC Capital Markets, Director General de WorldQuant, Vicepresidente Ejecutivo y accionista sustancial de Revere Data (ahora parte de FactSet), y Profesor Adjunto de la Universidad de Connecticut, Estados Unidos. Actualmente, es el Presidente y CEO de Quantigic$^\circledR$ Solutions y Profesor Titular en la Free University de Tbilisi, Georgia. Cuenta con m\'as de 17 a\~{n}os de experiencia pr\'actica en el trading cuantitativo y en las finanzas cuantitativas, m\'as de 130 publicaciones de f\'isica, finanzas, investigaci\'on del c\'ancer y otros campos, m\'as de 3,400 citas y el \'indice h mayor a 30, m\'as de 160,000 descargas en la SSRN y m\'as de un cuarto de mill\'on de seguidores en LinkedIn.\\

\noindent {\bf{Juan Andr\'es Serur}} tiene una Maestr\'ia en Finanzas de la Universidad del CEMA, Argentina. Con m\'as de 7 a\~{n}os de experiencia en el trading en el mercado de acciones, trabaja como analista cuantitativo y estratega en una firma argentina de gesti\'on de activos y como consultor financiero para grandes corporaciones. Adem\'as, se desempe\~{n}a como Secretario Acad\'emico del Programa de la Maestr\'ia en Finanzas de la Universidad del CEMA, en donde imparte cursos de finanzas computacionales de grado y posgrado como Profesor Asistente. En el a\~{n}o 2016 gan\'o el primer puesto en una Competencia de Simulaci\'on en el Mercado de Capitales Argentino para la categor\'ia de Universidades e Instituciones Profesionales.

\newpage

\newpage

\section{Introducci\'on y resumen}

{}Una estrategia de trading\index{estrategia de trading} puede ser definida como un conjunto de instrucciones cuyo objetivo es lograr ciertas tenencias de activos en momentos determinados $t_1,t_2,\dots$ Estas tenencias pueden ser nulas (pero no necesariamente) en uno o m\'as momentos en el tiempo. En muchos casos, el objetivo principal de una estrategia\index{estrategia} es obtener un beneficio directo, es decir, generar retornos positivos sobre la inversi\'on\index{inversi\'on}. Sin embargo, existen estrategias de trading\index{estrategia de trading} que no siempre son rentables de manera independiente. Por ejemplo, una estrategia de cobertura\index{estrategia de cobertura} puede formar parte de un plan global, el cual puede ser o no una estrategia de trading\index{estrategia de trading}. En este sentido, una aerol\'inea cubriendo sus costos ante aumentos en el precio del combustible con futuros de commodities\index{futuros de commodities} constituye una estrategia de trading\index{estrategia de trading} que, al mismo tiempo, forma parte de la etapa de gesti\'on del riesgo de la estrategia de negocios global de la aerol\'inea, cuyo fin es generar ganancias a trav\'es de la venta de sus servicios.

{}En el caso de estrategias\index{estrategia} que se encuentran enfocadas a generar rentabilidad de manera independiente, uno podr\'ia decir que la frase ``comprar barato, vender caro'' captura su esencia. Sin embargo, este punto de vista es un tanto superfluo, ya que solo aplica a aquellas estrategias\index{estrategia} que consisten en comprar y vender un activo individual (por ejemplo, una acci\'on), pero excluye un gran n\'umero de estrategias que funcionan de otra forma. Por ejemplo, una estrategia de cobertura\index{estrategia de cobertura} utilizada en el proceso de gesti\'on del riesgo\index{gesti\'on de riesgo} puede no siempre implicar ``comprar barato, vender caro''. Esto se debe a que cubrir riesgos\index{riesgo} -- o, b\'asicamente, transferir el riesgo\index{riesgo} (o parte de \'este) a otros traders -- no es gratuito y en muchos casos el trader pagar\'a una prima\index{prima} por la cobertura\index{cobertura} de los mismos. El ``arbitraje estad\'istico''\index{arbitraje estad\'istico}, una popular estrategia\index{estrategia} entre los fondos de cobertura constituye otro claro ejemplo de esto, en donde el portafolio\index{portafolio} de inversiones puede consistir en miles de acciones y la rentabilidad no proviene de comprar barata y vender cara cada acci\'on o grupo de acciones, sino que lo esperado es que, estad\'isticamente, algunas generen p\'erdidas y otras ganancias. Esto puede volverse complejo r\'apidamente.

{}El prop\'osito de este trabajo es recolectar una gran variedad de estrategias de trading\index{estrategia de trading} en el contexto de finanzas (que es opuesto al trading de tarjetas de b\'eisbol, autos cl\'asicos, etc.) a trav\'es de todas las clases de activos\index{clases de activo} (o al menos de aquellas que son m\'as conocidas). En este trabajo utilizamos intencionalmente el t\'ermino ``clase de activo''\index{clases de activo} de forma flexible e incluimos lo que puede ser referido como ``subclases de activos''. De esta forma, una definici\'on m\'as acotada podr\'ia incluir acciones\index{acci\'on}, bonos\index{bono}, efectivo\index{efectivo}, divisas\index{divisas (FX)}, bienes ra\'ices\index{bienes ra\'ices}, commodities\index{commodity} e infraestructura\index{infraestructura}. Sin embargo, esta definici\'on ser\'ia muy estrecha para los fines de este trabajo. Es por ello que tambi\'en consideramos: derivados\index{derivado} tales como opciones\index{opci\'on} y futuros\index{futuro}; fondos de inversi\'on cotizados (ETFs, por sus siglas en ingl\'es);\index{fondo de inversi\'on cotizado (ETF)} \'indices\index{indice @ \'indice} (usualmente comerciados a trav\'es de ETFs\index{fondo de inversi\'on cotizado (ETF)} y futuros\index{futuro}); volatilidad, la cual puede ser tratada como una clase de activo\index{clases de activo} (y comerciada a trav\'es de, por ejemplo, notas de intercambio cotizadas\index{nota de intercambio cotizada (ETN)}); activos estructurados\index{activos estructurados} (tales como obligaciones de deuda colateralizadas\index{obligaci\'on de deuda garantizada (CDO)} y activos respaldados por hipotecas\index{valores respaldados por hipotecas (MBS)}); bonos convertibles\index{bono convertible} (representando un h\'ibrido entre bonos\index{bono} y acciones\index{acci\'on}); activos en distress\index{activo en distress} (los cuales no son una clase de activos\index{clases de activo} diferente, pero las estrategias de trading\index{estrategia de trading} merecen un tratamiento distinto); criptomonedas\index{criptomonedas}; activos miscel\'aneos tales como clima\index{clima} y energ\'ia\index{energ\'ia} (operados con derivados\index{derivado}); y tambi\'en estrategias de trading\index{estrategia de trading} tales como arbitraje impositivo\index{arbitraje impositivo} y macro global\index{macro global} (las cuales utilizan algunos de los activos antes mencionados). Algunas de las estrategias son simples y pueden ser descriptas en unas pocas palabras, mientras que otras (de hecho, la mayor\'ia) requieren una descripci\'on matem\'atica mucho m\'as detallada, la cual es proporcionada de manera expl\'icita.

{}Es importante tener en cuenta que, a diferencia de las leyes de la naturaleza (f\'isica), que est\'an (aparentemente) talladas sobre una piedra y no cambian con el tiempo, los mercados financieros\index{mercados financieros} son un resultado del hombre y cambian casi de forma continua, siendo esto en ciertas ocasiones bastante dram\'atico. Una de las consecuencias de esta transitoriedad es que muchas estrategias de trading\index{estrategia de trading} que pueden haber funcionado bien durante un tiempo, pueden dejar de hacerlo, a veces de forma muy abrupta. Un claro ejemplo de esto es la evoluci\'on que experiment\'o la Bolsa de Nueva York (NYSE, por sus siglas en ingl\'es)\index{Bolsa de Nueva York (NYSE)} cuando comenz\'o a migrar de su sistema de los especialistas\index{sistema de los especialistas} operado por humanos a un sistema electr\'onico\index{sistema electr\'onico}, a fines de 2006.\footnote{\, El NYSE\index{Bolsa de Nueva York (NYSE)} comenz\'o con su ``Mercado H\'ibrido''\index{Mercado H\'ibrido} (v\'ease, por ejemplo, \cite{HendershottMoulton2011}). Sin embargo, las condiciones para la desaparici\'on definitiva del sistema de los especialistas\index{sistema de los especialistas} parecen haber estado dadas durante un per\'iodo de tiempo considerable. Para ver una l\'inea de tiempo, consulte, por ejemplo, \cite{Pisani2010}.} Una de las consecuencias inmediatas fue que muchas estrategias de arbitraje estad\'istico\index{estrategias de arbitraje estad\'istico} que fueron rentables durante a\~{n}os, dejaron de serlo de la noche a la ma\~{n}ana a causa de un aumento en la volatilidad. Eventualmente, el mercado\index{mercado} se inund\'o con estrategias de trading de alta frecuencia\index{trading de alta frecuencia (HFT)} (HFT, por sus siglas en ingl\'es)\footnote{\, V\'ease, por ejemplo, \cite{Aldridge2013}, \cite{Lewis2014}.} disminuyendo a\'un m\'as los m\'argenes de rentabilidad de muchas estrategias\index{estrategia} antes consideradas ``muy buenas''.

{}Sin embargo, los avances tecnol\'ogicos dieron lugar a la aparici\'on de nuevos tipos de estrategias\index{estrategia}, incluyendo estrategias\index{estrategia} basadas en la miner\'ia de datos\index{miner\'ia de datos} y aprendizaje autom\'atico\index{aprendizaje autom\'atico}, las cuales apuntan a identificar -- generalmente bastante ef\'imeras -- se\~{n}ales\index{senzal, efimera @ se\~{n}al, ef\'imera} o tendencias\index{tendencia, ef\'imera} analizando grandes vol\'umenes de distintos tipos de datos.  Muchas de estas se\~{n}ales de trading son tan d\'ebiles\index{senzal de trading, debil @ se\~{n}al de trading, d\'ebil} que no pueden operarse por s\'i mismas, por lo que se combinan miles, de hecho, decenas o incluso cientos de miles, sino millones, de tales se\~{n}ales\index{senzal @ se\~{n}al} con ponderaciones no triviales para amplificar y mejorar la se\~{n}al\index{senzal @ se\~{n}al} general de forma tal que se puedan operar por s\'i solas y generen rentabilidad incluso despu\'es de los costos de transacci\'on\index{costos de transacci\'on} y el slippage\index{slippage}, incluyendo aquel causado por el HFT\index{trading de alta frecuencia (HFT)}.\footnote{\, V\'ease, por ejemplo, \cite{KakushadzeTulchinsky2016}, \cite{KakushadzeYu2017b}.}

{}Considerando la naturaleza intr\'insecamente ef\'imera de los mercados financieros\index{mercados financieros} y las estrategias de trading\index{estrategia de trading} dise\~{n}adas para obtener beneficios de \'estos, el prop\'osito de estas notas {\em no} es ense\~{n}ar al lector c\'omo hacer dinero utilizando estrategias de trading\index{estrategia de trading}, sino que simplemente proporcionar informaci\'on acerca de las estrategias\index{estrategia} que las personas han considerado en un amplio abanico de clases de activos\index{clases de activo} y estilos de trading. Teniendo en cuenta lo anterior, hacemos el siguiente DESCARGO: {\em Cualquier informaci\'on u opiniones aqu\'i provistas tienen una finalidad meramente informativa y no est\'an destinadas, ni deben ser interpretadas, como un consejo de inversi\'on, o un asesoramiento legal, tributario o de otro tipo, o una oferta, solicitud, recomendaci\'on o endoso de cualquier estrategia de trading, activo, producto o servicio.} Para m\'as informaci\'on acerca de descargos de responsabilidad legal, consulte el Ap\'endice \ref{app.B}.

{}Esperamos que estas notas sean \'utiles para acad\'emicos, profesionales, estudiantes y aspirantes a investigadores/traders en los a\~{n}os por venir. Intencionalmente estas notas -- para no duplicar la literatura anterior y para evitar que este manuscrito abarque miles de p\'aginas -- no contienen ninguna simulaci\'on num\'erica, backtests\index{backtest}, estudios emp\'iricos, etc. Sin embargo, proporcionamos una cornucopia ecl\'ectica de referencias, incluidas algunas con an\'alisis emp\'iricos muy detallados. Nuestro prop\'osito aqu\'i es describir, en muchos casos de forma muy detallada, diversas estrategias de trading\index{estrategia de trading}. Adicionalmente, el Ap\'endice \ref{app.A} proporciona un c\'odigo fuente\index{codigo fuente @ c\'odigo fuente} para ilustrar un backtesting fuera de la muestra\index{backtesting fuera de la muestra} (consulte el Ap\'endice \ref{app.B} para obtener m\'as informaci\'on sobre el descargo legal).\footnote{\, El c\'odigo en el Ap\'endice \ref{app.A} no est\'a escrito para ser ``elegante'' u \'optimo en t\'erminos de velocidad.} !`Esperamos que lo disfrute!

\newpage

\section{Opciones\index{opci\'on}}\label{sec.options}

\subsection{Generalidades}

{}Una opci\'on\index{opci\'on} es una clase de derivado financiero\index{derivado financiero}. Se trata de un contrato que el lanzador de opciones\index{lanzador de opciones} vende al comprador. Generalmente, una opci\'on\index{opci\'on} otorga el derecho, pero no la obligaci\'on, de comprar o vender un activo financiero, conocido como activo subyacente\index{activo subyacente} (por ejemplo, acciones ordinarias\index{acci\'on ordinaria}) a un precio predeterminado (conocido como precio de ejercicio\index{precio de ejercicio}) durante un per\'iodo de tiempo o una fecha espec\'ifica (conocida como fecha de ejercicio\index{fecha de ejercicio}). El comprador de una opci\'on debe pagar una prima\index{prima} al vendedor. Para m\'as informaci\'on acerca de la valuaci\'on de opciones\index{valuaci\'on de opciones}, v\'ease, por ejemplo, \cite{Harrison1981}, \cite{Baxter1996}, \cite{Hull2012}, \cite{Kakushadze2015a}.

{}Una opci\'on de compra (call) europea\index{opci\'on de compra europea} es el derecho (pero no la obligaci\'on) de comprar una acci\'on en la fecha de vencimiento\index{fecha de vencimiento} $T$ a un precio de ejercicio\index{precio de ejercicio} $K$ acordado en el momento $t=0$. El valor\index{valor} de un call\index{opci\'on call} al vencimiento es $f^{call}(S_T,k)=(S_T-k)^+$. Aqu\'i $(x)^+=x$ si $x>0$ y $(x)^+=0$ si $x\leq 0$. Si el precio de la acci\'on al vencimiento\index{vencimiento} es $S_T > k$, el comprador de la opci\'on\index{comprador de la opci\'on} gana $S_T - k$ (excluyendo el costo de la opci\'on\index{opci\'on} en $t=0$). Si el precio al vencimiento\index{vencimiento} es $S_T \leq k$, las ganancias son nulas dado que no tendr\'ia sentido ejercer la opci\'on\index{opci\'on} si $S_T < k$ (ya que ser\'ia m\'as barato adquirirla directamente en el mercado\index{mercado}) y por \'ultimo si $S_T = k$, el comprador es indiferente entre ejercer y no ejercer -- todo esto es v\'alido si no consideramos costos transaccionales\index{costos transaccionales}. De forma similar, una opci\'on de venta (put) europea\index{opci\'on de venta europea} es el derecho (pero no la obligaci\'on) de vender una acci\'on en la fecha de vencimiento\index{fecha de vencimiento} $T$. El valor\index{valor} de un put\index{opci\'on put} al vencimiento\index{vencimiento} es $f^{put}(S_T,k)=(k-S_T)^+$.

{}Una opci\'on\index{opci\'on} puede ser emitida sobre una gran variedad de activos subyacentes\index{activo subyacente}, por ejemplo, acciones\index{acci\'on}, bonos\index{bono}, futuros\index{futuro}, \'indices\index{indice @ \'indice}, commodities\index{commodity}, divisas\index{divisas (FX)}, etc. Por conveniencia terminol\'ogica, en las siguientes hojas nos referiremos con frecuencia al activo subyacente\index{activo subyacente} como ``acciones'', aunque en muchos casos la discusi\'on puede generalizarse f\'acilmente a otros activos. Adem\'as, existe una gran variedad de estilos de opciones\index{estilo de opciones} (m\'as all\'a de las opciones europeas\index{opci\'on europea} -- para m\'as informaci\'on sobre opciones europeas\index{opci\'on europea}, v\'ease, por ejemplo, \cite{Black1973}), por ejemplo, opciones americanas\index{opci\'on americana} (que pueden ser ejercidas en cualquier momento antes de la fecha de ejercicio\index{fecha de ejercicio} -- v\'ease, por ejemplo, \cite{Kim1990}), opciones bermuda\index{opci\'on bermuda} (que pueden ejercerse solo en fechas espec\'ificas en o antes de la expiraci\'on\index{expiraci\'on} -- v\'ease, por ejemplo, \cite{Andersen1999}), opciones canarias\index{opci\'on canaria} (que se pueden ejercer, por ejemplo, trimestralmente, pero no antes de que haya transcurrido un per\'iodo de tiempo determinado, por ejemplo, 1 a\~{n}o; v\'ease, por ejemplo, \cite{Henrard2006}), opciones asi\'aticas\index{opci\'on asi\'atica} (cuyo pago\index{pago} est\'a determinado por el precio promedio del activo subyacente\index{promedio del activo subyacente} durante un per\'iodo de tiempo preestablecido - v\'ease, por ejemplo, \cite{Rogers1995}), opciones con barrera\index{opci\'on con barrera} (que pueden ejercerse solo si el precio del valor del activo subyacente pasa un cierto nivel o ``barrera'' -- v\'ease, por ejemplo, \cite{Haug2001}), otras opciones ex\'oticas\index{opci\'on ex\'otica} (existe una amplia categor\'ia de opciones\index{opci\'on} que, por lo general, est\'an estructuradas de forma compleja, v\'ease, por ejemplo, \cite{Fabozzi2002}), etc. Mencionemos tambi\'en opciones binarias\index{opci\'on, binaria} (tambi\'en conocidas como todo o nada\index{opci\'on, todo o nada} -- \textit{all-or-nothing} -- u opciones digitales\index{opci\'on digital}) que pagan una cantidad preestablecida, por ejemplo, \$1, si el valor del activo subyacente\index{activo subyacente} cumple una condici\'on predefinida al vencimiento\index{vencimiento}, de lo contrario, simplemente caducan sin pagar nada al titular; v\'ease, por ejemplo, \cite{Breeden1978}.

{}Algunas estrategias de trading\index{estrategia de trading} pueden ser construidas utilizando distintas combinaciones de opciones\index{opci\'on}. A grandes rasgos, se pueden dividir en dos tipos de estrategias\index{estrategia}: direccionales y no direccionales. Las estrategias direccionales\index{estrategia direccional} implican una expectativa sobre la direcci\'on de los movimientos futuros del precio de las acciones. Las estrategias no direccionales\index{estrategia no direccional} (tambi\'en conocidas como neutrales) no se basan en la direcci\'on futura, es decir, el trader es indiferente con respecto a que el precio de la acci\'on suba o baje.

{}Al mismo tiempo, las estrategias direccionales\index{estrategia direccional} se pueden dividir en dos subgrupos: (i) estrategias alcistas\index{estrategia alcista}, en las cuales el trader se beneficia si el precio de las acciones aumenta; y (ii) estrategias bajistas\index{estrategia bajista}, en donde el trader se beneficia si el precio de las acciones disminuye. Las estrategias no direccionales\index{estrategia no direccional} se pueden dividir en dos subgrupos: (a) estrategias de volatilidad\index{estrategia de volatilidad}, que se benefician si el precio de las acciones experimenta grandes movimientos (es decir, entorno de alta volatilidad); y (b) estrategias laterales\index{estrategia lateral}, que se benefician si el precio de las acciones se mantiene estable (es decir, entorno de baja volatilidad). Adem\'as, se pueden distinguir estrategias destinadas a la generaci\'on de ingresos, estrategias cuyo objetivo es la generaci\'on de ganancias de capital, estrategias de cobertura\index{estrategia de cobertura}\index{estrategia, generaci\'on de ingresos}\index{estrategia, ganancia de capital}, etc. (v\'ease, por ejemplo, \cite{Cohen2005}).

{}En el resto de esta secci\'on, a menos que se indique lo contrario, todas las opciones\index{opci\'on} son para la misma acci\'on y tienen el mismo tiempo al vencimiento (TTM, por sus siglas en ingl\'es)\index{tiempo al vencimiento (TTM)}. Las abreviaturas del grado del dinero\index{abreviaturas del grado del dinero} (``moneyness'') son: ATM = at-the-money (en el dinero), ITM = in-the-money (dentro del dinero), OTM = out-of-the-money (fuera del dinero). Tambi\'en: $f_T$ es el pago\index{pago} al vencimiento\index{vencimiento} $T$; $S_0$ es el precio de la acci\'on en el momento $t=0$ de establecer\index{establecer} la operaci\'on (es decir, iniciar la posici\'on); $S_T$ es el precio de las acciones al vencimiento\index{vencimiento}; $C$ es el cr\'edito neto recibido en $t=0$ y $D$ es el d\'ebito neto requerido en $t=0$, seg\'un corresponda; $H=D$ (para una operaci\'on de d\'ebito neto) o $H=-C$ (para una operaci\'on de cr\'edito neto);\footnote{\, $H$ es el d\'ebito neto de todas las primas de las opciones\index{prima de la opci\'on} compradas menos el cr\'edito neto de todas las primas de las opciones\index{prima de la opci\'on} vendidas.} $S_{*sup}$ y $S_{*inf}$ son los precios de equilibrio\index{precios de equilibrio} superiores e inferiores (es decir, un precio tal que $f_T=0$), respectivamente, al vencimiento\index{vencimiento}; si solo hay un precio de equilibrio\index{precios de equilibrio}, se denota con $S_*$; $P_{max}$ es la ganancia m\'axima al vencimiento\index{vencimiento}; $L_{max}$ es la p\'erdida m\'axima al vencimiento\index{vencimiento}.

\subsection{Estrategia: Call cubierta\index{call cubierta} (Covered call)}

{}Esta estrategia consiste en comprar una acci\'on\index{acci\'on} y lanzar una opci\'on call\index{opci\'on call} con un precio de ejercicio\index{precio de ejercicio} $K$ contra la posici\'on larga en la acci\'on. La perspectiva del trader sobre el precio de las acciones es neutral a alcista\index{perspectiva del trader, neutral a alcista}. Esta estrategia (de cobertura) tiene una funci\'on de pagos\index{pago} (es decir, el valor al vencimiento, o payoff en ingl\'es) igual a la de lanzar una opci\'on put\index{opci\'on put} (put corta\index{put corta}/descubierta\index{put, descubierta}).\footnote{\, Esto est\'a relacionado con la paridad put-call\index{paridad put-call} (v\'ease, por ejemplo, \cite{Stoll1969}, \cite{Hull2012}).} Mientras mantiene la posici\'on larga en el activo subyacente, el trader puede generar ingresos vendiendo peri\'odicamente opciones call\index{opci\'on call} OTM\index{opci\'on, OTM (fuera del dinero)}. Tenemos:\footnote{\, Para obtener m\'as literatura sobre estrategias de cobertura con opciones call\index{cobertura con opciones call}, v\'ease, por ejemplo, \cite{Pounds1978}, \cite{Whaley2002}, \cite{Feldman2004}, \cite{Hill2006}, \cite{Kapadia2007}, \cite{Che2011}, \cite{Mugwagwa2012}, \cite{Israelov2014}, \cite{Israelov2015a}, \cite{Hemler2015}.}
\begin{eqnarray}
 && f_T = S_T - S_0 - (S_T - K)^+ + C = K - S_0 - (K - S_T)^+ + C\\
 && S_* = S_0 - C\\
 && P_{max} = K - S_0 + C\\
 && L_{max} = S_0 - C
\end{eqnarray}

\subsection{Estrategia: Put cubierta\index{put cubierta} (Covered put)}

{}Esta estrategia consiste en vender una acci\'on y lanzar una opci\'on put\index{opci\'on put} con un precio de ejercicio\index{precio de ejercicio} $K$ contra la posici\'on corta en la acci\'on. La perspectiva del trader es neutral a bajista\index{perspectiva del trader, neutral a bajista}. Esta estrategia tiene la misma funci\'on de pagos\index{pago} (es decir, el valor al vencimiento, o payoff en ingl\'es) que lanzar una opci\'on call\index{opci\'on call} (call corta\index{call corta}/descubierta\index{call, descubierta}). Mientras mantiene la posici\'on corta en el activo subyacente\index{activo subyacente}, el trader puede generar ingresos vendiendo peri\'odicamente opciones put\index{opci\'on put} OTM\index{opci\'on, OTM (fuera del dinero)}. Tenemos:\footnote{\, La estrategia de put cubierta\index{estrategia de put cubierta} es sim\'etrica a la estrategia de call cubierta\index{estrategia de call cubierta}. La literatura acad\'emica sobre esta estrategia es escasa. V\'ease, por ejemplo, \cite{Che2016}.}
\begin{eqnarray}
 && f_T = S_0 - S_T - (K - S_T)^+ + C = S_0 - K - (S_T - K)^+ + C\\
 && S_* = S_0 + C\\
 && P_{max} = S_0 - K + C\\
 && L_{max} = \mbox{ilimitado}
\end{eqnarray}

\subsection{Estrategia: Put protectora\index{put protectora} (Protective put)}

{}Esta estrategia consiste en la compra de una acci\'on y de una opci\'on put\index{opci\'on put} ATM\index{opci\'on, ATM (en el dinero)} u OTM\index{opci\'on, OTM (fuera del dinero)} con un precio de ejercicio\index{precio de ejercicio} $K\leq S_0$. La perspectiva del trader es alcista\index{perspectiva del trader, alcista}. Esta estrategia es de cobertura\index{estrategia, cobertura}: la opci\'on de venta\index{opci\'on de venta} cubre el riesgo\index{riesgo} de una ca\'ida en el precio de la acci\'on. Tenemos:\footnote{\, Para algunos estudios sobre estrategias de protecci\'on con opciones put\index{estrategia de protecci\'on con opciones put}, v\'ease, por ejemplo, \cite{Figlewski1993}, \cite{Israelov2015b}, \cite{Israelov2017.1}, \cite{Israelov2017}.}
\begin{eqnarray}
 && f_T =  S_T  - S_0 + (K - S_T)^+ - D = K - S_0 + (S_T - K)^+ - D\\
 && S_* = S_0 + D\\
 && P_{max} =\mbox{ilimitado}\\
 && L_{max} = S_0 - K + D
\end{eqnarray}

\subsection{Estrategia: Call protectora\index{call protectora} (Protective call)}

{}Esta estrategia consiste en vender una acci\'on\index{vender una acci\'on} y comprar una opci\'on call\index{opci\'on call} ATM\index{opci\'on, ATM (en el dinero)} u OTM\index{opci\'on, OTM (fuera del dinero)} con un precio de ejercicio\index{precio de ejercicio} $K\geq S_0$. La perspectiva del trader es bajista\index{perspectiva del trader, bajista}. Esta estrategia es de cobertura\index{estrategia, cobertura}: la opci\'on de compra\index{opci\'on de compra} cubre el riesgo\index{riesgo} de una suba en el precio de la acci\'on. Tenemos:\footnote{\, Esta estrategia es sim\'etrica a la put cubierta\index{put cubierta}. La literatura acad\'emica parece ser escasa. V\'ease, por ejemplo, \cite{Jabbour2010}, \cite{Tokic2013}.}
\begin{eqnarray}
 && f_T =  S_0  - S_T + (S_T - K)^+ - D = S_0 - K + (K - S_T)^+ - D\\
 && S_* = S_0 - D\\
 && P_{max} = S_0 - D\\
 && L_{max} = K - S_0 + D
\end{eqnarray}

\subsection{Estrategia: Diferencial alcista con call\index{diferencial alcista con call} (Bull call spread)}

{}Este diferencial vertical\index{diferencial vertical} consiste en una posici\'on larga en una opci\'on call\index{opci\'on call} cercana a ATM\index{opci\'on, ATM (en el dinero)} con un precio de ejercicio\index{precio de ejercicio} $K_1$ y una posici\'on corta en otra opci\'on call\index{opci\'on call} OTM\index{opci\'on, OTM (fuera del dinero)} con un precio de ejercicio\index{precio de ejercicio} m\'as alto $K_2$. Para crear este trade se requiere un d\'ebito neto. La perspectiva del trader es alcista\index{perspectiva del trader, alcista}: la estrategia se beneficia si el precio de la acci\'on sube. Esta estrategia es de ganancia de capital\index{estrategia, ganancia de capital}. Tenemos:\footnote{\, Para m\'as informaci\'on sobre diferenciales verticales\index{diferencial vertical} alcistas/bajistas con calls/puts, v\'ease, por ejemplo, \cite{Cartea2012}, \cite{Chaput2003}, \cite{Chaput2005}, \cite{Chen1999}, \cite{Cong2013}, \cite{Cong2014}, \cite{Matsypura2010}, \cite{Shah2017}, \cite{Wong2011}, \cite{Zhang2015}. V\'ease tambi\'en \cite{Clarke2013}, \cite{Cohen2005}, \cite{Jabbour2010}, \cite{McMillan2002}, \cite{OI1995}.}
\begin{eqnarray}
 && f_T = (S_T - K_1)^+ - (S_T - K_2)^+ - D\\
 && S_* = K_1 + D\\
 && P_{max} = K_2 - K_1 -D\\
 && L_{max} = D
\end{eqnarray}

\subsection{Estrategia: Diferencial alcista con put\index{diferencial alcista con put} (Bull put spread)}

{}Este diferencial vertical\index{diferencial vertical} consiste en una posici\'on larga en una opci\'on put\index{opci\'on put} OTM\index{opci\'on, OTM (fuera del dinero)} con un precio de ejercicio\index{precio de ejercicio} $K_1$ y una posici\'on corta en otra opci\'on put\index{opci\'on put} OTM\index{opci\'on, OTM (fuera del dinero)} con un precio de ejercicio\index{precio de ejercicio} mayor $K_2$. Este trade genera un cr\'edito neto. La perspectiva del trader es alcista\index{perspectiva del trader, alcista}. Esta estrategia es de generaci\'on de ingresos\index{estrategia, generaci\'on de ingresos}. Tenemos:
\begin{eqnarray}
 && f_T = (K_1 - S_T)^+ - (K_2 - S_T)^+ + C\\
 && S_* = K_2 - C\\
 && P_{max} = C\\
 && L_{max} = K_2 - K_1 - C\
\end{eqnarray}

\subsection{Estrategia: Diferencial bajista con call\index{diferencial bajista con call} (Bear call spread)}

{}Este diferencial vertical\index{diferencial vertical} consiste en una posici\'on larga en una opci\'on call\index{opci\'on call} OTM\index{opci\'on, OTM (fuera del dinero)} con un precio de ejercicio\index{precio de ejercicio} $K_1$ y una posici\'on corta en otra opci\'on call\index{opci\'on call} OTM\index{opci\'on, OTM (fuera del dinero)} con un precio de ejercicio\index{precio de ejercicio} menor $K_2$. Este trade genera un cr\'edito neto. La perspectiva del trader es bajista\index{perspectiva del trader, bajista}. Esta estrategia es de generaci\'on de ingresos\index{estrategia, generaci\'on de ingresos}. Tenemos:
\begin{eqnarray}
 && f_T = (S_T - K_1 )^+ - ( S_T- K_2 )^+ + C\\
 && S_* = K_2 + C\\
 && P_{max} = C\\
 && L_{max} = K_1 -  K_2  - C\
\end{eqnarray}

\subsection{Estrategia: Diferencial bajista con put\index{diferencial bajista con put} (Bear put spread)}

{}Este diferencial vertical\index{diferencial vertical} consiste en una posici\'on larga en una opci\'on put\index{opci\'on put} cercana a ATM\index{opci\'on, ATM (en el dinero)} con un precio de ejercicio\index{precio de ejercicio} $K_1$ y una posici\'on corta en una opci\'on put\index{opci\'on put} OTM\index{opci\'on, OTM (fuera del dinero)} con un precio de ejercicio\index{precio de ejercicio} menor $K_2$. Para crear este trade se requiere un d\'ebito neto. La perspectiva del trader es bajista\index{perspectiva del trader, bajista}: esta estrategia genera beneficios si el precio de la acci\'on baja. Esta estrategia es de ganancia de capital\index{estrategia, ganancia de capital}. Tenemos:
\begin{eqnarray}
 && f_T = (K_1 - S_T )^+ - ( K_2 - S_T )^+ - D\\
 && S_* = K_1 - D\\
 && P_{max} = K_1 - K_2 - D\\
 && L_{max} = D\
\end{eqnarray}

\subsection{Estrategia: Forward sint\'etico largo\index{forward sint\'etico}}

{}Esta estrategia consiste en la compra de una opci\'on call\index{opci\'on call} ATM\index{opci\'on, ATM (en el dinero)} y la venta de una opci\'on put\index{opci\'on put} ATM\index{opci\'on, ATM (en el dinero)} con el precio de ejercicio\index{precio de ejercicio} $K = S_0$. Este trade puede ser de d\'ebito neto o cr\'edito neto. Generalmente, $|H| \ll S_0$. La perspectiva del trader es alcista\index{perspectiva del trader, alcista}: esta estrategia imita una posici\'on larga en una acci\'on o en un futuro\index{futuro}; replica una posici\'on larga en un contrato forward\index{contrato forward} con el precio de entrega\index{precio de entrega} $K$ y la madurez\index{madurez} igual a la de las opciones\index{opci\'on}. Esta estrategia es de ganancia de capital\index{estrategia, ganancia de capital}. Tenemos:\footnote{\, Para m\'as informaci\'on sobre contratos forward sint\'eticos\index{contrato forward sint\'etico} largos/cortos, v\'ease, por ejemplo, \cite{Benavides2009}, \cite{Bozic2012}, \cite{DeMaskey1995}, \cite{Ebrahim2005}, \cite{Nandy2016}.}
\begin{eqnarray}
 && f_T = (S_T  -  K)^+ - ( K -  S_T )^+ - H = S_T - K - H\\
 && S_* = K + H \\
 && P_{max} =  \mbox{ilimitado}\\
 && L_{max} = K + H
\end{eqnarray}

\subsection{Estrategia: Forward sint\'etico corto\index{forward sint\'etico}}

{}Esta estrategia consiste en la compra de una opci\'on put\index{opci\'on put} ATM\index{opci\'on, ATM (en el dinero)} y la venta de una opci\'on call\index{opci\'on call} \index{opci\'on, ATM (en el dinero)} con el precio de ejercicio\index{precio de ejercicio} $K = S_0$. Este trade puede ser de d\'ebito neto o cr\'edito neto. Generalmente, $|H| \ll S_0$. La perspectiva del trader es bajista\index{perspectiva del trader, bajista}: esta estrategia imita una posici\'on corta en una acci\'on o en un futuro\index{futuro}; replica una posici\'on corta en un contrato forward\index{contrato forward} con el precio de entrega\index{precio de entrega} $K$ y la madurez\index{madurez} igual a la de las opciones\index{opci\'on}. Esta estrategia es de ganancia de capital\index{estrategia, ganancia de capital}. Tenemos:
\begin{eqnarray}
 && f_T = ( K -  S_T )^+ - (S_T  -  K)^+  - H = K - S_T - H\\
 && S_* = K - H \\
 && P_{max} =  K - H \\
 && L_{max} = \mbox{ilimitado}
\end{eqnarray}

\subsection{Estrategia: Combo\index{combo} largo (Long risk reversal)}\label{sub.long.risk.rev}

{}Esta estrategia consiste en la compra de una opci\'on call\index{opci\'on call} OTM\index{opci\'on, OTM (fuera del dinero)} con un precio de ejercicio\index{precio de ejercicio} $K_1$ y en la venta de una opci\'on put\index{opci\'on put} OTM\index{opci\'on, OTM (fuera del dinero)} con un precio de ejercicio\index{precio de ejercicio} $K_2$. La perspectiva del trader es alcista\index{perspectiva del trader, alcista}. Esta estrategia es de ganancia de capital\index{estrategia, ganancia de capital}.\footnote{\, Para m\'as informaci\'on sobre estrategias de combos\index{estrategia de combo} largos/cortos, v\'ease, por ejemplo, \cite{Rusnakova2015}, \cite{Soltes2011}, \cite{Soltes2012}. V\'ease tambi\'en, por ejemplo, \cite{Chaput2003}.} Tenemos ($K_1 > K_2$):
\begin{eqnarray}
 && f_T = (S_T - K_1)^+ - (K_2 - S_T)^+ - H\\
 && S_* = K_1 + H,~~~H > 0\\
 && S_* = K_2 + H,~~~H < 0\\
 && K_2 \leq S_* \leq K_1,~~~H = 0\\
 && P_{max} = \mbox{ilimitado}\\
 && L_{max} = K_2 + H
\end{eqnarray}

\subsection{Estrategia: Combo\index{combo} corto (Short risk reversal)}

{}Esta estrategia consiste en la compra de una opci\'on put\index{opci\'on put} OTM\index{opci\'on, OTM (fuera del dinero)} con un precio de ejercicio\index{precio de ejercicio} $K_1$ y en la venta de una opci\'on call\index{opci\'on call} OTM\index{opci\'on, OTM (fuera del dinero)} con un precio de ejercicio\index{precio de ejercicio} $K_2$. La perspectiva del trader es bajista\index{perspectiva del trader, bajista}. Esta estrategia es de ganancia de capital\index{estrategia, ganancia de capital}. Tenemos ($K_2 > K_1$):
\begin{eqnarray}
 && f_T = (K_1 - S_T )^+ - ( S_T -  K_2 )^+ - H\\
 && S_* = K_1 - H,~~~H > 0\\
 && S_* = K_2 - H,~~~H < 0\\
 && K_1 \leq S_* \leq K_2,~~~H = 0\\
 && P_{max} = K_1 - H \\
 && L_{max} = \mbox{ilimitado}\
\end{eqnarray}

\subsection{Estrategia: Escalera\index{escalera} alcista con call (Bull call ladder)}

{}Este diferencial vertical\index{diferencial vertical} consiste en una posici\'on larga en una opci\'on call\index{opci\'on call} cercana (usualmente) a ATM\index{opci\'on, ATM (en el dinero)} con un precio de ejercicio\index{precio de ejercicio} $K_1$, una posici\'on corta en una opci\'on call\index{opci\'on call} OTM\index{opci\'on, OTM (fuera del dinero)} con un precio de ejercicio\index{precio de ejercicio} $K_2$ y una posici\'on corta en otra opci\'on call\index{opci\'on call} OTM\index{opci\'on, OTM (fuera del dinero)} con un precio de ejercicio\index{precio de ejercicio} mayor $K_3$. Esta estrategia es equivalente a un diferencial alcista construido con opciones call\index{opci\'on call} financiado con la venta de otra opci\'on call OTM\index{opci\'on, OTM (fuera del dinero)} (con el precio de ejercicio\index{precio de ejercicio} $K_3$).\footnote{\, En este sentido, esta es una estrategia de ``ingresos''\index{estrategia, generaci\'on de ingresos}.} Esta posici\'on ajusta las perspectivas del trader de alcista\index{perspectiva del trader, alcista} (diferencial alcista\index{diferencial alcista}) a conservadoramente alcista\index{perspectiva del trader, conservadoramente alcista} o incluso no direccional\index{perspectiva del trader, no direccional} (con una expectativa de baja volatilidad). Tenemos:
\begin{eqnarray}
 && f_T = (S_T - K_1)^+ -  (S_T - K_2 )^+ - (S_T - K_3)^+ - H\\
 && S_{*inf} = K_1 + H,~~~H > 0 \\
 && S_{*sup} = K_3 + K_2 - K_1 - H\\
 && P_{max} = K_2 - K_1 - H\\
 && L_{max} = \mbox{ilimitado}
\end{eqnarray}

\subsection{Estrategia: Escalera\index{escalera} alcista con put (Bull put ladder)}

{}Este diferencial vertical\index{diferencial vertical} consiste en una posici\'on corta en una opci\'on put\index{opci\'on put} cercana (usualmente) a ATM\index{opci\'on, ATM (en el dinero)} con un precio de ejercicio\index{precio de ejercicio} $K_1$, una posici\'on larga en una opci\'on put\index{opci\'on put} OTM\index{opci\'on, OTM (fuera del dinero)} con un precio de ejercicio\index{precio de ejercicio} $K_2$ y una posici\'on larga en otra opci\'on put\index{opci\'on put} OTM\index{opci\'on, OTM (fuera del dinero)} con un precio de ejercicio\index{precio de ejercicio} menor $K_3$. Esta estrategia generalmente es ejecutada cuando un diferencial alcista\index{diferencial alcista} construido con opciones put\index{opci\'on put} sale mal (las acciones bajan), entonces el trader compra otra opci\'on put\index{opci\'on put} OTM\index{opci\'on, OTM (fuera del dinero)} (con el precio de ejercicio\index{precio de ejercicio} $K_3$) para ajustar la posici\'on a bajista\index{posici\'on, bajista}. Tenemos:\footnote{\, Para m\'as informaci\'on sobre estas estrategias, v\'ease, por ejemplo, \cite{Amaitiek2010}, \cite{Harcarikova2016}, \cite{He2016}, \cite{Soltes2010a}.}
\begin{eqnarray}
 && f_T = (K_3  -  S_T )^+ +  (K_2  -  S_T )^+ - (K_1  -  S_T )^+ - H\\
 && S_{*sup} = K_1 + H,~~~H < 0 \\
 && S_{*inf} = K_3 + K_2 - K_1 - H\\
 && P_{max} = K_3 + K_2 - K_1 - H\\
 && L_{max} = K_1 - K_2 + H \
\end{eqnarray}

\subsection{Estrategia: Escalera\index{escalera} bajista con call (Bear call ladder)}

{}Este diferencial vertical\index{diferencial vertical} consiste en una posici\'on corta en una opci\'on call\index{opci\'on call} cercana (usualmente) a ATM\index{opci\'on, ATM (en el dinero)} con un precio de ejercicio\index{precio de ejercicio} $K_1$, una posici\'on larga en una opci\'on call\index{opci\'on call} OTM\index{opci\'on, OTM (fuera del dinero)} con un precio de ejercicio\index{precio de ejercicio} $K_2$ y una posici\'on larga en otra opci\'on call\index{opci\'on call} OTM\index{opci\'on, OTM (fuera del dinero)} con un precio de ejercicio\index{precio de ejercicio} mayor $K_3$. Esta estrategia generalmente es ejecutada cuando un diferencial bajista\index{diferencial bajista} construido con opciones call\index{opci\'on call} sale mal (las acciones suben), entonces el trader compra otra opci\'on call\index{opci\'on call} OTM\index{opci\'on, OTM (fuera del dinero)} (con el precio de ejercicio\index{precio de ejercicio} $K_3$) para ajustar la posici\'on a alcista\index{posici\'on, alcista}. Tenemos:
\begin{eqnarray}
 && f_T = (S_T - K_3 )^+ +  (S_T - K_2)^+ - (S_T - K_1)^+ - H\\
 && S_{*inf} = K_1 - H,~~~H < 0 \\
 && S_{*sup} = K_3 + K_2 - K_1 + H\\
 && P_{max} = \mbox{ilimitado}\\
 && L_{max} = K_2 - K_1 + H \
\end{eqnarray}

\subsection{Estrategia: Escalera\index{escalera} bajista con put (Bear put ladder)}

{}Este diferencial vertical\index{diferencial vertical} consiste en una posici\'on larga en una opci\'on put\index{opci\'on put} cercana (usualmente) a ATM\index{opci\'on, ATM (en el dinero)} con un precio de ejercicio\index{precio de ejercicio} $K_1$, una posici\'on corta en una opci\'on put\index{opci\'on put} OTM\index{opci\'on, OTM (fuera del dinero)} con un precio de ejercicio\index{precio de ejercicio} $K_2$ y una posici\'on corta en otra opci\'on put\index{opci\'on put} OTM\index{opci\'on, OTM (fuera del dinero)} con un precio de ejercicio\index{precio de ejercicio} menor $K_3$. Esta estrategia es equivalente a un diferencial bajista\index{diferencial bajista} construido con opciones put\index{opci\'on put} financiado con la venta de otra opci\'on put\index{opci\'on put} OTM\index{opci\'on, OTM (fuera del dinero)} (con el precio de ejercicio\index{precio de ejercicio} $K_3$).\footnote{\, En este sentido, esta es una estrategia de ``ingresos''\index{estrategia, generaci\'on de ingresos}.} Esta posici\'on ajusta las perspectivas del trader de bajista\index{perspectiva del trader, bajista} (diferencial bajista\index{diferencial bajista}) a conservadoramente bajista\index{perspectiva del trader, conservadoramente bajista} o incluso no direccional\index{perspectiva del trader, no direccional} (con una expectativa de baja volatilidad). Tenemos (asumiendo $K_3 + K_2 - K_1 + H > \mbox{max}(H, 0)$):
\begin{eqnarray}
 && f_T = (K_1 - S_T)^+ -  (K_2 - S_T)^+ - (K_3 - S_T)^+ - H\\
 && S_{*sup} = K_1 - H,~~~H > 0 \\
 && S_{*inf} = K_3 + K_2 - K_1 + H\\
 && P_{max} = K_1 - K_2 - H\\
 && L_{max} = K_3 + K_2 - K_1 + H
\end{eqnarray}

\subsection{Estrategia: Diferencial temporal con call\index{diferencial temporal con call} (Calendar call spread)}

{}Este es un diferencial horizontal\index{diferencial horizontal} que consiste en una posici\'on larga en una opci\'on call\index{opci\'on call} cercana a ATM\index{opci\'on, ATM (en el dinero)} con un TTM\index{tiempo al vencimiento (TTM)} $T^\prime$ y una posici\'on corta en otra opci\'on call con el mismo precio de ejercicio\index{precio de ejercicio} $K$, pero un TTM\index{tiempo al vencimiento (TTM)} m\'as corto $T < T^\prime$. Para crear este trade se requiere un d\'ebito neto. La perspectiva del trader es neutral a alcista\index{perspectiva del trader, neutral a alcista}. Al vencimiento\index{vencimiento} de la opci\'on call\index{opci\'on call} corta ($t = T$), el mejor escenario es que el precio de la acci\'on sea igual al precio de ejercicio\index{precio de ejercicio} ($S_T = K$). En $t=T$, sea $V$ el valor de la opci\'on\index{opci\'on call} call larga (que vence en $t=T^\prime$) asumiendo $S_T = K$. Tenemos:\footnote{\, Para m\'as informaci\'on sobre diferenciales temporales/diagonales con call/put\index{diferencial, temporal}\index{diferencial, diagonal}, v\'ease, por ejemplo, \cite{Carmona2003}, \cite{Carr2005}, \cite{Dale2015}, \cite{Gatheral2014}, \cite{Kawaller2002}, \cite{Liu2010}, \cite{Manoliu2004}, \cite{Pirrong2017}, \cite{Till2008}.}
\begin{eqnarray}
 && P_{max} = V - D\\
 && L_{max} = D\
\end{eqnarray}
Si al vencimiento\index{vencimiento} de la opci\'on call\index{opci\'on call} corta el precio de las acciones es $S_{stop-loss} \leq S_T \leq K$, en donde $S_{stop-loss}$ es el precio de stop-loss\index{precio de stop-loss} por debajo del cual el trader desarmar\'ia toda la posici\'on, entonces el trader puede lanzar otra opci\'on call\index{opci\'on call} con el precio de ejercicio\index{precio de ejercicio} $K$ y un TTM\index{tiempo al vencimiento (TTM)} $T_1 <T^\prime$. Mientras mantiene la posici\'on larga en la opci\'on call con el TTM\index{tiempo al vencimiento (TTM)} $T^\prime$, el trader puede generar ingresos mediante la venta peri\'odica de opciones call\index{opci\'on call} con vencimientos\index{vencimiento} m\'as cortos. En este sentido, esta estrategia se asemeja a la call cubierta\index{call cubierta}.

\subsection{Estrategia: Diferencial temporal con put\index{diferencial temporal con put} (Calendar put spread)}

{}Este es un diferencial horizontal\index{diferencial horizontal} que consiste en una posici\'on larga en una opci\'on put\index{opci\'on put} cercana a ATM\index{opci\'on, ATM (en el dinero)} con un TTM\index{tiempo al vencimiento (TTM)} $T^\prime$ y una posici\'on corta en otra opci\'on put\index{opci\'on put} con el mismo precio de ejercicio\index{precio de ejercicio} $K$, pero un TTM\index{tiempo al vencimiento (TTM)} m\'as corto $T < T^\prime$. Para crear este trade se requiere un d\'ebito neto. La perspectiva del trader es neutral a bajista\index{perspectiva del trader, neutral a bajista}. Al vencimiento\index{vencimiento} de la opci\'on put\index{opci\'on put} corta ($t = T$), el mejor escenario es que el precio de la acci\'on sea igual al precio de ejercicio\index{precio de ejercicio} ($S_T = K$). En $t=T$, sea $V$ el valor de la opci\'on put\index{opci\'on put} larga (que vence en $t=T^\prime$) asumiendo $S_T = K$. Tenemos:
\begin{eqnarray}
 && P_{max} = V - D\\
 && L_{max} = D\
\end{eqnarray}
Si al vencimiento\index{vencimiento} del put\index{opci\'on put} corto el precio de las acciones $K \leq S_T \leq S_{stop-loss}$, en donde $S_{stop-loss}$ es el precio de stop-loss\index{precio de stop-loss} por encima del cual el trader desarmar\'ia toda la posici\'on, entonces el trader puede lanzar otra opci\'on put\index{opci\'on put} con el precio de ejercicio\index{precio de ejercicio} $K$ y un TTM\index{tiempo al vencimiento (TTM)} $T_1 <T^\prime$. Mientras mantiene la posici\'on larga en la opci\'on put\index{opci\'on put} con el TTM\index{tiempo al vencimiento (TTM)} $T^\prime$, el trader puede generar ingresos mediante la venta peri\'odica de opciones put\index{opci\'on put} con vencimientos\index{vencimiento} m\'as cortos. En este sentido, esta estrategia se asemeja a la put cubierta\index{put cubierta}.

\subsection{Estrategia: Diferencial diagonal con call\index{diferencial diagonal con call} (Diagonal call spread)}

{}Este diferencial diagonal\index{diferencial diagonal} consiste en una posici\'on larga en una opci\'on call\index{opci\'on call} muy ITM\index{opci\'on, ITM (dentro del dinero)} con un precio de ejercicio\index{precio de ejercicio} $K_1$ y un TTM\index{tiempo al vencimiento (TTM)} $T^\prime$ y una posici\'on corta en una opci\'on call\index{opci\'on call} OTM\index{opci\'on, OTM (fuera del dinero)} con un precio de ejercicio\index{precio de ejercicio} $K_2$ y un TTM\index{tiempo al vencimiento (TTM)} m\'as corto $T < T^\prime$. Para crear este trade se requiere un d\'ebito neto. La perspectiva del trader es alcista\index{perspectiva del trader, alcista}. En $t=T$, sea $V$ el valor de la opci\'on call\index{opci\'on call} larga (que vence en $t=T^\prime$) asumiendo $S_T = K$. Tenemos:
\begin{eqnarray}
 && P_{max} = V - D\\
 && L_{max} = D\
\end{eqnarray}
Si al vencimiento\index{vencimiento} de la opci\'on call\index{opci\'on call} corta el precio de las acciones es $S_{stop-loss} \leq S_T \leq K_2$, en donde $S_{stop-loss}$ es el precio de stop-loss\index{precio de stop-loss} por debajo del cual el trader desarmar\'ia toda la posici\'on, entonces el trader puede lanzar otra opci\'on call\index{opci\'on call} OTM\index{opci\'on, OTM (fuera del dinero)} con un TTM\index{tiempo al vencimiento (TTM)} $T_1 <T^\prime$. Mientras mantiene la posici\'on larga en la opci\'on call\index{opci\'on call} con el TTM\index{tiempo al vencimiento (TTM)} $T^\prime$, el trader puede generar ingresos mediante la venta peri\'odica de opciones call\index{opci\'on call} OTM\index{opci\'on, OTM (fuera del dinero)} con vencimientos\index{vencimiento} m\'as cortos. En este sentido, esta estrategia se asemeja a un diferencial temporal con opciones call\index{diferencial temporal con call}. La principal diferencia radica en que, en esta estrategia, la opci\'on call\index{opci\'on call} (profundamente) ITM\index{opci\'on, ITM (dentro del dinero)} (a diferencia de la opci\'on call\index{opci\'on call} cercana a ATM\index{opci\'on, ATM (en el dinero)} en el diferencial temporal\index{diferencial temporal}) se asemeja m\'as a la acci\'on\index{acci\'on}, por lo que la posici\'on est\'a m\'as protegida contra un fuerte aumento en el precio de \'esta.

\subsection{Estrategia: Diferencial diagonal con put\index{diferencial diagonal con put} (Diagonal put spread)}

{}Este diferencial diagonal\index{diferencial diagonal} consiste en una posici\'on larga en una opci\'on put\index{opci\'on put} muy ITM\index{opci\'on, ITM (dentro del dinero)} con un precio de ejercicio\index{precio de ejercicio} $K_1$ y un TTM $T^\prime$ y una posici\'on corta en una opci\'on put\index{opci\'on put} OTM\index{opci\'on, OTM (fuera del dinero)} con un precio de ejercicio\index{precio de ejercicio} $K_2$ y un TTM\index{tiempo al vencimiento (TTM)} m\'as corto $T < T^\prime$. Para crear este trade se requiere un d\'ebito neto. La perspectiva del trader es bajista\index{perspectiva del trader, bajista}. En $t=T$, sea $V$ el valor de la opci\'on put\index{opci\'on put} larga (que vence en $t=T^\prime$) asumiendo $S_T = K$. Tenemos:
\begin{eqnarray}
 && P_{max} = V - D\\
 && L_{max} = D\
\end{eqnarray}
Si al vencimiento\index{vencimiento} de la opci\'on put\index{opci\'on put} corta el precio de las acciones es $K_2 \leq S_T \leq S_{stop-loss}$, en donde $S_{stop-loss}$ es el precio de stop-loss\index{precio de stop-loss} por encima del cual el trader desarmar\'ia toda la posici\'on, entonces el trader puede lanzar otra opci\'on put\index{opci\'on put} OTM\index{opci\'on, OTM (fuera del dinero)} con un TTM\index{tiempo al vencimiento (TTM)} $T_1 <T^\prime$. Mientras mantiene la posici\'on larga en la opci\'on put\index{opci\'on put} con el TTM\index{tiempo al vencimiento (TTM)} $T^\prime$, el trader puede generar ingresos mediante la venta peri\'odica de opciones put\index{opci\'on put} OTM\index{opci\'on, OTM (fuera del dinero)} con vencimientos\index{vencimiento} m\'as cercanos. En este sentido, esta estrategia se asemeja a un diferencial temporal con put\index{diferencial temporal con put}. La principal diferencia radica en que, en esta estrategia, la opci\'on put\index{opci\'on put} (profundamente) ITM\index{opci\'on, ITM (dentro del dinero)} (a diferencia de la opci\'on put\index{opci\'on put} cercana a ATM\index{opci\'on, ATM (en el dinero)} en el diferencial temporal) se asemeja m\'as a la acci\'on\index{acci\'on}, por lo que la posici\'on est\'a m\'as protegida contra una fuerte ca\'ida en el precio de \'esta.

\subsection{Estrategia: Cono\index{cono} largo (Long straddle)}

{}Esta estrategia de volatilidad\index{estrategia de volatilidad} consiste en una posici\'on larga en una opci\'on call\index{opci\'on call} ATM\index{opci\'on, ATM (en el dinero)} y una posici\'on larga en una opci\'on put\index{opci\'on put} ATM\index{opci\'on, ATM (en el dinero)} con un precio de ejercicio\index{precio de ejercicio} $K$. Para crear este trade se requiere un d\'ebito neto. La perspectiva del trader es neutral\index{perspectiva del trader, neutral}. Esta estrategia es de ganancia de capital\index{estrategia, ganancia de capital}. Tenemos:\footnote{\, Para m\'as informaci\'on sobre conos/cunas\index{estrategia, cono}\index{estrategia, cuna}, v\'ease, por ejemplo, \cite{Copeland1983}, \cite{Coval2001}, \cite{Engle2000}, \cite{Gao2017}, \cite{Goltz2009}, \cite{Guo2000}, \cite{Hansch1998}, \cite{Noh1994}, \cite{Rusnakova2012}, \cite{Suresh2015}. La literatura acad\'emica sobre guts\index{guts} largos/cortos (que pueden ser pensados como una variaci\'on de los conos\index{cono}) parece ser m\'as escasa. Por un libro de referencia, consulte, por ejemplo, \cite{Cohen2005}. Por estrategias de conos cubiertos\index{cono cubierto}, consulte, por ejemplo, \cite{Johnson1979}.}
\begin{eqnarray}
 && f_T = ( S_T  -  K)^+ +  (K - S_T)^+ - D\\
 && S_{*sup} = K + D\\
 && S_{*inf} = K -  D\\
 && P_{max} = \mbox{ilimitado}\\
 && L_{max} = D\
\end{eqnarray}

\subsection{Estrategia: Cuna\index{cuna} larga (Long strangle)}

{}Esta estrategia de volatilidad\index{estrategia de volatilidad} consiste en una posici\'on larga en una opci\'on call\index{opci\'on call} OTM\index{opci\'on, OTM (fuera del dinero)} con un precio de ejercicio\index{precio de ejercicio} $K_1$ y una posici\'on larga en una opci\'on put\index{opci\'on put} OTM\index{opci\'on, OTM (fuera del dinero)} con un precio de ejercicio\index{precio de ejercicio} $K_2$. Para crear este trade se requiere un d\'ebito neto. Debido a que las opciones call\index{opci\'on call} y las opciones put\index{opci\'on put} son OTM\index{opci\'on, OTM (fuera del dinero)}, esta estrategia es menos costosa de establecer que un cono\index{cono} largo. La contracara es que el movimiento en el precio de las acciones requerido para alcanzar uno de los puntos de equilibrio\index{punto de equilibrio} tambi\'en es m\'as significativo. La perspectiva del trader es neutral\index{perspectiva del trader, neutral}. Esta estrategia es de ganancia de capital\index{estrategia, ganancia de capital}. Tenemos:
\begin{eqnarray}
 && f_T = (S_T - K_1)^+ + (K_2  -  S_T )^+ - D\\
 && S_{*sup} = K_1 + D\\
 && S_{*inf} = K_2 -  D\\
 && P_{max} = \mbox{ilimitado}\\
 && L_{max} = D\
\end{eqnarray}

\subsection{Estrategia: Guts\index{guts} largos}

{}Esta estrategia de volatilidad\index{estrategia de volatilidad} consiste en una posici\'on larga en una opci\'on call\index{opci\'on call} ITM\index{opci\'on, ITM (dentro del dinero)} con un precio de ejercicio\index{precio de ejercicio} $K_1$ y una posici\'on larga en una opci\'on put\index{opci\'on put} ITM\index{opci\'on, ITM (dentro del dinero)} con un precio de ejercicio\index{precio de ejercicio} $K_2$. Para crear este trade se requiere un d\'ebito neto. Debido a que las opciones call\index{opci\'on call} y las opciones put\index{opci\'on put} son ITM\index{opci\'on, ITM (dentro del dinero)}, esta estrategia es m\'as costosa de establecer que un cono\index{cono} largo. La perspectiva del trader es neutral\index{perspectiva del trader, neutral}. Esta estrategia es de ganancia de capital\index{estrategia, ganancia de capital}. Tenemos (asumiendo $D > K_2 - K_1$):\footnote{\, De otra forma esta estrategia podr\'ia generar retornos libres de riesgo\index{retorno libre de riesgo}.}
\begin{eqnarray}
 && f_T = (S_T - K_1)^+ + (K_2  -  S_T )^+ - D\\
 && S_{*sup} = K_1 + D\\
 && S_{*inf} = K_2 - D\\
 && P_{max} = \mbox{ilimitado}\\
 && L_{max} = D - (K_2 - K_1)\
\end{eqnarray}

\subsection{Estrategia: Cono\index{cono} corto (Short straddle)}

{}Esta es una estrategia lateral\index{estrategia lateral} que consiste en una posici\'on corta en una opci\'on call\index{opci\'on call} ATM\index{opci\'on, ATM (en el dinero)} y una posici\'on corta en una opci\'on put\index{opci\'on put} ATM\index{opci\'on, ATM (en el dinero)} con un precio de ejercicio\index{precio de ejercicio} $K$. Este trade genera un cr\'edito neto. La perspectiva del trader es neutral\index{perspectiva del trader, neutral}. Esta es una estrategia de generaci\'on de ingresos\index{estrategia, generaci\'on de ingresos}. Tenemos:
\begin{eqnarray}
 && f_T = -(S_T - K)^+ - (K  -  S_T )^+ + C\\
 && S_{*sup} = K + C\\
 && S_{*inf} = K - C \\
 && P_{max} = C\\
 && L_{max} =  \mbox{ilimitado}\
\end{eqnarray}

\subsection{Estrategia: Cuna\index{cuna} corta (Short strangle)}

{}Esta es una estrategia lateral\index{estrategia lateral} que consiste en una posici\'on corta en una opci\'on call\index{opci\'on call} OTM\index{opci\'on, OTM (fuera del dinero)} con un precio de ejercicio\index{precio de ejercicio} $K_1$ y una posici\'on corta en una opci\'on put\index{opci\'on put} OTM\index{opci\'on, OTM (fuera del dinero)} con un precio de ejercicio\index{precio de ejercicio} $K_2$. Este trade genera un cr\'edito neto. Debido a que ambas posiciones son OTM\index{opci\'on, OTM (fuera del dinero)}, esta estrategia es menos riesgosa que un cono\index{cono} corto. La contracara es que el cr\'edito inicial es tambi\'en m\'as bajo. La perspectiva del trader es neutral\index{perspectiva del trader, neutral}. Esta es una estrategia de generaci\'on de ingresos\index{estrategia, generaci\'on de ingresos}. Tenemos:
\begin{eqnarray}
 && f_T = -(S_T - K_1)^+ - (K_2  -  S_T )^+ + C\\
 && S_{*sup} = K_1 + C\\
 && S_{*inf} = K_2 - C \\
 && P_{max} = C\\
 && L_{max} =  \mbox{ilimitado}\
\end{eqnarray}

\subsection{Estrategia: Guts\index{guts} cortos}

{}Esta es una estrategia lateral\index{estrategia lateral} que consiste en una posici\'on corta en una opci\'on call\index{opci\'on call} ITM\index{opci\'on, ITM (dentro del dinero)} con un precio de ejercicio\index{precio de ejercicio} $K_1$ y una posici\'on corta en una opci\'on put\index{opci\'on put} ITM\index{opci\'on, ITM (dentro del dinero)} con un precio de ejercicio\index{precio de ejercicio} $K_2$. Este trade genera un cr\'edito neto. Debido a que ambas posiciones son ITM\index{opci\'on, ITM (dentro del dinero)}, el cr\'edito inicial es mayor al generado por un cono\index{cono} corto. La contracara es que el riesgo de esta estrategia es m\'as alto. La perspectiva del trader es neutral\index{perspectiva del trader, neutral}. Esta es una estrategia de generaci\'on de ingresos\index{estrategia, generaci\'on de ingresos}. Tenemos:\footnote{\, Similar a los guts\index{guts} largos, aqu\'i asumimos que $C > K_2 - K_1$.}
\begin{eqnarray}
 && f_T = -(S_T - K_1)^+ - (K_2  -  S_T )^+ + C\\
 && S_{*sup} = K_1 + C\\
 && S_{*inf} = K_2 - C\\
 && P_{max} = C - (K_2 - K_1) \\
 && L_{max} = \mbox{ilimitado}\
\end{eqnarray}

\subsection{Estrategia: Cono sint\'etico\index{cono sint\'etico} largo con call (Long call synthetic straddle)}

{}Esta estrategia de volatilidad\index{estrategia de volatilidad} (que es la misma que un cono\index{cono} largo con la opci\'on put\index{opci\'on put} reemplazada por una opci\'on put sint\'etica\index{opci\'on put sint\'etica}) consiste en vender una acci\'on\index{vender una acci\'on} y comprar {\em dos} opciones call\index{opci\'on call} ATM\index{opci\'on, ATM (en el dinero)} (o las ITM\index{opci\'on, ITM (dentro del dinero)} m\'as cercanas) con un precio de ejercicio\index{precio de ejercicio} $K$. La perspectiva del trader es neutral\index{perspectiva del trader, neutral}. Esta estrategia es de ganancia de capital\index{estrategia, ganancia de capital}.\footnote{\, La literatura acad\'emica sobre conos sint\'eticos\index{cono sint\'etico} parece ser escasa. V\'ease, por ejemplo, \cite{Trifonov2011}, \cite{Trifonov2014}.} Tenemos (asumiendo $S_0\geq K$ y $D > S_0 - K$):
\begin{eqnarray}
 && f_T = S_0 - S_T + 2\times{(S_T - K)^+} - D \\
 && S_{*sup} = 2\times K - S_0 + D \\
 && S_{*inf} = S_0 - D \\
 && P_{max} = \mbox{ilimitado} \\
 && L_{max} = D - (S_0 - K) \
\end{eqnarray}

\subsection{Estrategia: Cono sint\'etico\index{cono sint\'etico} largo con put (Long put synthetic straddle)}

{}Esta estrategia de volatilidad\index{estrategia de volatilidad} (que es la misma que un cono\index{cono} largo con la opci\'on call\index{opci\'on call} reemplazada por una opci\'on call sint\'etica\index{opci\'on call sint\'etica}) consiste en comprar una acci\'on y comprar {\em dos} opciones put\index{opci\'on put} ATM\index{opci\'on, ATM (en el dinero)} (o las ITM\index{opci\'on, ITM (dentro del dinero)} m\'as cercanas) con un precio de ejercicio\index{precio de ejercicio} $K$. La perspectiva del trader es neutral\index{perspectiva del trader, neutral}. Esta estrategia es de ganancia de capital\index{estrategia, ganancia de capital}. Tenemos (asumiendo $S_0\leq K$ y $D > K - S_0$):
\begin{eqnarray}
 && f_T = S_T - S_0 + 2\times{(K - S_T)^+} - D \\
 && S_{*sup} = S_0 + D \\
 && S_{*inf} = 2\times K - S_0 - D \\
 && P_{max} = \mbox{ilimitado} \\
 && L_{max} = D - (K - S_0) \
\end{eqnarray}

\subsection{Estrategia: Cono sint\'etico\index{cono sint\'etico} corto con call (Short call synthetic straddle)}

{}Esta estrategia lateral\index{estrategia lateral} (que es la misma que un cono\index{cono} corto con la opci\'on put\index{opci\'on put} reemplazada por una opci\'on put sint\'etica\index{opci\'on put sint\'etica}) consiste en comprar una acci\'on y vender {\em dos} opciones call\index{opci\'on call} ATM\index{opci\'on, ATM (en el dinero)} (o las OTM\index{opci\'on, OTM (fuera del dinero)} m\'as cercanas) con un precio de ejercicio\index{precio de ejercicio} $K$.  La perspectiva del trader es neutral\index{perspectiva del trader, neutral}. Esta estrategia es de ganancia de capital\index{estrategia, ganancia de capital}. Tenemos (asumiendo $S_0\leq K$):
\begin{eqnarray}
 && f_T = S_T - S_0 - 2\times{(S_T - K)^+} + C \\
 && S_{*sup} = 2\times K - S_0 + C \\
 && S_{*inf} = S_0 - C \\
 && P_{max} = K - S_0 + C \\
 && L_{max} = \mbox{ilimitado} \
\end{eqnarray}

\subsection{Estrategia: Cono sint\'etico\index{cono sint\'etico} corto con put (Short put synthetic straddle)}

{}Esta estrategia lateral\index{estrategia lateral} (que es la misma que un cono\index{cono} corto con la opci\'on call\index{opci\'on call} reemplazada por una opci\'on call sint\'etica\index{opci\'on call sint\'etica}) consiste en vender una acci\'on\index{vender una acci\'on} y vender {\em dos} opciones put\index{opci\'on put} ATM\index{opci\'on, ATM (en el dinero)} (o las OTM\index{opci\'on, OTM (fuera del dinero)} m\'as cercanas) con un precio de ejercicio\index{precio de ejercicio} $K$. La perspectiva del trader es neutral\index{perspectiva del trader, neutral}. Esta estrategia es de ganancia de capital\index{estrategia, ganancia de capital}. Tenemos (asumiendo $S_0\geq K$):
\begin{eqnarray}
 && f_T = S_0 - S_T - 2\times{(K - S_T)^+} + C \\
 && S_{*sup} = S_0 + C \\
 && S_{*inf} = 2\times K - S_0  - C \\
 && P_{max} = S_0 - K + C \\
 && L_{max} = \mbox{ilimitado} \
\end{eqnarray}

\subsection{Estrategia: Cono cubierto corto\index{cono cubierto corto} (Covered short straddle)}

{}Esta estrategia consiste en aumentar una call cubierta\index{call cubierta} lanzando una opci\'on put\index{opci\'on put} con el mismo precio de ejercicio\index{precio de ejercicio} $K$ y el mismo TTM\index{tiempo al vencimiento (TTM)} que la opci\'on call\index{opci\'on call} vendida y, por lo tanto, incrementando los ingresos. La perspectiva del trader es alcista\index{perspectiva del trader, alcista}. Tenemos:
\begin{eqnarray}
 && f_T = S_T - S_0 - (S_T - K)^+ - (K - S_T)^+ + C\\
 && S_* = \frac{1}{2}\left(S_0 + K - C\right)\\
 && P_{max} = K - S_0 + C\\
 && L_{max} = S_0 + K - C\
\end{eqnarray}

\subsection{Estrategia: Cuna cubierta corta\index{cuna cubierta corta} (Covered short strangle)}

{}Esta estrategia consiste en aumentar una call cubierta\index{call cubierta} lanzando una opci\'on put\index{opci\'on put} OTM\index{opci\'on, OTM (fuera del dinero)} con un precio de ejercicio\index{precio de ejercicio} $K^\prime$ y el mismo TTM\index{tiempo al vencimiento (TTM)} que la opci\'on call\index{opci\'on call} vendida (cuyo precio de ejercicio\index{precio de ejercicio} es $K$) y, por lo tanto, incrementando los ingresos. La perspectiva del trader es alcista\index{perspectiva del trader, alcista}. Tenemos:
\begin{eqnarray}
 && f_T = S_T - S_0 - (S_T - K)^+ - (K^\prime - S_T)^+ + C\\
 && P_{max} = K - S_0 + C\\
 && L_{max} = S_0 + K^\prime - C\
\end{eqnarray}

\subsection{Estrategia: Correa\index{correa} (Strap)}

{}Esta es una estrategia de volatilidad\index{estrategia de volatilidad} que consiste en una posici\'on larga en {\em dos} opciones call\index{opci\'on call} ATM\index{opci\'on, ATM (en el dinero)} y una posici\'on larga en una opci\'on put\index{opci\'on put} ATM\index{opci\'on, ATM (en el dinero)} con un precio de ejercicio\index{precio de ejercicio} $K$. Para crear este trade se requiere un d\'ebito neto. La perspectiva del trader es alcista\index{perspectiva del trader, alcista}. Esta estrategia es de ganancia de capital\index{estrategia, ganancia de capital}. Tenemos:\footnote{\, Para m\'as informaci\'on sobre correas\index{correa} y bandas\index{banda}, v\'ease, por ejemplo, \cite{Jha2010}, \cite{Topaloglou2011}.}
\begin{eqnarray}
 && f_T = 2\times ( S_T  -  K)^+ + (K - S_T)^+ - D\\
 && S_{*sup} = K + \frac{D}{2}\\
 && S_{*inf} = K -  D\\
 && P_{max} = \mbox{ilimitado}\\
 && L_{max} = D\
\end{eqnarray}

\subsection{Estrategia: Banda\index{banda} (Strip)}

{}Esta es una estrategia de volatilidad\index{estrategia de volatilidad} que consiste en una posici\'on larga en una opci\'on call\index{opci\'on call} ATM\index{opci\'on, ATM (en el dinero)} y una posici\'on larga en {\em dos} opciones put\index{opci\'on put} ATM\index{opci\'on, ATM (en el dinero)} con un precio de ejercicio\index{precio de ejercicio} $K$. Para crear este trade se requiere un d\'ebito neto. La perspectiva del trader es bajista\index{perspectiva del trader, bajista}. Esta estrategia es de ganancia de capital\index{estrategia, ganancia de capital}. Tenemos:
\begin{eqnarray}
 && f_T = ( S_T  -  K)^+ +  2\times (K - S_T)^+ - D\\
 && S_{*sup} = K + D\\
 && S_{*inf} = K -  \frac{D}{2}\\
 && P_{max} = \mbox{ilimitado}\\
 && L_{max} = D\
\end{eqnarray}

\subsection{Estrategia: Diferencial ratio inverso\index{diferencial ratio inverso} con call (Call ratio backspread)}

{}Esta estrategia consiste en una posici\'on corta en $N_C$ opciones call\index{opci\'on call} cercanas a ATM\index{opci\'on, ATM (en el dinero)} con un precio de ejercicio\index{precio de ejercicio} $K_1$ y una posici\'on larga en $N_L$ opciones call\index{opci\'on call} OTM\index{opci\'on, OTM (fuera del dinero)} con un precio de ejercicio\index{precio de ejercicio} $K_2$, en donde $N_L > N_C$. Generalmente, $N_L = 2$ y $N_C = 1$, o $N_L = 3$ y $N_C = 2$. La perspectiva del trader es fuertemente alcista\index{perspectiva del trader, fuertemente alcista}. Esta estrategia es de ganancia de capital\index{estrategia, ganancia de capital}. Tenemos:\footnote{\, Para m\'as informaci\'on sobre diferenciales ratios (inversos) con call/put\index{diferencial ratio, con call}\index{diferencial ratio, con put}, v\'ease, por ejemplo, \cite{Augustin2015}, \cite{Chaput2008}, \cite{Soltes2010}, \cite{Soltes2010b}, \cite{Soltes2013}.}
\begin{eqnarray}
 && f_T = N_L \times (S_T - K_2)^+ - N_C\times (S_T - K_1)^+ - H\\
 && S_{*inf} = K_1 - H / N_C,~~~H < 0 \\
 && S_{*sup} = (N_L \times K_2 - N_C \times K_1 + H)/(N_{L} - N_{C})  \\
 && P_{max} = \mbox{ilimitado} \\
 && L_{max} = N_C\times (K_2 - K_1) + H\
\end{eqnarray}

\subsection{Estrategia: Diferencial ratio inverso\index{diferencial ratio inverso} con put (Put ratio backspread)}

{}Esta estrategia consiste en una posici\'on corta en $N_C$ opciones put\index{opci\'on put} cercanas a ATM\index{opci\'on, ATM (en el dinero)} con un precio de ejercicio\index{precio de ejercicio} $K_1$ y una posici\'on larga en $N_L$ opciones put\index{opci\'on put} OTM\index{opci\'on, OTM (fuera del dinero)} con un precio de ejercicio\index{precio de ejercicio} $K_2$, en donde $N_L > N_C$. Generalmente, $N_L = 2$ y $N_C = 1$, o $N_L = 3$ y $N_C = 2$. La perspectiva del trader es fuertemente bajista\index{perspectiva del trader, fuertemente bajista}. Esta estrategia es de ganancia de capital\index{estrategia, ganancia de capital}. Tenemos:
\begin{eqnarray}
 && f_T = N_L \times (K_2  -  S_T )^+ - N_C\times ( K_1  -   S_T)^+ - H\\
 && S_{*sup} = K_1 + H / N_C,~~~H < 0 \\
 && S_{*inf} = (N_L \times K_2 - N_C \times K_1 - H)/(N_{L} - N_{C})  \\
 && P_{max} = N_L \times K_2 - N_C \times K_1 - H \\
 && L_{max} = N_C \times (K_1 - K_2) + H \
\end{eqnarray}

\subsection{Estrategia: Diferencial ratio con call\index{diferencial ratio con call} (Ratio call spread)}

{}Esta estrategia consiste en una posici\'on corta en $N_C$ opciones call\index{opci\'on call} cercanas a ATM\index{opci\'on, ATM (en el dinero)} con un precio de ejercicio\index{precio de ejercicio} $K_1$ y una posici\'on larga en $N_L$ opciones call\index{opci\'on call} ITM\index{opci\'on, ITM (dentro del dinero)} con un precio de ejercicio\index{precio de ejercicio} $K_2$, en donde $N_L < N_C$. Generalmente, $N_L = 1$ y $N_C = 2$, o $N_L = 2$ y $N_C = 3$. Esta estrategia es de generaci\'on de ingresos\index{estrategia, generaci\'on de ingresos} si es estructurada de forma tal que genere un cr\'edito neto. La perspectiva del trader es neutral a bajista\index{perspectiva del trader, neutral a bajista}. Tenemos:\footnote{\, Entonces, puede apreciarse que la diferencia entre un diferencial ratio inverso\index{diferencial ratio inverso} con call/put y un diferencial ratio con call/put\index{diferencial ratio con call}\index{diferencial ratio, con put}, es que en el primero $N_L > N_C$, mientras que en el \'ultimo $N_L < N_C$.}
\begin{eqnarray}
 && f_T = N_L \times (S_T - K_2)^+ - N_C\times (S_T - K_1)^+ - H\\
 && S_{*inf} = K_2 + H / N_L,~~~H > 0 \\
 && S_{*sup} = (N_C \times K_1 - N_L \times K_2 - H)/(N_{C} - N_{L})  \\
 && P_{max} = N_L \times (K_1 - K_2) - H \\
 && L_{max} =  \mbox{ilimitado}\
\end{eqnarray}

\subsection{Estrategia: Diferencial ratio con put\index{diferencial ratio con put} (Ratio put spread)}

{}Esta estrategia consiste en una posici\'on corta en $N_C$ opciones put\index{opci\'on put} cercanas a ATM\index{opci\'on, ATM (en el dinero)} con un precio de ejercicio\index{precio de ejercicio} $K_1$ y una posici\'on larga en $N_L$ opciones put\index{opci\'on put} ITM\index{opci\'on, ITM (dentro del dinero)} con un precio de ejercicio\index{precio de ejercicio} $K_2$, en donde $N_L < N_C$. Generalmente, $N_L = 1$ y $N_C = 2$, o $N_L = 2$ y $N_C = 3$. Esta estrategia es de generaci\'on de ingresos\index{estrategia, generaci\'on de ingresos} si es estructurada de forma tal que genere un cr\'edito neto. La perspectiva del trader es neutral a alcista\index{perspectiva del trader, neutral a alcista}. Tenemos:
\begin{eqnarray}
 && f_T = N_L \times (K_2 - S_T)^+ - N_C\times (K_1 - S_T)^+ - H\\
 && S_{*sup} = K_2 - H / N_L,~~~H > 0 \\
 && S_{*inf} = (N_C \times K_1 - N_L \times K_2 + H)/(N_{C} - N_{L})  \\
 && P_{max} = N_L \times (K_2 - K_1) - H \\
 && L_{max} =  N_C \times K_1 - N_L \times K_2 + H\
\end{eqnarray}

\subsection{Estrategia: Mariposa\index{mariposa} larga con call (Long call butterfly)}

{}Esta es una estrategia lateral\index{estrategia lateral} que consiste en una posici\'on larga en una opci\'on call\index{opci\'on call} OTM\index{opci\'on, OTM (fuera del dinero)} con un precio de ejercicio\index{precio de ejercicio} $K_1$, una posici\'on corta en {\em dos} opciones call\index{opci\'on call} ATM\index{opci\'on, ATM (en el dinero)} con un precio de ejercicio\index{precio de ejercicio} $K_2$ y una posici\'on larga en una opci\'on call\index{opci\'on call} ITM\index{opci\'on, ITM (dentro del dinero)} con un precio de ejercicio\index{precio de ejercicio} $K_3$. Los precios de ejercicio\index{precio de ejercicio} son equidistantes: $K_2 - K_3 = K_1 - K_2 = \kappa$. Para crear este trade se requiere un d\'ebito neto relativamente bajo. La perspectiva del trader es neutral\index{perspectiva del trader, neutral}. Esta estrategia es de ganancia de capital\index{estrategia, ganancia de capital}. Tenemos:\footnote{\, Para m\'as informaci\'on sobre mariposas\index{mariposa} (incluyendo mariposas de hierro\index{mariposa de hierro}), v\'ease, por ejemplo, \cite{Balbas1999}, \cite{Howison2013}, \cite{Jongadsayakul2017}, \cite{Matsypura2010}, \cite{Youbi2017}, \cite{Wolf2014}, \cite{Wystup2017}. La literatura acad\'emica sobre estrategias c\'ondor\index{estrategia condor @ estrategia c\'ondor} (que pueden ser pensadas como una variaci\'on de mariposas\index{mariposa}) parece ser m\'as escasa. V\'ease, por ejemplo, \cite{Niblock2017}.}
\begin{eqnarray}
 && f_T = (S_T - K_1)^+ + (S_T - K_3 )^+ - 2\times (S_T - K_2  )^+  - D\\
 && S_{*inf} = K_3 + D\\
 && S_{*sup} = K_1 - D \\
 && P_{max} = \kappa - D\\
 && L_{max} =  D\
\end{eqnarray}

\subsubsection{Estrategia: Mariposa\index{mariposa} modificada con call (Modified call butterfly)}

{}Esta es una variaci\'on de la mariposa\index{mariposa} larga con call, cuya diferencia radica en que los precios de ejercicio\index{precio de ejercicio} no son equidistantes; en cambio tenemos $K_1 - K_2 < K_2 - K_3$. Esto resulta en una estrategia lateral\index{estrategia lateral} con un sesgo alcista. Tenemos:
\begin{eqnarray}
 && f_T = (S_T - K_1)^+ + (S_T - K_3 )^+ - 2\times (S_T - K_2  )^+  - D\\
 && S_{*} = K_3 + D\\
 && P_{max} = K_2 - K_3 - D\\
 && L_{max} =  D\
\end{eqnarray}

\subsection{Estrategia: Mariposa\index{mariposa} larga con put (Long put butterfly)}

{}Esta es una estrategia lateral\index{estrategia lateral} que consiste en una posici\'on larga en una opci\'on put\index{opci\'on put} OTM\index{opci\'on, OTM (fuera del dinero)} con un precio de ejercicio\index{precio de ejercicio} $K_1$, una posici\'on corta en {\em dos} opciones put\index{opci\'on put} ATM\index{opci\'on, ATM (en el dinero)} con un precio de ejercicio\index{precio de ejercicio} $K_2$ y una posici\'on larga en una opci\'on put\index{opci\'on put} ITM\index{opci\'on, ITM (dentro del dinero)} con un precio de ejercicio\index{precio de ejercicio} $K_3$. Los precios de ejercicio\index{precio de ejercicio} son equidistantes: $K_3 - K_2 = K_2 - K_1 = \kappa$. Para crear este trade se requiere un d\'ebito neto relativamente bajo. La perspectiva del trader es neutral\index{perspectiva del trader, neutral}. Esta estrategia es de ganancia de capital\index{estrategia, ganancia de capital}. Tenemos:
\begin{eqnarray}
 && f_T = (K_1 - S_T)^+ + (K_3 - S_T )^+ - 2\times (K_2  -  S_T )^+  - D\\
 && S_{*sup} = K_3 - D \\
 && S_{*inf} = K_1 + D \\
 && P_{max} = \kappa - D\\
 && L_{max} =  D
\end{eqnarray}

\subsubsection{Estrategia: Mariposa\index{mariposa} modificada con put (Modified put butterfly)}

{}Esta es una variaci\'on de la mariposa\index{mariposa} larga con put, cuya diferencia radica en que los precios de ejercicio\index{precio de ejercicio} no son equidistantes; en cambio tenemos $K_3 - K_2 < K_2 - K_1$. Esto resulta en una estrategia lateral\index{estrategia lateral} con un sesgo alcista. Tenemos (para $H > 0$ tenemos tambi\'en $S_{*sup} = K_3 - H$):\footnote{\, Idealmente, este trade deber\'ia ser estructurado de forma tal que genere un cr\'edito neto, aunque esto no siempre es posible.}
\begin{eqnarray}
 && f_T = (K_1 - S_T)^+ + (K_3 - S_T)^+ - 2\times (K_2 - S_T)^+  - H\\
 && S_{*inf} = 2\times K_2 - K_3 + H\\
 && P_{max} = K_3 - K_2 - H\\
 && L_{max} =  2\times K_2 - K_1 - K_3 + H\
\end{eqnarray}

\subsection{Estrategia: Mariposa\index{mariposa} corta con call (Short call butterfly)}

{}Esta es una estrategia de volatilidad\index{estrategia de volatilidad} que consiste en una posici\'on corta en una opci\'on call\index{opci\'on call} ITM\index{opci\'on, ITM (dentro del dinero)} con un precio de ejercicio\index{precio de ejercicio} $K_1$, una posici\'on larga en {\em dos} opciones call\index{opci\'on call} ATM\index{opci\'on, ATM (en el dinero)} con un precio de ejercicio\index{precio de ejercicio} $K_2$ y una posici\'on corta en una opci\'on call\index{opci\'on call} OTM\index{opci\'on, OTM (fuera del dinero)} con un precio de ejercicio\index{precio de ejercicio} $K_3$. Los precios de ejercicio\index{precio de ejercicio} son equidistantes: $K_3 - K_2 = K_2 - K_1 = \kappa$. Este trade genera un cr\'edito neto. En este sentido, esta es una estrategia de generaci\'on de ingresos\index{estrategia, generaci\'on de ingresos}. Sin embargo, la ganancia potencial es sustancialmente menor que la de un cono\index{cono} corto o una cuna\index{cuna} corta (aunque con menor riesgo\index{riesgo}). La perspectiva del trader es neutral\index{perspectiva del trader, neutral}. Tenemos:
\begin{eqnarray}
 && f_T = 2 \times (S_T - K_2)^+ - (S_T - K_1 )^+ - ( S_T - K_3 )^+ + C\\
 && S_{*sup} = K_3 - C\\
 && S_{*inf} = K_1 + C \\
 && P_{max} = C\\
 && L_{max} =  \kappa - C \
\end{eqnarray}

\subsection{Estrategia: Mariposa\index{mariposa} corta con put (Short put butterfly)}

{}Esta es una estrategia de volatilidad\index{estrategia de volatilidad} que consiste en una posici\'on corta en una opci\'on put\index{opci\'on put} ITM\index{opci\'on, ITM (dentro del dinero)} con un precio de ejercicio\index{precio de ejercicio} $K_1$, una posici\'on larga en {\em dos} opciones put\index{opci\'on put} ATM\index{opci\'on, ATM (en el dinero)} con un precio de ejercicio\index{precio de ejercicio} $K_2$ y una posici\'on corta en una opci\'on put\index{opci\'on put} OTM\index{opci\'on, OTM (fuera del dinero)} con un precio de ejercicio\index{precio de ejercicio} $K_3$. Los precios de ejercicio\index{precio de ejercicio} son equidistantes: $K_3 - K_2 = K_2 - K_1 = \kappa$. Este trade genera un cr\'edito neto. En este sentido, esta es una estrategia de generaci\'on de ingresos\index{estrategia, generaci\'on de ingresos}. Sin embargo, la ganancia potencial es sustancialmente menor que la de un cono\index{cono} corto o una cuna\index{cuna} corta (aunque con menor riesgo\index{riesgo}). La perspectiva del trader es neutral\index{perspectiva del trader, neutral}. Tenemos:
\begin{eqnarray}
 && f_T = 2 \times (K_2 - S_T)^+ - (K_1 - S_T)^+ - (K_3 - S_T)^+ + C\\
 && S_{*inf} = K_3 + C\\
 && S_{*sup} = K_1 - C \\
 && P_{max} = C\\
 && L_{max} =  \kappa - C \
\end{eqnarray}

\subsection{Estrategia: Mariposa de hierro\index{mariposa de hierro} larga (``Long'' iron butterfly)}

{}Esta estrategia lateral\index{estrategia lateral} es una combinaci\'on de una estrategia diferencial alcista con put\index{diferencial alcista con put} y una estrategia diferencial bajista con call\index{diferencial bajista con call}. Consiste en una posici\'on larga en una opci\'on put\index{opci\'on put} OTM\index{opci\'on, OTM (fuera del dinero)} con un precio de ejercicio\index{precio de ejercicio} $K_1$, una posici\'on corta en una opci\'on put\index{opci\'on put} ATM\index{opci\'on, ATM (en el dinero)} y en una opci\'on call\index{opci\'on call} ATM\index{opci\'on, ATM (en el dinero)} con un precio de ejercicio\index{precio de ejercicio} $K_2$, y una posici\'on larga en una opci\'on call OTM con un precio de ejercicio\index{precio de ejercicio} $K_3$. Los precios de ejercicio\index{precio de ejercicio} son equidistantes: $K_2 - K_1 = K_3 - K_2 = \kappa$. Este trade es de cr\'edito neto. La perspectiva del trader es neutral\index{perspectiva del trader, neutral}. Esta es una estrategia de generaci\'on de ingresos\index{estrategia, generaci\'on de ingresos}. Tenemos:
\begin{eqnarray}
 && f_T = (K_1 - S_T)^+ - (K_2 - S_T)^+ - (S_T - K_2)^+ + (S_T - K_3 )^+ + C\\
 && S_{*sup} = K_2 + C\\
 && S_{*inf} = K_2 - C \\
 && P_{max} = C\\
 && L_{max} =  \kappa - C\
\end{eqnarray}

\subsection{Estrategia: Mariposa de hierro\index{mariposa de hierro} corta (``Short'' iron butterfly)}

{}Esta estrategia de volatilidad\index{estrategia de volatilidad} es una combinaci\'on de una estrategia diferencial bajista con put\index{diferencial bajista con put} y una estrategia diferencial alcista con call\index{diferencial alcista con call}. Consiste en una posici\'on corta en una opci\'on put\index{opci\'on put} OTM\index{opci\'on, OTM (fuera del dinero)} con un precio de ejercicio\index{precio de ejercicio} $K_1$, una posici\'on larga en una opci\'on put\index{opci\'on put} ATM\index{opci\'on, ATM (en el dinero)} y en una opci\'on call\index{opci\'on call} ATM\index{opci\'on, ATM (en el dinero)} con un precio de ejercicio\index{precio de ejercicio} $K_2$, y una posici\'on corta en una opci\'on call\index{opci\'on call} OTM\index{opci\'on, OTM (fuera del dinero)} con un precio de ejercicio\index{precio de ejercicio} $K_3$. Los precios de ejercicio\index{precio de ejercicio} son equidistantes: $K_2 - K_1 = K_3 - K_2 = \kappa$. Este trade es de d\'ebito neto. La perspectiva del trader es neutral\index{perspectiva del trader, neutral}. Esta estrategia es de ganancia de capital\index{estrategia, ganancia de capital}. Tenemos:
\begin{eqnarray}
 && f_T = (K_2 - S_T)^+ + (S_T - K_2)^+ - (K_1 - S_T)^+ - (S_T - K_3 )^+ - D\\
 && S_{*sup} = K_2 + D\\
 && S_{*inf} = K_2 - D \\
 && P_{max} = \kappa - D\\
 && L_{max} =  D\
\end{eqnarray}

\subsection{Estrategia: C\'ondor\index{condor @ c\'ondor} largo con call (Long call condor)}

{}Esta es una estrategia lateral\index{estrategia lateral} que consiste en una posici\'on larga en una opci\'on call\index{opci\'on call} ITM\index{opci\'on, ITM (dentro del dinero)} con un precio de ejercicio\index{precio de ejercicio} $K_1$, una posici\'on corta en una opci\'on call\index{opci\'on call} ITM\index{opci\'on, ITM (dentro del dinero)} con un precio de ejercicio\index{precio de ejercicio} mayor $K_2$, una posici\'on corta en una opci\'on call\index{opci\'on call} OTM\index{opci\'on, OTM (fuera del dinero)} con un precio de ejercicio\index{precio de ejercicio} $K_3$ y una posici\'on larga en una opci\'on call\index{opci\'on call} OTM\index{opci\'on, OTM (fuera del dinero)} con un precio de ejercicio\index{precio de ejercicio} mayor $K_4$. Todos los precios de ejercicio\index{precio de ejercicio} son equidistantes: $K_4 - K_3 = K_3 - K_2 = K_2 - K_1 = \kappa$. Para crear este trade se requiere un d\'ebito neto relativamente bajo. La perspectiva del trader es neutral\index{perspectiva del trader, neutral}. Esta estrategia es de ganancia de capital\index{estrategia, ganancia de capital}. Tenemos:
\begin{eqnarray}
 && f_T = (S_T - K_1)^+ - (S_T - K_2)^+ - (S_T - K_3)^+ + (S_T - K_4)^+ - D\\
 && S_{*sup} = K_4 - D\\
 && S_{*inf} = K_1 + D\\
 && P_{max} = \kappa - D\\
 && L_{max} =  D\
\end{eqnarray}

\subsection{Estrategia: C\'ondor\index{condor @ c\'ondor} largo con put (Long put condor)}

{}Esta es una estrategia lateral\index{estrategia lateral} que consiste en una posici\'on larga en una opci\'on put\index{opci\'on put} OTM\index{opci\'on, OTM (fuera del dinero)} con un precio de ejercicio\index{precio de ejercicio} $K_1$, una posici\'on corta en una opci\'on put\index{opci\'on put} OTM\index{opci\'on, OTM (fuera del dinero)} con un precio de ejercicio\index{precio de ejercicio} mayor $K_2$, una posici\'on corta en una opci\'on put\index{opci\'on put} ITM\index{opci\'on, ITM (dentro del dinero)} con un precio de ejercicio\index{precio de ejercicio} $K_3$ y una posici\'on larga en una opci\'on put\index{opci\'on put} ITM\index{opci\'on, ITM (dentro del dinero)} con un precio de ejercicio\index{precio de ejercicio} mayor $K_4$. Todos los precios de ejercicio\index{precio de ejercicio} son equidistantes: $K_4 - K_3 = K_3 - K_2 = K_2 - K_1 = \kappa$. Para crear este trade se requiere un d\'ebito neto relativamente bajo. La perspectiva del trader es neutral\index{perspectiva del trader, neutral}. Esta estrategia es de ganancia de capital\index{estrategia, ganancia de capital}. Tenemos:
\begin{eqnarray}
 && f_T = (K_1 - S_T)^+ - (K_2  -  S_T )^+ - (K_3 - S_T )^+ + (K_4 - S_T)^+ - D\\
 && S_{*sup} = K_4 - D\\
 && S_{*inf} = K_1 + D \\
 && P_{max} = \kappa - D\\
 && L_{max} =  D\
\end{eqnarray}

\subsection{Estrategia: C\'ondor\index{condor @ c\'ondor} corto con call (Short call condor)}

{}Esta estrategia de volatilidad\index{estrategia de volatilidad} consiste en una posici\'on corta en una opci\'on call\index{opci\'on call} ITM\index{opci\'on, ITM (dentro del dinero)} con un precio de ejercicio\index{precio de ejercicio} $K_1$, una posici\'on larga en una opci\'on call\index{opci\'on call} ITM\index{opci\'on, ITM (dentro del dinero)} con un precio de ejercicio\index{precio de ejercicio} mayor $K_2$, una posici\'on larga en una opci\'on call\index{opci\'on call} OTM\index{opci\'on, OTM (fuera del dinero)} con un precio de ejercicio\index{precio de ejercicio} $K_3$ y una posici\'on corta en una opci\'on call\index{opci\'on call} OTM\index{opci\'on, OTM (fuera del dinero)} con un precio de ejercicio\index{precio de ejercicio} mayor $K_4$. Todos los precios de ejercicio\index{precio de ejercicio} son equidistantes: $K_4 - K_3 = K_3 - K_2 = K_2 - K_1 = \kappa$. Este trade genera un cr\'edito neto relativamente bajo. Al igual que en el caso de la mariposa\index{mariposa} corta con call, el beneficio potencial es sustancialmente m\'as bajo que en el caso de un cono\index{cono} corto o una cuna\index{cuna} corta (aunque con menor riesgo\index{riesgo}). Entonces, esta estrategia es de ganancia de capital y no de generaci\'on de ingresos\index{estrategia, ganancia de capital}\index{estrategia, generaci\'on de ingresos}. La perspectiva del trader es neutral\index{perspectiva del trader, neutral}. Tenemos:
\begin{eqnarray}
 && f_T = (S_T - K_2)^+ + (S_T - K_3)^+ -(S_T - K_1)^+ - (S_T - K_4)^+ + C\\
 && S_{*sup} = K_4 - C\\
 && S_{*inf} = K_1 + C\\
 && P_{max} = C\\
 && L_{max} =  \kappa - C\
\end{eqnarray}

\subsection{Estrategia: C\'ondor\index{condor @ c\'ondor} corto con put (Short put condor)}

{}Esta es una estrategia de volatilidad\index{estrategia de volatilidad} que consiste en una posici\'on corta en una opci\'on put\index{opci\'on put} OTM\index{opci\'on, OTM (fuera del dinero)} con un precio de ejercicio\index{precio de ejercicio} $K_1$, una posici\'on larga en una opci\'on put\index{opci\'on put} OTM\index{opci\'on, OTM (fuera del dinero)} con un precio de ejercicio\index{precio de ejercicio} mayor $K_2$, una posici\'on larga en una opci\'on put\index{opci\'on put} ITM\index{opci\'on, ITM (dentro del dinero)} con un precio de ejercicio\index{precio de ejercicio} $K_3$ y una posici\'on corta en una opci\'on put\index{opci\'on put} ITM\index{opci\'on, ITM (dentro del dinero)} con un precio de ejercicio\index{precio de ejercicio} mayor $K_4$. Todos los precios de ejercicio\index{precio de ejercicio} son equidistantes: $K_4 - K_3 = K_3 - K_2 = K_2 - K_1 = \kappa$. Este trade genera un cr\'edito neto relativamente bajo. Al igual que en el caso de la mariposa\index{mariposa} corta con put, el beneficio potencial es sustancialmente m\'as bajo que en el caso de un cono corto\index{cono} o una cuna\index{cuna} corta (aunque con menor riesgo\index{riesgo}). Entonces, esta estrategia es de ganancia de capital y no de generaci\'on de ingresos\index{estrategia, ganancia de capital}\index{estrategia, generaci\'on de ingresos}. La perspectiva del trader es neutral\index{perspectiva del trader, neutral}. Tenemos:
\begin{eqnarray}
 && f_T = (K_2  -  S_T )^+ + (K_3 - S_T )^+ -(K_1 - S_T)^+ - (K_4 - S_T)^+ + C\\
 && S_{*sup} = K_4 - C\\
 && S_{*inf} = K_1 + C \\
 && P_{max} = C\\
 && L_{max} =  \kappa - C\
\end{eqnarray}

\subsection{Estrategia: C\'ondor de hierro\index{condor de hierro @ c\'ondor de hierro} largo (Long iron condor)}

{}Esta estrategia lateral\index{estrategia lateral} es una combinaci\'on de una estrategia diferencial alcista con put\index{diferencial alcista con put} y una estrategia diferencial bajista con call\index{diferencial bajista con call}. Consiste en una posici\'on larga en una opci\'on put\index{opci\'on put} OTM\index{opci\'on, OTM (fuera del dinero)} con un precio de ejercicio\index{precio de ejercicio} $K_1$, una posici\'on corta en una opci\'on put\index{opci\'on put} OTM\index{opci\'on, OTM (fuera del dinero)} con un precio de ejercicio\index{precio de ejercicio} mayor $K_2$, una posici\'on corta en una opci\'on call\index{opci\'on call} OTM\index{opci\'on, OTM (fuera del dinero)} con un precio de ejercicio\index{precio de ejercicio} $K_3$ y una posici\'on larga en una opci\'on call\index{opci\'on call} OTM\index{opci\'on, OTM (fuera del dinero)} con un precio de ejercicio\index{precio de ejercicio} mayor $K_4$. Los precios de ejercicio\index{precio de ejercicio} son equidistantes: $K_4 - K_3 = K_3 - K_2 = K_2 - K_1 = \kappa$. Este trade genera un cr\'edito neto. La perspectiva del trader es neutral\index{perspectiva del trader, neutral}. Esta es una estrategia de generaci\'on de ingresos\index{estrategia, generaci\'on de ingresos}. Tenemos:
\begin{eqnarray}
 && f_T = (K_1 - S_T)^+ + (S_T - K_4 )^+ - (K_2 - S_T)^+ - (S_T - K_3)^+ + C\\
 && S_{*sup} = K_3 + C\\
 && S_{*inf} = K_2 - C \\
 && P_{max} = C\\
 && L_{max} =  \kappa - C\
\end{eqnarray}

\subsection{Estrategia: C\'ondor de hierro\index{condor de hierro @ c\'ondor de hierro} corto (Short iron condor)}

{}Esta estrategia de volatilidad\index{estrategia de volatilidad} es una combinaci\'on de una estrategia diferencial bajista con put\index{diferencial bajista con put} y una estrategia diferencial alcista con call\index{diferencial alcista con call}. Consiste en una posici\'on corta en una opci\'on put\index{opci\'on put} OTM\index{opci\'on, OTM (fuera del dinero)} con un precio de ejercicio\index{precio de ejercicio} $K_1$, una posici\'on larga en una opci\'on put\index{opci\'on put} OTM\index{opci\'on, OTM (fuera del dinero)} con un precio de ejercicio\index{precio de ejercicio} mayor $K_2$, una posici\'on larga en una opci\'on call\index{opci\'on call} OTM\index{opci\'on, OTM (fuera del dinero)} con un precio de ejercicio\index{precio de ejercicio} $K_3$ y una posici\'on corta en una opci\'on call\index{opci\'on call} OTM\index{opci\'on, OTM (fuera del dinero)} con un precio de ejercicio\index{precio de ejercicio} mayor $K_4$. Los precios de ejercicio\index{precio de ejercicio} son equidistantes: $K_4 - K_3 = K_3 - K_2 = K_2 - K_1 = \kappa$. Para crear este trade se requiere un d\'ebito neto. La perspectiva del trader es neutral\index{perspectiva del trader, neutral}. Esta estrategia es de ganancia de capital\index{estrategia, ganancia de capital}. Tenemos:
\begin{eqnarray}
 && f_T = (K_2 - S_T)^+ + (S_T - K_3)^+ -(K_1 - S_T)^+ - (S_T - K_4 )^+ - D\\
 && S_{*sup} = K_3 + D\\
 && S_{*inf} = K_2 - D \\
 && P_{max} = \kappa - D\\
 && L_{max} =  D\
\end{eqnarray}

\subsection{Estrategia: Conversi\'on\index{conversi\'on} larga (Long box)}

{}Esta estrategia de volatilidad\index{estrategia de volatilidad} puede ser vista como una combinaci\'on de un forward sint\'etico\index{forward sint\'etico} largo y un forward sint\'etico\index{forward sint\'etico} corto, o bien como una combinaci\'on de una estrategia diferencial alcista con call\index{diferencial alcista con call} y una estrategia diferencial bajista con put\index{diferencial bajista con put}. Consiste en una posici\'on larga en una opci\'on put\index{opci\'on put} ITM\index{opci\'on, ITM (dentro del dinero)} con un precio de ejercicio\index{precio de ejercicio} $K_1$, una posici\'on corta en una opci\'on put\index{opci\'on put} OTM\index{opci\'on, OTM (fuera del dinero)} con un precio de ejercicio\index{precio de ejercicio} menor $K_2$, una posici\'on larga en una opci\'on call\index{opci\'on call} ITM\index{opci\'on, ITM (dentro del dinero)} con el precio de ejercicio\index{precio de ejercicio} $K_2$ y una posici\'on corta en una opci\'on call\index{opci\'on call} OTM\index{opci\'on, OTM (fuera del dinero)} con el precio de ejercicio\index{precio de ejercicio} $K_1$. La perspectiva del trader es neutral\index{perspectiva del trader, neutral}. Esta estrategia es de ganancia de capital\index{estrategia, ganancia de capital}.\footnote{\, En ciertos casos puede ser utilizada como una estrategia impositiva\index{estrategia impositiva} -- v\'ease, por ejemplo,  \cite{Cohen2005}. Para m\'as informaci\'on sobre estrategias de este tipo, v\'ease, por ejemplo, \cite{BenZion2005}, \cite{Bharadwaj2001}, \cite{Billingsley1985}, \cite{Clarke2013}, \cite{Fung2004}, \cite{Hemler1997}, \cite{Jongadsayakul2016}, \cite{Ronn1989}, \cite{Vipul2009}.} Tenemos (asumiendo $K_1 \geq K_2 + D$):
\begin{eqnarray}
 f_T &=& (K_1 - S_T)^+ - (K_2 - S_T)^+ + (S_T - K_2)^+ - (S_T - K_1)^+ - D \nonumber \\
    &=& K_1 - K_2 - D \\
  P_{max} &=& (K_1 - K_2) - D \
\end{eqnarray}

\subsection{Estrategia: Collar\index{collar}}

{}Esta estrategia (tambi\'en conocida como ``cerca\index{cerca}'') es una opci\'on call cubierta\index{call cubierta} aumentada por una pocisi\'on larga en una opci\'on put\index{opci\'on put}, la cual sirve como seguro ante una ca\'ida en el precio de la acci\'on.\footnote{\, De forma similar, un collar corto es una opci\'on put cubierta\index{put cubierta} aumentada por una posici\'on larga en una opci\'on call\index{opci\'on call}.} Consiste en la compra de una acci\'on, una posici\'on larga en una opci\'on put\index{opci\'on put} OTM\index{opci\'on, OTM (fuera del dinero)} con un precio de ejercicio\index{precio de ejercicio} $K_1$ y una posici\'on corta en una opci\'on call\index{opci\'on call} OTM\index{opci\'on, OTM (fuera del dinero)} con un precio de ejercicio\index{precio de ejercicio} mayor $K_2$. La perspectiva del trader es moderadamente alcista\index{perspectiva del trader, moderadamente alcista}. Esta estrategia es de ganancia de capital\index{estrategia, ganancia de capital}. Tenemos:\footnote{\, Para m\'as informaci\'on sobre estrategias collar\index{estrategia collar}\index{collar}, v\'ease, por ejemplo, \cite{Bartonova2012}, \cite{Burnside2011}, \cite{DAntonio2008}, \cite{Israelov2016}, \cite{Li2017}, \cite{Officer2004}, \cite{Officer2006}, \cite{Shan2010}, \cite{Szado2010}, \cite{Szado2011}, \cite{Timmermans2017}, \cite{Yim2011}.}
\begin{eqnarray}
 && f_T = S_T - S_0 +(K_1 - S_T)^+ - (S_T - K_2)^+ - H\\
 && S_{*} = S_0 + H \\
 && P_{max} = K_2 - S_0 - H \\
 && L_{max} =  S_0 - K_1 + H \
\end{eqnarray}

\subsection{Estrategia: Gaviota\index{gaviota} corta alcista (Bullish short seagull spread)}

{}Esta estrategia es un diferencial alcista con call\index{diferencial alcista con call} financiado con la venta de una opci\'on put\index{opci\'on put} OTM\index{opci\'on, OTM (fuera del dinero)}. Consiste en una posici\'on corta en una opci\'on put\index{opci\'on put} OTM\index{opci\'on, OTM (fuera del dinero)} con un precio de ejercicio\index{precio de ejercicio} $K_1$, una posici\'on larga en una opci\'on call\index{opci\'on call} ATM\index{opci\'on, ATM (en el dinero)} con un precio de ejercicio\index{precio de ejercicio} $K_2$ y una posici\'on corta en una opci\'on call\index{opci\'on call} OTM\index{opci\'on, OTM (fuera del dinero)} con un precio de ejercicio\index{precio de ejercicio} $K_3$. Idealmente, el trade deber\'ia ser estructurado de forma tal que el costo inicial sea cero. La perspectiva del trader es alcista\index{perspectiva del trader, alcista}. Esta estrategia es de ganancia de capital\index{estrategia, ganancia de capital}. Tenemos:\footnote{\, La literatura acad\'emica sobre diferenciales gaviota\index{diferencial gaviota} parece ser escasa. Por un libro de referencia, v\'ease, por ejemplo, \cite{Wystup2017}.}
\begin{eqnarray}
 && f_T = -(K_1 - S_T)^+ + (S_T - K_2)^+ - (S_T - K_3)^+ - H\\
 && S_{*} = K_2 + H,~~~H > 0\\
 && S_{*} = K_1 + H,~~~H < 0\\
 && K_1 \leq S_{*} \leq K_2,~~~H = 0\\
 && P_{max} = K_3 - K_2 - H \\
 && L_{max} =  K_1 + H \
\end{eqnarray}

\subsection{Estrategia: Gaviota\index{gaviota} larga bajista (Bearish long seagull spread)}

{}Esta estrategia es un combo\index{combo} corto cubierto ante una suba en el precio de la acci\'on con la compra de una opci\'on call\index{opci\'on call} OTM\index{opci\'on, OTM (fuera del dinero)}. Consiste en una posici\'on larga en una opci\'on put\index{opci\'on put} OTM\index{opci\'on, OTM (fuera del dinero)} con un precio de ejercicio\index{precio de ejercicio} $K_1$, una posici\'on corta en una opci\'on call\index{opci\'on call} ATM\index{opci\'on, ATM (en el dinero)} con un precio de ejercicio\index{precio de ejercicio} $K_2$ y una posici\'on larga en una opci\'on call\index{opci\'on call} OTM\index{opci\'on, OTM (fuera del dinero)} con un precio de ejercicio\index{precio de ejercicio} $K_3$. Idealmente, el trade deber\'ia ser estructurado de forma tal que el costo inicial sea cero. La perspectiva del trader es bajista\index{perspectiva del trader, bajista}. Esta estrategia es de ganancia de capital\index{estrategia, ganancia de capital}. Tenemos:
\begin{eqnarray}
 && f_T = (K_1 - S_T)^+ -(S_T - K_2)^+ + (S_T - K_3)^+ - H\\
 && S_{*} = K_1 - H,~~~H > 0\\
 && S_{*} = K_2 - H,~~~H < 0\\
 && K_1 \leq S_{*} \leq K_2,~~~H = 0\\
 && P_{max} = K_1 - H \\
 && L_{max} =  K_3 - K_2 + H \
\end{eqnarray}

\subsection{Estrategia: Gaviota\index{gaviota} corta bajista (Bearish short seagull spread)}

{}Esta estrategia es un diferencial bajista con put\index{diferencial bajista con put} financiado con la venta de una opci\'on call\index{opci\'on call} OTM\index{opci\'on, OTM (fuera del dinero)}. Consiste en una posici\'on corta en una opci\'on put\index{opci\'on put} OTM\index{opci\'on, OTM (fuera del dinero)} con un precio de ejercicio\index{precio de ejercicio} $K_1$, una posici\'on larga en una opci\'on put\index{opci\'on put} ATM\index{opci\'on, ATM (en el dinero)} con un precio de ejercicio\index{precio de ejercicio} $K_2$ y una posici\'on corta en una opci\'on call\index{opci\'on call} OTM\index{opci\'on, OTM (fuera del dinero)} con un precio de ejercicio\index{precio de ejercicio} $K_3$. Idealmente, el trade deber\'ia ser estructurado de forma tal que el costo inicial sea cero. La perspectiva del trader es bajista\index{perspectiva del trader, bajista}. Esta estrategia es de ganancia de capital\index{estrategia, ganancia de capital}. Tenemos:
\begin{eqnarray}
 && f_T = -(K_1 - S_T)^+ + (K_2 - S_T)^+ - (S_T - K_3)^+ - H\\
 && S_{*} = K_2 - H,~~~H > 0\\
 && S_{*} = K_3 - H,~~~H < 0\\
 && K_2 \leq S_{*} \leq K_3,~~~H = 0\\
 && P_{max} = K_2 - K_1 - H \\
 && L_{max} =  \mbox{ilimitado} \
\end{eqnarray}

\subsection{Estrategia: Gaviota\index{gaviota} larga alcista (Bullish long seagull spread)}

{}Esta estrategia es un combo\index{combo} largo cubierto ante una ca\'ida en el precio de la acci\'on con la compra de una opci\'on put\index{opci\'on put} OTM\index{opci\'on, OTM (fuera del dinero)}. Consiste en una posici\'on larga en una opci\'on put\index{opci\'on put} OTM\index{opci\'on, OTM (fuera del dinero)} con un precio de ejercicio\index{precio de ejercicio} $K_1$, una posici\'on corta en una opci\'on put\index{opci\'on put} ATM\index{opci\'on, ATM (en el dinero)} con un precio de ejercicio\index{precio de ejercicio} $K_2$ y una posici\'on larga en una opci\'on call\index{opci\'on call} OTM\index{opci\'on, OTM (fuera del dinero)} con un precio de ejercicio\index{precio de ejercicio} $K_3$. Idealmente, el trade deber\'ia ser estructurado de forma tal que el costo inicial sea cero. La perspectiva del trader es alcista\index{perspectiva del trader, alcista}. Esta estrategia es de ganancia de capital\index{estrategia, ganancia de capital}. Tenemos:
\begin{eqnarray}
 && f_T = (K_1 - S_T)^+ - (K_2 - S_T)^+ + (S_T - K_3)^+ - H\\
 && S_{*} = K_3 + H,~~~H > 0\\
 && S_{*} = K_2 + H,~~~H < 0\\
 && K_2 \leq S_{*} \leq K_3,~~~H = 0\\
 && P_{max} = \mbox{ilimitado} \\
 && L_{max} =  K_2 - K_1 + H \
\end{eqnarray}

\newpage

\section{Acciones}\label{sec.stocks}

\subsection{Estrategia: Precio-momentum\index{precio-momentum}}\label{sub.prc.mom}

{}Emp\'iricamente, los rendimientos de las acciones presentan cierta ``inercia'', dando lugar a un fen\'onemo conocido como ``efecto momentum''\index{efecto de momentum}, en donde los retornos futuros se encuentran correlacionados con los retornos pasados (v\'ease, por ejemplo, \cite{Asness1994}, \cite{Asness2014}, \cite{Asness2013}, \cite{Grinblatt2004}, \cite{Jegadeesh1993}). Sea $t$ el tiempo medido en unidades de 1 mes, con $t=0$ correspondiente al tiempo m\'as cercano. Sea $P_i(t)$ las series de tiempo\index{serie de tiempo} de los precios (totalmente ajustados\index{precio, totalmente ajustado} por splits\index{splits} y dividendos\index{dividendo}) de la acci\'on $i$ ($i=1,\dots,N$, en donde $N$ es el n\'umero de acciones dentro del universo de trading\index{universo de trading}). Sea
\begin{eqnarray}
 && R_i(t) = \frac{P_i(t)}{P_i(t+1)} - 1\\
 && R_i^{acum} = \frac{P_i(S)}{P_i(S + T)} - 1 \label{cum.ret}\\
 && R_i^{media} = \frac{1}{T}~\sum_{t = S}^{S + T - 1} R_i(t)\\
 && R_i^{riesg.ajust} =  \frac{R_i^{media}}{\sigma_i}\label{riesg.ajust}\\
 && \sigma_i^2 = \frac{1}{T-1}~\sum_{t = S}^{S + T - 1} \left(R_i(t) - R_i^{media}\right)^2\label{hist.vol}
\end{eqnarray}
Aqu\'i: $R_i(t)$ es el retorno mensual; $R_i^{acum}$ es el retorno acumulado\index{retorno acumulado} calculado sobre $T$ meses, lo cual se conoce como ``per\'iodo de formaci\'on\index{periodo de formacion @ per\'iodo de formaci\'on}'' (usualmente $T=12$) salteando los $S$ meses m\'as recientes, lo cual se conoce como ``per\'iodo de omisi\'on\index{periodo de omision @ per\'iodo de omisi\'on}'' (usualmente $S=1$);\footnote{\, Usualmente, el mes m\'as reciente es salteado debido a que, emp\'iricamente, se ha observado un efecto de reversi\'on a la media (es decir, efecto contrario)\index{efecto, contrario}\index{efecto, reversi\'on a la media} en retornos mensuales, posiblemente debido a problemas de liquidez\index{liquidez}/microestructura -- v\'ease, por ejemplo, \cite{Asness1994}, \cite{Boudoukh1994}, \cite{Grinblatt2004}, \cite{Jegadeesh1990}, \cite{Lo1990}.} $R_i^{media}$ es la media mensual de los retornos calculada sobre el per\'iodo de formaci\'on\index{periodo de formacion @ per\'iodo de formaci\'on}; $R_i^{riesg.ajust}$ es la media de los retornos ajustada por riesgo\index{retornos ajustados por riesgo} sobre el per\'iodo de formaci\'on\index{periodo de formacion @ per\'iodo de formaci\'on}; y $\sigma_i$ es la volatilidad mensual calculada sobre el per\'iodo de formaci\'on\index{periodo de formacion @ per\'iodo de formaci\'on}.

{}La estrategia de precio-momentum\index{estrategia de precio-momentum} consiste en comprar las acciones con mejor desempe\~{n}o y vender aquellas con el peor desempe\~{n}o, en donde el ``desempe\~{n}o'' es medido por un criterio de selecci\'on basado en $R_i^{acum}$, $R_i^{media}$, $R_i^{riesg.ajust}$ o alg\'un otro criterio. Por ejemplo, una vez que las acciones son ordenadas en funci\'on de $R_i^{acum}$ (de forma decreciente), el trader puede comprar las acciones en el decil\index{decil} superior (ganadoras\index{ganadores}) y vender las acciones en el decil\index{decil} inferior (perdedoras\index{perdedores}).\footnote{\, Existe cierto grado de arbitrariedad al momento de definir los ganadores\index{ganadores} y perdedores\index{perdedores}.} Esta puede ser una estrategia de costo cero\index{estrategia de costo cero}, es decir, el portafolio\index{portafolio} correspondiente es d\'olar-neutral. De forma alternativa, un portafolio solo con posiciones largas\index{portafolio solo con posiciones largas} puede ser construido comprando aquellas acciones que se encuentran en, por ejemplo, el decil\index{decil} superior. Una vez que el portafolio\index{portafolio} se establece en $t = 0$, se mantiene inalterado durante un tiempo predefinido, conocido como ``per\'iodo de tenencia\index{periodo de tenencia @ per\'iodo de tenencia}'',\footnote{\, Sin embargo, por ejemplo, un portafolio solo con posiciones largas\index{portafolio solo con posiciones largas} puede ser liquidado antes del per\'iodo de tenencia\index{periodo de tenencia @ per\'iodo de tenencia} debido a eventos inesperados, tales como ca\'idas fuertes en el mercado\index{ca\'ida fuerte en el mercado}.} el cual puede ser 1 mes o m\'as (portafolios\index{portafolio} con un per\'iodo de tenencia m\'as largo, generalmente presentan una disminuci\'on en los retornos netos de costos transaccionales\index{costos transaccionales} ya que el efecto momentum\index{efecto de momentum} tiende a desvanecerse con el tiempo). Portafolios\index{portafolio} con m\'ultiple per\'iodos de tenencia se pueden construir mediante la superposici\'on de portafolios\index{portafolio} con 1 mes de per\'iodo de tenencia (v\'ease, por ejemplo, \cite{Jegadeesh1993}).

{}La prescripci\'on anterior no fija las ponderaciones relativas $w_i$ de las acciones dentro del portafolio\index{portafolio}. Para portafolios solo con posiciones largas\index{portafolio solo con posiciones largas} tenemos $w_i \geq 0$ y
\begin{equation}
\sum_{i=1}^N w_i = 1
\end{equation}
Entonces, si el nivel de inversi\'on\index{nivel de inversi\'on} total deseado es $I$, la acci\'on $i$ tiene $I \times w_i$ d\'olares invertidos. Esto, redondeando, se traduce en $Q_i = I \times w_i / P_i(0)$ acciones\index{acci\'on}.\footnote{\, Es decir, asumiendo que las acci\'ones son compradas al precio $P_i(0)$, lo cual no considera el impacto del slippage\index{slippage}.} Uno simplemente puede tomar ponderaciones uniformes, $w_i = 1/N$ para todas las acciones, aunque otros esquemas de ponderaci\'on\index{esquema de ponderaci\'on} son posibles. Por ejemplo, podemos tener ponderaciones no uniformes $w_i \propto 1/\sigma_i$, o $w_i \propto 1/\sigma_i^2$, etc.

{}Para un portafolio d\'olar-neutral\index{portafolio dolar-neutral @ portafolio d\'olar-neutral} podemos tener ponderaciones $w_i$ negativas y
\begin{eqnarray}
 && \sum_{i=1}^N |w_i| = 1\\
 && \sum_{i=1}^N w_i = 0
\end{eqnarray}
Entonces, si el nivel de inversi\'on\index{nivel de inversi\'on} total deseado es $I = I_L + I_C$, en donde $I_L$ es la inversi\'on\index{inversi\'on} larga total, y $I_C$ es el valor absoluto de la inversi\'on\index{inversi\'on} corta total,\footnote{\, Para portafolios d\'olar-neutral\index{portafolio dolar-neutral @ portafolio d\'olar-neutral} $I_L = I_C$ y $I = 2\times I_L$.} luego la acci\'on $i$ tiene $I \times w_i$ d\'olares invertidos, en donde $w_i > 0$ para las posiciones largas, y $w_i < 0$ para las posiciones cortas. Uno puede simplemente tomar ponderaciones m\'odulo-uniformes, en donde $w_i = 1/2N_L$ para todas las $N_L$ posiciones largas, y $w_i = -1/2N_C$ para todas las $N_C$ posiciones cortas. Sin embargo, otros esquemas de ponderaci\'on\index{esquema de ponderaci\'on} son posibles, por ejemplo, como arriba, ponderaciones ajustadas por $\sigma_i$, $\sigma_i^2$, etc.\footnote{\, Para m\'as literatura sobre estrategias de momentum\index{estrategia de momentum}, v\'ease, por ejemplo, \cite{Antonacci2017}, \cite{Asem2010}, \cite{Barroso2014}, \cite{Bhojraj2006}, \cite{Chordia2002}, \cite{Chuang2014}, \cite{Cooper2004}, \cite{Daniel2016}, \cite{Geczy2016}, \cite{Griffin2003}, \cite{Grundy2001}, \cite{Hwang2004}, \cite{Jegadeesh2001}, \cite{Karolyi2004}, \cite{Korajczyk2004}, \cite{Liu2008}, \cite{Moskowitz1999}, \cite{Rouwenhorst1998}, \cite{Sadka2002}, \cite{Siganos2006}, \cite{Stivers2010}.}

\subsection{Estrategia: Ganancias-momentum\index{ganancias-momentum}}

{}Esta estrategia consiste en comprar los ganadores\index{ganadores} y vender los perdedores\index{perdedores} como en el caso de la estrategia de precio-momentum\index{estrategia de precio-momentum}, pero utilizando las ganancias\index{ganancias} como el criterio de selecci\'on. Una forma de definir tal criterio, es mediante las ganancias inesperadas estandarizadas (SUE, por sus siglas en ingl\'es)\index{ganancias inesperadas estandarizadas (SUE)} \cite{Chan1996}:\footnote{\, Tambi\'en v\'ease, por ejemplo, \cite{Bartov2005}, \cite{Battalio2007}, \cite{Bernard1989}, \cite{Bernard1990}, \cite{Bhushan1994}, \cite{Chordia2009}, \cite{Chordia2006}, \cite{Czaja2013}, \cite{Doyle2006}, \cite{Foster1984}, \cite{Hew1996}, \cite{Hirshleifer2009}, \cite{Jansen2016},  \cite{Livnat2006}, \cite{Loh2012}, \cite{Mendenhall2004}, \cite{Ng2008}, \cite{Rendleman1982}, \cite{Stickel1991}, \cite{Watts1978}.}
\begin{eqnarray}
 && \mbox{SUE}_i = \frac{E_i - E^\prime_i}{\sigma_i}
\end{eqnarray}
Aqu\'i: $E_i$ es la ganancia\index{ganancias} por acci\'on\index{acci\'on} anunciada por la firma $i$ en el trimestre m\'as reciente; $E_i^\prime$ es la ganancia\index{ganancias} por acci\'on\index{acci\'on} anunciada 4 trimestres atr\'as; $\sigma_i$ es el desv\'io est\'andar\index{desvio estandar @ desv\'io est\'andar} de las ganancias inesperadas\index{ganancias inesperadas} $E_i - E^\prime_i$ sobre los \'ultimos 8 trimestres. Similar que para la estrategia de precio-momentum\index{estrategia de precio-momentum}, el trader puede, por ejemplo, construir un portafolio d\'olar-neutral\index{portafolio dolar-neutral @ portafolio d\'olar-neutral} comprando las acciones en el decil\index{decil} superior seg\'un SUE\index{ganancias inesperadas estandarizadas (SUE)}, y vendiendo las acciones\index{vender acciones} en el decil\index{decil} inferior.\footnote{\, Generalmente, el per\'iodo de tenencia\index{periodo de tenencia @ per\'iodo de tenencia} es de 6 meses. Los retornos tienden a disminuir cuando el per\'iodo de tenencia\index{periodo de tenencia @ per\'iodo de tenencia} es m\'as largo.}

\subsection{Estrategia: Valor\index{valor} (Value)\index{value}}\label{sub.value}

{}Esta estrategia consiste en comprar ganadores\index{ganadores} y vender perdedores\index{perdedores} como en las estrategias de precio-momentum y ganancias-momentum\index{estrategia de ganancias-momentum}\index{estrategia, ganancias-momentum}\index{estrategia, precio-momentum}, pero el criterio de selecci\'on se basa en el value\index{value} de la compa\~{n}\'ia. Value\index{value} puede ser definido con el ratio book-to-price (B/P, por sus siglas en ingl\'es)\index{ratio book-to-price (B/P)} (v\'ease, por ejemplo, \cite{Rosenberg1985}). Aqu\'i ``book'' es el valor de libros\index{valor de libros} de la compa\~{n}\'ia {\em por las acciones\index{acci\'on} en circulaci\'on} (entonces el ratio B/P\index{ratio B/P} es lo mismo que el ratio book-to-market\index{ratio book-to-market}, en donde ahora ``book'' representa su valor total de libros, {\em no} por las acciones\index{acci\'on} en circulaci\'on, y ``market'' (``mercado'' en ingl\'es) es su capitalizaci\'on burs\'atil\index{capitalizaci\'on burs\'atil}). El trader puede, por ejemplo, construir un portafolio de costo cero\index{portafolio de costo cero} comprando las acciones en el decil\index{decil} superior seg\'un el ratio B/P\index{ratio B/P}, y vendiendo las acciones\index{vender acciones} en el decil\index{decil} inferior. Puede haber variaciones en la definici\'on del ratio B/P\index{ratio B/P}. As\'i, por ejemplo, \cite{Asness2013} usa precios actuales (es decir, los m\'as recientes), mientras que \cite{Fama1992} y otros autores utilizan precios contempor\'aneos con el valor de libros\index{valor de libros}.\footnote{\, El per\'iodo de tenencia\index{periodo de tenencia @ per\'iodo de tenencia} generalmente es de 1 a 6 meses. Para m\'as literatura sobre estrategias de value\index{estrategia de value}, v\'ease, por ejemplo, \cite{Erb2006}, \cite{Fama1993}, \cite{Fama1996}, \cite{Fama1998}, \cite{Fama2012}, \cite{Fisher2016}, \cite{Gerakos2012}, \cite{Novy-Marx2013}, \cite{Piotroski2000}, \cite{Piotroski2012}, \cite{Stattman1980}, \cite{Suhonen2017}, \cite{Zhang2005}.}

\subsection{Estrategia: Anomal\'ia de baja volatilidad\index{anomal\'ia de baja volatilidad}}\label{vol.anomaly}

{}Esta estrategia tiene sus ra\'ices en la evidencia emp\'irica que muestra que los rendimientos futuros de portafolios con retornos pasados de baja volatilidad\index{portafolios con retornos pasados de baja volatilidad} son mayores a los de aquellos portafolios con retornos pasados de alta volatilidad\index{portafolios con retornos pasados de alta volatilidad},\footnote{\, V\'ease, por ejemplo, \cite{Ang2006}, \cite{Ang2009}, \cite{Baker2011}, \cite{Black1972}, \cite{Blitz2007}, \cite{Clarke2006}, \cite{Clarke2010}, \cite{Frazzini2014}, \cite{Fu2009}, \cite{Garcia-Feijoo2015}, \cite{Li2014}, \cite{Li2016}, \cite{Merton1987}.} lo que va en contra de la ``ingenua'' expectativa de que activos\index{activo} con mayor riesgo\index{riesgo} deber\'ian proporcionar mayores rendimientos. De esta forma, si $\sigma_i$ es definida como la volatilidad hist\'orica\index{volatilidad hist\'orica} (calculada sobre las series de tiempo\index{serie de tiempo} de los retornos hist\'oricos\index{retornos hist\'oricos}, como en la Ecuaci\'on (\ref{hist.vol})), el trader puede, por ejemplo, construir un portafolio d\'olar-neutral\index{portafolio dolar-neutral @ portafolio d\'olar-neutral} comprando acciones en el decil\index{decil} inferior seg\'un $\sigma_i$ (acciones de baja volatilidad), y vendiendo acciones\index{vender acciones} en el decil\index{decil} superior (acciones de alta volatilidad). La longitud de la muestra utilizada para calcular la volatilidad hist\'orica\index{volatilidad hist\'orica} puede ser desde, por ejemplo, 6 meses (126 d\'ias de trading\index{dias de trading @ d\'ias de trading}) a 1 a\~{n}o (252 d\'ias de trading\index{dias de trading @ d\'ias de trading}), con una duraci\'on similar para el per\'iodo de tenencia\index{periodo de tenencia @ per\'iodo de tenencia} (en este caso el ``per\'iodo de omisi\'on\index{periodo de omision @ per\'iodo de omisi\'on}'' no es requerido).

\subsection{Estrategia: Volatilidad impl\'icita\index{volatilidad impl\'icita}}

{}Esta estrategia se basa en la observaci\'on emp\'irica de que las acciones con mayores incrementos en las volatilidades impl\'icitas de las opciones call en el mes anterior presentan en promedio mayores rendimientos futuros, mientras que las acciones con mayores aumentos en las volatilidades impl\'icitas de las opciones put en el mes anterior tienen en promedio rendimientos futuros m\'as bajos (v\'ease, por ejemplo, \cite{An2014}, \cite{Chen2016}).\footnote{\, V\'ease tambi\'en, por ejemplo, \cite{Bali2009}, \cite{Bollen2004}, \cite{Busch2011}, \cite{Chakravarty2004}, \cite{Conrad2013}, \cite{Cremers2010}, \cite{Pan2006}, \cite{Xing2010}.} Por lo tanto, el trader puede, por ejemplo, construir un portafolio d\'olar-neutral\index{portafolio dolar-neutral @ portafolio d\'olar-neutral} comprando las acciones en el decil\index{decil} superior seg\'un el incremento en la volatilidad impl\'icita de las opciones call, y vendiendo acciones\index{vender acciones} en el decil\index{decil} superior seg\'un el incremento en la volatilidad impl\'icita de las opciones put. Distintas alternativas pueden ser consideradas, por ejemplo, comprar las acciones en el decil\index{decil} superior seg\'un la {\em diferencia} entre el cambio de la volatilidad impl\'icita de las opciones call y el cambio de la volatilidad impl\'icita de las opciones put.

\subsection{Estrategia: Portafolio multifactor\index{portafolio multifactor}}\label{sub.multifactor}

{}Esta estrategia consiste en comprar y vender acciones\index{vender acciones} en funci\'on de m\'ultiples factores\index{factor} tales como el value\index{value}, momentum\index{momentum}, etc. Por ejemplo, usualmente el value\index{value} y el momentum\index{momentum} se encuentran negativamente correlacionados, y, por lo tanto, combinarlos puede agregar valor (v\'ease, por ejemplo, \cite{Asness2013}). Hay una variedad de maneras en que los $F>1$ factores\index{factor} pueden ser combinados.\footnote{\, Y el per\'iodo de tenencia\index{periodo de tenencia @ per\'iodo de tenencia} depende de qu\'e factores\index{factor} son combinados.} Una de las formas m\'as simples es diversificar la exposici\'on\index{exposici\'on} a los $F$ factores\index{factor} con ciertas ponderaciones $w_A$, en donde $A=1,\dots,F$ etiqueta los distintos factores\index{factor}. Esto es, si $I$ es el nivel de inversi\'on\index{nivel de inversi\'on} total, entonces los $F$ portafolios\index{portafolio} (cada uno construido en funci\'on de un factor\index{factor}) tiene un nivel de inversi\'on\index{nivel de inversi\'on} $I_A = w_A \times I$, en donde (asumiendo que todo $w_A > 0$)
\begin{equation}
 \sum_{A = 1}^F w_A = 1
\end{equation}
As\'i, uno puede simplemente tomar ponderaciones uniformes $w_A = 1/F$, aunque este puede no ser el esquema de ponderaci\'on \'optimo\index{esquema de ponderaci\'on \'optimo}. Por ejemplo, similar a la Subsecci\'on \ref{sub.prc.mom}, hay esquemas de ponderaci\'on\index{esquema de ponderaci\'on} con $w_A \propto 1/\sigma_A$, $w_A\propto 1/\sigma_A^2$, etc., en donde $\sigma_A$ es la volatilidad hist\'orica\index{volatilidad hist\'orica} del portafolio del factor\index{portafolio del factor} correspondiente (con todos los factores uniformemente normalizados, por ejemplo, en funci\'on de los d\'olares invertidos).\footnote{\, Otro enfoque es asignar las ponderaciones $w_A$ optimizando un portafolio\index{portafolio} de $F$ retornos esperados\index{retorno esperado} correspondientes a los $F$ factores\index{factor} (utilizando una matriz de covarianza\index{matriz de covarianza} invertible con dimensiones $F\times F$ sobre los retornos).}

{}De forma alternativa, se pueden considerar $F$ rankings\index{ranking} de las acciones basados en $F$ factores\index{factor}. Luego, uno puede combinarlos de varias maneras. Por ejemplo, en el caso de dos factores\index{factor}, momentum\index{momentum} y value\index{value}, uno puede tomar las acciones de un quintil\index{quintil} superior (ganadoras\index{ganadores}) e inferior (perdedoras\index{perdedores}) seg\'un momentum\index{momentum} y luego dividirlas en la mitad superior y la mitad inferior, respectivamente, en funci\'on del value\index{value}. O uno puede tomar los quintiles\index{quintil} superior e inferior seg\'un value\index{value} y luego dividirlos seg\'un momentum\index{momentum}.\footnote{\, Estas dos formas generalmente no producen los mismos portafolios\index{portafolio}.} Otra forma consiste en definir los rankings netos de su promedio\index{ranking neto de la media}, es decir, netos de la media de todos los rankings\index{ranking}
\begin{eqnarray}
 s_{Ai} = \mbox{rank}(f_{Ai}) - \frac{1}{N}~\sum_{j=1}^N \mbox{rank}(f_{Aj})
\end{eqnarray}
en donde $f_{Ai}$ es el valor num\'erico del factor\index{factor} etiquetado por $A$ (por ejemplo, momentum\index{momentum}) para la acci\'on etiquetada por $i$ ($i=1,\dots,N$). Uno puede entonces simplemente promediar los rankings\index{ranking}:
\begin{equation}\label{comb.rank}
 s_i = \frac{1}{F}~\sum_{A=1}^F s_{Ai}
\end{equation}
El ``puntaje'' $s_i$ promediado puede generar confusiones al momento de definir los portafolios (por ejemplo, si hay una ambig\"uedad en el borde del decil\index{decil} superior) que pueden ser resueltas simplemente dando preferencia a uno de los rankings factoriales\index{ranking factorial}. Promediar sobre $s_{Ai}$ simplemente minimiza la suma de los cuadrados de las distancias euclidianas\index{distancia euclidiana} entre el $N$-vector $s_i$ y los $K$ $N$-vectores $s_{Ai}$. Uno puede introducir ponderaciones no uniformes en esta suma (lo que equivaldr\'ia a un promedio ponderado\index{promedio ponderado} en la Ecuaci\'on (\ref{comb.rank})), o incluso usar una definici\'on diferente de la distancia (por ejemplo, la distancia de Manhattan\index{distancia de Manhattan}), lo que complicar\'ia el problema computacionalmente. Etc.\footnote{\, Para literatura adicional sobre estrategias multifactoriales\index{estrategia multifactorial}, v\'ease, por ejemplo, \cite{Amenc2016}, \cite{Amenc2015}, \cite{Arnott2013}, \cite{Asness1997}, \cite{Barber2015}, \cite{Cochrane1999}, \cite{FamaEF1996}, \cite{Grinold2000}, \cite{Hsu2018}, \cite{Kahn2015}, \cite{Kahn2016}, \cite{Kozlov2013}, \cite{Malkiel2014}, \cite{Wang2005}.}

\subsection{Estrategia: Momentum residual\index{momentum residual}}\label{res.mom}

{}Esta es la misma que la estrategia de precio-momentum\index{estrategia de precio-momentum}, pero aqu\'i los retornos de las acciones $R_i(t)$ son reemplazados por los residuos $\epsilon_i(t)$ de una regresi\'on {\em serial}\index{regresi\'on serial} de los retornos de las acciones $R_i(t)$ sobre, por ejemplo, los 3 factores de Fama-French\index{factores de Fama-French} $\mbox{MKT}(t)$, $\mbox{SMB}(t)$, $\mbox{HML}(t)$,\footnote{\, Los retornos de las acciones $R_i$ se definen como los excesos de estos sobre la tasa libre de riesgo\index{tasa libre de riesgo} (la tasa de inter\'es del Tesoro\index{tasa de inter\'es del Tesoro} a un mes); MKT\index{portafolio, MKT} es el exceso de retorno\index{exceso de retorno} del portafolio del mercado\index{portafolio de mercado}; SMB\index{portafolio, SMB} es el exceso de retorno\index{exceso de retorno} del portafolio ``compa\~{n}\'ias peque\~{n}as menos compa\~{n}\'ias grandes'' (Small minus Big)\index{Small minus Big (SMB)} (seg\'un su capitalizaci\'on burs\'atil\index{capitalizaci\'on burs\'atil}); HML\index{portafolio, HML} es el exceso de retorno\index{exceso de retorno} del portafolio ``alto menos bajo''  (High minus Low)\index{High minus Low (HML)} (seg\'un book-to-market). V\'ease, por ejemplo, \cite{Carhart1997}, \cite{Fama1993} por m\'as detalles.\label{fn.FF}} con el intercepto\index{intercepto} (v\'ease, por ejemplo, \cite{Blitz2011}):\footnote{\, Para algunos estudios adicionales relacionados con la estrategia de momentum residual\index{estrategia de momentum residual}, v\'ease, por ejemplo, \cite{Blitz2013}, \cite{Chang2016}, \cite{Chaves2012}, \cite{Chuang2015}, \cite{Grundy2001}, \cite{Gutierrez2007}, \cite{Huhn2017}, \cite{Huij2017}, \cite{VanOord2016}.}
\begin{equation}
 R_i(t) =  \alpha_i + \beta_{1,i}~\mbox{MKT}(t) + \beta_{2,i}~\mbox{SMB}(t) + \beta_{3,i}~\mbox{HML}(t) + \epsilon_i(t) \
\end{equation}
La regresi\'on\index{regresi\'on} se corre sobre un per\'iodo de 36 meses\cite{Blitz2011} (con un per\'iodo de omisi\'on de 1 mes) para estimar los coeficientes de la regresi\'on\index{coeficiente de la regresi\'on} $\alpha_i$, $\beta_{1,i}$, $\beta_{2,i}$, $\beta_{3,i}$. Una vez que se estiman los coeficientes, los residuos se pueden calcular para el per\'iodo de formaci\'on\index{periodo de formacion @ per\'iodo de formaci\'on} de 12 meses (nuevamente, con un per\'iodo de omisi\'on de 1 mes):
\begin{equation}
 \epsilon_i(t) = R_i(t) - \beta_{1,i}~\mbox{MKT}(t) - \beta_{2,i}~\mbox{SMB}(t) - \beta_{3,i}~\mbox{HML}(t)  \
\end{equation}
Tenga en cuenta que $\alpha_i$ no se incluye en el c\'alculo de los residuos para el per\'iodo de formaci\'on\index{periodo de formacion @ per\'iodo de formaci\'on} de 12 meses ya que $\alpha_i$ fue computada por el per\'iodo de 36 meses. Los residuos $\epsilon_i(t)$ son utilizados luego para, por ejemplo, calcular los retornos de los residuos ajustados por riesgo\index{retornos ajustados por riesgo} ${\widetilde R}_i^{riesg.ajust}$ (aqu\'i $S=1$ y $T=12$; el per\'iodo de tenencia\index{periodo de tenencia @ per\'iodo de tenencia} generalmente es 1 mes, pero puede ser m\'as largo):
\begin{eqnarray}
 && \epsilon_i^{media} = \frac{1}{T}~\sum_{t = S}^{S + T - 1} \epsilon_i(t)\\
 && {\widetilde R}_i^{riesg.ajust} =  \frac{\epsilon_i^{media}}{{\widetilde \sigma}_i}\\
 && {\widetilde \sigma}_i^2 = \frac{1}{T-1}~\sum_{t = S}^{S + T - 1} \left(\epsilon_i(t) - \epsilon_i^{media}\right)^2
\end{eqnarray}
De esta forma un portafolio d\'olar-neutral\index{portafolio dolar-neutral @ portafolio d\'olar-neutral} se puede construir comprando las acciones en el decil\index{decil} superior seg\'un ${\widetilde R}_i^{riesg.ajust}$, y vendiendo las acciones\index{vender acciones} en el decil\index{decil} inferior (con ponderaciones (no) uniformes).

\subsection{Estrategia: Trading de pares\index{trading de pares}}\label{pairs.trading}

{}Esta estrategia d\'olar-neutral\index{estrategia dolar-neutral @ estrategia d\'olar-neutral} consiste en identificar acciones que hist\'oricamente presentan alta correlaci\'on (llam\'emoslas acci\'on A y acci\'on B) y, cuando una valoraci\'on err\'onea\index{valoraci\'on err\'onea} (es decir, una desviaci\'on de la correlaci\'on hist\'orica\index{correlaci\'on hist\'orica}) ocurre, vender la acci\'on ``cara''\index{acci\'on cara} y comprar la acci\'on ``barata''\index{acci\'on barata}. Este es un ejemplo de una estrategia de reversi\'on a la media\index{estrategia de reversi\'on a la media}. Sean $P_A(t_1)$ y $P_B(t_1)$ los precios de la acci\'on A y de la acci\'on B en el tiempo $t_1$, y sean $P_A(t_2)$ y $P_B(t_2)$ los precios de la acci\'on A y de la acci\'on B en un tiempo m\'as lejano $t_2$. Todos los precios son totalmente ajustados\index{precio, totalmente ajustado} por splits\index{splits} y dividendos\index{dividendo}. Los retornos correspondientes (desde $t_1$ a $t_2$) son
\begin{eqnarray}
 &&R_A = {P_A(t_2)\over P_A(t_1)} - 1\\
 &&R_B = {P_B(t_2)\over P_B(t_1)} - 1
\end{eqnarray}
Dado que normalmente los retornos son peque\~{n}os, podemos usar una definici\'on alternativa:
\begin{eqnarray}
 &&R_A = \ln\left({P_A(t_2)\over P_A(t_1)}\right)\\
 &&R_B = \ln\left({P_B(t_2)\over P_B(t_1)}\right)
\end{eqnarray}
Luego, sean ${\widetilde R}_A$ y ${\widetilde R}_B$ los retornos netos de la media\index{retorno neto de la media}:
\begin{eqnarray}
 &&{\overline R} = {1\over 2}\left(R_A + R_B\right)\\
 &&{\widetilde R}_A = R_A - {\overline R}\\
 &&{\widetilde R}_B = R_B - {\overline R}
\end{eqnarray}
en donde ${\overline R}$ es el retorno medio. Una acci\'on se encuentra ``cara\index{acci\'on, cara}'' si su retorno neto de la media\index{retorno neto de la media} es positivo, y se encuentra ``barata\index{acci\'on, barata}'' si su retorno neto de la media\index{retorno neto de la media} es negativo. Los n\'umeros de acciones\index{acci\'on} $Q_A$, $Q_B$ a vender/comprar se determina seg\'un la inversi\'on\index{inversi\'on} total deseada en d\'olares $I$ (Ecuaci\'on (\ref{tot.inv.pt})) y el requisito de d\'olar-neutralidad\index{dolar-neutralidad @ d\'olar-neutralidad} (Ecuaci\'on (\ref{dol.neu})):
\begin{eqnarray}
 &&P_A~\left|Q_A\right| + P_B~\left|Q_B\right| = I\label{tot.inv.pt}\\
 &&P_A~Q_A + P_B~Q_B = 0\label{dol.neu}
\end{eqnarray}
en donde $P_A$, $P_B$ son los precios de las acciones al momento $t_*$ en que la posici\'on es establecida ($t_*\geq t_2$).\footnote{\, Por algunas publicaciones sobre el trading de pares\index{trading de pares}, v\'ease, por ejemplo, \cite{Bogomolov2013}, \cite{Bowen2016}, \cite{Bowen2010}, \cite{Caldeira2013}, \cite{Chen2017}, \cite{Do2010}, \cite{Do2012}, \cite{Elliott2005}, \cite{Engle1987}, \cite{Gatev2006}, \cite{Huck2009}, \cite{Huck2015}, \cite{Huck2014}, \cite{Jacobs2015}, \cite{Kakushadze2015b}, \cite{Kim2011}, \cite{Kishore2012}, \cite{Krauss2017}, \cite{KraussStubinger2017}, \cite{Liew2013}, \cite{Lin2006}, \cite{Liu2017}, \cite{Miao2014}, \cite{Perlin2009}, \cite{Pizzutilo2013}, \cite{Rad2016}, \cite{StuingerBredthauer2017}, \cite{Stubinger2017}, \cite{Vaitonis2016}, \cite{Vidyamurthy2004}, \cite{Xie2014}, \cite{Yoshikawa2017}, \cite{Zeng2014}.}

\subsection{Estrategia: Reversi\'on a la media\index{reversi\'on a la media} -- grupo\index{grupo} \'unico}\label{sub1.3}

{}Esta es una generalizaci\'on de la estrategia de trading de pares\index{estrategia de trading de pares} de $N>2$ acciones que hist\'oricamente est\'an altamente correlacionadas (por ejemplo, acciones que pertenecen a la misma industria\index{industria} o sector\index{sector}). Sean $R_i$, $i=1,\dots,N$, los retornos de las $N$ acciones:
\begin{eqnarray}
 &&R_i = \ln\left({P_i(t_2)\over P_i(t_1)}\right)\\
 &&{\overline R} = {1\over N}\sum_{i=1}^N R_i\\
 &&{\widetilde R}_i = R_i - {\overline R}
\end{eqnarray}
Siguiendo la intuici\'on de la estrategia de trading de pares\index{trading de pares}, podemos vender las acciones con ${\widetilde R}_i$ positivo y comprar aquellas con ${\widetilde R}_i$ negativo. Tenemos las siguientes condiciones:
\begin{eqnarray}\label{tot.inv}
 &&\sum_{i=1}^N P_i~\left|Q_i\right| = I\\
 &&\sum_{i=1}^N P_i~Q_i = 0\label{dollar.neutral}
\end{eqnarray}
Aqu\'i: $I$ es la inversi\'on\index{inversi\'on} total deseada en d\'olares; la Ecuaci\'on (\ref{dollar.neutral}) es la restricci\'on de d\'olar-neutralidad\index{restricci\'on de d\'olar-neutralidad}; $Q_i<0$ para las ventas en corto\index{ventas en corto}; $Q_i>0$ para las posiciones largas; $P_i$ son los precios en el momento en que las posiciones se establecen. Tenemos 2 ecuaciones y $N > 2$ inc\'ognitas. Una forma simple (que es una de las muchas posibilidades) de definir $Q_i$ es tener las posiciones en d\'olares\index{posici\'on en d\'olares} $D_i = P_i~Q_i$ proporcionales a los retornos netos de la media\index{retorno neto de la media}:
\begin{equation}\label{D.i}
 D_i = -\gamma~{\widetilde R}_i
\end{equation}
en donde $\gamma > 0$ (recordar que vendemos las acciones con ${\widetilde R}_i>0$ y compramos aquellas con ${\widetilde R}_i<0$). Luego, la Ecuaci\'on (\ref{dollar.neutral}) se satisface autom\'aticamente, mientras que la Ecuaci\'on (\ref{tot.inv}) soluciona $\gamma$:
\begin{equation}
 \gamma = {I \over \sum_{i=1}^N \left|{\widetilde R}_i\right|}
\end{equation}

\subsubsection{Estrategia: Reversi\'on a la media\index{reversi\'on a la media} -- m\'ultiples grupos\index{grupo}}\label{sub1.4}

{}La estrategia de reversi\'on a la media\index{estrategia de reversi\'on a la media} de la Subsecci\'on \ref{sub1.3} puede ser f\'acilmente generalizada al caso en donde tenemos $K>1$ grupos\index{grupo} tal que las acciones pertenecientes a cada grupo\index{grupo} se encuentran hist\'oricamente altamente correlacionados.\footnote{\, Por ejemplo, esos grupos\index{grupo} pueden ser sectores\index{sector}, tales como energ\'ia\index{energ\'ia}, tecnolog\'ia, salud, etc.} Podemos simplemente tratar los diferentes grupos\index{grupo} independientemente unos de otros y construir una estrategia de reversi\'on a la media\index{estrategia de reversi\'on a la media} siguiendo el procedimiento anterior en cada grupo\index{grupo}. Luego, por ejemplo, podemos asignar inversiones\index{inversi\'on} a esas $K$ estrategias independientes de forma uniforme.

{}Hay una forma conveniente de tratar a todos los grupos\index{grupo} en una forma ``unificada'' utilizando una regresi\'on lineal\index{regresi\'on lineal}. Sean los $K$ grupos\index{grupo} llamados $A=1,\dots,K$. Sea $\Lambda_{iA}$ una matriz con dimensiones $N\times K$ tal que si la acci\'on etiquetada por $i$ ($i=1,\dots,N$) pertenece al grupo\index{grupo} llamado $A$, luego $\Lambda_{iA}=1$; en caso contrario, $\Lambda_{iA}=0$. Supondremos que todas y cada una de las acciones pertenecen a un solo grupo\index{grupo} (por lo tanto, no hay grupos\index{grupo} vac\'ios):
\begin{eqnarray}
 &&N_A = \sum_{i=1}^N \Lambda_{iA} > 0\\
 &&N = \sum_{A=1}^K N_A
\end{eqnarray}
Tenemos
\begin{eqnarray}
 &&\Lambda_{iA} = \delta_{G(i), A}\label{bin.lam}\\
 &&G:\{1,\dots, N\} \mapsto \{1,\dots, K\}
\end{eqnarray}
Aqu\'i: $G$ es el mapa entre las acciones y los grupos\index{grupo}; y $\Lambda_{iA}$ es la matriz de cargas\index{matriz de cargas}.

{}Ahora consideramos una regresi\'on lineal\index{regresi\'on lineal} de los retornos $R_i$ sobre $\Lambda_{iA}$ (sin el intercepto\index{intercepto} y con ponderaciones unitarias):
\begin{equation}\label{R.Lambda}
 R_i = \sum_{A=1}^K \Lambda_{iA}~f_A + \varepsilon_i
\end{equation}
en donde $f_A$ son los coeficientes de la regresi\'on\index{coeficiente de la regresi\'on} dados por (en notaci\'on matricial, en donde $R$ es el $N$-vector $R_i$, $f$ es el $K$-vector $f_A$, y $\Lambda$ es la matriz con dimensiones $N\times K$ $\Lambda_{iA}$)
\begin{eqnarray}
 &&f = Q^{-1}~\Lambda^T~R\\
 &&Q = \Lambda^T~\Lambda
\end{eqnarray}
y $\varepsilon_i$ son los residuos de la regresi\'on\index{residuos de la regresi\'on}. Cuando $\Lambda_{iA}$ es binario, dado por la Ecuaci\'on (\ref{bin.lam}), estos residuos no son m\'as que los retornos $R_i$ netos de la media con respecto a cada uno de los correspondientes grupos\index{grupo}:
\begin{eqnarray}
 &&\varepsilon = R - \Lambda~Q^{-1}~\Lambda^T~R\\
 &&Q_{AB} = N_A~\delta_{AB}\\
 &&{\overline R}_A = {1\over N_A}\sum_{j\in J_A} R_j\\
 &&\varepsilon_i = R_i - {\overline R}_{G(i)} = {\widetilde R}_i
\end{eqnarray}
en donde ${\overline R}_A$ es la media de los retornos del grupo\index{grupo} etiquetado por $A$, y ${\widetilde R}_i$ es el retorno neto de la media\index{retorno neto de la media} que se obtiene de sustraer de $R_i$ la media de los retornos del grupo\index{grupo} etiquetado por $A=G(i)$ al que las acciones etiquetadas por $i$ pertenecen: $J_A = \{i| G(i) = A\}\subset \{1,\dots,N\}$.

{}Los retornos netos de la media\index{retorno neto de la media} son grupo neutral, es decir,
\begin{equation}\label{ind.neutral}
 \sum_{i=1}^N {\widetilde R}_i~\Lambda_{iA} = 0,~~~A=1,\dots,K
\end{equation}
Tambi\'en, tenga en cuenta que autom\'aticamente tenemos (entonces $D_i$ dado por la Ecuaci\'on (\ref{D.i}) satisface la Ecuaci\'on (\ref{dollar.neutral}))
\begin{equation}\label{d.n.1}
 \sum_{i=1}^N {\widetilde R}_i~\nu_i = 0
\end{equation}
en donde $\nu_i\equiv 1$, $i=1,\dots, N$, es decir, el $N$-vector $\nu$ es el vector unitario. En t\'erminos de una regresi\'on\index{regresi\'on}, $\nu$ equivale al intercepto\index{intercepto}. No tuvimos que a\~{n}adir el intercepto\index{intercepto} a la matriz de cargas\index{matriz de cargas} $\Lambda$ dado que ya est\'a subsumido en ella:
\begin{equation}
 \sum_{A=1}^K \Lambda_{iA} = \nu_i
\end{equation}

\subsection{Reversi\'on a la media\index{reversi\'on a la media} -- regresi\'on ponderada\index{regresi\'on ponderada}}\label{sub.mean.rev.w.reg}

{}Las condiciones (\ref{ind.neutral}) satisfechas por los retornos netos de la media\index{retorno neto de la media} cuando la matriz de cargas\index{matriz de cargas} es binaria, simplemente implican que esos retornos son grupo-neutrales, es decir, ortogonales a los $K$ $N$-vectores $v^{(A)}$ que comprenden las columnas de $\Lambda_{iA}$. Dicha ortogonalidad puede definirse para cualquier matriz de cargas\index{matriz de cargas}, no solo para el caso binario. Entonces, podemos considerar una generalizaci\'on en donde la matriz de cargas\index{matriz de cargas}, llamada $\Omega_{iA}$, puede tener algunas columnas binarias, pero generalmente no es necesario. Las columnas binarias, si las hay, pueden, por ejemplo, ser factores de riesgo basados en una industria\index{industria} (o sector\index{sector}); las columnas no binarias se interpretan como factores de riesgo\index{factor de riesgo} no basados en la industria; y la condici\'on de ortogonalidad\index{condici\'on de ortogonalidad}
\begin{equation}
 \sum_{i=1}^N {\widetilde R}_i~\Omega_{iA},~~~A=1,\dots,K
\end{equation}
puede ser satisfecha si los retornos ${\widetilde R}_i$ est\'an relacionados con los residuos $\varepsilon_i$ de la regresi\'on\index{regresi\'on} de $R_i$ sobre $\Omega_{iA}$ con algunas ponderaciones de la regresi\'on\index{ponderaciones de la regresi\'on} (generalmente no uniformes) $z_i$ v\'ia
\begin{eqnarray}
 \label{w.reg.4}
 &&{\widetilde R} = Z~\varepsilon\\
 \label{w.reg.1}
 &&\varepsilon = R - \Omega~Q^{-1}~\Omega^T~Z~R\\
 \label{w.reg.2}
 &&Z = \mbox{diag}(z_i)\\
 \label{w.reg.3}
 &&Q = \Omega^T~Z~\Omega
\end{eqnarray}
Si el intercepto\index{intercepto} se incluye en $\Omega_{iA}$ (es decir, una combinaci\'on lineal de las columnas de $\Omega_{iA}$ es igual al $N$-vector $\nu$ unitario), luego autom\'aticamente tenemos
\begin{equation}
 \sum_{i=1}^N {\widetilde R}_i = 0
\end{equation}
Las ponderaciones $z_i$ pueden ser, por ejemplo, tomadas como $z_i = 1/\sigma_i^2$, en donde $\sigma_i$ son las volatilidades hist\'oricas\index{volatilidad hist\'orica}.\footnote{\, Por algunas publicaciones sobre estrategias de reversi\'on a la media\index{estrategia, reversi\'on a la media}\index{estrategia, reversi\'on a la media}, v\'ease, por ejemplo, \cite{Avellaneda2010}, \cite{Black1991}, \cite{Black1992}, \cite{Cheung2010}, \cite{Chin2002}, \cite{Conrad1998}, \cite{Daniel2001}, \cite{DaSilva2009}, \cite{Doan2014}, \cite{Drobetz2001}, \cite{Hodges1993}, \cite{Idzorek2007}, \cite{Jansen2016}, \cite{Jegadeesh1995},  \cite{Kakushadze2015b}, \cite{Kang2002}, \cite{Kudryavtsev2012}, \cite{Lakonishok1994}, \cite{Lehmann1990}, \cite{Li2012}, \cite{LiewRoberts2013},  \cite{Lo1990}, \cite{Mun2000}, \cite{OTool2013}, \cite{Pole2007}, \cite{Poterba1988}, \cite{Satchell2000}, \cite{Schiereck1999}, \cite{Shi2015}, \cite{Yao2012}.}

\subsection{Estrategia: Media m\'ovil\index{media m\'ovil}}

{}Esta estrategia se basa en encontrar puntos en qu\'e los precios de las acciones cruzan su media m\'ovil\index{media m\'ovil}. Uno puede usar diferentes tipos de medias m\'oviles\index{media m\'ovil} (MAs, por sus siglas en ingl\'es), como una media m\'ovil simple (SMA, por sus siglas en ingl\'es)\index{media m\'ovil simple (SMA)}, o una media m\'ovil exponencial (EMA, por sus siglas en ingl\'es)\index{media m\'ovil exponencial (EMA)}:\footnote{\, Para $T\gg 1$ tenemos $\lambda^T\ll 1$ y $\mbox{EMA}(T, \lambda) \approx (1-\lambda)~P(1) + \lambda~\mbox{EMA}(T - 1, \lambda)$, en donde $\mbox{EMA}(T - 1, \lambda)$ est\'a basada en $P(2),P(3),\dots, P(T)$. Adem\'as, por algunas publicaciones sobre estrategias basadas en medias m\'oviles\index{media m\'ovil}, v\'ease, por ejemplo, \cite{BenZion2003}, \cite{Brock1992}, \cite{Dzikevicius2010}, \cite{Edwards1992}, \cite{Faber2007}, \cite{Felix2008}, \cite{Fifield2008}, \cite{Fong2005}, \cite{Gencay1996}, \cite{Gencay1998}, \cite{GencayStengos1998}, \cite{Glabadanidis2015}, \cite{Gunasekarage2001}, \cite{Hung2016}, \cite{James1968}, \cite{Jasemi2012}, \cite{Kilgallen2012}, \cite{Li2015}, \cite{Lo2000}, \cite{Metghalchi2012}, \cite{Patari2014}, \cite{Taylor1992}, \cite{Weller2009}, \cite{Zakamulin2014a}, \cite{Zakamulin2015}.}
\begin{eqnarray}
&& \mbox{SMA}(T) = \frac{1}{T}\sum_{t = 1}^{T} P(t) \\
&& \mbox{EMA}(T, \lambda) = {\sum_{t=1}^T \lambda^{t-1}~P(t) \over \sum_{t=1}^T \lambda^{t-1}} = {{1-\lambda}\over{1-\lambda^T}}~\sum_{t=1}^T \lambda^{t-1}~P(t)\
\end{eqnarray}
Aqu\'i: $t=1$ corresponde al tiempo m\'as reciente de la serie de tiempo\index{serie de tiempo} de los precios hist\'oricos $P(t)$ de la acci\'on\index{precio hist\'orico de la acci\'on}; $T$ es la longitud de la MA ($t$ y $T$ son usualmente medidos en d\'ias de trading\index{dias de trading @ d\'ias de trading}); y $\lambda < 1$ es el factor que suprime las contribuciones de datos pasados. Abajo MA se referir\'a a SMA\index{media m\'ovil simple (SMA)} o EMA\index{media m\'ovil exponencial (EMA)}. Una estrategia simple se define de la siguiente forma ($P$ es el precio en $t=0$, en el d\'ia de trading\index{dias de trading @ d\'ias de trading} inmediatamente despu\'es del \'ultimo d\'ia de trading\index{dias de trading @ d\'ias de trading} $t=1$ en la serie de tiempo\index{serie de tiempo} $P(t)$):
\begin{eqnarray}\label{MA1}
\mbox{Se\~{n}al} = \begin{cases}
     \mbox{Establecer posici\'on larga/liquidar posici\'on corta}    \mbox{ si } P > \mbox{MA}(T)	\\
     \mbox{Establecer posici\'on corta/liquidar posici\'on larga}    \mbox{ si } P < \mbox{MA}(T)
\end{cases}
\end{eqnarray}
Esta estrategia puede ser ejecutada tomando solo posiciones largas\index{solo posiciones largas}, solo posiciones cortas, o ambas, largas y cortas. Se puede aplicar f\'acilmente a m\'ultiples acciones (siguiendo la l\'ogica de acciones individuales, sin interacci\'on transversal entre las se\~{n}ales\index{senzal @ se\~{n}al} para acciones individuales). Si se utiliza un gran n\'umero de acciones, puede ser posible construir portafolios d\'olar-neutral\index{portafolio dolar-neutral @ portafolio d\'olar-neutral} (o cercanos a d\'olar-neutral).

\subsection{Estrategia: Dos medias m\'oviles\index{media m\'ovil}}\label{sub.2.mov.avg}

{}La variante m\'as simple de esta estrategia reemplaza al precio de las acciones $P$ en la Ecuaci\'on (\ref{MA1}) por otra media m\'ovil\index{media m\'ovil}. Es decir, ahora tenemos 2 medias m\'oviles\index{media m\'ovil} con longitudes $T^\prime$ y $T$, en donde $T^\prime < T$ (por ejemplo, $T^\prime = 10$ y $T=30$), y la se\~{n}al\index{senzal @ se\~{n}al} est\'a dada por:
\begin{eqnarray}
\mbox{Se\~{n}al} = \begin{cases}
     \mbox{Establecer/liquidar posiciones largas/cortas}    \mbox{ si } \mbox{MA}(T^\prime) > \mbox{MA}(T)	\\
     \mbox{Establecer/liquidar posiciones cortas/largas}    \mbox{ si } \mbox{MA}(T^\prime) < \mbox{MA}(T)
\end{cases}
\end{eqnarray}
La se\~{n}al\index{senzal @ se\~{n}al} se puede refinar, por ejemplo, incorporando reglas de ``stop-loss''\index{regla de stop-loss} para proteger las ganancias realizadas\index{ganancia realizada}. Por ejemplo, si se ha establecido una posici\'on larga, el trader puede definir un umbral para liquidar dicha posici\'on si la acci\'on comienza a caer (incluso si la media m\'ovil\index{media m\'ovil} m\'as corta no ha cruzado a la media m\'ovil\index{media m\'ovil} m\'as larga a\'un):
\begin{eqnarray}
\mbox{Se\~{n}al} = \begin{cases}
     \mbox{Establecer posiciones largas}    \mbox{ si } \mbox{MA}(T^\prime) > \mbox{MA}(T)	\\
     \mbox{Liquidar posiciones largas}    \mbox{ si } P < (1 - \Delta) \times P_1\\
     \mbox{Establecer posiciones cortas}    \mbox{ si } \mbox{MA}(T^\prime) < \mbox{MA}(T)	\\
     \mbox{Liquidar posiciones cortas}    \mbox{ si } P > (1 + \Delta) \times P_1
\end{cases}
\end{eqnarray}
Aqu\'i $\Delta$ es un porcentaje predefinido, por ejemplo, $\Delta = 2\%$. Por lo tanto, una posici\'on larga se liquida si el precio actual $P$ cae m\'as de 2\% debajo del precio del d\'ia anterior $P_1$; y una posici\'on corta se liquida si $P$ sube 2\% por encima de $P_1$. Otras variaciones pueden ser usadas.

\subsection{Estrategia: Tres medias m\'oviles\index{media m\'ovil}}

{}En algunos casos, utilizar 3 medias m\'oviles\index{media m\'ovil} con longitudes $T_1 < T_2 < T_3$ (por ejemplo, $T_1 = 3$, $T_2 = 10$, $T_3 = 21$) puede ayudar a filtrar se\~{n}ales falsas\index{senzal falsa @ se\~{n}al falsa}:
\begin{eqnarray}
\mbox{Se\~{n}al} = \begin{cases}
     \mbox{Establecer posici\'on larga}    \mbox{ si } \mbox{MA}(T_1) > \mbox{MA}(T_2) > \mbox{MA}(T_3)	\\
     \mbox{Liquidar posici\'on larga}    \mbox{ si } \mbox{MA}(T_1) \leq \mbox{MA}(T_2)\\
     \mbox{Establecer posici\'on corta}    \mbox{ si } \mbox{MA}(T_1) < \mbox{MA}(T_2) < \mbox{MA}(T_3)	\\
     \mbox{Liquidar posici\'on corta}    \mbox{ si } \mbox{MA}(T_1) \geq \mbox{MA}(T_2)
\end{cases}
\end{eqnarray}

\subsection{Estrategia: Soporte\index{soporte} y resistencia\index{resistencia}}

{}Esta estrategia utiliza los niveles de ``soporte'' $S$ y ``resistencia'' $R$\index{nivel, de resistencia}\index{nivel, de soporte}, que se pueden calcular utilizando el ``punto de pivote\index{punto de pivote}'' (conocido como el ``centro'') $C$ de la siguiente forma:\footnote{\, Existen otras definiciones del punto de pivote\index{punto de pivote} (por ejemplo, utilizando el precio de apertura\index{precio de apertura} del d\'ia de trading actual) y un nivel mayor/menor de soporte/resistencia\index{nivel, de resistencia}\index{nivel, de soporte}. Por algunas publicaciones sobre estrategias de soporte y resistencia\index{estrategia, resistencia}\index{estrategia, soporte}, v\'ease, por ejemplo, \cite{Amiri2010}, \cite{Brock1992}, \cite{Garzarelli2014}, \cite{Hardy1978}, \cite{Kahneman1979}, \cite{Murphy1986}, \cite{Osler2000}, \cite{Osler2003}, \cite{Person2007}, \cite{Pring1985}, \cite{Shiu2011}, \cite{Thomsett2003}, \cite{Zapranis2012}.}
\begin{eqnarray}
&& C = \frac{P_{MA} + P_{MI} + P_C}{3} \\
&& R = 2\times C - P_{MI} \\
&& S = 2\times C - P_{MA} \
\end{eqnarray}
Aqu\'i $P_{MA}$, $P_{MI}$ y $P_C$ son el precio m\'aximo, m\'inimo\index{precio, m\'aximo}\index{precio, m\'inimo} y de cierre\index{precio de cierre} del d\'ia anterior. Una forma de definir una se\~{n}al de trading\index{senzal de trading @ se\~{n}al de trading} es la siguiente (tal como arriba, $P$ es el precio actual):
\begin{eqnarray}
\mbox{Se\~{n}al} = \begin{cases}
      \mbox{Establecer posici\'on larga}    \mbox{ si } P > C \\
      \mbox{Liquidar posici\'on larga}    \mbox{ si } P \geq R \\
      \mbox{Establecer posici\'on corta}   \mbox{ si } P < C \\
      \mbox{Liquidar posici\'on corta}    \mbox{ si } P \leq S\
\end{cases}
\end{eqnarray}

\subsection{Estrategia: Canal\index{canal}}

{}Esta estrategia consiste en comprar y vender una acci\'on cuando alcanza el piso y el techo de un canal\index{canal}, respectivamente. Un canal\index{canal} es un rango/banda, limitado por un techo y un piso, dentro de los cuales el precio de las acciones fluct\'ua. La expectativa del trader es que, si se alcanza el piso o el techo, el precio de las acciones rebotar\'a en la direcci\'on opuesta. Por otro lado, si el precio de las acciones rompe el techo o el piso, el trader puede concluir que ha surgido una nueva tendencia\index{tendencia} y, entonces, sigue esta nueva tendencia\index{tendencia}. Una definici\'on simple y com\'un de un canal\index{canal} es el canal de Donchian\index{canal de Donchian} \cite{Donchian1960}, en donde el techo $B_{techo}$ y el piso $B_{piso}$ se definen de la siguiente forma (con las mismas notaciones que arriba):\footnote{\, Para obtener m\'as literatura sobre estrategias de trading con canales\index{estrategia de trading con canal}, v\'ease, por ejemplo, \cite{Batten1996}, \cite{Birari2014}, \cite{Dempster2002}, \cite{DeZwart2009}, \cite{Elder2014}, \cite{Sullivan1999}.}
\begin{eqnarray}
&& B_{techo} = \mbox{max}(P(1), P(2), \dots, P(T)) \\
&& B_{piso} = \mbox{min}(P(1), P(2), \dots, P(T)) \
\end{eqnarray}
Luego, una estrategia de trading\index{estrategia de trading} simple se define de la siguiente forma:
\begin{eqnarray}
\mbox{Se\~{n}al} = \begin{cases}
     \mbox{Establecer posici\'on larga/liquidar posici\'on corta}    \mbox{ si } P = B_{piso}	\\
     \mbox{Establecer posici\'on corta/liquidar posici\'on larga}    \mbox{ si } P = B_{techo}
\end{cases}
\end{eqnarray}
Cuanto m\'as ancho sea el canal\index{canal}, mayor es la volatilidad. Por lo general, un indicador de canal\index{indicador de canal} se utiliza junto con otros indicadores. Por ejemplo, la se\~{n}al\index{senzal @ se\~{n}al} puede ser m\'as robusta cuando un cambio en la direcci\'on del precio\index{cambio en la direcci\'on del precio} (o un quiebre del canal\index{quiebre del canal}) ocurre con un aumento en el volumen comercializado\index{volumen comercializado}.

\subsection{Estrategia: Eventos -- Fusiones y Adquisiciones\index{fusiones y adquisiciones (M\&A)}}

{}Esta estrategia, conocida como ``arbitraje de fusiones\index{arbitraje de fusiones}'' o ``arbitraje de riesgo\index{arbitraje de riesgo}'', intenta capturar los  excesos de retornos\index{exceso de retorno} generados por actividades corporativas\index{actividades corporativas} tales como fusiones y adquisiciones (M\&A, por sus siglas en ingl\'es)\index{fusiones y adquisiciones (M\&A)}. Una oportunidad de arbitraje de fusi\'on\index{oportunidad de arbitraje de fusi\'on} surge cuando una compa\~{n}\'ia que cotiza en la bolsa\index{compa\~{n}\'ia que cotiza en la bolsa} pretende adquirir otra compa\~{n}\'ia que tambi\'en cotiza en la bolsa\index{compa\~{n}\'ia que cotiza en la bolsa} a un precio que difiere del precio de mercado\index{precio de mercado} de esta \'ultima. En este sentido, existen dos tipos principales de transacciones: fusiones en efectivo\index{fusi\'on en efectivo} y fusiones por acciones\index{fusi\'on de acciones}. En el caso de una fusi\'on en efectivo\index{fusi\'on en efectivo}, el trader establece una posici\'on larga en las acciones de la empresa objetivo\index{empresa objetivo}. En el caso de una fusi\'on por acciones\index{fusi\'on de acciones}, el trader establece una posici\'on larga en las acciones de la empresa objetivo\index{empresa objetivo} (ll\'amese A) y una posici\'on corta en las acciones de la compa\~{n}\'ia adquirente\index{compa\~{n}\'ia adquirente} (ll\'amese B). Por ejemplo, si el precio actual de A es \$67, el precio actual de B es \$35, y bajo el acuerdo de fusi\'on de acciones\index{fusi\'on de acciones} propuesto cada acci\'on\index{acci\'on} de A es intercambiada por 2 acciones\index{acci\'on} de B, entonces el trader compra 1 acci\'on\index{acci\'on} de A y vende 2 acciones\index{acci\'on} de B generando un cr\'edito neto inicial de $\$3 = 2 \times \$35 - \$67$, que es el beneficio por cada acci\'on\index{acci\'on} de A comprada si el trato se cierra. El riesgo\index{riesgo} del trader est\'a en que, si el trato no se cierra, entonces de forma muy probable perder\'a dinero con este trade.\footnote{\, Por algunas publicaciones sobre arbitraje de fusiones\index{arbitraje de fusiones}, v\'ease, por ejemplo, \cite{Andrade2001}, \cite{Andries2017}, \cite{Baker2012}, \cite{Baker2002}, \cite{Bester2017}, \cite{Brown1986}, \cite{Cao2016}, \cite{Cornelli2002}, \cite{Dukes1992}, \cite{Hall2013}, \cite{Harford2005}, \cite{Hsieh2005}, \cite{Huston2000}, \cite{Jetley2010}, \cite{Karolyi1999}, \cite{Khan2002}, \cite{Larker1987}, \cite{Lin2013}, \cite{Maheswaran2005}, \cite{Mitchell2001}, \cite{Officer2004}, \cite{Officer2006}, \cite{Samuelson1986}, \cite{Subramanian2004}, \cite{VanTassel2016}, \cite{Walkling1985}.}

\subsection{Estrategia: Aprendizaje autom\'atico\index{aprendizaje autom\'atico} -- acci\'on individual con KNN\index{acci\'on individual con KNN}}\label{MLKNN}

{}Algunas estrategias se basan en t\'ecnicas de aprendizaje autom\'atico\index{tecnicas de aprendizaje automatico @ t\'ecnicas de aprendizaje autom\'atico}, tales como el algoritmo de los k vecinos m\'as cercanos (KNN, por sus siglas en ingl\'es)\index{algoritmo de los k vecinos m\'as cercanos (KNN)} (v\'ease, por ejemplo, \cite{Altman1992}, \cite{Samworth2012}), para predecir los retornos futuros de las acciones (la variable objetivo\index{variable objetivo}) en funci\'on de un conjunto de variables predictivas (caracter\'isticas)\index{variable, caracter\'istica}\index{variable, predictor}, que pueden basarse en datos t\'ecnicos, fundamentales y/o en otros tipos de datos\index{datos, fundamental}\index{datos, t\'ecnico}. La estrategia que describimos aqu\'i es una estrategia de una sola acci\'on\index{estrategia de una sola acci\'on}, es decir, para cada acci\'on la variable objetivo\index{variable objetivo} se predice utilizando los datos de precio y volumen\index{datos de precio y volumen} solo de esta acci\'on (pero no datos de corte transversal, es decir, no hay datos de otras acciones). La variable objetivo\index{variable objetivo} $Y(t)$ se define como el retorno acumulado\index{retorno acumulado} sobre los pr\'oximos $T$ d\'ias de trading\index{dias de trading @ d\'ias de trading} (como arriba, los valores enteros ascendentes de $t$, que se mide en d\'ias de trading\index{dias de trading @ d\'ias de trading}, corresponden a retroceder en el tiempo):
\begin{equation}
 Y(t) = {P(t - T)\over P(t)} - 1
\end{equation}
Las variables predictivas\index{variable predictiva} $X_a(t)$, $a=1,\dots,m$, se definen utilizando precio $P(t^\prime)$ y volumen\index{volumen} $V(t^\prime)$ en los tiempos $t^\prime$ antes de $t$ (es decir, $t^\prime > t$), por lo tanto, estos son datos ``fuera de la muestra''. Ejemplos de tales variables son medias m\'oviles\index{media m\'ovil} del precio y volumen\index{volumen} con distintas longitudes:
\begin{eqnarray}
 && X_1(t) = \frac{1}{T_1}~\sum_{s = 1}^{T_1} V(t + s) \\
 && X_2(t) = \frac{1}{T_2}~\sum_{s = 1}^{T_2} P(t + s) \\
 && X_3(t) = \frac{1}{T_3}~\sum_{s = 1}^{T_3} P(t + s) \\
 &&\dots
\end{eqnarray}
Adem\'as, las variables predictivas\index{variable predictiva} se normalizan entre 0 y 1:
\begin{equation}
 {\widetilde X}_a(t) =\frac{X_a(t) - X^-_a}{X_a^+ - X_a^-}
\end{equation}
en donde $X_a^+$ y $X_a^-$ son los valores m\'aximos y m\'inimos de $X_a(t)$ durante el per\'iodo de entrenamiento\index{periodo de entrenamiento @ per\'iodo de entrenamiento}. El ingrediente final es el n\'umero $k$ de los vecinos m\'as cercanos (v\'ease abajo). Para un valor dado de $t$ podemos tomar $k$ vecinos m\'as cercanos del $m$-vector ${\widetilde X}_a(t)$ entre los $m$-vectores ${\widetilde X}_a(t^\prime)$, $t^\prime = t +1, t+2,\dots,t+T_*$, utilizando el algoritmo KNN\index{algoritmo de los k vecinos m\'as cercanos (KNN)} (aqu\'i $T_*$ es el tama\~{n}o de la muestra). Para ejecutar el KNN\index{algoritmo de los k vecinos m\'as cercanos (KNN)} podemos usar la distancia euclidiana\index{distancia euclidiana} $D(t,t^\prime)$ entre ${\widetilde X}_a(t)$ y ${\widetilde X}_a(t^\prime)$ definida como
\begin{equation}
 [D(t,t^\prime)]^2 = \sum_{a=1}^m ({\widetilde X}_a(t) - {\widetilde X}_a(t^\prime))^2
\end{equation}
Sin embargo, podemos usar otra medida de distancia (por ejemplo, la distancia de Manhattan\index{distancia de Manhattan}). Asumamos que los $k$ vecinos m\'as cercanos de ${\widetilde X}_a(t)$ son ${\widetilde X}_a(t^\prime_\alpha(t))$, $\alpha=1,\dots,k$. (Tenga en cuenta que los $k$ valores $t^\prime_\alpha(t)$ dependen de $t$.) Luego podemos {\em definir} el valor {\em predicho}\index{valor predicho} ${\cal Y}(t)$ simplemente como el promedio de los valores realizados correspondientes $Y(t^\prime_\alpha(t))$:
\begin{equation}\label{y.mean}
 {\cal Y}(t) = {1\over k}~\sum_{\alpha=1}^k Y(t^\prime_\alpha(t))
\end{equation}
Alternativamente, podemos, por ejemplo, considerar un modelo lineal\index{modelo lineal}
\begin{equation}\label{y.reg}
 {\cal Y}(t) = \sum_{\alpha=1}^k Y(t^\prime_\alpha(t))~w_\alpha + v
\end{equation}
y estimar los coeficientes $w_\alpha$ y $v$ corriendo una regresi\'on\index{regresi\'on}\footnote{\, Podemos correr esta regresi\'on\index{regresi\'on} sin el intercepto\index{intercepto}, en cuyo caso solo tenemos los coeficientes $w_\alpha$, o con el intercepto\index{intercepto}, en cuyo caso, tambi\'en tenemos el coeficiente $v$.} de los valores realizados $Y(t)$ sobre $Y(t^\prime_\alpha(t))$ para alg\'un n\'umero -- ll\'amelo $M$ -- de valores de $t$. Es decir, colocamos $Y(t)$ para cada valor de $t$ en un $M$-vector y lo regresamos sobre la matriz con dimensiones $M \times k$ de los valores correspondientes $Y(t^\prime_\alpha(t))$. Los coeficientes de esta regresi\'on\index{regresi\'on} son $w_\alpha$ y $v$.

{}La ventaja de usar la Ecuaci\'on (\ref{y.mean}) es simplicidad -- no hay par\'ametros para entrenar en este caso. Sin embargo, todav\'ia tenemos que hacer el backtesting de la estrategia (v\'ease abajo) fuera de la muestra. La desventaja es que las contribuciones igualmente ponderadas de todos los $k$ vecinos m\'as cercanos podr\'ian ser sub\'optimas. En este sentido, hay varios esquemas de ponderaci\'on\index{esquema de ponderaci\'on} (por ejemplo, basados en la distancia) que podr\'ian ser considerados. Ponderaciones no triviales es precisamente lo que la Ecuaci\'on (\ref{y.reg}) intenta capturar. Sin embargo, esto requiere entrenamiento\index{entrenamiento} y validaci\'on cruzada\index{validaci\'on cruzada} (utilizando m\'etricas tales como la ra\'iz cuadrada del error cuadr\'atico medio\index{ra\'iz cuadrada del error cuadr\'atico medio (RMSE)}), y los par\'ametros ajustados $w_\alpha$ y $v$ pueden ser (y muchas veces son) inestables fuera de muestra. Los datos se pueden dividir en, por ejemplo, 60\% para entrenamiento\index{entrenamiento} y 40\% para validaci\'on cruzada\index{validaci\'on cruzada}. En \'ultima instancia, la estrategia debe tener un buen rendimiento fuera de la muestra.

{}La se\~{n}al\index{senzal @ se\~{n}al} en $t=0$ se puede definir usando el valor predicho\index{valor predicho} ${\cal Y} = {\cal Y}(0)$, el cual es el retorno esperado\index{retorno esperado} para los pr\'oximos $T$ d\'ias. Para el trading de una sola acci\'on\index{trading de una sola acci\'on}\footnote{\, Alternativamente, uno puede usar los retornos esperados\index{retorno esperado} ${\cal Y}_i$ computados para $N$ acciones (en donde $N \gg 1$) utilizando un algoritmo de aprendizaje autom\'atico\index{algoritmo de aprendizaje autom\'atico} como arriba y luego usar estos retornos esperados\index{retorno esperado} en estrategias de corte transversal\index{estrategia de corte transversal} con m\'ultiples activos tales como reversi\'on a la media\index{reversi\'on a la media}/arbitraje estad\'istico\index{arbitraje estad\'istico}.} uno puede simplemente definir umbrales para establecer\index{establecer} posiciones largas y cortas, y liquidar\index{liquidar} posiciones existentes, por ejemplo, de la siguiente forma:\footnote{\, Por algunas publicaciones sobre el uso de aprendizaje autom\'atico\index{aprendizaje autom\'atico} para predecir los retornos de las acciones, v\'ease, por ejemplo, \cite{Adam2001}, \cite{AngQuek2006}, \cite{Chen2014}, \cite{Chen2003}, \cite{Creamer2007}, \cite{Creamer2010a}, \cite{Gestel2001}, \cite{Grudnitski1993}, \cite{Huang2005}, \cite{Huang2009}, \cite{Huerta2013}, \cite{Kablan2009}, \cite{KakushadzeYu2016b}, \cite{KakushadzeYu2017c}, \cite{KakushadzeYu2018}, \cite{Kara2011}, \cite{Kim2003}, \cite{Kim2006}, \cite{Kim2000}, \cite{Kordos2011}, \cite{Kryzanowski1993}, \cite{Kumar2001}, \cite{Liew2018}, \cite{Lu2009}, \cite{Milosevic2016}, \cite{Novak2016}, \cite{Ou2009}, \cite{Refenes1994}, \cite{RodriguezGonzalez2011}, \cite{Saad1998}, \cite{Schumaker2010}, \cite{Subha2012}, \cite{Tay2001a}, \cite{Teixeira2010}, \cite{Tsai2010}, \cite{Vanstone2009}, \cite{Yao2000}, \cite{Yao1999}, \cite{Yu2005}. \label{fn.KNN.ML}}
\begin{eqnarray}
\mbox{Se\~{n}al} = \begin{cases}
       \mbox{Establecer posici\'on larga}    \mbox{ si } {\cal Y} > z_1 \\
        \mbox{Liquidar posici\'on larga}    \mbox{ si } {\cal Y} \leq z_2 \\
        \mbox{Establecer posici\'on corta}    \mbox{ si } {\cal Y} < -z_1 \\
        \mbox{Liquidar posici\'on corta}    \mbox{ si } {\cal Y} \geq -z_2
        \end{cases}
\end{eqnarray}
Aqu\'i, $z_1$ y $z_2$ son los umbrales definidos por el trader. Esta se\~{n}al\index{senzal @ se\~{n}al} debe ser probada fuera de muestra. El n\'umero $k$ de vecinos m\'as cercanos se puede optimizar utilizando un backtest\index{backtest} (probando un grupo de valores de $k$). Alternativamente, uno puede usar una heur\'istica com\'un, por ejemplo, $k = \mbox{floor}(\sqrt{T_*})$ o $k = \mbox{ceiling}(\sqrt{T_*})$. V\'ease tambi\'en, por ejemplo, \cite{Hall2008}.

\subsection{Estrategia: Arbitraje estad\'istico\index{arbitraje estad\'istico} -- Optimizaci\'on\index{optimizaci\'on}}\label{sub.opt}

{}Sea $C_{ij}$ la matriz de covarianza muestral\index{matriz de covarianza, muestral} para las $N$ acciones en un portafolio\index{portafolio}.\footnote{\, La matriz de covarianza muestral\index{matriz de covarianza muestral} basada en las series de tiempo\index{serie de tiempo} de los retornos hist\'oricos\index{retornos hist\'oricos} es singular si $T\leq N + 1$, en donde $T$ es el n\'umero de observaciones en la serie de tiempo\index{serie de tiempo}. Incluso si no es singular, a no ser que $T \gg N$, lo que es muy raro, los elementos fuera de la diagonal de la matriz de covarianza muestral\index{matriz de covarianza muestral} generalmente son inestables fuera de muestra. Por lo tanto, en la pr\'actica, t\'ipicamente una matriz de covarianza del modelo\index{matriz de covarianza del modelo} (que es positiva-definida y deber\'ia ser estable fuera de la muestra) es usada (v\'ease abajo).} Sea $D_i$ las tenencias en d\'olares\index{tenencia en d\'olares} en el portafolio\index{portafolio}. El P\&L\index{P\&L del portafolio} $P$, la volatilidad $V$ y el ratio de Sharpe\index{ratio de Sharpe} $S$ del portafolio {\em esperados} son dados por
\begin{eqnarray}
 &&P = \sum_{i=1}^N E_i~D_i\\
 &&V^2 = \sum_{i,j=1}^N C_{ij}~D_i~D_j\\
 &&S = {P / V}
\end{eqnarray}
Aqu\'i $E_i$ son los retornos esperados de las acciones\index{retornos esperados de las acciones}. En lugar de las tenencias en d\'olares\index{tenencia en d\'olares} $D_i$, es m\'as conveniente trabajar con ponderaciones de tenencias\index{ponderaciones de tenencias} sin dimensiones (que son positivas/negativas para posiciones largas/cortas)
\begin{equation}
 w_i = {D_i / I}
\end{equation}
en donde $I$ es el nivel de inversi\'on\index{nivel de inversi\'on} total. Las ponderaciones de tenencias\index{ponderaciones de tenencias} satisfacen la condici\'on
\begin{equation}\label{w.norm}
 \sum_{i=1}^N \left|w_i\right| = 1
\end{equation}
Tenemos $P = I\times {\widetilde P}$, $V = I\times {\widetilde V}$ y $S = {\widetilde P}/{\widetilde V}$, en donde
\begin{eqnarray}
 &&{\widetilde P} = \sum_{i=1}^N E_i~w_i\\
 &&{\widetilde V}^2 = \sum_{i,j=1}^N C_{ij}~w_i~w_j
\end{eqnarray}
Para determinar las ponderaciones del portafolio\index{ponderaciones del portafolio} $w_i$, a menudo se requiere que el ratio de Sharpe\index{ratio de Sharpe} \cite{Sharpe1966}, \cite{Sharpe1994} sea maximizado:
\begin{equation}\label{max.sharpe}
 S \rightarrow \mbox{max}
\end{equation}
Suponiendo que no hay condiciones adicionales en $w_i$ (por ejemplo, l\'imites superiores o inferiores), la soluci\'on a la Ecuaci\'on (\ref{max.sharpe}) en ausencia de costos transaccionales\index{costos transaccionales} es dada por
\begin{equation}\label{w.max.sharpe}
 w_i = \gamma~ \sum_{j=1}^N C^{-1}_{ij}E_j
\end{equation}
en donde $C^{-1}$ es la inversa\index{inversa} de $C$, y el coeficiente de normalizaci\'on $\gamma$ es determinado por la Ecuaci\'on (\ref{w.norm}) (y $\gamma>0$ entonces ${\widetilde P}>0$). Las ponderaciones dadas por la Ecuaci\'on (\ref{w.max.sharpe}) gen\'ericamente no corresponden a un portafolio d\'olar-neutral\index{portafolio dolar-neutral @ portafolio d\'olar-neutral}. Para tener un portafolio d\'olar-neutral\index{portafolio dolar-neutral @ portafolio d\'olar-neutral}, necesitamos maximizar el ratio de Sharpe\index{ratio de Sharpe} sujeto a la restricci\'on de d\'olar-neutralidad\index{restricci\'on de d\'olar-neutralidad}.

\subsubsection{D\'olar-neutralidad\index{dolar-neutralidad @ d\'olar-neutralidad}}

{}Podemos lograr la d\'olar-neutralidad\index{dolar-neutralidad @ d\'olar-neutralidad} de la siguiente forma. En ausencia de l\'imites, costos transaccionales\index{costos transaccionales}, etc., el ratio de Sharpe\index{ratio de Sharpe} es invariante bajo simult\'aneas reescalaciones de todas las ponderaciones de tenencia\index{ponderaciones de tenencias} $w_i\rightarrow \zeta~w_i$, en donde $\zeta >0$. Debido a esta invariancia de escala\index{invariancia de escala}, el problema de la maximizaci\'on del ratio de Sharpe\index{maximizaci\'on del ratio de Sharpe} se puede reformular como un problema de minimizaci\'on de una funci\'on objetivo\index{funci\'on objetivo} cuadr\'atica:
\begin{eqnarray}
 &&g(w, \lambda) = {\lambda\over 2} \sum_{i,j=1}^N C_{ij}~w_i~ w_j - \sum_{i=1}^N E_i~w_i\\
 &&g(w, \lambda)\rightarrow\mbox{min}
\end{eqnarray}
en donde $\lambda > 0$ es un par\'ametro, y la minimizaci\'on es con respecto a $w_i$. La soluci\'on est\'a dada por
\begin{equation}
 w_i = {1\over\lambda}~\sum_{j=1}^N C^{-1}_{ij} E_j
\end{equation}
y $\lambda$ se fija a trav\'es de la Ecuaci\'on (\ref{w.norm}). El enfoque de la funci\'on objetivo\index{funci\'on objetivo} -- que es la optimizaci\'on de media-varianza\index{optimizaci\'on de media-varianza} \cite{Markowitz1952} -- es conveniente si queremos imponer restricciones {\em lineales homog\'eneas}\index{restricciones lineales homog\'eneas} (las cuales no atentan contra la invariancia de escala\index{invariancia de escala} antes mencionada) sobre $w_i$, por ejemplo, la restricci\'on de d\'olar-neutralidad\index{restricci\'on de d\'olar-neutralidad}. Introducimos el multiplicador de Lagrange\index{multiplicador de Lagrange} $\mu$:\footnote{\, Introduciendo m\'ultiples multiplicadores de Lagrange\index{multiplicador de Lagrange}, podemos tener m\'ultiples restricciones homog\'eneas lineales\index{restricciones lineales homog\'eneas} (v\'ease, por ejemplo,\cite{Kakushadze2015b}).}
\begin{eqnarray}\label{obj.fun.dn}
 &&g(w, \mu, \lambda) = {\lambda\over 2} \sum_{i,j=1}^N C_{ij}~w_i~ w_j - \sum_{i=1}^N E_i~w_i - \mu~\sum_{i=1}^N w_i\\
 &&g(w, \mu, \lambda)\rightarrow\mbox{min}
\end{eqnarray}
La minimizaci\'on con respecto a $w_i$ y $\mu$ ahora conduce a las siguientes ecuaciones:
\begin{eqnarray}\label{opt.w.7}
 &&\lambda~\sum_{j=1}^N C_{ij}~w_j = E_i + \mu\\
 &&\sum_{i=1}^N w_i = 0\label{opt.mu.7}
\end{eqnarray}
Entonces tenemos d\'olar-neutralidad\index{dolar-neutralidad @ d\'olar-neutralidad}. La soluci\'on a las Ecuaciones (\ref{opt.w.7}) y (\ref{opt.mu.7}) est\'a dada por:
\begin{equation}
 w_i = {1\over\lambda}\left[\sum_{j = 1}^N C^{-1}_{ij} E_j - \sum_{j=1}^N C^{-1}_{ij}~{{\sum_{k,l=1}^N C^{-1}_{kl} E_l}\over{\sum_{k,l = 1}^N C^{-1}_{kl}}}\right]
 \label{w.opt.const}
\end{equation}
Por construcci\'on, $w_i$ satisface la restricci\'on de d\'olar-neutralidad\index{restricci\'on de d\'olar-neutralidad} (\ref{opt.mu.7}), y $\lambda$ se fija a trav\'es de la Ecuaci\'on (\ref{w.norm}). Los retornos esperados\index{retorno esperado} $E_i$ pueden basarse en reversi\'on a la media, momentum, aprendizaje autom\'atico u otras se\~{n}ales\index{senzal, de aprendizaje automatico @ se\~{n}al, de aprendizaje autom\'atico}\index{senzal, reversion a la media @ se\~{n}al, reversi\'on a la media}\index{senzal, de momentum @ se\~{n}al, de momentum}. La Ecuaci\'on (\ref{w.opt.const}) construye un portafolio d\'olar-neutral\index{portafolio dolar-neutral @ portafolio d\'olar-neutral} con ``gesti\'on de riesgos\index{gesti\'on de riesgo}''. Por ejemplo, las ponderaciones $w_i$ (aproximadamente) son suprimidas por las volatilidades de las acciones $\sigma_i$ (en donde $\sigma_i^2 = C_{ii}$) asumiendo que en promedio $|E_i|$ son de orden $\sigma_i$.\footnote{\, Generalmente, $C_{ij}$ es una matriz de covarianza del modelo\index{matriz de covarianza del modelo} multifactorial de riesgo\index{modelo multifactorial de riesgo}. Para una discusi\'on general, v\'ease, por ejemplo, \cite{Grinold2000}. Para implementaciones expl\'icitas (incluyendo el c\'odigo fuente\index{codigo fuente @ c\'odigo fuente}), v\'ease, por ejemplo, \cite{Kakushadze2015e}, \cite{KakushadzeYu2016a}, \cite{KakushadzeYu2017a}. Para modelos multifactoriales\index{modelo multifactorial}, las ponderaciones son {\em aproximadamente} neutrales con respecto a las columnas de la matriz de cargas factoriales\index{matriz de cargas factorial}. La neutralidad exacta se alcanza en el l\'imite de riesgo espec\'ifico\index{riesgo espec\'ifico} cero, en donde la optimizaci\'on\index{optimizaci\'on} se reduce a una regresi\'on ponderada\index{regresi\'on ponderada} (v\'ease, por ejemplo, \cite{Kakushadze2015b}).}

{}La implementaci\'on anterior de la restricci\'on de d\'olar-neutralidad\index{restricci\'on de d\'olar-neutralidad} a trav\'es de la minimizaci\'on de la funci\'on objetivo\index{funci\'on objetivo} cuadr\'atica (\ref{obj.fun.dn}) es equivalente a imponer esta restricci\'on en la maximizaci\'on del ratio de Sharpe\index{maximizaci\'on del ratio de Sharpe} dado que costos transaccionales\index{costos transaccionales}, l\'imites de posici\'on/trading\index{limites, de posicion @ l\'imites, de posici\'on}\index{limites, de trading @ l\'imites, de trading}, restricciones no lineales/no homog\'eneas\index{restricciones, no homog\'eneas}\index{restricciones, no lineales}, etc., no est\'an presentes. M\'as generalmente, la maximizaci\'on del ratio de Sharpe\index{maximizaci\'on del ratio de Sharpe} no es equivalente a minimizar una funci\'on objetivo\index{funci\'on objetivo} cuadr\'atica (v\'ease, por ejemplo, \cite{Kakushadze2015b}), aunque en la pr\'actica este \'ultimo enfoque suele usarse.

\subsection{Estrategia: Market-making\index{market-making}}

{}Siendo muy simplista, esta estrategia consiste en capturar la diferencia entre el precio m\'aximo de compra (bid) y precio m\'inimo de venta (ask)\index{diferencia bid-ask} para una acci\'on dada. Esto se puede resumir (nuevamente, de manera simplista) de la siguiente manera:
\begin{eqnarray}
\mbox{Regla} = \begin{cases}
      \mbox{Comprar al bid}  \\
      \mbox{Vender al ask}
\end{cases}
\end{eqnarray}
En un mercado\index{mercado} en donde la mayor parte del flujo de \'ordenes es ``tonto''\index{flujo de ordenes, tonto @ flujo de \'ordenes, tonto} (o desinformado), esta estrategia en promedio funcionar\'ia muy bien. Sin embargo, en un mercado\index{mercado} en donde la mayor parte del flujo de \'ordenes es ``inteligente'' (o informado, es decir, ``t\'oxico'')\index{flujo de ordenes, informado @ flujo de \'ordenes, informado}\index{flujo de ordenes, inteligente @ flujo de \'ordenes, inteligente}\index{flujo de ordenes, toxico @ flujo de \'ordenes, t\'oxico}, esta estrategia, tal como fue explicada, perder\'ia dinero. Esto se debe a la {\em selecci\'on adversa\index{selecci\'on adversa}}, en donde, precisamente porque la mayor parte del flujo de \'ordenes es inteligente\index{flujo de ordenes, inteligente @ flujo de \'ordenes, inteligente}, la mayor parte de las operaciones se completan al precio bid\index{bid} (ask\index{ask}) cuando el mercado\index{mercado} est\'a operando a hacia abajo (hacia arriba), por lo que estas operaciones perder\'ian dinero. Adem\'as, la mayor\'ia de las \'ordenes l\'imites\index{ordenes l\'imites @ \'ordenes l\'imites} a comprar (vender) al bid\index{bid} (ask\index{ask}) nunca se completar\'an dado que el precio se alejar\'a de estos valores, es decir, incrementar\'a (disminuir\'a). Por lo tanto, idealmente, esta estrategia deber\'ia estar estructurada de tal manera que capture el flujo de \'ordenes tonto\index{flujo de ordenes tonto @ flujo de \'ordenes tonto} y evite el flujo de \'ordenes inteligente\index{flujo de ordenes inteligente @ flujo de \'ordenes inteligente}, algo que no es tan simple.

{}Un enfoque es, en un momento dado, dentro de un horizonte\index{horizonte} de tiempo corto, estar en el lado ``correcto'' del mercado\index{mercado}, es decir, tener una se\~{n}al\index{senzal @ se\~{n}al} de horizonte corto indicando la direcci\'on del mercado\index{mercado} y colocar \'ordenes l\'imites\index{ordenes l\'imites @ \'ordenes l\'imites} en consecuencia (comprar al bid\index{bid} si la se\~{n}al\index{senzal @ se\~{n}al} indica un incremento en el precio, y vender al ask\index{ask} si la se\~{n}al\index{senzal @ se\~{n}al} indica una disminuci\'on en el precio). Si la se\~{n}al\index{senzal @ se\~{n}al} fuera (por arte de magia) 100\% correcta, esto capturar\'ia el flujo de \'ordenes tonto\index{flujo de ordenes tonto @ flujo de \'ordenes tonto} suponiendo que las \'ordenes\index{orden} se completan. Este es un gran supuesto dado que para que esto est\'e garantizado, el trader tendr\'ia que ser el \#1 en la cola entre muchos otros participantes del mercado colocando \'ordenes l\'imites\index{ordenes l\'imites @ \'ordenes l\'imites} al mismo precio. Aqu\'i es donde el trading de alta frecuencia\index{trading de alta frecuencia (HFT)} entra en escena -- b\'asicamente, se trata de la velocidad con la que se inician, cancelan y reemplazan las \'ordenes\index{orden, cancelada}\index{orden, cancelada y reemplazada}\index{orden, iniciada}. La infraestructura y la tecnolog\'ia son claves en esto.

{}Otra posibilidad es modular la se\~{n}al\index{senzal @ se\~{n}al} de corto plazo con otra se\~{n}al\index{senzal @ se\~{n}al} de largo plazo (que, de todas formas, puede ser otra se\~{n}al intrad\'ia\index{senzal intradia @ se\~{n}al intrad\'ia}). La se\~{n}al\index{senzal @ se\~{n}al} de largo plazo normalmente obtendr\'a m\'as ``centavos por acci\'on'' (CPS, por sus siglas en ingl\'es)\index{centavos por acci\'on}\footnote{\, ``Centavos por acci\'on''\index{centavos por acci\'on} es definido como el P\&L realizado\index{P\&L realizado} en centavos (y no en d\'olares) dividido por el total de acciones\index{acci\'on} negociadas (que incluye tanto el establecimiento como la liquidaci\'on de posiciones\index{establecimiento, de posici\'on}\index{liquidaci\'on, de posici\'on}). Tenga en cuenta que la se\~{n}al\index{senzal @ se\~{n}al} de largo plazo generalmente tiene un ratio de Sharpe\index{ratio de Sharpe} m\'as bajo que la se\~{n}al\index{senzal @ se\~{n}al} de corto plazo.} que la se\~{n}al\index{senzal @ se\~{n}al} de corto plazo. De esta forma, ahora ciertas operaciones pueden ser rentables incluso en presencia de selecci\'on adversa\index{selecci\'on adversa}, porque se establecen en base a la se\~{n}al\index{senzal @ se\~{n}al} de largo plazo. Es decir, ``pierden dinero'' en el corto plazo debido a selecci\'on adversa\index{selecci\'on adversa} (dado que el mercado\index{mercado} opera a trav\'es de las \'ordenes l\'imites\index{ordenes l\'imites @ \'ordenes l\'imites} correspondientes), pero generan dinero en el largo plazo. El aspecto de market-making\index{market-making} de esto es valioso mientras se establece una orden l\'imite pasiva\index{orden l\'imite pasiva} ya que a diferencia de una orden de mercado\index{orden de mercado} o una orden l\'imite agresiva\index{orden l\'imite, agresiva}, ahorra dinero. Por otro lado, en algunos casos, si la se\~{n}al\index{senzal @ se\~{n}al} de largo plazo es lo suficientemente fuerte y la se\~{n}al\index{senzal @ se\~{n}al} de corto plazo es en la misma direcci\'on, una orden l\'imite pasiva\index{orden l\'imite pasiva} probablemente no se completar\'ia y tendr\'ia m\'as sentido colocar una orden agresiva\index{orden agresiva}. Tal flujo de \'ordenes agresivo\index{flujo de ordenes agresivo @ flujo de \'ordenes agresivo} no es tonto sino inteligente, dado que est\'a basado en se\~{n}ales\index{senzal @ se\~{n}al} no triviales de corto y largo plazo con un retorno esperado\index{retorno esperado} positivo.\footnote{\, El flujo de \'ordenes tonto\index{flujo de ordenes tonto @ flujo de \'ordenes tonto} puede provenir, por ejemplo, de traders minoristas\index{trader minorista} desinformados. Tambi\'en puede provenir de los traders institucionales\index{trader institucional} con un horizonte ultra largo (fondos mutuos\index{fondo mutuo}, fondos de pensiones\index{fondo de pensi\'on}, etc.), cuyas perspectivas pueden ser de meses o a\~{n}os y no est\'an preocupados por la diferencia de unos pocos centavos en el precio de ejecuci\'on\index{precio de ejecuci\'on} en el corto plazo (es decir, esto es solo flujo de \'ordenes ``tonto de corto plazo''\index{flujo de ordenes tonto @ flujo de \'ordenes tonto}). Para una discusi\'on m\'as detallada, v\'ease, por ejemplo, \cite{Kakushadze2015d}, \cite{Lo2008}. Por algunas publicaciones sobre el trading de alta frecuencia\index{trading de alta frecuencia (HFT)} y market-making\index{market-making}, v\'ease, por ejemplo, \cite{Aldridge2013}, \cite{Anand2016}, \cite{Avellaneda2008}, \cite{Baron2014}, \cite{Benos2017}, \cite{Benos2016}, \cite{Biais2014}, \cite{BiaisMoinas2014}, \cite{Bowen2010}, \cite{Bozdog2011}, \cite{Brogaard2018}, \cite{Brogaard2015}, \cite{Brogaard2014},  \cite{Budish2015}, \cite{Carrion2013}, \cite{Carrion2017}, \cite{Easley2011}, \cite{Easley2012}, \cite{Egginton2016}, \cite{Hagstromer2013}, \cite{Hagstromer2014}, \cite{Harris2016}, \cite{Hasbrouck2013}, \cite{Hendershott2011}, \cite{Hendershott2013}, \cite{HendershottRiordan2013}, \cite{Hirschey2018}, \cite{Holden2014}, \cite{Jarrow2012}, \cite{Khandani2011}, \cite{Kirilenko2017}, \cite{Korajczyk2017}, \cite{Kozhan2012}, \cite{LiDeng2014}, \cite{Madhavan2012}, \cite{Menkveld2013}, \cite{Menkveld2016}, \cite{Muthuswamy2011}, \cite{OHara2015}, \cite{Pagnotta2012}, \cite{Riordan2012}, \cite{VanKervel2017}.} Y la velocidad a\'un sigue siendo importante.

\subsection{Estrategia: Combo alfa\index{combo alfa}}

{}Con los avances tecnol\'ogicos -- hardware cada vez m\'as barato y m\'as potente -- ahora a trav\'es de la miner\'ia de datos es posible extraer cientos de miles e incluso millones de alfas\index{alfa} utilizando m\'etodos de aprendizaje autom\'atico\index{metodos de aprendizaje automatico @ m\'etodos de aprendizaje autom\'atico}. Aqu\'i el t\'ermino ``alfa''\index{alfa} -- siguiendo la jerga com\'un del trader -- generalmente significa cualquier ``retorno esperado\index{retorno esperado} razonable'' que uno puede desear operar y no es necesariamente lo mismo que el alfa ``acad\'emico''\index{alfa acad\'emico}.\footnote{\, Por el alfa ``acad\'emico''\index{alfa acad\'emico} nos referimos al alfa de Jensen\index{alfa de Jensen} \cite{Jensen1968} o un \'indice similar de rendimiento.} En la pr\'actica, a menudo la informaci\'on sobre c\'omo se construyen los alfas\index{alfa} puede que ni siquiera est\'e disponible, por ejemplo, los \'unicos datos disponibles podr\'ian ser los datos de posiciones, entonces ``alfa''\index{alfa} luego es un conjunto de instrucciones para lograr ciertas tenencias de acciones (o alg\'un otro instrumento) en ciertos momentos $t_1,t_2,\dots$ Adem\'as, ``aprendizaje autom\'atico\index{aprendizaje autom\'atico}'' aqu\'i se refiere a m\'etodos sofisticados que van m\'as all\'a de los m\'etodos para una sola acci\'on\index{metodos para una sola accion @ m\'etodos para una sola acci\'on} demostrados en la Subsecci\'on \ref{MLKNN} e implican an\'alisis de corte transversal\index{an\'alisis de corte transversal} basado en precio-volumen\index{precio-volumen} as\'i como en otros tipos de datos (por ejemplo, capitalizaci\'on de mercado\index{capitalizaci\'on de mercado}, algunos otros datos fundamentales\index{datos fundamentales} tales como ganancias\index{ganancias}, datos de clasificaci\'on de la industria\index{datos de clasificaci\'on de la industria}, datos de sentimiento\index{datos de sentimiento}, etc.) para una gran cantidad de acciones (t\'ipicamente, algunas miles y m\'as). 101 ejemplos expl\'icitos de tales alfas de trading cuantitativos\index{alfas de trading cuantitativos} se exponen en \cite{Kakushadze2016}.\footnote{\, Este es un campo secreto, por lo que la literatura es muy escasa. V\'ease tambi\'en, por ejemplo, \cite{KakushadzeTulchinsky2016}, \cite{Tulchinsky2015}.} La otra cara es que estos alfas\index{alfa} ubicuos son d\'ebiles, ef\'imeros y no se pueden operar por s\'i solos, ya que cualquier ganancia se erosionar\'ia por los costos transaccionales\index{costos transaccionales}. Para mitigar esto, uno combina un gran n\'umero de tales alfas\index{alfa} y opera un ``mega-alfa''. Por lo tanto, estrategias de ``combo alfa''\index{estrategia de combo alfa}.

{}Esto no es cr\'itico, pero para mantenerlo simple, asumamos que todos los alfas\index{alfa} operan los mismos instrumentos subyacentes\index{instrumento subyacente}, para ser m\'as concretos, el mismo universo de (digamos, 2,500) las acciones m\'as l\'iquidas de los Estados Unidos\index{acciones m\'as l\'iquidas de los Estados Unidos}. Cada alfa\index{alfa} produce tenencias deseadas\index{tenencias deseadas} para este universo de trading\index{universo de trading}. Lo que necesitamos aqu\'i son las ponderaciones con las que combinamos los alfas\index{alfa} individuales, cuyo n\'umero $N$ puede ser muy grande (cientos de miles e incluso millones).\footnote{\, Tenga en cuenta que $N$ aqu\'i se refiere al n\'umero de alfas\index{alfa}, y no al n\'umero de acciones subyacentes\index{acci\'on subyacente}.} Aqu\'i se presenta un procedimiento para fijar las ponderaciones de los alfas\index{ponderaciones de los alfas} $w_i$, $i=1,\dots,N$ \cite{KakushadzeYu2017b} (v\'ease tambi\'en \cite{KakushadzeYu2018}):\\

{}$\bullet$ 1) Comenzar con las series de tiempo\index{serie de tiempo} de los retornos realizados de los {\em alfas}\index{retornos realizados de los alfas}\footnote{\, Aqu\'i $s=1,\dots,T = M+1$ etiqueta los tiempos $t_s$, en donde, tal como antes, $t_1$ corresponde al tiempo m\'as reciente (aunque la direcci\'on del tiempo no es crucial a continuaci\'on), y los retornos de los alfas\index{retornos de los alfas} son $R_{is} = R_i(t_s)$. Generalmente, los retornos de los alfas\index{retornos de los alfas} se computan de forma diaria, de cierre\index{cierre} a cierre\index{cierre}.} $R_{is}$, $i=1,\dots,N$, $s=1,\dots,M+1$.

{}$\bullet$ 2) Calcular los retornos netos de la media\index{retorno neto de la media} serial $X_{is} = R_{is} - {1\over {M+1}}\sum_{s=1}^{M+1} R_{is}$.

{}$\bullet$ 3) Calcular varianzas muestrales\index{varianza muestral} de los retornos de los alfas:\index{retornos de los alfas}\footnote{\, Su normalizaci\'on es irrelevante.}\\ $\sigma_i^2 = {1\over M}\sum_{s=1}^{M+1} X^2_{is}$.

{}$\bullet$ 4) Calcular los retornos netos de la media\index{retorno neto de la media} normalizados $Y_{is} = X_{is} /\sigma_i$.

{}$\bullet$ 5) Mantener solo las primeras $M$ columnas en $Y_{is}$: $s=1,\dots, M$.

{}$\bullet$ 6) Ajustar por la media de corte transversal $Y_{is}$: $\Lambda_{is} = Y_{is} - {1\over N}\sum_{j=1}^N Y_{js}$.

{}$\bullet$ 7) Mantener solo las primeras $M-1$ columnas en $\Lambda_{is}$: $s=1,\dots,M-1$.

{}$\bullet$ 8) Tomar los retornos esperados de los alfas\index{retornos esperados de los alfas} $E_i$ y normalizarlos: ${\widetilde E}_i = E_i/\sigma_i$. Una (pero por lejos no es el \'unica) forma de computar los retornos esperados de los alfas\index{retornos esperados de los alfas} es mediante medias m\'oviles\index{media m\'ovil} de $d$ d\'ias (tenga en cuenta que $d$ no tiene que ser lo mismo que $T$):
\begin{equation}
 E_i = {1\over d}~\sum_{s=1}^d R_{is}
\end{equation}

{}$\bullet$ 9) Calcular los residuos ${\widetilde \varepsilon}_i$ de la regresi\'on\index{regresi\'on} (sin el intercepto\index{intercepto} y con las ponderaciones unitarias) de ${\widetilde E}_i$ sobre $\Lambda_{is}$.

{}$\bullet$ 10) Establecer las ponderaciones del portafolio de alfas\index{ponderaciones del portafolio de alfas} como $w_i = \eta~ {\widetilde\varepsilon}_i/\sigma_i$.

{}$\bullet$ 11) Establecer el coeficiente de normalizaci\'on $\eta$ tal que $\sum_{i=1}^N \left|w_i\right| = 1$.

\subsection{Comentarios adicionales}

{}Finalizamos esta secci\'on con breves comentarios sobre algunas de las estrategias de trading de acciones\index{estrategia de trading de acciones} discutidas anteriormente. Primero, las estrategias de an\'alisis t\'ecnico de una sola acci\'on\index{estrategias de an\'alisis t\'ecnico de una sola acci\'on} (es decir, aquellas basadas \'unicamente en datos de acciones individuales en lugar de datos de corte transversal\index{datos, corte transversal}\index{datos, acci\'on individual}) tales como aquellas basadas en medias m\'oviles\index{media m\'ovil}, soporte\index{soporte} y resistencia\index{resistencia}, canal\index{canal} e incluso la estrategia de acci\'on individual con KNN\index{acci\'on individual con KNN}, son consideradas ``no cient\'ificas'' por muchos profesionales y acad\'emicos. A primera vista, ``fundamentalmente'' hablando (no debe confundirse con an\'alisis fundamental\index{an\'alisis fundamental}), no hay ninguna raz\'on por la cual, por ejemplo, una media m\'ovil\index{media m\'ovil} corta cruzando a una media m\'ovil\index{media m\'ovil} larga deber\'ia tener alg\'un poder predictivo.\footnote{\, {\em Arguendo}, el efecto momentum\index{efecto de momentum} parece proporcionar una base para tal poder de predicci\'on en algunos casos. Sin embargo, entonces se podr\'ia argumentar que, por ejemplo, estas son estrategias de momentum\index{estrategia de momentum} disfrazadas.}  Esto no quiere decir que las medias m\'oviles\index{media m\'ovil} son ``no cient\'ificas'' o que no deben ser utilizadas. Despu\'es de todo, por ejemplo, las estrategias de seguimiento de la tendencia/momentum\index{estrategia, momentum}\index{estrategia, seguimiento de la tendencia} est\'an basadas en medias m\'oviles\index{media m\'ovil}, es decir, los retornos esperados\index{retorno esperado} se calculan mediante medias m\'oviles\index{media m\'ovil}. Sin embargo, si se observa un gran n\'umero de acciones de forma transversal, se introduce un elemento {\em estad\'istico} en el juego. Estrategias de reversi\'on a la media\index{reversi\'on a la media} se espera que funcionen porque se espera que las acciones est\'en correlacionadas si pertenecen a la misma industria\index{industria}, etc. Esto se relaciona con el an\'alisis fundamental\index{an\'alisis fundamental} y -- m\'as importante -- con la {\em percepci\'on} de los traders acerca de c\'omo los precios/retornos de las acciones ``deber\'ian'' comportarse en base a los fundamentos\index{fundamentos} de las empresas. Sin embargo, aqu\'i tambi\'en es importante tener en cuenta que el mercado de valores\index{mercado de valores} -- una construcci\'on imperfecta creada por el hombre -- no se rige por las leyes de la naturaleza de la misma manera que, por ejemplo, el movimiento de los planetas en el sistema solar se rige por leyes fundamentales de la gravedad (v\'ease, por ejemplo, \cite{Kakushadze2015d}). Los mercados\index{mercado} se comportan como lo hacen porque sus participantes se comportan de cierta manera, que a veces son irracionales y ciertamente no siempre eficientes. En este sentido, la diferencia clave entre las estrategias de an\'alisis t\'ecnico\index{estrategias de an\'alisis t\'ecnico} y estrategias de arbitraje estad\'istico\index{estrategias de arbitraje estad\'istico} es que estas \'ultimas se basan en ciertas percepciones que se derivan de horizontes de tenencia\index{horizonte de tenencia} m\'as largos (estrategias basadas en an\'alisis fundamental\index{an\'alisis fundamental}) a horizontes\index{horizonte} m\'as cortos (arbitraje estad\'istico\index{arbitraje estad\'istico}) mejorado a\'un m\'as por las estad\'isticas, es decir, el hecho de que estas estrategias se basen en un gran n\'umero de acciones cuyas propiedades est\'an ``estratificadas'' de acuerdo con algunas caracter\'isticas estad\'isticas y de otros tipos.

{}Esto nos lleva al segundo punto sobre estas ``estratificaciones'' en el contexto del arbitraje estad\'istico\index{arbitraje estad\'istico}. As\'i, en la Subsecci\'on \ref{sub.mean.rev.w.reg} podemos utilizar una matriz de clasificaci\'on binaria de la industria\index{clasificaci\'on binaria de la industria} como la matriz de cargas\index{matriz de cargas} $\Omega_{iA}$. Tales clasificaciones de la industria\index{clasificaci\'on de la industria} se basan en datos fundamentales/econ\'omicos\index{datos, econ\'omicos}\index{datos, fundamental} relevantes, tales como los productos y servicios de las empresas, fuentes de ingresos, proveedores, competidores, socios, etc. Son independientes de los datos de precios\index{datos de precios} y, si est\'an bien construidos, tienden a ser bastante estables fuera de la muestra, ya que las empresas rara vez saltan de industria\index{industria}. Sin embargo, las clasificaciones binarias tambi\'en pueden construirse bas\'andose \'unicamente en datos de precios\index{datos de precios}, a trav\'es de algoritmos de clustering\index{algoritmo de clustering} (v\'ease, por ejemplo, \cite{KakushadzeYu2016b}). Alternativamente, la matriz $\Omega_{iA}$ puede ser no binaria y construida usando, por ejemplo, componentes principales\index{componentes principales} (v\'ease, por ejemplo, \cite{KakushadzeYu2017a}). Algunas de las columnas de $\Omega_{iA}$ pueden estar basadas en factores de estilo de riesgo\index{factores de estilo de riesgo} de horizonte m\'as largo tales como value\index{value}, crecimiento, tama\~{n}o, momentum\index{momentum}, liquidez\index{liquidez} y volatilidad (v\'ease, por ejemplo, \cite{Ang2006}, \cite{Anson2013}, \cite{Asness1995}, \cite{Asness2001}, \cite{Asness2000}, \cite{Banz1981}, \cite{Basu1977}, \cite{Fama1992}, \cite{Fama1993}, \cite{Haugen1995}, \cite{Jegadeesh1993}, \cite{Lakonishok1994}, \cite{Liew2000}, \cite{Pastor2003}, \cite{Scholes1977}),\footnote{\, Para consideraciones sobre (i)liquidez\index{liquidez}, v\'ease tambi\'en, por ejemplo, \cite{Amihud2002}.} o en factores de estilo\index{factores de estilo} de horizonte m\'as corto \cite{Kakushadze2015c}.

\newpage

\section{Fondos de Inversi\'on Cotizados (ETFs)\index{fondo de inversi\'on cotizado (ETF)}}\label{sec.etfs}

\subsection{Estrategia: Rotaci\'on de momentum sectorial\index{rotaci\'on de momentum sectorial}}

{}La evidencia emp\'irica sugiere que el efecto de momentum\index{efecto de momentum} existe no solo para acciones individuales sino que tambi\'en para sectores\index{sector} e industrias\index{industria}.\footnote{\, Para literatura pertinente, v\'ease, por ejemplo,\cite{Cavaglia2002}, \cite{Conover2008}, \cite{Doeswijk2011}, \cite{Dolvin2011}, \cite{Gao2015}, \cite{Hong2007}, \cite{Levis1999}, \cite{Moskowitz1999}, \cite{ONeal2000}, \cite{Sefton2005}, \cite{Simpson2016}, \cite{Sorensen1986}, \cite{Stoval1996}, \cite{Swinkels2002}, \cite{Szakmary2015}, \cite{Wang2017}.} Una estrategia de rotaci\'on de momentum sectorial\index{estrategia de rotaci\'on de momentum sectorial} consiste en asignar mayor ponderaci\'on a los sectores\index{sector} que presentan rendimientos superiores y menor ponderaci\'on a aquellos sectores\index{sector} que presentan rendimientos menores, en donde el rendimiento se basa en el momentum\index{momentum} durante los $T$ meses del per\'iodo de formaci\'on\index{periodo de formacion @ per\'iodo de formaci\'on} (que generalmente var\'ia de 6 a 12 meses). ETFs\index{fondo de inversi\'on cotizado (ETF)} (por sus siglas en ingl\'es) concentrados en sectores\index{sector}/industrias\index{industria} espec\'ificas ofrecen una simple forma de implementar una estrategia de rotaci\'on sectorial/por industria\index{rotaci\'on, por industria}\index{rotaci\'on sectorial} sin tener que comprar o vender una gran cantidad de acciones subyacentes\index{acci\'on subyacente}. Similar a la Subsecci\'on \ref{sub.prc.mom}, como medida de momentum por sector/industria\index{momentum, por sector}\index{momentum, por industria}, podemos utilizar los retornos acumulados\index{retorno acumulado} de los ETFs:
\begin{equation}
 R_i^{acum}(t) = \frac{P_i(t)}{P_i(t + T)} - 1
\end{equation}
Aqu\'i $P_i(t)$ es el precio del ETF\index{fondo de inversi\'on cotizado (ETF)} etiquetado por $i$. (Tal como antes, $t+T$ es $T$ meses en el {\em pasado} con respecto a $t$.) Justo despu\'es del tiempo $t$, el trader puede, por ejemplo, comprar los ETFs\index{fondo de inversi\'on cotizado (ETF)} en el decil\index{decil} superior seg\'un $R_i^{acum}(t)$ y mantener el portafolio\index{portafolio} durante el per\'iodo de tenencia\index{periodo de tenencia @ per\'iodo de tenencia} (generalmente, de 1 a 3 meses). Estrategias d\'olar-neutral\index{estrategia dolar-neutral @ estrategia d\'olar-neutral} tambi\'en se pueden construir, por ejemplo, comprando los ETFs\index{fondo de inversi\'on cotizado (ETF)} en el decil\index{decil} superior y vendiendo los ETFs\index{vender ETFs} en el decil\index{decil} inferior (como para el caso de las acciones, los ETFs\index{fondo de inversi\'on cotizado (ETF)} se pueden vender en corto).\footnote{\, Por algunas publicaciones sobre ETFs\index{fondo de inversi\'on cotizado (ETF)}, v\'ease, por ejemplo, \cite{Agapova2011a}, \cite{Aldridge2016}, \cite{Ben-David2017}, \cite{Bhattacharya2017}, \cite{Buetow2012}, \cite{Clifford2014}, \cite{Hill2015}, \cite{Krause2014}, \cite{Madhavan2016}, \cite{Madura2008}, \cite{Nyaradi2010}, \cite{Oztekin2017}.}

\subsubsection{Estrategia: Rotaci\'on de momentum sectorial\index{rotaci\'on de momentum sectorial} con filtro de media m\'ovil}

{}Esta es una variaci\'on/refinamiento de la estrategia de rotaci\'on de momentum sectorial\index{estrategia de rotaci\'on de momentum sectorial}. Un ETF\index{fondo de inversi\'on cotizado (ETF)} en el decil\index{decil} superior (inferior) es comprado (vendido) solo si pasa un filtro adicional basado en una media m\'ovil\index{media m\'ovil} $\mbox{MA}(T^\prime)$ del precio del ETF:
\begin{eqnarray}
\mbox{Regla} = \begin{cases}
     \mbox{Comprar ETFs en el decil superior solo si } P > \mbox{MA}(T^\prime) \\
     \mbox{Vender ETFs en el decil inferior solo si } P < \mbox{MA}(T^\prime)
\end{cases}
\end{eqnarray}
Aqu\'i $P$ es el precio del ETF al momento de la transacci\'on y $\mbox{MA}(T^\prime)$ se calcula utilizando precios {\em diarios} ($T^\prime$ puede, pero no necesita ser igual a $T$; por ejemplo, $T^\prime$ puede ser de 100 a 200 d\'ias).

\subsubsection{Estrategia: Rotaci\'on sectorial con doble momentum\index{rotaci\'on sectorial con doble momentum}}

{}Para disminuir el riesgo\index{riesgo} de comprar ETFs\index{fondo de inversi\'on cotizado (ETF)} cuando el mercado general\index{mercado general} presenta una tendencia negativa -- para el caso de las estrategias que solo consisten en posiciones largas\index{estrategias solo con posiciones largas} -- el momentum relativo (es decir, de corte transversal)\index{momentum, relativo}\index{momentum, corte transversal} de ETFs sectoriales\index{ETF sectorial} puede ser refinado con el momentum absoluto (es decir, de series de tiempo)\index{momentum, absoluto}\index{momentum, de series de tiempo} de, por ejemplo, un ETF del \'indice de mercado\index{ETF del \'indice de mercado} (v\'ease, por ejemplo, \cite{Antonacci2014}, \cite{Antonacci2017}).\footnote{\, Para algunos estudios adicionales sobre el momentum relativo\index{momentum relativo}, momentum absoluto\index{momentum absoluto} y t\'opicos relacionados, v\'ease, por ejemplo, \cite{Ahn2003}, \cite{Bandarchuk2013}, \cite{Berk1999}, \cite{Cooper2004}, \cite{Fama2008}, \cite{Hurst2017}, \cite{Johnson2002}, \cite{Liu2008}, \cite{Moskowitz2012}, \cite{Sagi2007}, \cite{Schwert2003}, \cite{Zhang2006}.} Por lo tanto, una posici\'on larga basada en la se\~{n}al de rotaci\'on sectorial\index{senzal de rotaci\'on sectorial @ se\~{n}al de rotaci\'on sectorial} (antes discutida) se establece solo si el \'indice de mercado general\index{indice de mercado general @ \'indice de mercado general} presenta una tendencia\index{tendencia} alcista; de lo contrario, los fondos disponibles se invierten en un ETF\index{fondo de inversi\'on cotizado (ETF)} (por ejemplo, un ETF de oro o un ETF del Tesoro\index{ETF del Tesoro}) no correlacionado con el \'indice de mercado general\index{indice de mercado general @ \'indice de mercado general}:
\begin{eqnarray}
\mbox{Regla} = \begin{cases}
     \mbox{Comprar ETFs en el decil superior si } P > \mbox{MA}(T^\prime) \\
     \mbox{Comprar un ETF no correlacionado si } P \leq \mbox{MA}(T^\prime)
\end{cases}
\end{eqnarray}
Aqu\'i $P$ es el precio del ETF del \'indice de mercado\index{ETF del \'indice de mercado} al momento de la transacci\'on y $\mbox{MA}(T^\prime)$ es la media m\'ovil\index{media m\'ovil} del precio de dicho ETF. Generalmente, $T^\prime$ es de 100 a 200 d\'ias.

\subsection{Estrategia: Rotaci\'on de alfa\index{rotaci\'on de alfa}}\label{sub.alpha.rot}

{}Esta estrategia es la misma que la estrategia de rotaci\'on de momentum sectorial\index{estrategia de rotaci\'on de momentum sectorial} con los retornos acumulados de los ETFs\index{retornos acumulados de ETFs} $R^{acum}_i$ reemplazados por los alfas de los ETFs\index{alfas de los ETFs} $\alpha_i$, siendo estos los coeficientes de la regresi\'on\index{coeficiente de la regresi\'on} que corresponden al intercepto\index{intercepto} en una regresi\'on {\em serial}\index{regresi\'on serial} de los retornos de los ETFs\index{retornos de los ETFs}\footnote{\, Generalmente, el per\'iodo de estimaci\'on\index{periodo de estimacion @ per\'iodo de estimaci\'on} es 1 a\~{n}o, y $R_i(t)$ son retornos diarios o semanales.\label{fn.alpha.rot}} $R_i(t)$ sobre, por ejemplo, los 3 factores de Fama-French\index{factores de Fama-French} $\mbox{MKT}(t)$, $\mbox{SMB}(t)$, $\mbox{HML}(t)$ (v\'ease la nota al pie \ref{fn.FF}):\footnote{\, Alfa\index{alfa} aqu\'i es el alfa de Jensen\index{alfa de Jensen} definido por los retornos de los ETFs\index{retornos de los ETFs} a diferencia de los retornos de fondos mutuos\index{retornos de los fondos mutuos} utilizados en \cite{Jensen1968}. Para m\'as literatura relacionada con el alfa de Jensen\index{alfa de Jensen}, v\'ease, por ejemplo, \cite{Bollen2005}, \cite{Droms2001}, \cite{Elton1996}, \cite{Goetzmann1994}, \cite{Grinblatt1992}, \cite{Jan2004}.}
\begin{equation}
 R_i(t) =  \alpha_i + \beta_{1,i}~\mbox{MKT}(t) + \beta_{2,i}~\mbox{SMB}(t) + \beta_{3,i}~\mbox{HML}(t) + \epsilon_i(t) \
\end{equation}

\subsection{Estrategia: R-cuadrado\index{R-cuadrado}}

{}Estudios emp\'iricos sobre fondos mutuos\index{fondo mutuo} (v\'ease, por ejemplo, \cite{Amihud2013}, \cite{Ferson2016}) e ETFs\index{fondo de inversi\'on cotizado (ETF)} (v\'ease, por ejemplo, \cite{GarynTal2014a}, \cite{GarynTal2014b}) han demostrado que combinando el alfa\index{alfa} con un indicador basado en $R^2$ de una regresi\'on {\em serial}\index{regresi\'on serial} de los retornos $R_i(t)$ sobre m\'ultiples factores\index{factor}, por ejemplo, los 3 factores de Fama-French\index{factores de Fama-French} $\mbox{MKT}(t)$, $\mbox{SMB}(t)$, $\mbox{HML}(t)$ m\'as el factor de momentum de Carhart\index{factor de momentum de Carhart (MOM)} $\mbox{MOM}(t)$ (v\'ease la nota al pie \ref{fn.FF}), agrega valor en la predicci\'on de los retornos futuros\index{predicci\'on de retornos futuros}. As\'i, a partir de la regresi\'on serial\index{regresi\'on serial}
\begin{equation}
 R_i(t) =  \alpha_i + \beta_{1,i}~\mbox{MKT}(t) + \beta_{2,i}~\mbox{SMB}(t) + \beta_{3,i}~\mbox{HML}(t) + \beta_{4,i}~\mbox{MOM}(t) + \epsilon_i(t) \
\end{equation}
podemos estimar $\alpha_i$ (los coeficientes de la regresi\'on\index{coeficiente de la regresi\'on} que corresponden al intercepto\index{intercepto}) y el $R^2$ de la regresi\'on\index{R-cuadrado de la regresi\'on}, que se define como (``SS\index{suma de cuadrados (SS)}'' que significa ``suma de cuadrados\index{suma de cuadrados (SS)}''  por sus siglas en ingl\'es):
\begin{eqnarray}
 &&R^2 = 1 - {\mbox{SS}_{res}\over\mbox{SS}_{tot}}\\
 &&\mbox{SS}_{res} = \sum_{i=1}^N\epsilon^2_i(t)\\
 &&\mbox{SS}_{tot} = \sum_{i=1}^N (R_i(t) - \overline{R}(t))^2\\
 &&{\overline R}(t) = {1\over N}~\sum_{i=1}^N R_i(t)
\end{eqnarray}
Una estrategia R-cuadrado\index{estrategia R-cuadrado} entonces consiste en dar mayor ponderaci\'on a los ETFs\index{fondo de inversi\'on cotizado (ETF)} con mayor ``selectividad\index{selectividad}'' (definida como $1-R^2$ \cite{Amihud2013}) y dar menor ponderaci\'on a aquellos ETFs\index{fondo de inversi\'on cotizado (ETF)} con menor ``selectividad\index{selectividad}''. Por ejemplo, uno puede primero ordenar los ETFs\index{fondo de inversi\'on cotizado (ETF)} en quintiles\index{quintil} seg\'un el $R^2$, y luego ordenar los ETFs\index{fondo de inversi\'on cotizado (ETF)} en cada uno de tales quintiles\index{quintil} en otros sub quintiles seg\'un el alfa\index{alfa} (resultando en 25 grupos de ETFs\index{fondo de inversi\'on cotizado (ETF)}). Uno puede entonces, por ejemplo, comprar ETFs\index{fondo de inversi\'on cotizado (ETF)} en el grupo correspondiente al quintil\index{quintil} $R^2$ m\'as bajo y en el sub quintil\index{quintil} con alfa\index{alfa} m\'as alto y vender ETFs\index{fondo de inversi\'on cotizado (ETF)} en el grupo correspondiente al quintil\index{quintil} $R^2$ m\'as alto y en el sub quintil con alfa\index{alfa} m\'as bajo. Otras variaciones son posibles. Finalmente, tanto el per\'iodo de estimaci\'on\index{periodo de estimacion @ per\'iodo de estimaci\'on} como los retornos para estimar $R^2$ pueden ser los mismos que en la estrategia de rotaci\'on de alfa\index{estrategia de rotaci\'on de alfa} (v\'ease Subsecci\'on \ref{sub.alpha.rot} y la nota al pie \ref{fn.alpha.rot}). Sin embargo, per\'iodos de estimaci\'on\index{periodo de estimacion @ per\'iodo de estimaci\'on} m\'as largos pueden ser considerados, especialmente si $R_i(t)$ son retornos mensuales.\footnote{\, Adem\'as, tenga en cuenta que en \cite{Amihud2013} $R^2$ es una medida de cu\'an activa es la gesti\'on\index{gesti\'on activa} de un fondo mutuo\index{fondo mutuo}. En \cite{GarynTal2014a}, \cite{GarynTal2014b} $R^2$ se aplica a ETFs gestionados activamente\index{ETF gestionado activamente}. Para m\'as literatura sobre ETFs gestionados activamente\index{ETF gestionado activamente}, v\'ease, por ejemplo, \cite{Mackintosh2017}, \cite{Meziani2015}, \cite{Rompotis2011a}, \cite{Rompotis2011b}, \cite{Schizas2014}, \cite{Sherrill2018}.}

\subsection{Estrategia: Reversi\'on a la media\index{reversi\'on a la media}}

{}Una forma (entre otras tantas) de construir una estrategia de reversi\'on a la media\index{estrategia de reversi\'on a la media} para ETFs\index{fondo de inversi\'on cotizado (ETF)} es utilizando un indicador de la ``fuerza interna de las barras de precios'' (IBS, por sus siglas en ingl\'es)\index{fuerza interna de las barras de precios (IBS)} basado en los precios de cierre $P_C$, m\'aximo $P_{MA}$ y m\'inimo $P_{MI}$ del d\'ia anterior\index{precio de cierre}\index{precio, m\'aximo}\index{precio, m\'inimo}:\footnote{\, V\'ease, por ejemplo, \cite{Pagonidis2014}. Por alguna literatura adicional relacionada, v\'ease, por ejemplo, \cite{Brown2018}, \cite{Caginalp2014}, \cite{Chan2013}, \cite{Dunis2013}, \cite{Lai2016}, \cite{Levy2013}, \cite{Marshall2013}, \cite{Rudy2010}, \cite{Schizas2011}, \cite{Smith2015}, \cite{Yu2014}.}
\begin{equation}	
 \mbox{IBS} =  \frac{P_C - P_{MI}}{P_{MA} - P_{MI}} \
\end{equation}
Tenga en cuenta que IBS\index{fuerza interna de las barras de precios (IBS)} oscila entre 0 y 1.\footnote{\, Una medida equivalente pero m\'as sim\'etrica es $Y = \mbox{IBS} - 1/2 = (P_C - P_*)/(P_{MA} - P_{MI})$, en donde $P_* = (P_{MA} + P_{MI})/2$. Tenga en cuenta que $Y$ oscila desde $1/2$ para $P_C = P_{MA}$ a $-1/2$ para $P_C = P_{MI}$.} Un ETF\index{fondo de inversi\'on cotizado (ETF)} se puede considerar como ``caro'' si su IBS\index{fuerza interna de las barras de precios (IBS)} es cercano a 1, y como ``barato'' si su IBS\index{fuerza interna de las barras de precios (IBS)} es cercano 0. Al ordenar\index{ordenar} un universo de ETFs\index{fondo de inversi\'on cotizado (ETF)} transversalmente de acuerdo con IBS\index{fuerza interna de las barras de precios (IBS)}, una estrategia d\'olar-neutral\index{estrategia dolar-neutral @ estrategia d\'olar-neutral} puede construirse, por ejemplo, vendiendo los ETFs\index{fondo de inversi\'on cotizado (ETF)} en el decil\index{decil} superior y comprando los ETFs\index{fondo de inversi\'on cotizado (ETF)} en el decil\index{decil} inferior. Al igual que con las estrategias de acciones discutidas anteriormente, las ponderaciones pueden ser uniformes o no uniformes, por ejemplo, uno puede basarse en las volatilidades hist\'oricas de los ETFs\index{fondo de inversi\'on cotizado (ETF)}. Adem\'as, las estrategias de reversi\'on a la media\index{estrategia de reversi\'on a la media} que hemos discutido anteriormente para las acciones tambi\'en se pueden adaptar a los ETFs\index{fondo de inversi\'on cotizado (ETF)}.

\subsection{Estrategia: ETFs apalancados (LETFs)\index{ETF apalancado (LEFT)}}

{}Un ETF apalancado (LETF, por sus siglas en ingl\'es) (inverso)\index{ETF, apalancado}\index{ETF, apalancado inverso} apunta a duplicar o triplicar (el inverso de) el rendimiento diario de su \'indice subyacente\index{indice subyacente @ \'indice subyacente}.\footnote{\, Para algunos estudios sobre ETFs apalancados\index{ETF apalancado (LEFT)}, v\'ease por ejemplo, \cite{AvellanedaZhang2010}, \cite{Bai2015}, \cite{Charupat2011}, \cite{Cheng2010}, \cite{Ivanov2014}, \cite{Jarrow2010}, \cite{Jiang2017}, \cite{Lu2012}, \cite{Shum2016}, \cite{Tang2013}, \cite{Trainor2010}, \cite{Tuzun2013}.} Para mantener un apalancamiento\index{apalancamiento} diario de $2\times$ o $3\times$, los LETFs\index{ETF apalancado (LEFT)} requieren un rebalanceo diario, comprando en los d\'ias en que el mercado\index{mercado} est\'a en alza y vendiendo cuando el mercado\index{mercado} se encuentra a la baja. Esto puede resultar en una deriva\index{deriva} negativa a largo plazo, que puede explotarse vendiendo un ETF apalancado\index{ETF apalancado (LEFT)} y un ETF apalancado inverso\index{ETF apalancado inverso}\index{ETF inverso} (ambos con el mismo nivel de apalancamiento\index{apalancamiento} y para el mismo \'indice subyacente\index{indice subyacente @ \'indice subyacente}) e invirtiendo los ingresos en, por ejemplo, un ETF del Tesoro\index{ETF del Tesoro}. Esta estrategia puede tener un fuerte riesgo de baja\index{riesgo de baja} a corto plazo si una de las piernas vendidas en corto tiene un rendimiento sustancialmente positivo.

\subsection{Estrategia: Seguimiento de la tendencia con m\'ultiples activos\index{seguimiento de la tendencia con m\'ultiples activos}}\label{sub.multiasset.ETF}

{}Uno de los atributos de los ETFs\index{fondo de inversi\'on cotizado (ETF)} es su poder de diversificaci\'on\index{poder de diversificaci\'on}: los ETFs\index{fondo de inversi\'on cotizado (ETF)} permiten obtener exposici\'on\index{exposici\'on} a diferentes sectores\index{sector}, pa\'ises, clases de activos\index{clases de activo}, factores\index{factor}, etc., tomando posiciones en un n\'umero relativamente peque\~{n}o de ETFs\index{fondo de inversi\'on cotizado (ETF)} (en lugar de tomar posiciones en un gran n\'umero de instrumentos subyacentes\index{instrumento subyacente}). Aqu\'i nos centramos en portafolios de seguimiento de la tendencia con posiciones largas\index{portafolio de seguimiento de la tendencia con posiciones largas}. Uno tiene que determinar las ponderaciones $w_i$ de cada ETF\index{fondo de inversi\'on cotizado (ETF)}. Una (pero por lejos no es la \'unica) forma de definir estas ponderaciones es la siguiente. Primero, como en la estrategia de rotaci\'on de momentum sectorial\index{estrategia de rotaci\'on de momentum sectorial}, computamos los retornos acumulados\index{retorno acumulado} $R^{acum}_i$ (sobre un per\'iodo $T$ de, por ejemplo, 6-12 meses). Solo tomamos los ETFs\index{fondo de inversi\'on cotizado (ETF)} con $R^{acum}_i$ positivo. Si se desea, opcionalmente, podemos filtrar a\'un m\'as los ETFs\index{fondo de inversi\'on cotizado (ETF)} como en la estrategia de rotaci\'on de momentum sectorial\index{estrategia de rotaci\'on de momentum sectorial} con un filtro de media m\'ovil, manteniendo solo los ETFs\index{fondo de inversi\'on cotizado (ETF)} cuyos \'ultimos precios de cierre\index{precio de cierre} $P_i$ son m\'as altos que sus medias m\'oviles\index{media m\'ovil} a largo plazo correspondientes $\mbox{MA}_i(T^\prime)$ (generalmente, la longitud $T^\prime$ es de 100 a 200 d\'ias). Ahora, en lugar de tomar los ETFs\index{fondo de inversi\'on cotizado (ETF)} en el decil\index{decil} superior acorde a $R^{acum}_i$ (como en la estrategia de rotaci\'on de momentum sectorial\index{estrategia de rotaci\'on de momentum sectorial}), podemos asignar ponderaciones $w_i$ distintas de cero para todos los ETF restantes\index{fondo de inversi\'on cotizado (ETF)}, cuyo n\'umero en este contexto es relativamente peque\~{n}o por naturaleza. Las ponderaciones pueden, por ejemplo, asignarse de la siguiente manera:
\begin{eqnarray}	
 && w_i = \gamma_1~R^{acum}_i \label{etf.w1}\\
 && w_i = \gamma_2~R^{acum}_i / \sigma_i \label{etf.w2}\\
 && w_i = \gamma_3~R^{acum}_i / \sigma_i^2 \label{etf.w3}
\end{eqnarray}
Aqu\'i: $\sigma_i$ es la volatilidad hist\'orica\index{volatilidad hist\'orica}; y los coeficientes generales de normalizaci\'on $\gamma_1,\gamma_2,\gamma_3$ en cada caso se calculan en base al requisito de que $\sum_{i=1}^N w_i = 1$ (en donde $N$ es el n\'umero de los ETFs\index{fondo de inversi\'on cotizado (ETF)} en el portafolio\index{portafolio} despu\'es de que todos los filtros son aplicados, es decir, aquellos con ponderaciones distintas de cero). As\'i, las ponderaciones en la Ecuaci\'on (\ref{etf.w1}) son simplemente proporcionales a los retornos acumulados\index{retorno acumulado} pasados $R^{acum}_i$, que se toman como la medida de momentum\index{momentum}, entonces los retornos esperados\index{retorno esperado} tambi\'en est\'an dados por (o, m\'as precisamente, proporcionales a) $R^{acum}_i$. El problema con este esquema de ponderaci\'on\index{esquema de ponderaci\'on} es que asigna mayor ponderaci\'on a los ETFs m\'as vol\'atiles\index{fondo de inversi\'on cotizado (ETF)} dado que en promedio $R^{acum}_i \propto\sigma_i$. Las ponderaciones en la Ecuaci\'on (\ref{etf.w2}) mitigan esto, mientras que las ponderaciones en la Ecuaci\'on (\ref{etf.w3}) optimizan el ratio de Sharpe\index{ratio de Sharpe} del portafolio de los ETFs\index{portafolio de ETFs} asumiendo una matriz de covarianza diagonal\index{matriz de covarianza} $C_{ij} = \mbox{diag}(\sigma_i^2)$ para los retornos de los ETFs\index{retornos de los ETFs}, es decir, ignorando sus correlaciones\index{correlaci\'on}.\footnote{\, Por algunas publicaciones sobre portafolios de m\'ultiples activos\index{portafolio de multiples activos @ portafolio de m\'ultiples activos}, asignaci\'on din\'amica de activos\index{asignaci\'on din\'amica de activos} y temas relacionados, v\'ease, por ejemplo, \cite{Bekkers2009}, \cite{Black1992}, \cite{Detemple2010}, \cite{Doeswijk2014}, \cite{Faber2015}, \cite{Faber2016}, \cite{Mladina2014}, \cite{Petre2015}, \cite{Sassetti2006}, \cite{Sharpe2009}, \cite{Sharpe1988}, \cite{Sorensen1999}, \cite{Tripathi2016}, \cite{Wu2003}, \cite{Zakamulin2014b}.} Imponiendo l\'imites $w_i \leq w^{max}_i$, uno puede mitigar a\'un m\'as el problema de la ponderaci\'on excesiva en ciertos ETFs\index{fondo de inversi\'on cotizado (ETF)}.

\newpage

\section{Renta Fija\index{renta fija}}\label{sec.fixed.income}

\subsection{Generalidades}

\subsubsection{Bonos con cup\'on cero\index{bono con cup\'on cero}}

{}Una promesa de pago de \$1 al vencimiento\index{vencimiento} $T$ puede considerarse como un activo, que tiene alg\'un valor en el momento $t$ antes de $T$. Este activo es llamado {\em bono de descuento\index{bono de descuento}} (con cup\'on cero). Sea su precio $P(t,T)$ en el momento $0\leq t\leq T$. Luego, $P(T,T)=1$. El {\em rendimiento\index{rendimiento}} de un bono de descuento\index{bono de descuento} es definido como\footnote{\, M\'as precisamente, esta definici\'on supone una composici\'on continua\index{composici\'on continua}. Para composici\'on peri\'odica\index{composici\'on peri\'odica} en $n$ tiempos discretos $T_i = T_0 + i\delta$, $i=1,\dots,n$, el rendimiento\index{rendimiento} entre $t=T_0$ y $t=T_n$ se encuentra dado por $R(T_0,T_n) =\delta^{-1}\left([P(T_0, T_n)]^{-1/n} - 1\right)$ asumiendo $P(T_n, T_n) = 1$, es decir, $T_n$ es la fecha de madurez\index{madurez}. La Ecuaci\'on (\ref{yield.cont}) se obtiene en el l\'imite donde $n\rightarrow\infty$, $\delta\rightarrow 0$, $n\delta=\mbox{fijo}$ (e igual a $T-t$ en la Ecuaci\'on (\ref{yield.cont})).\label{fn.periodic.yield}}
\begin{equation}\label{yield.cont}
 R(t,T)=-{\ln(P(t,T))\over {T-t}}
\end{equation}
y tiene el significado de una tasa de inter\'es\index{tasa de inter\'es} promedio sobre el per\'iodo de tiempo $T - t$. Cuanto mayor sea el precio del bono\index{precio del bono} en el momento $t$, menor es el rendimiento\index{rendimiento} $R(t,T)$ y viceversa. A continuaci\'on nos referimos a un bono con cup\'on cero\index{bono con cup\'on cero} con un principal\index{principal} de \$1 y una madurez\index{madurez} $T$ como $T$-bono.

\subsubsection{Bonos\index{bono} con cupones\index{cup\'on}}

{}En la pr\'actica, un bono\index{bono} suele pagar no solo su principal\index{principal} al vencimiento\index{vencimiento} $T$, sino que tambi\'en hace peque\~{n}os pagos de cupones\index{pago de cup\'on} antes del mismo. Considere un bono\index{bono} que realiza $n$ pagos de cupones\index{pago de cup\'on} regulares a una tasa {\em no compuesta}\index{tasa no compuesta} fija $k$ en los momentos $T_i=T_0+i\delta$, $i=1,2,\dots,n$, y tambi\'en paga un principal\index{principal} \$1 al vencimiento\index{vencimiento} $T$. El importe de cada pago de cup\'on\index{pago de cup\'on} es $k\delta$, en donde $\delta$ es el per\'iodo de pago\index{periodo de pago @ per\'iodo de pago}. Este flujo de ingresos es equivalente a poseer un $T$-bono m\'as $k\delta$ unidades de cada $T_i$-bono, $i=1,\dots,
n$. El precio del bono con cup\'on\index{bono con cup\'on} al momento $t$ es entonces
\begin{equation}
 P_c (t,T)=P(t,T)+k\delta\sum_{i=I(t)}^n P(t,T_i)
\end{equation}
en donde $I(t) =  \mbox{min}(i:~t<T_i)$. Al momento $t=T_0$ tenemos
\begin{equation}\label{fixed.coupon}
 P_c (T_0,T)=P(T_0,T)+k\delta\sum_{i=1}^n P(T_0,T_i)
\end{equation}
Si deseamos que el bono con cup\'on\index{bono con cup\'on} empiece con su valor nominal\index{valor nominal} ($P_c(T_0,T)=1$), entonces la tasa de cup\'on\index{tasa de cup\'on} correspondiente se encuentra dada por
\begin{equation}
 k={{1-P(T_0,T)}\over\delta~{\sum_{i=1}^n P(T_0,T_i)}}
\end{equation}

\subsubsection{Bonos con tasa flotante\index{bono con tasa flotante}}

{}Un bono\index{bono} puede tambi\'en tener pagos de cupones {\em flotantes}\index{pago de cup\'on flotante}. Considere un
bono\index{bono} que paga \$1 al vencimiento\index{vencimiento} $T$, y que tambi\'en hace pagos de cupones\index{pago de cup\'on} en los momentos
$T_i=T_0+i\delta$, $i=1,2,\dots,n$, con importes basados en una tasa variable\index{tasa variable} (usualmente LIBOR\index{Tasa Interbancaria de Oferta de Londres (LIBOR)} (por sus siglas en ingl\'es) -- v\'ease la Subsecci\'on \ref{sub.swap.spread})
\begin{equation}
 L(T_{i-1})={1\over \delta}\left[{1\over P(T_{i-1},T_i)}-1\right]
\end{equation}
El pago de cup\'on\index{pago de cup\'on} al momento $T_i$ es
\begin{equation}\label{float.coupon}
 X_i= L(T_{i-1})\delta={1\over P(T_{i-1},T_i)}-1
\end{equation}
que son los intereses\index{interes @ inter\'es} que obtendr\'iamos comprando \$1 de valor de un $T_i$-bono al momento $T_{i-1}$. De hecho, un $T_i$-bono vale $P(T_{i-1}, T_i)$ al momento $t = T_{i-1}$, entonces \$1 de valor de $T_i$-bono en $t = T_{i-1}$ vale $1/P(T_{i-1}, T_i)$ en $t=T_i$, entonces los intereses\index{interes @ inter\'es} ganados se encuentran dados por la Ecuaci\'on (\ref{float.coupon}). El valor total del bono con cup\'on variable\index{bono con cup\'on variable} en $t=T_0$ est\'a dado por:
\begin{equation}\label{var.coupon}
 V_0=1 - \left[P(T_0,T_n)-P(T_0,T)\right]
\end{equation}
Si $T = T_n$, entonces tenemos $V_0=1$.
Esto se debe a que este bono\index{bono} es equivalente a la siguiente secuencia de operaciones. Al momento $t=T_0$ tomar \$1 y comprar $T_1$-bonos con ello.
Al momento $t=T_1$ tomar el inter\'es\index{interes @ inter\'es} de los $T_1$-bonos como el $T_1$-cup\'on, y comprar $T_2$-bonos con el principal\index{principal} restante de \$1. Repetir esto hasta que nos quede \$1 al momento $T_n$. Esto tiene exactamente el mismo flujo de efectivo\index{flujo de efectivo} que un bono con cup\'on variable\index{bono con cup\'on variable}, por lo que los precios iniciales deben coincidir. Si $T > T_n$, entonces $V_0 < 1$ y se puede determinar de la siguiente manera. En primer lugar, tenga en cuenta que
\begin{equation}
 V_0 = P(T_0, T) + V_0^{cupones}
\end{equation}
en donde $V_0^{cupones}$ es el valor total de todos los $n$ pagos de cupones\index{pago de cup\'on} en $t=T_0$. Este valor es independiente de $T$ y es determinado desde
\begin{equation}
 P(T_0,T_n) + V_0^{cupones} = 1
\end{equation}
que es el valor del bono con cup\'on variable\index{bono con cup\'on variable} con la madurez\index{madurez} $T_n$. Por lo tanto, equivale a la Ecuaci\'on (\ref{var.coupon}).

\subsubsection{Swaps\index{swap}}\label{sub.swaps}

{}Los swaps\index{swap} son contratos que intercambian un flujo de pagos de tasa flotante\index{pago de tasa flotante} por un flujo de pagos de tasa fija\index{pago de tasa fija} o viceversa. Un swap\index{swap} en donde recibimos un flujo de pagos de tasa fija\index{pago de tasa fija} a cambio de pagos de tasa flotante\index{pago de tasa flotante} es simplemente un portafolio\index{portafolio} con una posici\'on larga en un bono con cupones fijos\index{bono con cup\'on fijo} y una posici\'on corta en un bono con cupones variables\index{bono con cup\'on variable}. El precio del primero en $t=T_0$ est\'a dado por la ecuaci\'on (\ref{fixed.coupon}), mientras que el del \'ultimo viene dado por la Ecuaci\'on (\ref{var.coupon}). La tasa fija\index{tasa fija} que da al swap\index{swap} un valor inicial nulo es independiente de la madurez\index{madurez} $T$ y est\'a dada por
\begin{equation}
 k={{1-P(T_0,T_n)}\over \delta\sum_{i=1}^n P(T_0,T_i)}
\end{equation}

\subsubsection{Duraci\'on\index{duraci\'on} y convexidad\index{convexidad}}

{}La duraci\'on de Macaulay\index{duraci\'on de Macaulay} de un bono\index{bono} es un promedio ponderado\index{promedio ponderado} de la madurez\index{madurez} de sus flujos de efectivo\index{flujo de efectivo}, en donde las ponderaciones son los valores presentes de dichos flujos de efectivo\index{flujo de efectivo}. Por ejemplo, para un bono con cup\'on de tasa fija\index{bono con cup\'on de tasa fija} tenemos (v\'ease la Ecuaci\'on (\ref{fixed.coupon}))
\begin{equation}
 \mbox{MacD}(t, T) = {1\over P_c(t, T)}~\left[(T-t)~P(t,T) + k\delta~\sum_{i=I(t)}^n (T_i - t)~P(t, T_i) \right]
\end{equation}
La duraci\'on modificada\index{duraci\'on modificada} se define como (asumiendo {\em desplazamientos paralelos\index{desplazamiento paralelo}} en la curva de rendimientos\index{curva de rendimientos})\footnote{\, Es decir, $\partial R(t,\tau)/\partial R(t, T) = 1$ para todo $t < \tau < T$. Para cambios no uniformes esto se complica.}
\begin{equation}\label{mod.dur}
 \mbox{ModD}(t, T) = -{\partial \ln\left(P_c(t,T)\right)\over\partial R(t,T)}
\end{equation}
Para composici\'on continua\index{composici\'on continua}, la duraci\'on de Macaulay\index{duraci\'on de Macaulay} y la duraci\'on modificada\index{duraci\'on modificada} son lo mismo (v\'ease la Ecuaci\'on (\ref{yield.cont})). Para composici\'on peri\'odica\index{composici\'on peri\'odica}, estas difieren. Para un rendimiento\index{rendimiento} {\em constante} $R(t,\tau) = Y = \mbox{constante}$ (para todo $t < \tau < T$), est\'an relacionadas a trav\'es de (v\'ease la nota al pie  \ref{fn.periodic.yield}):
\begin{equation}
  \mbox{ModD}(t, T) = \mbox{MacD}(t, T) / (1 + Y\delta)
\end{equation}
La duraci\'on modificada\index{duraci\'on modificada} es una medida relativa de la sensibilidad del precio de los bonos\index{precio del bono} a los cambios en las tasas de inter\'es\index{tasa de inter\'es}: $\Delta P_c(t,T) / P_c(t,T) \approx - \mbox{ModD}(t, T)~\Delta R(t,T)$ (para desplazamientos paralelos\index{desplazamiento paralelo} $\Delta R(t,\tau) = \Delta R = \mbox{constante}$, para todo $t < \tau < T$). De forma similar, la duraci\'on d\'olar\index{duraci\'on dolar @ duraci\'on d\'olar} definida como
\begin{equation}
 \mbox{DD}(t, T) = -{\partial P_c(t,T)\over\partial R(t,T)} = \mbox{ModD}(t, T)~P_c(t, T)
\end{equation}
es una medida absoluta de la sensibilidad del precio de los bonos\index{precio del bono} a los cambios en las tasas de inter\'es\index{tasa de inter\'es}.

{}La convexidad\index{convexidad} de un bono\index{bono} se define como (nuevamente, asumiendo desplazamientos paralelos\index{desplazamiento paralelo})\footnote{\, Por alguna literatura sobre las propiedades de los bonos\index{bono}, v\'ease, por ejemplo, \cite{Baxter1996}, \cite{Bessembinder2008}, \cite{Cerovic2014}, \cite{Chance1996}, \cite{Chen2007}, \cite{Chen2010}, \cite{Christensen1999}, \cite{Cole1995}, \cite{Fabozzi2006a}, \cite{Fabozzi2012a}, \cite{Fabozzi2012b}, \cite{Fabozzi2010}, \cite{Henderson2003}, \cite{Horvath1998}, \cite{Hotchkiss2002}, \cite{Hull2012}, \cite{Hull2005}, \cite{Jostova2013}, \cite{Kakushadze2015a}, \cite{Leland1997}, \cite{Litterman1991}, \cite{Macaulay1938}, \cite{Martellini2003}, \cite{Osborne2005}, \cite{Samuelson1945}, \cite{Stulz2010}, \cite{Tuckman2011}.}
\begin{equation}
 C(t, T) = -{1\over P_c(t,T)}~{\partial^2 P_c(t,T)\over\partial R(t,T)^2}
\end{equation}
y corresponde a los efectos no lineales en los cambios del precio de los bonos\index{precio del bono} a los cambios en la tasa de inter\'es\index{tasa de inter\'es}:
\begin{equation}
 \Delta P_c(t,T) / P_c(t,T) \approx - \mbox{ModD}(t, T)~\Delta R(t,T) + {1\over 2}~C(t, T)~[\Delta R(t,T)]^2
\end{equation}

\subsection{Estrategia: Bullets\index{bullet}}

{}En un portafolio bullet\index{portafolio bullet}, todos los bonos\index{bono} tienen la misma fecha de vencimiento\index{fecha de vencimiento} $T$, por lo tanto, esta estrategia se concentra en un segmento espec\'ifico de la curva de rendimientos\index{curva de rendimientos}. La madurez\index{madurez} se puede elegir en funci\'on de las perspectivas del trader\index{perspectiva del trader} sobre las tasas de inter\'es\index{tasa de inter\'es} futuras: si se espera que las tasas de inter\'es\index{tasa de inter\'es} bajen (es decir, el precio de los bonos\index{precio del bono} suba), concentrar la estrategia en bonos de mayor madurez\index{madurez} tendr\'ia m\'as sentido; si se espera que las tasas de inter\'es\index{tasa de inter\'es} suban (es decir, el precio de los bonos\index{precio del bono} baje), concentrar la estrategia en bonos de menor madurez\index{madurez} estar\'ia m\'as justificado; sin embargo, si el trader no est\'a seguro acerca de las tasas de inter\'es\index{tasa de inter\'es} futuras, un portafolio diversificado\index{portafolio diversificado} (por ejemplo, un portafolio barbell/ladder\index{portafolio, barbell}\index{portafolio, ladder} -- v\'ease abajo) ser\'ia m\'as apropiado (y no un portafolio bullet\index{portafolio bullet}). Generalmente, los bonos\index{bono} en un portafolio bullet\index{portafolio bullet} se compran a lo largo del tiempo, lo que disminuye el riesgo de tasa de inter\'es\index{riesgo de tasa de inter\'es} en cierta forma: si las tasas de inter\'es\index{tasa de inter\'es} suben, las \'ultimas compras de bonos se realizar\'an a tasas m\'as altas; si las tasas de inter\'es\index{tasa de inter\'es} bajan, las primeras compras de bonos realizadas presentar\'an mayores rendimientos\index{rendimiento}.\footnote{\, Por literatura sobre estrategias bullet\index{bullet} y barbell\index{estrategia, barbell}\index{estrategia, bullet} (v\'ease abajo), v\'ease, por ejemplo, \cite{FabbozziMP2006}, \cite{Grantier1988}, \cite{Jones1991}, \cite{Mann1997}, \cite{Pascalau2015}, \cite{Su2010}, \cite{Wilner1996}, \cite{Yamada1999}.}

\subsection{Estrategia: Barbells\index{barbell}}

{}En esta estrategia, todos los bonos comprados\index{bono} se concentran en dos fechas de madurez\index{madurez} $T_1$ (madurez\index{madurez} corta) y $T_2$ (madurez\index{madurez} larga), entonces este portafolio\index{portafolio} es una combinaci\'on de dos estrategias bullet\index{estrategia bullet}. Esta estrategia aprovecha los mayores rendimientos\index{rendimiento} de los bonos\index{bono} a largo plazo mientras que cubre el riesgo de tasa de inter\'es\index{riesgo de tasa de inter\'es} con bonos\index{bono} de corto plazo: si las tasas de inter\'es\index{tasa de inter\'es} suben, los bonos\index{bono} de largo plazo van a perder valor, pero los ingresos de los bonos\index{bono} de corto plazo se puede reinvertir a mayores tasas.\footnote{\, Aplanamiento/empinamiento de la curva de rendimientos\index{curva de rendimientos} (el margen\index{margen} entre las tasas de inter\'es\index{tasa de inter\'es} de bonos de largo plazo y corto plazo disminuye/aumenta) tiene un impacto positivo/negativo en el valor del portafolio\index{portafolio}.} La duraci\'on modificada\index{duraci\'on modificada} (ll\'amese $D$) de una estrategia barbell\index{estrategia barbell} es la misma que la duraci\'on modificada\index{duraci\'on modificada} (ll\'amese $D_*$) de una estrategia bullet\index{estrategia bullet} con una madurez\index{madurez} de rango medio (ll\'amese $T_*$, $T_1 < T_* < T_2$). Sin embargo, la convexidad\index{convexidad} (ll\'amese $C$) de una estrategia barbell\index{estrategia barbell} es mayor que la convexidad\index{convexidad} (ll\'amese $C_*$) de la estrategia bullet\index{estrategia bullet}. Intuitivamente, esto se puede entender al observar que la duraci\'on modificada\index{duraci\'on modificada} aumenta aproximadamente de forma lineal con la madurez\index{madurez}, mientras que la convexidad\index{convexidad} aumenta aproximadamente de forma cuadr\'atica con la madurez\index{madurez}. Con fines ilustrativos y para mantenerlo simple, consideremos una estrategia barbell\index{estrategia barbell} que consiste en $w_1$ d\'olares de bonos con cup\'on cero\index{bono con cup\'on cero} con una madurez\index{madurez} corta $T_1$ y $w_2$ d\'olares de bonos con cup\'on cero\index{bono con cup\'on cero} con una madurez\index{madurez} larga $T_2$ (cada bono\index{bono} tiene \$1 de valor nominal\index{valor nominal}). Adem\'as, asumamos una composici\'on continua\index{composici\'on continua} y una tasa de rendimiento\index{rendimiento} constante $Y$. Entonces tenemos
\begin{eqnarray}	
&& D = {{{\widetilde w}_1~T_1 + {\widetilde w}_2~T_2}\over{{\widetilde w}_1 + {\widetilde w}_2}} \\
&& T_* = D_* = D\\
&& C = {{{\widetilde w}_1~T_1^2 + {\widetilde w}_2~T_2^2}\over{{\widetilde w}_1 + {\widetilde w}_2}}\\
&& C_* = T_*^2
\end{eqnarray}
en donde ${\widetilde w}_1 = w_1~\exp(-T_1~Y)$ y ${\widetilde w}_2 = w_2~\exp(-T_2~Y)$. Una simple transformaci\'on algebraica nos da
\begin{equation}
 C - C_* = {{\widetilde w}_1{\widetilde w}_2\over ({\widetilde w}_1 + {\widetilde w}_2)^2}~(T_2 - T_1)^2 > 0
\end{equation}
Mayor convexidad\index{convexidad} de un portafolio barbell\index{portafolio barbell} proporciona una mejor protecci\'on contra los desplazamientos paralelos\index{desplazamiento paralelo} en la curva de rendimientos\index{curva de rendimientos}. Sin embargo, esto viene a expensas de un rendimiento\index{rendimiento} general m\'as bajo.

\subsection{Estrategia: Escaleras\index{escalera} (Ladders)\index{ladder}}

{}Un ladder\index{ladder} es un portafolio de bonos\index{portafolio de bonos} con las asignaciones de capital\index{asignaci\'on de capital} (aproximadamente) iguales en bonos\index{bono} con $n$ fechas de vencimiento\index{vencimiento} $T_i$, $i=1,\dots,n$ (en donde el n\'umero de escalones $n$ es considerable, por ejemplo, $n=10$). Las fechas de vencimiento\index{vencimiento} son equidistantes: $T_{i+1} = T_i + \delta$. Esta es una estrategia de ``duration-targeting''\index{estrategia de duration-targeting},\footnote{\, Por alguna literatura sobre estrategias ladder\index{ladder} y estrategias de ``duration-targeting''\index{estrategia de duration-targeting}\index{estrategia, ladder}, v\'ease, por ejemplo, \cite{Bierwag1978}, \cite{Bohlin2004}, \cite{CheungKwan2010}, \cite{Dyl1986}, \cite{Fridson2014}, \cite{Judd2011}, \cite{Langetieg1990}, \cite{Leibowitz2013}, \cite{Leibowitz2014}, \cite{Leibowitz2015}.} que mantiene una duraci\'on\index{duraci\'on} aproximadamente constante vendiendo bonos\index{bono} de menor madurez\index{madurez} a medida que se acercan al vencimiento\index{vencimiento} y reemplaz\'andolos por nuevos bonos\index{bono} con vencimiento m\'as lejano. Un portafolio ladder\index{portafolio ladder} apunta a diversificar el riesgo de tasa de inter\'es y el riesgo de reinversi\'on\index{riesgo de reinversi\'on}\footnote{\, El riesgo de reinversi\'on\index{riesgo de reinversi\'on} es el riesgo\index{riesgo} de que los ingresos (provenientes de los pagos de cupones\index{pago de cup\'on} y/o el principal\index{principal}) ser\'an reinvertidos a una tasa menor que la inversi\'on\index{inversi\'on} original.} evitando la exposici\'on\index{exposici\'on} a bonos con solo unas pocas fechas de vencimiento\index{vencimiento} (como en el caso de bullets\index{bullet} y barbells\index{barbell}). Esta estrategia tambi\'en genera un flujo de ingresos regular provenientes de los cupones\index{cup\'on} de cada bono\index{bono}. La madurez\index{madurez} de un portafolio ladder\index{portafolio ladder} puede definirse como la madurez\index{madurez} promedio:
\begin{equation}
 T = {1\over n}~\sum_{i=1}^n T_i
\end{equation}
El ingreso es mayor para valores m\'as altos de $T$; sin embargo, as\'i tambi\'en es el riesgo de tasa de inter\'es\index{riesgo de tasa de inter\'es}.

\subsection{Estrategia: Inmunizaci\'on de bonos\index{inmunizaci\'on de bonos}}

{}La inmunizaci\'on de bonos\index{inmunizaci\'on de bonos} se utiliza en casos tales como cuando una obligaci\'on de efectivo futura est\'a predeterminada. Una soluci\'on simple ser\'ia comprar un bono con cup\'on cero\index{bono con cup\'on cero} con la madurez\index{madurez} requerida (y un rendimiento\index{rendimiento} deseable/aceptable). Sin embargo, tal bono\index{bono} puede que no siempre est\'e disponible en el mercado\index{mercado}, entonces un portafolio\index{portafolio} de bonos\index{bono} con diferentes fechas de vencimiento\index{vencimiento} debe ser utilizado en su lugar. Tal portafolio\index{portafolio} est\'a sujeto al riesgo de tasa de inter\'es y riesgo de reinversi\'on\index{riesgo de reinversi\'on}. Una forma de mitigar estos riesgos\index{riesgo} es construir un portafolio\index{portafolio} cuya duraci\'on\index{duraci\'on} coincide con la madurez\index{madurez} de la futura obligaci\'on en efectivo (y, por lo tanto, ``inmunizando'' el portafolio de bonos\index{portafolio de bonos} ante desplazamientos paralelos\index{desplazamiento paralelo} en la curva de rendimientos\index{curva de rendimientos}). Considere un portafolio\index{portafolio} de bonos\index{bono} con 2 fechas de vencimiento\index{vencimiento} distintas $T_1, T_2$ y sus correspondientes duraciones\index{duraci\'on} $D_1,D_2$ (en donde ``duraci\'on\index{duraci\'on}'' se refiere a duraci\'on modificada\index{duraci\'on modificada}). Sean los montos en d\'olares invertidos en estos bonos\index{bono} $P_1,P_2$; el monto total a invertir sea $P$; la duraci\'on\index{duraci\'on} deseada del portafolio\index{portafolio} sea $D$ (que est\'a relacionada con la madurez\index{madurez} $T_*$ de la futura obligaci\'on en efectivo -- v\'ease abajo); y el rendimiento\index{rendimiento} {\em constante} (que se supone que es {\em el mismo para todos los bonos\index{bono}} -- v\'ease abajo) sea $Y$. Entonces $P$ se fija utilizando $Y$ y el importe de la obligaci\'on futura $F$:
\begin{equation}\label{PF}
 P = F / (1 + Y\delta)^{T_*/\delta}
\end{equation}
en donde asumimos composici\'on peri\'odica\index{composici\'on peri\'odica} y $\delta$ es la longitud de cada per\'iodo de composici\'on\index{periodo de composicion @ per\'iodo de composici\'on} (por ejemplo, 1 a\~{n}o).\footnote{\, Para asegurarnos la simplicidad, en la Ecuaci\'on (\ref{PF}) el n\'umero $n=T_*/\delta$ de per\'iodos de composici\'on\index{periodo de composicion @ per\'iodo de composici\'on} se supone que es un n\'umero entero. La extensi\'on a un n\'umero no entero $T_*/\delta$ es simple.} Entonces tenemos:
\begin{eqnarray}
 &&P_1 + P_2 = P\\
 &&P_1~D_1 + P_2~D_2 = P~D
\end{eqnarray}
en donde
\begin{equation}
 D = T_* / (1 + Y\delta)
\end{equation}
Con 3 bonos\index{bono}, tambi\'en podemos igualar la convexidad\index{convexidad}:
\begin{eqnarray}
 &&P_1 + P_2 + P_3 = P\\
 &&P_1~D_1 + P_2~D_2 + P_3~D_3 = P~D\\
 &&P_1~C_1 + P_2~C_2 + P_3~C_3 = P~C
\end{eqnarray}
en donde $C_1,C_2,C_3$ son las convexidades de los 3 bonos\index{bono} y
\begin{equation}
 C = T_*(T_*+\delta)/(1+Y\delta)^2
\end{equation}
En la pr\'actica, la curva de rendimientos\index{curva de rendimientos} cambia a lo largo del tiempo, lo que (entre otras cosas) requiere que el portafolio\index{portafolio} sea balanceado peri\'odicamente. Esto introduce costos de transacci\'on\index{costos de transacci\'on} no triviales, que tambi\'en deben tenerse en cuenta. Adem\'as, los rendimientos\index{rendimiento} no son los mismos para todos los bonos\index{bono} en el portafolio\index{portafolio}, lo que introduce complejidad adicional en el problema.\footnote{\, Por algunas publicaciones sobre inmunizaci\'on de bonos\index{inmunizaci\'on de bonos}, incluyendo t\'ecnicas de optimizaci\'on\index{tecnicas de optimizacion @ t\'ecnicas de optimizaci\'on} m\'as sofisticadas, v\'ease, por ejemplo, \cite{Albrecht1985}, \cite{Alexander1985}, \cite{Bierwag1979}, \cite{Bodie1996}, \cite{Boyle1978}, \cite{Christensen1985}, \cite{DeLaPena2017}, \cite{Fisher1971}, \cite{Fong1983}, \cite{Fong1984}, \cite{Hurlimann2002}, \cite{Hurlimann2012}, \cite{Iturricastillo2010}, \cite{Khang1983}, \cite{Kocherlakota1988}, \cite{Kocherlakota1990}, \cite{Montrucchio1991}, \cite{Nawalkha1996}, \cite{Reddington1952}, \cite{Reitano1996}, \cite{Shiu1987}, \cite{Shiu1988}, \cite{Zheng2003}.}

\subsection{Estrategia: Mariposa duraci\'on-d\'olar-neutral\index{mariposa duraci\'on-d\'olar-neutral}}

{}Esta es una combinaci\'on de costo cero\index{combinaci\'on de costo cero} de una posici\'on larga en un portafolio barbell\index{portafolio barbell} (con fechas de vencimiento\index{vencimiento} $T_1$ (m\'as corta) y $T_3$ (m\'as larga)) y una posici\'on corta en un portafolio bullet\index{portafolio bullet} (con un vencimiento\index{vencimiento} medio $T_2$, en donde $T_1 < T_2 < T_3$). Sean los montos en d\'olares invertidos en los 3 bonos\index{bono} $P_1,P_2, P_3$; y sean las correspondientes duraciones modificadas\index{duraci\'on modificada} $D_1,D_2,D_3$. Luego, el costo cero (es decir, la d\'olar-neutralidad\index{dolar-neutralidad @ d\'olar-neutralidad}) y la duraci\'on-d\'olar-neutralidad\index{duracion-dolar-neutralidad @ duraci\'on-d\'olar-neutralidad} (el \'ultimo protege al portafolio\index{portafolio} de desplazamientos paralelos\index{desplazamiento paralelo} en la curva de rendimientos\index{curva de rendimientos}) implican que
\begin{eqnarray}	
&& P_1 + P_3 = P_2 \\
&& P_1~D_1 + P_3~D_3 = P_2~D_2
\end{eqnarray}
Esto fija $P_1,P_3$ mediante $P_2$. Mientras que el portafolio\index{portafolio} es inmune a desplazamientos paralelos\index{desplazamiento paralelo} en la curva de rendimientos\index{curva de rendimientos}, no es inmune a los cambios en la pendiente o la curvatura\index{curvatura} de la curva de rendimientos\index{curva de rendimientos}.\footnote{\, Por alguna literatura sobre varias estrategias mariposas con bonos\index{estrategia mariposa con bonos}, v\'ease, por ejemplo, \cite{Bedendo2007}, \cite{Brooks2017}, \cite{Christiansen2005}, \cite{Fontaine2017}, \cite{Gibson2000}, \cite{Grieves1999}, \cite{Heidari2003}, \cite{Martellini2002}.}

\subsection{Estrategia: Mariposa cincuenta-cincuenta\index{mariposa cincuenta-cincuenta}}

{}Esta es una variaci\'on de la mariposa est\'andar\index{mariposa}. Siguiendo con la notaci\'on anterior para la mariposa duraci\'on-d\'olar-neutral\index{mariposa duraci\'on-d\'olar-neutral}, tenemos
\begin{equation}\label{curve.neutral}
 P_1~D_1 = P_3~D_3 = {1\over 2}~P_2~D_2
\end{equation}
Entonces, la mariposa cincuenta-cincuenta\index{mariposa cincuenta-cincuenta} sigue siendo duraci\'on-d\'olar-neutral, pero ya no es d\'olar-neutral (es decir, no es una estrategia de costo cero\index{estrategia de costo cero}). En cambio, la duraci\'on d\'olar\index{duraci\'on dolar @ duraci\'on d\'olar} de ambas alas\index{ala} es la misma (de ah\'i el t\'ermino ``cincuenta-cincuenta''). Como resultado, la estrategia es (aproximadamente) neutral a peque\~{n}as inclinaciones y aplanamientos de la curva de rendimientos\index{curva de rendimientos}, a saber, si el margen de tasas de inter\'es\index{diferencial de la tasa de inter\'es} entre el cuerpo\index{cuerpo} y el ala de corto plazo\index{ala de corto plazo} es igual al cambio del margen\index{cambio del margen} de tasas de inter\'es entre el cuerpo\index{cuerpo} y el ala de largo plazo\index{ala de largo plazo}. Es por eso que esta estrategia es tambi\'en conocida como ``mariposa neutral a la curva\index{mariposa neutral a la curva}'' (cuyo costo es la no d\'olar-neutralidad).

\subsection{Estrategia: Mariposa regresi\'on-ponderada\index{mariposa regresi\'on-ponderada}}

{}Emp\'iricamente, las tasas de inter\'es\index{tasa de inter\'es} de corto plazo son considerablemente m\'as vol\'atiles que las tasas de inter\'es\index{tasa de inter\'es} de largo plazo.\footnote{\, V\'ease, por ejemplo, \cite{Edwards2003}, \cite{Joslin2017}, \cite{Mankiw1984}, \cite{Shiller1979}, \cite{Sill1996}, \cite{Turnovsky1989}.} Por lo tanto, el cambio en el margen de tasas de inter\'es\index{diferencial de la tasa de inter\'es} entre el cuerpo\index{cuerpo} y el ala de corto plazo\index{ala de corto plazo} (de la mariposa\index{mariposa} -- v\'ease arriba) se puede esperar que sea mayor por alg\'un factor -- ll\'amese $\beta$ -- al cambio del margen\index{cambio del margen} de tasas de inter\'es entre el cuerpo\index{cuerpo} y el ala de largo plazo\index{ala de largo plazo} (entonces, generalmente $\beta > 1$). Este factor se puede obtener a partir de datos hist\'oricos\index{datos hist\'oricos} mediante, por ejemplo, una regresi\'on\index{regresi\'on} lineal del cambio del margen\index{cambio del margen} entre el cuerpo\index{cuerpo} y el ala de corto plazo\index{ala de corto plazo} sobre el cambio del margen\index{cambio del margen} entre el cuerpo\index{cuerpo} y el ala de largo plazo\index{ala de largo plazo}. Entonces, en lugar de la Ecuaci\'on (\ref{curve.neutral}), tenemos las siguientes condiciones de duraci\'on-d\'olar-neutralidad\index{duracion-dolar-neutralidad @ duraci\'on-d\'olar-neutralidad} y ``curva-neutralidad\index{curva-neutralidad}'':
\begin{eqnarray}
 &&P_1~D_1 + P_3~D_3 = P_2~D_2\\
 &&P_1~D_1 = \beta~P_3~D_3 \label{reg.but}
\end{eqnarray}

\subsubsection{Estrategia: Mariposa madurez-ponderada\index{mariposa madurez-ponderada}}

{}Esta es una variaci\'on de la mariposa regresi\'on-ponderada\index{mariposa regresi\'on-ponderada}, en donde en lugar de fijar $\beta$ en la Ecuaci\'on (\ref{reg.but}) mediante una regresi\'on\index{regresi\'on} basada en datos hist\'oricos\index{datos hist\'oricos}, este coeficiente se basa en los vencimientos de los 3 bonos\index{vencimiento del bono}:
\begin{equation}
 \beta = {{T_2 - T_1}\over{T_3 - T_2}}
\end{equation}

\subsection{Estrategia: Factor de bajo riesgo\index{factor de bajo riesgo}}

{}Al igual que en las acciones, la evidencia emp\'irica sugiere que los bonos de menor riesgo\index{bono} tienden a superar a los bonos de mayor riesgo\index{bono} en t\'erminos de retornos ajustados por riesgo (``anomal\'ia de bajo riesgo\index{anomal\'ia de bajo riesgo}'').\footnote{\, Por alguna literatura, v\'ease, por ejemplo, \cite{DeCarvalho2014}, \cite{Derwall2009}, \cite{Frazzini2014}, \cite{Houweling2017}, \cite{Ilmanen2011}, \cite{Ilmanen2004},  \cite{Kozhemiakin2007}, \cite{Ng2015}.} Uno puede definir el ``riesgo'' de un bono\index{bono} utilizando diferentes m\'etricas, por ejemplo, la calificaci\'on crediticia\index{calificaci\'on crediticia} y la madurez\index{madurez}. Por ejemplo, un portafolio largo\index{portafolio largo} se puede construir (v\'ease, por ejemplo, \cite{Houweling2017}) tomando los Bonos de Grado de Inversi\'on\index{Bonos de Grado de Inversi\'on} con las calificaciones crediticias\index{calificaci\'on crediticia} que van desde AAA a A-, y luego tomando el decil\index{decil} inferior por madurez\index{madurez}. Del mismo modo, uno puede tomar los Bonos de Alto Rendimiento\index{Bonos de Alto Rendimiento} con las calificaciones crediticias\index{calificaci\'on crediticia} que van desde BB+ a B-, y luego tomar el decil\index{decil} inferior por madurez\index{madurez}.

\subsection{Estrategia: Factor de value\index{factor de value}}

{}El ``value\index{value}'' para bonos\index{bono} (v\'ease, por ejemplo, \cite{Correia2012}, \cite{Houweling2017}, \cite{LHoir2010}) es m\'as dif\'icil de definir que para las acciones. Una forma es comparar el margen crediticio\index{margen crediticio}\footnote{\, El margen crediticio\index{margen crediticio} es la diferencia entre el rendimiento de los bonos\index{rendimientos del bono} y la tasa libre de riesgo\index{tasa libre de riesgo}.} observado y una predicci\'on te\'orica de dicho margen. Una forma de estimar esto \'ultimo es, por ejemplo, a trav\'es de una regresi\'on lineal de corte transversal\index{regresi\'on, de corte transversal}  (entre $N$ bonos\index{bono} etiquetados por $i=1,\dots,N$) \cite{Houweling2017}:
\begin{eqnarray}\label{cr.spr.reg}	
&& S_i = \sum_{r =1}^K \beta_r~I_{ir} + \gamma~T_i + \epsilon_i \\
&& S^*_i = S_i - \epsilon_i
\end{eqnarray}
Aqu\'i: $S_i$ es el margen crediticio\index{margen crediticio}; $I_{ir}$ es una variable dummy\index{variable dummy} ($I_{ir} = 1$ si el bono\index{bono} etiquetado por $i$ tiene calificaci\'on crediticia\index{calificaci\'on crediticia} $r$; de otra manera, $I_{ir}=0$) para la calificaci\'on crediticia de los bonos\index{calificaci\'on crediticia de los bonos} $r$ (que etiqueta las $K$ calificaciones crediticias\index{calificaci\'on crediticia} presentes entre los $N$ bonos\index{bono}, pudiendo ser una de las 21 calificaciones crediticias\index{calificaci\'on crediticia});\footnote{\, Estas calificaciones crediticias\index{calificaci\'on crediticia} son AAA, AA+, AA, AA-, A+, A, A-, BBB+, BBB, BBB-, BB+, BB, BB-, B+, B, B-, CCC+, CCC, CCC-, CC, C.} $T_i$ son los vencimientos de los bonos\index{vencimiento del bono}; $\beta_r,\gamma$ son los coeficientes de la regresi\'on\index{coeficiente de la regresi\'on}; $\epsilon_i$ son los residuos de la regresi\'on\index{residuos de la regresi\'on}; y $S^*_i$ es el valor ajustado (te\'orico) del margen crediticio\index{margen crediticio}. La matriz con dimensiones $N\times K$ $I_{ir}$ no tiene ninguna columna con todos los valores iguales a cero (entonces $K$ puede ser menor de 21). Tenga en cuenta que por definici\'on, ya que cada bono\index{bono} tiene una y solo una calificaci\'on crediticia\index{calificaci\'on crediticia}, tenemos
\begin{equation}
 \sum_{r = 1}^K I_{ir} = 1
\end{equation}
entonces el intercepto\index{intercepto} se subsume en $I_{ir}$ (es por eso por lo que no hay un coeficiente de la regresi\'on\index{coeficiente de la regresi\'on} separado para el intercepto\index{intercepto}). Luego, el value\index{value} se define como $V_i = \ln(S_i/S^*_i)$ o $V_i = \epsilon_i/S^*_i = S_i/S^*_i - 1$, y los bonos\index{bono} en el decil\index{decil} superior seg\'un $V_i$ son seleccionados para el portafolio\index{portafolio}.

\subsection{Estrategia: Factor carry\index{carry factor}}

{}El carry\index{carry} se define como el retorno proveniente de la apreciaci\'on del valor del bono\index{valor del bono} cuando este rueda hacia abajo por la curva de rendimientos\index{curva de rendimientos} (v\'ease, por ejemplo, \cite{Beekhuizen2016}, \cite{Koijen2018}):\footnote{\, Aqu\'i, por simplicidad, consideramos bonos con cup\'on cero\index{bono con cup\'on cero}. El resultado final a continuaci\'on tambi\'en es v\'alido para bonos con cup\'on\index{bono con cup\'on}.}
\begin{equation}
 C(t, t+\Delta t, T) = {{P(t + \Delta t, T) - P(t, T)}\over P(t, T)}
\end{equation}
Aqu\'i $\Delta t$ es el per\'iodo sobre el cual se calcula el carry\index{carry}. Surge una simplificaci\'on si asumimos que toda la estructura temporal\index{estructura temporal} de tasas de inter\'es\index{tasa de inter\'es} se mantiene constante, es decir, el rendimiento\index{rendimiento} $R(t, T) = f(T-t)$ es una funci\'on solo de $T-t$ (es decir, el tiempo a la  madurez\index{madurez}). Luego, al momento $t+\Delta t$ el rendimiento\index{rendimiento} es $R(t+\Delta t, T) = R(t, T - \Delta t)$. Entonces, tenemos\footnote{\, Para portafolios financiados\index{portafolio financiado}, $R(t, T)$ en la segunda l\'inea de la Ecuaci\'on (\ref{carry.eq}) es reemplazada por $R(t, T) - r_f$, en donde $r_f$ es la tasa libre de riesgo\index{tasa libre de riesgo}. Sin embargo, este cambio general no afecta las tenencias actuales en la estrategia de carry\index{estrategia de carry}.}
\begin{eqnarray}
 C(t, t+\Delta t, T) &=& {{\left.P(t + \Delta t, T)\right|_{R(t+\Delta t, T)} - \left.P(t, T)\right|_{R(t, T)}}\over \left.P(t, T)\right|_{R(t, T)}} = \nonumber\\
 &=& R(t, T)~\Delta t + C_{roll}(t, t+\Delta t, T)\label{carry.eq}
\end{eqnarray}
en donde (teniendo en cuenta la definici\'on de la duraci\'on modificada\index{duraci\'on modificada}, la Ecuaci\'on (\ref{mod.dur}))
\begin{eqnarray}
 C_{roll}(t, t+\Delta t, T) &=& {{\left.P(t + \Delta t, T)\right|_{R(t, T - \Delta t)} - \left.P(t + \Delta t, T)\right|_{R(t, T)}}\over \left.P(t, T)\right|_{R(t, T)}}\approx\nonumber\\
 &\approx & -\mbox{ModD}(t, T)\left[R(t, T - \Delta t) - R(t, T)\right]
\end{eqnarray}
Por lo tanto, si la estructura temporal\index{estructura temporal} de las tasas de inter\'es\index{tasa de inter\'es} es constante, luego el carry\index{carry} $C(t, t+\Delta t, T)$ recibe dos contribuciones: i) $R(t, T)~\Delta t$ del rendimiento del bono\index{rendimientos del bono}; y ii) $C_{roll}(t, t+\Delta t, T)$ del bono\index{bono} rodando hacia abajo por la curva de rendimientos\index{rodando hacia abajo por la curva de rendimientos}. Una estrategia de costo cero\index{estrategia de costo cero} se puede construir, por ejemplo, comprando bonos\index{bono} en el decil\index{decil} superior seg\'un el carry\index{carry} y vendiendo bonos\index{bono} en el decil\index{decil} inferior.

\subsection{Estrategia: Rodando hacia abajo por la curva de rendimientos\index{rodando hacia abajo por la curva de rendimientos}}

{}El objetivo de esta estrategia es capturar el componente ``roll-down''\index{componente roll-down} $C_{roll}(t, t+\Delta t, T)$ de los rendimientos de los bonos\index{rendimientos del bono}. Estos retornos se maximizan en los segmentos m\'as inclinados de la curva de rendimientos\index{curva de rendimientos}. Por lo tanto, el trader puede, por ejemplo, comprar bonos a largo o mediano plazo\index{bono} en dichos segmentos y mantenerlos mientras est\'an ``rodando hacia abajo por la curva''\index{rodando hacia abajo por la curva}.\footnote{\, Por algunos estudios sobre las estrategias relacionadas a los bonos ``rondando hacia abajo por la curva de rendimientos''\index{estrategia, rondando hacia abajo por la curva de rendimientos}, v\'ease, por ejemplo, \cite{Ang1998}, \cite{Bieri2004}, \cite{Bieri2005}, \cite{Dyl1981}, \cite{GrievesMann1999}, \cite{Grieves1992}, \cite{Osteryoung1981}, \cite{Pantalone1984}, \cite{Pelaez1997}.} Los bonos\index{bono}, idealmente, deben venderse mientras se acercan al vencimiento\index{vencimiento} y los ingresos pueden utilizarse para comprar nuevos bonos a largo/mediano plazo\index{bono} del segmento m\'as empinado de la curva de rendimientos\index{curva de rendimientos} en ese momento.

\subsection{Estrategia: Margen de la curva de rendimientos\index{margen de la curva de rendimientos} (Flattener\index{flattener} \& Steepener\index{steepener})}\label{sub.yield.curve}

{}Esta estrategia consiste en comprar o vender el margen de la curva de rendimientos\index{margen de la curva de rendimientos}.\footnote{\, Por alguna literatura sobre la  estrategia sobre el margen de la curva de rendimientos\index{estrategia sobre el margen de la curva de rendimientos}, la din\'amica de la curva de rendimientos\index{din\'amica de la curva de rendimientos} y t\'opicos relacionados, v\'ease, por ejemplo, \cite{Bernadell2005}, \cite{Boyd2010}, \cite{Chua2006}, \cite{Diebold2002}, \cite{Diebold2006}, \cite{Dolan1999}, \cite{Evans2007}, \cite{Fuss2011}, \cite{Jones1991}, \cite{Kalev2006}, \cite{Krishnamurthy2002}, \cite{ShillerMod1979}.} El margen de la curva de rendimientos\index{margen de la curva de rendimientos} se define como la diferencia entre los rendimientos\index{rendimiento} de dos bonos\index{bono} del mismo emisor con diferentes fechas de vencimiento\index{vencimiento}. Si se espera que las tasas de inter\'es\index{tasa de inter\'es} caigan, es de esperar que la curva de rendimientos\index{curva de rendimientos} se empine. Si se espera que las tasas de inter\'es\index{tasa de inter\'es} suban, es de esperar que la curva de rendimientos\index{curva de rendimientos} se aplane. La estrategia sobre el margen de la curva de rendimientos\index{estrategia sobre el margen de la curva de rendimientos} se puede resumir a trav\'es de la siguiente regla:
\begin{eqnarray}
\mbox{Regla} = \begin{cases}
        \mbox{Flattener: Vender margen si se espera que las tasas aumenten} \\
        \mbox{Steepener: Comprar margen si se espera que las tasas disminuyan}
        \end{cases}
\end{eqnarray}
Vender el margen\index{margen} es equivalente a vender los bonos\index{bono} de vencimiento m\'as corto (tambi\'en conocidos como la pierna delantera\index{pierna delantera}) y comprar los bonos\index{bono} de mayor vencimiento (tambi\'en conocidos como la pierna trasera\index{pierna trasera}). Comprar el margen\index{margen} es la operaci\'on opuesta: comprar la pierna delantera\index{pierna delantera} y vender la pierna trasera\index{pierna trasera}. Si la curva de rendimientos\index{curva de rendimientos} experimenta desplazamientos paralelos\index{desplazamiento paralelo}, esta estrategia puede generar p\'erdidas. Si la duraci\'on d\'olar\index{duraci\'on dolar @ duraci\'on d\'olar} de las piernas delanteras y traseras\index{pierna delantera}\index{pierna trasera} coincide, el portafolio\index{portafolio} es inmune a peque\~{n}os desplazamientos paralelos\index{desplazamiento paralelo} en la curva de rendimientos\index{curva de rendimientos}.

\subsection{Estrategia: Arbitraje de la base del CDS\index{arbitraje de la base del CDS}}\label{sub.cds}

{}Un swap de incumplimiento crediticio (CDS, por sus siglas en ingl\'es)\index{swap de incumplimiento crediticio (CDS)} es un seguro contra el incumplimiento\index{incumplimiento} de un bono\index{bono}.\footnote{\, Por literatura sobre el arbitraje de la base del CDS\index{arbitraje de la base del CDS} y t\'opicos relacionados, v\'ease, por ejemplo, \cite{Bai2013}, \cite{Choudhry2004}, \cite{Choudhry2006}, \cite{Choudhry2007}, \cite{DeWit2006}, \cite{Fontana2010}, \cite{Fontana2016}, \cite{Kim2016}, \cite{Kim2017}, \cite{Nashikkar2011}, \cite{Rajan2007}, \cite{Wang2014}, \cite{Zhu2006}.} El precio de los CDS\index{precio del CDS}, conocido como margen, es una prima peri\'odica (por ejemplo, anual)\index{prima, anual}\index{prima, peri\'odica} por d\'olar de la deuda asegurada. Los CDS\index{swap de incumplimiento crediticio (CDS)} b\'asicamente convierten al bono\index{bono} en un instrumento libre de riesgo\index{instrumento libre de riesgo}. Por lo tanto, el margen de los CDS\index{margen del CDS} deber\'ia igualar al margen del rendimiento del bono\index{margen del rendimiento del bono}, es decir, al margen\index{margen} entre el rendimiento del bono\index{rendimientos del bono} y la tasa libre de riesgo\index{tasa libre de riesgo}. La diferencia entre el margen del CDS\index{margen del CDS} y el margen del bono\index{margen del bono} es conocida como la base del CDS\index{base de CDS}:
\begin{equation}
 \mbox{base del CDS} = \mbox{margen del CDS} - \mbox{margen del bono}
\end{equation}
Una base negativa indica que el margen del bono\index{margen del bono} es demasiado alto en relaci\'on con el margen del CDS\index{margen del CDS}, es decir, el bono\index{bono} se encuentra relativamente barato. La operaci\'on de arbitraje de CDS\index{operaci\'on de arbitraje de CDS} entonces consiste en comprar el bono\index{bono} y asegurarlo con un CDS\index{swap de incumplimiento crediticio (CDS)}\footnote{\, Tenga en cuenta que un CDS\index{swap de incumplimiento crediticio (CDS)} es equivalente a una posici\'on sint\'etica corta en un bono\index{posici\'on sint\'etica corta en un bono}.} generando as\'i una ganancia libre de riesgo\index{ganancia libre de riesgo}.\footnote{\, En el caso de una base positiva, te\'oricamente uno entrar\'ia en la posici\'on opuesta, es decir, vender\'ia el bono\index{bono} y vender\'ia un CDS\index{swap de incumplimiento crediticio (CDS)}. Sin embargo, en la pr\'actica, esto generalmente implicar\'ia que el trader ya posee el bono\index{bono} y el CDS\index{swap de incumplimiento crediticio (CDS)}, es decir, esto equivaldr\'ia a desarmar una posici\'on existente.}

\subsection{Estrategia: Arbitraje del margen del swap\index{arbitraje del margen del swap}}\label{sub.swap.spread}

{}Esta estrategia d\'olar-neutral\index{estrategia dolar-neutral @ estrategia d\'olar-neutral} consiste en una posici\'on larga (corta) en un swap de tasas de inter\'es\index{swap de tasas de inter\'es} (v\'ease la Subsecci\'on \ref{sub.swaps}) y una posici\'on corta (larga) en un bono del Tesoro\index{bono del Tesoro} (con rendimiento\index{rendimiento} constante $Y_{Tesoro}$) con la misma madurez\index{madurez} que el swap\index{swap}. Una posici\'on larga (corta) en un swap\index{swap} implica recibir (proveer) los pagos de cupones a la tasa fija $r_{swap}$\index{pago de cup\'on, a la tasa fija} a cambio de proveer (recibir) los pagos de cupones a la tasa variable\index{pago de cup\'on a la tasa variable} igual a la Tasa Interbancaria de Oferta de Londres (LIBOR, por sus siglas en ingl\'es)\index{Tasa Interbancaria de Oferta de Londres (LIBOR)} $L(t)$. La posici\'on corta (larga) en el bono del Tesoro\index{bono del Tesoro} genera (se financia a) la ``tasa repo\index{tasa repo}'' (la tasa de descuento\index{tasa de descuento} a la cual el banco central recompra activos gubernamentales\index{activo gubernamental} de los bancos comerciales) $r(t)$ en una cuenta a margen\index{cuenta al margen}. La tasa por d\'olar invertido $C(t)$ a la que esta estrategia genera P\&L\index{P\&L} est\'a dada por
\begin{eqnarray}	
 && C(t) = \pm[C_1 - C_2(t)]\\
 && C_1 = r_{swap} - Y_{Tesoro}\\
 && C_2(t) =  L(t) - r(t)
\end{eqnarray}
en donde el signo mas (menos) corresponde a la estrategia de swap\index{estrategia de swap} larga (corta). La estrategia larga (corta) de swap\index{estrategia de swap} es rentable si la LIBOR\index{Tasa Interbancaria de Oferta de Londres (LIBOR)} baja (sube). Entonces, esto es b\'asicamente una apuesta sobre la tasa LIBOR\index{Tasa Interbancaria de Oferta de Londres (LIBOR)}.\footnote{\, Por algunas publicaciones sobre m\'argenes de swaps\index{margen del swap} y t\'opicos relacionados, v\'ease, por ejemplo, \cite{Asgharian2008}, \cite{Aussenegg2014}, \cite{Chen1994}, \cite{Collin-Dufresne2001}, \cite{Duarte2006}, \cite{Dubil2011}, \cite{Duffie1996}, \cite{Duffie1997b}, \cite{Feldhutter2008}, \cite{Fisher2002}, \cite{Jermann2016}, \cite{Jordan1997}, \cite{Kambhu2006}, \cite{Keane1996}, \cite{Klinger2016}, \cite{Kobor2005}, \cite{Lang1998}, \cite{Liu2006}, \cite{Minton1997}.}

\newpage

\section{\'Indices\index{indice @ \'indice}}\label{sec.indexes}

\subsection{Generalidades}

{}Un \'indice\index{indice @ \'indice} es un portafolio diversificado\index{portafolio diversificado} de activos combinados con ciertas ponderaciones. Los activos subyacentes\index{activo subyacente} son a menudo acciones, por ejemplo, en \'indices\index{indice @ \'indice} tales como el DJIA\index{Dow Jones Industrial Average (DJIA)}, S\&P 500\index{S\&P 500}, Russell 3000\index{Russell 3000}, etc. Las ponderaciones del DJIA\index{Dow Jones Industrial Average (DJIA)} se basan en los precios, mientras que las ponderaciones del S\&P 500\index{S\&P 500} y las del Russell 3000\index{Russell 3000} se basan en la capitalizaci\'on burs\'atil\index{capitalizaci\'on burs\'atil}. Veh\'iculos de inversi\'on\index{veh\'iculo de inversi\'on} tales como futuros del \'indice\index{futuros del \'indice}, ETFs basados en \'indices\index{ETF basado en \'indice}, etc., permiten ganar exposici\'on\index{exposici\'on} a un \'indice general\index{indice general @ \'indice general} tomando solo una posici\'on.\footnote{\, Por algunas publicaciones sobre \'indices\index{indice @ \'indice}, v\'ease, por ejemplo, \cite{Antoniou1995}, \cite{Beneish1996}, \cite{Bologna2002}, \cite{Bos2000}, \cite{Chang1999}, \cite{Chiang2002}, \cite{Edwards1988}, \cite{Frino2004}, \cite{Graham1994}, \cite{Hautcoeur2006}, \cite{Illueca2003}, \cite{Kenett2013}, \cite{Lamoureux1987}, \cite{Larsen1998}, \cite{Lo2016}, \cite{Schwartz1991}, \cite{Spyrou2005}, \cite{Yo2001}.}

\subsection{Estrategia: Arbitraje ``Cash \& Carry''\index{arbitraje Cash \& Carry}}

{}Esta estrategia (tambi\'en conocida como ``arbitraje de \'indice\index{arbitraje de indice @ arbitraje de \'indice}'') pretende explotar las ineficiencias entre el precio spot\footnote{\, ``Spot\index{spot}'' se refiere al valor actual del \'indice\index{indice @ \'indice} basado en los precios actuales de sus constituyentes. ``Cash''\index{efectivo} se refiere al portafolio subyacente del \'indice\index{portafolio subyacente del \'indice}. Esta es la jerga com\'un utilizada por los traders.} del \'indice\index{precio spot del \'indice} y los precios de los futuros del \'indice\index{precios de los futuros del \'indice}.\footnote{\, V\'ease, por ejemplo, \cite{Brenner1989}, \cite{Buhler1995}, \cite{Butterworth2010}, \cite{Chan1993}, \cite{Cornell1983}, \cite{Dwyer1996}, \cite{Fassas2011}, \cite{Puttonen1993}, \cite{Richie2008}, \cite{Yadav1990}, \cite{Yadav1994}.} Te\'oricamente, el precio de los futuros del \'indice\index{futuros del \'indice} debe ser igual al precio spot\index{precio spot} teniendo en cuenta el costo del carry\index{carry} durante la vida del contrato de futuros\index{contrato de futuros}:
\begin{equation}\label{FuturesEq}	
 F^{*}(t, T) = \left[S(t) - D(t, T)\right]\exp\left(r\left(T-t\right)\right)
\end{equation}
Aqu\'i: $F^{*}(t, T)$ es el precio te\'orico (``justo'') al momento $t$ del contrato de futuros\index{contrato de futuros} con el tiempo de entrega\index{tiempo de entrega} $T$; $S(t)$ es el valor spot\index{valor spot} al momento $t$; $D(t, T)$ es la suma de los (valores descontados de) dividendos\index{dividendo} pagados por las acciones subyacentes\index{acci\'on subyacente} entre el tiempo $t$ y el tiempo de entrega\index{entrega}; y $r$ es la tasa libre de riesgo\index{tasa libre de riesgo}, la que, en aras de simplicidad, se supone que es constante desde $t$ hasta el tiempo de entrega\index{entrega}.\footnote{\, La Ecuaci\'on (\ref{FuturesEq}), adem\'as, ignora algunos otros aspectos pertinentes, como los impuestos, la asimetr\'ia de las tasas de inter\'es\index{tasa de inter\'es} (para posiciones largas y cortas), costos de transacci\'on\index{costos de transacci\'on}, etc.} La base se define como
\begin{eqnarray}\label{Basis}	
 && B(t, T) =  \frac{F(t, T) - F^{*}(t, T)}{S(t)}\
\end{eqnarray}
en donde $F(t,T)$ es el precio actual del contrato de futuros\index{contrato de futuros} con el tiempo de entrega\index{tiempo de entrega} $T$. Si $B(t, T) \neq 0$, m\'as precisamente, si $|B(t, T)|$ supera los costos de transacci\'on\index{costos de transacci\'on} pertinentes a ejecutar la operaci\'on de arbitraje\index{operaci\'on de arbitraje}, entonces hay una oportunidad de arbitraje. Si la base es positiva (negativa), el precio de los futuros\index{precio de los futuros} es caro (barato) comparado con el precio spot\index{precio spot}, entonces la operaci\'on de arbitraje\index{operaci\'on de arbitraje} consiste en vender (comprar) los futuros\index{futuro} y comprar (vender) el cash\index{efectivo} (es decir, la cesta del \'indice\index{cesta del \'indice}).\footnote{\, Vender los futuros\index{futuro} no plantea problemas. Sin embargo, vender el cash\index{efectivo} puede ser problem\'atico cuando hay problemas con las ventas en corto\index{problemas con las ventas en corto} tales como el caso de las acciones dif\'iciles de pedir prestadas (acciones ``hard-to-borrow'' en ingl\'es)\index{activo hard-to-borrow}, etc. Continuamente manteniendo un libro d\'olar-neutral\index{libro d\'olar-neutral} considerable con posiciones largas en cash\index{efectivo} y cortas en futuros\index{futuro} puede ayudar a sortear tales problemas.} La posici\'on se cierra cuando la base vuelve a cero, es decir, el precio de los futuros\index{precio de los futuros} converge a su valor justo\index{valor justo}. Tales oportunidades de arbitraje son de corta duraci\'on y con el advenimiento del trading de alta frecuencia\index{trading de alta frecuencia (HFT)} requieren una ejecuci\'on extremadamente r\'apida. En muchos casos, el slippage\index{slippage} puede ser prohibitivo para ejecutar la operaci\'on.\footnote{\, En algunos casos, cestas incompletas\index{cestas incompletas} aproximando el \'indice\index{indice @ \'indice} pueden ser ejecutadas para reducir los costos de transacci\'on\index{costos de transacci\'on}, por ejemplo, en los \'indices ponderados por la capitalizaci\'on de mercado\index{indice ponderado por capitalizacion de mercado @ \'indice ponderado por capitalizaci\'on de mercado}, al omitir las acciones con las capitalizaciones m\'as bajas (y por lo tanto menos l\'iquidas). Sin embargo, tales coberturas imperfectas\index{cobertura imperfecta} tambi\'en aumentan el riesgo\index{riesgo} de perder dinero en la operaci\'on.}

\subsection{Estrategia: Trading de dispersi\'on\index{trading de dispersi\'on} en \'indices de acciones\index{indice de acciones @ \'indice de acciones}}\label{sub.dispersion}

{}Esta estrategia consiste en tomar posiciones largas en las volatilidades de los constituyentes del \'indice\index{constituyentes del \'indice}\index{volatilidades de los constituyentes del \'indice} y una posici\'on corta en la volatilidad del \'indice\index{volatilidad del \'indice}. Se basa en la observaci\'on emp\'irica de que, la mayor parte del tiempo,\footnote{\, Pero no siempre -- v\'ease abajo. Por algunas publicaciones sobre la volatilidad del \'indice\index{volatilidad del \'indice} versus las de los constituyentes y el trading de dispersi\'on y correlaci\'on\index{trading de correlaci\'on}, v\'ease, por ejemplo, \cite{Carrasco2007}, \cite{Deng2008}, \cite{Lozovaia2005}, \cite{Marshall2008}, \cite{Marshall2009}, \cite{Maze2012}, \cite{Meissner2016}, \cite{Nelken2006}.} la volatilidad impl\'icita\index{volatilidad impl\'icita} ${\widetilde\sigma}_I$ de las opciones del \'indice\index{opci\'on de indice @ opci\'on de \'indice} es considerablemente m\'as alta que su volatilidad te\'orica $\sigma_I$ dada por
\begin{equation}\label{vol.index}
 \sigma_I^2 = \sum_{i,j = 1}^{N} w_{i} w_{j} \sigma_{i} \sigma_{j} \rho_{ij}
\end{equation}
en donde $w_i$ son las ponderaciones de las acciones en el \'indice\index{indice @ \'indice}, $\sigma_i$ son las volatilidades impl\'icitas de las opciones de acciones individuales\index{opci\'on de acci\'on individual}, y $\rho_{ij}$ es la matriz de correlaci\'on muestral\index{matriz de correlaci\'on muestral} ($\rho_{ii}=1$)\footnote{\, Tenga en cuenta que las correlaciones\index{correlaci\'on} por pares $\rho_{ij}$, $i\neq j$, son inestables fuera de la muestra, lo que puede introducir un error considerable en este c\'alculo. \label{fn.corr}}, la cual es computada en funci\'on de series de tiempo\index{serie de tiempo} de los retornos hist\'oricos\index{retornos hist\'oricos}.\footnote{\, Por algunos estudios pertinentes, v\'ease, por ejemplo,\cite{Bakshi2003a}, \cite{Bakshi2003b}, \cite{BakshiKM2003}, \cite{Bollen2004}, \cite{Branger2004}, \cite{Coval2001}, \cite{Dennis2002}, \cite{Dennis2006}, \cite{Driessen2009}, \cite{Garleanu2009}, \cite{Lakonishok2007}.} Dicho de otra manera, las opciones del \'indice\index{opci\'on de indice @ opci\'on de \'indice} tienen un precio m\'as alto que el precio correspondiente a la volatilidad te\'orica antes mencionada. Entonces, una estrategia b\'asica se puede estructurar de la siguiente manera. Para cada acci\'on en el \'indice\index{indice @ \'indice} tenemos una posici\'on larga en $n_i$ (cercanos a ATM\index{opci\'on, ATM (en el dinero)}) conos sobre opciones de acciones individuales\index{cono sobre opciones de acci\'on individual} (cuyos pagos\index{pago} se basan en los precios de las acciones $P_i$), y tenemos una posici\'on corta en un (cercano a ATM\index{opci\'on, ATM (en el dinero)}) cono sobre opciones del \'indice\index{cono sobre opciones del \'indice}\index{indice @ \'indice} (cuyo pago\index{pago} se basa en el nivel del \'indice\index{nivel del \'indice} $P_I$ -- v\'ease abajo), en donde
\begin{eqnarray}	
 && n_{i} = \frac{S_{i} P_I}{\sum_{i = 1}^{N} S_{i} P_{i}}
\end{eqnarray}
Aqu\'i: $S_i$ es el n\'umero de acciones en circulaci\'on\index{acciones en circulaci\'on} para la acci\'on $i$ (estamos asumiendo que el \'indice es ponderado por la capitalizaci\'on de mercado\index{indice, ponderado por capitalizacion de mercado @ \'indice, ponderado por capitalizaci\'on de mercado}); y $P_I$ es el nivel del \'indice\index{nivel del \'indice}. Con esta definici\'on de $n_i$, tenemos $P_I = \sum_{i=1}^N n_i P_i$, por lo que el pago\index{pago} del cono sobre opciones del \'indice\index{cono sobre opciones del \'indice} coincide con los pagos\index{pago} de los conos sobre opciones de acciones individuales\index{cono sobre opciones de acci\'on individual} lo m\'as posible.\footnote{\, Si las opciones ATM\index{opci\'on ATM} no est\'an disponibles para una acci\'on dada, opciones OTM\index{opci\'on OTM} (cercanas a ATM\index{opci\'on, ATM (en el dinero)}) pueden ser usadas.} Todas las opciones\index{opci\'on} tienen aproximadamente 1 mes hasta el vencimiento\index{vencimiento} y todas las posiciones permanecen abiertas hasta el vencimiento\index{vencimiento}.\footnote{\, Se puede argumentar que esta estrategia es una estrategia de volatilidad\index{estrategia de volatilidad}. Sin embargo, tambi\'en se puede argumentar que es una estrategia de correlaci\'on\index{trading de correlaci\'on} dado que la volatilidad del portafolio\index{portafolio} depende de las correlaciones\index{correlaci\'on} entre sus componentes (v\'ease la Ecuaci\'on (\ref{vol.index})). As\'i, cuando la volatilidad impl\'icita del \'indice\index{volatilidad impl\'icita del \'indice} ${\widetilde \sigma}_I$ es m\'as alta que el valor te\'orico $\sigma_I$, esto se puede interpretar (posiblemente) como que la correlaci\'on de pares media impl\'icita\index{correlaci\'on, impl\'icita} es m\'as alta que el promedio de correlaci\'on\index{correlaci\'on} de pares basado en $\rho_{ij}$. En este sentido, a veces la volatilidad impl\'icita del \'indice\index{volatilidad impl\'icita del \'indice} puede ser inferior a su valor te\'orico, por lo que la estrategia de dispersi\'on\index{estrategia de dispersi\'on} que consiste en vender la volatilidad del \'indice\index{volatilidad del \'indice} perder\'ia dinero y la estrategia inversa podr\'ia ser la m\'as adecuada. V\'ease, por ejemplo, \cite{Deng2008}.}

\subsubsection{Estrategia: Trading de dispersi\'on\index{trading de dispersi\'on} -- subconjunto del portafolio\index{subconjunto del portafolio}}

{}Para algunos \'indices\index{indice @ \'indice}, algunas de las acciones que lo componen pueden no tener opciones. A menudo, estas ser\'ian aquellas acciones menos l\'iquidas, con menor capitalizaci\'on de mercado\index{capitalizaci\'on de mercado}. Entonces \'estas tendr\'ian que ser excluidas del portafolio\index{portafolio} adquirido. Reducir el n\'umero comprado de opciones de acciones individuales\index{opci\'on de acci\'on individual} subyacentes\index{opci\'on de acci\'on subyacente individual} tambi\'en es conveniente para reducir los costos de transacci\'on\index{costos de transacci\'on}. Adem\'as, la matriz de correlaci\'on muestral\index{matriz de correlaci\'on muestral} $\rho_{ij}$ es singular para una t\'ipica ventana de estimaci\'on (por ejemplo, retornos de cierre contra cierre\index{retornos de cierre contra cierre} diarios, retrocediendo 1 a\~{n}o, que es equivalente a unos 252 d\'ias de trading\index{dias de trading @ d\'ias de trading}) cuando el n\'umero de activos es grande (500 para el S\&P 500\index{S\&P 500} e incluso m\'as grande para otros \'indices\index{indice @ \'indice}). Adem\'as, tal como se mencion\'o anteriormente, las correlaciones\index{correlaci\'on} de pares son inestables fuera de muestra, lo que aumenta los errores en el valor te\'orico $\sigma_I$ computado mediante la Ecuaci\'on (\ref{vol.index}). Esto se puede mitigar de la siguiente manera.\footnote{\, La variaci\'on de la estrategia de trading de dispersi\'on\index{estrategia de trading de dispersi\'on} que discutimos aqu\'i es similar pero no id\'entica a la estrategia basada en el PCA (an\'alisis de componentes principales)\index{an\'alisis de componentes principales (PCA)} discutida en \cite{Deng2008}, \cite{Larsson2011}, \cite{Su2006}. La construcci\'on (v\'ease abajo) de un modelo estad\'istico de riesgo\index{modelo estad\'istico de riesgo} es m\'as eficiente.}

{}La matriz de correlaci\'on\index{matriz de correlaci\'on} singular e inestable se puede hacer no singular y m\'as estable al reemplazarla con un modelo estad\'istico de riesgo\index{modelo estad\'istico de riesgo} \cite{KakushadzeYu2017a}. Sean $V^{(A)}_i$ los componentes principales\index{componentes principales} de $\rho_{ij}$ con los eigenvalores\index{eigenvalor} $\lambda^{(A)}$ en orden decreciente, $\lambda^{(1)} > \lambda^{(2)} > \lambda^{(r)}$, en donde $r$ es el rango\index{rango} de $\rho_{ij}$ (si $r<N$, los otros eigenvalores\index{eigenvalor} son nulos: $\lambda^{(A)} = 0$, $A>r$). La matriz de correlaci\'on\index{matriz de correlaci\'on} basada en el modelo estad\'istico de riesgo\index{modelo estad\'istico de riesgo} est\'a dada por
\begin{eqnarray}
 &&\psi_{ij} = \xi_i^2~\delta_{ij} + \sum_{A=1}^K \lambda^{(A)}~V^{(A)}_i~V^{(A)}_j\\
 &&\xi_i^2 = 1 - \sum_{A=1}^K \lambda^{(A)}\left[V^{(A)}_i\right]^2
\end{eqnarray}
en donde $K < r$ es el n\'umero de factores de riesgo\index{factor de riesgo} basados en los primeros $K$ componentes principales\index{componentes principales} que se seleccionan para explicar el riesgo sistem\'atico\index{riesgo sistem\'atico}, y $\xi_i$ es el riesgo espec\'ifico (tambi\'en conocido como idiosincr\'atico)\index{riesgo, idiosincr\'atico}\index{riesgo, espec\'ifico}. La forma simple de fijar $K$ es a trav\'es de eRank (rango efectivo)\index{eRank (rango efectivo)} \cite{Roy2007} -- v\'ease \cite{KakushadzeYu2017a} para detalles y c\'odigo fuente\index{codigo fuente @ c\'odigo fuente} completo para construir $\psi_{ij}$ y fijar $K$. Entonces, ahora podemos usar $\psi_{ij}$ (en lugar de $\rho_{ij}$) para calcular la volatilidad te\'orica $\sigma_I$:
\begin{equation}\label{vol.index.stat}
 \sigma_I^2 = \sum_{i,j=1}^N w_{i} w_{j} \sigma_{i} \sigma_{j} \psi_{ij} = \sum_{i=1}^N w_i^2\sigma_i^2\xi_i^2 +
 \sum_{A=1}^K\left[\sum_{i=1}^N \lambda^{(A)}V^{(A)}_i w_i\sigma_i\right]^2
\end{equation}
El primer t\'ermino en el lado derecho de la Ecuaci\'on (\ref{vol.index.stat}) se debe al riesgo espec\'ifico\index{riesgo espec\'ifico}. El portafolio largo\index{portafolio largo} entonces contiene solo conos\index{cono} que corresponden a las $N_*$ opciones de acciones individuales\index{opci\'on de acci\'on individual} con los $N_*$ valores m\'as bajos de $w_i^2\sigma_i^2\xi_i^2$. Por ejemplo, para el S\&P 500\index{S\&P 500} podemos tomar $N_*=100$.

\subsection{Estrategia: Arbitraje intrad\'ia\index{arbitraje intrad\'ia} entre ETFs de \'indices\index{ETF de \'indice}}

{}Esta estrategia consiste en explotar las valoraciones err\'oneas\index{valoraci\'on err\'onea} a corto plazo entre dos ETFs\index{fondo de inversi\'on cotizado (ETF)} (ll\'amelos ETF1 y ETF2) sobre el mismo \'indice subyacente\index{indice subyacente @ \'indice subyacente}.\footnote{\, Por ejemplo, los ETFs del S\&P 500\index{ETFs del S\&P 500}, SPDR Trust\index{SPDR Trust} (tablero de cotizaciones\index{tablero de cotizaciones} SPY) y iShares\index{iShares} (tablero de cotizaciones\index{tablero de cotizaciones} IVV). V\'ease, por ejemplo, \cite{Marshall2013}. Para m\'as literatura sobre arbitraje de ETFs\index{arbitraje de ETF} y t\'opicos relacionados, v\'ease, por ejemplo, \cite{Abreu2002}, \cite{Ackert2000}, \cite{Ben-David2012}, \cite{Brown2018}, \cite{Cherry2004}, \cite{Dolvin2009}, \cite{Garvey2009}, \cite{HendershottMoulton2011}, \cite{Johnson2008}, \cite{Maluf2013}.} Se puede resumir como sigue:
\begin{eqnarray}
 \mbox{Regla} = \begin{cases}
          \text{Comprar ETF2, vender ETF1}  &  \mbox{ si } P^{Bid}_1 \geq P^{Ask}_2\times{\kappa}  \\
          \text{Liquidar posici\'on}  &  \mbox{ si } P^{Bid}_2 \geq P^{Ask}_1  \\
          \text{Comprar ETF1, vender ETF2}  &  \mbox{ si } P^{Bid}_2 \geq P^{Ask}_1\times{\kappa} \\
          \text{Liquidar posici\'on}  &  \mbox{ si } P^{Bid}_1 \geq P^{Ask}_2
        \end{cases}
\end{eqnarray}
Aqu\'i: $\kappa$ es un umbral predefinido, que est\'a cerca de 1, por ejemplo, $\kappa=1.002$ (v\'ease, por ejemplo, \cite{Marshall2013}); $P^{Bid}_1$ y $P^{Bid}_2$ son los precios bid\index{precios bid} para ETF1 y ETF2, y $P^{Ask}_1$ y $P^{Ask}_2$ son los precios ask\index{precios ask}. \'Ordenes l\'imites ``completar o matar''\index{orden l\'imite completar o matar} (en ingl\'es, \'ordenes ``fill-or-kill'') pueden ser utilizadas para ejecutar las operaciones. Tales oportunidades de arbitraje son ef\'imeras y requieren de un sistema de ejecuci\'on de \'ordenes\index{sistema de ejecuci\'on de \'ordenes} muy r\'apido o bien el slippage\index{slippage} erosionar\'a con rapidez las ganancias.

\subsection{Estrategia: Targeting de volatilidad sobre \'indice\index{targeting de volatilidad sobre \'indice} con activo libre de riesgo\index{activo libre de riesgo}}\label{vol.targ}

{}Una estrategia de targeting de volatilidad\index{estrategia de targeting de volatilidad} apunta a mantener un nivel de volatilidad constante, que se puede lograr rebalanceando\index{rebalanceo} de forma peri\'odica (semanal, mensual, etc.) entre un activo de riesgo -- en este caso un \'indice\index{indice @ \'indice} -- y un activo sin riesgo\index{activo sin riesgo} (por ejemplo, bonos del Tesoro\index{bono del Tesoro}).\footnote{\, Por algunos estudios pertinentes, v\'ease, por ejemplo, \cite{Albeverio2013}, \cite{Anderson2014}, \cite{Cirelli2017}, \cite{Cooper2010}, \cite{Giese2012}, \cite{Khuzwayo2014}, \cite{KimEnke2016}, \cite{Kirby2012}, \cite{Papageorgiou2017}, \cite{Perchet2014}, \cite{Torricelli2018}, \cite{Zakamulin2014b}.} Si $\sigma$ es la volatilidad del activo de riesgo\footnote{\, Por lo general, esta es la volatilidad impl\'icita\index{volatilidad impl\'icita} a diferencia de la volatilidad hist\'orica\index{volatilidad hist\'orica}, teniendo en cuenta que la primera se considera una medida de la volatilidad futura (``forward-looking''). Alternativamente, puede basarse en varias t\'ecnicas de predicci\'on de volatilidad.} y la volatilidad objetivo (target en ingl\'es)\index{volatilidad objetivo} es $\sigma_*$, entonces la ponderaci\'on de la asignaci\'on\index{ponderaci\'on de la asignaci\'on} en el activo de riesgo est\'a dada por\footnote{\, Si hay un apalancamiento\index{apalancamiento} m\'aximo preestablecido $L$, entonces $w$ est\'a limitado a $L$.} $w = \sigma_*/\sigma$, y la ponderaci\'on de la asignaci\'on\index{ponderaci\'on de la asignaci\'on} en el activo libre de riesgo\index{activo libre de riesgo} es $1-w$. Para evitar el exceso de operaciones y reducir los costos de transacci\'on\index{costos de transacci\'on}, el rebalanceo\index{rebalanceo} (en lugar de hacerlo peri\'odicamente) se puede hacer en base a, por ejemplo, un umbral preestablecido $\kappa$, es decir, solo se procede a rebalancear si el cambio porcentual $|\Delta w| / w$ desde el \'ultimo rebalanceo\index{rebalanceo} supera $\kappa$.

\newpage

\section{Volatilidad}\label{sec.volatility}

\subsection{Generalidades}

{}Algunas estrategias de trading de opciones\index{estrategia de trading de opciones} discutidas en la Secci\'on \ref{sec.options} son estrategias de volatilidad\index{estrategia de volatilidad}, en el sentido en que apuestan a una volatilidad futura alta o baja.\footnote{\, Por ejemplo, conos\index{cono} largos (cortos) apuestan a aumentos (disminuciones) en la volatilidad.} Hay varias formas de hacer apuestas de volatilidad, y, por lo tanto, la volatilidad se puede ver como una clase de activo\index{clases de activo} por s\'i misma. La volatilidad hist\'orica\index{volatilidad hist\'orica} se basa en una serie de tiempo\index{serie de tiempo} de retornos pasados. En cambio, la volatilidad impl\'icita\index{volatilidad impl\'icita} extra\'ida de las opciones\index{opci\'on} se considera una medida de volatilidad futura.\footnote{\, V\'ease, por ejemplo, \cite{Abken1996}, \cite{Ane2001}, \cite{Canina1993}, \cite{Christensen1998}, \cite{Derman1994}, \cite{Dumas1998}, \cite{Dupire1994}, \cite{Glasserman2010}, \cite{He2015}, \cite{Lamoureux1993}, \cite{Mayhew1995}, \cite{Skiadopoulos1999}.} El \'indice de volatilidad VIX\index{VIX (CBOE Volatility Index)} (CBOE\index{Chicago Board Options Exchange (CBOE)} Volatility Index (en ingl\'es), o el \'Indice de Volatilidad del CBOE (en espa\~{n}ol), tambi\'en conocido como ``\'indice de incertidumbre\index{indice de incertidumbre @ \'indice de incertidumbre}'' o ``\'indice de medida de miedo\index{indice de medida de miedo @ \'indice de medida de miedo}'')\footnote{\, V\'ease, por ejemplo, \cite{Aijo2008}, \cite{Corrado2005}, \cite{Fleming1995}, \cite{Maghrebi2007}, \cite{Shaikh2015}, \cite{Siriopoulos2009}, \cite{Skiadopoulos2004}, \cite{Whaley2000}, \cite{Whaley2009}.} y otros \'indices de volatilidad\index{indice de volatilidad @ \'indice de volatilidad}\footnote{\, Por ejemplo, RVX (CBOE Russell 2000 Volatility Index), VXEEM (CBOE Emerging Markets ETF Volatility Index), TYVIX (CBOE/CBOT 10-year U.S. Treasury Note Volatility Index), GVZ (CBOE Gold ETF Volatility Index), EUVIX (CBOE/CME FX Euro Volatility Index), VXGOG (CBOE Equity VIX on Google), VVIX (CBOE VIX of VIX Index), etc. (todos los nombres aqu\'i se dan en ingles).} y derivados\index{derivado} (opciones\index{opci\'on} y futuros\index{futuro}) sobre \'indices de volatilidad\index{indice de volatilidad @ \'indice de volatilidad} tales como el VIX\index{VIX (CBOE Volatility Index)} proporcionan medios para el trading de volatilidad.

\subsection{Estrategia: Trading de la base de futuros del VIX\index{trading de la base de futuros del VIX}}

{}Esta es esencialmente una estrategia de reversi\'on a la media\index{estrategia de reversi\'on a la media}. Est\'a enraizada en la observaci\'on emp\'irica (v\'ease, por ejemplo, \cite{Mixon2007}, \cite{Nossman2009}, \cite{Simon2014})\footnote{\, Para algunos estudios adicionales sobre el trading de la base de futuros del VIX\index{base de futuros del VIX} y temas relacionados, v\'ease, por ejemplo, \cite{Buetow2016}, \cite{Donninger2014}, \cite{Fu2016}, \cite{Lee2017}, \cite{Zhang2010}, \cite{ZhangZhu2006}.} que la base de los futuros del VIX\index{base de futuros del VIX} (definida abajo) no tiene poder de pron\'ostico para los cambios posteriores del VIX pero tiene un poder de pron\'ostico sustancial para los cambios posteriores en el precio de futuros del VIX\index{precio de futuros del VIX}. La base de los futuros del VIX\index{base de futuros del VIX} $B_{VIX}$ (para nuestros prop\'ositos aqu\'i) se define como
\begin{eqnarray}
&& B_{VIX} = P_{UX1} - P_{VIX} \\
&& D = \frac{B_{VIX}}{T} \
\end{eqnarray}
Aqu\'i: $P_{UX1}$ es el precio del contrato de futuros del primer mes\index{futuros del primer mes} del VIX;\footnote{\, UX1 tiene aproximadamente 1 mes hasta su madurez\index{madurez}, UX2 tiene aproximadamente 2 meses, etc.} $P_{VIX}$ es el precio del VIX\index{precio del VIX}; $D$ es el valor del roll diario\index{valor del roll diario}; y $T$ es el n\'umero de d\'ias h\'abiles hasta la liquidaci\'on\index{liquidaci\'on} (que se supone que es al menos 10). Emp\'iricamente, los precios de los futuros\index{precio de los futuros} tienden a caer cuando la base es positiva y suben cuando la base es negativa (reversi\'on a la media\index{reversi\'on a la media}). Entonces, la estrategia consiste en vender los futuros del VIX\index{vender futuros del VIX} cuando la curva de futuros del VIX\index{curva de futuros del VIX} tiene pendiente positiva (esto es tambi\'en conocido como ``contango''\index{contango}, entonces la base es positiva), y comprar los futuros del VIX\index{futuros del VIX} cuando la curva de futuros del VIX\index{curva de futuros del VIX} presenta pendiente negativa (esto es tambi\'en conocido como ``backwardation''\index{backwardation}, entonces la base es negativa). Aqu\'i se presenta una simple regla de trading\index{regla de trading} (v\'ease, por ejemplo, \cite{Simon2014}):
\begin{eqnarray}
\mbox{Regla} = \begin{cases}
          \text{Abrir posici\'on larga en UX1}  &  \mbox{ si }  D < -0.10   \\
          \text{Cerrar posici\'on larga en UX1}  &  \mbox{ si }  D > -0.05   \\
          \text{Abrir posici\'on corta en UX1}  &  \mbox{ si } D > 0.10 \\
          \text{Cerrar posici\'on corta en UX1}  &  \mbox{ si } D < 0.05 \\
        \end{cases}
\end{eqnarray}
Una posici\'on en UX1 corta (larga) est\'a expuesta a un riesgo\index{riesgo} de un aumento (disminuci\'on) repentino en la volatilidad, que normalmente se produce durante una venta masiva\index{venta masiva} (un rally\index{rally}) en el mercado accionario\index{mercado accionario}, entonces, el riesgo\index{riesgo} puede ser cubierto, por ejemplo, vendiendo (comprando) futuros mini-S\&P 500\index{futuros mini-S\&P 500}.\footnote{\, Generalmente, el VIX\index{VIX (CBOE Volatility Index)} y el mercado accionario\index{mercado accionario} son anti-correlacionados.} El ratio de cobertura\index{ratio de cobertura} se puede estimar, por ejemplo, en base a una regresi\'on serial\index{regresi\'on serial} hist\'orica de los cambios en el precio de futuros del VIX\index{precio de futuros del VIX} sobre los retornos del contrato de futuros mini-S\&P\index{futuros mini-S\&P 500} del primer mes\index{contrato de futuros, primer mes}.\footnote{\, Por detalles, v\'ease, por ejemplo, \cite{Simon2014}.}

\subsection{Estrategia: Carry de volatilidad\index{carry de volatilidad} con dos ETNs\index{nota de intercambio cotizada (ETN)}}\label{ETNs.carry}

{}El VXX es una nota de intercambio cotizada (ETN, por sus siglas en ingl\'es)\index{nota de intercambio cotizada (ETN)} que ``replica'' al VIX\index{VIX (CBOE Volatility Index)} mediante un portafolio\index{portafolio} de contratos de futuros del VIX\index{contrato de futuros del VIX} con vencimientos a corto plazo (meses 1 y 2). Para mantener una madurez\index{madurez} constante, al cierre\index{cierre} de cada d\'ia, una porci\'on de los futuros\index{futuro} de vencimiento m\'as corto se vende y con los ingresos se compran los futuros\index{futuro} de vencimiento m\'as largo. Dado que la curva de futuros del VIX\index{curva de futuros del VIX} se encuentra en contango\index{contango} la mayor parte del tiempo, los futuros\index{futuro} de vencimiento m\'as largo tienen un precio m\'as alto que los futuros\index{futuro} de vencimiento m\'as corto, por lo que este rebalanceo\index{rebalanceo} genera una disminuci\'on en el valor del VXX a lo largo del tiempo, lo que se conoce como la p\'erdida por roll (o contango)\index{perdida, de contango @ p\'erdida, de contango}\index{perdida, de roll @ p\'erdida, de roll}. Adem\'as, a medida que pasa el tiempo, los futuros\index{futuro} convergen al spot\index{spot} (VIX\index{VIX (CBOE Volatility Index)}), por lo tanto, el VXX pierde valor mientras la curva de futuros del VIX\index{curva de futuros del VIX} se encuentra en contango\index{contango}. El VXZ es otro ETN\index{nota de intercambio cotizada (ETN)} que replica al VIX\index{VIX (CBOE Volatility Index)} mediante un portafolio\index{portafolio} de futuros del VIX\index{futuros del VIX} de vencimiento medio (de 4 a 7 meses). El VXZ tambi\'en sufre la p\'erdida de roll\index{perdida de roll @ p\'erdida de roll}, pero en menor grado que el VXX, dado que la pendiente de la curva de los futuros del VIX\index{futuros del VIX} en contango\index{contango} disminuye con la madurez\index{madurez}.\footnote{\, Para algunos estudios sobre ETNs de volatilidad\index{ETN de volatilidad} y temas relacionados, v\'ease, por ejemplo,\cite{Alexander2012}, \cite{Avellaneda2018}, \cite{DeLisle2014}, \cite{Deng2012}, \cite{Eraker2014}, \cite{Gehricke2018}, \cite{Grasselli2018}, \cite{Hancock2013}, \cite{Husson2011}, \cite{Liu2012}, \cite{Liu2018}, \cite{Moran2007}.} Una estrategia b\'asica entonces es vender el VXX y comprar el VXZ con un ratio de cobertura\index{ratio de cobertura} que se puede determinar mediante una regresi\'on serial\index{regresi\'on serial}.\footnote{\, Tenemos $h = \beta =  \rho\sigma_{X}/\sigma_{Z}$, en donde: $h$ (conocido como el ratio de cobertura \'optimo\index{ratio de cobertura \'optimo}) es el n\'umero del VXZ para comprar por cada VXX vendido en corto; $\beta$ es el coeficiente (para los retornos del VXZ) de la regresi\'on serial\index{regresi\'on serial} (con el intercepto\index{intercepto}) de los retornos del VXX sobre los retornos del VXZ; $\sigma_{X}$ y $\sigma_{Z}$ son las volatilidades hist\'oricas\index{volatilidad hist\'orica} del VXX y del VXZ, respectivamente; y $\rho$ es la correlaci\'on hist\'orica\index{correlaci\'on hist\'orica} de pares entre el VXX y el VXZ.} Sin embargo, esta estrategia no est\'a exenta de riesgos\index{riesgo}. Puede haber picos\index{pico} a corto plazo en el VXX (los picos\index{pico} correspondientes en el VXZ usualmente son considerablemente m\'as peque\~{n}os), lo que puede conducir en corto plazo a sustanciales drawdowns en el P\&L\index{drawdown en el P\&L}, incluso si la estrategia es rentable a nivel general en el largo plazo.

\subsubsection{Estrategia: Cobertura\index{cobertura} del VXX con futuros del VIX\index{futuros del VIX}}\label{vol.hedge}

{}En lugar de usar una posici\'on larga en el VXZ para cubrir la posici\'on corta en el VXX, uno puede usar directamente una cesta\index{cesta} de, por ejemplo, futuros del VIX\index{futuros del VIX} de mediana madurez.\footnote{\, \'Estos pueden tener fechas de madurez\index{madurez} de, por ejemplo, 4 a 7 meses (imitando as\'i la composici\'on del VXZ).} Los $N$ futuros del VIX\index{futuros del VIX} tienen ciertas ponderaciones $w_i$. Estas ponderaciones se pueden fijar de varias maneras, por ejemplo, minimizando el error de rastreo\index{error de rastreo}, es decir, corriendo una regresi\'on serial\index{regresi\'on serial} (con el intercepto\index{intercepto}) de los retornos del VXX sobre los retornos de los $N$ futuros\index{retorno, de los futuros}. Entonces tenemos:
\begin{equation}\label{opt.hedge}
 w_i = \sigma_X~\sum_{j=1}^N C^{-1}_{ij}\sigma_j \rho_j
\end{equation}
Aqu\'i: $\rho_i$ es la correlaci\'on hist\'orica\index{correlaci\'on hist\'orica} de pares entre los futuros\index{futuro} etiquetados por $i$ y el VXX; $C_{ij}$ es la matriz de covarianza muestral\index{matriz de covarianza muestral} con dimensiones $N\times N$ de los $N$ futuros\index{futuro} ($\sigma_i^2 = C_{ii}$ es la varianza hist\'orica\index{varianza hist\'orica} para los futuros\index{futuro} etiquetados por $i$); y $\sigma_X$ es la volatilidad hist\'orica\index{volatilidad hist\'orica} del VXX. Ciertos $w_i$ pueden resultar negativos. Esto no es necesariamente un problema, pero uno puede desear imponer l\'imites as\'i $w_i\geq 0$. Adem\'as, uno puede desear que la estrategia sea d\'olar-neutral, lo que equivaldr\'ia a imponer la restricci\'on
\begin{eqnarray}
 && \sum_{i = 1}^{N} w_i = 1
\end{eqnarray}
que los ratios de cobertura \'optimos\index{ratio de cobertura \'optimo} (\ref{opt.hedge}) generalmente no satisfacen. Adem\'as, en lugar de minimizar el error de rastreo\index{error de rastreo}, uno puede desear minimizar la varianza\index{varianza} de todo el portafolio\index{portafolio}. Entre otras. El portafolio\index{portafolio} se puede rebalancear mensualmente o con mayor frecuencia.

\subsection{Estrategia: Prima de riesgo de volatilidad\index{prima de riesgo de volatilidad}}

{}La evidencia emp\'irica indica que la volatilidad impl\'icita\index{volatilidad impl\'icita} tiende a ser m\'as alta que la volatilidad realizada\index{volatilidad realizada} la mayor parte del tiempo, lo cual se conoce como la ``prima de riesgo de volatilidad\index{prima de riesgo de volatilidad}''.\footnote{\, Por alguna literatura pertinente, v\'ease, por ejemplo, \cite{Bakshi2003a}, \cite{Bollerslev2011}, \cite{Carr2009}, \cite{Carr2016}, \cite{Christensen1998}, \cite{Eraker2009}, \cite{Ge2016}, \cite{Miao2012}, \cite{Prokopczuk2014}, \cite{Saretto2009}, \cite{Todorov2010}.} En pocas palabras, la mayor parte del tiempo, las opciones\index{opci\'on} tienen un precio m\'as alto que los precios que uno esperar\'ia en funci\'on de la volatilidad realizada\index{volatilidad realizada}, entonces la idea es vender la volatilidad. Por ejemplo, el trader puede vender conos\index{cono} sobre opciones del S\&P 500\index{opciones del S\&P 500}. Como un posible proxy de la prima de riesgo de volatilidad\index{prima de riesgo de volatilidad}, el trader puede, por ejemplo, usar la diferencia entre el VIX\index{VIX (CBOE Volatility Index)} al inicio del mes en curso y la volatilidad realizada\index{volatilidad realizada} (en \%, dado que el VIX\index{VIX (CBOE Volatility Index)} es expresado en \%) de los retornos diarios del S\&P 500\index{S\&P 500} desde el inicio del mes en curso. Si el margen\index{margen} es positivo, el trader vende conos\index{cono}. Si hay un pico de volatilidad\index{pico de volatilidad} (lo que suele suceder si el mercado\index{mercado} se vuelve muy bajista), la estrategia perder\'a dinero. Esta estrategia es rentable en mercados laterales\index{mercado lateral}.\footnote{\, Tambi\'en, las opciones de \'indices\index{opci\'on de indice @ opci\'on de \'indice} son m\'as adecuadas para esta estrategia que las opciones de acciones individuales\index{opci\'on de acci\'on individual} dado que las primeras normalmente tienen mayor prima de riesgo de volatilidad (v\'ease Subsecci\'on \ref{sub.dispersion}).}

\subsubsection{Estrategia: Prima de riesgo de volatilidad\index{prima de riesgo de volatilidad} con cobertura de Gamma\index{cobertura de Gamma}}\label{sub.Gamma.scalping}

{}Los conos ATM\index{cono ATM} en la estrategia anterior son Delta-neutral\index{estrategia, Delta-neutral}.\footnote{\, Algunas de las Griegas\index{Griegas} de las opciones\index{opci\'on} son: $\Theta = {\partial V/\partial t}$ (Theta\index{Theta}), $\Delta = {\partial V/ \partial S}$ (Delta\index{Delta}), $\Gamma = {\partial^2 V/ \partial S^2}$ (Gamma\index{Gamma}), $\nu =  {\partial V/ \partial \sigma}$ (Vega\index{Vega}). Aqu\'i: $V$ es el valor de la opci\'on\index{opci\'on}; $t$ es el tiempo; $S$ es el precio del subyacente\index{subyacente}; $\sigma$ es la volatilidad impl\'icita\index{volatilidad impl\'icita}.} Entonces, esta estrategia es una ``juego a Vega\index{juego a Vega}'', es decir, el trader est\'a vendiendo Vega\index{vendiendo Vega}. Si el subyacente\index{subyacente} (S\&P 500\index{S\&P 500}) se mueve, el cono\index{cono} corto ya no es Delta-neutral: si el subyacente\index{subyacente} sube (baja), el Delta\index{Delta} se vuelve negativo (positivo). Entonces una variaci\'on de esta estrategia es utilizar la cobertura de Gamma\index{cobertura de Gamma} para mantener la estrategia cercana a Delta-neutral, que se logra comprando (vendiendo) el subyacente\index{subyacente} si este se mueve hacia arriba (abajo). Entonces esto se convierte en un ``juego a Theta\index{juego a Theta}'', es decir, la estrategia ahora apunta a capitalizar el decaimiento de Theta\index{decaimiento de Theta} en el valor de las opciones\index{opci\'on} vendidas. Entonces, el precio de esto es el costo de la cobertura de Gamma\index{cobertura de Gamma}, que reduce el P\&L\index{P\&L}. En tanto y cuanto el subyacente\index{subyacente} se aleja cada vez m\'as del precio de ejercicio\index{precio de ejercicio} de las opciones put y opciones call vendidas\index{opci\'on call}\index{opci\'on, call}\index{opci\'on, put}, la cobertura de Gamma\index{cobertura de Gamma} se vuelve m\'as y m\'as cara y eventualmente superar\'a el cr\'edito de las opciones vendidas, al punto al cual la estrategia empieza a perder dinero.

\subsection{Estrategia: Asimetr\'ia de volatilidad\index{asimetr\'ia de volatilidad} -- combo largo}

{}Las opciones put\index{opci\'on put} OTM\index{opci\'on, OTM (fuera del dinero)} con el precio del subyacente\index{subyacente} dado por $S_0 = K + \kappa$ tienden a tener un precio m\'as alto que las opciones call\index{opci\'on call} OTM\index{opci\'on, OTM (fuera del dinero)} con el precio del subyacente\index{subyacente} dado por $S_0 = K - \kappa$ (aqu\'i $K$ es el precio de ejercicio\index{precio de ejercicio}, y $\kappa > 0$ es la distancia desde el precio de ejercicio\index{precio de ejercicio}). Es decir, con todo lo dem\'as igual, la volatilidad impl\'icita\index{volatilidad impl\'icita} de las opciones put\index{opci\'on put} es mayor que la de las opciones call\index{opci\'on, call}.\footnote{\, Para algunos estudios pertinentes, v\'ease, por ejemplo,\cite{Bondarenko2014}, \cite{Chambers2014}, \cite{Corrado1997}, \cite{Damghani2013}, \cite{DeMiguel2013}, \cite{Doran2010}, \cite{Doran2007}, \cite{Fengler2012}, \cite{Flint2017}, \cite{Jackwerth2000}, \cite{Kozhan2013}, \cite{Liu2016}, \cite{Mixon2011}, \cite{Zhang2008}.} La estrategia de combo largo\index{combo largo} (v\'ease la Subsecci\'on \ref{sub.long.risk.rev}), en donde el trader compra opciones call\index{opci\'on call} OTM\index{opci\'on, OTM (fuera del dinero)} y vende opciones put\index{opci\'on put} OTM\index{opci\'on, OTM (fuera del dinero)}, captura esta asimetr\'ia\index{asimetr\'ia}. Sin embargo, esta es una estrategia direccional\index{estrategia direccional} -- pierde dinero si el precio del subyacente\index{subyacente} cae por debajo de $K_{put} - C$, en donde $K_{put}$ es el precio de ejercicio\index{precio de ejercicio} del put, y $C > 0$ es la prima\index{prima} de las opciones put\index{opci\'on put} menos la prima\index{prima} de las opciones call\index{opci\'on, call}.

\subsection{Estrategia: Trading de volatilidad con swaps de varianza\index{swap de varianza}}

{}Un problema con el trading de volatilidad utilizando opciones\index{opci\'on} es la necesidad de (casi continuamente) rebalancear la posici\'on para evitar una exposici\'on direccional\index{exposici\'on direccional},\footnote{\, V\'ease la Subsecci\'on \ref{sub.Gamma.scalping} para una estrategia de cobertura de Delta\index{estrategia de cobertura de Delta} (tambi\'en conocida como ``Gamma scalping\index{Gamma scalping}'').} lo que en la pr\'actica puede ser engorroso y costoso. Para evitar la necesidad de cobertura de Delta\index{cobertura de Delta}, uno puede hacer apuestas de volatilidad usando swaps de varianza\index{swap de varianza}. Un swap de varianza\index{swap de varianza} es un contrato de derivados\index{contrato de derivado} cuyo pago\index{pago} $P(T)$ en la madurez\index{madurez} $T$ es proporcional a la diferencia entre la varianza realizada\index{varianza realizada} $v(T)$ del subyacente\index{subyacente} y el nivel de ejercicio de la varianza\index{nivel de ejercicio de la varianza} preestablecida $K$:
\begin{eqnarray}	
 && P(T) = N \times \left(v(T) - K \right) \\
 && v(T) = \frac{F}{T} \sum_{t = 1}^{T} R^2(t) \label{var.swap}\\
 && R(t) = \ln \left[\frac{S(t)}{S(t-1)} \right]
\end{eqnarray}
Aqu\'i: $t=0,1,\dots,T$ etiqueta los puntos de la muestra (por ejemplo, d\'ias de trading\index{dias de trading @ d\'ias de trading}); $S(t)$ es el precio del subyacente\index{subyacente} al momento $t$; $R(t)$ es el retorno logar\'itmico\index{retorno logar\'itmico} desde $t-1$ a $t$; $F$ es el factor de anualizaci\'on\index{factor de anualizaci\'on} (as\'i, si $t$ etiqueta d\'ias de trading\index{dias de trading @ d\'ias de trading}, entonces $F = 252$); y $N$ es el ``nocional de varianza\index{nocional de varianza}'', el cual es preestablecido. Tenga en cuenta que en la Ecuaci\'on (\ref{var.swap}), la media de $R(t)$ sobre el per\'iodo $t=1$ a $t=T$ no se resta, y por lo tanto, tenemos $T$ en el denominador.\footnote{\, Si la media se resta, entonces el denominador ser\'ia $T-1$.} Un swap de varianza\index{swap de varianza} largo (corto) es una apuesta a que la volatilidad futura realizada\index{volatilidad realizada} ser\'a mayor (menor) que la volatilidad impl\'icita\index{volatilidad impl\'icita} actual. Swaps de varianza\index{swap de varianza} largos (cortos) por lo tanto, se pueden usar en lugar de, por ejemplo, conos\index{cono} largos (cortos) para tomar posiciones largas (cortas) sobre la volatilidad. Por ejemplo, la estrategia de dispersi\'on\index{estrategia de dispersi\'on} en la Subsecci\'on \ref{sub.dispersion} puede ser ejecutada vendiendo un swap de varianza\index{swap de varianza} sobre un \'indice\index{indice @ \'indice} y comprando swaps de varianza\index{swap de varianza} sobre los constituyentes del \'indice\index{constituyentes del \'indice} (comparado con vender y comprar conos\index{cono}).\footnote{\, Por algunas publicaciones sobre swaps de varianza\index{swap de varianza}, v\'ease, por ejemplo, \cite{Ait-Sahalia2015}, \cite{Bernard2014}, \cite{Broadie2008}, \cite{Bossu2006}, \cite{Carr2007}, \cite{CarrLee2009}, \cite{Carr2012}, \cite{Demeterfi1999}, \cite{Elliott2007}, \cite{Filipovic2016}, \cite{Hafner2007}, \cite{Hardle2010}, \cite{Jarrow2013}, \cite{Konstantinidi2016}, \cite{Leontsinis2016}, \cite{Liverance2010}, \cite{Martin2011}, \cite{Rujivan2012}, \cite{Schoutens2005}, \cite{Wystup2014}, \cite{Zhang2014}, \cite{Zheng2014}.}

\newpage

\section{Divisas (FX)\index{divisas (FX)}}\label{sec.FX}

{}\subsection{Estrategia: Medias m\'oviles\index{media m\'ovil} con filtro HP\index{filtro HP}}\label{sub.fx.momentum}

{}En la Subsecci\'on \ref{sub.2.mov.avg} hablamos de una estrategia de trading\index{estrategia de trading} para acciones en la cual la se\~{n}al de trading\index{senzal de trading @ se\~{n}al de trading} se basa en la intersecci\'on de 2 medias m\'oviles\index{media m\'ovil} (una m\'as corta y otra m\'as larga). Un enfoque similar puede aplicarse a FX (por sus siglas en ingl\'es)\index{divisas (FX)} tambi\'en. Sin embargo, las series de tiempo de las tasas spot de FX\index{series de tiempo de las tasas spot de FX} tienden a ser bastante ruidosas, lo que puede conducir a se\~{n}ales falsas\index{senzal falsa @ se\~{n}al falsa} basadas en medias m\'oviles\index{media m\'ovil}. Para mitigar esto, antes de calcular las medias m\'oviles\index{media m\'ovil}, primero se puede filtrar el ruido\index{ruido} de alta frecuencia utilizando, por ejemplo, el conocido filtro de Hodrick-Prescott (HP)\index{filtro, Hodrick-Prescott}.\footnote{\, Tambi\'en conocido como el m\'etodo de Whittaker-Henderson\index{metodo de Whittaker-Henderson @ m\'etodo de Whittaker-Henderson} en las ciencias actuariales. Para algunos estudios pertinentes, v\'ease, por ejemplo, \cite{Baxter1999}, \cite{Bruder2013}, \cite{Dao2014}, \cite{Ehlgen1998}, \cite{Harris2009}, \cite{Harvey2008}, \cite{Henderson1924}, \cite{Henderson1925}, \cite{Henderson1938}, \cite{Hodrick1997}, \cite{Joseph1952}, \cite{Lahmiri2014}, \cite{Mcelroy2008}, \cite{Weinert2007}, \cite{Whittaker1923}, \cite{Whittaker1924}.} Entonces, el componente de la tendencia\index{componente de la tendencia} de baja frecuencia restante (a diferencia de la tasa spot sin procesar\index{tasa spot sin procesar}) se puede utilizar para calcular las medias m\'oviles\index{media m\'ovil} y generar la se\~{n}al de trading\index{senzal de trading @ se\~{n}al de trading} (v\'ease, por ejemplo, \cite{Harris2009}). El filtro HP\index{filtro HP} est\'a dado por:
 \begin{eqnarray}	
  && S(t) = S^*(t) + \nu(t)\\
  && g = \sum_{t = 1}^T \left[S(t) - S^*(t)\right]^2 + \lambda~\sum_{t = 2}^{T-1} \left[S^*(t+1) - 2S^*(t) + S^*(t - 1)\right]^2 \label{HP}\\
  && g \rightarrow\mbox{min}\label{HP1}
\end{eqnarray}
Aqu\'i: la funci\'on objetivo\index{funci\'on objetivo} $g$ se minimiza con respecto al conjunto de $T$ valores de $S^*(t)$, $t=1,\dots,T$; $S(t)$ es la tasa spot de FX\index{tasa spot de FX} al momento $t$; $S^*(t)$ es el componente de frecuencia m\'as baja (``regular''); $\nu(t)$ es el componente de frecuencia m\'as alta (``irregular''), que es tratado como ruido\index{ruido}; el primer t\'ermino en la Ecuaci\'on (\ref{HP}) minimiza el ruido\index{ruido}, mientras que el segundo t\'ermino (basado en la segunda derivada discretizada de $S^*(t)$) penaliza la variaci\'on en $S^*(t)$; y $\lambda$ es el par\'ametro de suavizaci\'on\index{par\'ametro de suavizaci\'on}. No hay un m\'etodo ``fundamental'' para fijar $\lambda$. A veces (pero no siempre) se establece como $\lambda = 100 \times n^2$, en donde $n$ es la frecuencia de datos medida en base anual (v\'ease, por ejemplo, \cite{Baxter1999} para m\'as detalles). Por lo tanto, para los datos mensuales, que es lo que normalmente se utiliza en este contexto, $n=12$. El per\'iodo de estimaci\'on\index{periodo de estimacion @ per\'iodo de estimaci\'on} generalmente abarca varios a\~{n}os (de datos mensuales). Una vez que $S^*(t)$ se determina, dos medias m\'oviles\index{media m\'ovil} $\mbox{MA}(T_1)$ y $\mbox{MA}(T_2)$, $T_1 < T_2$, se calculan en base a $S^*(t)$. Entonces, como antes, $\mbox{MA}(T_1) > \mbox{MA}(T_2)$ es una se\~{n}al de compra\index{senzal de compra @ se\~{n}al de compra}, y $\mbox{MA}(T_1) < \mbox{MA}(T_2)$ es una se\~{n}al de venta\index{senzal de venta @ se\~{n}al de venta}.

\subsection{Estrategia: Carry trade\index{carry trade}}\label{sub.fx.carry}

{}De conformidad con la ``Paridad de Tasas de Inter\'es no Cubierta\index{Paridad de Tasas de Inter\'es no Cubierta (UIRP)}'' (UIRP\index{Paridad de Tasas de Inter\'es no Cubierta (UIRP)}, por sus siglas en ingl\'es), cualquier exceso de inter\'es\index{interes @ inter\'es} obtenido en un pa\'is en comparaci\'on con otro debido a una diferencia entre las tasas de inter\'es libre de riesgo\index{tasa de inter\'es libre de riesgo}, ser\'ia compensado precisamente por la depreciaci\'on en la tasa de FX\index{tasa de FX} entre sus monedas\index{moneda}:
\begin{eqnarray}\label{UIP}	
 && (1 + r_d) = {E_t(S(t + T))\over S(t)}~(1 + r_f)
\end{eqnarray}
Aqu\'i: $r_d$ es la tasa de inter\'es dom\'estica\index{tasa de inter\'es dom\'estica}; $r_f$ es la tasa de inter\'es extranjera\index{tasa de inter\'es extranjera}; se supone que tanto $r_d$ como $r_f$ son constantes durante el per\'iodo de composici\'on\index{periodo de composicion @ per\'iodo de composici\'on} $T$; $S(t)$ es la tasa de FX spot\index{tasa de FX spot} al momento $t$, que es el valor de 1 unidad de la moneda extranjera\index{moneda extranjera} en unidades de la moneda dom\'estica\index{moneda dom\'estica}; y $E_t(S(t + T))$ es la tasa de FX spot\index{tasa de FX spot} futura (al momento $t + T$) esperada al momento $t$.\footnote{\, De esta forma, 1 USD\index{USD (d\'olar estadounidense)} invertido en el tiempo $t$ en un activo libre de riesgo\index{activo libre de riesgo} en los Estados Unidos pagar\'ia $(1+r_d)$ USD\index{USD (d\'olar estadounidense)} en el tiempo $t + T$. De forma alternativa, 1 USD\index{USD (d\'olar estadounidense)} comprar\'ia $1/S(t)$ JPY\index{JPY (yen japon\'es)} en el tiempo $t$, cuya suma podr\'ia invertirse en un activo libre de riesgo\index{activo libre de riesgo} en Jap\'on en el momento $t$, que valdr\'ia $(1/S(t)) \times (1+r_f)$ JPY\index{JPY (yen japon\'es)} al momento $t+T$, que a su vez podr\'ia ser cambiado por $(E_t(S(t + T))/S(t)) \times (1+r_f)$ USD\index{USD (d\'olar estadounidense)} al momento $t+T$. Exigiendo que las inversiones\index{inversi\'on} de los Estados Unidos y Jap\'on proporcionen el mismo rendimiento conduce a la Ecuaci\'on (\ref{UIP}).} La UIRP\index{Paridad de Tasas de Inter\'es no Cubierta (UIRP)} no siempre se mantiene, dando lugar a oportunidades de trading -- las cuales {\em no} son oportunidades de arbitraje sin riesgo\index{oportunidad de arbitraje sin riesgo} (v\'ease abajo). De esta forma, la UIRP\index{Paridad de Tasas de Inter\'es no Cubierta (UIRP)} implica que las monedas\index{moneda} de alta tasa de inter\'es\index{tasa de inter\'es} deber\'ian depreciarse con respecto a las monedas\index{moneda} de baja tasa de inter\'es\index{tasa de inter\'es}, aunque emp\'iricamente en promedio, lo opuesto tiende a ocurrir, es decir, tales monedas\index{moneda} tienden a apreciarse (hasta cierto punto).\footnote{\, Esto se conoce como ``rompecabezas/anomal\'ia de premio/descuento a plazo''\index{anomal\'ia, descuento a plazo} o ``enigma de Fama\index{enigma de Fama}''. Por alguna literatura sobre la UIRP\index{Paridad de Tasas de Inter\'es no Cubierta (UIRP)} y t\'opicos relacionados, v\'ease, por ejemplo, \cite{Anker1999}, \cite{Ayuso1996}, \cite{Bacchetta2006}, \cite{Bacchetta2010}, \cite{Baillie2000}, \cite{Bekaert2007}, \cite{Beyaert2007}, \cite{Bilson1981}, \cite{Chaboud2005}, \cite{Engel1996}, \cite{Fama1984}, \cite{Frachot1996}, \cite{Froot1990}, \cite{Hansen1980}, \cite{Harvey2015}, \cite{Hodrick1987}, \cite{Ilut2012}, \cite{Lewis1995}, \cite{Lustig2007}, \cite{Mark2001}, \cite{Roll2008}.} As\'i, la estrategia b\'asica de carry\index{estrategia de carry} consiste en vender los forwards\index{forward} sobre las monedas\index{moneda} que presentan un premio a plazo\index{premio a plazo}, es decir, la tasa de FX a plazo\index{tasa de FX a plazo} $F(t, T)$ excede a la tasa de FX spot\index{tasa de FX spot} $S(t)$, y en comprar los forwards\index{forward} sobre las monedas\index{moneda} que se encuentran con un descuento a plazo\index{descuento a plazo}, es decir, la tasa de FX a plazo\index{tasa de FX a plazo} $F(t, T)$ es menor que la tasa de FX spot\index{tasa de FX spot} $S(t)$.\footnote{\, Ignorando los costos de transacci\'on\index{costos de transacci\'on}, esto es equivalente a pedir prestado (prestar) monedas\index{moneda} con tasas de inter\'es\index{tasa de inter\'es} bajas (altas) sin cubrir el riesgo de tasa de FX\index{riesgo de tasa de FX}.} La tasa de FX a plazo\index{tasa de FX a plazo} es dada por\footnote{\, Esto se conoce como ``Paridad de Tasas de Inter\'es Cubierta'' (CIRP, por sus siglas en ingl\'es)\index{Paridad de Tasas de Inter\'es Cubierta (CIRP)}. Tenga en cuenta que, suponiendo que la Ecuaci\'on (\ref{CIRP}) se mantiene (v\'ease abajo), cuando la UIRP\index{Paridad de Tasas de Inter\'es no Cubierta (UIRP)} (es decir, la Ecuaci\'on (\ref{UIP})) no se sostiene, $F(t, T) \neq E_t(S(t + T))$.}
\begin{equation}\label{CIRP}
 F(t, T) = S(t)~{{1 + r_d}\over{1 + r_f}}
\end{equation}
Como se mencion\'o anteriormente, la estrategia de carry\index{estrategia de carry}\footnote{\, Por algunas publicaciones sobre el carry trade de divisas\index{carry trading de divisas} y t\'opicos relacionados, v\'ease, por ejemplo, \cite{Bakshi2013}, \cite{Brunnermeier2008}, \cite{Burnside2011}, \cite{Burnside2007}, \cite{Burnside2008}, \cite{Clarida2009}, \cite{Deardorff1979}, \cite{Doskov2015}, \cite{Hau2014}, \cite{Jurek2014}, \cite{Lustig2011}, \cite{Lustig2014}, \cite{Olmo2009}, \cite{Ready2017}, \cite{Rhee1992}.} no est\'a exenta de riesgos\index{riesgo}: esta operaci\'on puede generar p\'erdidas si la moneda\index{moneda} a la cual se pide el pr\'estamo (se presta) de repente se aprecia (se deprecia) con respecto a su contraparte, es decir, est\'a expuesta al riesgo de tasa de FX\index{riesgo de tasa de FX}. Por otro lado, si pedimos prestada la moneda\index{moneda} de baja tasa de inter\'es\index{tasa de inter\'es} con fecha de vencimiento\index{fecha de vencimiento} $T$, e invertimos los fondos en la moneda\index{moneda} de alta tasa de inter\'es\index{tasa de inter\'es}, y cubrimos esta posici\'on con un contrato a plazo\index{contrato a plazo} para intercambiar la moneda\index{moneda} de alta tasa de inter\'es\index{tasa de inter\'es} por la moneda\index{moneda} de baja tasa de inter\'es\index{tasa de inter\'es} en la fecha de vencimiento\index{vencimiento} $T$ (para que podamos cubrir el pr\'estamo\index{prestamo @ pr\'estamo}), ignorando los costos de transacci\'on\index{costos de transacci\'on} (y otras sutilezas como impuestos, etc.), esta es una posici\'on libre de riesgo\index{posici\'on libre de riesgo} y cualquier ganancia de los mismos equivaldr\'ia a un arbitraje sin riesgo\index{arbitraje sin riesgo}. Entonces tenemos la Ecuaci\'on (\ref{CIRP}), que es una condici\'on de no arbitraje libre de riesgo\index{condici\'on de no arbitraje libre de riesgo}.\footnote{\, No obstante, las desviaciones de la CIRP\index{Paridad de Tasas de Inter\'es Cubierta (CIRP)} (es decir, la Ecuaci\'on (\ref{CIRP})) ocurren, lo que da lugar a arbitraje de inter\'es cubierto\index{arbitraje de inter\'es cubierto}. V\'ease, por ejemplo, \cite{Akram2008}, \cite{Avdjiev2016}, \cite{Baba2009}, \cite{Boulos1994}, \cite{Clinton1988}, \cite{Coffey2009}, \cite{Cosandier1981}, \cite{Du2018}, \cite{Duffie2017}, \cite{Frenkel1975}, \cite{Frenkel1981}, \cite{Liao2016}, \cite{Mancini-Griffoli2011}, \cite{Popper1993}, \cite{Rime2017}.}

\subsubsection{Estrategia: Carry alto-menos-bajo\index{carry alto-menos-bajo}}

{}La estrategia de carry\index{estrategia de carry} discutida anteriormente se puede aplicar a monedas extranjeras\index{moneda extranjera} individuales. Tambi\'en se puede aplicar de forma transversal a m\'ultiples monedas extranjeras\index{moneda extranjera}. Sea $s(t) = \ln(S(t))$ (tasa de FX spot logar\'itmica\index{tasa de FX spot logar\'itmica}) y $f(t, T) = \ln(F(t, T))$ (tasa de FX a plazo logar\'itmica\index{tasa de FX a plazo logar\'itmica}). El descuento a plazo\index{descuento a plazo} $D(t, T)$ se define como
\begin{equation}
 D(t, T) = s(t) - f(t, T)
\end{equation}
De conformidad con la CIRP\index{Paridad de Tasas de Inter\'es Cubierta (CIRP)}, en la Ecuaci\'on (\ref{CIRP}), tenemos
\begin{equation}
 D(t, T) = \ln\left({{1 + r_f}\over{1+r_d}}\right) \approx r_f - r_d
\end{equation}
Cuando el descuento a plazo\index{descuento a plazo} es positivo, compramos un forward\index{forward} (es decir, pedimos prestada la moneda dom\'estica\index{moneda dom\'estica} e invertimos en la moneda extranjera\index{moneda extranjera}), y cuanto mayor sea el descuento a plazo\index{descuento a plazo}, m\'as rentable es la estrategia. Para un descuento a plazo\index{descuento a plazo} negativo, vendemos un forward\index{forward} (es decir, pedimos prestada la moneda extranjera\index{moneda extranjera} e invertimos en la moneda dom\'estica\index{moneda dom\'estica}), y cuanto menor sea el descuento a plazo\index{descuento a plazo}, m\'as rentable es la estrategia. Entonces, podemos construir una operaci\'on de corte transversal\index{operaci\'on de corte transversal} (incluyendo una estrategia de costo cero, es decir, un trade d\'olar-neutral\index{trade dolar-neutral @ trade d\'olar-neutral} -- v\'ease, por ejemplo, \cite{Lustig2011}) comprando los forwards\index{forward} sobre las monedas\index{moneda} en alg\'un cuantil\index{cuantil} superior\footnote{\, A diferencia de las acciones, que hay miles, existe un n\'umero limitado de monedas\index{moneda}. Por lo tanto, uno no necesariamente tiene el lujo de tomar deciles\index{decil} superiores e inferiores seg\'un el descuento a plazo\index{descuento a plazo}. Entonces, este cuantil\index{cuantil} puede ser la mitad, un tercio, etc., dependiendo del n\'umero de monedas\index{moneda}.} seg\'un el descuento a plazo\index{descuento a plazo} y vendiendo los forwards\index{forward} sobre las monedas\index{moneda} en el cuantil\index{cuantil} inferior correspondiente. Los forwards\index{forward} pueden, por ejemplo, tener una vida de un mes.

\subsection{Estrategia: Carry trade sobre el d\'olar\index{carry trade sobre el d\'olar}}

{}Esta estrategia se basa en el promedio de descuento a plazo\index{descuento a plazo} transversal ${\overline D}(t, T)$ (v\'ease, por ejemplo, \cite{Lustig2014}) para una cesta\index{cesta} de $N$ monedas extranjeras\index{moneda extranjera}:
\begin{eqnarray}	
 && {\overline D}(t, T) = \frac{1}{N} \sum_{i = 1}^{N} D_i(t, T) \
\end{eqnarray}
en donde $D_i(t, T)$ es el descuento a plazo\index{descuento a plazo} de la moneda\index{moneda} etiquetada por $i = 1,\dots,N$. Esta estrategia entonces consiste en tomar posiciones largas (cortas), con ponderaciones iguales en los forwards de todas las $N$ monedas extranjeras\index{moneda extranjera}\index{forward, moneda extranjera}, cuando ${\overline D}(t, T)$ es positivo (negativo), en donde $T$ puede ser $1,2,3,6,12$ meses. La evidencia emp\'irica sugiere que esta estrategia se relaciona con el estado de la econom\'ia de los Estados Unidos, a saber, cuando \'esta es d\'ebil, el promedio de los descuentos a plazo\index{descuento a plazo} tiende a ser positivo.\footnote{\, V\'ease, por ejemplo, \cite{Cooper2008}, \cite{Joslin2017}, \cite{Joslin2014}, \cite{Lustig2014}, \cite{Stambaugh1988}, \cite{Tille2001}.}

\subsection{Estrategia: Combo de momentum \& carry\index{combo de momentum \& carry}}

{}Esta es una combinaci\'on de la estrategia de momentum\index{estrategia de momentum} (la Subsecci\'on \ref{sub.fx.momentum})\footnote{\, Para literatura adicional sobre estrategias de momentum con FX\index{estrategia de momentum con FX} y t\'opicos relacionados, v\'ease, por ejemplo, \cite{Accominotti2014}, \cite{Ahmerkamp2013}, \cite{BurnsideEich2011}, \cite{Chiang1995}, \cite{Grobys2016}, \cite{Menkhoff2012}, \cite{Okunev2003}, \cite{Serban2010}.} y la estrategia de carry\index{estrategia de carry} (la Subsecci\'on \ref{sub.fx.carry}), o sus variaciones. Hay una gran variedad de formas en las que se pueden combinar estas estrategias (incluyendo un combo igualmente ponderado, o algunas ideas discutidas en, por ejemplo, la Subsecci\'on \ref{sub.multifactor} y la Subsecci\'on \ref{sub.multiasset.ETF}). Una combinaci\'on simple se basa en minimizar la varianza\index{varianza} de la estrategia de combo utilizando la matriz de covarianza muestral\index{matriz de covarianza muestral} de los retornos hist\'oricos\index{retornos hist\'oricos} $R_1(t_s)$ y $R_2(t_s)$ de las dos estrategias (v\'ease, por ejemplo, \cite{Olszweski2013}). Sea (aqu\'i Var y Cor son la varianza serial\index{varianza serial} y la correlaci\'on serial\index{correlaci\'on serial}, respectivamente)
\begin{eqnarray}
 &&\sigma_1^2 = \mbox{Var}(R_1(t_s))\\
 &&\sigma_2^2 = \mbox{Var}(R_2(t_s))\\
 &&\rho = \mbox{Cor}(R_1(t_s), R_2(t_s))
\end{eqnarray}
Luego, minimizando la varianza hist\'orica\index{varianza hist\'orica} de los retornos combinados $R(t_s)$, se fijan las ponderaciones $w_1$ y $w_2$ de la estrategia:
\begin{eqnarray}	
    && R(t_s) = w_1~R_1(t_s) + w_2~R_2(t_s)  \\
    && w_1 + w_2 = 1\\
    && \mbox{Var}(R(t_s))\rightarrow \mbox{min}\\
    && w_1 = \frac{\sigma_2^2 - \sigma_1\sigma_2\rho}{\sigma_1^2 + \sigma_2^2 - 2\sigma_1\sigma_2\rho}  \\
    && w_2 = \frac{\sigma_1^2 - \sigma_1\sigma_2\rho}{\sigma_1^2 + \sigma_2^2 - 2\sigma_1\sigma_2\rho}
\end{eqnarray}

\subsection{Estrategia: Arbitraje triangular con FX\index{arbitraje triangular con FX}}

{}Esta estrategia se basa en 3 pares de divisas\index{par de divisas}.\footnote{\, Aunque uno tambi\'en puede considerar m\'as de 3 pares, lo que se conoce como arbitraje multidivisa\index{arbitraje multidivisa} (v\'ease, por ejemplo, \cite{Moosa2003}).} Sean esas monedas\index{moneda} A, B y C. Entonces tenemos 2 cadenas: i) intercambio A por B, intercambio B por C, e intercambio C por A; y ii) intercambio A por C, intercambio C por B, e intercambio B por A. Nos centraremos en la primera cadena, ya que la segunda se obtiene cambiando B por C. Cada par de divisas\index{par de divisas} tiene el bid\index{bid} y el ask\index{ask}; por ejemplo, $Bid(A\rightarrow B)$ y $Ask(B\rightarrow A)$ para el par A-B. Entonces, la tasa a la que A se intercambia por B es $Bid(A\rightarrow B)$, mientras que la tasa a la que B se intercambia por A es $1/Ask(B\rightarrow A)$. Por lo tanto, $Bid(B\rightarrow A) = 1/Ask(B\rightarrow A)$, y $Ask(A\rightarrow B) = 1/Bid(A\rightarrow B)$. En la cadena i) el trader comienza con A y vuelve a A con el tipo de cambio\index{tipo de cambio} global
\begin{equation}
 R(A\rightarrow B \rightarrow C\rightarrow A) = Bid(A\rightarrow B) \times Bid(B\rightarrow C) \times {1\over Ask(C\rightarrow A)}
\end{equation}
Si esta cantidad es mayor que 1, entonces el trader obtiene una ganancia. Tales oportunidades son ef\'imeras, por lo que los datos de mercado\index{datos del mercado} y los sistemas de ejecuci\'on\index{sistemas de ejecuci\'on} r\'apidos son cr\'iticos aqu\'i.\footnote{\, Para obtener informaci\'on sobre el arbitraje triangular y temas relacionados, v\'ease, por ejemplo, \cite{Aiba2006}, \cite{Aiba2002}, \cite{Aiba2003}, \cite{Akram2008}, \cite{Choi2011}, \cite{Cross2015}, \cite{Fenn2009}, \cite{Goldstein1964}, \cite{Gradojevic2017}, \cite{Ito2012}, \cite{Moosa2001}, \cite{Morisawa2009}, \cite{Mwangi2012}, \cite{Osu2010}.}

\newpage

\section{Commodities\index{commodity}}\label{sec.commodities}

{}\subsection{Estrategia: Roll yields\index{roll yield}}

{}Cuando los futuros de commodities\index{futuros de commodities} se encuentran en backwardation\index{backwardation} (contango\index{contango}), es decir, cuando la estructura temporal\index{estructura temporal} de los precios de los futuros\index{precio de los futuros} presenta una pendiente negativa (positiva), posiciones largas (cortas) en futuros\index{posici\'on, larga en futuros}\index{posici\'on, corta en futuros}, en promedio, generan retornos positivos debido al roll yield\index{roll yield}. El roll yield\index{roll yield} se genera al rebalancear\index{rebalanceo} las posiciones de futuros: cuando los contratos de futuros\index{contrato de futuros} largos (cortos) est\'an a punto de expirar, se venden (se cubren) y otros contratos de futuros\index{contrato de futuros} con fecha de vencimiento\index{vencimiento} m\'as lejana se compran (se venden). Sea
\begin{equation}
 \phi = P_1 / P_2
\end{equation}
en donde $P_1$ es el precio de futuros del primer mes\index{precio de futuros del primer mes}, y $P_2$ es el precio de futuros del segundo mes\index{precio de futuros del segundo mes}. El ratio $\phi$ es una medida de backwardation\index{backwardation} ($\phi > 1$) y contango\index{contango} ($\phi < 1$). Un portafolio con posiciones largas y cortas de costo cero\index{portafolio con posiciones largas y cortas de costo cero} se puede construir basado en $\phi$, por ejemplo, comprando futuros de commodities\index{futuros de commodities} con valores m\'as altos de $\phi$ y vendiendo futuros\index{futuro} con valores m\'as bajos de los mismos.\footnote{\, Para algunos estudios pertinentes, v\'ease, por ejemplo, \cite{Anson1998}, \cite{Arnott2014}, \cite{Erb2006}, \cite{Fama1987}, \cite{Fama1988}, \cite{Feldman2006}, \cite{Fuertes2015}, \cite{Gorton2013}, \cite{Gorton2006}, \cite{Greer2000}, \cite{Leung2016}, \cite{Ma1992}, \cite{Mou2010}, \cite{Mouakhar2010}, \cite{Symeonidis2012}, \cite{Taylor2016}, \cite{Telser1958}.}

\subsection{Estrategia: Trading basado en la presi\'on de cobertura\index{presi\'on de cobertura (HP)}}

{}Esta estrategia se basa en los datos de las posiciones de los especuladores y hedgers proporcionados (semanalmente) por la Comisi\'on de Negociaci\'on de Futuros de Mercanc\'ias de los Estados Unidos (CFTC, por sus siglas en ingl\'es) en el reporte de los Compromisos de los Comerciantes (COT, por sus siglas en ingl\'es)\index{Compromisos de los Comerciantes (COT)}. Para cada commodity\index{commodity}, la ``presi\'on de cobertura'' (HP, por sus siglas en ingl\'es)\index{presi\'on de cobertura (HP)}, por separado para los hedgers\index{hedgers} y los especuladores\index{especulador}, se calcula como el n\'umero de contratos largos dividido por el n\'umero total de contratos (largos y cortos). Entonces, HP\index{presi\'on de cobertura (HP)} est\'a entre 0 y 1. Un alto (bajo) HP de los hedgers es indicativo de contango\index{contango} (backwardation\index{backwardation}), mientras que un HP alto (bajo) de los especuladores es indicativo de backwardation\index{backwardation} (contango\index{contango}). Un portafolio de costo cero\index{portafolio de costo cero} se puede construir, por ejemplo, de la siguiente forma. En primer lugar, el universo de los commodities\index{commodity} se divide en la mitad superior y la mitad inferior seg\'un el HP de los especuladores. Luego, los futuros de commodities\index{futuros de commodities} en la mitad superior se compran si est\'an en el quintil\index{quintil} inferior seg\'un el HP de los hedgers, y  los futuros de commodities\index{futuros de commodities} en la mitad inferior se venden si est\'an en el quintil\index{quintil} superior seg\'un el HP de los hedgers. Generalmente, los per\'iodos de formaci\'on y de tenencia\index{periodo de tenencia @ per\'iodo de tenencia}\index{periodo, de formacion @ per\'iodo, de formaci\'on} son de 6 meses.\footnote{\, Por algunas publicaciones sobre estrategias de trading\index{estrategia de trading} basadas en tales datos y temas relacionados, v\'ease, por ejemplo,\cite{Basu2013}, \cite{Bessembinder1992}, \cite{Carter1983}, \cite{ChengXiong2013}, \cite{DeRoon2000}, \cite{Dewally2013}, \cite{Fernandez-Perez2016}, \cite{Fishe2014}, \cite{Fuertes2015}, \cite{Hirshleifer1990}, \cite{Lehecka2013}, \cite{Miffre2012}, \cite{Switzer2010}.}

\subsection{Estrategia: Diversificaci\'on de portafolios\index{diversificaci\'on de portafolio} con commodities\index{commodity}}

{}Los mercados de commodities\index{mercado de commodities} t\'ipicamente tienen una baja correlaci\'on\index{correlaci\'on} con los mercados accionarios\index{mercado accionario}, lo cual puede ser utilizado para mejorar las caracter\'isticas de rendimientos\index{caracter\'isticas de rendimiento} de portafolios de acciones\index{portafolio de acciones}. Hay maneras diferentes de hacer esto. Un ``enfoque pasivo\index{enfoque pasivo}'' consistir\'ia en comprar commodities\index{commodity} con una porci\'on preestablecida de los fondos disponibles, manteni\'endolos y rebalanceando\index{rebalanceo} el portafolio\index{portafolio} con cierta periodicidad (por ejemplo, mensual o anual). Un ``enfoque activo''\index{enfoque activo} consistir\'ia en una asignaci\'on t\'actica de activos\index{asignaci\'on t\'actica de activos} aumentando/disminuyendo la exposici\'on\index{exposici\'on} a commodities\index{commodity} en funci\'on de un aumento/disminuci\'on en la tasa de descuento de la Fed\index{tasa de descuento de la Fed} (emp\'iricamente, los retornos de commodities\index{retornos de commodities} tienden a estar correlacionadas de forma considerable con la pol\'itica monetaria de la Fed\index{pol\'itica monetaria de la Fed}) o alguna otra metodolog\'ia.\footnote{\, Por algunas publicaciones sobre estrategias de diversificaci\'on\index{estrategia de diversificaci\'on} utilizando commodities\index{commodity} y t\'opicos relacionados, v\'ease, por ejemplo, \cite{Adams2015}, \cite{Bernardi2018}, \cite{Bjornson1997}, \cite{Blitz2008}, \cite{Bodie1983}, \cite{Bodie1980}, \cite{Chan2011}, \cite{Chance1994}, \cite{Chong2010}, \cite{Conover2010}, \cite{Creti2013}, \cite{Daumas2017}, \cite{Draper2006}, \cite{Edwards1996}, \cite{Elton1987}, \cite{Frankel2006}, \cite{Gorton2006}, \cite{Greer1978}, \cite{Greer2007}, \cite{Hess2008}, \cite{Jensen2000}, \cite{Jensen2002}, \cite{Kaplan1998}, \cite{Lummer1993},  \cite{MarshallCahan2008}, \cite{Miffre2007}, \cite{Nguyen2010}, \cite{Taylor2004}, \cite{Vrugt2007}, \cite{Wang2004}, \cite{Weiser2003}.}

\subsection{Estrategia: Valor\index{valor} (Value)\index{value}}

{}Esta estrategia es similar a la estrategia de value\index{estrategia de value} para acciones (v\'ease la Subsecci\'on \ref{sub.value}). El value\index{value} para commodities\index{commodity} se puede definir como, por ejemplo, el ratio (v\'ease, por ejemplo, \cite{Asness2013})
\begin{equation}
 v = P_5 / P_0
\end{equation}
en donde $P_5$ es el precio spot\index{precio spot} de 5 a\~{n}os atr\'as,\footnote{\, O el precio spot\index{precio spot} promedio de los precios entre 5.5 y 4.5 a\~{n}os atr\'as.} y $P_0$ es el precio spot\index{precio spot} actual. Entonces uno puede construir un portafolio de costo cero\index{portafolio de costo cero} comprando, por ejemplo, los commodities\index{commodity} en el tercil\index{tercil} superior seg\'un value\index{value}, y vendiendo los del tercil\index{tercil} inferior. El portafolio\index{portafolio} es rebalanceado mensualmente.

\subsection{Estrategia: Prima de asimetr\'ia\index{prima de asimetr\'ia}}

{}Esta estrategia se basa en la correlaci\'on\index{correlaci\'on} negativa observada emp\'iricamente entre la asimetr\'ia\index{asimetr\'ia} de los retornos hist\'oricos\index{retornos hist\'oricos} y los retornos esperados\index{retorno esperado} futuros de los futuros de commodities\index{futuros de commodities}. La asimetr\'ia\index{asimetr\'ia} $S_i$ se define como ($i=1,\dots,N$ etiqueta diferentes commodities\index{commodity}):
\begin{eqnarray}
 &&S_i = {1\over \sigma^3_i~T}~\sum_{s=1}^T \left[R_{is} - {\overline R}_{i}\right]^3\\
 &&{\overline R}_i = {1\over T}~\sum_{s=1}^T R_{is}\\
 &&\sigma_i^2 = {1\over {T-1}}~\sum_{s=1}^T \left[R_{is} - {\overline R}_{i}\right]^2
\end{eqnarray}
en donde $R_{is}$ son las series de tiempo\index{serie de tiempo} de los retornos hist\'oricos\index{retornos hist\'oricos} (con $T$ observaciones en cada serie de tiempo\index{serie de tiempo}). Una estrategia de costo cero\index{estrategia de costo cero} se puede construir, por ejemplo, comprando los futuros de commodities\index{futuros de commodities} en el quintil\index{quintil} inferior seg\'un la asimetr\'ia\index{asimetr\'ia}, y vendiendo los futuros\index{futuro} en el quintil\index{quintil} superior.\footnote{\, V\'ease, por ejemplo, \cite{Fernandez-Perez2018}. Para obtener informaci\'on adicional pertinente, v\'ease, por ejemplo, \cite{Barberis2008}, \cite{ChristieDavid2001}, \cite{Eastman2008}, \cite{Gilbert2006}, \cite{Junkus1991}, \cite{Kumar2009}, \cite{Lien2010}, \cite{Lien2015}, \cite{Mitton2007}, \cite{Stulz1996}, \cite{Tversky1992}.}

\subsection{Estrategia: Trading con modelos de valuaci\'on\index{modelo de valuaci\'on}}

{}La estructura a plazos de futuros de commodities\index{estructura a plazos de futuros de commodities} no es trivial. Una forma de modelarla es a trav\'es de procesos estoc\'asticos\index{proceso estoc\'astico}. Sea $S(t)$ el precio spot\index{precio spot}, y sea $X(t) = \ln(S(t))$. Luego, $X(t)$ se puede modelar utilizando, por ejemplo, un movimiento Browniano\index{movimiento Browniano} con reversi\'on a la media (es decir, un proceso de Ornstein-Uhlenbeck\index{proceso de Ornstein-Uhlenbeck} \cite{Uhlenbeck1930}):\footnote{\, Este es un modelo de un factor. Modelos m\'as complejos incluyendo modelos multifactoriales\index{modelo multifactorial}, modelos de volatilidad no constante/estoc\'astica\index{modelos de volatilidad estoc\'astica}, etc., se pueden considerar en su lugar. Para algunos estudios sobre la modelaci\'on de precios de los futuros\index{precio de los futuros} a trav\'es de procesos estoc\'asticos\index{proceso estoc\'astico} y t\'opicos relacionados, v\'ease, por ejemplo, \cite{Andersen2010}, \cite{Bessembinder1995}, \cite{Borovkova2006}, \cite{Casassus2005}, \cite{Chaiyapo2017}, \cite{Choi2014}, \cite{Geman2006}, \cite{Gibson1990}, \cite{Hilliard1998}, \cite{Jankowitsch2008}, \cite{Litzenberger1995}, \cite{Liu2011}, \cite{Milonas1991}, \cite{Miltersen1998}, \cite{Ng1994}, \cite{Nielsen2044}, \cite{Paschke2012}, \cite{Pindyck2001}, \cite{Routledge2000}, \cite{Schwartz1997}, \cite{Schwartz1998}, \cite{Schwartz2000}.}
\begin{eqnarray}
    && dX(t) = \kappa \left[a - X(t)\right] dt + \sigma~dW(t) \
\end{eqnarray}
Aqu\'i los par\'ametros $\kappa$ (par\'ametro de reversi\'on a la media\index{par\'ametro de reversi\'on a la media}), $a$ (media de largo plazo\index{media de largo plazo}) y $\sigma$ (volatilidad logar\'itmica\index{volatilidad logar\'itmica}) se asume que son constantes; y $W(t)$ es un movimiento Browniano\index{movimiento Browniano} bajo una medida de probabilidad neutral al riesgo\index{medida de probabilidad neutral al riesgo}, llam\'emosla ${\bf Q}$.\footnote{\, Tenga en cuenta que este modelo se reduce al modelo de Black-Scholes\index{modelo de Black-Scholes} \cite{Black1973} en el l\'imite $\kappa\rightarrow 0$, $a\rightarrow \infty$, $\kappa~a = \mbox{fijo}$.} Con el argumento est\'andar de valuaci\'on\index{argumento est\'andar de valuaci\'on} (v\'ease, por ejemplo, \cite{Baxter1996}, \cite{Hull2012}, \cite{Kakushadze2015a}) tenemos que el precio de los futuros\index{precio de los futuros} $F(t,T)$ est\'a dado por (que es el precio en el momento $t$ del contrato de futuros\index{contrato de futuros} con fecha de entrega\index{fecha de entrega} $T$)
\begin{eqnarray}
 &&F(t, T) = E_t(S(T))\\
 &&\ln(F(t,T)) = E_t(X(T)) + {1\over 2}V_t(X(T))
\end{eqnarray}
Aqu\'i $E_t(\cdot)$ y $V_t(\cdot)$ son la expectativa condicional\index{expectativa condicional} y la varianza condicional\index{varianza, condicional}, respectivamente, al momento $t$. Esto da:
\begin{eqnarray}	
 \ln(F(t, T)) &=& \exp\left(-\kappa(T - t)\right) X(t) + a\left[1 - \exp\left(-\kappa(T - t)\right)\right] + \nonumber\\
 &+& \frac{\sigma^2}{4 \kappa}\left[1-\exp\left(-2\kappa(T - t)\right)\right]
\end{eqnarray}
Los par\'ametros $\kappa, a, \sigma$ se pueden ajustar utilizando datos hist\'oricos\index{datos hist\'oricos} (por ejemplo, utilizando m\'inimos cuadrados no lineales\index{minimos cuadrados no lineales @ m\'inimos cuadrados no lineales}). Entonces, el precio de mercado\index{precio de mercado} actual se compara con el precio estimado por el modelo para identificar los futuros\index{futuro} que est\'an caros (se\~{n}al de venta\index{senzal de venta @ se\~{n}al de venta}) y los que est\'an baratos (se\~{n}al de compra\index{senzal de compra @ se\~{n}al de compra}) en comparaci\'on con la predicci\'on del modelo. Aqu\'i hay dos comentarios de precauci\'on que deben tenerse en cuenta. En primer lugar, el modelo ajustado podr\'ia funcionar dentro de la muestra pero no tener poder predictivo fuera de la muestra, por lo que el poder de predicci\'on debe ser comprobado (v\'ease, por ejemplo, \cite{Paschke2012}). Segundo, a priori podr\'iamos escribir cualquier modelo de estructura temporal\index{estructura temporal} razonable con propiedades cualitativas deseables (por ejemplo, reversi\'on a la media\index{reversi\'on a la media}) y ajustar los par\'ametros utilizando datos hist\'oricos\index{datos hist\'oricos} sin considerar ninguna referencia a una din\'amica estoc\'astica\index{din\'amica estoc\'astica} subyacente, incluido el uso de, por ejemplo, t\'ecnicas de aprendizaje autom\'atico conocidas como ``caja negra''\index{tecnicas de aprendizaje automatico ``caja negra'' @ t\'ecnicas de aprendizaje autom\'atico ``caja negra''}. Entonces, mientras el modelo funcione fuera de muestra, aqu\'i no hay una bala m\'agica y ``el lujo'' no equivale a ``mejor''.

\newpage

\section{Futuros\index{futuro}}\label{sec.futures}

\subsection{Estrategia: Cobertura con futuros\index{futuro}}

{}La exposici\'on\index{exposici\'on} a ciertos riesgos\index{riesgo} se puede mitigar mediante la cobertura con futuros\index{futuro}. Por ejemplo, un trader de granos que al momento $t$ anticipa que \'el o ella tendr\'a que comprar (vender) $X$ toneladas de soja en un momento posterior $T$ puede cubrir el riesgo\index{riesgo} de un aumento (disminuci\'on) en los precios de la soja entre $t$ y $T$ comprando (vendiendo) en el momento $t$ un contrato de futuros\index{contrato de futuros} con fecha de entrega\index{fecha de entrega} $T$ por la cantidad deseada de soja. Esta estrategia simple puede tener diversos ajustes y variaciones.\footnote{\, Por algunas publicaciones sobre cobertura con futuros\index{futuro}, v\'ease, por ejemplo, \cite{Ahmadi1986}, \cite{Cheung1990}, \cite{Ederington1979}, \cite{Geczy1997}, \cite{Ghosh1993}, \cite{Grant2016}, \cite{Hanly2018}, \cite{Lebeck1978}, \cite{LienTse2000}, \cite{Mun2016},  \cite{Wolf1987}, \cite{Working1953}.}

\subsubsection{Estrategia: Cobertura cruzada\index{cobertura cruzada}}

{}En ciertas ocasiones, puede que no haya disponible un contrato de futuros\index{contrato de futuros} para el activo que el trader desea cubrir. En tales casos, el trader puede ser capaz de obtener cobertura mediante un contrato de futuros\index{contrato de futuros} para otro activo con caracter\'isticas similares.\footnote{\, Por algunas publicaciones sobre la cobertura cruzada\index{cobertura cruzada} con futuros\index{futuro}, v\'ease, por ejemplo, \cite{Anderson1981}, \cite{Ankirchner2012}, \cite{AnkirchnerHeyne2012}, \cite{Benet1990}, \cite{Blake1984}, \cite{Blank1984}, \cite{BrooksDK2007}, \cite{ChenS2007}, \cite{Dahlgran2000}, \cite{DeMaskey1997}, \cite{DeMaskey1998}, \cite{Foster2002}, \cite{Franken2003}, \cite{Hartzog1982}, \cite{Lafuente2013}, \cite{McEnally1979}, \cite{Mun1997}.} A la madurez\index{madurez} $T$, el pago\index{pago} de la posici\'on de cobertura cruzada establecida en el momento $t$ (asumiendo que la posici\'on corta en los futuros\index{posici\'on, corta en futuros} tiene un ratio de cobertura\index{ratio de cobertura} unitario) est\'a dado por:
\begin{eqnarray}
 && S(T) - F(T,T) + F(t, T) = \nonumber\\
 && [S_*(T) - F(T, T)] + [S(T) - S_*(T)] + F(t, T)
\end{eqnarray}
Aqu\'i: el sub\'indice $*$ indica que el activo subyacente\index{activo subyacente} del contrato de futuros\index{contrato de futuros} es diferente del activo a cubrir; $S(t)$ es el precio spot\index{precio spot}; $F(t,T)$ es el precio de los futuros\index{precio de los futuros}; el primer t\'ermino en el lado derecho representa el riesgo de base\index{riesgo de base} causado por la diferencia en la entrega\index{entrega} entre los precios de los futuros\index{futuro} y el spot\index{precio spot}; y el segundo t\'ermino representa la diferencia entre los dos precios subyacentes\index{precio subyacente}. En la pr\'actica, el ratio \'optimo de cobertura\index{ratio de cobertura \'optimo} puede no ser 1 y puede estimarse mediante, por ejemplo, una regresi\'on serial\index{regresi\'on serial} u otros m\'etodos.\footnote{\, Para diversas t\'ecnicas sobre el c\'alculo del ratio de cobertura \'optimo\index{ratio de cobertura \'optimo}, v\'ease, por ejemplo, \cite{Baillie1991}, \cite{Brooks2001}, \cite{Brooks2002}, \cite{Cecchetti1988}, \cite{Davis2006}, \cite{Holmes1996}, \cite{Lien1992}, \cite{Lien2004}, \cite{Lien1993}, \cite{Lindahl1992}, \cite{Low2002}, \cite{Kroner1993}, \cite{Monoyios2004}, \cite{Moosa2003b}, \cite{Myers1991}.}

\subsubsection{Estrategia: Cobertura de riesgo de tasa de inter\'es\index{cobertura de riesgo de tasa de inter\'es}}

{}Los activos de renta fija\index{activo de renta fija} son sensibles a las variaciones de la tasa de inter\'es\index{tasa de inter\'es} (v\'ease la Secci\'on \ref{sec.fixed.income}) y por esto, los traders suelen utilizar contratos de futuros\index{contrato de futuros} para cubrir el riesgo de tasa de inter\'es\index{riesgo de tasa de inter\'es}. Una posici\'on de cobertura\index{posici\'on de cobertura} larga (corta) consiste en comprar (vender) los futuros de tasas de inter\'es\index{futuros de tasas de inter\'es} para protegerse contra un aumento (disminuci\'on) en el precio del activo subyacente\index{activo subyacente}, es decir, una disminuci\'on (aumento) en las tasas de inter\'es\index{tasa de inter\'es}.\footnote{\, Para algunos estudios sobre la cobertura del riesgo de tasa de inter\'es\index{riesgo de tasa de inter\'es} con futuros\index{futuro}, v\'ease, por ejemplo, \cite{Booth1984}, \cite{Briys1992}, \cite{Cerovic2011}, \cite{Clare2000}, \cite{Fortin1984}, \cite{Gay1983}, \cite{Hilliard1984}, \cite{Hilliard1989}, \cite{Ho1983}, \cite{Kolb1982}, \cite{Lee1993}, \cite{Pepic2014}, \cite{Picou1981}, \cite{Trainer1983}, \cite{Yawitz1985}, \cite{Yeutter1982}.} El correspondiente P\&L\index{P\&L} ($P_L(t,T)$ para la cobertura\index{cobertura} larga y $P_C(t,T)$ para la cobertura\index{cobertura} corta, asumiendo que la posici\'on se establece en $t=0$ con un ratio de cobertura\index{ratio de cobertura} unitario y madurez\index{madurez} $T$) est\'a dado por:
\begin{eqnarray}
 && P_L(t,T) = B(0,T) - B(t,T) \\
 && P_C(t,T) = B(t, T) - B(0, T) \\
 && B(t,T) = S(t) - F(t,T) \
\end{eqnarray}
en donde $B(t,T)$ es la base de los futuros\index{base del futuro}. En la pr\'actica, el ratio de cobertura\index{ratio de cobertura} puede no ser 1. Si la cobertura\index{cobertura} es contra un bono\index{bono} en la cesta de entrega de los futuros\index{cesta de entrega de los futuros},\footnote{\, T\'ipicamente, un contrato de futuros de tasas de inter\'es\index{contrato de futuros de tasas de inter\'es} permite que sea entregado no solo uno sino que cualquier bono\index{bono} a partir de una matriz predefinida de bonos\index{bono} (con diferentes fechas de madurez\index{madurez}, cupones\index{cup\'on}, etc.). De ah\'i el uso del factor de conversi\'on\index{factor de conversi\'on} (v\'ease abajo), que se define de la siguiente forma\cite{Hull2012}: ``El factor de conversi\'on\index{factor de conversi\'on} de un bono\index{bono} se establece igual al precio cotizado que tendr\'ia el bono\index{bono} por d\'olar de principal\index{principal} en el primer d\'ia del mes de entrega\index{mes de entrega} con el supuesto de que la tasa de inter\'es\index{tasa de inter\'es} para todas las fechas de madurez\index{madurez} es 6\% por a\~{n}o (con la composici\'on semestral\index{composici\'on semestral}).''} entonces el modelo de factor de conversi\'on\index{modelo de factor de conversi\'on}\footnote{\, El modelo de factor de conversi\'on\index{modelo de factor de conversi\'on} se aplica solo a los contratos de futuros\index{contrato de futuros} que utilizan factores de conversi\'on\index{factor de conversi\'on}, tales como los futuros sobre T-bond\index{T-bond} y T-note\index{T-note}\index{futuros, T-bond}\index{futuros, T-note}.} se utiliza com\'unmente para calcular el ratio de cobertura\index{ratio de cobertura} $h_C$:
\begin{equation}
 h_C = C ~{M_B\over M_F}
\end{equation}
en donde $M_B$ es el valor nominal del bono\index{bono}, $M_F$ es el valor nominal de los futuros\index{futuro}, y $C$ es el factor de conversi\'on\index{factor de conversi\'on}. A diferencia del modelo de factor de conversi\'on\index{modelo de factor de conversi\'on}, el ratio de cobertura\index{ratio de cobertura} de la duraci\'on modificada\index{duraci\'on modificada} $h_D$ se puede utilizar para bonos entregables\index{bono, entregable} y no entregables\index{bono, no entregable}:
\begin{eqnarray}
 && h_D = \beta~\frac{D_B}{D_F}\
\end{eqnarray}
en donde $D_B$ es la duraci\'on d\'olar\index{duraci\'on dolar @ duraci\'on d\'olar}\footnote{\, Recordar que la duraci\'on d\'olar\index{duraci\'on dolar @ duraci\'on d\'olar} es igual al precio multiplicado por la duraci\'on modificada\index{duraci\'on modificada}.} del bono\index{bono}, $D_F$ es la duraci\'on d\'olar\index{duraci\'on dolar @ duraci\'on d\'olar} de los futuros\index{futuro}, y $\beta$ (que a menudo se establece que sea igual 1) es el cambio en el rendimiento de los bonos\index{rendimientos del bono} en relaci\'on con el cambio en el rendimiento de los futuros\index{rendimiento de los futuros}, ambos tomados para un cambio dado en la tasa libre de riesgo\index{tasa libre de riesgo}.\footnote{\, El factor $\beta$ se puede estimar en base a los datos hist\'oricos\index{datos hist\'oricos}. Para algunos estudios sobre ratios de cobertura de futuros de tasas de inter\'es\index{ratio de cobertura de futuros de tasas de inter\'es} y temas relacionados, v\'ease, por ejemplo, \cite{Chang1990}, \cite{Chen2005}, \cite{Daigler1998}, \cite{Falkenstein1996}, \cite{Fisher1971}, \cite{GayKolb1983}, \cite{Geske1987}, \cite{Grieves2004}, \cite{Grieves2005}, \cite{Hegde1982}, \cite{Kolb1981}, \cite{Kuberek1983}, \cite{Landes1985}, \cite{Pitts1985}, \cite{Rendleman1999}, \cite{Toevs1986}, \cite{Viswanath1993}.}

\subsection{Estrategia: Diferenciales de calendario\index{diferencial de calendario}}

{}Un diferencial de futuros alcista (bajista)\index{diferencial de futuros, bajista}\index{diferencial de futuros, alcista} consiste en comprar (vender) un futuro al mes m\'as cercano\index{futuro al mes m\'as cercano} y vender (comprar) un futuro a un mes diferido\index{futuro a un mes diferido}. Esto reduce la exposici\'on\index{exposici\'on} a la volatilidad del mercado\index{volatilidad del mercado} y permite centrarse m\'as en los fundamentos\index{fundamentos}. As\'i, para el caso de futuros de commodities\index{futuros de commodities}, generalmente, aquellos contratos con vencimiento al mes m\'as cercano\index{contrato al mes m\'as cercano} tienden a reaccionar m\'as frente a cambios en la oferta\index{oferta} y la demanda\index{demanda} que los contratos a un mes diferido\index{contratos a un mes diferido}. Por lo tanto, si el trader espera una oferta\index{oferta} baja (alta) y una demanda\index{demanda} alta (baja), entonces puede hacer una apuesta con un diferencial alcista (bajista).\footnote{\, Por algunas publicaciones sobre diferenciales de calendario con futuros\index{diferencial de calendario con futuros} y t\'opicos relacionados, v\'ease, por ejemplo, \cite{Abken1989}, \cite{Adrangi2006}, \cite{Barrett1995}, \cite{Bernstein1990}, \cite{Bessembinder1992}, \cite{Bessembinder1993}, \cite{BessembinderChan1992}, \cite{Billingsley1988}, \cite{Castelino1984}, \cite{Cole1999}, \cite{Daigler2007}, \cite{DeRoon1998}, \cite{DeRoon2000}, \cite{Dunis2006}, \cite{Dunis2010}, \cite{Dutt1997}, \cite{Frino2002}, \cite{Girma1998}, \cite{Hou2017}, \cite{Kawaller2002}, \cite{Kim1997}, \cite{McComas2003}, \cite{Moore2006}, \cite{Ng1994}, \cite{Pechanok2012}, \cite{Perchanok2013}, \cite{Poitras1990},  \cite{Ross2006}, \cite{Salcedo2004}, \cite{Schap2005}, \cite{Shimko1994}, \cite{Till2017}, \cite{vandenGoorbergh2004}.}

\subsection{Estrategia: Trading contrario\index{trading contrario} (reversi\'on a la media\index{reversi\'on a la media})}

{}Esta estrategia es similar a la estrategia de reversi\'on a la media\index{estrategia de reversi\'on a la media} discutida en la Subsecci\'on \ref{sub1.3}. Dentro de un universo dado de futuros\index{futuro} etiquetados por $i=1,\dots,N$, el retorno del ``\'indice de mercado''\index{retorno del \'indice de mercado} se calcula como un promedio igualmente ponderado\index{promedio igualmente ponderado}:
\begin{equation}\label{fut.mkt.ix}
 R_m = {1\over N}~\sum_{i=1}^N R_i
\end{equation}
en donde $R_i$ son los retornos de los futuros\index{retornos de los futuros} individuales (t\'ipicamente durante la \'ultima semana). Las ponderaciones de asignaci\'on de capital\index{ponderaciones de asignaci\'on de capital} $w_i$ entonces est\'an dadas por
\begin{eqnarray}\label{w.fut}
 &&w_i = -\gamma \left[R_i - R_m\right]
\end{eqnarray}
en donde $\gamma > 0$ se fija a trav\'es de la condici\'on de normalizaci\'on
\begin{equation}\label{norm.fut}
 \sum_{i=1}^N |w_i| = 1
\end{equation}
Tenga en cuenta que la estrategia es autom\'aticamente neutral en d\'olares. Consiste en comprar los perdedores\index{perdedores} y vender los ganadores\index{ganadores} con respecto al \'indice de mercado\index{indice de mercado @ \'indice de mercado} (v\'ease, por ejemplo, \cite{Wang2004}).\footnote{\, Para obtener informaci\'on adicional pertinente, v\'ease, por ejemplo, \cite{Bali2008}, \cite{Bessembinder1995}, \cite{Bianchi2015}, \cite{Chaves2016}, \cite{Fuertes2015}, \cite{Irwin1996}, \cite{Julio2013}, \cite{Leung2016}, \cite{Monoyios2002}, \cite{Rao2011}, \cite{Rosales2017}, \cite{Tse2017}.} Como en el caso de las acciones\index{acci\'on}, el esquema de ponderaci\'on\index{esquema de ponderaci\'on} simple dado por la Ecuaci\'on (\ref{w.fut}) es propenso a sobre invertir en futuros\index{futuro} vol\'atiles, lo cual se puede mitigar suprimiendo $w_i$ por $1/\sigma_i$ o $1/\sigma_i^2$, en donde $\sigma_i$ son las volatilidades hist\'oricas\index{volatilidad hist\'orica}. El portafolio\index{portafolio} es rebalanceado semanalmente.

\subsubsection{Estrategia: Trading contrario\index{trading contrario} -- actividad de mercado\index{actividad de mercado}}

{}Se pueden agregar campanas y silbidos a la estrategia de reversi\'on a la media\index{estrategia de reversi\'on a la media} ``b\'asica'' antes descripta incorporando filtros de volumen\index{volumen} e inter\'es abierto\index{interes abierto @ inter\'es abierto}. Sea $V_i$ el volumen\index{volumen} total de los futuros\index{futuro} etiquetados por $i$ durante la \'ultima semana (es decir, la suma de los vol\'umenes\index{volumen} diarios durante la \'ultima semana), y $V_i^\prime$ sea el volumen total\index{volumen} durante la semana anterior. Sean $U_i$ y $U^\prime_i$ las cantidades an\'alogas para el inter\'es abierto\index{interes abierto @ inter\'es abierto}. Sea
\begin{eqnarray}
 && v_i = \ln(V_i / V_i^\prime)\\
 && u_i = \ln(U_i / U_i^\prime)
\end{eqnarray}
Entonces la estrategia se puede construir, por ejemplo, tomando la mitad superior de los futuros\index{futuro} por el factor de volumen\index{volumen} $v_i$, tomando la mitad inferior de estos futuros\index{futuro} por el factor de inter\'es abierto\index{interes abierto @ inter\'es abierto} $u_i$, y aplicando la estrategia definida por la Ecuaci\'on (\ref{w.fut}) a este subconjunto de los futuros\index{futuro}.\footnote{\, La raz\'on detr\'as de esto es: i) es probable que mayores cambios en el volumen\index{volumen} sean indicativos de una mayor sobre-reacci\'on\index{sobre-reacci\'on} (v\'ease, por ejemplo, \cite{Bloom1994}, \cite{Conrad1994}, \cite{DeBondt1985}, \cite{Gervais2001}, \cite{Odean2002}, \cite{Statman2006}), entonces se puede esperar un mayor efecto ``snap-back'' (en ingl\'es) (es decir, reversi\'on a la media)\index{efecto, reversi\'on a la media}; y ii) el inter\'es abierto\index{interes abierto @ inter\'es abierto} se relaciona con la actividad por parte de los hedgers\index{hedgers} y es un proxy de la profundidad del mercado (v\'ease, por ejemplo, \cite{BessembinderS1993}), por lo que un aumento en el inter\'es abierto\index{interes abierto @ inter\'es abierto} es indicativo de un mercado m\'as profundo, en donde los incrementos en el volumen\index{volumen} tienen menores efectos en los precios en comparaci\'on a cuando hay una disminuci\'on en el inter\'es abierto\index{interes abierto @ inter\'es abierto}.}

\subsection{Estrategia: Seguimiento de la tendencia\index{seguimiento de la tendencia} (momentum\index{momentum})}

{}Varias estrategias de momentum\index{estrategia de momentum} con futuros\index{futuro} se pueden construir de manera similar a las de las acciones. Aqu\'i hay un ejemplo simple (v\'ease, por ejemplo, \cite{Balta2013}, \cite{Moskowitz2012}).\footnote{\, Por alguna literatura adicional pertinente, v\'ease, por ejemplo, \cite{Ahn2002}, \cite{Bianchi2015}, \cite{Dusak1973}, \cite{Fuertes2015}, \cite{Fuertes2010}, \cite{Hayes2011}, \cite{Kazemi2009}, \cite{Miffre2007}, \cite{Pirrong2005}, \cite{Reynauld1984}, \cite{Schneeweis2006}, \cite{Szakmary2010}.} Sean $R_i$ los retornos de los futuros\index{retornos de los futuros} etiquetados por $i=1,\dots,N$ durante el per\'iodo pasado $T$ (que se puede medir en, por ejemplo, d\'ias, semanas o meses). Luego las ponderaciones $w_i$ del portafolio de trading\index{portafolio de trading} est\'an dadas por
\begin{eqnarray}
 &&w_i = \gamma~{\eta_i\over\sigma_i}\label{fut.mom}\\
 &&\eta_i = \mbox{sign}(R_i)
\end{eqnarray}
en donde $\sigma_i$ son las volatilidades hist\'oricas\index{volatilidad hist\'orica} (computadas durante el per\'iodo $T$ u otro per\'iodo), y $\gamma > 0$ se fija mediante la condici\'on de normalizaci\'on
\begin{equation}\label{fut.trend.norm}
 \sum_{i=1}^N |w_i| = 1
\end{equation}
Tenga en cuenta que esta estrategia es equivalente a la estrategia de optimizaci\'on\index{estrategia de optimizaci\'on} (v\'ease la Subsecci\'on \ref{sub.opt}, Ecuaci\'on (\ref{w.max.sharpe})) con una matriz de covarianza\index{matriz de covarianza} diagonal $C_{ij} = \sigma_i^2~\delta_{ij}$ (es decir, las correlaciones\index{correlaci\'on} entre  los diferentes futuros\index{futuro} son ignoradas) y los retornos esperados\index{retorno esperado} $E_i = \eta_i\sigma_i$. Esto debe contrastarse con los retornos esperados\index{retorno esperado} basados en los retornos acumulados\index{retorno acumulado} (Ecuaci\'on (\ref{cum.ret})), que en este caso es igual a $R_i$. Un problema con el uso de $E_i = \eta_i\sigma_i$ a diferencia de $E_i = R_i$ es que, para peque\~{n}os $|R_i|$ (por ejemplo, comparado con $\sigma_i$), $\eta_i$ puede voltearse f\'acilmente a pesar de que el cambio en $R_i$ sea peque\~{n}o. Esto se traduce en una inestabilidad indeseable en la estrategia. Hay formas de mitigar esto, por ejemplo, suavizando $\eta_i = \tanh(R_i/\kappa)$, en donde $\kappa$ es alg\'un par\'ametro, por ejemplo, la desviaci\'on est\'andar de corte transversal\index{desviaci\'on est\'andar de corte transversal} de $R_i$ (v\'ease, por ejemplo, \cite{Kakushadze2015b}). Alternativamente, uno puede simplemente tomar $E_i = R_i$ (y adem\'as usar una $C_{ij}$ no diagonal). Adem\'as, tenga en cuenta que las ponderaciones definidas por la Ecuaci\'on (\ref{fut.mom}) no son d\'olar-neutrales. Esto puede ser rectificado mediante la sustracci\'on de la media:
\begin{equation}\label{fut.w.neut}
 w_i = \gamma\left[{\eta_i\over\sigma_i} - {1\over N}~\sum_{j=1}^N {\eta_j\over\sigma_j}\right]
\end{equation}
Una desventaja de esto es que ahora algunos futuros\index{futuro} con $\eta_i > 0$ pueden ser vendidos, y algunos futuros\index{futuro} con $\eta_i < 0$ pueden ser comprados. Para evitar esto, si el n\'umero $N_+ = |H_+|$ de los futuros\index{futuro} con $\eta_i >0$ no es dram\'aticamente diferente del n\'umero $N_- = |H_-|$ de futuros\index{futuro} con $\eta_i < 0$ (aqu\'i $H_\pm = \{i| \pm\eta_i > 0\}$), podemos hacer que las ponderaciones sean
\begin{eqnarray}
 &&w_i = \gamma_+~{\eta_i\over\sigma_i},~~~i\in H_+\\\
 &&w_i = \gamma_-~{\eta_i\over\sigma_i},~~~i\in H_-
\end{eqnarray}
Entonces, ahora tenemos dos par\'ametros $\gamma_\pm$, que pueden fijarse para satisfacer la Ecuaci\'on (\ref{fut.trend.norm}) y la condici\'on de d\'olar-neutralidad\index{condici\'on de d\'olar-neutralidad}
\begin{equation}
 \sum_{i=1}^N w_i = 0
\end{equation}
Sin embargo, si la mayor parte de $\eta_i$ son positivos (negativos), es decir, tenemos retornos asim\'etricos, luego las posiciones largas (cortas) estar\'an bien diversificadas, mientras que las posiciones cortas (largas) no lo estar\'an. Esto puede suceder, por ejemplo, si el mercado general\index{mercado general} es muy alcista (bajista). La Ecuaci\'on (\ref{fut.w.neut}) mitiga esto hasta cierto punto. Sin embargo, $\eta_i$ todav\'ia puede estar sesgado en este caso. Una forma sencilla de evitar esto por completo es utilizar los retornos netos de la media\index{retorno neto de la media} ${\widetilde R}_i$ en lugar de $R_i$, en donde ${\widetilde R}_i = R_i - R_m$, y el retorno del ``\'indice de mercado''\index{retorno del \'indice de mercado} $R_m$ es definido por la Ecuaci\'on (\ref{fut.mkt.ix}).\footnote{\, Es decir, en este caso los ganadores\index{ganadores} y perdedores\index{perdedores} seg\'un el momentum\index{momentum} se definen con respecto al \'indice de mercado\index{indice de mercado @ \'indice de mercado}, y los ganadores\index{ganadores} as\'i definidos se compran, mientras que los perdedores\index{perdedores} se venden.} Luego $\eta_i = \mbox{sign}({\widetilde R}_i$) ya no est\'an sesgados y la d\'olar-neutralidad\index{dolar-neutralidad @ d\'olar-neutralidad} se puede lograr como se indica arriba.\footnote{\, Adem\'as, en lugar de utilizar retornos acumulados\index{retorno acumulado} $R_i$, uno puede usar medias m\'oviles exponenciales\index{media m\'ovil exponencial (EMA)} (para suprimir las contribuciones pasadas -- v\'ease la Secci\'on \ref{sec.stocks}), el filtro de Hodrick-Prescott\index{filtro de Hodrick-Prescott} (para quitar el ruido\index{ruido} e identificar la tendencia\index{tendencia} -- v\'ease la Secci\'on  \ref{sec.FX}), el filtro de Kalman\index{filtro de Kalman} (v\'ease, por ejemplo, \cite{Babbs1999}, \cite{Benhamou2016}, \cite{Bruder2013}, \cite{deMoura2016}, \cite{Elliott2005}, \cite{EngleW1987}, \cite{Harvey1984}, \cite{Harvey1990}, \cite{HatemiJ2006}, \cite{Kalman1960}, \cite{Lautier2004}, \cite{Levine2016}, \cite{Martinelli2010}, \cite{Vidyamurthy2004}), o algunos otros filtros de series de tiempo\index{filtro de series de tiempo}.}

\newpage

\section{Activos Estructurados\index{activos estructurados}}\label{sec.str}

\subsection{Generalidades: Obligaciones de deuda garantizadas (CDOs)\index{obligaci\'on de deuda garantizada (CDO)}}\label{sub.hedgingCDOs}

{}Un CDO\index{obligaci\'on de deuda garantizada (CDO)} (por sus siglas en ingl\'es) es un valor respaldado por activos (ABS, por sus siglas en ingl\'es)\index{valor respaldado por activos (ABS)} que consiste en una cesta\index{cesta} de activos tales como bonos\index{bono}, swaps de incumplimiento crediticio\index{swap de incumplimiento crediticio (CDS)}, etc. Se dividen en m\'ultiples tramos\index{tramo}, que consisten en activos con diferentes calificaciones crediticias\index{calificaci\'on crediticia} y tasas de inter\'es\index{tasa de inter\'es}. Cada tramo\index{tramo} tiene un punto de intervenci\'on\index{punto de intervenci\'on} $a$ y un punto de desprendimiento\index{punto de desprendimiento} $d$. Por ejemplo, un tramo\index{tramo} de 3-8\% (para cual $a = 3\%$ y $d = 8\%$) significa que comienza a perder su valor cuando la p\'erdida de la cartera subyacente excede el 3\%; y cuando la p\'erdida de la cartera subyacente supera el 8\%, el valor del tramo\index{valor del tramo} pasa a ser cero.\footnote{\, Ejemplos de tramos\index{tramo} son (en orden decreciente de riesgo de incumplimiento\index{riesgo de incumplimiento} y de la tasa de pago de la prima peri\'odica\index{tasa de pago de la prima peri\'odica}): tramo de capital 0-3\%\index{tramo de capital}; tramo junior mezzanine 3-7\%\index{tramo junior mezzanine}; tramo senior mezzanine 7-10\%\index{tramo senior mezzanine}; tramo senior 10-15\%\index{tramo senior}; y tramo super senior 15-30\%\index{tramo super senior}.} Un comprador (con una posici\'on larga) de un tramo de un CDO\index{tramo de un CDO} es un vendedor de protecci\'on\index{vendedor de protecci\'on}: a cambio de recibir pagos peri\'odicos de primas\index{pago peri\'odico de prima}, en caso de un evento de incumplimiento\index{incumplimiento}, el comprador tiene la obligaci\'on de cubrir el incumplimiento\index{incumplimiento} correspondiente al tama\~{n}o del tramo\index{tramo}. Un vendedor (posici\'on corta) de un tramo de un CDO\index{tramo de un CDO} es un comprador de protecci\'on\index{comprador de protecci\'on}: a cambio de hacer pagos peri\'odicos de primas\index{pago peri\'odico de prima}, el vendedor recibe un pago en caso de incumplimiento\index{incumplimiento}. Los CDOs sint\'eticos\index{CDO sint\'etico} son ``sintetizados'' a trav\'es de derivados de cr\'edito\index{derivados de cr\'edito} tales como CDS\index{swap de incumplimiento crediticio (CDS)} (swap de incumplimiento crediticio\index{swap de incumplimiento crediticio (CDS)} -- v\'ease la Subsecci\'on \ref{sub.cds}) en un conjunto de entidades de referencia\index{entidad de referencia} (por ejemplo, bonos\index{bono}, pr\'estamos\index{prestamo @ pr\'estamo}, nombres de empresas o pa\'ises). Los conjuntos de referencias\index{conjunto de referencia} para CDOs de un solo tramo\index{CDO de un solo tramo} negociados en la bolsa son los \'indices de CDS\index{indice de CDS @ \'indice de CDS} tales como el CDX y el iTraxx.\footnote{\, Por algunas publicaciones sobre CDOs\index{obligaci\'on de deuda garantizada (CDO)} y temas relacionados, v\'ease, por ejemplo, \cite{Altman2005}, \cite{Amato2005}, \cite{Amato2003}, \cite{Andersen2005}, \cite{Andersen2003}, \cite{Belkin1998}, \cite{Bielecki2011}, \cite{Bol2009}, \cite{Boscher2002}, \cite{Cousin2012}, \cite{Das2005}, \cite{Davis2001}, \cite{Ding2011}, \cite{Douglas2007}, \cite{Duffie2004}, \cite{Duffie2001}, \cite{DuffieHuang1996}, \cite{Duffie1997a}, \cite{Duffie1997b}, \cite{Fabozzi2006a},  \cite{Finger1999}, \cite{Frey2001}, \cite{Gibson2004}, \cite{Goodman2002}, \cite{GoodmanLucas2002}, \cite{Houdain2006}, \cite{Hull2006}, \cite{Hull2010}, \cite{Jarrow1997}, \cite{Jarrow1995}, \cite{Jobst2005}, \cite{Jobst2006a}, \cite{Jobst2006b}, \cite{Jobst2006c}, \cite{Jobst2007}, \cite{Laurent2005}, \cite{Li2000}, \cite{Lucas2006}, \cite{Meissner2008}, \cite{Packer2005}, \cite{Prince2005}, \cite{Schmidt2002}, \cite{Schonbucher2003}, \cite{Tavakoli1998}, \cite{Vasicek2015}.}

{}Sean $t_i$, $i=1,\dots,n$ los tiempos en que se hicieron los pagos peri\'odicos de la prima\index{pago peri\'odico de prima}.\footnote{\, Por simplicidad, podemos asumir que cualquier pago por incumplimiento\index{pago por incumplimiento} tambi\'en se hace en esos tiempos.} Sea $H(t)$ el conjunto de posibles incumplimientos\index{incumplimiento} $\ell_\alpha$, $\alpha = 1,\dots, K$, que pueden ocurrir en tiempo $t$, y que $p_\alpha(t)$ denote las probabilidades correspondientes (que dependen de un modelo). Aqu\'i $\ell_\alpha$ son los montos en d\'olares de los incumplimientos\index{incumplimiento}.\footnote{\, Si la cantidad nocional\index{cantidad nocional} del cr\'edito en incumplimiento\index{credito en incumplimiento @ cr\'edito en incumplimiento} etiquetado por $\alpha$ es $M_{\alpha}$, entonces $\ell_\alpha = M_{\alpha}(1 - R_\alpha)$, en donde $R_\alpha$ es la tasa de recuperaci\'on\index{tasa de recuperaci\'on} (que puede ser distinta de cero) de dicho cr\'edito.} La p\'erdida esperada $L(t)$ se puede calcular como
\begin{equation}
 L(t) = \sum_{\alpha = 1}^K p_\alpha(t)~\mbox{max}(\mbox{min}(\ell_\alpha, L_d) - L_a, 0)
\end{equation}
en donde $L_a = a~M_{CDO}$, $L_d = d~M_{CDO}$, y $M_{CDO}$ es el nocional del CDO\index{nocional del CDO} en d\'olares.\footnote{\, Recordemos que el punto de intervenci\'on\index{punto de intervenci\'on} $a$ y el punto de desprendimiento\index{punto de desprendimiento} $d$ se miden en \%.} Desde la perspectiva del trader largo en el tramo\index{tramo}, el valor de mercado (MTM, por sus siglas en ingl\'es)\index{valor de mercado (MTM)} del tramo\index{tramo}, ll\'amese ${\cal M}$, es dado por
\begin{eqnarray}
 &&{\cal M} = P - C\\
 &&P = S~\sum_{i=1}^n D_i~\Delta_i \left[M_{tr} - L(t_i)\right]\\
 &&C = \sum_{i=1}^n D_i \left[L(t_i) - L(t_{i-1})\right]
\end{eqnarray}
Aqu\'i: $P$ es la pierna premium\index{pierna premium}; $C$ es la pierna contingente (de incumplimiento)\index{pierna contingente}\index{pierna, de incumplimiento}; $S$ es el margen\index{margen, de tramo de CDO}; $\Delta_i = t_i - t_{i-1}$; $D_i$ es el factor de descuento libre de riesgo\index{factor de descuento libre de riesgo} para la fecha de pago $t_i$; y $M_{tr} = L_d - L_a$ es el nocional del tramo\index{nocional del tramo}. (Tambi\'en, $t$ se mide en a\~{n}os, $t_0$ es el momento inicial, y $L(t_0) = 0$). Estableciendo el MTM\index{valor de mercado (MTM)} ${\cal M} = 0$, se fija el valor del margen\index{margen, de tramo de CDO} $S=S_*$.

{}Podemos definir adem\'as la ``duraci\'on arriesgada\index{duraci\'on arriesgada}'' ${\cal D}$ del tramo\index{tramo} como la primera derivada del MTM\index{valor de mercado (MTM)} con respecto al margen\index{margen, de tramo de CDO}:
\begin{eqnarray}
 &&{\cal M}(S) = \left(S - S_*\right)\sum_{i=1}^n D_i~\Delta_i \left[M_{tr} - L(t_i)\right]\label{MTM}\\
 &&{\cal D} = \partial{\cal M}/\partial S = \sum_{i=1}^n D_i~\Delta_i \left[M_{tr} - L(t_i)\right]\label{risky-duration}
\end{eqnarray}
La duraci\'on arriesgada\index{duraci\'on arriesgada} ${\cal D}_{ix}$ tambi\'en se puede definir de manera similar para un \'indice de CDS\index{indice de CDS @ \'indice de CDS}.

\subsection{Estrategia: Carry\index{carry}, tramo de capital\index{tramo de capital} -- cobertura con \'indice\index{cobertura del \'indice}}

{}Esta estrategia consiste en comprar el tramo de capital\index{tramo de capital} (de calidad m\'as baja) y cubrir el Delta vendiendo el \'indice\index{indice @ \'indice}. El Delta\index{Delta} (es decir, el ratio de cobertura\index{ratio de cobertura}) est\'a dado por\footnote{\, Para algunos estudios sobre la cobertura de tramos de CDOs\index{cobertura de tramo del CDO} y t\'opicos relacionados, v\'ease, por ejemplo, \cite{Arnsdorf2007}, \cite{Bielecki2007}, \cite{Bielecki2008}, \cite{Carmona2010}, \cite{Cont2013}, \cite{Frey2008}, \cite{Frey2010}, \cite{Giesecke2006}, \cite{Herbertsson2008}, \cite{Houdain2006}, \cite{Laurent2011}, \cite{Walker2008}.}
\begin{eqnarray}\label{delta-ix}	
 \Delta_{ix} = \frac{ {\cal D}}{{\cal D}_{ix}} \
\end{eqnarray}
Las primas\index{prima} recibidas del tramo de capital\index{tramo de capital} son m\'as altas que las primas\index{prima} pagadas en la posici\'on en el \'indice\index{posici\'on en el \'indice}. El riesgo\index{riesgo} es la exposici\'on\index{exposici\'on} a los eventos de cr\'edito del tramo de capital\index{eventos de cr\'edito del tramo de capital}.

\subsection{Estrategia: Carry\index{carry}, senior/mezzanine -- cobertura con \'indice\index{cobertura del \'indice}}

{}Esta estrategia consiste en {\em vender} un tramo\index{tramo} de alta calidad (por ejemplo, senior/mezzanine) y cubrir la posici\'on para lograr la Delta-neutralidad {\em comprando} el \'indice\index{indice @ \'indice}.\footnote{\, Las primas\index{prima} recibidas del \'indice\index{indice @ \'indice} son m\'as altas que las primas\index{prima} que se pagan por la posici\'on corta en el tramo\index{tramo}. Por lo tanto, este trade es ``opuesto'' al trade que establece una posici\'on larga en el tramo de capital\index{tramo de capital} y la cubre con el \'indice\index{indice @ \'indice}.} El Delta\index{Delta} est\'a dado por la Ecuaci\'on (\ref{delta-ix}).

\subsection{Estrategia: Carry\index{carry} -- cobertura con tramo\index{cobertura de tramo}}\label{sub.tranche.hedge}

{}Esta estrategia consiste en comprar un tramo\index{tramo} de baja calidad y cubrir el Delta de la posici\'on vendiendo un tramo\index{tramo} de alta calidad. El ratio de cobertura\index{ratio de cobertura} est\'a dado por:
\begin{eqnarray}	
&& \Delta_{alto} = \frac{{\cal D}_{bajo}}{{\cal D}_{alto}} \
\end{eqnarray}
Aqu\'i ${\cal D}_{bajo}$ y ${\cal D}_{alto}$ son las duraciones arriesgadas\index{duraci\'on arriesgada} de los tramos\index{tramo} de baja y alta calidad, respectivamente.

\subsection{Estrategia: Carry\index{carry} -- cobertura con CDS\index{cobertura con CDS}}

{}Esta estrategia consiste en comprar un tramo\index{tramo} de baja calidad y cubrir el Delta de la posici\'on vendiendo un CDS\index{swap de incumplimiento crediticio (CDS)} de referencia individual con menores pagos de primas\index{pago de prima} que el tramo\index{tramo} largo (en lugar del \'indice\index{indice @ \'indice} o un tramo\index{tramo} de mayor calidad). El ratio de cobertura\index{ratio de cobertura} est\'a dado por la Ecuaci\'on (\ref{delta-ix}) con ${\cal D}_{ix}$ reemplazado por la duraci\'on arriesgada\index{duraci\'on arriesgada} ${\cal D}_{CDS}$ del CDS\index{swap de incumplimiento crediticio (CDS)}:
\begin{eqnarray}	
&& \Delta_{CDS} = \frac{{\cal D}}{{\cal D}_{CDS}} \
\end{eqnarray}

\subsection{Estrategia: CDOs\index{obligaci\'on de deuda garantizada (CDO)} -- trade de la curva\index{trade de la curva}}

{}Como en el caso de los bonos\index{bono} (v\'ease la Subsecci\'on \ref{sub.yield.curve}), un trade de la curva\index{trade de la curva} flattener\index{flattener} (steepener\index{steepener}) implica la venta (compra) de un tramo\index{tramo} a corto plazo y la compra (venta) simult\'anea de un tramo\index{tramo} a largo plazo. Dicho de otra manera, con un flattener\index{flattener} (steepener\index{steepener}), el trader est\'a comprando (vendiendo) la protecci\'on a corto plazo y vendiendo (comprando) la protecci\'on a largo plazo, es decir, el trader espera que el margen de la curva\index{margen de la curva} se aplane (empine), por lo que el margen\index{margen} entre los tramos\index{tramo} a largo y a corto plazo disminuya (aumente). El carry\index{carry} del trade de la curva\index{trade de la curva} durante el per\'iodo de tiempo $t$ hasta el tiempo $t+\Delta t$ se puede definir de la siguiente manera
\begin{eqnarray}	
&& C(t, t+\Delta t) = \left(M_{largo}~S_{largo} - M_{corto}~S_{corto}\right)\Delta t
\end{eqnarray}
en donde $M_{largo}$ y $M_{corto}$ son los nocionales del tramo largo y corto\index{nocional del tramo}, y $S_{largo}$ y $S_{corto}$ son los m\'argenes\index{margen, de tramo de CDO} correspondientes. La operaci\'on puede ser estructurada para ser d\'olar-neutral (es decir, nocional neutral, $M_{largo}=M_{corto}$),\footnote{\, En este caso, para una curva de pendiente ascendente, un flattener\index{flattener} (steepener\index{steepener}) tiene un carry\index{carry} positivo (negativo) ya que $S_{largo} > S_{corto}$ ($S_{largo} < S_{corto}$).} duraci\'on arriesgada neutral (${\cal D}_{largo} = {\cal D}_{corto}$, v\'ease la Ecuaci\'on (\ref{risky-duration})), carry neutral ($M_{largo}S_{largo} = M_{corto}S_{corto}$), etc.\footnote{\, Por algunas publicaciones sobre el trade de la curva\index{trade de la curva} y t\'opicos relacionados, v\'ease, por ejemplo, \cite{Bobey2010}, \cite{Burtshell2009}, \cite{Choros-Tomczyk2016}, \cite{Crabbe2002}, \cite{Detlefsen2013}, \cite{Hagenstein2004}, \cite{Hamerle2012}, \cite{Hull2004}, \cite{Kakodkar2006}, \cite{Koopman2012}, \cite{Lin2008}, \cite{Rajan2007}.} El P\&L\index{P\&L} de la estrategia est\'a dado por (${\cal M}_{largo}$ y ${\cal M}_{corto}$ son los MTM de la posici\'on larga y corta en los tramos\index{tramo}, v\'ease la Ecuaci\'on (\ref{MTM})):
\begin{equation}
 \mbox{P\&L} = {\cal M}_{largo} - {\cal M}_{corto}
\end{equation}

\subsection{Estrategia: Valores respaldados por hipotecas (MBS)\index{valores respaldados por hipotecas (MBS)}}

{}Esta estrategia consiste en comprar un MBS passthrough\index{MBS passthrough} (MBS viene de sus siglas en ingl\'es)\footnote{\, Un MBS\index{valores respaldados por hipotecas (MBS)} es un activo respaldado por un conjunto de hipotecas\index{hipotecas}. En un MBS passthrough\index{MBS passthrough}, que es el tipo de MBS\index{valores respaldados por hipotecas (MBS)} m\'as com\'un, los flujos de efectivo\index{flujo de efectivo} se pasan de los deudores a los traders a trav\'es de un intermediario.} y cubrir la exposici\'on a la tasa de inter\'es\index{exposici\'on a la tasa de inter\'es} con swaps de tasas de inter\'es\index{swap de tasas de inter\'es}. El principal riesgo\index{riesgo} de un MBS passthrough\index{MBS passthrough} es el riesgo de prepago\index{riesgo de prepago}, ya que los propietarios tienen la opci\'on de prepagar sus hipotecas\index{hipotecas}. Los propietarios de viviendas refinancian sus hipotecas\index{hipotecas} cuando las tasas de inter\'es\index{tasa de inter\'es} bajan, lo que resulta en una convexidad\index{convexidad} negativa en el precio del MBS\index{precio del MBS} en funci\'on de las tasas de inter\'es\index{tasa de inter\'es} (por ejemplo, la tasa de swap\index{tasa de swap} a 5 a\~{n}os). Los ratios de cobertura\index{ratio de cobertura} son dependientes de un modelo y una variedad de modelos de prepago\index{modelo de prepago} se pueden construir. Alternativamente, se puede seguir un enfoque no param\'etrico mediante el cual, usando datos hist\'oricos\index{datos hist\'oricos}, se estima la primera derivada del precio del MBS\index{precio del MBS} passthrough\index{precio del MBS passthrough} $P$ con respecto a la tasa de swap\index{tasa de swap} a 5 a\~{n}os $R$ con la restricci\'on de que $P$ es una funci\'on no creciente de $R$ (v\'ease, por ejemplo, \cite{Duarte2006}),\footnote{\, Para obtener informaci\'on adicional pertinente, v\'ease, por ejemplo, \cite{Ambrose2004}, \cite{Biby2001}, \cite{Bielecki2011}, \cite{Boudoukh1997}, \cite{Brazil1988}, \cite{Brennan1985}, \cite{Carron1988}, \cite{Chinloy1989}, \cite{Davidson1988}, \cite{Dechario2010}, \cite{Downing2009}, \cite{Dunn1981a}, \cite{Dunn1981b}, \cite{Dynkin2001}, \cite{Fabozzi2006b}, \cite{Gabaix2007}, \cite{Glaeser1997}, \cite{Hu2001}, \cite{Longstaff2005}, \cite{Kau1995}, \cite{McConnell2011}, \cite{McKenzie2002}, \cite{Nothaft1995}, \cite{Passmore2005}, \cite{Richard1989}, \cite{Schultz2016}, \cite{Schwartz1989}, \cite{Schwartz1992}, \cite{Stanton1995}, \cite{Thibodeau1989}, \cite{Vickery2010}.} empleando, por ejemplo, una regresi\'on restringida\index{regresi\'on restringida} (v\'ease, por ejemplo, \cite{Ait-Sahalia2003}).

\newpage

\section{Convertibles\index{convertibles}}\label{sec.convertible}

\subsection{Estrategia: Arbitraje de convertible\index{arbitraje de convertible}}

{}Un bono convertible\index{bono convertible} es un activo h\'ibrido\index{activo hibrido @ activo h\'ibrido} con una opci\'on embebida\index{opci\'on embebida} para convertir el bono\index{bono} (un instrumento de renta fija\index{instrumento de renta fija}) en un n\'umero preestablecido (conocido como el ratio de conversi\'on\index{ratio de conversi\'on}) de acciones del mismo emisor (un instrumento de capital) cuando, por ejemplo, el precio de la acci\'on alcanza un nivel preestablecido (conocido como el precio de conversi\'on\index{precio de conversi\'on}). Emp\'iricamente, los convertibles\index{convertibles} al momento de la emisi\'on tienden a estar subvalorados en relaci\'on con su valor ``justo''\index{valor justo}.\footnote{\, Por algunas publicaciones sobre bonos convertibles\index{bono convertible} y t\'opicos relacionados, v\'ease, por ejemplo, \cite{Agarwal2011}, \cite{Ammann2010}, \cite{Ammann2003}, \cite{Batta2010}, \cite{Brennan1988}, \cite{Brown2012}, \cite{Calamos2003}, \cite{Chan2007}, \cite{Choi2010}, \cite{Choi2009}, \cite{DeJong2011}, \cite{Duca2012}, \cite{Dutordoir2014}, \cite{Grundy2016}, \cite{Henderson2005}, \cite{Henderson2012}, \cite{Ingersoll1977}, \cite{Kang1996}, \cite{King1986}, \cite{King2014}, \cite{Korkeamaki2013}, \cite{Lewis1999}, \cite{Lewis2011}, \cite{Loncarski2006}, \cite{Loncarski2009}, \cite{Mayers1998}, \cite{Ryabkov2015}, \cite{Stein1992}, \cite{Tsiveriotis1998}, \cite{Marle2017}, \cite{Zabolotnyuk2010}.} Esto da lugar a oportunidades de arbitraje. Una estrategia de arbitraje de convertible\index{estrategia de arbitraje de convertible} consiste en comprar un bono convertible\index{bono convertible} y al mismo tiempo vender $h$ unidades de las acciones subyacentes\index{acci\'on subyacente}, en donde el ratio de cobertura\index{ratio de cobertura} est\'a dado por
\begin{eqnarray}
 &&h = \Delta \times C\\
 &&\Delta = \partial V/\partial S
\end{eqnarray}
Aqu\'i: $C$ es el ratio de conversi\'on\index{ratio de conversi\'on}; $V$ es el valor de la opci\'on de conversi\'on\index{opci\'on de conversi\'on} (que depende de un modelo); $S$ es el precio de las acciones subyacentes\index{precio de la acci\'on subyacente}; y $\Delta$ es el Delta\index{Delta} (dependiente de un modelo) de la opci\'on de conversi\'on\index{opci\'on de conversi\'on}.\footnote{\, El Delta\index{Delta} cambia con el precio de las acciones $S$. Para tener en cuenta esto, el Gamma de la opci\'on\index{Gamma de la opci\'on} puede ser utilizado como en la Subsecci\'on \ref{sub.Gamma.scalping} (la cobertura de Gamma\index{cobertura de Gamma}).} Normalmente, la posici\'on se mantiene durante 6 a 12 meses a partir de la fecha de emisi\'on del convertible y el ratio de cobertura\index{ratio de cobertura} se actualiza diariamente.

\subsection{Estrategia: Diferencial ajustado por opciones de convertible\index{diferencial ajustado por opciones de convertible}}

{}Esta estrategia consiste en comprar y vender simult\'aneamente dos bonos convertibles\index{bono convertible} diferentes del mismo emisor. La posici\'on larga es en un bono\index{bono} con un diferencial ajustado por opciones (OAS, por sus siglas en ingl\'es)\index{diferencial ajustado por opciones (OAS)} m\'as alto, y la posici\'on corta es en un bono\index{bono} con un OAS\index{diferencial ajustado por opciones (OAS)} menor (v\'ease, por ejemplo, \cite{Calamos2003}). Entonces la operaci\'on es rentable si estos dos m\'argenes\index{margen} convergen.

{}El OAS\index{diferencial ajustado por opciones (OAS)} se puede calcular de la siguiente manera (v\'ease, por ejemplo, \cite{Hull2012}).\footnote{\, Para m\'as literatura relacionada con el OAS\index{diferencial ajustado por opciones (OAS)} (principalmente enfocada en aplicaciones a MBS\index{valores respaldados por hipotecas (MBS)}), v\'ease, por ejemplo, \cite{Boyarchenko2014}, \cite{Brazil1988}, \cite{Brown1999}, \cite{Cerrato2008}, \cite{Dong2009}, \cite{Hayre1990}, \cite{Huang2003}, \cite{Levin2005}, \cite{LiuXu1998}, \cite{Stroebel2012}, \cite{Windas2007}.} Una forma sencilla (pero no la \'unica)\footnote{\, Por algunas publicaciones sobre la valuaci\'on de bonos convertibles\index{valuaci\'on de bonos convertibles}, v\'ease, por ejemplo, \cite{Ayache2013}, \cite{Batten2014}, \cite{Brennan1977}, \cite{Finnerty2017}, \cite{Ingersoll1977}, \cite{Kang1996}, \cite{King1986}, \cite{Kwok2014}, \cite{McConnell1986}, \cite{Milanov2013}, \cite{Park2017}, \cite{Sorensson1993}, \cite{Tsiveriotis1998}, \cite{Xiao2013}, \cite{Zabolotnyuk2010}.} para calcular el precio $P_C$ del bono convertible\index{bono convertible} es asumir que
\begin{equation}
 P_C = P_B + V
\end{equation}
en donde $P_B$ es el precio del bono por s\'i solo\index{bono} (es decir, sin la opci\'on embebida\index{opci\'on embebida}), y $V$ es el valor de la opci\'on de conversi\'on\index{opci\'on de conversi\'on}, que es una opci\'on call\index{opci\'on call}. $P_B$ se computa descontando los flujos de efectivo\index{flujo de efectivo} futuros del bono\index{bono}. Por otro lado, $V$ depende de la curva de tasas de inter\'es libre de riesgo\index{curva de tasas de inter\'es libre de riesgo}. En la iteraci\'on inicial, se calcula $V$ (utilizando un modelo de valuaci\'on\index{modelo de valuaci\'on} para la opci\'on call\index{opci\'on call}) asumiendo que la curva de bonos con cup\'on cero del Tesoro\index{curva del Tesoro}\index{curva de bonos con cup\'on cero del Tesoro} es la curva de tasas de inter\'es libre de riesgo\index{curva de tasas de inter\'es libre de riesgo}. Esta iteraci\'on inicial $V ^ {(0)}$ puede no coincidir con $P^{mkt}_C - P_B$, en donde $P^{mkt}_C$ es el {\em precio de mercado\index{precio de mercado}} del bono convertible\index{bono convertible}. Luego, iterativamente (por ejemplo, utilizando el m\'etodo de bisecci\'on\index{metodo de biseccion @ m\'etodo de bisecci\'on}) {\em se realizan desplazamientos paralelos\index{desplazamiento paralelo}} de la curva del Tesoro\index{curva del Tesoro} hasta que $V$, que se calcula utilizando la curva desplazada, sea tal que $V = P^{mkt}_C - P_B$. El desplazamiento paralelo\index{desplazamiento paralelo} de la curva obtenido a trav\'es de este procedimiento iterativo es el OAS\index{diferencial ajustado por opciones (OAS)}.

\newpage

\section{Arbitraje Fiscal\index{arbitraje fiscal}}\label{sec.tax}

\subsection{Estrategia: Arbitraje fiscal con bonos municipales\index{arbitraje fiscal con bonos municipales}}

{}Esta estrategia es una de las formas m\'as comunes y simples de realizar un arbitraje fiscal\index{arbitraje fiscal}. Se trata de pedir prestado dinero y comprar bonos municipales exentos de impuestos\index{bonos municipales exentos de impuestos}.\footnote{\, Por algunas publicaciones sobre arbitraje fiscal con bonos municipales\index{arbitraje fiscal con bonos municipales} y t\'opicos relacionados, v\'ease, por ejemplo, \cite{Ang2017}, \cite{Buser1986}, \cite{Chalmers1998}, \cite{Erickson2003}, \cite{Heaton1988}, \cite{Kochin1988}, \cite{Longstaff2011}, \cite{Miller1977}, \cite{Poterba1986}, \cite{Poterba1989}, \cite{Skelton1983}, \cite{Trzcinka1982}, \cite{YawitzMaloney1985}.} El retorno de la estrategia est\'a dado por
\begin{eqnarray}	
 && R = r_{largo} - r_{corto}\left(1 - \tau\right) \
\end{eqnarray}
Aqu\'i: $r_{largo}$ es la tasa de inter\'es\index{tasa de inter\'es} de los bonos municipales\index{bono municipal} comprados, $r_{corto}$ es la tasa de inter\'es\index{tasa de inter\'es} del pr\'estamo\index{prestamo @ pr\'estamo}, y $\tau$ es la tasa de impuesto corporativa. Esta estrategia es atractiva para las empresas en jurisdicciones en donde las normas tributarias les permiten comprar bonos municipales exentos de impuestos\index{bonos municipales exentos de impuestos} y deducir los gastos por intereses de su ingreso imponible (esto tambi\'en se conoce como ``escudo fiscal\index{escudo fiscal}'').

\subsection{Estrategia: Arbitraje fiscal transfronterizo\index{arbitraje fiscal transfronterizo}}

{}Las ganancias corporativas en los Estados Unidos son gravadas impositivamente dos veces. La ganancia se grava primero a nivel corporativo. Luego, se grava de nuevo cuando los accionistas\index{accionista} reciben los dividendos\index{dividendo}. En otros pa\'ises, los sistemas tributarios est\'an dise\~{n}ados para aliviar la carga tributaria, por ejemplo, al no gravar los dividendos\index{dividendo} (como, por ejemplo, en Singapur), o dando a los accionistas\index{accionista} cr\'editos fiscales\index{credito fiscal @ cr\'edito fiscal} adjuntos a los pagos de dividendos\index{pago de dividendo} (como, por ejemplo, en Australia). En el caso de que este sistema tributario corporativo de ``imputaci\'on de dividendos\index{imputaci\'on de dividendos}'' otorgue el cr\'edito fiscal\index{credito fiscal @ cr\'edito fiscal} completo a los accionistas\index{accionista}, esto se puede describir esquem\'aticamente de la siguiente manera (v\'ease, por ejemplo, \cite{McDonald2001}):\footnote{\, Sin embargo, puede haber limitaciones en el cr\'edito fiscal\index{credito fiscal @ cr\'edito fiscal} y otras sutilezas presentes dependiendo de la jurisdicci\'on, diversas circunstancias, etc.}
\begin{eqnarray}
 \begin{cases}
      \mbox{Tasa de impuesto corporativa}  = \tau_{c} \\
      \mbox{Dividendo pagado en efectivo}  = D \\
      \mbox{Cr\'edito tributario por dividendos} = C = D~{\tau_{c}\over{1 - \tau_{c}}} \\
      \mbox{Ganancia imponible} = I_t = D + C = {D\over {1 - \tau_{c}} } \\
      \mbox{Tasa impositiva personal}   = \tau_{p}\\
      \mbox{Impuestos personales} = T = I_t~\tau_{p}\\
      \mbox{Dividendos despu\'es de cr\'edito e impuestos} = I = D + C - T = D~{{1 - \tau_{p}}\over{1 - \tau_{c}}}
\end{cases}
\end{eqnarray}
Entonces, si la ganancia corporativa es $P$ y la corporaci\'on distribuye todas sus ganancias despu\'es de impuestos como dividendos\index{dividendo}, entonces $D = P\left(1-\tau_c\right)$ y $I = P\left(1-\tau_p\right)$, es decir, no hay doble imposici\'on fiscal.\footnote{\, Por el contrario, en el sistema de doble imposici\'on\index{sistema de doble imposici\'on} tendr\'iamos en su lugar: $D = P\left(1-\tau_c\right)$, $I_t = D$, $T = I_t~\tau_{p}$, $I = I_t - T = P\left(1-\tau_c\right)\left(1-\tau_p\right)$.}

{}Mientras que en pa\'ises con sistemas de imputaci\'on\index{sistema de imputaci\'on} los traders nacionales disfrutan de cr\'editos fiscales\index{credito fiscal @ cr\'edito fiscal}, en general, los traders extranjeros no. Si no hubiera cr\'editos fiscales\index{credito fiscal @ cr\'edito fiscal}, se esperar\'ia que la ca\'ida de precios entre cum-dividendo\index{cum-dividendo} y ex-dividendo\index{ex-dividendo}\footnote{\, Cum-dividendo\index{cum-dividendo} significa que el comprador de acciones tiene el derecho a recibir un dividendo\index{dividendo} que ha sido declarado, pero no pagado. Ex-dividendo\index{ex-dividendo} significa que el vendedor de acciones tiene el derecho al dividendo\index{dividendo}, no el comprador.} refleje el dividendo\index{dividendo}. En presencia de cr\'editos fiscales\index{credito fiscal @ cr\'edito fiscal}, se espera que la ca\'ida sea mayor: si refleja plenamente el cr\'edito fiscal\index{credito fiscal @ cr\'edito fiscal}, entonces es $D\left(1+\kappa\right)$, en donde $\kappa$ es la tasa de cr\'edito fiscal\index{tasa de cr\'edito fiscal}. (En la nomenclatura anterior, $1 + \kappa = 1/(1+\tau_c)$). Entonces, un inversionista extranjero est\'a efectivamente penalizado por mantener las acciones. Para evitar esto, el inversionista extranjero puede vender las acciones cum-dividendo\index{cum-dividendo} y volver a comprarlas ex-dividendo\index{ex-dividendo}.\footnote{\, Asumiendo que los costos de transacci\'on\index{costos de transacci\'on} no son prohibitivamente altos.} Alternativamente, el inversionista extranjero puede prestar las acciones a un inversionista nacional cum-dividendo\index{cum-dividendo} y recibir las acciones de vuelta ex-dividendo\index{ex-dividendo} junto con una parte preestablecida del cr\'edito fiscal\index{credito fiscal @ cr\'edito fiscal} -- asumiendo que no hay restricciones en tal arbitraje fiscal transfronterizo\index{arbitraje fiscal transfronterizo}. Un acuerdo de swap\index{acuerdo de swap} tambi\'en lograr\'ia el mismo resultado.\footnote{\, Por algunas publicaciones sobre arbitraje fiscal transfronterizo\index{arbitraje fiscal transfronterizo} y t\'opicos relacionados, v\'ease, por ejemplo, \cite{Allen1995}, \cite{Amihud1997}, \cite{Bellamy1994}, \cite{Booth1987}, \cite{BoothJ1984}, \cite{Brown1993}, \cite{Bundgaard2013}, \cite{Callaghan2003}, \cite{Christoffersen2005}, \cite{Christoffersen2003}, \cite{Eun2003}, \cite{Green1999}, \cite{Harris2001}, \cite{Lakonishok1986}, \cite{Lasfer1995}, \cite{Lessambo2016}, \cite{McDonald2001}, \cite{Monkhouse1993}, \cite{Shaviro2002}, \cite{Wells2016}, \cite{Wood1997}.}

\subsubsection{Estrategia: Arbitraje fiscal transfronterizo\index{arbitraje fiscal transfronterizo} con opciones\index{opci\'on}}

{}En ausencia de cr\'edito fiscal\index{credito fiscal @ cr\'edito fiscal}, existe un l\'imite superior te\'orico sobre el valor de una opci\'on put americana\index{opci\'on put americana} (v\'ease, por ejemplo, \cite{Hull2012}):
\begin{eqnarray}	
&& V_{put}(K, T) \leq V_{call}(K, T) - S_0 + K + D \
\end{eqnarray}
Aqu\'i: $V_{put}$ ($V_{call}$) es el precio de la opci\'on put (call)\index{opci\'on, call}\index{opci\'on, put} al tiempo $t=0$; $K$ es el precio de ejercicio\index{precio de ejercicio}; $S_0$ es el precio de las acciones en $t=0$; $T$ es la madurez\index{madurez}; y $D$ es el valor presente de los dividendos\index{dividendo} durante la vida de la opci\'on\index{opci\'on}. Las opciones put\index{opci\'on put} se ejercen de manera \'optima ex-dividendo\index{ex-dividendo}. Por lo tanto, en presencia de un cr\'edito fiscal\index{credito fiscal @ cr\'edito fiscal}, se espera que los precios de las opciones put reflejen el cr\'edito fiscal\index{credito fiscal @ cr\'edito fiscal}, es decir, deber\'ian ser m\'as altos que en ausencia del cr\'edito fiscal\index{credito fiscal @ cr\'edito fiscal} (v\'ease, por ejemplo, \cite{McDonald2001}). Entonces, el trader extranjero puede vender las acciones cum-dividendo\index{cum-dividendo} (al precio $S_0$) y vender una opci\'on put\index{opci\'on put} muy ITM\index{opci\'on, ITM (dentro del dinero)}, cuyo valor cercano al vencimiento\index{vencimiento} es aproximadamente (aqu\'i $\kappa$ es la tasa de cr\'edito fiscal\index{tasa de cr\'edito fiscal} antes definida)
\begin{equation}
 V_{put}(K,T) = K - \left[S_0 - D\left(1+\kappa\right)\right]
\end{equation}
Una vez ejercido el put ex-dividendo\index{ex-dividendo} al precio de ejercicio\index{precio de ejercicio} $K$, el P\&L\index{P\&L} es el mismo que con las estrategias antes discutidas, que consisten en realizar un pr\'estamo de las acciones\index{prestamo de acciones @ pr\'estamo de acciones} o utilizar un swap\index{estrategia de swap}:
\begin{eqnarray}
&& \mbox{P\&L} = S_0 + V_{put}(K,T) - K = D\left(1+\kappa\right) \
\end{eqnarray}

\newpage

\section{Activos Miscel\'aneos}\label{se.misc}

\subsection{Estrategia: Cobertura contra la inflaci\'on\index{cobertura contra la inflaci\'on} -- swaps de inflaci\'on\index{swap de inflaci\'on}}

{}Esta estrategia consiste en comprar (vender) swaps de inflaci\'on\index{swap de inflaci\'on} para canjear una tasa fija (flotante)\index{tasa, fija}\index{tasa, flotante} de inflaci\'on\index{inflaci\'on} por una tasa flotante (fija)\index{tasa, fija}\index{tasa, flotante}. Los swaps de inflaci\'on\index{swap de inflaci\'on} conceptualmente son similares a los swaps de tasas de inter\'es\index{swap de tasas de inter\'es} (v\'ease la Subsecci\'on \ref{sub.swaps}). Un comprador (vendedor) de un swap de inflaci\'on\index{swap de inflaci\'on} tiene una posici\'on larga (corta) en la inflaci\'on\index{inflaci\'on} y recibe una tasa flotante (fija)\index{tasa, fija}\index{tasa, flotante}. El comprador tiene un retorno positivo si la inflaci\'on\index{inflaci\'on} supera a la inflaci\'on\index{inflaci\'on} esperada (es decir, la tasa fija del swap\index{tasa fija del swap}, tambi\'en conocida como ``tasa de equilibrio''\index{tasa de equilibrio}). La tasa fija\index{tasa fija} normalmente se calcula como el diferencial de la tasa de inter\'es\index{diferencial de la tasa de inter\'es} entre las notas/bonos del Tesoro\index{bono del Tesoro}\index{nota, del Tesoro} (seg\'un corresponda) y los Valores del Tesoro Protegidos Contra la Inflaci\'on (conocidos como TIPS, por sus siglas en ingl\'es)\index{Valores del Tesoro Protegidos Contra la Inflaci\'on (TIPS)} con la misma madurez\index{madurez} que la del swap\index{swap}. La tasa flotante\index{tasa flotante} normalmente se basa en un \'indice de inflaci\'on\index{indice de inflacion @ \'indice de inflaci\'on} tal como el \'Indice de Precios al Consumidor (CPI, por sus siglas en ingl\'es)\index{Indice de Precios al Consumidor (CPI) @ \'Indice de Precios al Consumidor (CPI)}. El tipo m\'as com\'un de swap de inflaci\'on\index{swap de inflaci\'on} es el swap de inflaci\'on con cup\'on cero\index{swap de inflaci\'on con cup\'on cero}, que tiene un solo flujo de efectivo\index{flujo de efectivo} en la fecha de vencimiento\index{vencimiento} $T$ (medida en a\~{n}os). Este flujo de efectivo\index{flujo de efectivo} es la diferencia entre el flujo de efectivo\index{flujo de efectivo} $C_{fijo}$ a la tasa fija y el flujo de efectivo\index{flujo de efectivo} $C_{flotante}$ a la tasa flotante. Estos flujos de efectivo\index{flujo de efectivo}, por \$1 de nocional, est\'an dados por:
\begin{eqnarray}
 &&C_{fija} = (1 + K)^T - 1\\\
 &&C_{flotante} = I(T) / I(0) - 1
\end{eqnarray}
Aqu\'i: $K$ es la tasa fija\index{tasa fija}; y $I(t)$ es el valor del CPI\index{Indice de Precios al Consumidor (CPI) @ \'Indice de Precios al Consumidor (CPI)} al momento $t$ ($t=0$ es el momento en que se establece el contrato de swap\index{contrato de swap}). Otro tipo de swap de inflaci\'on\index{swap de inflaci\'on} es el swap de inflaci\'on interanual (year-on-year\index{swap de inflaci\'on interanual} o YoY\index{anzo a anzo @ a\~{n}o a a\~{n}o}, por sus siglas en ingl\'es), que hace referencia a la inflaci\'on anual\index{inflaci\'on anual} (a diferencia de la inflaci\'on acumulada\index{inflaci\'on acumulada} para el caso del swap con cup\'on cero\index{swap con cup\'on cero}). Por lo tanto, asumiendo por simplicidad pagos anuales, tenemos (aqu\'i $t=1,\dots,T$ es medido en a\~{n}os):\footnote{\, Por algunas publicaciones sobre swaps de inflaci\'on\index{swap de inflaci\'on} y t\'opicos relacionados, v\'ease, por ejemplo, \cite{Belgrade2004}, \cite{BelgradeBK2004}, \cite{Bouzoubaa2010}, \cite{Christensen2010}, \cite{Deacon2004}, \cite{Fleming2013}, \cite{Haubrich2012}, \cite{Hinnerich2008}, \cite{Jarrow2003}, \cite{Kenyon2008}, \cite{Lioui2005}, \cite{Martellini2015}, \cite{Mercurio2005}, \cite{Mercurio2006}, \cite{Mercurio2009}, \cite{Mercurio2008}.}
\begin{eqnarray}
 &&C_{fija}(t) =  K\\\
 &&C_{flotante}(t) = I(t) / I(t-1) - 1
\end{eqnarray}

\subsection{Estrategia: Arbitraje con TIPS del Tesoro\index{arbitraje con TIPS del Tesoro}}

{}Esta estrategia se basa en la observaci\'on emp\'irica de que los bonos del Tesoro\index{bono del Tesoro} tienden a estar sobrevalorados en relaci\'on con los TIPS\index{Valores del Tesoro Protegidos Contra la Inflaci\'on (TIPS)}\footnote{\, Los TIPS\index{Valores del Tesoro Protegidos Contra la Inflaci\'on (TIPS)} pagan cupones fijos\index{cup\'on fijo} semestrales\index{cupones fijos semestrales} a una tasa fija\index{tasa fija}, pero los pagos de cupones\index{pago de cup\'on} (y del principal\index{principal}) se ajustan en funci\'on de la inflaci\'on\index{inflaci\'on}. Por algunas publicaciones sobre TIPS\index{Valores del Tesoro Protegidos Contra la Inflaci\'on (TIPS)}, productos indexados a la inflaci\'on\index{productos indexados a la inflaci\'on} y t\'opicos relacionados, v\'ease, por ejemplo, \cite{Adrian2010}, \cite{AngBW2008}, \cite{Bardong2004}, \cite{Barnes2010}, \cite{Barr1997}, \cite{Bekaert2010}, \cite{Buraschi2005}, \cite{Campbell2017}, \cite{ChenLC2010}, \cite{Chernov2012}, \cite{Christensen2012}, \cite{DAmico2018}, \cite{Deacon2004}, \cite{Dudley2009}, \cite{Evans1998}, \cite{Fleckenstein2017}, \cite{Fleming2012}, \cite{Grishchenko2013}, \cite{Grishchenko2016}, \cite{Gurkaynak2010}, \cite{Hordahl2012}, \cite{Hordahl2014}, \cite{Hunter2005}, \cite{Jacoby2008}, \cite{Joyce2010}, \cite{Kandel1996}, \cite{Kitsul2013}, \cite{Kozicki2012}, \cite{Mehra2002}, \cite{Pennacchi1991}, \cite{Pflueger2011}, \cite{Remolona1998}, \cite{Roll1996}, \cite{Roll2004}, \cite{Sack2004}, \cite{Seppala2004}, \cite{Shen2006}, \cite{Shen2001}, \cite{Woodward1990}, \cite{Veronesi1999}.} casi todo el tiempo (v\'ease, por ejemplo, \cite{Campbell2009}, \cite{Driessen2017}, \cite{Fleckenstein2012}, \cite{Haubrich2012}). La estrategia consiste en vender un bono del Tesoro\index{bono del Tesoro} (cuyo precio es $P_{Tesoro}$, la tasa de cup\'on fija\index{tasa de cup\'on fija} es $r_{Tesoro}$, y la madurez\index{madurez} es $T$) y compensar esta posici\'on corta con un portafolio sint\'etico\index{portafolio sint\'etico}, que reproduce exactamente los pagos de los cupones y del principal del bono del Tesoso\index{bono del Tesoro}\index{pago, de cup\'on}\index{pago, de principal} a un menor costo. Este portafolio sint\'etico\index{portafolio sint\'etico} se construye comprando los TIPS\index{Valores del Tesoro Protegidos Contra la Inflaci\'on (TIPS)} (cuyo precio es $P_{TIPS}$ y la madurez\index{madurez} $T$ es la misma que la del bono del Tesoro\index{bono del Tesoro}) con una tasa de cup\'on fija\index{tasa de cup\'on fija} $r$ y $n$ pagos de cupones\index{pago de cup\'on} en los tiempos $t_i$, $i=1,\dots,n$ (con $t_n = T$), y simult\'aneamente vendiendo $n$ swaps de inflaci\'on con cup\'on cero\index{swap de inflaci\'on con cup\'on cero} con las fechas de vencimiento\index{vencimiento} $t_i$, la tasa fija\index{tasa fija} $K$, y los nocionales\index{nocional} $N_i = r + \delta_{t_i, T}$ por \$1 de principal de los TIPS\index{principal de TIPS}. Los flujos de efectivo\index{flujo de efectivo} (por \$1 de nocional\index{nocional}) en $t=t_i$ son dados por (como arriba, $I(t)$ es el valor del CPI\index{Indice de Precios al Consumidor (CPI) @ \'Indice de Precios al Consumidor (CPI)} al momento $t$; tambi\'en, el tiempo se mide en las unidades de los per\'iodos de composici\'on\index{periodo de composicion @ per\'iodo de composici\'on} (t\'ipicamente, semestrales)):
\begin{eqnarray}
&& C_{TIPS}(t_i) = N_i~ I(t_i)/I(0) \\
&& C_{swap}(t_i) = N_i\left[\left(1 + K\right)^{t_i} - I(t_i)/I(0)\right] \\
&& C_{total}(t_i) = C_{swap}(t_i) + C_{TIPS}(t_i) =  N_i\left(1 + K\right)^{t_i} \label{totalCF}\
\end{eqnarray}
Entonces, el portafolio sint\'etico\index{portafolio sint\'etico} convierte los pagos indexados\index{pago indexado} de los TIPS\index{Valores del Tesoro Protegidos Contra la Inflaci\'on (TIPS)} en pagos fijos con las tasas de cup\'on\index{tasa de cup\'on} efectivas $r_{eff}(t_i) = r \left(1 + K\right)^{t_i}$. Estos pagos de cupones sint\'eticos\index{pago de cup\'on sint\'etico} casi replican a los cupones de los bonos del Tesoro\index{cupones de bonos del Tesoro} $r_{Tesoro}$. La coincidencia exacta implica peque\~{n}as posiciones largas o cortas en STRIPS\index{Transacciones Separadas de Intereses y Principales Nominativos (STRIPS)}\footnote{\, Los STRIPS\index{Transacciones Separadas de Intereses y Principales Nominativos (STRIPS)} = ``Transacciones Separadas de Intereses y Principales Nominativos''. Esencialmente, los STRIPS\index{Transacciones Separadas de Intereses y Principales Nominativos (STRIPS)} son bonos de descuento con cup\'on cero\index{bono de descuento con cup\'on cero}.} (por sus siglas en ingl\'es), que est\'an dadas por (v\'ease, por ejemplo, \cite{Fleckenstein2013} por m\'as detalles)
\begin{equation}\label{strips}
 S(t_i) = D(t_i)\left\{\left[r_{Tesoro} - r_{eff}(t_i)\right]  + \delta_{t_i, T}\left[1 - \left(1 + K\right)^{t_i} \right] \right\}
\end{equation}
en donde $D(\tau)$ es el valor de los STRIPS\index{Transacciones Separadas de Intereses y Principales Nominativos (STRIPS)} con madurez\index{madurez} $\tau$ al momento $t=0$ (es decir, $D(\tau)$ es un factor de descuento\index{factor de descuento}). En la Ecuaci\'on (\ref{strips}), el segundo t\'ermino en los corchetes (que es proporcional a $\delta_{t_i, T}$ y es distinto de cero solo para $i=n$, es decir, en la madurez\index{madurez} $T$) se incluye dado que tambi\'en los principales\index{principal} deben coincidir en la fecha de madurez\index{madurez}. Tenga en cuenta que las posiciones en los STRIPS\index{Transacciones Separadas de Intereses y Principales Nominativos (STRIPS)} se establecen en $t=0$. El flujo de efectivo\index{flujo de efectivo} neto $C(0)$ en $t=0$ viene dado por (n\'otese que los flujos de efectivo\index{flujo de efectivo} netos en $t > 0$ son nulos por la replicaci\'on\index{replicaci\'on})
\begin{equation}
 C(0) = P_{Tesoro} - P_{TIPS} - \sum_{i=1}^n S(t_i)
\end{equation}
Emp\'iricamente $C(0)$ tiende a ser positivo (incluso despu\'es de costos de transacci\'on\index{costos de transacci\'on}). De ah\'i viene el arbitraje\index{arbitraje}.

\subsection{Estrategia: Riesgo del clima\index{riesgo del clima} -- cobertura de la demanda\index{cobertura de la demanda}}

{}Diversas empresas y sectores\index{sector} de la econom\'ia pueden verse afectados por las condiciones clim\'aticas\index{condiciones clim\'aticas}, tanto directa como indirectamente. El riesgo del clima\index{riesgo del clima} se puede cubrir con derivados del clima\index{derivados del clima}. No hay \'indices del clima\index{indices del clima @ \'indices del clima} ``negociables'', entonces varios \'indices sint\'eticos\index{indice sintetico @ \'indice sint\'etico} han sido creados. Los m\'as comunes se basan en la temperatura. Los grados-d\'ia de refrigeraci\'on (CDD, por sus siglas en ingl\'es)\index{grados-d\'ia de refrigeraci\'on (CDD)} y los grados-d\'ia de calefacci\'on (HDD, por sus siglas en ingl\'es)\index{grados-d\'ia de calefacci\'on (HDD)} miden temperaturas extremadamente altas y temperaturas extremadamente bajas, respectivamente:\footnote{\, Por algunas publicaciones sobre derivados del clima\index{derivados del clima}, \'indices del clima\index{indices del clima @ \'indices del clima} y t\'opicos relacionados, v\'ease, por ejemplo, \cite{Alaton2010}, \cite{Barrieu2002}, \cite{Barrieu2010}, \cite{Benth2003}, \cite{Benth2005}, \cite{Benth2007}, \cite{BenthSK2007}, \cite{Bloesch2015}, \cite{Brocket2010}, \cite{Brockett2005}, \cite{Brody2002}, \cite{Campbell2005}, \cite{Cao2000}, \cite{Cao2004}, \cite{Cartea2005}, \cite{Chaumont2006}, \cite{ChenRT2006}, \cite{Corbally2002}, \cite{DavisM2001}, \cite{Dischel1998a}, \cite{Dischel1998b}, \cite{Dischel1999}, \cite{Dorfleitner2010}, \cite{Dornier2000}, \cite{Ederington1979}, \cite{Geman1998}, \cite{Geman2005}, \cite{Ghiulnara2010}, \cite{Golden2007}, \cite{Goncu2012}, \cite{Hamisultane2009}, \cite{Hanley1999}, \cite{Hardle2011}, \cite{HuangSL2008}, \cite{Huault2011}, \cite{Hunter1999}, \cite{Jain2000}, \cite{Jewson2004a}, \cite{Jewson2004b}, \cite{Jewson2005}, \cite{Jewson2003}, \cite{Lazo2011}, \cite{Lee2009}, \cite{Leggio2002}, \cite{Mraoua2007}, \cite{Muller2000}, \cite{Oetomo2005}, \cite{Parnaudeau2018}, \cite{PerezGonzalez2010}, \cite{Richards2004}, \cite{SaltyteBenth2012}, \cite{Schiller2010}, \cite{Svec2007}, \cite{Swishchuk2013}, \cite{Tang2011}, \cite{Thornes2006}, \cite{Vedenov2004}, \cite{Wilson2016}, \cite{Woodard2008}, \cite{Yang2009}, \cite{Zapranis2008}, \cite{Zapranis2009}, \cite{Zeng2000}.}
\pagebreak
\begin{eqnarray}
&& I_{CDD} = \sum_{i=1}^n \mbox{max}(0, T_i - T_{base}) \label{temprature} \\
&& I_{HDD} = \sum_{i=1}^n \mbox{max}(0, T_{base} - T_i) \label{temprature1}\\
&& T_i = {{T_i^{min} + T_i^{max}}\over 2}
\end{eqnarray}
Aqu\'i: $i=1,\dots,n$ etiqueta los d\'ias; $n$ es la vida del contrato (una semana, un mes o una temporada) medida en d\'ias; $T^{min}_i$ y $T^{max}_i$ son las temperaturas m\'inimas y m\'aximas registradas en los d\'ias marcados por $i$; y $T_{base} = 65^{\circ}\mbox{F}$. Entonces, el riesgo de demanda\index{riesgo de demanda} para los d\'ias de calor, por ejemplo, puede cubrirse con una posici\'on corta en los futuros\index{posici\'on, corta en futuros} o una posici\'on larga en una opci\'on put\index{opci\'on put} con los ratios de cobertura\index{ratio de cobertura} dados por (aqu\'i (Cov) Var es la (co)varianza serial):
\begin{eqnarray}
&& h^{HDD}_{futuros} = \mbox{Cov}(q_w, I_{HDD})/\mbox{Var}(I_{HDD}) \\
&& h^{HDD}_{put} = -\mbox{Cov}\left(q_w, \mbox{max}(K - I_{HDD}, 0)\right)/\mbox{Var}\left(\mbox{max}(K - I_{HDD}, 0)\right) \
\end{eqnarray}
Aqu\'i: $q_w$ es la parte de la demanda\index{demanda} que se ve afectada por las condiciones clim\'aticas\index{condiciones clim\'aticas} (es importante considerar que puede haber otros componentes ex\'ogenos que afectan a la demanda\index{demanda}, y que van m\'as all\'a del clima); y $K$ es el precio de ejercicio\index{precio de ejercicio}. De modo similar, el riesgo de demanda\index{riesgo de demanda} para los d\'ias de frio puede, por ejemplo, cubrirse con una posici\'on larga en los futuros\index{posici\'on, larga en futuros} o una posici\'on larga en una opci\'on call\index{opci\'on call} con los ratios de cobertura\index{ratio de cobertura} dados por:
\begin{eqnarray}
&& h^{CDD}_{futuros} = \mbox{Cov}(q_w, I_{CDD})/\mbox{Var}(I_{CDD}) \\
&& h^{CDD}_{call} = \mbox{Cov}\left(q_w, \mbox{max}(I_{CDD} - K, 0)\right)/\mbox{Var}\left(\mbox{max}(I_{CDD} - K, 0)\right) \
\end{eqnarray}

\subsection{Estrategia: Energ\'ia\index{energ\'ia} -- diferencial spark\index{diferencial spark}}

{}El diferencial spark\index{diferencial spark} es la diferencia entre el precio mayorista de la electricidad y el precio del gas natural requerido para producirla.\footnote{\, Entonces, el diferencial spark\index{diferencial spark} mide el margen bruto de una central el\'ectrica que produce con gas, excluyendo todos los dem\'as costos de operaci\'on, mantenimiento, capital, etc. Adem\'as, si la planta de energ\'ia utiliza un combustible distinto al gas natural, entonces el margen\index{margen} correspondiente tiene un nombre diferente. Para el carb\'on se llama ``diferencial oscuro''\index{diferencial oscuro}; para la energ\'ia nuclear se llama ``diferencial cuarc\index{diferencial cuarc}''; etc. Por literatura sobre diferenciales de energ\'ia\index{diferencial de energ\'ia}, cobertura de energ\'ia\index{cobertura de energ\'ia} y t\'opicos relacionados, v\'ease, por ejemplo, \cite{Benth2011}, \cite{Benth2014}, \cite{Carmona2003}, \cite{Cassano2013}, \cite{Deng2001}, \cite{Edwards2009},  \cite{Elias2016}, \cite{Emery2002}, \cite{Fiorenzani2006}, \cite{Fusaro2005}, \cite{Hsu1998}, \cite{James2003}, \cite{Kaminski2004}, \cite{Li2009}, \cite{Maribu2007}, \cite{Martinez2018}, \cite{Wang2013}.} Un diferencial spark\index{diferencial spark} se puede construir, por ejemplo, tomando una posici\'on corta en futuros de electricidad\index{futuros de electricidad} y una posici\'on larga en el n\'umero correspondiente de futuros de combustible\index{futuros de combustible}. Tales posiciones son utilizadas por los productores de electricidad para protegerse contra los cambios en el precio de la electricidad o en el costo del combustible, as\'i como tambi\'en por traders o especuladores\index{especulador} quienes quieren hacer una apuesta a una central el\'ectrica. El n\'umero de futuros de combustible\index{futuros de combustible} est\'a determinado por la {\em tasa de calentamiento\index{tasa de calentamiento}} $H$, que mide la eficiencia con la que la planta convierte el combustible en electricidad:
\begin{equation}
 H = Q_F/Q_E
\end{equation}
Aqu\'i: $Q_F$ es la cantidad de combustible utilizado para producir la cantidad de electricidad $Q_E$; $Q_F$ se mide en MMBtu\index{MMBtu}; Btu = unidad t\'ermica brit\'anica\index{Btu (unidad t\'ermica brit\'anica)}, que es aproximadamente 1,055 Julios\index{Julio}; MBtu\index{MBtu} = 1,000 Btu\index{Btu (unidad t\'ermica brit\'anica)}; MMBtu\index{MMBtu} = 1,000,000 Btu\index{Btu (unidad t\'ermica brit\'anica)}; $Q_E$ se mide en Mwh = Megavatio hora\index{Megavatio hora (Mwh)}; la tasa de calentamiento\index{tasa de calentamiento} $H$ se mide en MMBtu\index{MMBtu}/Mwh\index{Megavatio hora (Mwh)}. El diferencial spark\index{diferencial spark} se mide en \$/Mwh\index{Megavatio hora (Mwh)}. Entonces, si el precio de la electricidad es $P_E$ (medido en \$/Mwh\index{Megavatio hora (Mwh)}) y el precio del combustible es $P_F$ (medido en \$/MMBtu\index{MMBtu}), luego el diferencial spark\index{diferencial spark} est\'a dado por
\begin{equation}
 S = P_E - H~P_F
\end{equation}
El ratio de cobertura\index{ratio de cobertura} para los futuros\index{futuro} se ve afectado por los tama\~{n}os de los contratos de futuros\index{tama\~{n}o de los contratos de futuros} disponibles. As\'i, un contrato de futuros de electricidad\index{contrato de futuros de electricidad} es $F_E = 736$ Mwh\index{Megavatio hora (Mwh)}, y un contrato de futuros de gas\index{contrato de futuros de gas} es $F_F = 10,000$ MMBtu\index{MMBtu}. Entonces, el ratio de cobertura\index{ratio de cobertura} est\'a dado por
\begin{equation}
 h = H~{F_E / F_F}
\end{equation}
que generalmente no es un n\'umero entero. Por lo tanto, se representa (aproximadamente, dentro de la precisi\'on deseada) como un ratio $h \approx N_F / N_E$ con el m\'inimo denominador posible $N_E$, en donde $N_F$ y $N_E$ son n\'umeros enteros. Entonces, la cobertura\index{cobertura} consiste en comprar $N_F$ contratos de futuros de gas\index{contrato de futuros de gas} para cada conjunto de $N_E$ contratos de futuros de electricidad\index{contrato de futuros de electricidad} vendidos.

\newpage

\section{Activos en Distress\index{activo en distress}}\label{sec.distressed}

\subsection{Estrategia: Inversi\'on pasiva en deuda en distress\index{deuda en distress}}\label{passivedistr}

{}Los activos en distress\index{activo en distress} (este t\'ermino en ingl\'es significa ``dificultades'' en espa\~{n}ol)  son aquellos cuyos emisores est\'an sufriendo dificultades financieras/operacionales\index{dificultad, financiera}\index{dificultad, operacional}\index{dificultad operacional}, incumplimiento\index{incumplimiento} o bancarrota\index{bancarrota}. Una de las formas de definir la deuda en distress\index{deuda en distress} es considerar el margen\index{margen} entre los rendimientos\index{rendimiento} de bonos del Tesoro\index{bono del Tesoro} y los del emisor. Si estos son mayores que alg\'un n\'umero preestablecido, por ejemplo, 1,000 puntos b\'asicos (bps, por sus siglas en ingl\'es)\index{punto b\'asico (bps)}, uno podr\'ia considerar que efectivamente, la deuda del emisor se encuentra en dificultades (v\'ease, por ejemplo, \cite{Harner2008}). Una simple estrategia de trading pasiva\index{estrategia de trading pasiva} en la deuda en distress\index{estrategia de trading pasiva en deuda en distress} consiste en comprar la deuda de una compa\~{n}\'ia en distress\index{compa\~{n}\'ia en distress} con un gran descuento,\footnote{\, Por literatura pertinente, v\'ease, por ejemplo, \cite{Altman1998}, \cite{Clark1983}, \cite{Eberhart1999}, \cite{Friewald2012}, \cite{Gande2010}, \cite{Gilson2010}, \cite{Gilson2012}, \cite{Harner2011}, \cite{Hotchkiss1997}, \cite{Jiang2012}, \cite{Lhabitant2002}, \cite{Morse1988}, \cite{Moyer2012}, \cite{Putnam1991}, \cite{Quintero1989}, \cite{Reiss1991}, \cite{Volpert1991}.} esperando que la compa\~{n}\'ia pague su deuda. T\'ipicamente, un portafolio de deuda en distress se diversifica\index{portafolio de deuda en distress, diversificado} en diversas industrias\index{industria}, entidades y nivel de prioridad\index{deuda, nivel de prioridad} y calidad de la deuda. Se anticipa que solo una peque\~{n}a fracci\'on de los activos que se mantienen en el portafolio\index{portafolio} tendr\'an rendimientos positivos, pero aquellos que lo hagan, proporcionar\'an altas tasas de rendimiento (v\'ease, por ejempo, \cite{Greenhaus1991}). Existen dos categor\'ias generales de estrategias pasivas en deuda en distress\index{estrategias pasivas en deuda en distress} (v\'ease, por ejemplo, \cite{Altman2006}). En primer lugar, usando distintos modelos (v\'ease la Subsecci\'on \ref{distressstrat}), uno puede tratar de predecir si una empresa se declarar\'a en bancarrota\index{bancarrota}. En segundo lugar, algunas estrategias se centran en los activos de empresas en incumplimiento\index{incumplimiento} o bancarrota\index{bancarrota}, en las cuales un proceso de reorganizaci\'on\index{reorganizaci\'on} exitoso es el motor de los retornos. Por lo general, las posiciones se establecen en fechas clave, como al final del mes de incumplimiento\index{mes de incumplimiento} o al final del mes de la presentaci\'on de bancarrota\index{presentaci\'on de bancarrota}, con el fin de explotar la sobre-reacci\'on\index{sobre-reacci\'on} en el mercado de deuda en distress\index{mercado de deuda en distress}\index{deuda en distress} (v\'ease, por ejemplo, \cite{Eberhart1992}, \cite{Gilson1995}).

\subsection{Estrategia: Inversi\'on activa en activos en distress\index{inversi\'on activa en activos en distress}}

{}Esta estrategia consiste en comprar activos en distress\index{activo en distress} con la visi\'on (a diferencia de la estrategia pasiva discutida anteriormente) de adquirir cierto grado de control en la gesti\'on y direcci\'on de la empresa. Ante una situaci\'on de dificultades\index{situaci\'on de dificultades}, una empresa tiene varias opciones para llevar adelante su proceso de reorganizaci\'on\index{proceso de reorganizaci\'on}. Puede solicitar la protecci\'on por bancarrota\index{protecci\'on por bancarrota} bajo el Cap\'itulo 11\index{Capitulo 11 @ Cap\'itulo 11} de la Ley de Quiebras de los Estados Unidos para reorganizarse. O puede trabajar directamente con sus acreedores fuera del Tribunal.\footnote{\, Por alguna literatura, v\'ease, por ejemplo, \cite{Altman2006}, \cite{Chatterjee1996}, \cite{Gilson1995}, \cite{Gilson1990}, \cite{Jostarndt2010}, \cite{Levy1991}, \cite{Markwardt2016}, \cite{Peric2015}, \cite{Rosenberg1992}, \cite{Swank1995}, \cite{Ward1993}.} A continuaci\'on se presentan algunos escenarios para la inversi\'on activa\index{inversi\'on activa}.

\subsubsection{Estrategia: Planificaci\'on de una reorganizaci\'on\index{reorganizaci\'on}}

{}Un inversor puede presentar un plan de reorganizaci\'on\index{plan de reorganizaci\'on} al Tribunal con el objetivo de obtener alguna participaci\'on en la gesti\'on de la empresa, intentar aumentar su valor y generar ganancias. Los planes por parte de los tenedores sustanciales de deuda tienden a ser m\'as competitivos.

\subsubsection{Estrategia: Compra de deuda en circulaci\'on}

{}Esta estrategia consiste en comprar la deuda en circulaci\'on de una compa\~{n}\'ia en distress\index{compa\~{n}\'ia en distress} con un sustancial descuento bajo la premisa de que, despu\'es de la reorganizaci\'on\index{reorganizaci\'on}, una parte de esta deuda se convertir\'a en capital de la compa\~{n}\'ia\index{capital de la compa\~{n}\'ia}, dando as\'i al inversor un cierto nivel de control sobre la misma.

\subsubsection{Estrategia: Prestar a poseer\index{prestar a poseer}}

{}Esta estrategia consiste en financiar (a trav\'es de pr\'estamos asegurados\index{prestamo asegurado @ pr\'estamo asegurado}) una compa\~{n}\'ia en distress\index{compa\~{n}\'ia en distress} que no est\'a a\'un en bancarrota con la visi\'on de que i) supere la situaci\'on de dificultades\index{situaci\'on de dificultades}, evite la bancarrota\index{bancarrota} y aumente su valor de capital\index{valor de capital}, o ii) solicite protecci\'on bajo el C\'apitulo 11\index{Capitulo 11 @ Cap\'itulo 11} y, en la reorganizaci\'on\index{reorganizaci\'on}, el pr\'estamo asegurado\index{prestamo asegurado @ pr\'estamo asegurado} se convierta en capital de la compa\~{n}\'ia\index{capital de la compa\~{n}\'ia} con derechos de control\index{derechos de control}.

\subsection{Estrategia: Rompecabezas de riesgo de dificultades\index{rompecabezas de riesgo de dificultades}}\label{distressstrat}

{}Algunos estudios sugieren que las empresas m\'as propensas a la bancarrota\index{bancarrota} ofrecen mayores rendimientos, lo cual podr\'ia considerarse de cierta manera una prima de riesgo\index{prima de riesgo} (v\'ease, por ejemplo, \cite{Chan1991}, \cite{Fama1992}, \cite{Fama1996}, \cite{Vassalou2004}). Sin embargo, estudios m\'as recientes sugieren lo opuesto, es decir, que tales compa\~{n}\'ias no superan a aquellas m\'as saludables y que, al contrario, estas \'ultimas ofrecen mayores rendimientos. Esto se conoce como ``rompecabezas de riesgo de dificultades\index{rompecabezas de riesgo de dificultades}'' (v\'ease, por ejemplo, \cite{George2010}, \cite{Godfrey2015}, \cite{Griffin2002}, \cite{Ozdagli2010}). Entonces, esta estrategia consiste en comprar las compa\~{n}\'ias m\'as seguras y vender las m\'as riesgosas. Como un proxy, uno puede usar la probabilidad de bancarrota\index{bancarrota} $P_i$, $i=1,\dots,N$ ($N$ es el n\'umero de acciones), que puede, por ejemplo, ser modelada a trav\'es de una regresi\'on log\'istica\index{regresi\'on log\'istica} (v\'ease, por ejemplo, \cite{Campbell2008}).\footnote{\, Para algunos estudios sobre modelos para estimar las probabilidades de bancarrota\index{probabilidad de bancarrota}, variables explicativas\index{variable explicativa} y t\'opicos relacionados, v\'ease, por ejemplo, \cite{Alaminos2016}, \cite{Altman1968}, \cite{Altman1993}, \cite{Aretz2013}, \cite{Beaver1966}, \cite{Beaver2005}, \cite{Bellovary2007}, \cite{Brezigar2012}, \cite{Callejon2013}, \cite{Chaudhuri2011}, \cite{Chava2004}, \cite{ChenHL2011}, \cite{Cultrera2015}, \cite{Dichev1998}, \cite{Duffie2007}, \cite{DuJardin2015}, \cite{ElKalak2016}, \cite{Fedorova2013}, \cite{Ferreira2016}, \cite{Gordini2014}, \cite{Griffin2002}, \cite{Hensher2007}, \cite{Hillegeist2004}, \cite{Jo1997}, \cite{Jonsson1996}, \cite{Korol2013}, \cite{Laitinen2000}, \cite{McKey2002}, \cite{Min2006}, \cite{Mossman1998}, \cite{Odom1990}, \cite{Ohlson1980}, \cite{Philosophov2005}, \cite{Pindado2008}, \cite{Podobnik2010}, \cite{Ribeiro2012}, \cite{Shin2002}, \cite{Shumway2001}, \cite{Slowinski1995}, \cite{Tinoco2013}, \cite{Tsai2014}, \cite{Wilson1994}, \cite{Woodlock2014}, \cite{Yang2011}, \cite{Zhou2013}, \cite{Zmijewski1984}.} Un portafolio de costo cero\index{portafolio de costo cero} se puede construir, por ejemplo, vendiendo las acciones en el decil\index{decil} superior seg\'un $P_i$, y comprando las acciones en el decil\index{decil} inferior. Generalmente, el portafolio\index{portafolio} se reequilibra mensualmente, aunque un rebalanceo\index{rebalanceo} anual es tambi\'en posible (con retornos similares).

\subsubsection{Estrategia: Rompecabezas de riesgo de dificultades\index{rompecabezas de riesgo de dificultades} -- gesti\'on del riesgo\index{gesti\'on de riesgo}}

{}Esta estrategia es una variaci\'on de la estrategia de rompecabezas de riesgo de dificultades\index{estrategia de rompecabezas de riesgo de dificultades} presentada en la Subsecci\'on \ref{distressstrat}. Los estudios emp\'iricos sugieren que las estrategias de costo cero basadas en ``healthy-minus-distressed'' (HMD, por sus siglas en ingl\'es)\index{healthy-minus-distressed (HMD)}\index{estrategia, HMD}\index{estrategia, costo cero}, o ``compa\~{n}\'ias saludables menos compa\~{n}\'ias en distress'', tienden a tener un beta de mercado\index{beta de mercado} muy variable, que se vuelve significativamente negativo despu\'es de fuertes ca\'idas del mercado\index{ca\'ida del mercado} (generalmente se asocia con un incremento en la volatilidad), que puede causar grandes p\'erdidas si el mercado\index{mercado} rebota abruptamente (v\'ease, por ejemplo, \cite{Garlappi2011}, \cite{ODoherty2012}, \cite{Opp2017}). Esto es similar a lo que sucede en otras estrategias basadas en factores\index{estrategias basadas en factores}.\footnote{\, V\'ease, por ejemplo, \cite{Barroso2014}, \cite{Blitz2011}, \cite{Daniel2016}.} Para mitigar esto, la estrategia se puede modificar de la siguiente manera (v\'ease, por ejemplo, \cite{Eisdorfer2015}):
\begin{eqnarray}\label{HMD}
&& \mbox{HMD}_* = \frac{\sigma_{objetivo}}{\widehat{\sigma}}~\mbox{HMD} \
\end{eqnarray}
Aqu\'i: HMD\index{healthy-minus-distressed (HMD)} es el mismo que aquel de la estrategia est\'andar HMD\index{estrategia, HMD} en la Subsecci\'on \ref{distressstrat}; $\sigma_{objetivo}$ es el nivel de volatilidad objetivo\index{volatilidad objetivo} (generalmente, entre 10\% y 15\%, dependiendo de las preferencias del trader); y $\widehat{\sigma}$ es la volatilidad realizada\index{volatilidad realizada} estimada durante el a\~{n}o anterior utilizando datos diarios. Entonces, 100\% de la inversi\'on\index{inversi\'on} se asigna solo si $\widehat{\sigma} = \sigma_{objetivo}$, y una cantidad menor se asigna cuando $\widehat{\sigma} > \sigma_{objetivo}$. Cuando $\widehat{\sigma} < \sigma_{objetivo}$, la estrategia podr\'ia apalancarse.\footnote{\, O, simplemente, 100\% de la inversi\'on\index{inversi\'on} podr\'ia asignarse sin apalancamiento\index{apalancamiento}, en cuyo caso el prefactor en la Ecuaci\'on (\ref{HMD}) es en cambio $\mbox{min}(\sigma_{objetivo}/\widehat{\sigma}, 1)$.}

\newpage

\section{Bienes Ra\'ices\index{bienes ra\'ices}}\label{sec.realestate}

\subsection{Generalidades}

{}Los bienes ra\'ices\index{bienes ra\'ices} (o bienes inmuebles), a diferencia de la mayor\'ia de los otros activos financieros, son tangibles. Se pueden dividir en dos grupos principales: bienes ra\'ices comerciales (oficinas, centros comerciales, etc.) y residenciales (casas, apartamentos, etc.)\index{bienes ra\'ices, comerciales}\index{bienes ra\'ices, residenciales}. Hay varias maneras de obtener cierta exposici\'on\index{exposici\'on} a los bienes ra\'ices\index{bienes ra\'ices}, por ejemplo, a trav\'es de fideicomisos de inversi\'on inmobiliaria (REITs, por sus siglas en ingl\'es)\index{fideicomiso de inversi\'on inmobiliaria (REIT)}, que a menudo cotizan en las principales bolsas y permiten a los traders tomar una participaci\'on l\'iquida\index{participaci\'on l\'iquida} en este tipo de activos.\footnote{\, Los REITs\index{fideicomiso de inversi\'on inmobiliaria (REIT)} son en cierto sentido similares a los fondos mutuos\index{fondo mutuo}, ya que proporcionan una manera para que los inversionistas individuales adquieran una propiedad en los portafolios de bienes ra\'ices que generan ingresos\index{portafolio de bienes ra\'ices que genera ingresos}.} Hay varias formas de medir el rendimiento de una inversi\'on inmobiliaria\index{inversi\'on inmobiliaria}. Una forma com\'un y sencilla es la siguiente:
\begin{eqnarray}
&& R(t_1, t_2) = \frac{P(t_2) + C(t_1, t_2)}{P(t_1)} - 1
\end{eqnarray}
Aqu\'i: $R(t_1, t_2)$ es el retorno de la inversi\'on\index{inversi\'on} desde el inicio del per\'iodo de tenencia\index{periodo de tenencia @ per\'iodo de tenencia} $t_1$ hasta el final de per\'iodo de tenencia\index{periodo de tenencia @ per\'iodo de tenencia} $t_2$; $P(t_1)$ y $P(t_2)$ son los valores de mercado\index{valor de mercado} de la propiedad en dichos tiempos; $C(t_1, t_2)$ representa los flujos de efectivo\index{flujo de efectivo} recibidos, netos de costos.\footnote{\, Por alguna literatura, v\'ease, por ejemplo, \cite{Block2011}, \cite{Eldred2004}, \cite{Geltner1995}, \cite{Goetzmann1990}, \cite{Hoesli2008}, \cite{Hudson-Wilson2005}, \cite{Larkin2004}, \cite{Mazurczak2011}, \cite{Pivar2003}, \cite{Steinert2001}.}

\subsection{Estrategia: Diversificaci\'on de activos mixtos\index{diversificaci\'on de activos mixtos} con bienes ra\'ices\index{bienes ra\'ices}}\label{sub.realstatediv1}

{}Los bienes ra\'ices\index{bienes ra\'ices} son atractivos como una herramienta de diversificaci\'on\index{diversificaci\'on}. Los estudios emp\'iricos sugieren que su correlaci\'on\index{correlaci\'on} con activos tradicionales\index{activos tradicionales}, tales como bonos\index{bono} y acciones, es baja y se mantiene as\'i incluso en eventos extremos del mercado\index{eventos extremos del mercado} (por ejemplo, crisis financieras\index{crisis financiera}), cuando las correlaciones\index{correlaci\'on} entre activos tradicionales\index{activos tradicionales} tienden a incrementar. Adem\'as, la correlaci\'on\index{correlaci\'on} tiende a ser m\'as baja en horizontes\index{horizonte} de tiempo m\'as largos, por lo que los inversores a largo plazo pueden mejorar el rendimiento de su portafolio\index{rendimiento del portafolio} en t\'erminos de retornos ajustados por riesgo\index{retornos ajustados por riesgo} mediante la inclusi\'on de activos inmobiliarios\index{activos inmobiliarios} (v\'ease, por ejemplo, \cite{Feldman2003}, \cite{Geltner2006}, \cite{Seiler1999}, \cite{Webb1988}). Entonces, una estrategia simple consiste en comprar y mantener activos inmobiliarios\index{activos inmobiliarios} dentro de un portafolio tradicional\index{portafolio tradicional} que contiene, por ejemplo, bonos\index{bono}, acciones\index{acci\'on}, etc. La asignaci\'on \'optima var\'ia seg\'un las preferencias de los inversores (en t\'erminos de riesgo\index{riesgo} y retorno) y del horizonte\index{horizonte} temporal (v\'ease, por ejemplo, \cite{Geltner1995}, \cite{Lee2005}, \cite{Mueller2003}, \cite{Rehring2012}). Para obtener las ponderaciones, se pueden utilizar t\'ecnicas tales como optimizaci\'on de media-varianza\index{optimizaci\'on de media-varianza} o modelos vectoriales autorregresivos (VAR, por sus siglas en ingl\'es)\index{modelo vectorial autorregresivo (VAR)}\footnote{\, Por literatura sobre la metodolog\'ia del VAR\index{modelo vectorial autorregresivo (VAR)}, v\'ease, por ejemplo, \cite{Barberis2000}, \cite{Campbell1991}, \cite{Campbell2003}, \cite{CampbellViceira2004}, \cite{CampbellViceira2005}, \cite{Kandel1987}, \cite{Sorensen2005}.} para calcular la asignaci\'on \'optima condicional al horizonte\index{horizonte} temporal y las caracter\'isticas de rendimiento\index{caracter\'isticas de rendimiento} deseadas (v\'ease, por ejemplo, \cite{Fugazza2007}, \cite{Hoevenaars2008}, \cite{MacKinnon2009}).

\subsection{Estrategia: Diversificaci\'on intra-activos\index{diversificaci\'on intra-activos} con bienes ra\'ices\index{bienes ra\'ices}}\label{sub.realstatediv2}

{}Esta estrategia se trata de diversificar las tenencias inmobiliarias\index{tenencias inmobiliarias} (que pueden ser una parte de un portafolio\index{portafolio} m\'as grande como en la Subsecci\'on \ref{sub.realstatediv1}). Los activos inmobiliarios\index{activos inmobiliarios} se pueden diversificar por su \'area geogr\'afica, tipo de propiedad, tama\~{n}o, proximidad a un \'area metropolitana, regi\'on econ\'omica, etc. (v\'ease, por ejemplo, \cite{Eichholtz1995}, \cite{Hartzell1986}, \cite{Hartzell1987}, \cite{Hudson-Wilson1990}, \cite{Seiler1999}, \cite{Viezer2000}). Varias t\'ecnicas de construcci\'on de portafolios\index{tecnicas de construccion de portafolio @ t\'ecnicas de construcci\'on de portafolio} (tales como las mencionadas en la Subsecci\'on \ref{sub.realstatediv1}) se pueden aplicar para determinar las asignaciones.

\subsubsection{Estrategia: Diversificaci\'on por el tipo de propiedad\index{diversificaci\'on por el tipo de propiedad}}

{}Esta estrategia consiste en invertir en activos inmobiliarios\index{activos inmobiliarios} de diferentes tipos, por ejemplo, apartamentos, oficinas, propiedades industriales\index{propiedad industrial} (que incluyen tambi\'en los edificios y las propiedades para la fabricaci\'on), centros comerciales, etc. Estudios emp\'iricos sugieren que la diversificaci\'on por el tipo de propiedad\index{diversificaci\'on por el tipo de propiedad} puede ser beneficiosa para la reducci\'on del riesgo no sistem\'atico\index{riesgo no sistem\'atico}\index{reducci\'on del riesgo no sistem\'atico}, incluso despu\'es de tener en cuenta los costos de transacci\'on\index{costos de transacci\'on} (v\'ease, por ejemplo, \cite{Firstenberg1988}, \cite{Grissom1987}, \cite{Miles1984}, \cite{Mueller1995}).

\subsubsection{Estrategia: Diversificaci\'on por la regi\'on econ\'omica\index{diversificaci\'on econ\'omica}}

{}Esta estrategia consiste en diversificar las inversiones inmobiliarias\index{inversi\'on inmobiliaria} por diferentes regiones divididas seg\'un las caracter\'isticas econ\'omicas, pudiendo ser estas la actividad econ\'omica\index{actividad econ\'omica} principal, estad\'isticas de empleo, ingresos promedios, etc. Los estudios emp\'iricos sugieren que dicha diversificaci\'on\index{diversificaci\'on} puede reducir el riesgo no sistem\'atico\index{riesgo no sistem\'atico} y los costos de transacci\'on\index{costos de transacci\'on} (v\'ease, por ejemplo, \cite{Hartzell1987}, \cite{Malizia1991}, \cite{Mueller1993}).

\subsubsection{Estrategia: Diversificaci\'on por el tipo de propiedad\index{tipo de propiedad} y la regi\'on geogr\'afica}

{}Esta estrategia combina la diversificaci\'on\index{diversificaci\'on} basada en m\'as de un atributo, por ejemplo, el tipo de propiedad\index{tipo de propiedad} y la regi\'on. Por lo tanto, si consideramos cuatro tipos de propiedades\index{tipo de propiedad}: oficina, comercio minorista, industrial y residencial, y cuatro regiones de EE.UU.\index{regiones de EE.UU.}, a saber, Este, Medio Oeste, Sur y Oeste, podemos diversificar el portafolio en los 16 grupos resultantes (v\'ease, por ejemplo, \cite{Viezer2000}).\footnote{\, Para obtener informaci\'on adicional pertinente, v\'ease, por ejemplo, \cite{DeWit2010}, \cite{Ertugrul2011}, \cite{Gatzlaff1995}, \cite{Hartzell2007}, \cite{Hastings2007}, \cite{Ross1991}, \cite{Seiler1999}, \cite{Worzala1997}.}

\subsection{Estrategia: Momentum inmobiliario\index{momentum inmobiliario} -- enfoque regional}

{}Esta estrategia consiste en comprar propiedades inmobiliarias\index{propiedad inmobiliaria} en funci\'on de sus rendimientos pasados. La evidencia emp\'irica sugiere que hay un efecto de momentum\index{efecto de momentum} en las \'areas estad\'isticas metropolitanas (MSAs, por sus siglas en ingl\'es)\index{area estadistica metropolitana (MSA) @ \'area estad\'istica metropolitana (MSA)} de los Estados Unidos, es decir, las \'areas con rendimientos pasados m\'as altos (bajos) tienden a continuar ofreciendo rendimientos m\'as altos (bajos) en el futuro (v\'ease, por ejemplo, \cite{Beracha2015}, \cite{Beracha2011}). En algunos casos, una estrategia de costo cero\index{estrategia de costo cero} se puede construir, por ejemplo, utilizando veh\'iculos inmobiliarios alternativos\index{veh\'iculos inmobiliarios alternativos} tales como REITs\index{fideicomiso de inversi\'on inmobiliaria (REIT)}, futuros\index{futuro} y opciones\index{opci\'on} sobre los \'indices de vivienda de los Estados Unidos basados en diferentes \'areas geogr\'aficas.\footnote{\, Por algunas publicaciones sobre estrategias de momentum inmobiliario\index{estrategia de momentum inmobiliario} (incluyendo el uso de REITs\index{fideicomiso de inversi\'on inmobiliaria (REIT)} y otros veh\'iculos de inversi\'on\index{veh\'iculo de inversi\'on} mencionados anteriormente) y t\'opicos relacionados, v\'ease, por ejemplo, \cite{Abraham1993}, \cite{Abraham1996}, \cite{Anglin2003}, \cite{Buttimer2005}, \cite{Caplin2011}, \cite{Capozza2004}, \cite{Case1987}, \cite{Case1989}, \cite{Case1990}, \cite{ChanHS1990}, \cite{Chan1998}, \cite{ChenHVC1998}, \cite{Cho1996}, \cite{Chui2003a}, \cite{Chui2003b}, \cite{Cooper1999}, \cite{DerwallHBM2009}, \cite{deWit2013}, \cite{Genesove2012}, \cite{Genesove1997}, \cite{Genesove2001}, \cite{Goebel2013}, \cite{Graff1999}, \cite{Graff1997}, \cite{Gupta2012}, \cite{Guren2014}, \cite{Haurin2002}, \cite{Haurin2010}, \cite{Head2014}, \cite{Kallberg2000}, \cite{Kang1989}, \cite{Karolyi1998}, \cite{Knight2002}, \cite{Krainer2001}, \cite{Kuhle2000}, \cite{Lee2010}, \cite{Levitt2008}, \cite{LiWang1995}, \cite{LinY2004}, \cite{LiuMei1992}, \cite{Malpezzi1999}, \cite{Meen2002}, \cite{Mei1995}, \cite{Mei1998}, \cite{Moss2015}, \cite{Nelling1998}, \cite{NovyMarx2009}, \cite{OrtaloMagne2006}, \cite{Piazzesi2009}, \cite{Peterson1997}, \cite{PoterbaS2008}, \cite{Smith1976}, \cite{Stein1995}, \cite{Stevenson2001}, \cite{Stevenson2002}, \cite{Taylor1999}, \cite{Titman1986}, \cite{Wheaton1990}, \cite{Yavas1995}, \cite{Young1996}.}

\subsection{Estrategia: Cobertura contra la inflaci\'on\index{cobertura contra la inflaci\'on} con bienes ra\'ices\index{bienes ra\'ices}}

{}Los estudios emp\'iricos sugieren una fuerte relaci\'on entre los retornos de los bienes ra\'ices\index{retornos de los bienes ra\'ices} y la tasa de inflaci\'on\index{tasa de inflaci\'on}. Por lo tanto, los bienes ra\'ices\index{bienes ra\'ices} pueden ser utilizados como una cobertura\index{cobertura} contra inflaci\'on\index{inflaci\'on}. Adem\'as, emp\'iricamente, algunos tipos de propiedades\index{tipo de propiedad} (por ejemplo, los bienes ra\'ices comerciales\index{bienes ra\'ices comerciales}, que tienden a ajustarse m\'as r\'apido a los aumentos inflacionarios de precios) parecen proporcionar una mejor cobertura\index{cobertura} que otros, aunque esto puede depender de varios aspectos tales como la muestra, el mercado\index{mercado}, etc.\footnote{\, Por literatura pertinente, v\'ease, por ejemplo, \cite{Bond1998}, \cite{Fama1977}, \cite{Gunasekarage2008}, \cite{Hamelink1996}, \cite{HartzellHekman1987}, \cite{LeMoigne2008}, \cite{Mauer2002}, \cite{Miles1997}, \cite{Newell1996}, \cite{Sing2000}, \cite{Wurtzebach1991}.}

\subsection{Estrategia: Arreglar y vender (Fix-and-flip)\index{arreglar y vender}}

{}Esta es una estrategia de inversi\'on inmobiliaria\index{estrategia de inversi\'on inmobiliaria} a corto plazo. Consiste en comprar una propiedad, que normalmente se encuentra en una situaci\'on deteriorada y requiere renovaciones, con un descuento (sustancial) por debajo de los precios de mercado\index{precio de mercado}. El inversor renueva la propiedad y la revende a un precio lo suficientemente alto para cubrir los costos de renovaci\'on y obtener una ganancia.\footnote{\, Por literatura pertinente, v\'ease, por ejemplo, \cite{Anacker2009}, \cite{Anacker2015}, \cite{Bayer2015}, \cite{Chinco2012}, \cite{Corbett2006}, \cite{Depken2009}, \cite{Depken2011}, \cite{Fu2014}, \cite{Hagopian1999}, \cite{Kemp2007}, \cite{Leung2013}, \cite{Montelongo2008}, \cite{Villani2006}.}

\newpage

\section{Efectivo\index{efectivo}}\label{sec.cash}

\subsection{Generalidades}

{}El efectivo\index{efectivo} es un activo, aunque a veces su funci\'on como un activo puede pasarse por alto o darse por sentada. Como un activo, el efectivo\index{efectivo} se puede utilizar de varias maneras, por ejemplo, i) como una {\em herramienta de gesti\'on de riesgos\index{herramienta de gesti\'on de riesgo}}, dado que puede ayudar a mitigar los drawdowns\index{drawdown} y la volatilidad; ii) como una {\em herramienta de gesti\'on de oportunidades\index{herramienta de gesti\'on de oportunidades}}, dado que permite aprovechar situaciones espec\'ificas o inusuales; y iii) como una {\em herramienta de gesti\'on de liquidez\index{herramienta de gesti\'on de liquidez}} en situaciones inesperadas que requieren fondos l\'iquidos\index{fondos l\'iquidos}. Hay varias formas de incluir fondos l\'iquidos\index{fondos l\'iquidos} en un portafolio\index{portafolio}, por ejemplo, mediante las letras del Tesoro de los Estados Unidos a corto plazo, certificados de dep\'ositos bancarios (CDs, por sus siglas en ingl\'es)\index{certificado de dep\'osito bancario (CD)}, papeles comerciales\index{papel comercial}, aceptaciones bancarias (BAs, por sus siglas en ingl\'es)\index{aceptaci\'on bancaria}, eurod\'olares\index{eurod\'olares}, y acuerdos de recompra (REPO, por sus siglas en ingl\'es)\index{acuerdo de recompra (repo)} (tambi\'en conocidos como repos\index{acuerdo de recompra (repo)}), etc.\footnote{\, Por alguna literatura, v\'ease, por ejemplo, \cite{Cook1993}, \cite{Cook1986}, \cite{Damiani2012}, \cite{Duchin2010}, \cite{Goodfriend2011}, \cite{Schaede1990}, \cite{Summers1980}, \cite{Ysmailov2017}.}

\subsection{Estrategia: Lavado de dinero\index{lavado de dinero} -- el lado oscuro del efectivo\index{efectivo}}

{}El lavado de dinero\index{lavado de dinero}, en t\'erminos generales, es una actividad en donde el efectivo\index{efectivo} se utiliza como un veh\'iculo para transformar ganancias ilegales en activos de apariencia leg\'itima. Hay tres pasos principales en el proceso de lavado de dinero\index{lavado de dinero}. El primer (y m\'as riesgoso) paso es la {\em colocaci\'on\index{colocaci\'on}}, a trav\'es del cual se introducen fondos ilegales en la econom\'ia legal mediante medios fraudulentos, por ejemplo, dividiendo los fondos en peque\~{n}as cantidades y deposit\'andolos en varias cuentas bancarias, evitando de esta forma su detecci\'on. El segundo paso, {\em estratificaci\'on\index{estratificaci\'on}}, implica mover el dinero entre diferentes cuentas e incluso pa\'ises, creando de esta forma cierta complejidad y separando el dinero de su fuente original en varios grados. El tercer paso es la {\em integraci\'on\index{integraci\'on}}, mediante el cual quienes lavan el dinero lo recuperan a trav\'es de fuentes de apariencia leg\'itima, por ejemplo, negocios que requieren un uso intensivo de efectivo como bares y restaurantes, lavado de autos, hoteles (al menos en algunos pa\'ises), establecimientos de apuestas, garajes de estacionamiento, etc.\footnote{\, Por alguna literatura, v\'ease, por ejemplo, \cite{Ardizzi2014}, \cite{Cox2015}, \cite{Gilmour2015}, \cite{Hopton1999}, \cite{John2009}, \cite{Kumar2012}, \cite{Levi2006}, \cite{Schneider2008}, \cite{Seymour2008}, \cite{Soudijn2016}, \cite{Walker1999}, \cite{Wright2017}.}

\subsection{Estrategia: Gesti\'on de liquidez\index{gesti\'on de liquidez}}

{}Desde una perspectiva de gesti\'on de portafolios\index{gesti\'on de portafolio}, esta estrategia consiste en definir de forma \'optima la cantidad de efectivo\index{efectivo} a mantener en el portafolio\index{portafolio} para satisfacer las demandas de liquidez generadas por eventos imprevistos.\footnote{\, Tenga en cuenta que esta no es necesariamente la misma raz\'on por la cual se mantiene efectivo\index{efectivo} como en las estrategias de Kelly\index{estrategia de Kelly}. Por algunos estudios pertinentes, v\'ease, por ejemplo, \cite{Browne2000}, \cite{Cover1984}, \cite{Davis2012}, \cite{Hsieh2015}, \cite{Hsieh2016}, \cite{Kelly1956}, \cite{Laureti2010}, \cite{LoOZ2017}, \cite{Maslov1998}, \cite{Nekrasov2014}, \cite{Rising2012}, \cite{Samuelson1971}, \cite{Thorp2006}, \cite{Thorp1967}.} El efectivo\index{efectivo} proporciona liquidez\index{liquidez} inmediata, mientras que otros activos tendr\'ian que ser liquidados primero, lo cual puede estar asociado con costos de transacci\'on\index{costos de transacci\'on} sustanciales, especialmente si la liquidaci\'on\index{liquidaci\'on} es abrupta.\footnote{\, Por alguna literatura, v\'ease, por ejemplo, \cite{Agapova2011b}, \cite{Aragon2017}, \cite{Cao2013}, \cite{Chernenko2016}, \cite{Connor1995}, \cite{JiangLi2017}, \cite{Kruttli2018}, \cite{Leland1995}, \cite{Simutin2014}, \cite{Yan2006}.} Desde una perspectiva corporativa, mantener efectivo\index{efectivo} puede ser una medida de precauci\'on dirigida a evitar el d\'eficit de flujo de efectivo\index{deficit de flujo de efectivo @ d\'eficit de flujo de efectivo} que puede producir, entre otras cosas, la p\'erdida de oportunidades de inversi\'on\index{oportunidad de inversi\'on}, dificultades financieras\index{dificultad financiera}, etc.\footnote{\, Por algunas publicaciones sobre aspectos corporativos de la gesti\'on de liquidez\index{gesti\'on de liquidez} y t\'opicos relacionados, v\'ease, por ejemplo, \cite{Acharya2007}, \cite{Almeida2005}, \cite{Azmat2017}, \cite{Chidambaran2001}, \cite{Disatnik2014}, \cite{Froot1993}, \cite{Han2007}, \cite{Opler1999}, \cite{Sher2014}.}

\subsection{Estrategia: Acuerdo de recompra (REPO)\index{acuerdo de recompra (repo)}}

{}Un acuerdo de recompra (REPO)\index{acuerdo de recompra (repo)} es un activo equivalente a efectivo\index{activo equivalente a efectivo} que proporciona liquidez\index{liquidez} inmediata a una tasa de inter\'es\index{tasa de inter\'es} preestablecida por un per\'iodo espec\'ifico de tiempo a cambio de otro activo utilizado como colateral\index{colateral}. Un acuerdo de recompra inverso\index{acuerdo de recompra inverso} es lo opuesto. Por lo tanto, una estrategia REPO\index{estrategia REPO} consiste en pedir prestado (prestar) efectivo\index{efectivo} con intereses\index{interes @ inter\'es} a cambio de activos\index{activo} con el compromiso de recomprarlos de (revenderlos a) la contraparte\index{contraparte}. Este tipo de transacci\'on generalmente abarca desde 1 d\'ia hasta 6 meses.\footnote{\, V\'ease, por ejemplo, \cite{Adrian2013}, \cite{Bowsher1979}, \cite{Duffie1996}, \cite{Garbade2004}, \cite{Gorton2012}, \cite{Happ1986}, \cite{Kraenzlin2007}, \cite{Lumpkin1987}, \cite{Ruchin2011}, \cite{Schatz2012}, \cite{Simmons1954}, \cite{Sollinger1994}, \cite{Zhang2018}.}

\subsection{Estrategia: Empe\~{n}o\index{empe\~{n}o}}

{}REPOs\index{acuerdo de recompra (repo)} son en cierto sentido similares a las estrategias de empe\~{n}o\index{estrategia de empe\~{n}o} mucho m\'as antiguas. Un prestamista\index{prestamista} otorga un pr\'estamo asegurado en efectivo\index{prestamo asegurado en efectivo @ pr\'estamo asegurado en efectivo} con un inter\'es\index{interes @ inter\'es} y un per\'iodo pre-acordado (aunque a veces, este \'ultimo se puede extender). El pr\'estamo\index{prestamo @ pr\'estamo} se asegura con un colateral\index{colateral}, que es un art\'iculo(s) valioso(s), tales como joyer\'ia, electr\'onica, veh\'iculos, libros o instrumentos musicales raros, etc. Si el pr\'estamo\index{prestamo @ pr\'estamo} no se devuelve bajo los t\'erminos acordados, entonces el prestatario pierde el colateral\index{colateral} y el prestamista\index{prestamista} puede conservarlo o venderlo. El importe del pr\'estamo\index{prestamo @ pr\'estamo}, por lo general, se encuentra a un descuento significativo con respecto al valor de tasaci\'on\index{valor de tasaci\'on} del colateral\index{colateral}.\footnote{\, En la Secci\'on \ref{sec.commodities} discutimos estrategias de trading\index{estrategia de trading} basadas en futuros de commodities\index{futuros de commodities}. Los prestamistas\index{prestamista}, entre otras cosas, comercian commodities f\'isicos\index{commodities f\'isicos} tales como plata y oro. Por alguna literatura sobre empe\~{n}o\index{empe\~{n}o} y t\'opicos relacionados, v\'ease, por ejemplo, \cite{Bos2012}, \cite{Bouman1988}, \cite{Caskey1991}, \cite{DEste2014}, \cite{Fass2004}, \cite{Maaravi2017}, \cite{McCants2007}, \cite{Shackman2006}, \cite{Zhou2016}.}

\subsection{Estrategia: Usura\index{usura}}

{}A diferencia del empe\~{n}o\index{empe\~{n}o}, la usura\index{usura} en muchas jurisdicciones es ilegal. La usura\index{usura} consiste en ofrecer un pr\'estamo\index{prestamo @ pr\'estamo} a tasas de inter\'es\index{tasa de inter\'es} excesivamente altas. Tal pr\'estamo\index{prestamo @ pr\'estamo} no est\'a necesariamente asegurado por un colateral\index{colateral}. En cambio, el usurero\index{usurero} a veces puede recurrir al chantaje y/o a la violencia para hacer cumplir los t\'erminos de un pr\'estamo\index{prestamo @ pr\'estamo} (v\'ease, por ejemplo, \cite{Aldohni2013}).

\newpage

\section{Criptomonedas\index{criptomonedas}}

\subsection{Generalidades}

{}Las criptomonedas\index{criptomonedas}, tales como el Bitcoin (BTC)\index{Bitcoin (BTC)}, el Ethereum (ETH)\index{ETH (ether/Ethereum)}, etc., a diferencia de las monedas fiduciarias\index{moneda fiduciaria} tradicionales (USD\index{USD (d\'olar estadounidense)}, EUR\index{EUR (euro)}, etc.), son monedas digitales descentralizadas\index{moneda digital descentralizada} basadas en protocolos de internet con una fuente abierta de entre pares (P2P, por sus siglas en ingl\'es)\index{protocolo de internet con una fuente abierta de entre pares (P2P)}. Las criptomonedas\index{criptomonedas} tales como el BTC\index{Bitcoin (BTC)} y el ETH\index{ETH (ether/Ethereum)} utilizan la tecnolog\'ia blockchain\index{tecnolog\'ia blockchain} \cite{Nakamoto2008}.\footnote{\, El blockchain\index{blockchain} es un libro de registro distribuido\index{registro distribuido} que mantiene un registro de todas las transacciones. Es una cadena secuencial de bloques (que contiene registros de transacciones), los cuales est\'an vinculados mediante la criptograf\'ia\index{criptograf\'ia} y un marcador del tiempo\index{marcador del tiempo}. Ning\'un bloque puede modificarse retroactivamente sin alterar todos los bloques subsiguientes, lo que hace que el blockchain\index{blockchain} sea resistente a la modificaci\'on de datos debido a su propio dise\~{n}o. Para lograr una modificaci\'on del blockchain\index{blockchain}, el cual es mantenido por una gran red como un libro de registro distribuido\index{registro distribuido} con una actualizaci\'on constante en una gran cantidad de sistemas al mismo tiempo, se requerir\'ia una colusi\'on de la mayor parte de la red.} La capitalizaci\'on burs\'atil\index{capitalizaci\'on burs\'atil} total de criptomonedas\index{criptomonedas} se mide en cientos de miles de millones de d\'olares.\footnote{\, Las criptomonedas\index{criptomonedas} son altamente vol\'atiles, por lo que su capitalizaci\'on de mercado\index{capitalizaci\'on de mercado} var\'ia en el tiempo de forma sustancial.} Muchos inversores se sienten atra\'idos por las criptomonedas\index{criptomonedas} como activos especulativos para comprarlos y mantenerlos\index{activo especulativo de compra y mantenimiento}. Algunos los ven como diversificadores\index{diversificador} debido a su baja correlaci\'on\index{correlaci\'on} con activos tradicionales\index{activos tradicionales}. Otros los perciben como el futuro del dinero. Algunos inversores simplemente quieren hacer dinero r\'apido en una burbuja especulativa\index{burbuja especulativa}. Etc.\footnote{\, Por literatura pertinente, v\'ease, por ejemplo, \cite{Baek2014}, \cite{Bariviera2017}, \cite{Bouoiyour2015}, \cite{Bouoiyour2016}, \cite{BouriGupta2017}, \cite{Bouri2017}, \cite{Brandvold2015}, \cite{Briere2015}, \cite{Cheah2015}, \cite{CheungRS2015}, \cite{Ciaian2015}, \cite{Donier2015}, \cite{Dowd2015}, \cite{Dyhrberg2015}, \cite{Dyhrberg2016}, \cite{Eisl2015}, \cite{Fry2016}, \cite{Gajardo2018}, \cite{Garcia2015}, \cite{Garcia2014}, \cite{Harvey2014}, \cite{Harvey2016}, \cite{KimKim2016}, \cite{Kristoufek2015}, \cite{Chuen2018}, \cite{LiewLB2018}, \cite{Ortisi2016}, \cite{VanAlstyne2014}, \cite{WangVergne2017}, \cite{White2015}.} Sea como sea, a diferencia de, por ejemplo, las acciones, no hay ``fundamentos\index{fundamentos}'' evidentes para los criptoactivos\index{criptoactivos}, los cuales podr\'ian ser la base para construir estrategias de trading ``fundamentales''\index{estrategias de trading fundamentales} (por ejemplo, estrategias basadas en value\index{estrategia basada en value}). Por lo tanto, las estrategias de trading de criptomonedas\index{estrategia de trading de criptomonedas}, generalmente, se basan en la miner\'ia de datos para tendencias\index{miner\'ia de datos para tendencias} mediante t\'ecnicas de aprendizaje autom\'atico\index{tecnicas de aprendizaje automatico @ t\'ecnicas de aprendizaje autom\'atico}.

\subsection{Estrategia: Red neuronal artificial (ANN)\index{red neuronal artificial (ANN)}}\label{sub.ANN}

{}Esta estrategia utiliza una ANN\index{red neuronal artificial (ANN)} (por sus siglas en ingl\'es) para pronosticar los movimientos a corto plazo del BTC\index{Bitcoin (BTC)} utilizando, como variables de entrada, indicadores t\'ecnicos\index{indicador t\'ecnico}. En una ANN\index{red neuronal artificial (ANN)} tenemos una capa de entrada\index{capa de entrada} (input layer en ingl\'es), una capa de salida\index{capa de salida} (output layer en ingl\'es), y un cierto n\'umero de capas ocultas\index{capas ocultas} (hidden layers en ingl\'es). Entonces, en esta estrategia la capa de entrada\index{capa de entrada} se construye utilizando indicadores t\'ecnicos\index{indicador t\'ecnico}.\footnote{\, Por lo tanto, en esp\'iritu, es algo similar a la estrategia de trading de acci\'on individual con KNN\index{estrategia de trading de acci\'on individual con KNN}\index{acci\'on individual con KNN} discutida en la Subsecci\'on \ref{MLKNN}, que utiliza el algoritmo de los k vecinos m\'as cercanos (KNN)\index{algoritmo de los k vecinos m\'as cercanos (KNN)} (opuesto a la ANN\index{red neuronal artificial (ANN)}).} Por ejemplo, podemos usar medias m\'oviles (exponenciales) ((E)MAs, por sus siglas en ingl\'es)\index{media m\'ovil, exponencial}, desviaciones est\'andar m\'oviles (exponenciales) ((E)MSDs, por sus siglas en ingl\'es)\index{desviaci\'on est\'andar m\'ovil, exponencial}, el \'indice de fuerza relativa (RSI, por sus siglas en ingl\'es)\index{indice de fuerza relativa (RSI) @ \'indice de fuerza relativa (RSI)},\footnote{\, Generalmente, un RSI\index{indice de fuerza relativa (RSI) @ \'indice de fuerza relativa (RSI)} $> 0.7$ ($< 0.3$) se interpreta como una se\~{n}al de sobrecompra (sobreventa). V\'ease, por ejemplo, \cite{Wilder1978}.} etc. De forma m\'as concreta, podemos construir la capa de entrada\index{capa de entrada} de la siguiente forma (v\'ease, por ejemplo, \cite{Nakano2018}). Sea $P(t)$ el precio del BTC\index{precio del BTC} al momento $t$, en donde $t=1,2,\dots$ se mide en ciertas unidades (por ejemplo, intervalos de 15 minutos; tambi\'en, $t=1$ es el tiempo m\'as reciente). Consideremos:
\begin{eqnarray}
 &&R(t) = {P(t)\over P(t+1)} - 1\\
 &&{\widetilde R}(t, T_1) = R(t) - {\overline R}(t, T_1)\\
 &&{\overline R}(t, T_1) = {1\over T_1}\sum_{t^\prime = t+1}^{t+T_1} R(t^\prime)\\
 &&{\widehat R}(t, T_1) = {{\widetilde R}(t, T_1)\over\sigma(t, T_1)}\\
 &&[\sigma(t, T_1)]^2 = {1\over{T_1-1}}\sum_{t^\prime = t+1}^{t+T_1}[{\widetilde R}(t, T_1)]^2
\end{eqnarray}
Entonces: $R(t)$ es el retorno desde $t+1$ hasta $t$; ${\overline R}(t, T_1)$ es el retorno medio serial desde $t+T_1$ hasta $t+1$, es decir, sobre $T_1$ per\'iodos, en donde $T_1$ se puede elegir para que sea lo suficientemente largo como para proporcionar una estimaci\'on razonable de la volatilidad (v\'ease abajo); ${\widetilde R}(t, T_1)$ es el retorno neto de la media serial\index{retorno neto de la media serial}; $\sigma(t, T_1)$ es la volatilidad calculada a partir de $t+T_1$ hasta $t+1$; y ${\widehat R}(t, T_1)$ es el retorno ajustado de la media serial y normalizado. A continuaci\'on, para simplificar, omitiremos la referencia al par\'ametro $ T_1 $ y usaremos ${\widehat R}(t)$ para denotar los rendimientos normalizados.

{}Luego, podemos definir las EMAs\index{media m\'ovil exponencial (EMA)}, EMSDs\index{desviaci\'on est\'andar m\'ovil exponencial (EMSD)} y el RSI\index{indice de fuerza relativa (RSI) @ \'indice de fuerza relativa (RSI)} de la siguiente forma:\footnote{\, Tenga en cuenta que esto se puede hacer de m\'as de una manera.}
\begin{eqnarray}
 &&\mbox{EMA}(t, \lambda, \tau) = {{1-\lambda}\over{1-\lambda^\tau}}~\sum_{t^\prime=t+1}^{t+\tau} \lambda^{t^\prime - t - 1}~{\widehat R}(t^\prime)\\
 &&[\mbox{EMSD}(t, \lambda, \tau)]^2 = {{1-\lambda}\over{\lambda-\lambda^\tau}}~\sum_{t^\prime=t+1}^{t+\tau} \lambda^{t^\prime - t - 1}~[{\widehat R}(t^\prime) - \mbox{EMA}(t, \lambda, \tau)]^2\\
 &&\mbox{RSI}(t, \tau) = {\nu_+(t, \tau) \over {\nu_+(t, \tau) + \nu_-(t, \tau)}}\\
 &&\nu_{\pm}(t, \tau) = \sum_{t^\prime = t+1}^{t+\tau} \mbox{max}(\pm {\widehat R}(t^\prime), 0)
\end{eqnarray}
Aqu\'i: $\tau$ es la longitud de la media m\'ovil\index{longitud de la media m\'ovil}; $\lambda$ es el par\'ametro de suavizaci\'on exponencial\index{par\'ametro de suavizaci\'on exponencial}.\footnote{\, Para reducir el n\'umero de par\'ametros, podemos, por ejemplo, tomar $\lambda = (\tau-1)/(\tau+1)$.\label{fn.lambda.tau}}

{}La capa de entrada\index{capa de entrada} luego puede consistir en, por ejemplo, ${\widehat R}(t)$, $\mbox{EMA}(t, \lambda_a, \tau_a)$, $\mbox{EMSD}(t, \lambda_a, \tau_a)$, y $\mbox{RSI}(t, \tau^\prime_{a^\prime})$, en donde $a=1,\dots,m$, $a^\prime = 1,\dots,m^\prime$. Los valores $\tau_a$ pueden, por ejemplo, ser elegidos para corresponder a 30 minutos, 1 hora, 3 horas y 6 horas (entonces $m=4$; v\'ease la nota al pie \ref{fn.lambda.tau} para los valores de lambda $\lambda_a$). Los valores $\tau^\prime_{a^\prime}$ pueden, por ejemplo, ser elegidos para corresponder a 3 horas, 6 horas y 12 horas (entonces $m^\prime=3$). No existe una bala m\'agica aqu\'i. Estos valores se pueden elegir en funci\'on de un backtest fuera de la muestra\index{backtest fuera de la muestra} teniendo en cuenta, sin embargo, el peligro que siempre se encuentra presente de sobreajuste\index{sobreajuste}, al considerar tantos par\'ametros libres (v\'ease abajo), incluyendo $\tau_a$, $\lambda_a$ y $\tau^\prime_{a^\prime}$.

{}La capa de salida\index{capa de salida} se puede construir de la siguiente manera. Asumamos que el objetivo es pronosticar a qu\'e cuantil\index{cuantil} pertenecer\'an los retornos futuros normalizados. Sea el n\'umero de cuantiles\index{cuantil} $K$. As\'i, para los valores de $t$ correspondiente al conjunto de datos de entrenamiento\index{datos de entrenamiento} $D_{entrenamiento}$,\footnote{\, Idealmente, al calcular los cuantiles\index{cuantil}, un n\'umero apropiado $d_1$ de los valores de $t = t_d, t_{d-1},\dots, t_{d-d_1+1}$, $d = |D_{entrenamiento}|$, deber\'ian ser excluidos para asegurar de que todos los valores de las EMA\index{media m\'ovil exponencial (EMA)}, EMSD\index{desviaci\'on est\'andar m\'ovil exponencial (EMSD)} y el RSI\index{indice de fuerza relativa (RSI) @ \'indice de fuerza relativa (RSI)} se calculan utilizando los n\'umeros requeridos de puntos de datos.} tenemos los retornos normalizados ${\widehat R}(t)$, $t\in D_{entrenamiento}$. Los valores de los $(K-1)$ cuantiles\index{cuantil} de ${\widehat R}(t)$, $t\in D_{entrenamiento}$, sea $q_\alpha$, $\alpha=1,\dots,(K-1)$. Para cada valor de $t$, podemos definir $K$-vectores de supervisi\'on $S_{\alpha}(t)$, $\alpha = 1,\dots,K$, de la siguiente forma:
\begin{eqnarray}
\begin{cases}
    S_1(t) = 1,~~~{\widehat R}(t) \leq q_1\\
    S_\alpha(t) = 1,~~~q_{\alpha-1} \leq {\widehat R}(t) < q_\alpha,~~~1 < \alpha < K\\
    S_K(t) = 1,~~~q_{K-1} \leq {\widehat R}(t)\\
    S_\alpha(t) = 0,~~~\mbox{otros casos}
\end{cases}
\end{eqnarray}
La capa de salida\index{capa de salida} entonces puede ser un $K$-vector no negativo $p_\alpha(t)$, cuyos elementos se interpretan como las probabilidades de que los retornos normalizados futuros pertenezcan al $\alpha$-th cuantil\index{cuantil}. Entonces, tenemos
\begin{equation}\label{output.layer}
 \sum_{\alpha=1}^K p_\alpha(t) = 1
\end{equation}
La capa de salida\index{capa de salida} se construye desde la capa de entrada\index{capa de entrada} como alguna de sus funciones no lineales, con cierto n\'umero de par\'ametros a determinar a trav\'es del entrenamiento\index{entrenamiento}. En una ANN\index{red neuronal artificial (ANN)} tenemos $L$ capas etiquetadas con $\ell=1,\dots, L$, en donde $\ell = 1$ corresponde a la capa de entrada\index{capa de entrada}, y $\ell = L$ corresponde a la capa de salida\index{capa de salida}. En cada capa tenemos $N^{(\ell)}$ nodos\index{nodo} y los correspondientes $N^{(\ell)}$-vectores ${\vec X}^{(\ell)}$ con componentes $X^{(\ell)}_{i^{(\ell)}}$, $i^{(\ell)} = 1,\dots, N^{(\ell)}$:\footnote{\, Suprimimos la variable de tiempo $t$ para simplificar la notaci\'on.}
\begin{eqnarray}
 &&X^{(\ell)}_{i^{(\ell)}} = h^{(\ell)}_{i^{(\ell)}} ({\vec Y}^{(\ell)}),~~~\ell=2,\dots, L\\
 &&Y^{(\ell)}_{i^{(\ell)}} = \sum_{j^{(\ell-1)} = 1}^{N^{(\ell - 1)}} A^{(\ell)}_{i^{(\ell)} j^{(\ell-1)}}~X^{(\ell - 1)}_{j^{(\ell-1)}} + B^{(\ell)}_{i^{(\ell)}}
\end{eqnarray}
Aqu\'i: ${\vec Y}^{(\ell)}$ es un $N^{(\ell)}$-vector con componentes $Y^{(\ell)}_{i^{(\ell)}}$, $i^{(\ell)} = 1,\dots, N^{(\ell)}$; $X^{(1)}_{i^{(1)}}$ son los datos de entrada (para cada valor de $t$, es decir, ${\widehat R}(t)$, $\mbox{EMA}(t, \lambda_a, \tau_a)$, $\mbox{EMSD}(t, \lambda_a, \tau_a)$, y $\mbox{RSI}(t, \tau^\prime_{a^\prime})$ -- v\'ease arriba); $X^{(L)}_{i^{(L)}}$ son los datos de salida $p_\alpha(t)$ (es decir, $N^{(L)} = K$ y el \'indice $i^{(L)}$ es lo mismo que $\alpha$); los par\'ametros desconocidos $A^{(\ell)}_{i^{(\ell)} j^{(\ell-1)}}$ (llamados ponderaciones) y $B^{(\ell)}_{i^{(\ell)}}$ (llamado sesgo\index{sesgo}) se determinan a trav\'es del entrenamiento\index{entrenamiento} (v\'ease abajo); y hay mucha arbitrariedad al momento de escoger los valores de $N^{(\ell)}$ y las funciones de activaci\'on\index{funci\'on de activaci\'on} $h^{(\ell)}_{i^{(\ell)}}$. Una posible elecci\'on (entre muchas otras) es la siguiente (v\'ease, por ejemplo, \cite{Nakano2018}):\footnote{\, Nuevamente, no hay una bala m\'agica aqu\'i. A priori, una gran cantidad de funciones de activaci\'on\index{funci\'on de activaci\'on} se pueden utilizar, por ejemplo, sigmoide\index{sigmoide} (tambi\'en conocida como log\'istica), tanh (tangente hiperb\'olica)\index{tangente hiperb\'olica (tanh)}, unidad lineal rectificada (ReLU, por sus siglas en ingl\'es)\index{unidad lineal rectificada (ReLU)}, softmax\index{softmax}, etc. Por algunos estudios pertinentes, v\'ease, por ejemplo, \cite{Bengio2009}, \cite{Chandra2003}, \cite{daSGomes2011}, \cite{Glorot2011}, \cite{Goodfellow2013}, \cite{Karlik2011}, \cite{Mhaskar1993}, \cite{Singh2003}, \cite{Wu2009}.}
\begin{eqnarray}
 &&h^{(\ell)}_{i^{(\ell)}}({\vec Y}^{(\ell)}) = \mbox{max}\left(Y^{(\ell)}_{i^{(\ell)}}, 0\right),~~~\ell = 2,\dots,L-1~~~\mbox{(ReLU)}\\
 &&h^{(L)}_{i^{(L)}}({\vec Y}^{(L)}) = Y^{(L)}_{i^{(L)}} \left[\sum_{j^{(L)} = 1}^{N^{(L)}} Y^{(L)}_{j^{(L)}}\right]^{-1}~~~\mbox{(softmax)}
\end{eqnarray}
Es decir, ReLU\index{unidad lineal rectificada (ReLU)} se utiliza en las capas ocultas\index{capas ocultas} (y el algoritmo se mueve a la siguiente capa solo si algunas neuronas est\'an activadas (disparadas) en la capa $\ell$, es decir, al menos algunos $Y^{(\ell)}_{i^{(\ell)}} > 0$), y softmax\index{softmax} se utiliza en la capa de salida\index{capa de salida} (para que tengamos la condici\'on (\ref{output.layer}) por construcci\'on). Adem\'as, para entrenar el modelo, es decir, para determinar los par\'ametros desconocidos, alg\'un tipo de funci\'on de error\index{funci\'on de error} $E$ debe minimizarse, por ejemplo, la llamada entrop\'ia cruzada\index{entrop\'ia cruzada} (v\'ease, por ejemplo, \cite{deBoer2005}):
\begin{eqnarray}
&&  E = - \sum_{t \in D_{entrenamiento}} \sum_{\alpha=1}^K S_\alpha(t)~\ln(p_{\alpha}(t))
\end{eqnarray}
Para minimizar $E$, se puede, por ejemplo, utilizar el m\'etodo de descenso gradiente estoc\'astico (SGD, por sus siglas en ingl\'es)\index{descenso gradiente estoc\'astico (SGD)}, que minimiza la funci\'on de error\index{funci\'on de error} iterativamente hasta que el procedimiento converge.\footnote{\, Se puede utilizar una variedad de m\'etodos. Por algunos estudios pertinentes, v\'ease, por ejemplo, \cite{Denton1996}, \cite{Dong2008}, \cite{Dreyfus1990}, \cite{Ghosh2012}, \cite{Kingma2014}, \cite{Ruder2017}, \cite{Rumelhart1986}, \cite{Schmidhuber2015}, \cite{Wilson2018}.}

{}Por \'ultimo, debemos especificar las reglas de trading\index{regla de trading}. Hay una serie de posibilidades aqu\'i dependiendo del n\'umero de cuantiles\index{cuantil}, es decir, la selecci\'on de $K$. Una se\~{n}al de trading\index{senzal de trading @ se\~{n}al de trading} razonable est\'a dada por:
\begin{eqnarray}
\mbox{Se\~{n}al} = \begin{cases}
     \mbox{Comprar}, ~~   \mbox{ si y solo si} & \mbox{max}(p_\alpha(t)) = p_K(t) \\
     \mbox{Vender},  ~~~  \mbox{ si y solo si } & \mbox{max}(p_\alpha(t)) = p_1(t)
\end{cases}
\end{eqnarray}
Por lo tanto, el trader compra el BTC\index{Bitcoin (BTC)} si la clase predicha\index{clase predicha} es $p_K(t)$ (el cuantil\index{cuantil} superior), y vende si es $p_1(t)$ (el cuantil\index{cuantil} inferior). Esta regla de trading\index{regla de trading} puede ser modificada. Por ejemplo, la se\~{n}al de compra\index{senzal de compra @ se\~{n}al de compra} se puede basar en los 2 cuantiles\index{cuantil} superiores, y la se\~{n}al de venta\index{senzal de venta @ se\~{n}al de venta} se puede basar en los 2 cuantiles\index{cuantil} inferiores (v\'ease, por ejemplo, \cite{Nakano2018}).\footnote{\, Diversas t\'ecnicas utilizadas en la aplicaci\'on de las ANNs\index{red neuronal artificial (ANN)} a otras clases de activos\index{clases de activo}, tales como acciones\index{acci\'on}, tambi\'en pueden ser de utilidad para las criptomonedas\index{criptomonedas}. V\'ease, por ejemplo, \cite{Ballings2015}, \cite{Chong2017}, \cite{Dash2016}, \cite{Oliveira2013}, \cite{Sezer2017}, \cite{Yao1999}. Por alguna literatura adicional, v\'ease la nota al pie \ref{fn.KNN.ML}.}

\subsection{Estrategia: An\'alisis de sentimiento\index{an\'alisis del sentimiento} -- na\"{\i}ve Bayes Bernoulli}\label{Sentiment.a}

{}Las estrategias basadas en el an\'alisis de sentimiento\index{an\'alisis del sentimiento} de las redes sociales\index{an\'alisis del sentimiento de las redes sociales} se han utilizado en el trading de acciones\footnote{\, Por alguna literatura, v\'ease, por ejemplo, \cite{Bollen2011}, \cite{BollenMZ2011}, \cite{Kordonis2016}, \cite{LiewB2016}, \cite{Mittal2012}, \cite{Nisar2018}, \cite{Pagolu2016}, \cite{Rao2012}, \cite{Ruan2018}, \cite{Sprenger2014}, \cite{Sul2017}), \cite{ZhangFG2011}.} y tambi\'en se han aplicado al trading de criptomonedas\index{trading de criptomonedas}. La premisa es utilizar un esquema de clasificaci\'on de aprendizaje autom\'atico\index{esquema de clasificaci\'on de aprendizaje autom\'atico} para pronosticar, por ejemplo, la direcci\'on del movimiento en los precios del BTC\index{precio del BTC} en funci\'on de datos de tweets. Esto implica recopilar todos los tweets que contengan al menos una palabra clave\index{palabra clave} de un vocabulario de aprendizaje\index{vocabulario de aprendizaje} pertinente (para la predicci\'on de los precios del BTC\index{precio del BTC}) $V$ durante un per\'iodo de tiempo, y limpiar los datos.\footnote{\, Esto, entre otras cosas, incluye la eliminaci\'on de tweets duplicados que probablemente est\'an generados por robots de Twitter, la eliminaci\'on de {\em palabras vac\'ias\index{palabra vac\'ia}} (por ejemplo, (los equivalentes en ingl\'es de) ``el/la'', ``es/esta'', ``en'', ``que/cual'', etc.), que no agregan valor y luego, la aplicaci\'on de un proceso conocido ampliamente como {\em stemming\index{stemming}} para reducir las palabras a su ra\'iz o a su forma base\index{forma base} (por ejemplo, (los equivalentes en ingl\'es de) ``invirtiendo'' e ``invertido'' se reducen a ``invertir'', etc.). Esto \'ultimo se puede lograr utilizando, por ejemplo, el algoritmo de stemming de Porter\index{algoritmo de stemming de Porter} u otros algoritmos similares (por algunas publicaciones, v\'ease, por ejemplo, \cite{Hull1996}, \cite{Porter1980}, \cite{Raulji2016}, \cite{Willett2006}).} Los datos resultantes se procesan posteriormente asignando lo que se conoce como caracter\'istica (o feature)\index{caracter\'istica} ($M$-vector) $X_i$ a cada tweet etiquetado por $i=1,\dots,N$, en donde $N$ es el n\'umero de tweets en el conjunto de datos. Aqu\'i $M = |V|$ es el n\'umero de palabras clave\index{palabra clave} en el vocabulario de aprendizaje\index{vocabulario de aprendizaje} $V$. Por lo tanto, los componentes de cada vector $X_i$ son $X_{ia}$, en donde $a=1,\dots,M$ etiqueta las palabras\index{palabra} en $V$. As\'i, si la palabra\index{palabra} $w_a\in V$ etiquetada por $a$ no se encuentra presente en el tweet $T_i$ etiquetado por $i$, entonces $X_{ia} = 0$. Si $w_a$ est\'a presente en $T_i$, entonces podemos establecer que $X_{ia} = 1$ o $X_{ia} = n_{ia}$, en donde $n_{ia}$ cuenta el n\'umero de veces que $w_a$ aparece en $T_i$. En el primer caso (que es en lo que nos centramos en esta estrategia) tenemos una distribuci\'on de probabilidad\index{distribuci\'on de probabilidad} de Bernoulli\index{distribuci\'on de probabilidad de Bernoulli}, mientras que en el \'ultimo caso tenemos una distribuci\'on multinomial\index{distribuci\'on multinomial}.

{}Luego, debemos construir un modelo que, dados los $N$ vectores de caracter\'isticas\index{vectores de caracter\'isticas} $X_i$, predice uno de un n\'umero preestablecido $K$ de resultados\index{resultado} (llamados {\em clases\index{clase}}) $C_\alpha$, $\alpha=1,\dots,K$. Por ejemplo, podemos tener $K=2$ resultados\index{resultado}, en donde el BTC\index{Bitcoin (BTC)} est\'a previsto que suba o que baje, y que luego esto puede usarse como se\~{n}al de compra/venta\index{senzal de compra @ se\~{n}al de compra}\index{senzal, de venta @ se\~{n}al, de venta}. Alternativamente, como en la estrategia de ANN\index{estrategia, ANN} en la Subsecci\'on \ref{sub.ANN}, podemos tener $K$ cuantiles\index{cuantil} definidos para los rendimientos normalizados ${\widehat R}(t)$. Etc. Esto entonces define nuestras reglas de trading\index{regla de trading}. Una vez que las clases\index{clase} $C_\alpha$ son elegidas, una forma simple de pronosticarlas es mediante la construcci\'on de un modelo para estimar las probabilidades condicionales\index{probabilidad condicional} $P(C_\alpha | X_1,\dots,X_N)$. Aqu\'i, en general, $P(A|B)$ denota la probabilidad condicional\index{probabilidad condicional} de $A$ ocurriendo luego de asumir que $B$ es verdad. De acuerdo con el teorema de Bayes\index{teorema de Bayes}, tenemos
\begin{equation}
 P(A|B) = {P(B|A)~P(A) \over P(B)}
\end{equation}
en donde $P(A)$ y $P(B)$ son las probabilidades de los eventos $A$ y $B$ ocurriendo independientemente el uno del otro. Entonces, tenemos
\begin{equation}\label{Bayes}
 P(C_\alpha | X_1,\dots,X_N) = {P(X_1,\dots,X_N | C_\alpha)~P(C_\alpha) \over P(X_1,\dots,X_N)}
\end{equation}
Tenga en cuenta que $P(X_1,\dots,X_N)$ es independiente de $C_\alpha$ y no ser\'a importante a continuaci\'on. Ahora, $P(C_\alpha)$ se puede estimar a partir de los datos de entrenamiento\index{datos de entrenamiento}. La principal dificultad est\'a en la estimaci\'on de $P(X_1,\dots,X_N | C_\alpha)$. Aqu\'i logramos una simplificaci\'on si asumimos la independencia condicional\index{suposici\'on de la independencia condicional} ``na\"{\i}ve'' (de ah\'i el t\'ermino ``na\"{\i}ve Bayes''), es decir, que, dada la clase\index{clase} $C_\alpha$, para todo $i$ la caracter\'istica\index{caracter\'istica} $X_i$ es condicionalmente independiente de cualquier otra caracter\'istica\index{caracter\'istica} $X_j$, $j=1,\dots,N$ ($j\neq i$):
\begin{equation}
 P(X_i|C_\alpha, X_1,\dots,X_{i-1},X_{i+1},\dots,X_N) = P(X_i|C_\alpha)
\end{equation}
Luego, la Ecuaci\'on (\ref{Bayes}) se simplifica de la siguiente manera:
\begin{eqnarray}\label{naive.Bayes}
 &&P(C_\alpha | X_1,\dots,X_N) = \gamma ~ P(C_\alpha)~\prod_{i=1}^N P(X_i | C_\alpha)\\
 &&\gamma = 1 / P(X_1,\dots,X_N)
\end{eqnarray}
Las probabilidades condicionales\index{probabilidad condicional} $P(X_i | C_\alpha)$ se pueden estimar utilizando las probabilidades condicionales\index{probabilidad condicional} $P(w_a | C_\alpha)$ para las $M$ palabras\index{palabra} $w_a$ en el vocabulario de aprendizaje\index{vocabulario de aprendizaje} $V$:
\begin{eqnarray}
 &&P(X_i | C_\alpha) = \prod_{a=1}^M Q_{ia\alpha}\\
 &&Q_{ia\alpha} = P(w_a | C_\alpha),~~~X_{ia} = 1\\
 &&Q_{ia\alpha} = 1 - P(w_a | C_\alpha),~~~X_{ia} = 0
\end{eqnarray}
Las probabilidades condicionales\index{probabilidad condicional} $P(w_a | C_\alpha)$ simplemente pueden estimarse bas\'andose en las frecuencias de aparici\'on de las palabras\index{palabra} $w_a$ en los datos de entrenamiento\index{datos de entrenamiento}. Del mismo modo, las probabilidades $P(C_\alpha)$ se pueden estimar a partir de los datos de entrenamiento\index{datos de entrenamiento}.\footnote{\, Por algunas publicaciones sobre la aplicaci\'on del sentimiento de Twitter\index{sentimiento de Twitter} en el trading del Bitcoin\index{trading del Bitcoin}, v\'ease, por ejemplo, \cite{Colianni2015}, \cite{Georgoula2015}, que tambi\'en discuten otros m\'etodos de aprendizaje autom\'atico\index{metodos de aprendizaje automatico @ m\'etodos de aprendizaje autom\'atico} tales como m\'aquinas de vectores de soporte (SVM, por sus siglas en ingl\'es)\index{maquinas de vectores de soporte (SVM) @ m\'aquinas de vectores de soporte (SVM)} y regresi\'on log\'istica\index{regresi\'on log\'istica} (tambi\'en conocida como modelo logit\index{modelo logit}). Por alguna literatura sobre el trading del Bitcoin\index{trading del Bitcoin} utilizando otros datos de sentimiento\index{datos de sentimiento}, v\'ease, por ejemplo, \cite{Garcia2015}, \cite{Li2018}. Para algunos estudios sobre la aplicaci\'on del algoritmo tree boosting\index{algoritmo tree boosting} al trading de criptomonedas\index{trading de criptomonedas}, v\'ease, por ejemplo, \cite{Alessandretti2018}, \cite{Li2018}. Por algunas publicaciones pertinentes adicionales (que generalmente parecen ser relativamente escasas para el BTC\index{Bitcoin (BTC)} en comparaci\'on con la literatura similar para el trading de acciones), v\'ease, por ejemplo, \cite{Amjad2017}, \cite{JiangLiang2017}, \cite{Shah2014}.} Por lo tanto, si establecemos que el valor previsto $C_{pred}$ del resultado\index{resultado} coincida con el m\'aximo $P(C_\alpha | X_1,\dots,X_N)$, entonces tenemos
\begin{equation}
 C_{pred} = \mbox{argmax}_{\,C_{\alpha \in \{ 1,\dots, K\}}} P(C_\alpha)~\prod_{i=1}^N \prod_{a=1}^M [P(w_a | C_\alpha)]^{X_{ia}} [1 - P(w_a | C_\alpha)]^{1-X_{ia}}
\end{equation}

\newpage

\section{Macro Global\index{macro global}}

\subsection{Generalidades}

{}En realidad, las estrategias de trading macro\index{estrategia de trading macro} constituyen un estilo de inversi\'on\index{estilo de inversi\'on}, no una clase de activos\index{clases de activo}. Estos tipos de estrategias no se limitan a ninguna clase de activo\index{clases de activo} en particular o a una regi\'on geogr\'afica y pueden utilizar como un medio de inversi\'on acciones, bonos\index{bono}, monedas\index{moneda}, commodities\index{commodity}, derivados\index{derivado}, etc., apuntando a capitalizar los cambios regionales, econ\'omicos y pol\'iticos en todo el mundo\index{mundo}. Mientras que muchas estrategias macro\index{estrategia macro} se basan en opiniones subjetivas de los analistas (estas son estrategias discrecionales\index{estrategia discrecional}), el enfoque sistem\'atico\index{enfoque sistem\'atico} (estrategias no discrecionales\index{estrategia no discrecional}) tambi\'en juega un papel destacado. Las estrategias macro globales\index{estrategia macro global} pueden variar seg\'un su estilo, por ejemplo, existen estrategias direccionales\index{estrategia direccional}, estrategias con posiciones largas y cortas\index{estrategia con posiciones largas y cortas}, estrategias de valor relativo\index{estrategia de valor relativo}, etc.\footnote{\, Las estrategias macro\index{estrategia macro} se puede dividir en 3 clases: macro discrecional\index{macro discrecional}, macro sistem\'atico\index{macro sistem\'atico}, y CTA (por sus siglas en ingl\'es)/futuros gestionados\index{futuros, CTA}\index{futuros, gestionados}\index{asesor de trading de commodities (CTA)}\index{futuros gestionados}. Por algunas publicaciones sobre estrategias macro\index{estrategia macro} y t\'opicos relacionados, v\'ease, por ejemplo, \cite{Asgharian2004}, \cite{Chung2000}, \cite{Connor2004}, \cite{Dobson1984}, \cite{Drobny2006}, \cite{FabozziFocardi2010}, \cite{Fung1999}, \cite{Gliner2014}, \cite{Kidd2014}, \cite{Lambert2006}, \cite{Potjer2007}, \cite{Stefanini2006}, \cite{Zaremba2014}.}

\subsection{Estrategia: Macro momentum fundamental\index{macro momentum fundamental}}

{}Esta estrategia apunta a capturar los retornos generados por la infra-reacci\'on del mercado\index{infra-reacci\'on del mercado} ante cambios en tendencias macroecon\'omicas\index{tendencias macroecon\'omicas} comprando (vendiendo) los activos favorecidos (adversamente afectados) por las tendencias macroecon\'omicas\index{tendencias macroecon\'omicas} entrantes. Distintas clases de activos\index{clases de activo} pueden ser utilizadas para construir un portafolio de inversiones\index{portafolio de inversi\'on}, por ejemplo, \'indices globales de acciones\index{indice de acciones @ \'indice de acciones}, monedas\index{moneda}, bonos gubernamentales\index{bono gubernamental}, etc.\footnote{\, Diferentes clases de activos\index{clases de activo} se ven afectadas de forma diferente por las mismas tendencias macroecon\'omicas\index{tendencias macroecon\'omicas}. Por ejemplo, un aumento en el crecimiento es positivo para acciones\index{acci\'on} y monedas\index{moneda}, pero negativo para bonos\index{bono}.} Las ``variables de estado\index{variable de estado}'' a considerar son el ciclo econ\'omico\index{ciclo econ\'omico}, el comercio internacional\index{comercio internacional}, la pol\'itica monetaria\index{pol\'itica monetaria}, y las tendencias\index{tendencia} del sentimiento de riesgo\index{sentimiento de riesgo} (v\'ease, por ejemplo, \cite{BrooksAQR2017}).\footnote{\, Las tendencias del ciclo econ\'omico\index{tendencias del ciclo econ\'omico} se pueden estimar utilizando los cambios de 1 a\~{n}o en el crecimiento del GDP (por sus siglas en ingl\'es)\index{GDP (producto interno bruto)} real y el valor esperado del CPI\index{Indice de Precios al Consumidor (CPI) @ \'Indice de Precios al Consumidor (CPI)}, cada variable aportando un 50\% de peso. Las tendencias del comercio internacional\index{tendencias del comercio internacional} se pueden estimar usando los cambios de 1 a\~{n}o en las tasas de FX spot\index{tasa de FX spot} contra una cesta ponderada por exportaciones. Las tendencias de la pol\'itica monetaria\index{tendencias de la pol\'itica monetaria} se pueden estimar utilizando los cambios de 1 a\~{n}o en las tasas a corto plazo. Las tendencias del sentimiento de riesgo\index{tendencias del sentimiento de riesgo} se pueden estimar utilizando los excesos de retornos del mercado accionario\index{excesos de retornos del mercado accionario} a 1 a\~{n}o. Para obtener informaci\'on sobre los fundamentos de estas variables, consulte, por ejemplo, \cite{Bernanke2005}, \cite{Clarida2007}, \cite{Eichenbaum1995}.} Por ejemplo, algunos \'indices de acciones\index{indice de acciones @ \'indice de acciones} de algunos pa\'ises se clasifican de acuerdo con los valores de las 4 variables de estado\index{variable de estado} mencionadas anteriormente para cada pa\'is.\footnote{\, Hay una variedad de formas de hacer esta clasificaci\'on (ranking en ingl\'es)\index{ranking} utilizando las 4 variables. V\'ease, por ejemplo, la Subsecci\'on \ref{sub.multifactor}.} Un portafolio de costo cero\index{portafolio de costo cero} luego se puede construir, por ejemplo, tomando posiciones largas en los \'indices\index{indice @ \'indice} en el decil\index{decil} superior y posiciones cortas en aquellos que se encuentran en el decil\index{decil} inferior. Los portafolios\index{portafolio} as\'i construidos para varias clases de activos\index{clases de activo} pueden, por ejemplo, combinarse con ponderaciones iguales. T\'ipicamente, el per\'iodo de tenencia\index{periodo de tenencia @ per\'iodo de tenencia} var\'ia de tres a seis meses.

\subsection{Estrategia: Cobertura macro global contra la inflaci\'on\index{cobertura macro global contra la inflaci\'on}}

{}Los choques ex\'ogenos (como asuntos pol\'iticos o geopol\'iticos) pueden tener un impacto en los precios de commodities\index{precios de commodities} tales como el petr\'oleo, conduciendo a un aumento de los precios en las econom\'ias dependientes de \'este. Hay dos pasos en este proceso: (i) un traspaso de los precios de commodities\index{precios de commodities} a la inflaci\'on general (HI, por sus siglas en ingl\'es)\index{inflaci\'on general (HI)}, y (ii) luego, un traspaso desde HI\index{inflaci\'on general (HI)} a la inflaci\'on n\'ucleo (CI, por sus siglas en ingl\'es)\index{inflaci\'on n\'ucleo (CI)}.\footnote{\, HI\index{inflaci\'on general (HI)} es la inflaci\'on general\index{inflaci\'on general (HI)} medida por \'indices como el \'Indice de Precios al Consumidor (CPI, por sus siglas en ingl\'es)\index{Indice de Precios al Consumidor (CPI) @ \'Indice de Precios al Consumidor (CPI)}, basado en precios de una amplia cesta de bienes y servicios, mientras que CI\index{inflaci\'on n\'ucleo (CI)} excluye algunos productos tales como commodities\index{commodity}, que son altamente vol\'atiles y agregan un ruido\index{ruido} considerable al \'indice\index{indice @ \'indice}. Por estudios pertinentes, v\'ease, por ejemplo, \cite{Blanchard2007}, \cite{Blanchard2013}, \cite{Clark2010}, \cite{Hamilton2003}, \cite{Marques2003}, \cite{Trehan2005}, \cite{Noord2007}.} Es decir, HI\index{inflaci\'on general (HI)} refleja r\'apidamente algunos de los choques que ocurren en todo el mundo\index{mundo}. Entonces, la estrategia de cobertura macro global contra la inflaci\'on\index{estrategia de cobertura macro global contra la inflaci\'on} se basa en el margen\index{margen} entre HI\index{inflaci\'on general (HI)} y CI\index{inflaci\'on n\'ucleo (CI)} como indicador para cubrirse contra la inflaci\'on\index{inflaci\'on} utilizando commodities\index{commodity}:\footnote{\, Por algunos estudios sobre el uso de commodities\index{commodity} como cobertura contra la inflaci\'on\index{cobertura contra la inflaci\'on}, v\'ease, por ejemplo, \cite{Amenc2009}, \cite{Bodie1983}, \cite{Bodie1980}, \cite{Greer1978}, \cite{Hoevenaars2008}, \cite{Jensen2002}.}
\begin{eqnarray}
&& \mbox{CA} = \mbox{max}\left(0, \mbox{min}\left(\frac{\mbox{HI}_{YoY} - \mbox{CI}_{YoY}}{\mbox{HI}_{YoY}}, 1 \right) \right)
\end{eqnarray}
Aqu\'i: CA (por sus siglas en ingl\'es) es el porcentaje de asignaci\'on a commodities\index{porcentaje de asignaci\'on a commodities (CA)} dentro del portafolio\index{portafolio}, y ``YoY\index{anzo a anzo @ a\~{n}o a a\~{n}o}'' se refiere a ``a\~{n}o a a\~{n}o\index{anzo a anzo @ a\~{n}o a a\~{n}o}''. La operaci\'on de cobertura\index{cobertura} puede ser ejecutada, por ejemplo, comprando una cesta\index{cesta} de varios commodities\index{commodity} mediante ETFs\index{fondo de inversi\'on cotizado (ETF)}, futuros\index{futuro}, etc. (v\'ease, por ejemplo, \cite{Fulli-Lemaire2013}).

\subsection{Estrategia: Estrategia global de renta fija}

{}Esta estrategia de trading macro sistem\'atica\index{estrategia de trading macro sistem\'atica} se basa en un an\'alisis de corte transversal\index{an\'alisis de corte transversal} de bonos gubernamentales\index{bono gubernamental} de diversos pa\'ises utilizando variables como (v\'ease, por ejemplo, \cite{Fan2017}) el GDP\index{GDP (producto interno bruto)}, la inflaci\'on\index{inflaci\'on}, el riesgo soberano\index{riesgo soberano}, las tasas de inter\'es reales\index{tasa de inter\'es real}, la brecha de producci\'on\index{brecha de producci\'on} (output gap en ingl\'es), value\index{value}, momentum\index{momentum}, la diferencia temporal\index{diferencia temporal} (en la curva de rendimientos), y el ampliamente conocido predictor de Cochrane-Piazzesi\index{predictor de Cochrane-Piazzesi} \cite{Cochrane2005}. De esta forma, estos bonos\index{bono} se pueden clasificar en base a estos factores\index{factor} y un portafolio de costo cero\index{portafolio de costo cero} se puede construir comprando los bonos\index{bono} en el cuantil\index{cuantil} superior y vendiendo los bonos\index{bono} en el cuantil\index{cuantil} inferior. De forma similar a la Subsecci\'on \ref{sub.multifactor}, portafolios multifactoriales\index{portafolio multifactor} tambi\'en se pueden construir. Por lo general, ETFs de bonos de pa\'ises son utilizados en dichos portafolios\index{portafolio}.\footnote{\, Por algunos estudios sobre las inversiones con factores\index{inversi\'on con factores} en activos de renta fija\index{activo de renta fija}, v\'ease, por ejemplo, \cite{Beekhuizen2016}, \cite{Correia2012}, \cite{Houweling2017}, \cite{Koijen2018}, \cite{LHoir2010}, \cite{Staal2015}.}

\subsection{Estrategia: Trading con anuncios econ\'omicos\index{anuncio econ\'omico}}

{}La evidencia emp\'irica sugiere que las acciones tienden a obtener mayores retornos en las fechas de anuncios importantes tales como los anuncios del Comit\'e Federal de Mercado Abierto (FOMC, por sus siglas en ingl\'es)\index{Comit\'e Federal de Mercado Abierto (FOMC)}.\footnote{\, Por literatura pertinente, v\'ease, por ejemplo, \cite{Ai2016}, \cite{Bernanke2005}, \cite{Boyd2005}, \cite{Donninger2015}, \cite{Graham2003}, \cite{Jones1998}, \cite{Lucca2012}, \cite{Savor2013}.} Por lo tanto, una estrategia de trading macro\index{estrategia de trading macro} simple consiste en comprar las acciones en los d\'ias de anuncios\index{dias de anuncios @ d\'ias de anuncios} importantes, tales como los anuncios del FOMC\index{anuncios del FOMC}, y durante los d\'ias sin anuncios\index{dias sin anuncios @ d\'ias sin anuncios}, posicionarse en activos libres de riesgo\index{activo libre de riesgo}. Esto se hace a trav\'es de ETFs\index{fondo de inversi\'on cotizado (ETF)}, futuros\index{futuro}, etc., y no con las acciones individuales, ya que la estrategia consiste en pasar de un 100\% asignado a las acciones\index{acci\'on} a un 100\% asignado a los bonos del Tesoro\index{Tesoro} (v\'ease, por ejemplo, \cite{Stotz2016}).\footnote{\, Esta estrategia se puede sofisticar con varios filtros (por ejemplo, t\'ecnicos) (v\'ease, por ejemplo, \cite{Stotz2016}).}

\newpage

\section{Infraestructura\index{infraestructura}}

{}En general, invertir en infraestructura\index{infraestructura} incluye invertir en proyectos a largo plazo tales como transporte (carreteras, puentes, t\'uneles, ferrocarriles, puertos, aeropuertos, etc.), telecomunicaciones (cables de transmisi\'on, sat\'elites, torres, etc.), servicios p\'ublicos (generaci\'on de electricidad, transmisi\'on o distribuci\'on de gas o electricidad, suministro de agua, aguas residuales, residuos, etc.), energ\'ia\index{energ\'ia} (incluyendo pero no limitado a energ\'ia renovable), atenci\'on m\'edica (hospitales, cl\'inicas, hogares de ancianos, etc.), instalaciones educativas (escuelas, universidades, institutos de investigaci\'on, etc.), etc. Un inversionista puede ganar exposici\'on\index{exposici\'on} a los activos de infraestructura\index{activo de infraestructura} mediante inversiones\index{inversi\'on} directas o indirectas tales como inversiones de capital privado\index{inversiones de capital privado} (por ejemplo, a trav\'es de fondos de infraestructura no cotizados\index{fondo de infraestructura no cotizado}), fondos de infraestructura cotizados\index{fondo de infraestructura cotizado}, acciones de las compa\~{n}\'ias de infraestructura\index{compa\~{n}\'ia de infraestructura} que cotizan en la bolsa\index{compa\~{n}\'ia de infraestructura que cotiza en la bolsa}, bonos municipales\index{bono municipal} destinados a proyectos de infraestructura\index{proyecto de infraestructura}, etc.\footnote{\, Por algunas publicaciones sobre infraestructura\index{infraestructura} como una clase de activo\index{clases de activo} y t\'opicos relacionados, v\'ease, por ejemplo, \cite{Ansar2016}, \cite{Bitsch2010}, \cite{Blanc-Brude2016}, \cite{Blanc-Brude2017}, \cite{Blundell2006}, \cite{Clark2017}, \cite{Clark2012}, \cite{Finkenzeller2010}, \cite{Grigg2010}, \cite{Grimsey2002}, \cite{Hartigan2011}, \cite{Helm2009}, \cite{HelmTindall2009}, \cite{Herranz-Loncan2007}, \cite{Inderst2010}, \cite{McDevitt2008}, \cite{Newell2009}, \cite{Newell2008}, \cite{Peng2007}, \cite{Ramamurti2004}, \cite{Rickards2008}, \cite{Sanchez-Robles1998}, \cite{Sawant2010}, \cite{Sawant2010b}, \cite{Singhal2011}, \cite{Smit2009}, \cite{Torrance2007}, \cite{Vives1999}, \cite{Weber2016}, \cite{Wurstbauer2016}.}

{}Las inversiones en infraestructura\index{inversi\'on en infraestructura}, por su naturaleza, son inversiones de largo plazo, de compra y mantenimiento\index{estrategia de compra y mantenimiento}. Una estrategia de inversi\'on\index{estrategia de inversi\'on} consiste en utilizar activos de infraestructura\index{activo de infraestructura} para mejorar los retornos ajustados por riesgo\index{retornos ajustados por riesgo} de portafolios diversificados\index{portafolio diversificado}, por ejemplo, a trav\'es de ETFs de rastreo\index{ETF de rastreo}, fondos de infraestructura global\index{fondo de infraestructura global}, fondos de infraestructura no cotizados\index{fondo de infraestructura no cotizado}, etc.\footnote{\, V\'ease, por ejemplo, \cite{Dechant2013}, \cite{Haran2011}, \cite{Joshi2011}, \cite{Martin2010}, \cite{Nartea2010}, \cite{Newell2011}, \cite{Oyedele2014}, \cite{Panayiotou2016}, \cite{Rothballer2012}.} Otra estrategia de inversi\'on\index{estrategia de inversi\'on} es utilizar activos de infraestructura\index{activo de infraestructura} para obtener cobertura contra la inflaci\'on\index{cobertura contra la inflaci\'on}.\footnote{\, Infraestructura\index{infraestructura}, como bienes ra\'ices\index{bienes ra\'ices}, puede ser una inversi\'on de cobertura contra la inflaci\'on\index{inversi\'on de cobertura contra la inflaci\'on}, aunque aparentemente con cierta heterogeneidad. Por alguna literatura, v\'ease, por ejemplo, \cite{Armann2008}, \cite{Bird2014}, \cite{Inderst2010b}, \cite{Wurstbauer2015}, \cite{Rodel2012}.} Otra estrategia de inversi\'on\index{estrategia de inversi\'on} m\'as es generar flujos de efectivo\index{flujo de efectivo} estables mediante inversiones en infraestructura\index{inversi\'on en infraestructura}. Para este prop\'osito, proyectos ``brownfield''\index{proyecto brownfield} (asociados a activos establecidos que necesitan mejoras) son m\'as apropiados que los proyectos ``greenfield''\index{proyecto greenfield} (asociados a los activos por construir). Diversificaci\'on\index{diversificaci\'on} a trav\'es de diferentes sectores\index{sector} puede ser beneficioso en este sentido.\footnote{\, Por algunos estudios pertinentes, v\'ease, por ejemplo, \cite{Arezki2016}, \cite{Espinoza2002}, \cite{Leigland2018}, \cite{Panayiotou2014}, \cite{Weber2008}.}

\phantomsection
\addcontentsline{toc}{section}{Agradecimientos}
\section*{Agradecimientos}

{}JAS agradece a Juli\'{a}n R. Siri por sus valiosas discusiones. Los autores agradecen a Emiliano Serur por su ayuda con la correci\'on del manuscrito.

\appendix

\newpage

\section{C\'odigo Fuente\index{codigo fuente @ c\'odigo fuente} en R para Backtesting\index{backtesting}}\label{app.A}

{}En este ap\'endice brindamos el c\'odigo fuente\index{codigo fuente @ c\'odigo fuente} en R (R Package for Statistical Computing\index{R Package for Statistical Computing (R)}, \url{http://www.r-project.org}) para backtesting\index{backtesting} de estrategias intrad\'ia\index{estrategia intrad\'ia}, en donde la posici\'on se establece en la apertura\index{apertura} y se liquida al cierre\index{cierre} (del mercado) en el mismo d\'ia. El \'unico prop\'osito de este c\'odigo es ilustrar algunos trucos simples para realizar un backtesting fuera de muestra\index{backtesting fuera de la muestra}. En particular, este c\'odigo no trata el sesgo de supervivencia\index{sesgo de supervivencia} de ninguna manera,\footnote{\, Es decir, simplemente, no tiene en cuenta el hecho de que en el pasado hab\'ia tickers (tableros de cotizaciones)\index{ticker (tablero de cotizaciones)} que ya no existen en la actualidad, ya sea por bancarrota\index{bancarrota}, fusiones\index{fusi\'on}, adquisiciones\index{adquisici\'on}, etc. En su lugar, los datos de entrada se toman para los tickers (tableros de cotizaciones)\index{ticker (tablero de cotizaciones)} que existen en un d\'ia determinado mirando hacia atr\'as, digamos, algunos a\~{n}os.} aunque para este tipo de estrategias -- precisamente porque se trata de estrategias intrad\'ia\index{estrategia intrad\'ia} -- el sesgo de supervivencia\index{sesgo de supervivencia} no es perjudicial (v\'ease, por ejemplo, \cite{Kakushadze2015b}).\footnote{\, Por algunas publicaciones relacionadas con el sesgo de supervivencia\index{sesgo de supervivencia}, que es importante para las estrategias de horizonte m\'as largo, v\'ease, por ejemplo, \cite{Amin2003}, \cite{Brown1992}, \cite{Bu2007}, \cite{Carhart2002}, \cite{Davis1996}, \cite{Elton1996b}, \cite{Garcia1993}.}

{}La funci\'on principal (que internamente llama a algunas otras subfunciones) es {\tt{\small qrm.backtest()}} con las siguientes entradas: (i) {\tt{\small days}} es el lookback\index{lookback}; (ii) {\tt{\small d.r}} se utiliza para calcular el riesgo\index{riesgo}, tanto para la longitud de la desviaci\'on est\'andar m\'ovil\index{desviaci\'on est\'andar m\'ovil} {\tt{\small tr}} (computada internamente a trav\'es de las ventanas m\'oviles de {\tt{\small d.r}} d\'ias) as\'i como el lookback\index{lookback} para computar el modelo de riesgo\index{modelo de riesgo} (y, si es aplicable, una clasificaci\'on estad\'istica de la industria\index{clasificaci\'on estad\'istica de la industria}) -- v\'ease abajo; (iii) {\tt{\small d.addv}} se utiliza como el lookback\index{lookback} para el volumen diario promedio en d\'olares\index{volumen diario promedio en d\'olares} {\tt{\small addv}}, que es computado internamente; (iv) {\tt{\small n.addv}} es el n\'umero de los mejores tickers (tableros de cotizaciones)\index{ticker (tablero de cotizaciones)} seg\'un {\tt{\small addv}}, el cual es utilizado como el universo de trading\index{universo de trading} y se recalcula cada {\tt{\small d.r}} d\'ias; (v) {\tt{\small inv.lvl}} es el nivel de inversi\'on\index{nivel de inversi\'on} total (posiciones largas mas cortas, y la estrategia es d\'olar-neutral); (vi) {\tt{\small bnds}} controla los l\'imites de las posiciones\index{limites de la posicion @ l\'imites de la posici\'on} (que en esta estrategia son los mismos que los l\'imites de trading\index{limites de trading @ l\'imites de trading}), es decir, las tenencias en d\'olares\index{tenencia en d\'olares} $H_i$ para cada acci\'on est\'an delimitadas a trav\'es de ($B_i$ son los elementos {\tt{\small bnds}}, que pueden ser uniformes)
\begin{equation}
 |H_i| \leq B_i~A_i
\end{equation}
en donde $i=1,\dots,N$ etiqueta las acciones en el universo de trading\index{universo de trading}, y $A_i$ son los elementos correspondientes de {\tt{\small addv}}; (vii) {\tt{\small incl.cost}} es un Booleano\index{Booleano} para incluir costos transaccionales lineales\index{costos transaccionales lineales}, que se modelan de la siguiente manera.\footnote{\, Aqu\'i seguimos la discusi\'on en la Subsecci\'on 3.1 de \cite{KakushadzeYu2018b}.} Para la acci\'on etiquetada por $i$, sea $E_i$ su retorno esperado\index{retorno esperado} y $w_i$ sea su ponderaci\'on dentro del portafolio\index{portafolio}. El c\'odigo fuente\index{codigo fuente @ c\'odigo fuente} a continuaci\'on determina $w_i$ a trav\'es de la optimizaci\'on (de media-varianza)\index{optimizaci\'on, de media-varianza} (con l\'imites). Para la acci\'on etiquetada por $i$, sea $\tau_i$ el costo de trading lineal por {\em d\'olar} negociado. Incluir dichos costos en el problema de optimizaci\'on del portafolio\index{portafolio}\index{optimizaci\'on del portafolio} consiste en reemplazar el retorno esperado\index{retorno esperado} del portafolio\index{portafolio}
\begin{equation}
 E_{port} = \sum_{i=1}^N E_i~w_i
\end{equation}
por
\begin{equation}
 E_{port} = \sum_{i=1}^N \left[E_i~w_i - \tau_i~|w_i|\right]
\end{equation}
Un algoritmo completo para incluir los costos transaccionales lineales\index{costos transaccionales lineales} en la optimizaci\'on de media-varianza\index{optimizaci\'on de media-varianza} se presenta en, por ejemplo,\cite{Kakushadze2015b}. Sin embargo, para nuestros prop\'ositos aqu\'i el siguiente simple truco es suficiente. Podemos definir el retorno efectivo
\begin{equation}\label{E.tc}
 E_i^{eff} = \mbox{sign}(E_i)~\mbox{max}(|E_i| - \tau_i, 0)
\end{equation}
y simplemente establecer
\begin{equation}
 E_{port} = \sum_{i=1}^N E_i^{eff}~w_i
\end{equation}
Es decir, si la magnitud del retorno esperado\index{retorno esperado} para una acci\'on dada es menor que el costo esperado en el que se incurrir\'a, establecemos que el retorno esperado\index{retorno esperado} sea igual a cero, de lo contrario reducimos dicha magnitud por dicho costo. De esta manera podemos evitar un procedimiento iterativo no trivial (v\'ease, por ejemplo, \cite{Kakushadze2015b}), aunque esto es solo una aproximaci\'on.

{}Entonces, ?`qu\'e debemos usar como $\tau_i$ en (\ref{E.tc})? El modelo de \cite{Almgren2005} es razonable para nuestros prop\'ositos aqu\'i. Sea $H_i$ la cantidad de {\em d\'olares} negociados para la acci\'on etiquetada por $i$. Entonces para los costos transaccionales lineales\index{costos transaccionales lineales} tenemos
\begin{equation}
 T_i = \zeta~\sigma_i~{|H_i|\over A_i}
\end{equation}
en donde $\sigma_i$ es la volatilidad hist\'orica\index{volatilidad hist\'orica}, $A_i$ es el volumen promedio diario en d\'olares (ADDV, por sus siglas en ingl\'es)\index{volumen diario promedio en d\'olares}, y $\zeta$ es una constante de normalizaci\'on general que necesitamos fijar. Sin embargo, arriba trabajamos con las ponderaciones $w_i$, no con los montos en d\'olares negociados $H_i$. En nuestro caso de una estrategia de trading puramente intrad\'ia\index{estrategia de trading intrad\'ia}, las ponderaciones est\'an relacionadas simplemente a trav\'es de $H_i = I~w_i$, en donde $I$ es el nivel de inversi\'on\index{nivel de inversi\'on} total (es decir, las tenencias en d\'olares\index{tenencia en d\'olares} totales absolutas del portafolio\index{portafolio} una vez que \'este es establecido\index{establecer}). Por lo tanto, tenemos (note que $T_i = \tau_i~|H_i| = \tau_i~I~|w_i|$)
\begin{equation}
 \tau_i = \zeta~{\sigma_i~\over A_i}
\end{equation}
Para fijar la normalizaci\'on general $\zeta$, utilizaremos la siguiente heur\'istica. Asumiremos (de forma conservadora) que el costo lineal promedio por cada d\'olar negociado es igual a 10 bps (1 bps = 1 punto b\'asico\index{punto b\'asico (bps)} = 1/100 de 1\%),\footnote{\, Esto es equivalente a suponer que establecer un portafolio igualmente ponderado\index{portafolio igualmente ponderado} cuesta 10 bps.} es decir, $\mbox{media}(\tau_i) = 10^{-3}$ y $\zeta = 10^{-3} / \mbox{media}(\sigma_i / A_i)$.

{}Luego, internamente el c\'odigo obtiene datos de precios y vol\'umenes\index{datos de precio y volumen} ley\'endolos desde archivos delimitados por tabuladores\footnote{\, Este c\'odigo espec\'ifico no utiliza el precio m\'aximo, el precio m\'inimo,\index{precio, m\'aximo}\index{precio, m\'inimo} el precio promedio ponderado por volumen (VWAP, por sus siglas en ingl\'es), los precios intrad\'ia\index{precio, intrad\'ia} (por ejemplo, minuto a minuto), etc. Sin embargo, es sencillo modificarlo de tal manera que lo haga.} {\tt{\small nrm.ret.txt}} (retorno nocturno,\index{retorno nocturno} internamente referido como {\tt{\small ret}} -- v\'ease abajo), {\tt{\small nrm.open.txt}} (precio de apertura diario sin procesar, sin ajustar\index{precio de apertura, sin ajustar}, internamente referido como {\tt{\small open}}), {\tt{\small nrm.close.txt}} (precio de cierre diario sin procesar, sin ajustar\index{precio de cierre, sin ajustar}, internamente referido como {\tt{\small close}}), {\tt{\small nrm.vol.txt}} (volumen diario sin procesar, sin ajustar\index{volumen, sin ajustar}, internamente referido como {\tt{\small vol}}), {\tt{\small nrm.prc.txt}} (precio de cierre diario\index{precio de cierre} totalmente ajustado\index{precio, totalmente ajustado} por todos los splits\index{splits} y dividendos\index{dividendo}, internamente referido como {\tt{\small prc}}). Las filas de {\tt{\small ret}}, {\tt{\small open}}, {\tt{\small close}}, {\tt{\small vol}} y {\tt{\small prc}} corresponden a los $N$ tickers (tableros de cotizaciones)\index{ticker (tablero de cotizaciones)} (\'indice $i$). Sean los d\'ias de trading\index{dias de trading @ d\'ias de trading} etiquetados por $t=0,1,2,\dots, T$, en donde $t=0$ es el d\'ia m\'as reciente. Luego las columnas de {\tt{\small open}}, {\tt{\small close}}, {\tt{\small vol}} y {\tt{\small prc}} corresponden a los d\'ias de trading\index{dias de trading @ d\'ias de trading} $t=1,2,\dots, T$, es decir, el valor de $t$ es el mismo que el valor del \'indice de la columna. Por otro lado, las columnas de {\tt{\small ret}} corresponden a los retornos nocturnos\index{retorno nocturno} del cierre a la apertura\index{retornos de cierre contra apertura} desde el d\'ia de trading\index{dias de trading @ d\'ias de trading} $t$ hasta el d\'ia de trading\index{dias de trading @ d\'ias de trading} $t-1$. Es decir, la primera columna de {\tt{\small ret}} corresponde a los retornos nocturnos\index{retorno nocturno} del cierre a la apertura\index{retornos de cierre contra apertura} desde el d\'ia de trading\index{dias de trading @ d\'ias de trading} $t=1$ hasta el d\'ia de trading\index{dias de trading @ d\'ias de trading} $t=0$. Tambi\'en, {\tt{\small ret}}, ll\'amese $R_i(t)$, en donde $t=1,2,\dots, T$ etiqueta las columnas de {\tt{\small ret}}, se calcula de la siguiente manera:
\begin{eqnarray}
 && R_i(t) = \ln\left({P_i^{AO}(t-1) \over P_i^{AC}(t)}\right)\\
 && P_i^{AO}(t) = \gamma_i^{adj}(t)~P_i^{O}(t)\\
 &&\gamma_i^{adj}(t) = {P_i^{AC}(t) \over P_i^{C}(t)}
\end{eqnarray}
Aqu\'i: $P_i^{O}(t)$ es el precio de apertura sin procesar\index{precio de apertura sin procesar} (que es el elemento correspondiente a {\tt{\small open}} para $t=1,2,\dots, T$); $P_i^{C}(t)$ es el precio de cierre sin procesar\index{precio de cierre sin procesar} (que es el elemento correspondiente a {\tt{\small close}} para $t=1,2,\dots, T$); $P_i^{AC}(t)$ es el precio de cierre totalmente ajustado\index{precio de cierre ajustado}\index{precio, totalmente ajustado} (que es el elemento correspondiente a {\tt{\small prc}} para $t=1,2,\dots, T$); $\gamma_i^{adj}(t)$ es el factor de ajuste\index{factor de ajuste}, que se utiliza para calcular el precio de apertura totalmente ajustado\index{precio de apertura ajustado}\index{precio, totalmente ajustado} $P_i^{AO}(t)$; entonces $R_i(t)$ es el retorno nocturno\index{retorno nocturno} del cierre a la apertura\index{retornos de cierre contra apertura} basado en los precios totalmente ajustados\index{precio ajustado}\index{precio, totalmente ajustado}. Tenga en cuenta que los precios en $t=0$ requeridos para la computaci\'on de $R_i(1)$ {\em no} son parte de las matrices {\tt{\small open}}, {\tt{\small close}} y {\tt{\small prc}}. Adem\'as, el c\'odigo asume internamente que las matrices {\tt{\small ret}}, {\tt{\small open}}, {\tt{\small close}}, {\tt{\small vol}} y {\tt{\small prc}} est\'an todas alineadas, es decir, todos los tickers (tableros de cotizaciones)\index{ticker (tablero de cotizaciones)} y las fechas son iguales y est\'an en el mismo orden en cada uno de los 5 archivos {\tt{\small nrm.ret.txt}} (tenga en cuenta el etiquetado de los retornos descripto anteriormente), {\tt{\small nrm.open.txt}}, {\tt{\small nrm.close.txt}}, {\tt{\small nrm.vol.txt}} y {\tt{\small nrm.prc.txt}}. El ordenamiento de los tickers (tableros de cotizaciones)\index{ticker (tablero de cotizaciones)} en estos archivos es inmaterial, siempre y cuando sea el mismo en los 5 archivos ya que el c\'odigo es ajeno a dicho ordenamiento. Sin embargo, las fechas deben ordenarse de forma descendente, es decir, la primera columna corresponde a la fecha m\'as reciente, la segunda columna corresponde a la fecha anterior, etc. (aqu\'i ``fecha'' corresponde a un d\'ia de trading\index{dias de trading @ d\'ias de trading}). Por \'ultimo, tenga en cuenta que la funci\'on interna {\tt{\small read.x()}} lee estos archivos con el valor del par\'ametro {\tt{\small as.is = T}}. Esto significa que estos archivos est\'an en el formato delimitado por tabulaciones ``R-ready'', con $N+1$ l\'ineas delimitadas por tabulaciones. Las l\'ineas 2 hasta $N+1$ tienen $T+1$ elementos cada una. El primer elemento es un s\'imbolo de un ticker (tablero de cotizaciones)\index{ticker (tablero de cotizaciones)} (entonces los $N$ s\'imbolos comprenden {\tt{\small dimnames($\cdot$)[[1]]}} de la matriz correspondiente, por ejemplo, {\tt{\small open}} para los precios de apertura\index{precio de apertura}), y los otros $T$ elementos son los $T$ valores (por ejemplo, $P_i^O(t)$, $t=1,\dots,T$, para los precios de apertura\index{precio de apertura}). Sin embargo, la primera l\'inea tiene solo $T$ elementos, que son las etiquetas de los d\'ias de trading\index{dias de trading @ d\'ias de trading} (entonces estos comprenden {\tt{\small dimnames($\cdot$)[[2]]}} de la matriz correspondiente, por ejemplo, {\tt{\small open}} para los precios de apertura\index{precio de apertura}). Funciones internas que utilizan estos datos de entrada, tales como {\tt{\small calc.mv.avg()}} (que calcula las medias m\'oviles simples\index{media m\'ovil simple (SMA)}) y {\tt{\small calc.mv.sd()}} (que calcula las desviaciones est\'andar m\'oviles simples\index{desviaci\'on est\'andar m\'ovil simple}) son simples y se explican por s\'i mismas.

{}Como se mencion\'o anteriormente, el par\'ametro de entrada {\tt{\small d.r}} se utiliza para recomputar el universo de trading\index{universo de trading} y los modelos de riesgo\index{modelo de riesgo} (v\'ease abajo) cada {\tt{\small d.r}} d\'ias de trading\index{dias de trading @ d\'ias de trading}. Estos c\'alculos se realizan 100\% fuera de la muestra\index{calculo, fuera de la muestra @ c\'alculo, fuera de la muestra}, es decir, los datos utilizados en estos c\'alculos est\'an 100\% en el pasado con respecto al d\'ia de trading\index{dias de trading @ d\'ias de trading} en el que las cantidades resultantes se utilizan (de forma simulada). Esto se logra en parte mediante el uso de la funci\'on interna {\tt{\small calc.ix()}}. Tenga en cuenta que los datos de entrada descriptos anteriormente est\'an estructurados y se utilizan de tal manera que los backtests\index{backtest} son 100\% fuera de la muestra. Aqu\'i se deben distinguir dos aspectos conceptualmente diferentes. Por un lado, tenemos los retornos esperados\index{retorno esperado} y ``el resto''. Este \'ultimo -- que puede ser referido en cierta forma a la ``gesti\'on de riesgos\index{gesti\'on de riesgo}'' -- siendo la selecci\'on del universo, la computaci\'on del modelo de riesgo\index{modelo de riesgo}, etc., es decir, la maquinaria que nos conduce desde los retornos esperados\index{retorno esperado} a las tenencias deseadas\index{tenencias deseadas} (es decir, las posiciones de la estrategia). La parte de gesti\'on de riesgos\index{gesti\'on de riesgo} debe ser 100\% fuera de la muestra. En la vida real los retornos esperados\index{retorno esperado} son tambi\'en 100\% fuera de la muestra. Sin embargo, en el backtesting\index{backtesting}, mientras que los retornos esperados\index{retorno esperado} en ninguna circunstancia pueden mirar hacia el futuro, a veces pueden estar ``en el l\'imite de la muestra''. Por lo tanto, considere una estrategia que hoy opera sobre los retornos calculados con los precios del cierre de ayer y la apertura de hoy. Si asumimos que las posiciones se establecen en base a este retorno en alg\'un momento despu\'es de la apertura\index{apertura}, entonces el backtest\index{backtest} est\'a fuera de la muestra por el tiempo de ``retraso\index{retraso}'' que transcurre entre la apertura\index{apertura} y el momento en el que se establece la operaci\'on. Sin embargo, si asumimos que la posici\'on se establece en la apertura\index{apertura}, entonces esto es llamado estrategia con ``retraso-0''\index{estrategia con retraso-0}, y el backtest\index{backtest} es ``en el l\'imite de la muestra'', en el sentido de que en la vida real las \'ordenes\index{orden} ser\'an enviadas con alg\'un retraso\index{retraso}, aunque posiblemente peque\~{n}o, pero nunca podr\'ian ser ejecutadas exactamente en la apertura\index{apertura}. En este sentido, todav\'ia tiene sentido hacer un backtest de este tipo de estrategia para medir la intensidad de la se\~{n}al\index{senzal @ se\~{n}al}. Lo que no tendr\'ia sentido y nunca deber\'ia hacerse, es realizar un backtest\index{backtest} dentro de la muestra que mira hacia el futuro. Por ejemplo, usar los precios del cierre\index{precio de cierre} de hoy para computar los retornos esperados\index{retorno esperado} para operar en la apertura de hoy ser\'ia en gran medida dentro de la muestra. Por otro lado, el uso de los precios de ayer para operar en el mercado en la apertura de hoy se conoce como estrategia con ``retraso-1''\index{estrategia con retraso-1}, que es b\'asicamente 1 d\'ia fuera de la muestra (y, como es l\'ogico, se espera que el backtest sea mucho peor que una estrategia con retraso-0\index{estrategia con retraso-0}). El c\'odigo da ejemplos de ambas estrategias con retraso-0\index{estrategia, retraso-0} (reversi\'on a la media\index{reversi\'on a la media}) y con retraso-1 (momentum\index{momentum})\index{estrategia, retraso-1} (v\'ease los comentarios {\tt{\small DELAY-0}} y {\tt{\small DELAY-1}} en el c\'odigo).

{}El c\'odigo calcula internamente las tenencias deseadas\index{tenencias deseadas} a trav\'es de la optimizaci\'on\index{optimizaci\'on}. La funci\'on del optimizador\index{funci\'on del optimizador} (que incorpora l\'imites y restricciones lineales, tales como la d\'olar-neutralidad\index{dolar-neutralidad @ d\'olar-neutralidad}) {\tt{\small bopt.calc.opt()}} se puede encontrar en \cite{Kakushadze2015e}. Una de sus entradas es la inversa de la matriz de covarianza del modelo\index{inversa de la matriz de covarianza del modelo} para las acciones. Esta matriz se calcula internamente a trav\'es de las funciones tales como {\tt{\small qrm.cov.pc()}} y {\tt{\small qrm.erank.pc()}}, las cuales se dan en y utilizan la construcci\'on del modelo estad\'istico de riesgo\index{modelo estad\'istico de riesgo} de \cite{KakushadzeYu2017a}, o {\tt{\small qrm.gen.het()}}, el cual se da en y utiliza el modelo heter\'otico de riesgo\index{modelo heter\'otico de riesgo} de \cite{KakushadzeYu2016a}. Este \'ultimo requiere una clasificaci\'on binaria multinivel de la industria\index{clasificaci\'on binaria multinivel de la industria}. El siguiente c\'odigo crea una clasificaci\'on de este tipo a trav\'es de la funci\'on {\tt{\small qrm.stat.ind.class.all()}}, que se da en y utiliza la construcci\'on de la clasificaci\'on estad\'istica de la industria\index{clasificaci\'on estad\'istica de la industria} de \cite{KakushadzeYu2016b}. Sin embargo, el c\'odigo puede modificarse f\'acilmente para usar una clasificaci\'on fundamental de la industria\index{clasificaci\'on fundamental de la industria}, tal como GICS (Est\'andar Global de Clasificaci\'on de la Industria)\index{Estandar Global de Clasificacion de la Industria (GICS) @ Est\'andar Global de Clasificaci\'on de la Industria (GICS)}, BICS (Sistema de Bloomberg de Clasificaci\'on de la Industria)\index{Sistema de Bloomberg de Clasificaci\'on de la Industria (BICS)}, SIC (Clasificaci\'on Industrial Est\'andar)\index{Clasificaci\'on Industrial Est\'andar (SIC)}, etc. Un problema con esto es que pr\'acticamente es dif\'icil hacer esto 100\% fuera de la muestra. Sin embargo, ``in-sampleness'' (o el grado en que se encuentra dentro de la muestra) de una clasificaci\'on fundamental de la industria\index{clasificaci\'on fundamental de la industria} -- que es relativamente estable -- por lo general, no plantea un problema grave en tales backtests\index{backtest} dado que las acciones rara vez saltan de las industrias\index{industria}. Adem\'as, tenga en cuenta que las funciones ``externas'' mencionadas anteriormente tienen otros par\'ametros (que se establecen en sus valores predeterminados impl\'icitos en el c\'odigo a continuaci\'on), que se pueden modificar (consulte las referencias anteriores que proporcionan las funciones mencionadas).

{}Finalmente, el c\'odigo calcula internamente las tenencias deseadas\index{tenencias deseadas} y varias caracter\'isticas de rendimiento\index{caracter\'isticas de rendimiento} tales como el P\&L\index{P\&L} total durante el per\'iodo de backtesting\index{periodo de backtesting @ per\'iodo de backtesting}, el retorno anualizado\index{retorno anualizado}, el  ratio de Sharpe anualizado\index{ratio de Sharpe anualizado} y los centavos por acci\'on\index{centavos por acci\'on}. Estas y otras cantidades computadas internamente pueden obtenerse (por ejemplo, a trav\'es de entornos o listas), volcarse en archivos, imprimirse en una pantalla, etc. El c\'odigo es sencillo y puede modificarse seg\'un las necesidades/estrategias espec\'ificas del usuario. Su finalidad es ilustrativa/pedag\'ogica.\\
\\
{\tt{\small
\noindent qrm.backtest <- function (days = 252 * 5, d.r = 21, d.addv = 21,\\
\indent n.addv = 2000, inv.lvl = 2e+07, bnds = .01, incl.cost = F)\\
\{\\
\indent calc.ix <- function(i, d, d.r)\\
\indent \{\\
\indent \indent k1 <- d - i\\
\indent \indent k1 <- trunc(k1 / d.r)\\
\indent \indent ix <- d - k1 * d.r\\
\indent \indent return(ix)\\
\indent \}\\
\\
\indent calc.mv.avg <- function(x, days, d.r)\\
\indent \{\\
\indent \indent y <- matrix(0, nrow(x), days)\\
\indent \indent for(i in 1:days)\\
\indent \indent \indent y[, i] <- rowMeans(x[, i:(i + d.r - 1)])\\
\\
\indent \indent return(y)\\
\indent \}\\
\\
\indent calc.mv.sd <- function(x, days, d.r)\\
\indent \{\\
\indent \indent y <- matrix(0, nrow(x), days)\\
\indent \indent for(i in 1:days)\\
\indent \indent \indent y[, i] <- apply(x[, i:(i + d.r - 1)], 1, sd)\\
\\
\indent \indent return(y)\\
\indent \}\\
\\
\indent read.x <- function(file)\\
\indent \{\\
\indent \indent x <- read.delim(file, as.is = T)\\
\indent \indent x <- as.matrix(x)\\
\indent \indent mode(x) <- "numeric"\\
\indent \indent return(x)\\
\indent \}\\
\\
\indent calc.sharpe <- function (pnl, inv.lvl)\\
\indent \{\\
\indent \indent print(sum(pnl, na.rm = T))\\
\indent \indent print(mean(pnl, na.rm = T) * 252 / inv.lvl * 100)\\
\indent \indent print(mean(pnl, na.rm = T) / sd(pnl, na.rm = T) * sqrt(252))\\
\indent \}\\
\\
\indent ret <- read.x("nrm.ret.txt")\\
\indent open <- read.x("nrm.open.txt")\\
\indent close <- read.x("nrm.close.txt")\\
\indent vol <- read.x("nrm.vol.txt")\\
\indent prc <- read.x("nrm.prc.txt")\\
\\
\indent addv <- calc.mv.avg(vol * close, days, d.addv)\\
\indent ret.close <- log(prc[, -ncol(prc)]/prc[, -1])\\
\indent tr <- calc.mv.sd(ret.close, days, d.r)\\
\\
\indent ret <- ret[, 1:days]\\
\indent prc <- prc[, 1:days]\\
\indent close <- close[, 1:days]\\
\indent open <- open[, 1:days]\\
\indent close1 <- cbind(close[, 1], close[, -ncol(close)])\\
\indent open1 <- cbind(close[, 1], open[, -ncol(open)])\\
\\
\indent pnl <- matrix(0, nrow(ret), ncol(ret))\\
\indent des.hold <- matrix(0, nrow(ret), ncol(ret))\\
\\
\indent for(i in 1:ncol(ret))\\
\indent \{\\
\indent \indent ix <- calc.ix(i, ncol(ret), d.r)\\
\indent \indent if(i == 1)\\
\indent \indent \indent prev.ix <- 0\\
\\
\indent \indent if(ix != prev.ix)\\
\indent \indent \{\\
\indent \indent \indent liq <- addv[, ix]\\
\indent \indent \indent x <- sort(liq)\\
\indent \indent \indent x <- x[length(x):1]\\
\indent \indent \indent take <- liq >= x[n.addv]\\
\\
\indent \indent \indent r1 <- ret.close[take, (ix:(ix + d.r - 1))]\\
\\
\indent \indent \indent \#\#\# ind.list <- qrm.stat.ind.class.all(r1,\\
\indent \indent \indent \#\#\# \indent c(100, 30, 10), iter.max = 100)\\
\\
\indent \indent \indent \#\#\# rr <- qrm.gen.het(r1, ind.list)\\
\\
\indent \indent \indent rr <- qrm.cov.pc(r1)\\
\indent \indent \indent \#\#\# rr <- qrm.erank.pc(r1)\\
\\
\indent \indent \indent cov.mat <- rr\$inv.cov\\
\indent \indent \indent prev.ix <- ix\\
\indent \indent \}\\
\\
\indent \indent w.int <- rep(1, sum(take))\\\
\indent \indent ret.opt <- ret \#\#\# DELAY-0 MEAN-REVERSION\\
\indent \indent \#\#\# ret.opt <- -log(close/open) \#\#\# DELAY-1 MOMENTUM\\
\\
\indent \indent if(incl.cost)\\
\indent \indent \{\\
\indent \indent \indent lin.cost <- tr[take, i] / addv[take, i]\\
\indent \indent \indent lin.cost <- 1e-3 * lin.cost / mean(lin.cost)\\
\indent \indent \}\\
\indent \indent else\\
\indent \indent \indent lin.cost <- 0\\
\\
\indent \indent ret.lin.cost <- ret.opt[take, i]\\
\indent \indent ret.lin.cost <- sign(ret.lin.cost) *\\
\indent \indent \indent pmax(abs(ret.lin.cost) - lin.cost, 0)\\
\\
\indent \indent des.hold[take, i] <- as.vector(bopt.calc.opt(ret.lin.cost, w.int,\\
\indent \indent \indent cov.mat, bnds * liq[take]/inv.lvl, -bnds * liq[take]/inv.lvl))\\
\\
\indent \indent des.hold[take, i] <- -des.hold[take, i] *\\
\indent \indent \indent inv.lvl / sum(abs(des.hold[take, i]))\\
\\
\indent \indent pnl[take, i] <- des.hold[take, i] *\\
\indent \indent \indent (close1[take, i]/open1[take, i] - 1)\\
\\
\indent \indent pnl[take, i] <- pnl[take, i] - abs(des.hold[take, i]) * lin.cost\\
\indent \}\\
\\
\indent des.hold <- des.hold[, -1]\\
\indent pnl <- pnl[, -1]\\
\indent pnl <- colSums(pnl)\\
\indent calc.sharpe(pnl, inv.lvl)\\
\\
\indent trd.vol <- 2 * sum(abs(des.hold/open1[, -1]))\\
\indent cps <- 100 * sum(pnl) / trd.vol\\
\indent print(cps)\\
\}
}}

\newpage

\section{DESCARGOS DE RESPONSABILIDAD}\label{app.B}

{}Donde quiera que el contexto as\'i lo requiera, el g\'enero masculino incluye el femenino y/o el neutro, y la forma singular incluye el plural y {\em viceversa}. El autor de este libro (``Autor'') y sus afiliados incluyendo, sin limitaci\'on, Quantigic$^\circledR$ Solutions LLC (``Afiliados del Autor'' o ``sus Afiliados'') no otorgan de forma impl\'icita ni expresan ninguna garant\'ia o cualquier otra representaci\'on que sea, incluyendo, sin limitaci\'on, garant\'ias impl\'icitas de comercializaci\'on y adecuaci\'on para un prop\'osito particular, en relaci\'on con o con respecto al contenido de este documento, incluido, sin limitaci\'on, cualquier c\'odigo o algoritmo contenido en este documento (``Contenido'').

{}El lector puede usar el Contenido \'unicamente bajo su propio riesgo y el lector no tendr\'a derecho a ning\'un tipo de reclamo contra el Autor o sus Afiliados, y el Autor y sus Afiliados no tendr\'an responsabilidad alguna ante el lector o cualquier tercero por cualquier p\'erdida, gastos, costos de oportunidad, da\~{n}os o cualquier otro efecto adverso relacionado con el uso del Contenido por parte del lector, incluido sin limitaci\'on alguna: cualquier da\~{n}o directo, indirecto, incidental, especial, consecuencial o cualquier otro da\~{n}o en el que incurra el lector, sin importar la causa y bajo cualquier teor\'ia de responsabilidad; cualquier p\'erdida de ganancias (ya sea directa o indirectamente), p\'erdida de buena voluntad o reputaci\'on, p\'erdida de datos, costos de adquisici\'on de bienes o servicios sustitutos, o cualquier otra p\'erdida tangible o intangible; cualquier confianza depositada por el lector en la integridad, exactitud o existencia del Contenido o cualquier otro efecto del uso del Contenido; y cualquier y todas las dem\'as adversidades o efectos negativos que el lector pueda encontrar al usar el Contenido, independientemente de si el Autor o sus Afiliados es o son o deber\'ian haber sido conscientes de tales adversidades o efectos negativos.

{}Cualquier informaci\'on u opini\'on proporcionada en el presente documento tiene solo fines informativos y no est\'a destinada, ni debe ser interpretada, como un consejo de inversi\'on, asesoramiento legal, fiscal o cualquier otro tipo de asesoramiento, o una oferta, solicitud, recomendaci\'on o aprobaci\'on de cualquier estrategia de trading, activo, producto o servicio, o cualquier art\'iculo, libro o cualquier otra publicaci\'on a la que se haga referencia en este documento o cualquiera de sus contenidos.

{}El c\'odigo R incluido en el Ap\'endice \ref{app.A} del presente documento forma parte del c\'odigo R con derechos de autor de Quantigic$^\circledR$ Solutions LLC y se proporciona aqu\'i con el permiso expreso de Quantigic$^\circledR$ Solutions LLC. El propietario de los derechos de autor conserva todos los derechos, t\'itulos e intereses en y a su c\'odigo fuente con los derechos de autor incluido en el Ap\'endice \ref{app.A} de este documento, cualquiera y todos los derechos de autor correspondientes.

\newpage
\phantomsection

\phantomsection
\cleardoublepage
\addcontentsline{toc}{section}{Glosario}
\section*{Glosario}

\noindent
{\bf acci\'on:} \indent
un activo que representa la propiedad fraccional en una corporaci\'on.\\
\\
\noindent
{\bf acci\'on (o participaci\'on, o share en ingl\'es):} \indent
una unidad de participaci\'on accionaria en una corporaci\'on o activo financiero.\\
\\
\noindent
{\bf acci\'on barata:} \indent
una acci\'on que se percibe como subvaluada por alg\'un criterio.\\
\\
\noindent
{\bf acci\'on cara:} \indent
una acci\'on que se percibe como sobrevaluada por alg\'un criterio.\\
\\
\noindent
{\bf acci\'on ordinaria:} \indent
un activo que representa la propiedad en una corporaci\'on que le da derecho a su titular para ejercer control sobre los asuntos de la compa\~{n}\'ia (por ejemplo, a trav\'es de la votaci\'on en la elecci\'on de la junta directiva y la pol\'itica corporativa), con la prioridad m\'as baja (despu\'es de los tenedores de bonos, accionistas preferentes, etc.) para los derechos de los activos de la compa\~{n}\'ia en caso de liquidaci\'on.\\
\\
\noindent
{\bf acciones corporativas:} \indent
eventos iniciados por una empresa que cotiza en la bolsa, como divisiones de acciones, dividendos, fusiones y adquisiciones (M\&A, por sus siglas en ingl\'es), emisiones de derechos, escisiones, etc.\\
\\
\noindent
{\bf acciones en circulaci\'on:} \indent
el n\'umero total de acciones de una empresa en poder de todos sus accionistas.\\
\\
\noindent
{\bf acciones l\'iquidas de los Estados Unidos:} \indent
un subconjunto de acciones listadas en los Estados Unidos que generalmente se definen utilizando los filtros de volumen en d\'olares diario promedio (ADDV, por sus siglas en ingl\'es) y capitalizaci\'on burs\'atil (por ejemplo, las 2,000 acciones m\'as l\'iquidas seg\'un ADDV).\\
\\
\noindent
{\bf accionista:} \indent
un propietario de acciones en una empresa.\\
\\
\noindent
{\bf aceptaci\'on bancaria (BA, por sus siglas en ingl\'es):} \indent
un instrumento de deuda de corto plazo emitido por una empresa y garantizado por un banco comercial.\\
\\
\noindent
{\bf actividad econ\'omica:} \indent
producci\'on, distribuci\'on, intercambio y consumo de bienes y servicios.\\
\\
\noindent
{\bf activo (o valor):} \indent
en finanzas, usualmente un instrumento financiero fungible, negociable con valor monetario.\\
\\
\noindent
{\bf activo de renta fija:} \indent
un instrumento de deuda que genera rendimientos fijos mediante los pagos de intereses.\\
\\
\noindent
{\bf activo en distress:} \indent
un activo (por ejemplo, deuda) de una empresa en distress.\\
\\
\noindent
{\bf activo equivalente a efectivo:} \indent
un activo de inversi\'on a corto plazo altamente l\'iquido con alta calidad crediticia (por ejemplo, REPO).\\
\\
\noindent
{\bf activo especulativo:} \indent
un activo con poco o ning\'un valor intr\'inseco.\\
\\
\noindent
{\bf activo estructurado:} \indent
un instrumento (deuda) de estructura compleja como un CDO o ABS.\\
\\
\noindent
{\bf activo hard-to-borrow:} \indent
un activo en una ``Hard-to-Borrow List'', un registro de inventario utilizado por los corredores para activos que son dif\'iciles de tomar prestados para transacciones de venta en corto debido a la escasez de oferta o a la alta volatilidad.\\
\\
\noindent
{\bf activo h\'ibrido:} \indent
un activo con caracter\'isticas mixtas de dos clases de activos, por ejemplo, un bono convertible.\\
\\
\noindent
{\bf activo libre de riesgo (tambi\'en conocido como activo sin riesgo):}
un activo con una rentabilidad certera futura, por ejemplo, las letras del Tesoro.\\
\\
\noindent
{\bf activo l\'iquido:} \indent
un activo que se puede convertir en efectivo r\'apidamente con costos de transacci\'on m\'inimos.\\
\\
\noindent
{\bf activo sint\'etico (tambi\'en conocido como sint\'etico):} \indent
un instrumento financiero creado (a trav\'es de un portafolio de activos) para replicar (o reproducir de forma aproximada) los mismos flujos de efectivo que otro activo (ejemplos de activos sint\'eticos pueden ser put, call, cono, forward, futuros, etc.).\\
\\
\noindent
{\bf activos tradicionales:} \indent
acciones, bonos, efectivo, bienes ra\'ices y, en algunos casos, tambi\'en divisas y commodities.\\
\\
\noindent
{\bf acuerdo de recompra (tambi\'en conocido como REPO o repo, por sus siglas en ingl\'es):} \indent
un activo equivalente a efectivo que proporciona la liquidez inmediata a una tasa de inter\'es preestablecida por un per\'iodo espec\'ifico de tiempo a cambio de otro activo utilizado como un colateral.\\
\\
\noindent
{\bf acuerdo de recompra inverso:} \indent
un REPO desde el punto de vista del prestamista.\\
\\
\noindent
{\bf ajustar por la media (demeaning en ingl\'es):} \indent
restar de los elementos de una muestra su valor medio a trav\'es de dicha muestra.\\
\\
\noindent
{\bf ala:} \indent
una de las 2 piernas perif\'ericas (por vencimiento en portafolios de bonos y por precio de ejercicio en portafolios de opciones) de un portafolio mariposa.\\
\\
\noindent
{\bf alfa de Jensen:} \indent
un retorno anormal de un activo o portafolio, generalmente calculado como el coeficiente del intercepto en un modelo lineal, en donde los excesos de retornos de dicho activo o portafolio se regresan (serialmente) sobre los excesos de retornos de uno o m\'as portafolios de factores (por ejemplo, MKT).\\
\\
\noindent
{\bf algoritmo ``caja negra'':} \indent
un algoritmo que puede verse en t\'erminos de sus entradas y salidas, sin ning\'un conocimiento de su funcionamiento interno.\\
\\
\noindent
{\bf algoritmo de clustering:} \indent
agrupar objetos (en clusters) en funci\'on de alg\'un criterio (o criterios) de similitud.\\
\\
\noindent
{\bf algoritmo de los k vecinos m\'as cercanos (tambi\'en conocido como KNN o k-NN, por sus siglas en ingl\'es):} \indent
un algoritmo estad\'istico de clasificaci\'on basado en un criterio de similitud, como distancia, \'angulo, etc., entre vectores multidimensionales.\\
\\
\noindent
{\bf algoritmo de stemming de Porter:} \indent
un algoritmo para reducir palabras a su forma base (stemming en ingl\'es).\\
\\
\noindent
{\bf algoritmo de stemming:} \indent
v\'ease {\em algoritmo de stemming de Porter}.\\
\\
\noindent
{\bf alfa:} \indent
siguiendo la jerga del trader com\'un, cualquier ``retorno esperado'' razonable con el que uno quiere realizar trading.\\
\\
\noindent
{\bf alto (tambi\'en conocido como precio alto, o high en ingl\'es):} \indent
el precio m\'aximo alcanzado por una acci\'on (u otro activo) dentro de un d\'ia de trading determinado (o en alg\'un otro intervalo de tiempo); usamos ``precio m\'aximo'' en el texto principal.\\
\\
\noindent
{\bf an\'alisis de componentes principales (PCA, por sus siglas en ingl\'es):}
un procedimiento matem\'atico que transforma un n\'umero de variables (t\'ipicamente, correlacionadas) en un n\'umero (t\'ipicamente, m\'as peque\~{n}o) de variables no correlacionadas (componentes principales), con el primer componente principal representando la mayor variabilidad posible en los datos, y cada componente principal subsiguiente explicando la mayor cantidad posible de la variabilidad restante.\\
\\
\noindent
{\bf an\'alisis de sentimiento (tambi\'en conocido como miner\'ia de opini\'on):}
el uso del procesamiento de lenguaje natural y otras t\'ecnicas computacionales para extraer la informaci\'on de documentos (electr\'onicos) (por ejemplo, tweets), la cual es pertinente a un activo, por ejemplo, para pronosticar la direcci\'on de sus movimientos de precios.\\
\\
\noindent
{\bf an\'alisis fundamental:} \indent
evaluar los activos en funci\'on de datos fundamentales.\\
\\
\noindent
{\bf an\'alisis t\'ecnico:} \indent
una metodolog\'ia para pronosticar la direcci\'on de los precios utilizando los datos hist\'oricos del mercado, principalmente los datos de precio y volumen (comparado con {\em an\'alisis fundamental}).\\
\\
\noindent
{\bf anomal\'ia de baja volatilidad:} \indent
una observaci\'on emp\'irica de que los rendimientos futuros de los portafolios con retornos pasados que presentan baja volatilidad tienden a superar a los de portafolios con retornos pasados que presentan alta volatilidad.\\
\\
\noindent
{\bf anomal\'ia de descuento a plazo (tambi\'en conocida como anomal\'ia de premio a plazo, o rompecabezas de descuento a plazo, o rompecabezas de premio a plazo, o enigma de Fama):} \indent
un acontecimiento emp\'irico por el cual, en promedio, las monedas con tasas de inter\'es altas tienden a apreciarse (en cierto grado) con respecto a las monedas de tasas de inter\'es bajas.\\
\\
\noindent
{\bf anuncios del FOMC:} \indent
anuncios del Comit\'e Federal de Mercado Abierto (FOMC, por sus siglas en ingl\'es), tales como alzas en las tasas de inter\'es.\\
\\
\noindent
{\bf apalancamiento:} \indent
el uso de fondos prestados para comprar un activo.\\
\\
\noindent
{\bf apertura (tambi\'en conocida como precio de apertura, u open en ingl\'es):} \indent
el precio de apertura de una acci\'on en el NYSE (9:30 AM, hora del Este).\\
\\
\noindent
{\bf aprendizaje autom\'atico (ML, por sus siglas en ingl\'es):} \indent
un m\'etodo de an\'alisis de datos que automatiza la construcci\'on de modelos anal\'iticos predictivos basado en la premisa de que los sistemas computacionales pueden ``aprender'' de los datos, identificar patrones y tomar decisiones con m\'inima intervenci\'on humana.\\
\\
\noindent
{\bf arbitraje:} \indent
aprovechar una desviaci\'on (percibida) de los precios justos (esto es tambi\'en conocido como oportunidad de arbitraje) en uno o m\'as activos para obtener una ganancia.\\
\\
\noindent
{\bf arbitraje con TIPS del Tesoro:} \indent
una estrategia de trading que consiste en vender un T-bond y compensar la posici\'on corta mediante un portafolio de r\'eplica de menor costo que consiste en TIPS, swaps de inflaci\'on y STRIPS.\\
\\
\noindent
{\bf arbitraje convertible:} \indent
una estrategia de trading que involucra un bono convertible y acciones del mismo emisor.\\
\\
\noindent
{\bf arbitraje de fusiones (tambi\'en conocido como arbitraje de riesgo):}
una estrategia de trading cuyo objetivo es capturar los rendimientos en excesos generados en acciones corporativas tales como fusiones y adquisiciones (M\&A, por sus siglas en ingl\'es).\\
\\
\noindent
{\bf arbitraje de \'indice (tambi\'en conocido como arbitraje Cash \& Carry en ingl\'es):} \indent
una estrategia de arbitraje que explota las valoraciones err\'oneas entre el precio del \'indice spot y el precio de los futuros (es decir, la base de los futuros del \'indice).\\
\\
\noindent
{\bf arbitraje de intereses cubiertos:} \indent
una estrategia de trading que explota las desviaciones del CIRP.\\
\\
\noindent
{\bf arbitraje de la base del CDS (tambi\'en conocido como arbitraje de CDS ):} \indent
comprar un bono y asegurarlo con un CDS.\\
\\
\noindent
{\bf arbitraje de riesgo:} \indent
v\'ease {\em arbitraje de fusiones}.\\
\\
\noindent
{\bf arbitrage del margen del swap (tambi\'en conocido como arbitraje de swap-spread):} \indent
una estrategia d\'olar-neutral que consiste en una posici\'on larga (corta) en un swap de tasa de inter\'es y una posici\'on corta (larga) en un bono del Tesoro con el mismo vencimiento que el swap.\\
\\
\noindent
{\bf arbitraje estad\'istico (tambi\'en conocido como Stat Arb, o StatArb, por sus siglas en ingl\'es):} \indent
por lo general, se trata de estrategias de trading de corto plazo con universos de trading considerables (por ejemplo, miles de acciones) y se basa en complejas se\~{n}ales estad\'isticas de corte transversal (y serial) de reversi\'on a la media.\\
\\
\noindent
{\bf arbitraje fiscal:} \indent
obtener beneficios de las diferencias en c\'omo se gravan los ingresos, las ganancias de capital, las transacciones, etc.\\
\\
\noindent
{\bf arbitraje fiscal de bonos municipales:} \indent
una estrategia de trading basada en pedir prestado dinero y comprar bonos municipales exentos de impuestos.\\
\\
\noindent
{\bf arbitraje fiscal transfronterizo:} \indent
explotar las diferencias en los reg\'imenes tributarios de dos o m\'as pa\'ises.\\
\\
\noindent
{\bf arbitraje intradiario:} \indent
obtener beneficios de los errores de valoraci\'on intrad\'ia, por ejemplo, en ETFs y acciones.\\
\\
\noindent
{\bf arbitraje multidivisa:} \indent
arbitraje de 4 o m\'as pares de divisas.\\
\\
\noindent
{\bf arbitraje sin riesgo:} \indent
realizar ganancias sin ning\'un riesgo.\\
\\
\noindent
{\bf arbitraje triangular:} \indent
v\'ease {\em arbitraje triangular con FX}.\\
\\
\noindent
{\bf arbitraje triangular con FX:} \indent
arbitrar 3 pares de divisas.\\
\\
\noindent
{\bf \'area estad\'istica metropolitana (MSA, por sus siglas en ingl\'es):}
un \'area central que contiene un n\'ucleo de poblaci\'on sustancial, junto con las comunidades adyacentes que tienen un alto grado de integraci\'on econ\'omica y social con dicho n\'ucleo.\\
\\
\noindent
{\bf arreglar y vender:} \indent
una estrategia con bienes ra\'ices.\\
\\
\noindent
{\bf asignaci\'on de activos:} \indent
asignar las ponderaciones (porcentajes de asignaci\'on) a los activos en un portafolio, generalmente basados en consideraciones de riesgos y retornos.\\
\\
\noindent
{\bf asignaci\'on de capital:} \indent
v\'ease {\em asignaci\'on de activos}.\\
\\
\noindent
{\bf asignaci\'on din\'amica de activos:} \indent
ajustar con frecuencia las asignaciones de activos en un portafolio de acuerdo con las condiciones cambiantes del mercado.\\
\\
\noindent
{\bf asignaci\'on t\'actica de activos:} \indent
una estrategia de inversi\'on din\'amica que ajusta activamente las ponderaciones de asignaci\'on de los activos en un portafolio.\\
\\
\noindent
{\bf asimetr\'ia (skewness en ingl\'es):} \indent
una medida de asimetr\'ia en una distribuci\'on de probabilidad, definida como el valor medio de la potencia c\'ubica de las desviaciones de la media dividido por la potencia c\'ubica de la desviaci\'on est\'andar.\\
\\
\noindent
{\bf asimetr\'ia de volatilidad:} \indent
un suceso emp\'irico por el cual, con todo lo dem\'as igual, la volatilidad impl\'icita de las opciones de venta es mayor que la de las opciones de compra.\\
\\
\noindent
{\bf ask (tambi\'en conocido como precio ask, oferta o precio de oferta):}
el precio al cual un vendedor est\'a dispuesto (ofreciendo) a vender.\\
\\
\noindent
{\bf backtest:} \indent
una simulaci\'on del rendimiento de una estrategia utilizando datos hist\'oricos.\\
\\
\noindent
{\bf backtest con retraso-$d$:} \indent
un backtest en el que todas las cantidades utilizadas para establecer o liquidar las posiciones simuladas en un momento dado $t$ se calculan utilizando las cantidades hist\'oricas de al menos $d$ d\'ias de trading antes de $t$.\\
\\
\noindent
{\bf backtest fuera de la muestra (out-of-sample en ingl\'es):} \indent
un backtest en el que todas las cantidades utilizadas para establecer o liquidar las posiciones simuladas correspondientes a un momento dado $t$ se calculan utilizando las cantidades hist\'oricas de tiempos anteriores a $t$.\\
\\
\noindent
{\bf backwardation:} \indent
cuando la curva de futuros (estructura temporal) presenta una pendiente descendente.\\
\\
\noindent
{\bf bajo (tambi\'en conocido como precio bajo, o low en ingl\'es):} \indent
el precio m\'inimo alcanzado por una acci\'on (u otro activo) dentro de un d\'ia de trading determinado (o en alg\'un otro intervalo de tiempo); usamos ``precio m\'inimo'' en el texto principal.\\
\\
\noindent
{\bf bancarrota:} \indent
un estado legal (impuesto por una orden judicial) de una compa\~{n}\'ia que no puede pagar las deudas a sus acreedores.\\
\\
\noindent
{\bf banda (strip en ingl\'es):} \indent
una estrategia de trading con opciones.\\
\\
\noindent
{\bf barbell:} \indent
un portafolio de bonos que consiste en bonos con solo dos vencimientos (generalmente, cortos y largos).\\
\\
\noindent
{\bf base del CDS:} \indent
el margen del CDS menos el margen de rendimiento del bono asegurado.\\
\\
\noindent
{\bf base de futuros:} \indent
el precio de futuros menos el precio spot del activo subyacente.\\
\\
\noindent
{\bf beta de mercado:} \indent
una medida de la volatilidad (riesgo sistem\'atico) de un activo o portafolio en comparaci\'on con el mercado general.\\
\\
\noindent
{\bf bid (tambi\'en conocido como precio bid):} \indent
el precio al que el comprador est\'a dispuesto a comprar.\\
\\
\noindent
{\bf bienes ra\'ices (tambi\'en conocidos como bienes inmobiliarios):} \indent
bienes inmuebles tangibles, incluidos terrenos, estructuras construidas en \'el, etc.\\
\\
\noindent
{\bf bienes ra\'ices comerciales:} \indent
una propiedad inmobiliaria utilizada para fines comerciales (en lugar de espacio para vivir), por ejemplo, centros comerciales, tiendas minoristas, espacio de oficinas, etc.\\
\\
\noindent
{\bf Bitcoin (BTC):} \indent
la primera moneda digital descentralizada (criptomoneda) del mundo.\\
\\
\noindent
{\bf blockchain:} \indent
un libro de registro distribuido, que mantiene un registro de todas las transacciones y consiste en una cadena secuencial de bloques (que contiene los registros de las transacciones), que est\'an vinculados mediante la criptograf\'ia y un marcador del tiempo.\\
\\
\noindent
{\bf bono:} \indent
un instrumento de renta fija, con promesa de recibir alg\'un monto (principal) en alg\'un momento futuro $T$ (vencimiento), y posiblemente algunos montos menores (pagos de cupones) en algunos momentos antes de $T$.\\
\\
\noindent
{\bf bono con cup\'on:} \indent
un bono que realiza algunos pagos peri\'odicos de cupones antes del vencimiento.\\
\\
\noindent
{\bf bono con cup\'on cero:} \indent
v\'ease {\em bono de descuento}.\\
\\
\noindent
{\bf bono con cup\'on fijo (tambi\'en conocido como bono con tasa de c\'upon fija):} \indent
un bono con una tasa de cup\'on fija (en oposici\'on a una variable).\\
\\
\noindent
{\bf bono con cup\'on flotante (tambi\'en conocido como bono con cup\'on de tasa variable, o bono con cup\'on variable, o bono con cup\'on de tasa variable):} \indent
un bono con una tasa de cup\'on variable (en oposici\'on a una fija).\\
\\
\noindent
{\bf bono con cup\'on variable:} \indent
v\'ease {\em bono con cup\'on flotante}.\\
\\
\noindent
{\bf bono convertible:} \indent
un activo h\'ibrido con una opci\'on integrada para convertir a un n\'umero preestablecido (ratio de conversi\'on) de las acciones del emisor cuando, por ejemplo, el precio de la acci\'on alcanza un nivel preestablecido (precio de conversi\'on).\\
\\
\noindent
{\bf Bonos de Alto Rendimiento (tambi\'en conocidos como bonos basura):}
los bonos con las calificaciones crediticias del S\&P por debajo de BBB-.\\
\\
\noindent
{\bf bono de descuento (tambi\'en conocido como bono con cup\'on cero):}
un bono que paga solo su principal al vencimiento pero que no realiza ning\'un pago de cup\'on.\\
\\
\noindent
{\bf Bonos de Grado de Inversi\'on (tambi\'en conocidos como bonos IG, por sus siglas en ingl\'es):}
los bonos con las calificaciones crediticias del S\&P desde AAA a AA- (calidad crediticia alta) y desde A+ a BBB- (calidad crediticia media).\\
\\
\noindent
{\bf bono del Tesoro (tambi\'en conocido como T-bond en ingl\'es):} \indent
una obligaci\'on de deuda emitida por el Tesoro de los Estados Unidos con un vencimiento superior a 10 a\~{n}os.\\
\\
\noindent
{\bf bono entregable:} \indent
un bono en la cesta de entrega de un contrato de futuros de tasa de inter\'es.\\
\\
\noindent
{\bf bono municipal:} \indent
un bono emitido por un gobierno/territorio local o su agencia.\\
\\
\noindent
{\bf bono municipal exento de impuestos:} \indent
por ejemplo, los bonos municipales que no est\'an sujetos a los impuestos federales sobre la renta (sobre el inter\'es ganado) en los EE.UU.\\
\\
\noindent
{\bf bono no entregable:} \indent
un bono que no est\'a en la cesta de entrega de un contrato de futuros de tasa de inter\'es.\\
\\
\noindent
{\bf Booleano:} \indent
una variable binaria con solo dos valores posibles, {\tt{\small VERDADERO}} y {\tt{\small FALSO}} ({\tt{\small TRUE}} y {\tt{\small FALSE}} en ingl\'es).\\
\\
\noindent
{\bf brecha de producci\'on:} \indent
la diferencia entre la producci\'on real de una econom\'ia y su producci\'on potencial m\'axima como un porcentaje del GDP.\\
\\
\noindent
{\bf Btu:} \indent
la unidad t\'ermica brit\'anica, aproximadamente 1,055 Julios.\\
\\
\noindent
{\bf bullet:} \indent
un portafolio de bonos donde todos los bonos tienen el mismo vencimiento.\\
\\
\noindent
{\bf burbuja (tambi\'en conocida como burbuja econ\'omica, de activos, especulativa, de mercado, de precio o financiera):} \indent
un activo que cotiza a precios fuertemente inflados en comparaci\'on con su valor intr\'inseco.\\
\\
\noindent
{\bf burbuja especulativa:} \indent
v\'ease {\em burbuja}.\\
\\
\noindent
{\bf ca\'ida del mercado (o crash (en ingl\'es) del mercado):} \indent
una repentina disminuci\'on sustancial en los precios de los activos a trav\'es de su corte transversal.\\
\\
\noindent
{\bf calificaci\'on:} \indent
v\'ease {\em calificaci\'on crediticia}.\\
\\
\noindent
{\bf calificaci\'on crediticia (para bonos):} \indent
una medida de la solvencia de los bonos corporativos y gubernamentales (por ejemplo, las calificaciones crediticias del S\&P AAA, AA+, AA, AA-, A+, A, A-, BBB+, BBB, BBB-, BB+, BB, BB-, B+, B, B-, CCC+, CCC, CCC-, CC, C, D).\\
\\
\noindent
{\bf call casada:} \indent
v\'ease {\em call protectora}.\\
\\
\noindent
{\bf call cubierta:} \indent
v\'ease {\em estrategia de compra-lanzamiento}.\\
\\
\noindent
{\bf call desnuda:} \indent
una opci\'on call corta independiente (sin ninguna otra posici\'on).\\
\\
\noindent
{\bf call protectora (tambi\'en conocida como call casada, o put sint\'etico):}
la cobertura de una posici\'on corta en acciones con una posici\'on larga en una opci\'on call.\\
\\
\noindent
{\bf canal:} \indent
un rango/banda, limitado por un techo y un piso, dentro de los cuales el precio de las acciones fluct\'ua.\\
\\
\noindent
{\bf Canal de Donchian:} \indent
una definici\'on de canal com\'unmente utilizada en estrategias de trading de canal.\\
\\
\noindent
{\bf cantidad de corte transversal:} \indent
una cantidad (por ejemplo, media, desviaci\'on est\'andar, etc.) calculada a trav\'es de un conjunto de activos (por ejemplo, acciones en un portafolio) en lugar de en serie (es decir, a trav\'es de una serie de tiempo para cada activo).\\
\\
\noindent
{\bf cantidad hist\'orica:} \indent
una cantidad (por ejemplo, correlaci\'on, varianza, volatilidad, retorno, etc.) calculada en base a datos hist\'oricos.\\
\\
\noindent
{\bf cantidad no ajustada:} \indent
un precio o volumen no ajustado por splits y dividendos.\\
\\
\noindent
{\bf cantidad serial:} \indent
una cantidad (por ejemplo, media, desviaci\'on est\'andar, etc.) computada serialmente (es decir, a trav\'es de las series de tiempo para cada activo) en oposici\'on a aquellas de corte transversal (es decir, a trav\'es de un conjunto de activos).\\
\\
\noindent
{\bf capa:} \indent
v\'ease {\em capa de entrada, capa oculta, capa de salida}.\\
\\
\noindent
{\bf capa de entrada:} \indent
en una red neuronal artificial, la capa de nodos (neuronas artificiales) que procesa los datos de entrada.\\
\\
\noindent
{\bf capa de salida:} \indent
en una red neuronal artificial, la capa de nodos (neuronas artificiales) que genera los datos de salida (el resultado).\\
\\
\noindent
{\bf capas ocultas:} \indent
en una red neuronal artificial, las capas intermedias de nodos (neuronas artificiales) entre la capa de entrada y la capa de salida.\\
\\
\noindent
{\bf capitalizaci\'on burs\'atil (tambi\'en conocida como capitalizaci\'on de mercado):} \indent
el valor de mercado de las acciones en circulaci\'on de una empresa.\\
\\
\noindent
{\bf Cap\'itulo 11:} \indent
un cap\'itulo del T\'itulo 11, el c\'odigo de bancarrota de los Estados Unidos.\\
\\
\noindent
{\bf caracter\'istica (en aprendizaje autom\'atico):} \indent
un predictor, una variable de entrada.\\
\\
\noindent
{\bf caracter\'isticas de rendimiento:} \indent
para un portafolio o estrategia, caracter\'isticas tales como el retorno sobre el capital, el ratio de Sharpe, los centavos por acci\'on, la reducci\'on m\'axima, etc.\\
\\
\noindent
{\bf carry (tambi\'en conocido como costo de carry):} \indent
un retorno (positivo o negativo) de mantener un activo.\\
\\
\noindent
{\bf carry alto-menos-bajo:} \indent
una estrategia de trading de FX basada en la anomal\'ia de descuento a plazo.\\
\\
\noindent
{\bf carry trade (tambi\'en conocido como estrategia de carry):} \indent
una estrategia basada en ganar una diferencia entre pedir prestado un activo con carry bajo y prestar un activo con carry alto.\\
\\
\noindent
{\bf carry trade sobre el d\'olar:} \indent
una estrategia de trading de FX.\\
\\
\noindent
{\bf CDS de nombre individual:} \indent
un CDS en una sola entidad de referencia.\\
\\
\noindent
{\bf centavos por acci\'on (CPS, por sus siglas en ingl\'es):} \indent
el P\&L realizado en centavos (en lugar de d\'olares) dividido por el n\'umero total de acciones negociadas (que incluye tanto las operaciones del establecimiento como las de la liquidaci\'on).\\
\\
\noindent
{\bf certificado de dep\'osito bancario (tambi\'en conocido como certificado de dep\'osito, o CD, por sus siglas en ingl\'es):} \indent
un certificado de ahorro (un pagar\'e emitido por un banco) con una fecha de vencimiento y una tasa de inter\'es fijas.\\
\\
\noindent
{\bf cesta:} \indent
una cartera de activos combinada con ciertas ponderaciones.\\
\\
\noindent
{\bf cesta de entrega:} \indent
en futuros de tasa de inter\'es, el conjunto de bonos que se pueden entregar en la fecha de entrega.\\
\\
\noindent
{\bf cesta de entrega de los futuros:} \indent
v\'ease {\em cesta de entrega}.\\
\\
\noindent
{\bf cesta del \'indice:} \indent
un portafolio del \'indice.\\
\\
\noindent
{\bf cesta incompleta:} \indent
un subconjunto de un portafolio (o una cesta de activos) que idealmente se negociar\'ia, por ejemplo, en el arbitraje de \'indice.\\
\\
\noindent
{\bf cierre (tambi\'en conocido como precio de cierre, o close en ingl\'es):}
el precio de cierre de una acci\'on al cierre del NYSE (4:00 PM, hora del Este).\\
\\
\noindent
{\bf clase:} \indent
en aprendizaje autom\'atico, uno de los posibles resultados previstos de un algoritmo de aprendizaje autom\'atico.\\
\\
\noindent
{\bf clase de activo:} \indent
un grupo de activos con caracter\'isticas similares.\\
\\
\noindent
{\bf clase predicha:} \indent
en aprendizaje autom\'atico, el resultado predicho por un algoritmo.\\
\\
\noindent
{\bf clasificaci\'on de la industria:} \indent
una taxonom\'ia de empresas (acciones) basada en alg\'un criterio (o criterios) de similitud, por ejemplo, la principal fuente de ingresos de una empresa, la relaci\'on de los retornos de las acciones hist\'oricamente, etc.\\
\\
\noindent
{\bf clasificaci\'on binaria de la industria:} \indent
una clasificaci\'on de la industria donde cada empresa pertenece a una y solo una subindustria, industria, sector, etc.\\
\\
\noindent
{\bf clasificaci\'on estad\'istica de la industria:} \indent
un agrupamiento multinivel de empresas basado en t\'ecnicas puramente estad\'isticas, por ejemplo, agrupamiento basado en la distancia de los retornos de las empresas (comparado con {\em clasificaci\'on fundamental de la industria}).\\
\\
\noindent
{\bf clasificaci\'on fundamental de la industria:} \indent
una clasificaci\'on industrial de las empresas (en sectores, industrias, subindustrias, etc.) basada en datos fundamentales/econ\'omicos, tales como productos y servicios, distintas fuentes de ingresos, proveedores, competidores, socios, etc. (comparado con {\em clasificaci\'on estad\'istica de la industria}).\\
\\
\noindent
{\bf cobertura:} \indent
una inversi\'on (generalmente, a trav\'es de una posici\'on que compensa otra posici\'on en un activo relacionado) para reducir el riesgo de perder dinero en una posici\'on existente.\\
\\
\noindent
{\bf cobertura cruzada:} \indent
administrar el riesgo para un activo tomando una posici\'on opuesta (con alg\'un ratio de cobertura) en otro activo (o su derivado, por ejemplo, futuros), en donde los dos est\'an correlacionados positivamente y tienen movimientos similares.\\
\\
\noindent
{\bf cobertura de Delta:} \indent
cubrir una posici\'on larga (corta) en un activo derivado con una posici\'on corta (larga) en el activo subyacente con el ratio de cobertura igual al Delta del activo derivado.\\
\noindent
\\
{\bf cobertura de Gamma:} \indent
una estrategia de cobertura de opciones para eliminar o reducir la exposici\'on causada por los cambios en el Delta de una cartera de opciones resultantes de los movimientos de los precios del activo subyacente.\\
\\
\noindent
{\bf cobertura de la duraci\'on:} \indent
la cobertura del riesgo de duraci\'on (es decir, riesgo de tasa de inter\'es) con swaps de tasa de inter\'es o futuros de tasa de inter\'es.\\
\\
\noindent
{\bf cobertura del \'indice:} \indent
cubrir una posici\'on (por ejemplo, un tramo de un CDO) con el \'indice pertinente.\\
\\
\noindent
{\bf cobertura imperfecta:} \indent
cuando una cobertura se deshace en cierto grado (por ejemplo, debido a movimientos de los precios del activo subyacente).\\
\\
\noindent
{\bf c\'odigo fuente (tambi\'en conocido como c\'odigo):} \indent
c\'odigo escrito en alg\'un lenguaje de programaci\'on de computadora.\\
\\
\noindent
{\bf coeficiente de la regresi\'on:} \indent
la pendiente de una variable independiente en una regresi\'on lineal.\\
\\
\noindent
{\bf colateral:} \indent
algo de valor prometido como garant\'ia de pago de un pr\'estamo, que se perder\'a si el prestatario no cumple.\\
\\
\noindent
{\bf collar (tambi\'en conocido como cerca):} \indent
una estrategia de trading con opciones.\\
\\
\noindent
{\bf colocaci\'on:} \indent
la etapa inicial en el proceso de lavado de dinero, por medio del cual se introducen fondos ilegales en la econom\'ia legal por medios fraudulentos.\\
\\
\noindent
{\bf combo (para opciones):} \indent
un tipo de estrategias de trading con opciones.\\
\\
\noindent
{\bf completar:} \indent
cuando se completa una orden de compra o venta de un activo, con una finalizaci\'on parcial (por ejemplo, solo se ``llenan'' 100 acciones de una orden de compra de 200 acciones) conocida como ejecuci\'on parcial.\\
\\
\noindent
{\bf completar o matar una orden l\'imite (tambi\'en conocido como FOK, por sus siglas en ingl\'es):} \indent
una orden de l\'imite para comprar o vender acciones que debe ejecutarse de forma inmediata y completa o no se realiza (no se permiten ejecuciones parciales).\\
\\
\noindent
{\bf Compromisos de los Comerciantes (COT, por sus siglas en ingl\'es):}
los informes semanales proporcionados por la CFTC.\\
\\
\noindent
{\bf commodity:} \indent
una materia prima (por ejemplo, oro, plata, petr\'oleo, cobre) o un producto agr\'icola (por ejemplo, trigo, soja, arroz) que se puede comprar y vender.\\
\\
\noindent
{\bf commodity f\'isico:} \indent
el commodity real (por ejemplo, cobre) que se entrega al vencimiento a un comprador de contratos de futuros de commodities.\\
\\
\noindent
{\bf compa\~{n}\'ia adquirente:} \indent
la compa\~{n}\'ia que compra otra compa\~{n}\'ia (compa\~{n}\'ia objetivo) en una adquisici\'on corporativa.\\
\\
\noindent
{\bf componente principal:} \indent
para una matriz cuadrada sim\'etrica, un eigenvector de la misma normalizado de tal manera que la suma de los cuadrados de sus componentes es igual a 1, con diferentes componentes principales ordenados en orden descendente por los eigenvalores correspondientes.\\
\\
\noindent
{\bf composici\'on continua:} \indent
un l\'imite matem\'atico idealizado de composici\'on, en donde el n\'umero de per\'iodos de composici\'on $n$ va hasta el infinito, la longitud $\delta$ de cada per\'iodo de composici\'on va a cero, y el producto $n\times \delta$ se mantiene fijo y finito.\\
\\
\noindent
{\bf composici\'on peri\'odica:} \indent
una composici\'on con per\'iodos de capitalizaci\'on iguales, por ejemplo, la composici\'on trimestral, semestral o anual.\\
\\
\noindent
{\bf comprador de protecci\'on:} \indent
un comprador de seguros.\\
\\
\noindent
{\bf comprar y mantener activos/inversiones:} \indent
un activo/inversi\'on para una estrategia pasiva a largo plazo donde el inversionista mantiene una posici\'on larga independientemente de las fluctuaciones a corto plazo en el mercado.\\
\\
\noindent
{\bf computaci\'on fuera de muestra (out-of-sample en ingl\'es):} \indent
un c\'alculo en el que todas las cantidades que se utilizan para fines de pron\'osticar en cualquier momento simulado $t$ se calculan utilizando las cantidades hist\'oricas de tiempos anteriores a $t$.\\
\\
\noindent
{\bf condici\'on de no arbitraje libre de riesgo:} \indent
una condici\'on que garantiza que no se puedan obtener ganancias libres de riesgo mediante una estrategia de trading (al menos, en exceso a la tasa libre de riesgo).\\
\\
\noindent
{\bf c\'ondor:} \indent
un tipo de estrategias de opciones.\\
\\
\noindent
{\bf c\'ondor de hierro:} \indent
un tipo de estrategias de trading con opciones.\\
\\
\noindent
{\bf cono:} \indent
una estrategia de trading con opciones.\\
\\
\noindent
{\bf constituyentes del \'indice:} \indent
los activos en el portafolio del \'indice.\\
\\
\noindent
{\bf contango:} \indent
cuando la curva de futuros (estructura temporal) presenta una pendiente positiva.\\
\\
\noindent
{\bf contraparte:} \indent
la otra parte que participa en una transacci\'on financiera.\\
\\
\noindent
{\bf contrato al mes m\'as cercano:} \indent
v\'ease {\em futuro al mes m\'as cercano}.\\
\\
\noindent
{\bf contrato del primer mes:} \indent
v\'ease {\em futuro del primer mes}.\\
\\
\noindent
{\bf conversi\'on larga} \indent
una estrategia de trading con opciones.\\
\\
\noindent
{\bf convexidad (para bonos):} \indent
una medida de la dependencia no lineal de los precios de los bonos a los cambios en las tasas de inter\'es, que implica la segunda derivada del precio de los bonos con respecto a las tasas de inter\'es.\\
\\
\noindent
{\bf correa (strap en ingl\'es):} \indent
una estrategia de trading con opciones.\\
\\
\noindent
{\bf correlaci\'on:} \indent
una medida de cu\'an cerca se mueven dos activos entre s\'i, definida como la covarianza de sus retornos dividida por un producto de las desviaciones est\'andar de dichos retornos.\\
\\
\noindent
{\bf correlaci\'on serial:} \indent
una correlaci\'on por pares entre dos activos calculada en funci\'on de sus series de tiempo de los retornos hist\'oricos.\\
\\
\noindent
{\bf costos de trading (tambi\'en conocidos como costos de transacci\'on):}
los costos asociados con el trading de activos, incluidos (seg\'un corresponda) comisiones de operaciones, comisiones de corretaje, comisiones de la SEC, slippage, etc.\\
\\
\noindent
{\bf covarianza:} \indent
un valor medio del producto de las desviaciones de los retornos de dos activos de sus valores medios respectivos.\\
\\
\noindent
{\bf cr\'edito fiscal:} \indent
v\'ease {\em imputaci\'on de dividendos}.\\
\\
\noindent
{\bf criptoactivos:} \indent
criptomonedas y activos digitales similares.\\
\\
\noindent
{\bf criptograf\'ia:} \indent
la construcci\'on y el an\'alisis de protocolos que evitan que terceros lean mensajes privados.\\
\\
\noindent
{\bf criptomoneda:} \indent
un medio de intercambio digital que utiliza criptograf\'ia (por ejemplo, el BTC).\\
\\
\noindent
{\bf crisis financiera:} \indent
cuando algunos activos financieros pierden repentinamente una gran parte de su valor nominal.\\
\\
\noindent
{\bf cuantil:} \indent
cada una de las $n$ partes (aproximadamente) iguales de una muestra (por ejemplo, muestra de datos), en donde $n>1$.\\
\\
\noindent
{\bf cuenta al margen:} \indent
una cuenta de corretaje en la que el corredor presta dinero en efectivo al cliente para comprar activos.\\
\\
\noindent
{\bf cuerpo:} \indent
la pierna media (por vencimiento en portafolios de bonos y por precio de ejercicio en portafolios de opciones) de un portafolio mariposa.\\
\\
\noindent
{\bf cum-dividendo:} \indent
cuando el comprador de acciones tiene el derecho a recibir un dividendo que ha sido declarado, pero no pagado (comparado con {\em ex-dividendo}).\\
\\
\noindent
{\bf cuna:} \indent
una estrategia de trading con opciones.\\
\\
\noindent
{\bf curva de futuros (tambi\'en conocida como estructura de plazo de futuros):} \indent
la dependencia de los precios de los futuros al tiempo a la entrega.\\
\\
\noindent
{\bf curva de rendimientos (tambi\'en conocida como estructura temporal):} \indent
la dependencia de las tasas de inter\'es o de los rendimientos de los bonos a los vencimientos.\\
\\
\noindent
{\bf curva del Tesoro:} \indent
la curva de rendimientos de los valores del Tesoro.\\
\\
\noindent
{\bf curva-neutralidad:} \indent
la neutralidad aproximada de una cartera de bonos ante una peque\~{n}a inclinaci\'on y aplanamiento de la curva de rendimientos.\\
\\
\noindent
{\bf curvatura:} \indent
en una curva de rendimientos, el cambio en la pendiente de la misma en funci\'on de la madurez.\\
\\
\noindent
{\bf datos de entrenamiento (tambi\'en conocidos como conjunto de datos de formaci\'on):} \indent
en aprendizaje autom\'atico, un conjunto de datos de pares de entrada-salida conocidos de antemano, que se utilizan para entrenar un algoritmo de aprendizaje autom\'atico.\\
\\
\noindent
{\bf datos de precios:} \indent
los datos hist\'oricos y en tiempo real que contienen precios, vol\'umenes negociados y otras cantidades relacionadas (v\'ease {\em datos del mercado}).\\
\\
\noindent
{\bf datos de sentimiento:} \indent
los datos textuales utilizados en el an\'alisis de opiniones (por ejemplo, el contenido de los tweets).\\
\\
\noindent
{\bf datos del mercado:} \indent
los precios y datos relacionados con el trading de un activo financiero, reportado por una bolsa de comercio (o similar).\\
\\
\noindent
{\bf datos econ\'omicos:} \indent
los datos (t\'ipicamente, series de tiempo) que pertenecen a una econom\'ia.\\
\\
\noindent
{\bf datos fundamentales:} \indent
los datos pertenecientes a los fundamentos de acciones u otros activos, incluidas series de tiempo y/o datos de corte transversal.\\
\\
\noindent
{\bf decaimiento de Theta:} \indent
la disminuci\'on en el tiempo del valor de una opci\'on (u otro activo) a medida que el tiempo se acerca al vencimiento.\\
\\
\noindent
{\bf decil:} \indent
cada una de las 10 partes (aproximadamente) iguales de una muestra (por ejemplo, muestra de datos).\\
\\
\noindent
{\bf d\'eficit de flujo de efectivo:} \indent
la cantidad por la cual una obligaci\'on o un pasivo financiero excede la cantidad de efectivo (o, m\'as generalmente, fondos l\'iquidos) que est\'a disponible.\\
\\
\noindent
{\bf Delta:} \indent
la primera derivada del valor de un activo derivado (por ejemplo, una opci\'on) con respecto al precio del activo subyacente.\\
\\
\noindent
{\bf derechos de control:} \indent
los derechos legales otorgados a un inversionista (por ejemplo, un accionista que posee acciones ordinarias), como el derecho a transferir acciones, recibir informaci\'on financiera regular y precisa, votar sobre temas espec\'ificos en la compa\~{n}\'ia, etc.\\
\\
\noindent
{\bf deriva:} \indent
el cambio medio en una cantidad dependiente del tiempo durante un per\'iodo de tiempo, es decir, un promedio serial.\\
\\
\noindent
{\bf derivado (tambi\'en conocido como contrato derivado, o reclamaci\'on contingente):} \indent
un activo (por ejemplo, una opci\'on) cuyo pago futuro depende del valor de su activo subyacente (por ejemplo, acciones) y est\'a supeditado a alg\'un evento futuro incierto.\\
\\
\noindent
{\bf derivado crediticio:} \indent
contratos financieros (por ejemplo, CDS) que permiten a las partes transferir o recibir exposici\'on al riesgo de cr\'edito.\\
\\
\noindent
{\bf derivado del clima:} \indent
un derivado (por ejemplo, una opci\'on o un futuro) sobre un \'indice sint\'etico del clima.\\
\\
\noindent
{\bf descenso gradiente estoc\'astico (SGD, por sus siglas en ingl\'es):} \indent
un m\'etodo iterativo para optimizar una funci\'on objetivo que es diferenciable.\\
\\
\noindent
{\bf descuento a plazo:} \indent
cuando el forward de la tasa de FX es menor que la tasa de FX spot.\\
\\
\noindent
{\bf desplazamiento paralelo:} \indent
cuando en la curva de rendimientos, todas las tasas de inter\'es para todos los vencimientos cambian en la misma cantidad.\\
\\
\noindent
{\bf desprendimiento (tambi\'en conocido como punto de desprendimiento):}
el porcentaje de p\'erdida del portafolio subyacente en la que un tramo de un CDO (obligaci\'on de deuda garantizada) pierde todo su valor.\\
\\
\noindent
{\bf desviaci\'on est\'andar:} \indent
la ra\'iz cuadrada de la varianza.\\
\\
\noindent
{\bf desviaci\'on est\'andar m\'ovil:} \indent
en una serie de tiempo, una desviaci\'on est\'andar (posiblemente computada con algunas ponderaciones no triviales) sobre un intervalo de tiempo de longitud fija, en donde el tiempo m\'as reciente en dicho intervalo puede tomar varios valores en la serie de tiempo.\\
\\
\noindent
{\bf desviaci\'on est\'andar m\'ovil simple:} \indent
una desviaci\'on est\'andar m\'ovil sin suprimir las contribuciones pasadas (comparado con {\em desviaci\'on est\'andar m\'ovil exponencial o EMSD}).\\
\\
\noindent
{\bf deuda en distress:} \indent
v\'ease {\em activo en distress}.\\
\\
\noindent
{\bf d\'ias de anuncios:} \indent
los d\'ias con algunos anuncios econ\'omicos importantes tales como los del FOMC (comparado con {\em d\'ias sin anuncio}).\\
\\
\noindent
{\bf d\'ias de trading:} \indent
por lo general, los d\'ias en que el NYSE est\'a abierto.\\
\\
\noindent
{\bf d\'ias sin anuncios:} \indent
los d\'ias sin ning\'un anuncio econ\'omico importante, como, por ejemplo, los anuncios del FOMC (comparado con {\em d\'ias de anuncios}).\\
\\
\noindent
{\bf diferencia bid-ask:} \indent
el precio ask menos el precio bid.\\
\\
\noindent
{\bf diferencial ratio:} \indent
un tipo de estrategias con opciones.\\
\\
\noindent
{\bf diferencia temporal (en la curva de rendimientos):} \indent
un diferencial de tasas de inter\'es correspondientes a dos vencimientos diferentes.\\
\\
\noindent
{\bf diferencial (o margen, o spread en ingl\'es):} \indent
la diferencia entre dos cantidades, o un portafolio que consta de dos (o m\'as) piernas compuestas por el mismo tipo de activos, diferentes solo por una o m\'as cantidades espec\'ificas (por ejemplo, precio de ejercicio o precio de ejercicio y vencimiento).\\
\\
\noindent
{\bf diferencial ajustado por opciones (OAS, por sus siglas en ingl\'es):}
un desplazamiento paralelo en la curva del Tesoro (o alguna otra curva de rendimientos de referencia) que hace que el precio de un activo calculado en base a un modelo de la valuaci\'on coincida con su valor de mercado, con el objetivo de tener en cuenta las opciones embebidas del activo.\\
\\
\noindent
{\bf diferencial cuarc:} \indent
el an\'alogo del diferencial spark y del diferencial oscuro para las centrales el\'ectricas nucleares.\\
\\
\noindent
{\bf diferencial de calendario (para futuros):} \indent
comprar (vender) futuros en un mes cercano y vender (comprar) futuros en plazos diferidos.\\
\\
\noindent
{\bf diferencial de calendario (para opciones):} \indent
comprar una opci\'on con vencimiento m\'as largo (call o put) y vender una opci\'on con vencimiento m\'as corto (del mismo tipo, por el mismo subyacente y con el mismo precio de ejercicio).\\
\\
\noindent
{\bf diferencial de futuros:} \indent
v\'ease {\em diferencial de calendario (para futuros)}.\\
\\
\noindent
{\bf diferencial del tramo de un CDO:} \indent
para lograr un MTM nulo de un tramo de un CDO, el valor de la pierna de incumplimiento del tramo dividido por su duraci\'on arriesgada.\\
\\
\noindent
{\bf diferencial diagonal:} \indent
una estrategia de trading de opciones.\\
\\
\noindent
{\bf diferencial horizontal (tambi\'en conocido como diferencial de tiempo):}
v\'ease {\em diferencial de calendario}.\\
\\
\noindent
{\bf diferencial inverso:} \indent
un tipo de estrategias con opciones.\\
\\
\noindent
{\bf diferencial oscuro:} \indent
la diferencia entre el precio mayorista de la electricidad y el precio del carb\'on requerido para producirla por una central el\'ectrica de carb\'on.\\
\\
\noindent
{\bf diferencial spark:} \indent
la diferencia entre el precio mayorista de la electricidad y el precio del gas natural requerido para producirla por una central el\'ectrica que produce con gas.\\
\\
\noindent
{\bf diferencial vertical:} \indent
una estrategia con opciones que involucra todas las opciones call iguales o todas las opciones put iguales con la excepci\'on de sus precios de ejercicio.\\
\\
\noindent
{\bf dimnames:} \indent
un comando en R para los nombres de las etiquetas de las columnas y las filas de una matriz.\\
\\
\noindent
{\bf din\'amica estoc\'astica:} \indent
v\'ease {\em proceso estoc\'astico}.\\
\\
\noindent
{\bf distancia de Manhattan:} \indent
la distancia entre dos vectores, definida como la suma de los valores absolutos de las diferencias entre sus componentes.\\
\\
\noindent
{\bf distancia euclidiana:} \indent
la distancia entre dos vectores, definida como la ra\'iz de la suma de los cuadrados de las diferencias entre sus componentes.\\
\\
\noindent
{\bf distribuci\'on de probabilidad:} \indent
una funci\'on que proporciona las probabilidades de ocurrencia de diferentes resultados posibles.\\
\\
\noindent
{\bf distribuci\'on de probabilidad de Bernoulli:} \indent
una distribuci\'on de probabilidad discreta de una variable aleatoria que toma el valor 1 con probabilidad $p$ y el valor 0 con probabilidad $q=1-p$.\\
\\
\noindent
{\bf distribuci\'on de probabilidad multinomial:} \indent
una distribuci\'on de probabilidad discreta de una variable aleatoria que toma $k$ valores diferentes con probabilidades $p_1,\dots,p_k$.\\
\\
\noindent
{\bf diversificaci\'on:} \indent
asignar capital para reducir la exposici\'on a cualquier activo o riesgo en particular mediante la inversi\'on en una variedad de activos.\\
\\
\noindent
{\bf diversificaci\'on de portafolio:} \indent
v\'ease {\em diversificaci\'on}.\\
\\
\noindent
{\bf diversificaci\'on intra-activos:} \indent
en inversiones inmobiliarias, la diversificaci\'on por el tipo de propiedad (residencial, comercial, etc.), la diversificaci\'on econ\'omica (por diferentes regiones divididas seg\'un  las caracter\'isticas econ\'omicas), la diversificaci\'on geogr\'afica, etc.\\
\\
\noindent
{\bf dividendo:} \indent
una distribuci\'on de algunas de las ganancias de una compa\~{n}\'ia, seg\'un lo decidido por su junta directiva, a una clase de sus accionistas, generalmente (pero no siempre) trimestralmente.\\
\\
\noindent
{\bf doble imposici\'on:} \indent
un sistema de impuestos corporativos (por ejemplo, en los EE.UU.) donde los ingresos corporativos se gravan primero a nivel corporativo, y luego nuevamente cuando los accionistas reciben los dividendos.\\
\\
\noindent
{\bf d\'olar-neutralidad:} \indent
cuando la suma de tenencias en d\'olares en una cartera es nula (con tenencias de posiciones cortas en d\'olares definidas como negativas).\\
\\
\noindent
{\bf drawdown:} \indent
una disminuci\'on de un pico a un valle en el P\&L durante un per\'iodo determinado, en donde el pico (valle) se define como el m\'aximo (m\'inimo) del P\&L en dicho per\'iodo.\\
\\
\noindent
{\bf duraci\'on:} \indent
v\'ease {\em duraci\'on d\'olar, duraci\'on de Macaulay, duraci\'on modificada, duraci\'on arriesgada}.\\
\\
\noindent
{\bf duraci\'on arriesgada:} \indent
una suma ponderada (sobre las fechas de pago) de las diferencias (descontadas) entre el nocional (de un tramo de CDO o similar) y la p\'erdida esperada para cada una de esas fechas, en donde cada ponderaci\'on es el tiempo desde la fecha de pago anterior.\\
\\
\noindent
{\bf duraci\'on de Macaulay:} \indent
un vencimiento promedio ponderado de los flujos de efectivo de un bono, en donde las ponderaciones son los valores actuales de dichos flujos de efectivo.\\
\\
\noindent
{\bf duraci\'on d\'olar:} \indent
una medida de la sensibilidad absoluta del precio de los bonos a los cambios en las tasas de inter\'es, definida como la duraci\'on modificada multiplicada por el precio de los bonos.\\
\\
\noindent
{\bf duraci\'on-d\'olar-neutralidad:} \indent
cuando la suma de las duraciones d\'olar de una cartera de bonos es nula (con las duraciones d\'olar de las posiciones de bonos cortas definidas como negativas).\\
\\
\noindent
{\bf duraci\'on modificada:} \indent
una medida de la sensibilidad relativa en el precio de los bonos ante cambios en las tasas de inter\'es, definida como la primera derivada negativa del precio del bono con respecto al rendimiento del bono.\\
\\
\noindent
{\bf efectivo (para \'indices):} \indent
en la jerga del trader com\'un, ``efectivo'' (``cash'' en ingl\'es) se refiere a la cartera del \'indice subyacente (por ejemplo, las acciones del S\&P 500 para el \'indice S\&P 500).\\
\\
\noindent
{\bf efecto contrario:} \indent
v\'ease {\em efecto de reversi\'on a la media}.\\
\\
\noindent
{\bf efecto de reversi\'on a la media (tambi\'en conocido como reversi\'on a la media, o efecto contrario):} \indent
una tendencia de los precios de los activos y/o sus retornos a volver a sus valores medios, que pueden ser seriales y/o transversales, seg\'un el contexto.\\
\\
\noindent
{\bf eigenvalor (tambi\'en conocido como valor propio, o valor caracter\'istico):}
una ra\'iz de la ecuaci\'on caracter\'istica de una matriz (v\'ease {\em eigenvector)}.\\
\\
\noindent
{\bf eigenvector (tambi\'en conocido como vector propio, o vector caracter\'istico):}
para una matriz $A$ cuadrada y sim\'etrica con dimensiones $N\times N$, un $N$-vector $V$ que resuelve la ecuaci\'on caracter\'istica $A~V = \lambda~V$, en donde $\lambda$ es el eigenvalor correspondiente (que es un n\'umero).\\
\\
\noindent
{\bf EMA (por sus siglas en ingl\'es):} \indent
una media m\'ovil exponencial, una media m\'ovil serial con las contribuciones pasadas suprimidas con las ponderaciones que decrecen exponencialmente.\\
\\
\noindent
{\bf empresa en distress:} \indent
una compa\~{n}\'ia en crisis financiera u operativa.\\
\\
\noindent
{\bf empresa objetivo (o compa\~{n}\'ia objetivo):} \indent
la compa\~{n}\'ia elegida por la empresa adquirente para una potencial fusi\'on o adquisici\'on corporativa.\\
\\
\noindent
{\bf empresa que cotiza en la bolsa (tambi\'en conocida como empresa p\'ublica):}
una empresa cuyas acciones operan libremente en una bolsa de valores o en mercados extraburs\'atiles.\\
\\
\noindent
{\bf EMSD (por sus siglas en ingl\'es):} \indent
una desviaci\'on est\'andar m\'ovil exponencial, una desviaci\'on est\'andar m\'ovil serial con las contribuciones pasadas suprimidas con las ponderaciones que decrecen exponencialmente.\\
\\
\noindent
{\bf en la muestra:} \indent
cuando un c\'alculo o un backtest no se realiza con datos ``fuera de la muestra''.\\
\\
\noindent
{\bf enfoque sistem\'atico:} \indent
las estrategias de trading que son met\'odicas y est\'an basadas en reglas con objetivos bien definidos y con controles de riesgo (a diferencia de, por ejemplo, las opiniones subjetivas de los analistas).\\
\\
\noindent
{\bf enigma de Fama:} \indent
v\'ease {\em anomal\'ia de descuento a plazo}.\\
\\
\noindent
{\bf entidades de referencia:} \indent
CDS, bonos, pr\'estamos, nombres de empresas o pa\'ises, etc., sobre los cuales se proporciona protecci\'on de incumplimiento.\\
\\
\noindent
{\bf entrega:} \indent
transferir el instrumento subyacente (o commodity) en un contrato (por ejemplo, futuros o forwards) al comprador al vencimiento (fecha de entrega) a un precio previamente acordado (precio de entrega).\\
\\
\noindent
{\bf entrenamiento:} \indent
en aprendizaje autom\'atico, fijar par\'ametros libres en un algoritmo utilizando los datos de entrenamiento.\\
\\
\noindent
{\bf equity:} \indent
acciones u otro activo de una empresa que representa su propiedad.\\
\\
\noindent
{\bf eRank (tambi\'en conocido como rango efectivo, o effective rank en ingl\'es):} \indent
una medida de la dimensionalidad efectiva de una matriz.\\
\\
\noindent
{\bf error de rastreo:} \indent
la ra\'iz cuadrada de la varianza de las diferencias entre los retornos de una cartera y los del \'indice de referencia que dicha cartera pretende imitar o superar.\\
\\
\noindent
{\bf escalera (tambi\'en conocida como ladder en ingl\'es) (para bonos):}
un portafolio de bonos con asignaciones de capital (aproximadamente) iguales en bonos de un n\'umero considerable de vencimientos diferentes (y generalmente aproximadamente equidistantes).\\
\\
\noindent
{\bf escalera (tambi\'en conocida como ladder en ingl\'es) (para opciones):}
un diferencial vertical que consiste en 3 opciones, todas opciones call o todas opciones put, 2 de estas son posiciones largas y 1 es corta, o 1 es larga y 2 son cortas.\\
\\
\noindent
{\bf escal\'on:} \indent
en un portafolio ladder de bonos, los bonos con la misma madurez.\\
\\
\noindent
{\bf escudo fiscal:} \indent
la reducci\'on en los impuestos sobre la renta que resulta de tomar una deducci\'on permisible de la renta imponible.\\
\\
\noindent
{\bf especulador:} \indent
un participante del mercado que intenta beneficiarse del movimiento de precios de un activo (comparado con {\em hedger)}.\\
\\
\noindent
{\bf esquema de ponderaci\'on:} \indent
asignar ponderaciones a un portafolio de acuerdo con alguna regla, por ejemplo, al disminuir las contribuciones de las acciones vol\'atiles.\\
\\
\noindent
{\bf establecer:} \indent
comprar o vender en corto un activo o un portafolio desde una posici\'on nula.\\
\\
\noindent
{\bf estrategia:} \indent
v\'ease {\em estrategia de trading}.\\
\\
\noindent
{\bf estrategia alcista:} \indent
una estrategia direccional en la que el trader se beneficia si el precio del instrumento subyacente sube.\\
\\
\noindent
{\bf estrategia bajista:} \indent
una estrategia direccional en la que el trader se beneficia si el precio del instrumento subyacente baja.\\
\\
\noindent
{\bf estrategia de acci\'on individual:} \indent
una estrategia de trading que se deriva de una se\~{n}al de trading para cualquier acci\'on determinada utilizando los datos solo para esa acci\'on y no para otras acciones.\\
\\
\noindent
{\bf estrategia de carry de volatilidad:} \indent
una estrategia de trading que consiste en vender en corto el VXX y compensar la posici\'on corta comprando el VXZ (v\'ease {\em ETN de volatilidad}), generalmente con un ratio de cobertura no unitario.\\
\\
\noindent
{\bf estrategia de cobertura:} \indent
v\'ease {\em cobertura}.\\
\\
\noindent
{\bf estrategia de compra-lanzamiento:} \indent
comprar acciones y lanzar (vender) una opci\'on de compra contra la posici\'on de acciones.\\
\\
\noindent
{\bf estrategia de costo cero:} \indent
una estrategia d\'olar-neutral.\\
\\
\noindent
{\bf estrategia de deuda en distress:} \indent
una estrategia basada en la adquisici\'on de deuda de una empresa en distress.\\
\\
\noindent
{\bf estrategia de duration-targeting:} \indent
una estrategia (por ejemplo, un ladder de bonos) que mantiene una duraci\'on aproximadamente constante vendiendo los bonos de vencimiento m\'as corto a medida que se aproximan al vencimiento y comprando nuevos bonos de vencimiento m\'as largo.\\
\\
\noindent
{\bf estrategia de ganancia de capital:} \indent
una estrategia que se beneficia de la compra y venta de un activo (o, m\'as generalmente, de establecer y liquidar una posici\'on).\\
\\
\noindent
{\bf estrategia de ingresos:} \indent
una estrategia de trading que genera ingresos, generalmente mediante cierta exposici\'on a cierto tipo de riesgo.\\
\\
\noindent
{\bf estrategia de Kelly:} \indent
una estrategia de asignaci\'on (apuesta) basada en maximizar el valor esperado del logaritmo de la riqueza.\\
\\
\noindent
{\bf estrategia de momentum:} \indent
una estrategia de trading basada en el efecto momentum.\\
\\
\noindent
{\bf estrategia de precio-momentum:} \indent
una estrategia de momentum en donde el indicador de momentum se basa en los retornos pasados.\\
\\
\noindent
{\bf estrategia de reversi\'on a la media:} \indent
una estrategia de trading basada en el efecto de reversi\'on a la media.\\
\\
\noindent
{\bf estrategia de soporte y resistencia:} \indent
una estrategia de an\'alisis t\'ecnico basada en soporte y resistencia.\\
\\
\noindent
{\bf estrategia de targeting de volatilidad:} \indent
una estrategia de trading que apunta a mantener un nivel de volatilidad constante (objetivo de volatilidad o volatilidad objetivo) al reequilibrar entre un activo de riesgo y un activo sin riesgo.\\
\\
\noindent
{\bf estrategia de trading:} \indent
un conjunto de instrucciones para lograr ciertas tenencias de activos en ciertos tiempos predefinidos $t_1,t_2,\dots$, que pueden ser (pero no es necesario) nulas en una o m\'as de estas veces.\\
\\
\noindent
{\bf estrategia de trading fundamental:} \indent
una estrategia de trading basada en an\'alisis fundamental.\\
\\
\noindent
{\bf estrategia de trading pasiva:} \indent
una estrategia de trading basada en el enfoque de inversi\'on pasiva.\\
\\
\noindent
{\bf estrategia de valor relativo:} \indent
una estrategia que apunta a explotar las diferencias en los precios, rendimientos o tasas (por ejemplo, tasas de inter\'es) de activos relacionados (seg\'un alg\'un criterio) (por ejemplo, pares de acciones hist\'oricamente correlacionados en el trading de pares).\\
\\
\noindent
{\bf estrategia de value:} \indent
comprar acciones de alto value (alto ratio B/P) y vender acciones de bajo value (bajo ratio B/P).\\
\\
\noindent
{\bf estrategia de venta-lanzamiento:} \indent
vender acciones y lanzar (vender) una opci\'on de venta contra la posici\'on de acciones.\\
\\
\noindent
{\bf estrategia de volatilidad:} \indent
una estrategia de trading que apunta a capitalizar un entorno de alta volatilidad esperada, por ejemplo, comprando la volatilidad.\\
\\
\noindent
{\bf estrategia Delta-neutral:} \indent
una estrategia de trading que logra el Delta igual a cero mediante, por ejemplo, la cobertura de Delta.\\
\\
\noindent
{\bf estrategia direccional:} \indent
una estrategia que se beneficia en funci\'on de la direcci\'on futura del activo subyacente (o activos subyacentes) (comparado con {\em estrategia no direccional}).\\
\\
\noindent
{\bf estrategia discrecional:} \indent
una estrategia que se basa en las habilidades del gestor del fondo (comparado con {\em estrategia no discrecional}).\\
\\
\noindent
{\bf estrategia intrad\'ia:} \indent
una estrategia de trading que comienza con una posici\'on nula, compra y vende/va en corto los activos intrad\'ia, y termina el d\'ia con una posici\'on nula al cierre (en la jerga del trader en ingl\'es, ``goes home flat'').\\
\\
\noindent
{\bf estrategia lateral:} \indent
una estrategia de trading que apunta a capitalizar un entorno de baja volatilidad esperada, por ejemplo, vendiendo la volatilidad.\\
\\
\noindent
{\bf estrategia multifactorial:} \indent
una estrategia de trading basada en la exposici\'on a m\'ultiples factores, por ejemplo, momentum, value, etc. (portafolio multifactor).\\
\\
\noindent
{\bf estrategia no direccional (tambi\'en conocida como estrategia neutral):} \indent
una estrategia que no se basa en la direcci\'on futura del activo subyacente (o de los activos), por lo que el trader no tiene en cuenta si su precio sube o baja (comparado con {\em estrategia direccional}).\\
\\
\noindent
{\bf estrategia no discrecional:} \indent
una estrategia de trading basada en un enfoque sistem\'atico (a diferencia de discrecional).\\
\\
\noindent
{\bf estrategia ``prestar a poseer'' (loan-to-own en ingl\'es):} \indent
financiar una empresa en distress mediante pr\'estamos asegurados con el fin de adquirir su capital (con derechos de control) si la empresa se declara en quiebra.\\
\\
\noindent
{\bf estrategia sobre el margen de la curva de rendimientos:} \indent
una estrategia de bonos que hace una apuesta en el margen de la curva de rendimientos (flattener o steepener).\\
\\
\noindent
{\bf estratificaci\'on:} \indent
el paso intermedio en el proceso de lavado de dinero, que consiste en mover el dinero entre diferentes cuentas e incluso pa\'ises, creando de esta forma una red de transacciones compleja y separando el dinero de su fuente en varios grados.\\
\\
\noindent
{\bf estructura temporal (en futuros):} \indent
la dependencia de los precios de futuros al tiempo hasta su vencimiento.\\
\\
\noindent
{\bf estructura temporal (en tasas de inter\'es):} \indent
v\'ease {\em curva de rendimientos}.\\
\\
\noindent
{\bf ETF de \'indice (tambi\'en conocido como ETF de rastreo):} \indent
un ETF que rastrea a un \'indice.\\
\\
\noindent
{\bf ETF del Tesoro:} \indent
un ETF para un \'indice compuesto por las obligaciones de deuda del gobierno de EE.UU.\\
\\
\noindent
{\bf ETF gestionado activamente:} \indent
un fondo cotizado en la bolsa cuya asignaci\'on del portafolio subyacente se gestiona de forma activa.\\
\\
\noindent
{\bf ETF inverso:} \indent
un ETF dise\~{n}ado para rastrear el retorno inverso a su \'indice subyacente.\\
\\
\noindent
{\bf ETH:} \indent
ether/Ethereum, una criptomoneda.\\
\\
\noindent
{\bf ETN de volatilidad:} \indent
un ETN que rastrea al VIX, por ejemplo, el VXX o VXZ.\\
\\
\noindent
{\bf EUR:} \indent
euro, una unidad de la moneda de la eurozona.\\
\\
\noindent
{\bf eurod\'olar:} \indent
un dep\'osito en d\'olares en un banco fuera de los Estados Unidos.\\
\\
\noindent
{\bf exceso de retorno:} \indent
el retorno de un activo en exceso de alg\'un retorno de referencia (por ejemplo, tasa libre de riesgo).\\
\\
\noindent
{\bf ex-dividendo} \indent
cuando el vendedor de acciones tiene el derecho a recibir un dividendo que ha sido declarado, pero no pagado (comparado con {\em cum-dividendo}).\\
\\
\noindent
{\bf expectativa condicional (tambi\'en conocida como valor esperado condicional, o media condicional):} \indent
un valor promedio de una cantidad asumiendo que alguna condici\'on ocurre.\\
\\
\noindent
{\bf exposici\'on:} \indent
la cantidad que se puede perder (o ganar) en una inversi\'on.\\
\\
\noindent
{\bf factor (tambi\'en conocido como factor de riesgo):} \indent
una variable explicativa com\'un para un corte transversal de retornos de activos (por ejemplo, acciones).\\
\\
\noindent
{\bf factor de anualizaci\'on:} \indent
un factor multiplicativo para anualizar una cantidad.\\
\\
\noindent
{\bf factor de conversi\'on:} \indent
el precio cotizado que tendr\'ia un bono por d\'olar de principal el primer d\'ia del mes de entrega de un contrato de futuros de tasa de inter\'es, asumiendo que la tasa de inter\'es para todos los vencimientos es igual al 6\% anual con la composici\'on compuesta semestral.\\
\\
\noindent
{\bf factor de descuento:} \indent
el valor de un bono de descuento con \$1 de principal en el momento $t$ antes de su vencimiento $T$.\\
\\
\noindent
{\bf factor de descuento libre de riesgo:} \indent
un factor de descuento que utiliza una tasa libre de riesgo para descontar los flujos de efectivo futuros.\\
\\
\noindent
{\bf factor de momentum de Carhart (tambi\'en conocido como MOM):}
los ganadores menos los perdedores por momentum (de 12 meses).\\
\\
\noindent
{\bf factor de riesgo:} \indent
v\'ease {\em factor}.\\
\\
\noindent
{\bf factor de estilo de riesgo (tambi\'en conocido como factor de estilo):}
factores de riesgo tales como value, crecimiento, tama\~{n}o, momentum, liquidez y volatilidad.\\
\\
\noindent
{\bf factores de Fama-French:} \indent
MKT, el exceso de retorno (definido como el rendimiento en exceso de la tasa libre de riesgo, que a su vez, se define como la tasa del Tesoro a un mes) del portafolio del mercado; SMB, del portafolio ``compa\~{n}\'ias peque\~{n}as menos compa\~{n}\'ias grandes'' (por la capitalizaci\'on burs\'atil); HML, del portafolio ``alto menos bajo'' (por el ratio B/P).\\
\\
\noindent
{\bf fecha de ejercicio:} \indent
la fecha en la que se puede ejercer una opci\'on.\\
\\
\noindent
{\bf fideicomiso de inversiones inmobiliarias (REIT):} \indent
una empresa (que a menudo cotiza en grandes bolsas de valores y, por lo tanto, permite a los traders tomar una participaci\'on l\'iquida en bienes ra\'ices) que posee, opera o financia bienes ra\'ices que generan ingresos.\\
\\
\noindent
{\bf filtro de Hodrick-Prescott (tambi\'en conocido como filtro HP, o m\'etodo de Whittaker-Henderson en las ciencias actuariales):} \indent
un filtro de series de tiempo para separar el componente de baja frecuencia (``regular'') del componente de alta frecuencia (``irregular'') (ruido).\\
\\
\noindent
{\bf filtro de Kalman:} \indent
un filtro de series de tiempo para separar la se\~{n}al del ruido.\\
\\
\noindent
{\bf flattener:} \indent
una estrategia sobre el margen en la curva de rendimientos.\\
\\
\noindent
{\bf flujo de efectivo:} \indent
la cantidad neta de efectivo y activos equivalentes de efectivo que se transfieren hacia y desde una compa\~{n}\'ia (en el contexto corporativo) o una cartera (en el contexto de trading).\\
\\
\noindent
{\bf flujo de \'ordenes agresivo:} \indent
flujo de \'ordenes compuesto de \'ordenes agresivas.\\
\\
\noindent
{\bf flujo de \'ordenes informado:} \indent
v\'ease {\em flujo de \'ordenes inteligente}.\\
\\
\noindent
{\bf flujo de \'ordenes inteligente (tambi\'en conocido como flujo de \'ordenes t\'oxico):} \indent
flujo de \'ordenes basado en alg\'un retorno predictivo esperado.\\
\\
\noindent
{\bf flujo de \'ordenes tonto (tambi\'en conocido como flujo de \'ordenes no informado):} \indent
flujo de \'ordenes agresivo no basado en un retorno esperado predictivo.\\
\\
\noindent
{\bf fondo de infraestructura:} \indent
fondos de infraestructura no cotizados (inversiones de capital privado), fondos de infraestructura cotizados.\\
\\
\noindent
{\bf fondo de pensiones:} \indent
un conjunto de fondos que proporciona ingresos de jubilaci\'on.\\
\\
\noindent
{\bf fondo mutual:} \indent
un veh\'iculo de inversi\'on financiado por un conjunto de dinero recaudado de muchos inversionistas con el fin de comprar diversos activos (acciones, bonos, instrumentos del mercado monetario, etc.).\\
\\
\noindent
{\bf forma base (tambi\'en conocida como stem en ingl\'es):} \indent
en ling\"{u}\'istica, la parte de una palabra que es com\'un a todas sus variantes inflexas.\\
\\
\noindent
{\bf forward (tambi\'en conocido como contrato forward o contrato a plazo):}
un contrato establecido en el tiempo $t=0$, en el cual una de las dos partes acuerda vender al otro un activo determinado en alg\'un momento futuro $T$ (conocido como el vencimiento, fecha de entrega o madurez del contrato) a un precio de ejercicio previamente acordado $k$.\\
\\
\noindent
{\bf funci\'on de activaci\'on:} \indent
una funci\'on que define la salida de un nodo (neurona artificial) en una red neuronal artificial dada una entrada (o un conjunto de entradas).\\
\\
\noindent
{\bf funci\'on de error:} \indent
en aprendizaje autom\'atico, una funci\'on a minimizar que se construye a partir de los errores (o similar), por ejemplo, la suma de los cuadrados de los errores, o alguna otra funci\'on adecuada (no debe confundirse con la funci\'on de error de Gauss $\mbox{erf}(x)$).\\
\\
\noindent
{\bf funci\'on objetivo:} \indent
una funci\'on a ser maximizada o minimizada en optimizaci\'on.\\
\\
\noindent
{\bf fundamentos:} \indent
la informaci\'on cuantitativa y cualitativa sobre la salud financiera/econ\'omica y el valor de una empresa, activos, moneda, etc.\\
\\
\noindent
{\bf fusi\'on:} \indent
una consolidaci\'on de dos empresas en una sola.\\
\\
\noindent
{\bf fusi\'on de acciones:} \indent
una fusi\'on en donde cada acci\'on de la compa\~{n}\'ia objetivo se intercambia por un n\'umero (que puede ser fraccional) de las acciones de la compa\~{n}\'ia adquirente.\\
\\
\noindent
{\bf fusi\'on en efectivo:} \indent
una fusi\'on donde la compa\~{n}\'ia adquirente paga a los accionistas de la compa\~{n}\'ia objetivo en efectivo a cambio de sus acciones.\\
\\
\noindent
{\bf futuro (tambi\'en conocido como contrato de futuros):} \indent
un contrato a plazo estandarizado negociado en una bolsa de futuros.\\
\\
\noindent
{\bf futuro al mes m\'as cercano:} \indent
v\'ease {\em futuro del primer mes}.\\
\\
\noindent
{\bf \ futuro de \'indice:} \indent
un futuro basado en un \'indice.\\
\\
\noindent
{\bf futuro de commodities:} \indent
un contrato de futuros sobre commodities.\\
\\
\noindent
{\bf futuro de tasas de inter\'es:} \indent
un contrato de futuros t\'ipicamente con un conjunto (cesta de entrega) de instrumentos subyacentes (por ejemplo, bonos) que pagan intereses.\\
\\
\noindent
{\bf futuro de un mes diferido:} \indent
un contrato de futuros con la fecha de liquidaci\'on en los meses posteriores (comparado con {\em futuro del primer mes}).\\
\\
\noindent
{\bf futuro del primer mes:} \indent
un contrato de futuros con la fecha de liquidaci\'on m\'as cercana a la fecha actual (comparado con {\em futuro de un mes diferido}).\\
\\
\noindent
{\bf futuro del segundo mes:} \indent
un contrato de futuros con el vencimiento m\'as cercano despu\'es de los futuros del primer mes.\\
\\
\noindent
{\bf futuro mini-S\&P (tambi\'en conocido como e-mini):} \indent
un contrato de futuros sobre el S\&P 500 con un valor nocional de 50 veces el valor del \'indice burs\'atil S\&P 500.\\
\\
\noindent
{\bf Gamma:} \indent
la segunda derivada del valor de un activo de derivados (por ejemplo, una opci\'on) con respecto al precio del activo subyacente.\\
\\
\noindent
{\bf Gamma scalping:} \indent
la cobertura de Gamma alcanzada mediante la compra y venta del activo subyacente en respuesta a los movimientos de sus precios que causan cambios en el Delta de una cartera de opciones.\\
\\
\noindent
{\bf ganadores:} \indent
las acciones u otros activos en un portafolio o universo de trading que superan al resto en base a alg\'un criterio (\'indice de referencia).\\
\\
\noindent
{\bf ganancias:} \indent
el ingreso neto despu\'es de impuestos de una empresa.\\
\\
\noindent
{\bf ganancias inesperadas:} \indent
v\'ease {\em ganancias inesperadas estandarizadas}.\\
\\
\noindent
{\bf ganancias inesperadas estandarizadas (SUE, por sus siglas en ingl\'es):}
un ratio, cuyo numerador (ganancias inesperadas) es la diferencia entre las ganancias por acci\'on trimestrales anunciadas m\'as recientemente y aquellas anunciadas hace 4 trimestres, y cuyo denominador es la desviaci\'on est\'andar de las ganancias inesperadas en los \'ultimos 8 trimestres.\\
\\
\noindent
{\bf ganancias-momentum:} \indent
una estrategia de momentum basada en las ganancias.\\
\\
\noindent
{\bf ganancias realizadas:} \indent
v\'ease {\em P\&L realizado}.\\
\\
\noindent
{\bf gaviota:} \indent
un tipo de estrategias de opciones.\\
\\
\noindent
{\bf gesti\'on de riesgos:} \indent
identificar, analizar y mitigar riesgos potenciales.\\
\\
\noindent
{\bf grado del dinero (``moneyness'' en ingl\'es):} \indent
donde el precio de ejercicio de un contrato de derivados est\'a en relaci\'on con el precio actual de su activo subyacente, lo cual determina el valor intr\'inseco del derivado.\\
\\
\noindent
{\bf Griegas:} \indent
v\'ease {\em Delta, Gamma, Theta, Vega}.\\
\\
\noindent
{\bf guts:} \indent
una estrategia de trading con opciones.\\
\\
\noindent
{\bf hedger:} \indent
un participante del mercado que intenta reducir el riesgo asociado al movimiento del precio de un activo (comparado con {\em especulador}).\\
\\
\noindent
{\bf hipoteca:} \indent
un instrumento de deuda, asegurado por una propiedad inmobiliaria como una garant\'ia, en el que el prestatario debe pagar un conjunto predeterminado de pagos.\\
\\
\noindent
{\bf HMD (tambi\'en conocido como compa\~{n}\'ias saludables menos compa\~{n}\'ias en distress):} \indent
comprando las compa\~{n}\'ias m\'as seguras y vendiendo las m\'as riesgosas seg\'un su probabilidad de quiebra.\\
\\
\noindent
{\bf HML (tambi\'en conocido como Alto menos Bajo, o High minus Low en ingl\'es):} \indent
v\'ease {\em factores de Fama-French}.\\
\\
\noindent
{\bf horizonte (tambi\'en conocido como horizonte de inversi\'on):} \indent
v\'ease {\em per\'iodo de tenencia}.\\
\\
\noindent
{\bf IBS (por sus siglas en ingl\'es):} \indent
la fuerza interna de las barras de precios, definida como la diferencia entre el precio de cierre y el precio m\'inimo dividido por la diferencia entre el precio m\'aximo y el precio m\'inimo.\\
\\
\noindent
{\bf imputaci\'on de dividendos:} \indent
un sistema de impuestos corporativos en el que una parte o la totalidad del impuesto pagado por una empresa se puede atribuir o imputar a los accionistas a trav\'es de un cr\'edito fiscal para reducir el impuesto sobre la renta pagadero en una distribuci\'on a trav\'es de, por ejemplo, los dividendos.\\
\\
\noindent
{\bf incumplimiento:} \indent
no pagar un pr\'estamo/deuda.\\
\\
\noindent
{\bf independencia condicional:} \indent
$A$ y $B$ son condicionalmente independientes asumiendo que $C$ es cierto si y solo si la ocurrencia de $A$ asumiendo que $C$ es independiente de la ocurrencia de $B$ asumiendo $C$ y viceversa, es decir, $P(A\cap B | C) = P(A|C) \times P(B|C)$, en donde $P(A|B)$ es una probabilidad condicional.\\
\\
\noindent
{\bf indicador t\'ecnico:} \indent
una cantidad matem\'atica utilizada en el an\'alisis t\'ecnico.\\
\\
\noindent
{\bf \'indice:} \indent
una cartera diversificada de activos combinada con ciertas ponderaciones.\\
\\
\noindent
{\bf \'indice de CDS:} \indent
un \'indice de swaps de incumplimiento crediticio, tal como el CDX y el iTraxx.\\
\\
\noindent
{\bf \'indice de fuerza relativa (RSI, por sus siglas en ingl\'es):} \indent
durante un per\'iodo de tiempo especificado, la ganancia promedio de los per\'iodos alcistas dividida por la suma de la ganancia promedio de los per\'iodos alcistas y el valor absoluto de la p\'erdida promedio de los per\'iodos bajistas.\\
\\
\noindent
{\bf \'indice de inflaci\'on:} \indent
por ejemplo, el CPI.\\
\\
\noindent
{\bf \'indice de mercado general:} \indent
un \'indice basado en un amplio corte transversal de activos (por ejemplo, S\&P 500, Russell 3000, etc.).\\
\\
\noindent
{\bf \'Indice de Precios al Consumidor (CPI, por sus siglas en ingl\'es):}
una medida del nivel de precios de una cesta de mercado de bienes de consumo y servicios.\\
\\
\noindent
{\bf \'indice de volatilidad:} \indent
un \'indice (por ejemplo, el VIX) que mide la expectativa de volatilidad futura del mercado (30 d\'ias para el VIX) basado en las volatilidades impl\'icitas de los instrumentos subyacentes (el \'indice S\&P 500 para el VIX).\\
\\
\noindent
{\bf \'indice del clima:} \indent
un \'indice sint\'etico generalmente basado en la temperatura, utilizando, por ejemplo, grados-d\'ia de refrigeraci\'on (CDD, por sus siglas en ingl\'es) y grados-d\'ia de calefacci\'on (HDD, por sus siglas en ingl\'es).\\
\\
\noindent
{\bf industria (en clasificaci\'on de la industria):} \indent
un grupo de empresas basado (entre otras cosas) en la industria econ\'omica a la que pertenecen.\\
\\
\noindent
{\bf industria (en econom\'ia):} \indent
un grupo de empresas que se relacionan en funci\'on de sus actividades comerciales principales.\\
\\
\noindent
{\bf inflaci\'on:} \indent
un aumento sostenido en el nivel de precios de los bienes y servicios en una econom\'ia durante un per\'iodo de tiempo, el cual se mide como un cambio porcentual anual conocido como la tasa de inflaci\'on.\\
\\
\noindent
{\bf inflaci\'on acumulada:} \indent
la tasa de inflaci\'on medida desde el momento $t_1$ al tiempo $t_2$ (comparado con {\em inflaci\'on interanual}).\\
\\
\noindent
{\bf inflaci\'on general (HI, por sus siglas en ingl\'es):} \indent
una medida de la inflaci\'on total dentro de una econom\'ia, incluidos los precios de los commodities, tales como alimentos y energ\'ia (comparado con {\em inflaci\'on n\'ucleo}).\\
\\
\noindent
{\bf inflaci\'on interanual (YoY, por sus siglas en ingl\'es):} \indent
la inflaci\'on anual (comparado con {\em inflaci\'on acumulada}).\\
\\
\noindent
{\bf inflaci\'on n\'ucleo (CI, por sus siglas en ingl\'es):} \indent
la inflaci\'on a largo plazo, con art\'iculos sujetos a precios vol\'atiles (como alimentos y energ\'ia) excluidos (comparado con {\em inflaci\'on general}).\\
\\
\noindent
{\bf infra-reacci\'on:} \indent
en los mercados financieros, una respuesta insuficiente a las noticias, ya que algunos participantes del mercado tienden a ser conservadores y se aferran demasiado a sus creencias anteriores.\\
\\
\noindent
{\bf inmunizaci\'on de bonos:} \indent
hacer coincidir la duraci\'on de un portafolio de bonos con el vencimiento de una futura obligaci\'on en efectivo.\\
\\
\noindent
{\bf integraci\'on:} \indent
el \'ultimo paso del proceso de lavado de dinero mediante el cual quienes lavan dinero lo recuperan a trav\'es de fuentes de apariencia leg\'itima.\\
\\
\noindent
{\bf intercepto:} \indent
en una regresi\'on lineal, el coeficiente de la regresi\'on de la variable independiente (que tambi\'en se llama coloquialmente como el intercepto) cuyos elementos son todos iguales a 1.\\
\\
\noindent
{\bf inter\'es:} \indent
la cantidad pagada por el prestatario al prestamista por encima del capital (la cantidad real prestada).\\
\\
\noindent
{\bf inter\'es abierto (tambi\'en conocido como contratos abiertos, o compromisos abiertos):} \indent
el n\'umero total de contratos de futuros (u opciones) abiertos en un momento dado (es decir, aquellos contratos que no se han liquidado).\\
\\
\noindent
{\bf inter\'es compuesto (tambi\'en conocido como composici\'on, capitalizaci\'on):} \indent
reinversi\'on del inter\'es ganado para generar el inter\'es adicional en el futuro.\\
\\
\noindent
{\bf invariancia de escala:} \indent
una funci\'on $f(x_i)$ de $N$ variables $x_i$ ($i=1,\dots,N$) presenta una invariancia de escala si $f(\zeta x_i) = f(x_i)$ para un factor de escala arbitrario $\zeta$ toma valores en un intervalo continuo (por ejemplo, valores reales positivos).\\
\\
\noindent
{\bf inversi\'on:} \indent
asignaci\'on de dinero con la expectativa de obtener un retorno positivo.\\
\\
\noindent
{\bf inversi\'on activa (tambi\'en conocida como gesti\'on activa):} \indent
una estrategia de inversi\'on que involucra la compra y venta activa (frecuente) de activos en una cartera (comparado con {\em inversi\'on pasiva}) con el objetivo a explotar las oportunidades (percibidas) que pueden generar beneficios.\\
\\
\noindent
{\bf inversi\'on en infraestructura:} \indent
invertir en proyectos a largo plazo tales como transporte (carreteras, puentes, t\'uneles, ferrocarriles, puertos, aeropuertos, etc.), telecomunicaciones (cables de transmisi\'on, sat\'elites, torres, etc.), servicios p\'ublicos (generaci\'on de electricidad, transmisi\'on o distribuci\'on de gas o electricidad, suministro de agua, alcantarillado, residuos, etc.), energ\'ia (incluyendo pero no limitado a energ\'ia renovable), atenci\'on m\'edica (hospitales, cl\'inicas, hogares de ancianos, etc.), instalaciones educativas (escuelas, universidades, institutos de investigaci\'on, etc.), etc.\\
\\
\noindent
{\bf inversi\'on pasiva:} \indent
una estrategia de inversi\'on de horizonte largo, esencialmente de compra y mantenimiento, con el objetivo de minimizar los costos de transacci\'on y replicar el desempe\~{n}o de un portafolio de referencia (generalmente, bien diversificado).\\
\\
\noindent
{\bf iShares (tablero de cotizaciones IVV):} \indent
un ETF que rastrea al S\&P 500.\\
\\
\noindent
{\bf Julio:} \indent
una unidad de trabajo, calor y energ\'ia en el Sistema Internacional de Unidades (SI).\\
\\
\noindent
{\bf JPY:} \indent
yen japon\'es, una unidad de moneda japonesa.\\
\\
\noindent
{\bf lanzador de una opci\'on:} \indent
un vendedor de opciones.\\
\\
\noindent
{\bf largo-corto:} \indent
un portafolio o estrategia con tenencias (o posiciones) tanto largas como cortas.\\
\\
\noindent
{\bf lavado de dinero:} \indent
una actividad en la que el efectivo se utiliza como un veh\'iculo para transformar las ganancias ilegales en activos de apariencia leg\'itima.\\
\\
\noindent
{\bf LETF:} \indent
ETF apalancado (inverso), un ETF dise\~{n}ado para rastrear el retorno igual a (el inverso a) $n$ veces el retorno de su \'indice subyacente, en donde $n$ es el apalancamiento (generalmente, 2 o 3).\\
\\
\noindent
{\bf letra del Tesoro (tambi\'en conocida como T-bill en ingl\'es):} \indent
una obligaci\'on de deuda a corto plazo emitida por el Tesoro de los Estados Unidos con un vencimiento inferior a 1 a\~{n}o.\\
\\
\noindent
{\bf libro de registro distribuido:} \indent
una base de datos compartida y sincronizada a trav\'es de una red (generalmente grande, de par a par) que abarca varios sitios.\\
\\
\noindent
{\bf libro mayor (o libro de registro):} \indent
un libro u otra colecci\'on de cuentas financieras y registros de transacciones.\\
\\
\noindent
{\bf l\'imites de posici\'on:} \indent
l\'imites superiores o inferiores de las tenencias en d\'olares de diversos activos en un portafolio.\\
\\
\noindent
{\bf l\'imites de trading:} \indent
l\'imites superiores o inferiores sobre los montos en d\'olares de las operaciones permitidas para diversos activos en una cartera, al establecer, reequilibrar o liquidar.\\
\\
\noindent
{\bf liquidaci\'on:} \indent
el cumplimiento de las obligaciones bajo un futuro o contrato a plazo al momento del vencimiento.\\
\\
\noindent
{\bf liquidaci\'on (para activos o portafolios):} \indent
cerrar las posiciones abiertas.\\
\\
\noindent
{\bf liquidaci\'on (para empresas):} \indent
liquidar (poner fin a) el negocio de una empresa y distribuir sus activos a quienes tienen los derechos a reclamarlos, generalmente cuando la empresa es insolvente.\\
\\
\noindent
{\bf liquidez:} \indent
la disponibilidad de activos/fondos l\'iquidos.\\
\\
\noindent
{\bf log:} \indent
un logaritmo (normalmente el logaritmo natural, a menos que se especifique lo contrario).\\
\\
\noindent
{\bf lookback (tambi\'en conocido como per\'iodo lookback):} \indent
la longitud de una muestra de datos de una serie de tiempo utilizada en un backtest o alg\'un c\'omputo hist\'orico.\\
\\
\noindent
{\bf macro:} \indent
las estrategias de trading macro.\\
\\
\noindent
{\bf macro discrecional:} \indent
las estrategias macro globales discrecionales basadas en opiniones subjetivas de analistas.\\
\\
\noindent
{\bf macro global:} \indent
las estrategias de trading que apuntan a capitalizar los cambios regionales, econ\'omicos y pol\'iticos en todo el mundo.\\
\\
\noindent
{\bf macro sistem\'atico:} \indent
las estrategias de trading macro no discrecionales, sistem\'aticas.\\
\\
\noindent
{\bf madurez (tambi\'en conocida como fecha de madurez, o fecha de vencimiento):} \indent
el momento en que un instrumento financiero deja de existir y el principal y/o intereses se reembolsan en su totalidad.\\
\\
\noindent
{\bf m\'aquinas de vectores de soporte (SVM, por sus siglas en ingl\'es):}
en aprendizaje autom\'atico, un tipo de modelos de aprendizaje supervisado.\\
\\
\noindent
{\bf margen crediticio:} \indent
la diferencia entre el rendimiento de un bono y la tasa libre de riesgo (la misma que el margen de rendimiento de un bono).\\
\\
\noindent
{\bf margen de CDS:} \indent
una prima peri\'odica (por ejemplo, anual) por d\'olar de la deuda asegurada.\\
\\
\noindent
{\bf margen de mariposa:} \indent
una estrategia de opciones mariposa.\\
\\
\noindent
{\bf margen del swap:} \indent
la diferencia entre la tasa de inter\'es fija de un swap de tasa de inter\'es y el rendimiento de un activo del Tesoro con un vencimiento similar.\\
\\
\noindent
{\bf margen de tasas de inter\'es:} \indent
la diferencia entre las tasas de inter\'es pagadas por dos instrumentos.\\
\\
\noindent
{\bf margen del rendimiento de un bono (tambi\'en conocido como margen de un bono):} \indent
la diferencia entre el rendimiento de un bono y la tasa libre de riesgo.\\
\\
\noindent
{\bf margen en la curva de rendimientos:} \indent
el margen entre los bonos de vencimiento m\'as cortos y m\'as largos en la curva de rendimientos.\\
\\
\noindent
{\bf mariposa:} \indent
una cartera (de bonos u opciones) con 3 tramos, dos perif\'ericos (por vencimiento en portafolios de bonos y por precio de ejercicio en portafolios de opciones) alas y un cuerpo en el medio.\\
\\
\noindent
{\bf mariposa regresi\'on-ponderada:} \indent
un tipo de portafolio mariposa de bonos.\\
\\
\noindent
{\bf mariposa de hierro:} \indent
un tipo de estrategias de trading con opciones.\\
\\
\noindent
{\bf mariposa neutral a la curva:} \indent
una estrategia de bonos mariposa con la curva-neutralidad.\\
\\
\noindent
{\bf market-making:} \indent
proporcionar liquidez mediante la fijaci\'on simult\'anea de los precios de compra y venta en un instrumento financiero o commodity que se mantiene en el inventario con el fin de obtener una ganancia en el margen de oferta y demanda (o la diferencia bid-ask).\\
\\
\noindent
{\bf matriz de cargas:} \indent
v\'ease {\em matriz de cargas factoriales}.\\
\\
\noindent
{\bf matriz de cargas factoriales:} \indent
la matriz con las dimensiones $N\times K$ $\Omega_{iA}$ ($i=1,\dots,N$, $A=1,\dots,K$, generalmente $K\ll N$) en un modelo de $K$-factores $Y_i = \sum_{A=1}^K \Omega_{iA}~F_A + \epsilon_i$, en donde $Y_i$ son las variables observadas, $F_A$ son las variables no observadas (factores comunes), y $\epsilon_i$ son los t\'erminos de error no observados.\\
\\
\noindent
{\bf matriz de correlaci\'on:} \indent
una matriz con las dimensiones $N\times N$ con los elementos diagonales unitarios, cuyos elementos fuera de la diagonal son las correlaciones por pares de los $N$ activos diferentes.\\
\\
\noindent
{\bf matriz de correlaci\'on muestral:} \indent
una matriz de correlaci\'on para un conjunto de activos computada en funci\'on de las series de tiempo de sus retornos hist\'oricos.\\
\\
\noindent
{\bf matriz de covarianza:} \indent
una matriz con las dimensiones $N\times N$, cuyos elementos fuera de la diagonal son las covarianzas por pares de $N$ activos diferentes, y cuyos elementos en la diagonal son las varianzas correspondientes.\\
\\
\noindent
{\bf matriz de covarianza muestral:} \indent
una matriz de covarianza para un conjunto de activos computada en funci\'on de las series de tiempo de sus retornos hist\'oricos.\\
\\
\noindent
{\bf matriz inversa:} \indent
para una matriz cuadrada $A$ con las dimensiones $N\times N$, la matriz inversa $A^{-1}$ es la matriz cuadrada con las dimensiones $N\times N$ tal que $A~A^{-1} =A^{-1}~A = I$, en donde $I$ es la matriz identidad con las dimensiones $N\times N$ (cuyos elementos diagonales son iguales a 1, y cuyos elementos fuera de la diagonal son iguales a 0).\\
\\
\noindent
{\bf m\'aximo (tambi\'en conocido como precio m\'aximo, o high en ingl\'es):}
el precio m\'aximo alcanzado por una acci\'on (u otro activo) dentro de un d\'ia de trading determinado (o en alg\'un otro intervalo de tiempo).\\
\\
\noindent
{\bf MBS passthrough:} \indent
un MBS donde los flujos de efectivo se pasan de los deudores a los inversionistas a trav\'es de un intermediario.\\
\\
\noindent
{\bf media:} \indent
un valor promedio.\\
\\
\noindent
{\bf media de largo plazo:} \indent
en un proceso Ornstein-Uhlenbeck de reversi\'on a la media, el valor medio de la variable de estado en el l\'imite de tiempo infinito.\\
\\
\noindent
{\bf media m\'ovil:} \indent
en una serie de tiempo, un promedio (posiblemente computado con algunas ponderaciones no triviales) sobre un intervalo de tiempo de longitud fija (longitud de la media m\'ovil), en donde el tiempo m\'as reciente en dicho intervalo puede tomar varios valores en la serie de tiempo.\\
\\
\noindent
{\bf media m\'ovil simple (SMA, por sus siglas en ingl\'es):} \indent
una media m\'ovil sin suprimir las contribuciones pasadas (comparado con {\em media m\'ovil exponencial o EMA}).\\
\\
\noindent
{\bf medida de probabilidad:} \indent
una funci\'on real valorada en el intervalo entre 0 y 1 (0 corresponde al conjunto vac\'io y 1 corresponde al espacio completo) definida en un conjunto de eventos en un espacio de probabilidad que satisface la propiedad de aditividad contable, es decir, en pocas palabras, que la probabilidad de una uni\'on de eventos desunidos A y B es igual a la suma de sus probabilidades.\\
\\
\noindent
{\bf medida de probabilidad neutral al riesgo (tambi\'en conocida como medida de neutralidad al riesgo):} \indent
una medida de probabilidad te\'orica seg\'un la cual el precio actual de un activo es igual a su valor esperado futuro descontado por una tasa libre de riesgo.\\
\\
\noindent
{\bf Megavatio:} \indent
1,000,000 vatios.\\
\\
\noindent
{\bf Megavatio hora (Mwh, por sus siglas en ingl\'es):} \indent
1,000,000 vatios por 1 hora, que es igual a $3.6 \times 10^9$ Julios.\\
\\
\noindent
{\bf mercado:} \indent
un medio que permite a los compradores y vendedores interactuar y facilitar el intercambio de activos, productos, bienes, servicios, etc.\\
\\
\noindent
{\bf mercado de valores:} \indent
un mercado de acciones.\\
\\
\noindent
{\bf Mercado H\'ibrido:} \indent
una combinaci\'on de una plataforma de trading electr\'onica automatizada y un sistema de corredores tradicionales (operado por humanos).\\
\\
\noindent
{\bf mercado lateral:} \indent
cuando los precios se mantienen en un rango estrecho, sin tendencias al alza o a la baja.\\
\\
\noindent
{\bf mercado muy alcista (rally en ingl\'es):} \indent
en los mercados financieros, un per\'iodo de ganancias sostenidas.\\
\\
\noindent
{\bf mes de entrega:} \indent
el mes en que se produce la entrega en un contrato de futuros.\\
\\
\noindent
{\bf m\'etodo de bisecci\'on:} \indent
un m\'etodo de b\'usqueda de ra\'ices que divide repetidamente un intervalo y selecciona un subintervalo en el que una ra\'iz debe estar presente para una examinaci\'on m\'as detenida.\\
\\
\noindent
{\bf m\'etodo de Whittaker-Henderson:} \indent
v\'ease {\em filtro de Hodrick-Prescott}.\\
\\
\noindent
{\bf miner\'ia de datos:} \indent
un proceso de b\'usqueda de patrones/tendencias en grandes conjuntos de datos.\\
\\
\noindent
{\bf m\'inimo (tambi\'en conocido como precio m\'inimo, o low en ingl\'es):}
el precio m\'inimo alcanzado por una acci\'on (u otro activo) dentro de un d\'ia de trading determinado (o en alg\'un otro intervalo de tiempo).\\
\\
\noindent
{\bf m\'inimos cuadrados:} \indent
en el an\'alisis de regresi\'on, esto es la minimizaci\'on de la suma de los cuadrados de los residuos (posiblemente, con algunas ponderaciones no uniformes).\\
\\
\noindent
{\bf m\'inimos cuadrados no lineales:} \indent
los m\'inimos cuadrados utilizados para ajustar un conjunto de observaciones con un modelo que no es lineal en los par\'ametros desconocidos (o coeficientes de la regresi\'on).\\
\\
\noindent
{\bf MKT:} \indent
v\'ease {\em factores de Fama-French}.\\
\\
\noindent
{\bf modelo de Black-Scholes (tambi\'en conocido como modelo de Black-Scholes-Merton):} \indent
un modelo matem\'atico de la din\'amica de acciones (u otro activo subyacente) utilizado en la valuaci\'on de opciones y otros derivados, en donde el logaritmo del precio del activo subyacente se describe mediante un movimiento Browniano con una deriva constante.\\
\\
\noindent
{\bf modelo de factor de conversi\'on:} \indent
un modelo (basado en el factor de conversi\'on) com\'unmente utilizado para calcular los ratios de cobertura cuando se cubre el riesgo de tasa de inter\'es con los futuros de tasa de inter\'es.\\
\\
\noindent
{\bf modelo de riesgo:} \indent
un modelo matem\'atico para estimar el riesgo (por ejemplo, modelar una matriz de covarianza).\\
\\
\noindent
{\bf modelo de valuaci\'on:} \indent
un modelo para valuar (tasar) un activo o un conjunto de activos.\\
\\
\noindent
{\bf modelo estad\'istico de riesgo:} \indent
un modelo de riesgo creado utilizando solo los datos de precios (por ejemplo, utilizando los componentes principales de la matriz de correlaci\'on muestral de los retornos de las acciones), sin ninguna referencia a datos fundamentales (incluida cualquier clasificaci\'on industrial fundamental).\\
\\
\noindent
{\bf modelo heter\'otico de riesgo:} \indent
un modelo multifactorial de riesgo que combina una clasificaci\'on fundamental multinivel de la industria con an\'alisis de los principales componentes.\\
\\
\noindent
{\bf modelo multifactorial de riesgo:} \indent
un modelo de riesgo basado en un n\'umero (que puede ser considerable) de factores de riesgo.\\
\\
\noindent
{\bf MOM:} \indent
v\'ease {\em factor de momentum de Carhart}.\\
\\
\noindent
{\bf momentum (tambi\'en conocido como efecto momentum):} \indent
la observaci\'on emp\'irica de que los retornos futuros de los activos se correlacionan positivamente con sus retornos pasados.\\
\\
\noindent
{\bf momentum absoluto:} \indent
momentum de series de tiempo.\\
\\
\noindent
{\bf momentum relativo:} \indent
momentum de corte transversal.\\
\\
\noindent
{\bf moneda digital descentralizada:} \indent
una criptomoneda descentralizada como el BTC, el ETH, etc.\\
\\
\noindent
{\bf moneda dom\'estica:} \indent
la moneda del pa\'is de origen del trader.\\
\\
\noindent
{\bf moneda extranjera:} \indent
una moneda (que es diferente de la moneda dom\'estica) de un pa\'is que no es el pa\'is de origen del trader.\\
\\
\noindent
{\bf moneda fiduciaria:} \indent
una oferta legal declarada por un gobierno (por ejemplo, el d\'olar estadounidense) pero no respaldada por una mercanc\'ia f\'isica (como el oro).\\
\\
\noindent
{\bf movimiento Browniano (tambi\'en conocido como proceso de Wiener):}
un proceso estoc\'astico $W_t$ en tiempo continuo ($t$), en donde $W_0 = 0$, $W_t$ es una variable aleatoria normal con media 0 y varianza $t$, y el incremento $W_ {s + t} - W_s$ es una variable aleatoria normal con media 0 y varianza $t$ y es independiente de la historia de lo que hizo el proceso hasta el momento $s$.\\
\\
\noindent
{\bf multiplicador de Lagrange:} \indent
al minimizar una funci\'on (multivariada) $g(x)$ con respecto a $x$ sujeta a una restricci\'on $h(x)=0$, la variable adicional $\lambda$ en la funci\'on ${\widetilde g}(x, \lambda) = g(x) + \lambda~h(x)$, cuya minimizaci\'on (sin ningunas restricciones) con respecto a $x$ y $\lambda$ es equivalente al problema de minimizaci\'on restringido original.\\
\\
\noindent
{\bf nivel de prioridad de la deuda:} \indent
el orden en que se realizan los reembolsos de la deuda en caso de venta o quiebra del emisor de la misma.\\
\\
\noindent
{\bf nivel del \'indice:} \indent
para los \'indices ponderados por capitalizaci\'on burs\'atil, el valor actual del nivel del \'indice $I(t) = I(0) \times C(t) / C(0)$, en donde $I(0)$ es el valor inicial del nivel del \'indice (que se define, no se calcula), $C(t)$ es la capitalizaci\'on burs\'atil total actual de los constituyentes del \'indice, y $C(0)$ es su valor inicial.\\
\\
\noindent
{\bf nocional (tambi\'en conocido como valor nocional):} \indent
el valor total (en d\'olares) de una posici\'on.\\
\\
\noindent
{\bf nodo:} \indent
en una red neuronal artificial, una neurona artificial, que (mediante una funci\'on de activaci\'on) procesa un conjunto de datos de entrada y genera uno(s) de salida.\\
\\
\noindent
{\bf nota del Tesoro (tambi\'en conocida como T-note en ingl\'es):} \indent
un t\'itulo de deuda emitido por el Tesoro de los Estados Unidos con un vencimiento entre 1 y 10 a\~{n}os.\\
\\
\noindent
{\bf obligaci\'on de deuda garantizada (CDO, por sus siglas en ingl\'es):}
un valor respaldado por activos (ABS, por sus siglas en ingl\'es) que consiste en una cesta de activos tales como bonos, swaps de incumplimiento crediticio, etc.\\
\\
\noindent
{\bf opci\'on:} \indent
un contrato de derivados financieros que otorga al comprador (titular de la opci\'on) el derecho (pero no la obligaci\'on) de comprar (opci\'on de compra) o vender (opci\'on de venta) el activo subyacente a un precio previamente acordado y durante un per\'iodo de tiempo predefinido o en una fecha espec\'ifica.\\
\\
\noindent
{\bf opci\'on americana:} \indent
una opci\'on (por ejemplo, call o put) que se puede ejercer en cualquier d\'ia de trading en o antes del vencimiento.\\
\\
\noindent
{\bf opci\'on asi\'atica:} \indent
una opci\'on cuyo pago est\'a determinado por el precio promedio del subyacente durante un per\'iodo de tiempo predeterminado.\\
\\
\noindent
{\bf opci\'on bermuda:} \indent
una opci\'on que se puede ejercer solo en fechas especificadas en o antes del vencimiento.\\
\\
\noindent
{\bf opci\'on binaria (tambi\'en conocida como opci\'on digital, u opci\'on de todo o nada):} \indent
una opci\'on que paga un monto preestablecido, por ejemplo, \$1, si el activo subyacente cumple con una condici\'on predefinida al vencimiento, de lo contrario, simplemente caduca sin pagar nada al titular.\\
\\
\noindent
{\bf opci\'on call (tambi\'en conocida como call en ingl\'es, u opci\'on de compra):}
v\'ease {\em opci\'on de compra europea, opci\'on}.\\
\\
\noindent
{\bf opci\'on canaria:} \indent
una opci\'on que se puede ejercer, por ejemplo, trimestralmente, pero no antes de que haya transcurrido un per\'iodo de tiempo determinado, por ejemplo, 1 a\~{n}o.\\
\\
\noindent
{\bf opci\'on con barrera:} \indent
una opci\'on que se puede ejercer solo si el precio del activo subyacente pasa un cierto nivel o una ``barrera''.\\
\\
\noindent
{\bf opci\'on de acci\'on individual:} \indent
una opci\'on en una sola acci\'on subyacente (a diferencia de, por ejemplo, una opci\'on de una cesta de acciones tal como un \'indice).\\
\\
\noindent
{\bf opci\'on de compra europea (u opci\'on call europea):} \indent
un derecho (pero no una obligaci\'on) de comprar una acci\'on al vencimiento $T$ al precio de ejercicio $k$ acordado en el momento $t = 0$.\\
\\
\noindent
{\bf opci\'on de venta europea (u opci\'on put europea):} \indent
un derecho (pero no una obligaci\'on) de vender una acci\'on al vencimiento $T$ al precio de ejercicio $k$ acordado en el momento $t = 0$.\\
\\
\noindent
{\bf opci\'on dentro del dinero (ITM, por sus siglas en ingl\'es):} \indent
una opci\'on de compra (venta) cuyo precio de ejercicio est\'a por debajo (por encima) del precio actual del activo subyacente.\\
\\
\noindent
{\bf opci\'on embebida:} \indent
en un bono convertible, la opci\'on que permite convertir el bono a un n\'umero preestablecido de las acciones del emisor.\\
\\
\noindent
{\bf opci\'on en el dinero (ATM, por sus siglas en ingl\'es):} \indent
una opci\'on cuyo precio de ejercicio es el mismo que el precio actual del activo subyacente.\\
\\
\noindent
{\bf opci\'on fuera del dinero (OTM, por sus siglas en ingl\'es):} \indent
una opci\'on de compra (venta) cuyo precio de ejercicio est\'a por encima (por debajo) del precio actual del activo subyacente.\\
\\
\noindent
{\bf opci\'on put (tambi\'en conocida como put en ingl\'es, u opci\'on de venta):}
v\'ease {\em opci\'on de venta europea, opci\'on}.\\
\\
\noindent
{\bf opciones ex\'oticas:} \indent
una amplia categor\'ia de opciones que normalmente est\'an estructuradas de manera compleja.\\
\\
\noindent
{\bf operaci\'on de arbitraje:} \indent
un conjunto de transacciones ejecutadas con el fin de aprovechar una oportunidad de arbitraje.\\
\\
\noindent
{\bf optimizaci\'on:} \indent
v\'ease {\em optimizaci\'on de portafolio}.\\
\\
\noindent
{\bf optimizaci\'on de media-varianza:} \indent
una t\'ecnica de optimizaci\'on para construir un portafolio de activos de manera tal que su rendimiento esperado se maximice para un nivel dado de riesgo.\\
\\
\noindent
{\bf optimizaci\'on de portafolio:} \indent
seleccionar el mejor portafolio en funci\'on de alg\'un criterio (por ejemplo, maximizar el ratio de Sharpe).\\
\\
\noindent
{\bf orden:} \indent
las instrucciones de un inversionista a un corredor o firma de corretaje para comprar o vender un activo.\\
\\
\noindent
{\bf orden agresiva:} \indent
una orden de mercado, o una orden de l\'imite comercializable (para comprar en el ask o m\'as arriba, o para vender en el bid o m\'as abajo).\\
\\
\noindent
{\bf orden cancelada:} \indent
una orden realizada que ha sido cancelada posteriormente.\\
\\
\noindent
{\bf orden de cancelaci\'on-reemplazo:} \indent
una orden que se ha pedido pero que se cancel\'o posteriormente y se reemplaz\'o por otra.\\
\\
\noindent
{\bf orden l\'imite:} \indent
una orden para comprar o vender una acci\'on (u otro activo) a un precio espec\'ifico o mejor.\\
\\
\noindent
{\bf orden l\'imite pasiva:} \indent
una orden l\'imite de provisi\'on de liquidez para comprar en el bid (o inferior) o vender en el ask (o superior).\\
\\
\noindent
{\bf orden realizada:} \indent
una orden que se ha enviado a un mercado de valores y se ha colocado en una cola para su ejecuci\'on.\\
\\
\noindent
{\bf ordenar:} \indent
organizar un conjunto de forma ascendente o descendente en funci\'on de cierta cantidad (con ciertas instrucciones para resolver cualesquiera confusiones [to resolve ties, en ingl\'es]).\\
\\
\noindent
{\bf ortogonalidad:} \indent
los vectores $x_i$ y $y_i$ ($i=1,\dots,N$) son ortogonales si $\sum_{i=1}^N x_i~y_i = 0$.\\
\\
\noindent
{\bf P\&L realizado:} \indent
el P\&L en una operaci\'on completada, es decir, el P\&L resultante de establecer una posici\'on y luego liquidarla completamente.\\
\\
\noindent
{\bf pago:} \indent
la cantidad que el vendedor de la opci\'on paga al comprador siempre y cuando se ejerza la opci\'on.\\
\\
\noindent
{\bf pago de tasa fija:} \indent
un pago de cup\'on de un bono con cup\'on fijo.\\
\\
\noindent
{\bf pago de tasa flotante:} \indent
un pago de cup\'on de un bono con cup\'on flotante.\\
\\
\noindent
{\bf pago indexado:} \indent
los pagos ajustados seg\'un el valor de alg\'un \'indice (por ejemplo, el CPI) en swaps de inflaci\'on o TIPS.\\
\\
\noindent
{\bf palabra (tambi\'en conocida como palabra clave):} \indent
una palabra clave en un vocabulario de aprendizaje.\\
\\
\noindent
{\bf palabra clave:} \indent
en an\'alisis de sentimiento (por ejemplo, sentimiento de Twitter) usando t\'ecnicas de aprendizaje autom\'atico, una palabra en el vocabulario de aprendizaje que tiene una relaci\'on con el objetivo (por ejemplo, predecir los movimientos de precios de las acciones o criptomonedas).\\
\\
\noindent
{\bf palabra vac\'ia:} \indent
las palabras m\'as utilizadas en un idioma (por ejemplo, ``el'', ``es'', ``en'', ``cu\'al'', etc., o las palabras similares en ingl\'es u otra lengua) que no agregan ning\'un valor en un contexto particular y son ignorados por una herramienta de procesamiento de lenguaje natural.\\
\\
\noindent
{\bf papel comercial:} \indent
pagar\'es no asegurados a corto plazo emitidos por empresas.\\
\\
\noindent
{\bf par de FX:} \indent
monedas de 2 pa\'ises diferentes.\\
\\
\noindent
{\bf par\'ametro de reversi\'on a la media:} \indent
en un proceso Ornstein-Uhlenbeck que revierte a la media, el par\'ametro que controla la tasa de reversi\'on a la media.\\
\\
\noindent
{\bf par\'ametro de suavizaci\'on exponencial:} \indent
el factor de supresi\'on exponencial en una media m\'ovil exponencial.\\
\\
\noindent
{\bf paridad put-call:} \indent
la relaci\'on por la cual el pago de una opci\'on call europea (con un precio de ejercicio de $K$ y el vencimiento $T$) menos el pago de una opci\'on put europea (en el mismo subyacente y con el mismo precio de ejercicio y el mismo vencimiento) es igual al pago de un contrato forward (sobre el mismo subyacente) con el precio de ejercicio $K$ y el vencimiento $T$.\\
\\
\noindent
{\bf perdedores:} \indent
las acciones u otros activos en un portafolio o universo de trading que tienen un rendimiento inferior en funci\'on de alg\'un criterio (rendimiento de referencia).\\
\\
\noindent
{\bf p\'erdida de roll (tambi\'en conocida como p\'erdida de contango):} \indent
en ETNs como el VXX y el VXZ que consisten en portafolios de futuros del VIX, una ca\'ida en sus valores (cuando la curva de futuros del VIX est\'a en contango) debido a su reequilibrio diario requerido para mantener un vencimiento constante.\\
\\
\noindent
{\bf per\'iodo de backtesting:} \indent
el per\'iodo hist\'orico en el que se realiza un backtest.\\
\\
\noindent
{\bf per\'iodo de composici\'on:} \indent
el per\'iodo entre dos puntos consecutivos en el tiempo en el que los intereses se pagan o se suman al principal.\\
\\
\noindent
{\bf per\'iodo de entrenamiento:} \indent
en aprendizaje autom\'atico, el per\'iodo que abarcan los datos de entrenamiento cuando se trata de una serie de tiempo.\\
\\
\noindent
{\bf per\'iodo de estimaci\'on:} \indent
la longitud de una muestra de datos de una serie de tiempo utilizada para estimar algunos par\'ametros, por ejemplo, los coeficientes de una regresi\'on.\\
\\
\noindent
{\bf per\'iodo de formaci\'on:} \indent
en las estrategias de momentum, el per\'iodo durante el cual se calcula el indicador de momentum.\\
\\
\noindent
{\bf per\'iodo de omisi\'on:} \indent
en las estrategias de momentum de precios y estrategias similares, el per\'iodo (generalmente, el \'ultimo mes) que se omite antes del per\'iodo de formaci\'on (generalmente, los \'ultimos 12 meses antes del per\'iodo de omisi\'on).\\
\\
\noindent
{\bf per\'iodo de pago:} \indent
el per\'iodo entre dos pagos consecutivos de cup\'on de un bono.\\
\\
\noindent
{\bf per\'iodo de tenencia:} \indent
el per\'iodo durante el cual una posici\'on en un activo o un portafolio se mantiene despu\'es de establecerse y antes de liquidarse (o, de manera m\'as general, de rebalancearse).\\
\\
\noindent
{\bf permuta de incumplimiento crediticio (CDS, por sus siglas en ingl\'es):}
un swap que proporciona un seguro contra el incumplimiento de un bono.\\
\\
\noindent
{\bf perspectiva alcista:} \indent
cuando un trader espera que el mercado o un activo cotice al alza.\\
\\
\noindent
{\bf perspectiva bajista:} \indent
cuando un trader espera que el mercado o un activo cotice a la baja.\\
\\
\noindent
{\bf perspectiva neutral:} \indent
cuando un trader espera que el mercado o un activo cotice alrededor de su nivel actual.\\
\\
\noindent
{\bf pico:} \indent
un movimiento relativamente grande hacia arriba o hacia abajo del precio de un activo en un corto per\'iodo de tiempo.\\
\\
\noindent
{\bf pierna:} \indent
un componente en un portafolio de trading, generalmente cuando se puede pensar que dicho portafolio consiste en un n\'umero relativamente peque\~{n}o de grupos (por ejemplo, pierna larga y pierna corta, en referencia a posiciones largas y cortas, respectivamente).\\
\\
\noindent
{\bf pierna contingente:} \indent
en un CDO, la pierna de incumplimiento, la otra pierna es la pierna premium.\\
\\
\noindent
{\bf pierna delantera:} \indent
los bonos de vencimiento m\'as corto en una estrategia de margen de la curva de rendimientos (flattener o steepener).\\
\\
\noindent
{\bf pierna premium:} \indent
la pierna de un CDO correspondiente a las primas del CDO, la otra pierna es la pierna de incumplimiento.\\
\\
\noindent
{\bf pierna trasera:} \indent
los bonos de vencimiento m\'as largo en una estrategia de margen de la curva de rendimientos (flattener o steepener).\\
\\
\noindent
{\bf plan de reorganizaci\'on:} \indent
un plan de reorganizaci\'on de una empresa en quiebra que puede ser presentado (por ejemplo, por un acreedor con el fin de obtener una participaci\'on en la administraci\'on de la empresa) al Tribunal para su aprobaci\'on.\\
\\
\noindent
{\bf pol\'itica monetaria:} \indent
un proceso por el cual la autoridad monetaria (generalmente, un banco central) de un pa\'is controla el tama\~{n}o y la tasa de crecimiento de la oferta monetaria, modificando las tasas de inter\'es, comprando o vendiendo bonos gubernamentales y cambiando las reservas bancarias requeridas (la cantidad de dinero que los bancos deben mantener de forma l\'iquida).\\
\\
\noindent
{\bf ponderaciones (en portafolios):} \indent
v\'ease {\em ponderaciones de un portafolio}.\\
\\
\noindent
{\bf ponderaciones (en una ANN):} \indent
en una red neuronal artificial, los coeficientes de las entradas en el argumento de una funci\'on de activaci\'on.\\
\\
\noindent
{\bf ponderaciones de la regresi\'on:} \indent
las ponderaciones positivas $w_i$ (que no necesitan ser igual a 1) en la suma de los cuadrados $\sum_{i=1}^N w_i~\epsilon_i^2$, cuya minimizaci\'on determina los coeficientes de la regresi\'on y los residuos de la regresi\'on $\epsilon_i$.\\
\\
\noindent
{\bf ponderaciones de tenencias:} \indent
las ponderaciones que se asignan a los distintos activos en un portafolio.\\
\\
\noindent
{\bf ponderaciones de un portafolio:} \indent
los porcentajes relativos de las tenencias en d\'olares en un portafolio con respecto a su valor nocional total (definido como el valor nocional total de las posiciones largas mas el valor nocional absoluto total de las posiciones cortas).\\
\\
\noindent
{\bf porcentaje de asignaci\'on a commodities (CA, por sus siglas en ingl\'es):}
la ponderaci\'on de la asignaci\'on para commodities incluidos como la cobertura de inflaci\'on en un portafolio de otros activos.\\
\\
\noindent
{\bf portafolio:} \indent
una colecci\'on de activos en poder de una instituci\'on o inversionista individual.\\
\\
\noindent
{\bf portafolio de alfas (tambi\'en conocido como combo alfa):} \indent
un portafolio de (t\'ipicamente, un gran n\'umero de) alfas combinados con ciertas ponderaciones.\\
\\
\noindent
{\bf portafolio de factor:} \indent
un portafolio cuyo objetivo es alcanzar la exposici\'on a un factor dado.\\
\\
\noindent
{\bf portafolio igualmente ponderado:} \indent
un portafolio donde todos los activos tienen todas las tenencias en d\'olares iguales.\\
\\
\noindent
{\bf posici\'on:} \indent
la cantidad de acciones u de otro tipo de activo dentro de un portafolio, expresada en d\'olares, un n\'umero de acciones o algunas otras unidades, con las posiciones cortas teniendo posiblemente valores negativos seg\'un la convenci\'on utilizada.\\
\\
\noindent
{\bf posici\'on corta:} \indent
vender un activo sin poseerlo pidiendo prestado de otra persona o entitad, por lo general, una firma de corretaje.\\
\\
\noindent
{\bf precio ajustado:} \indent
el precio de una acci\'on ajustado por los splits y dividendos.\\
\\
\noindent
{\bf precio de conversi\'on:} \indent
el precio de las acciones subyacentes a las que un bono convertible se puede convertir en las acciones.\\
\\
\noindent
{\bf precio de ejecuci\'on:} \indent
el precio al que se llena (ejecuta) una orden (por ejemplo, para comprar acciones).\\
\\
\noindent
{\bf precio de ejercicio (o strike en ingl\'es):} \indent
el precio al que se puede ejercer un contrato derivado.\\
\\
\noindent
{\bf precio de equilibrio (tambi\'en conocido como punto de equilibrio):}
un precio del activo subyacente (por ejemplo, acciones) en una estrategia de trading con opciones en el cual se llega al equilibrio (es decir, cuando el P\&L es cero).\\
\\
\noindent
{\bf precio de stop-loss:} \indent
el precio de un activo en el cual una posici\'on en dicho activo se liquida (autom\'aticamente).\\
\\
\noindent
{\bf precio spot del \'indice:} \indent
el precio de mercado actual de una cesta del \'indice, en donde el n\'umero de unidades de cada activo constituyente est\'a determinado por el esquema de ponderaci\'on del \'indice (\'indice ponderado por capitalizaci\'on burs\'atil, \'indice ponderado por precios, etc.) con la constante de normalizaci\'on global fija dependiendo de un prop\'osito espec\'ifico, por ejemplo, para coincidir con la cesta del \'indice que se entrega al vencimiento de los futuros del \'indice en el caso de arbitraje de \'indice.\\
\\
\noindent
{\bf predictor (tambi\'en conocido como variable predictiva):} \indent
en aprendizaje autom\'atico, una variable de entrada.\\
\\
\noindent
{\bf predictor de Cochrane-Piazzesi:} \indent
un predictor de retornos de bonos.\\
\\
\noindent
{\bf premio a plazo:} \indent
cuando la tasa de FX a plazo es m\'as alta que la tasa de FX spot.\\
\\
\noindent
{\bf prepago:} \indent
liquidar una deuda o un pago a plazos en una fecha anterior a su fecha de vencimiento (por ejemplo, prepagar una hipoteca).\\
\\
\noindent
{\bf presi\'on de cobertura (HP, por sus siglas en ingl\'es):} \indent
en los mercados de futuros (de commodities), el n\'umero de posiciones largas dividido por el n\'umero total de contratos abiertos (posiciones largas m\'as posiciones cortas).\\
\\
\noindent
{\bf prestamista (en contexto de empe\~{n}o):} \indent
un prestamista que otorga un pr\'estamo asegurado en efectivo con la tasa de inter\'es y per\'iodo (que a veces se pueden extender) previamente acordados, en donde el pr\'estamo est\'a asegurado con un colateral (el prestatario lo pierde si no cumple), que es alg\'un art\'iculo(s) valioso(s), tales como joyer\'ia, electr\'onica, veh\'iculos, libros o instrumentos musicales raros, etc.\\
\\
\noindent
{\bf pr\'estamo:} \indent
prestar dinero u otro activo por una parte (prestamista) a otra (prestatario).\\
\\
\noindent
{\bf pr\'estamo asegurado:} \indent
un pr\'estamo asegurado con un colateral.\\
\\
\noindent
{\bf pr\'estamo garantizado:} \indent
un pr\'estamo garantizado por un tercero en caso de que el prestatario incumpla con el pr\'estamo.\\
\\
\noindent
{\bf previsi\'on de retornos futuros:} \indent
predecir retornos futuros.\\
\\
\noindent
{\bf prima (para opciones):} \indent
el costo de comprar una opci\'on.\\
\\
\noindent
{\bf prima (para productos tales como un seguro):} \indent
un pago peri\'odico para obtener la cobertura del seguro, por ejemplo, en un CDS, CDO, etc.\\
\\
\noindent
{\bf prima de asimetr\'ia:} \indent
en los futuros de commodities, un acontecimiento emp\'irico por el cual los rendimientos futuros esperados tienden a estar correlacionados negativamente con la asimetr\'ia de los rendimientos hist\'oricos.\\
\\
\noindent
{\bf prima de la opci\'on:} \indent
el costo que cobra el vendedor de la opci\'on al comprador.\\
\\
\noindent
{\bf prima de riesgo:} \indent
el rendimiento en exceso (esperado) de la tasa libre de riesgo obtenido de una inversi\'on.\\
\\
\noindent
{\bf prima de riesgo de volatilidad:} \indent
una ocurrencia emp\'irica que la volatilidad impl\'icita tiende a ser mayor que la volatilidad realizada la mayor parte del tiempo.\\
\\
\noindent
{\bf principal:} \indent
el monto que el emisor de la deuda (prestatario) debe al prestamista al vencimiento de la deuda.\\
\\
\noindent
{\bf principal de un bono:} \indent
la cantidad que el prestatario (el emisor del bono) debe al titular del bono en su totalidad al vencimiento.\\
\\
\noindent
{\bf probabilidad condicional:} \indent
$P(A|B)$, la probabilidad que $A$ ocurra asumiendo que $B$ es cierto.\\
\\
\noindent
{\bf proceso de Ornstein-Uhlenbeck:} \indent
un movimiento Browniano con reversi\'on a la media.\\
\\
\noindent
{\bf proceso estoc\'astico:} \indent
un grupo de variables aleatorias que cambian con el tiempo.\\
\\
\noindent
{\bf productos indexados a la inflaci\'on:} \indent
un activo (por ejemplo, TIPS) con pagos indexados basados en un \'indice de inflaci\'on.\\
\\
\noindent
{\bf promedio ponderado:} \indent
para los $N$ valores $x_i$ ($i=1,\dots,N$), la media ponderada dada por ${1\over N}\sum_{i=1}^N w_i~x_i$, en donde $w_i$ son las ponderaciones.\\
\\
\noindent
{\bf propiedades industriales:} \indent
propiedades inmobiliarias comerciales, incluyendo f\'abricas y otras propiedades, almacenes, etc.\\
\\
\noindent
{\bf proyecto brownfield:} \indent
un proyecto asociado con activos de infraestructura establecidos que necesitan mejoras.\\
\\
\noindent
{\bf proyecto greenfield:} \indent
un proyecto asociado a los activos de infraestructura a construir.\\
\\
\noindent
{\bf punto b\'asico (bps, por sus siglas en ingl\'es):} \indent
1/100 de 1\%.\\
\\
\noindent
{\bf punto de intervenci\'on:} \indent
el porcentaje de la p\'erdida del portafolio subyacente en el que un tramo de un CDO (obligaci\'on de deuda garantizada) comienza a perder su valor.\\
\\
\noindent
{\bf punto de pivote (tambi\'en conocido como centro):} \indent
en las estrategias de soporte y resistencia, una cantidad dependiente de su definici\'on, por ejemplo, puede estar definida como el promedio ponderado por igual de los precios m\'aximo, m\'inimo y del cierre del d\'ia de trading anterior.\\
\\
\noindent
{\bf put casada:} \indent
v\'ease {\em put protectora}.\\
\\
\noindent
{\bf put cubierta:} \indent
v\'ease {\em estrategia de venta-lanzamiento}.\\
\\
\noindent
{\bf put desnuda:} \indent
una opci\'on put corta independiente (sin ninguna otra posici\'on).\\
\\
\noindent
{\bf put protectora (tambi\'en conocida como put casada, o call sint\'etica):}
cubrir una posici\'on larga en una acci\'on con una posici\'on larga en una opci\'on put.\\
\\
\noindent
{\bf quintil:} \indent
cada una de las 5 partes (aproximadamente) iguales de una muestra (por ejemplo, muestra de datos).\\
\\
\noindent
{\bf R:} \indent
R Package for Statistical Computing.\\
\\
\noindent
{\bf ra\'iz cuadrada del error cuadr\'atico medio (RMSE, por sus siglas en ingl\'es):} \indent
la ra\'iz cuadrada del valor medio de los cuadrados de las diferencias entre los valores predichos y observados de una variable.\\
\\
\noindent
{\bf rango (para matrices):} \indent
el n\'umero m\'aximo de columnas linealmente independientes de una matriz.\\
\\
\noindent
{\bf rank (tambi\'en conocido como clasificaci\'on, o ranking en ingl\'es):}
la posici\'on de un elemento de un conjunto despu\'es de clasificarlo de acuerdo con alg\'un criterio (con ciertas instrucciones para resolver cualesquiera confusiones [to resolve ties, en ingl\'es]).\\
\\
\noindent
{\bf ratio de cobertura:} \indent
en una estrategia de cobertura, el n\'umero de unidades (o el nocional en d\'olares) del activo que cubre para cada unidad (o d\'olar) del activo a cubrir.\\
\\
\noindent
{\bf ratio de cobertura \'optimo:} \indent
un ratio de cobertura calculado mediante la minimizaci\'on de la varianza de un portafolio cubierto.\\
\\
\noindent
{\bf ratio de conversi\'on:} \indent
el n\'umero de acciones del emisor en las que se puede convertir un bono convertible.\\
\\
\noindent
{\bf ratio de Sharpe:} \indent
el retorno (en exceso) dividido por la volatilidad.\\
\\
\noindent
{\bf ratio de Sharpe anualizado:} \indent
el ratio de Sharpe diario multiplicado por la ra\'iz cuadrada de 252 (el n\'umero de d\'ias de trading en un a\~{n}o).\\
\\
\noindent
{\bf ratio del valor de libros sobre mercado:} \indent
el valor contable total de la compa\~{n}\'ia dividido por su capitalizaci\'on de mercado (igual al ratio B/P).\\
\\
\noindent
{\bf ratio del valor de libros sobre precio (tambi\'en conocido como ratio B/P, por sus siglas en ingl\'es):} \indent
el valor de libros de la compa\~{n}\'ia por las acciones en circulaci\'on dividido por el precio de sus acciones.\\
\\
\noindent
{\bf R-cuadrado:} \indent
en una regresi\'on, 1 menos un ratio, cuyo numerador es la suma de los cuadrados de los residuos, y cuyo denominador es la suma de los cuadrados de las desviaciones de los valores de la variable observable con respecto a su valor medio en la muestra de datos.\\
\\
\noindent
{\bf rebalancear:} \indent
cambiar las ponderaciones de tenencia en una cartera.\\
\\
\noindent
{\bf red de entre pares (P2P, por sus siglas en ingl\'es):} \indent
una arquitectura de aplicaci\'on inform\'atica distribuida con carga de trabajo particionada entre los pares igualmente privilegiados.\\
\\
\noindent
{\bf red neuronal artificial (ANN, por sus siglas en ingl\'es):} \indent
un sistema de computaci\'on (inspirado por la estructura neural de un cerebro) de nodos (neuronas artificiales) vinculados por las conexiones (similares a las sinapsis en un cerebro) que pueden transmitir las se\~{n}ales de un nodo a otro.\\
\\
\noindent
{\bf regiones de EE.UU.:} \indent
Este, Medio Oeste, Sur y Oeste.\\
\\
\noindent
{\bf regla de trading:} \indent
un conjunto de instrucciones de compra y venta, con las cantidades de los activos que se comprar\'an o vender\'an.\\
\\
\noindent
{\bf regresi\'on:} \indent
v\'ease {\em regresi\'on lineal}.\\
\\
\noindent
{\bf regresi\'on lineal (tambi\'en conocida como modelo lineal):} \indent
ajustar los datos para la variable observable mediante una combinaci\'on lineal de cierto n\'umero de funciones (lineales o no lineales) de las variables independientes, con o sin el intercepto.\\
\\
\noindent
{\bf regresi\'on log\'istica (tambi\'en conocida como modelo logit):} \indent
un modelo estad\'istico que generalmente se aplica a una variable dependiente binaria.\\
\\
\noindent
{\bf regresi\'on ponderada:} \indent
una regresi\'on lineal con las ponderaciones de regresi\'on no uniformes.\\
\\
\noindent
{\bf regresi\'on restringida:} \indent
una regresi\'on lineal sujeta a un conjunto de restricciones lineales o no lineales, por ejemplo, los m\'inimos cuadrados no negativos (NNLS, por sus siglas en ingl\'es), en donde se requiere que los coeficientes de la regresi\'on sean no negativos.\\
\\
\noindent
{\bf regresi\'on serial:} \indent
una regresi\'on en donde las variables independientes son series de tiempo (comparado con {\em regresi\'on transversal}).\\
\\
\noindent
{\bf regresi\'on transversal (la misma que regresi\'on de corte transversal):}
una regresi\'on en la que las variables independientes son los vectores cuyos elementos est\'an etiquetados por un \'indice de corte transversal, por ejemplo, el que marca las acciones en un portafolio (comparado con {\em regresi\'on serial}).\\
\\
\noindent
{\bf rendimiento:} \indent
v\'ease {\em rendimiento de un bono}.\\
\\
\noindent
{\bf rendimiento de un bono (aqu\'i, rendimiento al vencimiento):} \indent
la tasa de inter\'es general obtenida asumiendo que el bono se mantiene hasta el vencimiento y que todos los cupones y pagos de capital se realizan seg\'un lo acordado.\\
\\
\noindent
{\bf reorganizaci\'on:} \indent
un proceso de reestructuraci\'on de las finanzas de una compa\~{n}\'ia en bancarrota supervisado por el Tribunal.\\
\\
\noindent
{\bf replicaci\'on:} \indent
una estrategia mediante la cual un portafolio din\'amico de activos replica con precisi\'on los flujos de efectivo de otro activo o portafolio.\\
\\
\noindent
{\bf residuos de la regresi\'on:} \indent
las diferencias entre los valores observados y los valores ajustados (predichos por el modelo) de la variable dependiente en una regresi\'on lineal.\\
\\
\noindent
{\bf resistencia:} \indent
en an\'alisis t\'ecnico, el nivel de precios (percibido) en el que se espera que el precio de las acciones que est\'an subiendo rebote hacia abajo.\\
\\
\noindent
{\bf restricci\'on lineal homog\'enea:} \indent
para un $N$-vector $x_i$ ($i=1,\dots,N$), una restricci\'on de la forma $\sum_{i=1}^N a_i~x_i = 0$, en donde al menos algunos $a_i$ son distintos de cero.\\
\\
\noindent
{\bf resultado (tambi\'en conocido como clase):} \indent
en aprendizaje autom\'atico, uno de los posibles resultados (predicciones) de un algoritmo de aprendizaje autom\'atico.\\
\\
\noindent
{\bf retorno acumulado:} \indent
el retorno de un activo desde el momento $t_1$ al tiempo $t_2$.\\
\\
\noindent
{\bf retorno ajustado por riesgo:} \indent
el retorno dividido por la volatilidad.\\
\\
\noindent
{\bf retorno anualizado:} \indent
un retorno diario promedio multiplicado por 252 (el n\'umero de d\'ias de trading en un a\~{n}o).\\
\\
\noindent
{\bf retorno de cierre contra apertura:} \indent
el retorno desde el cierre del d\'ia de trading anterior a la apertura del d\'ia de trading actual.\\
\\
\noindent
{\bf retorno de cierre contra cierre:} \indent
el retorno desde el cierre del d\'ia de trading anterior al cierre del d\'ia de trading actual.\\
\\
\noindent
{\bf retorno esperado:} \indent
un retorno futuro esperado de un activo basado en alguna consideraci\'on razonable, por ejemplo, el retorno promedio realizado durante un per\'iodo de tiempo pasado.\\
\\
\noindent
{\bf retorno logar\'itmico:} \indent
el logaritmo natural del ratio entre el precio de un activo en el momento $t_2$ y su precio en el momento $t_1$ ($t_2 > t_1$).\\
\\
\noindent
{\bf retorno nocturno:} \indent
en t\'erminos generales, un retorno de alg\'un momento durante el d\'ia de trading anterior a alg\'un momento durante el d\'ia actual (por ejemplo, retorno de cierre a apertura, retorno de cierre a cierre); por lo general, retorno de cierre a apertura.\\
\\
\noindent
{\bf retorno realizado:} \indent
un retorno hist\'orico.\\
\\
\noindent
{\bf riesgo:} \indent
la posibilidad de que el retorno realizado difiera del retorno esperado.\\
\\
\noindent
{\bf riesgo a la baja:} \indent
el riesgo asociado a p\'erdidas.\\
\\
\noindent
{\bf riesgo de incumplimiento:} \indent
el riesgo (estimado/percibido) de un incumplimiento de pago de un prestatario.\\
\\
\noindent
{\bf riesgo de prepago:} \indent
el principal riesgo para los inversionistas en un MBS passthrough a trav\'es del cual los propietarios tienen la opci\'on de pagar por adelantado sus hipotecas (por ejemplo, refinanciando sus hipotecas a medida que caen las tasas de inter\'es).\\
\\
\noindent
{\bf riesgo de reinversi\'on:} \indent
el riesgo de que las ganancias (de los pagos de cup\'on y/o el principal de un bono o instrumento similar) se reinvertir\'an a una tasa m\'as baja que aquella de la inversi\'on original.\\
\\
\noindent
{\bf riesgo de tasa de inter\'es (tambi\'en conocido como exposici\'on a la tasa de inter\'es):} \indent
la exposici\'on a las fluctuaciones en las tasas de inter\'es, que afectan los precios de los bonos y otros activos de renta fija.\\
\\
\noindent
{\bf riesgo de tipo de cambio:} \indent
la exposici\'on a los cambios en la tasa de FX.\\
\\
\noindent
{\bf riesgo del clima:} \indent
el riesgo de que las empresas y varios sectores de la econom\'ia se vean afectados por las condiciones clim\'aticas.\\
\\
\noindent
{\bf riesgo espec\'ifico (tambi\'en conocido como riesgo idiosincr\'atico):} \indent
v\'ease {\em riesgo no sistem\'atico}.\\
\\
\noindent
{\bf riesgo no sistem\'atico:} \indent
el riesgo espec\'ifico (tambi\'en conocido como idiosincr\'atico), que es espec\'ifico para cada compa\~{n}\'ia, activo, etc., se puede reducir a trav\'es de la diversificaci\'on, aunque nunca se elimina por completo (comparado con {\em riesgo sistem\'atico}).\\
\\
\noindent
{\bf riesgo sistem\'atico:} \indent
el riesgo no diversificable inherente a todo el mercado o su segmento, como la exposici\'on a los amplios movimientos del mercado, que no se pueden diversificar en los portafolios solo con posiciones largas, pero que, sin embargo, puede reducirse sustancialmente o incluso eliminarse esencialmente en los portafolios con posiciones largas y cortas (por ejemplo, portafolios d\'olar-neutrales).\\
\\
\noindent
{\bf riesgo soberano:} \indent
el riesgo de que un gobierno pueda incumplir su deuda (deuda soberana, por ejemplo, bonos emitidos por el gobierno) u otras obligaciones, o que los cambios en la pol\'itica de un banco central puedan afectar adversamente los contratos de divisas.\\
\\
\noindent
{\bf rodando hacia abajo por la curva de rendimientos (tambi\'en conocido como rodando hacia abajo por la curva):} \indent
una estrategia de trading que consiste en comprar bonos a largo o mediano plazo del segmento m\'as empinado de la curva de rendimientos (suponiendo que tiene una pendiente ascendente) y venderlos a medida que se acercan al vencimiento.\\
\\
\noindent
{\bf roll:} \indent
en los contratos de futuros, reequilibrar las posiciones de futuros, es decir, el actual contrato de futuros largo (corto) que est\'a a punto de expirar se vende (se cubre) y se compra (se vende) otro contrato de futuros con un vencimiento m\'as largo.\\
\\
\noindent
{\bf roll yield:} \indent
en los futuros de commodities, los rendimientos positivos del roll generados por las posiciones largas (cortas) cuando la estructura temporal est\'a en backwardation (contango).\\
\\
\noindent
{\bf rompecabezas de riesgo de dificultades:} \indent
un suceso emp\'irico en el que las empresas con menor riesgo de quiebra tienden a generar mayores rendimientos que las m\'as riesgosas.\\
\\
\noindent
{\bf rotaci\'on:} \indent
v\'ease {\em rotaci\'on de alfa, rotaci\'on de momentum sectorial}.\\
\\
\noindent
{\bf rotaci\'on de alfa:} \indent
un tipo de estrategia de trading con ETFs.\\
\\
\noindent
{\bf rotaci\'on de momentum sectorial:} \indent
un tipo de estrategia de momentum con ETFs.\\
\\
\noindent
{\bf rotaci\'on sectorial con doble momentum:} \indent
una estrategia de momentum con ETFs.\\
\\
\noindent
{\bf ruido:} \indent
en una serie de tiempo financiera, las fluctuaciones aleatorias sin ninguna tendencia aparente.\\
\\
\noindent
{\bf Russell 3000:} \indent
un \'indice ponderado por la capitalizaci\'on de mercado de las 3,000 acciones m\'as grandes negociadas en los Estados Unidos por la capitalizaci\'on burs\'atil.\\
\\
\noindent
{\bf S\&P 500:} \indent
un \'indice ponderado por la capitalizaci\'on de mercado de las 500 acciones m\'as grandes negociadas en los Estados Unidos por la capitalizaci\'on burs\'atil.\\
\\
\noindent
{\bf sector (en una clasificaci\'on de la industria):} \indent
por lo general, el nivel menos granular en una clasificaci\'on industrial multinivel (por ejemplo, los sectores se dividen en industrias, las industrias se dividen en subindustrias).\\
\\
\noindent
{\bf sector (en una econom\'ia):} \indent
un \'area de la econom\'ia en la que las empresas comparten los productos o servicios similares.\\
\\
\noindent
{\bf seguimiento de la tendencia:} \indent
una estrategia de trading que apunta a capturar ganancias del momentum de un activo en una direcci\'on particular.\\
\\
\noindent
{\bf selecci\'on adversa:} \indent
un efecto causado por el flujo de \'ordenes inteligentes, por lo que la mayor\'ia de las \'ordenes limitadas para comprar en el bid (vender en el ask) se llenan cuando el mercado se mueve a trav\'es de ellas a la baja (al alza).\\
\\
\noindent
{\bf selectividad:} \indent
una medida cuantitativa de la gesti\'on activa de los fondos mutuos (as\'i como tambi\'en en los ETFs con gesti\'on activa).\\
\\
\noindent
{\bf sentimiento de las redes sociales:} \indent
el sentimiento sobre acciones u otros activos extra\'idos de las publicaciones o mensajes de las redes sociales (por ejemplo, en Twitter).\\
\\
\noindent
{\bf sentimiento de riesgo:} \indent
la tolerancia al riesgo de los traders en una respuesta a los patrones econ\'omicos mundiales, por lo que cuando el riesgo se percibe como bajo (alto), los traders tienden a participar en las inversiones de mayor (menor) riesgo (esto es tambi\'en conocido como ``risk-on/risk-off'' en ingl\'es).\\
\\
\noindent
{\bf sentimiento de Twitter:} \indent
el sentimiento sobre las acciones u otros activos extra\'idos de los tweets.\\
\\
\noindent
{\bf se\~{n}al:} \indent
una se\~{n}al de trading, por ejemplo, para comprar (se\~{n}al de compra) o vender (se\~{n}al de venta) un activo.\\
\\
\noindent
{\bf se\~{n}al de trading:} \indent
v\'ease {\em se\~{n}al}.\\
\\
\noindent
{\bf se\~{n}al intrad\'ia:} \indent
una se\~{n}al de trading utilizada por una estrategia intrad\'ia.\\
\\
\noindent
{\bf serie de tiempo:} \indent
una serie de puntos de datos indexados en un orden temporal, es decir, etiquetados por los valores de tiempo.\\
\\
\noindent
{\bf sesgo:} \indent
en una red neuronal artificial, el componente no homog\'eneo del argumento de una funci\'on de activaci\'on.\\
\\
\noindent
{\bf sigmoide:} \indent
la funci\'on de $x$ dada por $1/(1+\exp(-x))$.\\
\\
\noindent
{\bf sistema de ejecuci\'on de \'ordenes:} \indent
un componente de software que ejecuta las operaciones en base a las secuencias de \'ordenes compra y/o venta.\\
\\
\noindent
{\bf sistema de imputaci\'on:} \indent
v\'ease {\em imputaci\'on de dividendos}.\\
\\
\noindent
{\bf sistema de los especialistas:} \indent
un sistema de creaci\'on de mercado operado y controlado por humanos (en gran parte) en el NYSE antes de cambiar a transacciones electr\'onicas (principalmente).\\
\\
\noindent
{\bf slippage:} \indent
la diferencia entre el precio al que se coloca una orden (inicial) (o al que se espera que la orden sea ejecutada) y al que se llena (incluyendo despu\'es de cancelar-reemplazar la orden inicial cuando se persigue la oferta o demanda con \'ordenes l\'imites de compra o venta, respectivamente), a veces promediada sobre \'ordenes m\'ultiples (por ejemplo, cuando una orden grande se divide en m\'ultiples \'ordenes m\'as peque\~{n}as).\\
\\
\noindent
{\bf SMB (tambi\'en conocido como Small minus Big en ingl\'es):} \indent
v\'ease {\em factores de Fama-French}.\\
\\
\noindent
{\bf sobreajuste:} \indent
en un modelo estad\'istico, ajustar m\'as par\'ametros libres que los justificados por los datos, por lo que (a menudo sin darse cuenta) estos ajustan el ruido y hacen que el modelo sea impredecible fuera de la muestra.\\
\\
\noindent
{\bf sobre-reacci\'on:} \indent
en los mercados financieros, una respuesta irracional de los participantes del mercado (basada en la codicia o el miedo) a la nueva informaci\'on.\\
\\
\noindent
{\bf softmax:} \indent
la funci\'on $\exp(x_i) / \sum_{j=1}^N \exp(x_j)$ de un $N$-vector $x_i$ ($i=1,\dots,N$).\\
\\
\noindent
{\bf solo largo:} \indent
un portafolio o estrategia que solo posee tenencias (o posiciones) largas.\\
\\
\noindent
{\bf soporte:} \indent
en an\'alisis t\'ecnico, el nivel de precio (percibido) en el que se espera que el precio de una acci\'on con tendencia bajista comience a subir.\\
\\
\noindent
{\bf SPDR Trust (tablero de cotizaciones SPY):} \indent
un EFT que rastrea al S\&P 500.\\
\\
\noindent
{\bf split (tambi\'en conocido como divisi\'on de acciones):} \indent
una acci\'on corporativa en la que una empresa divide cada una de sus acciones existentes en acciones m\'ultiples (divisi\'on de acciones forward) o combina acciones m\'ultiples en una (divisi\'on inversa de acciones).\\
\\
\noindent
{\bf spot (tambi\'en conocido como precio al contado, o valor al contado):} \indent
el precio actual de un activo.\\
\\
\noindent
{\bf steepener:} \indent
una estrategia sobre el margen en la curva de rendimientos.\\
\\
\noindent
{\bf stemming:} \indent
reducir las palabras a su ra\'iz o a su forma base, es decir, dejar solo la parte de una palabra que es com\'un a todas sus variantes inflexas.\\
\\
\noindent
{\bf subindustria (en una clasificaci\'on de la industria):} \indent
por lo general, un subgrupo de empresas dentro de la misma industria se agrupan en base a un criterio m\'as detallado.\\
\\
\noindent
{\bf subyacente:} \indent
el instrumento subyacente (por ejemplo, las acciones en una opci\'on de acciones individuales).\\
\\
\noindent
{\bf swap (tambi\'en conocido como acuerdo de intercambio, o contrato de intercambio):} \indent
un contrato derivado a trav\'es del cual dos partes intercambian instrumentos financieros.\\
\\
\noindent
{\bf swap de inflaci\'on:} \indent
un swap cuyo comprador tiene una posici\'on larga en la inflaci\'on y recibe una tasa flotante (basada en un \'indice de la inflaci\'on) y paga una tasa fija (tasa de equilibrio).\\
\\
\noindent
{\bf swap de inflaci\'on con cup\'on cero:} \indent
un swap de inflaci\'on que tiene un solo flujo de efectivo al vencimiento y hace referencia a la inflaci\'on acumulada durante la vida del swap (comparado con {\em swap de inflaci\'on interanual}).\\
\\
\noindent
{\bf swap de inflaci\'on interanual:} \indent
un swap de inflaci\'on que hace referencia a la inflaci\'on anual (comparado con {\em swap de inflaci\'on con cup\'on cero}).\\
\\
\noindent
{\bf swap de tasas de inter\'es:} \indent
un contrato que intercambia un flujo de pagos de tasa flotante por un flujo de pagos de tasa fija o viceversa.\\
\\
\noindent
{\bf swap de varianza:} \indent
un contrato derivado cuyo pago al vencimiento es igual al producto de un coeficiente preestablecido (nocional de varianza) multiplicado por la diferencia entre el valor realizado de la varianza del subyacente a la madurez y la varianza de ejercicio preestablecida.\\
\\
\noindent
{\bf tablero de cotizaciones (o s\'imbolo, o ticker en ingl\'es):} \indent
una cadena de caracteres corta que representa un activo dado que cotiza en la bolsa.\\
\\
\noindent
{\bf tanh:} \indent
la tangente hiperb\'olica.\\
\\
\noindent
{\bf tasa:} \indent
v\'ease {\em tasa de inter\'es, inflaci\'on}.\\
\\
\noindent
{\bf tasa de calentamiento:} \indent
la eficiencia con la que una planta de producci\'on de electricidad convierte el combustible en la electricidad.\\
\\
\noindent
{\bf tasa de cup\'on:} \indent
una tasa no compuesta, fija o variable, a la cual un bono con cup\'on realiza pagos de cupones.\\
\\
\noindent
{\bf tasa de descuento (tambi\'en conocida como tasa de descuento de la Fed, o tasa de descuento Federal):} \indent
la tasa de inter\'es que se cobra a los bancos comerciales y otras instituciones de dep\'osito por los pr\'estamos recibidos de la Reserva Federal de los Estados Unidos.\\
\\
\noindent
{\bf tasa de equilibrio:} \indent
la tasa fijada de un swap de inflaci\'on.\\
\\
\noindent
{\bf tasa de FX:} \indent
v\'ease {\em tipo de cambio}.\\
\\
\noindent
{\bf tasa de FX spot (tambi\'en conocida como tasa spot de FX):} \indent
la tasa actual de FX.\\
\\
\noindent
{\bf tasa de intercambio forward (o tasa de FX forward):} \indent
la tasa de cambio de un contrato de divisas a plazo.\\
\\
\noindent
{\bf tasa de inter\'es:} \indent
el inter\'es por \$1 del principal.\\
\\
\noindent
{\bf tasa de inter\'es real:} \indent
la tasa de inter\'es ajustada por la inflaci\'on.\\
\\
\noindent
{\bf tasa de inter\'es flotante (tambi\'en conocida como tasa flotante, o tasa de inter\'es variable, o tasa variable):} \indent
la tasa de inter\'es sobre un pasivo que var\'ia durante el plazo del pr\'estamo.\\
\\
\noindent
{\bf tasa de recuperaci\'on:} \indent
el porcentaje del principal y los intereses devengados sobre una deuda en incumplimiento que se pueden recuperar.\\
\\
\noindent
{\bf tasa fija de inter\'es (tambi\'en conocida como tasa fija):} \indent
la tasa de inter\'es sobre un pasivo que permanece sin cambios ya sea por el plazo completo del pr\'estamo o por una parte de \'este.\\
\\
\noindent
{\bf tasa libre de riesgo:} \indent
la tasa de retorno de un activo libre de riesgo, muchas veces se considera la tasa del Tesoro a un mes.\\
\\
\noindent
{\bf tasa no compuesta:} \indent
una tasa de inter\'es aplicada al principal durante alg\'un per\'iodo sin ning\'un tipo de capitalizaci\'on.\\
\\
\noindent
{\bf tasa variable:} \indent
v\'ease {\em tasa de inter\'es flotante}.\\
\\
\noindent
{\bf tendencia:} \indent
la direcci\'on general del precio de un mercado o activo, esencialmente, el momentum.\\
\\
\noindent
{\bf tenedor de bonos:} \indent
propietario de bonos.\\
\\
\noindent
{\bf tenencia:} \indent
el contenido de un portafolio; tambi\'en, es una abreviaci\'on de, por ejemplo, tenencias en d\'olares.\\
\\
\noindent
{\bf tenencia en d\'olar (tambi\'en conocida como posici\'on en d\'olar):} \indent
el valor en d\'olares de la posici\'on de un activo en una cartera.\\
\\
\noindent
{\bf tenencias deseadas:} \indent
las tenencias en un portafolio a alcanzar por una estrategia de trading.\\
\\
\noindent
{\bf Teorema de Bayes:} \indent
$P(A|B) = P(B|A) \times P(A) / P(B)$, en donde $P(A|B)$ es la probabilidad condicional de que $A$ ocurra asumiendo que $B$ es verdad, y $P(A)$ y $P(B)$ son las probabilidades de que $A$ y $B$ ocurran independientemente el uno del otro.\\
\\
\noindent
{\bf tercil:} \indent
cada una de las 3 partes (aproximadamente) iguales de una muestra (por ejemplo, muestra de datos).\\
\\
\noindent
{\bf Tesoro:} \indent
el Departamento del Tesoro de los Estados Unidos.\\
\\
\noindent
{\bf Theta:} \indent
la primera derivada del valor de un activo derivado (por ejemplo, una opci\'on) con respecto al tiempo.\\
\\
\noindent
{\bf tiempo al vencimiento (TTM, por sus siglas en ingl\'es):} \indent
el tiempo restante antes de que caduque una opci\'on.\\
\\
\noindent
{\bf tipo de cambio (tambi\'en conocido como tasa de FX):} \indent
el tipo (o tasa) de cambio entre dos monedas diferentes.\\
\\
\noindent
{\bf trade de la curva:} \indent
un flattener o steepener (en bonos o CDOs).\\
\\
\noindent
{\bf trader:} \indent
una persona que compra y vende bienes, divisas, acciones, commodities, etc.\\
\\
\noindent
{\bf trader institucional:} \indent
un trader que compra y vende activos para una cuenta de un grupo o instituci\'on como, por ejemplo, un fondo de pensiones, un fondo mutuo, una compa\~{n}\'ia de seguros, un ETF, etc.\\
\\
\noindent
{\bf trader minorista:} \indent
un trader individual no profesional.\\
\\
\noindent
{\bf trading cuantitativo:} \indent
las estrategias de trading sistem\'aticas basadas en an\'alisis cuantitativos y c\'alculos matem\'aticos con poca o ninguna intervenci\'on humana m\'as all\'a del desarrollo de la estrategia (que incluye su codificaci\'on en un lenguaje inform\'atico de programaci\'on adecuado).\\
\\
\noindent
{\bf trading de correlaci\'on:} \indent
arbitrar la correlaci\'on de pares promedio de los constituyentes del \'indice versus su valor realizado futuro.\\
\\
\noindent
{\bf trading de dispersi\'on:} \indent
arbitrar la volatilidad impl\'icita del \'indice versus las volatilidades impl\'icitas de sus constituyentes.\\
\\
\noindent
{\bf trading de pares:} \indent
una estrategia de reversi\'on a la media que involucra dos activos hist\'oricamente correlacionados.\\
\\
\noindent
{\bf trading electr\'onico:} \indent
el trading de activos de forma electr\'onica, a diferencia de los traders humanos en los pisos de los mercados de valores.\\
\\
\noindent
{\bf trading en anuncios econ\'omicos:} \indent
una estrategia de trading que compra acciones en los d\'ias de anuncios importantes, como los anuncios del FOMC, mientras mantiene activos libres de riesgo en otros d\'ias.\\
\\
\noindent
{\bf tramo:} \indent
v\'ease {\em tramo de un CDO}.\\
\\
\noindent
{\bf tramo de capital:} \indent
el tramo de menor calidad de un CDO.\\
\\
\noindent
{\bf tramo de un CDO:} \indent
una parte de un CDO que consta de activos con diferentes calificaciones crediticias y tasas de inter\'es.\\
\\
\noindent
{\bf tramo junior mezzanine:} \indent
el siguiente tramo (con la calidad del mismo creciente) de un CDO despu\'es del tramo de capital.\\
\\
\noindent
{\bf tramo senior:} \indent
el siguiente tramo (con la calidad del mismo creciente) de un CDO despu\'es del tramo senior mezzanine.\\
\\
\noindent
{\bf tramo senior mezzanine:} \indent
el siguiente tramo (con la calidad del mismo creciente) de un CDO despu\'es del tramo junior mezzanine.\\
\\
\noindent
{\bf tramo super senior:} \indent
el tramo de mayor calidad de un CDO.\\
\\
\noindent
{\bf Transacciones Separadas de Intereses y Principales Nominativos (tambi\'en conocidas como STRIPS en ingl\'es):} \indent
los activos del Tesoro con cup\'on cero.\\
\\
\noindent
{\bf tree boosting:} \indent
una t\'ecnica de aprendizaje autom\'atico.\\
\\
\noindent
{\bf unidad lineal rectificada (ReLU, por sus siglas en ingl\'es):} \indent
la funci\'on de $x$ dada por $\mbox{max}(x, 0)$.\\
\\
\noindent
{\bf universo de trading (tambi\'en conocido como universo):} \indent
los tableros de cotizaciones de acciones (u otros activos) en un portafolio de trading.\\
\\
\noindent
{\bf usurero:} \indent
un prestamista que ofrece un pr\'estamo a tasas de inter\'es excesivamente altas.\\
\\
\noindent
{\bf validaci\'on cruzada (tambi\'en conocida como testeo fuera de la muestra):}
una t\'ecnica para evaluar modelos predictivos mediante la partici\'on de la muestra de datos original en un conjunto de entrenamiento para entrenar el modelo y un conjunto de pruebas para evaluarlo.\\
\\
\noindent
{\bf valor (o claim en ingl\'es):} \indent
el pago de una opci\'on (o alg\'un otro derivado).\\
\\
\noindent
{\bf valor de libros:} \indent
los activos totales de la empresa menos sus pasivos totales.\\
\\
\noindent
{\bf valor de mercado (mark-to-market (MTM) en ingl\'es):} \indent
valorar activos o portafolios en funci\'on de los precios de mercado pertinentes m\'as recientes.\\
\\
\noindent
{\bf valor de tasaci\'on:} \indent
la valuaci\'on del valor de una propiedad u otros objetos valiosos (por ejemplo, joyas) en un momento dado.\\
\\
\noindent
{\bf valor de un bono:} \indent
el valor de un bono en un momento dado antes del vencimiento.\\
\\
\noindent
{\bf valor justo (o razonable):} \indent
el valor de mercado de un activo o, en ausencia de un valor de mercado, un valor te\'orico basado en alg\'un modelo razonable.\\
\\
\noindent
{\bf valor nominal (tambi\'en conocido como principal):} \indent
el importe pagado al tenedor de bonos al vencimiento.\\
\\
\noindent
{\bf valor respaldado por activos (ABS, por sus siglas en ingl\'es):} \indent
un activo financiero garantizado por un conjunto de activos tales como pr\'estamos, hipotecas, regal\'ias, etc.\\
\\
\noindent
{\bf valor roll:} \indent
v\'ease {\em valor roll diario}.\\
\\
\noindent
{\bf valor roll diario:} \indent
la base de futuros dividida por el n\'umero de d\'ias h\'abiles hasta la liquidaci\'on.\\
\\
\noindent
{\bf valoraci\'on err\'onea:} \indent
una ineficiencia en el precio de un activo, cuando su precio no coincide con su valor intr\'inseco o valor justo (percibido).\\
\\
\noindent
{\bf valores del Tesoro:} \indent
los activos del Tesoro.\\
\\
\noindent
{\bf Valores del Tesoro Protegidos Contra la Inflaci\'on (TIPS, por sus siglas en ingl\'es):} \indent
los valores del tesoro que pagan los cupones fijos semestrales a una tasa fija, pero los pagos de cupones (y el principal) se ajustan en funci\'on de la inflaci\'on.\\
\\
\noindent
{\bf valor respaldado por hipotecas (MBS, por sus siglas en ingl\'es):}
un activo respaldado por un conjunto de hipotecas.\\
\\
\noindent
{\bf value (o valor):} \indent
un factor basado en el ratio book-to-price (B/P, por sus siglas en ingl\'es).\\
\\
\noindent
{\bf variable de estado:} \indent
una dentro de un conjunto de variables (que pueden o no ser observables) utilizadas para describir un sistema din\'amico.\\
\\
\noindent
{\bf variable dummy (tambi\'en conocida como variable binaria):} \indent
una variable predictiva que toma los valores binarios 0 o 1 para indicar la ausencia o presencia de alg\'un efecto o pertenencia o no pertenencia a alguna categor\'ia que pueda afectar el resultado (por ejemplo, si una empresa pertenece a un sector econ\'omico determinado).\\
\\
\noindent
{\bf variable explicativa:} \indent
una variable que tiene (o se espera que tenga) cierto poder explicativo para una variable observada (por ejemplo, se puede esperar que las horas estudiadas por un estudiante para un examen final sean una variable explicativa para la calificaci\'on/puntuaci\'on del examen final de dicho estudiante).\\
\\
\noindent
{\bf variable objetivo:} \indent
en aprendizaje autom\'atico, la variable cuyos valores deben modelarse y predecirse.\\
\\
\noindent
{\bf varianza:} \indent
un valor medio de los cuadrados de las desviaciones de los valores de una cantidad de su valor medio.\\
\\
\noindent
{\bf varianza muestral:} \indent
una varianza calculada en funci\'on de las series de tiempo de los retornos hist\'oricos de un activo.\\
\\
\noindent
{\bf vatio:} \indent
una unidad de potencia en el Sistema Internacional de Unidades (SI).\\
\\
\noindent
{\bf Vega:} \indent
la primera derivada del valor de un activo derivado (por ejemplo, una opci\'on) con respecto a la volatilidad impl\'icita del activo subyacente.\\
\\
\noindent
{\bf veh\'iculo de inversi\'on:} \indent
un producto de inversi\'on (por ejemplo, un ETF) utilizado por los traders para generar retornos positivos.\\
\\
\noindent
{\bf vencimiento:} \indent
la \'ultima fecha en la que un contrato de derivados (por ejemplo, un contrato de opciones o futuros) es v\'alido.\\
\\
\noindent
{\bf vencimiento de bonos:} \indent
el momento en que se paga el principal de un bono.\\
\\
\noindent
{\bf vendedor de protecci\'on:} \indent
un vendedor de seguros.\\
\\
\noindent
{\bf venta en corto:} \indent
establecer una posici\'on corta.\\
\\
\noindent
{\bf VIX:} \indent
el \'indice de volatilidad VIX del CBOE, tambi\'en conocido como el ``\'indice de incertidumbre'' o el ``\'indice de medida del miedo''.\\
\\
\noindent
{\bf vocabulario de aprendizaje:} \indent
en an\'alisis de sentimiento (por ejemplo, sentimiento de Twitter) usando t\'ecnicas de aprendizaje autom\'atico, un conjunto de palabras clave que est\'an relacionadas al objetivo (por ejemplo, predecir los movimientos de precios de las acciones o criptomonedas).\\
\\
\noindent
{\bf volatilidad:} \indent
una medida estad\'istica de la dispersi\'on de los retornos de un \'indice de mercado o un activo, que se expresa mediante la desviaci\'on est\'andar o la varianza de dichos retornos.\\
\\
\noindent
{\bf volatilidad impl\'icita:} \indent
en los precios de las opciones, la volatilidad del activo subyacente, que, cuando se utiliza como un dato de entrada en un modelo de valuaci\'on de opciones (como el modelo de Black-Scholes), produce un precio (seg\'un este modelo) de la opci\'on igual a su valor de mercado.\\
\\
\noindent
{\bf volatilidad logar\'itmica:} \indent
la desviaci\'on est\'andar de los logaritmos naturales de precios.\\
\\
\noindent
{\bf volatilidad realizada:} \indent
la volatilidad hist\'orica.\\
\\
\noindent
{\bf volumen:} \indent
el n\'umero de acciones o contratos negociados en un activo durante cierto per\'iodo.

\phantomsection
\cleardoublepage
\addcontentsline{toc}{section}{Acr\'onimos}
\section*{Acr\'onimos}

\noindent
{\bf ABS:} \indent
asset-backed security (valor respaldado por activos).\\
\\
\noindent
{\bf ADDV:} \indent
average daily dollar volume (volumen en d\'olares diario promedio).\\
\\
\noindent
{\bf ANN:} \indent
artificial neural network (red neuronal artificial).\\
\\
\noindent
{\bf ATM:} \indent
at-the-money (en el dinero).\\
\\
\noindent
{\bf B/P:} \indent
book-to-price.\\
\\
\noindent
{\bf BA:} \indent
banker's acceptance (aceptaci\'on bancaria).\\
\\
\noindent
{\bf BICS:} \indent
Bloomberg Industry Classification System (Sistema de Bloomberg de Clasificaci\'on de la Industria).\\
\\
\noindent
{\bf bps:} \indent
basis point (punto b\'asico).\\
\\
\noindent
{\bf BTC:} \indent
Bitcoin.\\
\\
\noindent
{\bf Btu:} \indent
British thermal unit (unidad t\'ermica brit\'anica).\\
\\
\noindent
{\bf CA:} \indent
commodity allocation percentage (porcentaje de asignaci\'on a commodities).\\
\\
\noindent
{\bf CBOE:} \indent
Chicago Board Options Exchange.\\
\\
\noindent
{\bf CD:} \indent
certificate of deposit (certificado de dep\'osito bancario).\\
\\
\noindent
{\bf CDD:} \indent
cooling-degree-days (grados-d\'ia de refrigeraci\'on).\\
\\
\noindent
{\bf CDO:} \indent
collateralized debt obligation (obligaci\'on de deuda garantizada).\\
\\
\noindent
{\bf CDS:} \indent
credit default swap (swap de incumplimiento crediticio).\\
\\
\noindent
{\bf CFTC:} \indent
U.S. Commodity Futures Trading Commission (Comisi\'on de Negociaci\'on de Futuros de Mercanc\'ias de los Estados Unidos).\\
\\
\noindent
{\bf CI:} \indent
core inflation (inflaci\'on n\'ucleo).\\
\\
\noindent
{\bf CIRP:} \indent
Covered Interest Rate Parity (Paridad de Tasas de Inter\'es Cubierta).\\
\\
\noindent
{\bf CME:} \indent
Chicago Mercantile Exchange.\\
\\
\noindent
{\bf COT:} \indent
Commitments of Traders (Compromisos de los Comerciantes).\\
\\
\noindent
{\bf CPI:} \indent
Consumer Price Index (\'Indice de Precios al Consumidor).\\
\\
\noindent
{\bf CPS:} \indent
cents-per-share (centavos por acci\'on).\\
\\
\noindent
{\bf CTA:} \indent
commodity trading advisor (asesor de trading de commodities).\\
\\
\noindent
{\bf DJIA:} \indent
Dow Jones Industrial Average.\\
\\
\noindent
{\bf EMA:} \indent
exponential moving average (media m\'ovil exponencial).\\
\\
\noindent
{\bf EMSD:} \indent
exponential moving standard deviation (desviaci\'on est\'andar m\'ovil exponencial).\\
\\
\noindent
{\bf ETF:} \indent
exchange-traded fund (fondo de inversi\'on cotizado).\\
\\
\noindent
{\bf ETH:} \indent
Ethereum.\\
\\
\noindent
{\bf ETN:} \indent
exchange-traded note (nota de intercambio cotizada).\\
\\
\noindent
{\bf EUR:} \indent
euro.\\
\\
\noindent
{\bf FOMC:} \indent
Federal Open Market Committee (Comit\'e Federal de Mercado Abierto).\\
\\
\noindent
{\bf FX:} \indent
foreign exchange (divisas).\\
\\
\noindent
{\bf GDP:} \indent
Gross Domestic Product (producto interno bruto, o PIB por sus siglas en espa\~{n}ol).\\
\\
\noindent
{\bf GICS:} \indent
Global Industry Classification Standard (Est\'andar Global de Clasificaci\'on de la Industria).\\
\\
\noindent
{\bf HDD:} \indent
heating-degree-days (grados-d\'ia de calefacci\'on).\\
\\
\noindent
{\bf HFT:} \indent
high frequency trading (trading de alta frecuencia).\\
\\
\noindent
{\bf HI:} \indent
headline inflation (inflaci\'on general).\\
\\
\noindent
{\bf HMD:} \indent
healthy-minus-distressed (compa\~{n}\'ias saludables menos compa\~{n}\'ias en distress).\\
\\
\noindent
{\bf HML:} \indent
High minus Low (alto menos bajo).\\
\\
\noindent
{\bf HP:} \indent
hedging pressure (presi\'on de cobertura); Hodrick-Prescott.\\
\\
\noindent
{\bf IBS:} \indent
internal bar strength (fuerza interna de las barras de precios).\\
\\
\noindent
{\bf ITM:} \indent
in-the-money (dentro del dinero).\\
\\
\noindent
{\bf JPY:} \indent
Japanese Yen (yen japon\'es).\\
\\
\noindent
{\bf LETF:} \indent
leveraged (inverse) ETF (ETF apalancado (inverso)).\\
\\
\noindent
{\bf LIBOR:} \indent
London Interbank Offer Rate (Tasa Interbancaria de Oferta de Londres).\\
\\
\noindent
{\bf M\&A:} \indent
mergers and acquisitions (fusiones y adquisiciones).\\
\\
\noindent
{\bf MA:} \indent
moving average (media m\'ovil).\\
\\
\noindent
{\bf ML:} \indent
machine learning (aprendizaje autom\'atico).\\
\\
\noindent
{\bf MBS:} \indent
mortgage-backed security (valor respaldado por hipotecas).\\
\\
\noindent
{\bf MBtu:} \indent
1,000 Btu.\\
\\
\noindent
{\bf MKT:} \indent
market (excess) return (retorno (en exceso) del portafolio del mercado).\\
\\
\noindent
{\bf MMBtu:} \indent
1,000,000 Btu.\\
\\
\noindent
{\bf MOM:} \indent
Carhart's momentum factor (factor de momentum de Carhart).\\
\\
\noindent
{\bf MSA:} \indent
metropolitan statistical area (\'area estad\'istica metropolitana).\\
\\
\noindent
{\bf MTM:} \indent
mark-to-market (valor de mercado).\\
\\
\noindent
{\bf Mwh:} \indent
Megawatt hour (Megavatio hora).\\
\\
\noindent
{\bf NYSE:} \indent
New York Stock Exchange (Bolsa de Nueva York).\\
\\
\noindent
{\bf OAS:} \indent
option adjusted spread (diferencial ajustado por opciones).\\
\\
\noindent
{\bf OTM:} \indent
out-of-the-money (fuera del dinero).\\
\\
\noindent
{\bf P\&L:} \indent
profit(s) and loss(es) (ganancia(s) y p\'erdida(s)).\\
\\
\noindent
{\bf P2P:} \indent
peer-to-peer (entre pares, par a par).\\
\\
\noindent
{\bf PCA:} \indent
principal component analysis (an\'alisis de componentes principales).\\
\\
\noindent
{\bf  REIT:} \indent
real estate investment trust (fideicomiso de inversi\'on inmobiliaria).\\
\\
\noindent
{\bf  ReLU:} \indent
rectified linear unit (unidad lineal rectificada).\\
\\
\noindent
{\bf  REPO/repo:} \indent
repurchase agreement (acuerdo de recompra).\\
\\
\noindent
{\bf  RMSE:} \indent
root mean square error (ra\'iz cuadrada del error cuadr\'atico medio).\\
\\
\noindent
{\bf  RSI:} \indent
relative strength index (\'indice de fuerza relativa).\\
\\
\noindent
{\bf S\&P:} \indent
Standard and Poor's.\\
\\
\noindent
{\bf SIC:} \indent
Standard Industrial Classification (Clasificaci\'on Industrial Est\'andar).\\
\\
\noindent
{\bf SMA:} \indent
simple moving average (media m\'ovil simple).\\
\\
\noindent
{\bf SMB:} \indent
Small minus Big (peque\~{n}as menos grandes).\\
\\
\noindent
{\bf SGD:} \indent
stochastic gradient descent (descenso gradiente estoc\'astico).\\
\\
\noindent
{\bf SS:} \indent
sum of squares (suma de cuadrados).\\
\\
\noindent
{\bf StatArb:} \indent
statistical arbitrage (arbitraje estad\'istico).\\
\\
\noindent
{\bf STRIPS:} \indent
Separate Trading of Registered Interest and Principal of Securities (Transacciones Separadas de Intereses y Principales Nominativos).\\
\\
\noindent
{\bf SUE:} \indent
standardized unexpected earnings (ganancias inesperadas estandarizadas).\\
\\
\noindent
{\bf SVM:} \indent
support vector machine (m\'aquina de vectores de soporte).\\
\\
\noindent
{\bf TTM:} \indent
time-to-maturity (tiempo al vencimiento).\\
\\
\noindent
{\bf TIPS:} \indent
Treasury Inflation-Protected Securities (Valores del Tesoro Protegidos Contra la Inflaci\'on).\\
\\
\noindent
{\bf UIRP:} \indent
Uncovered Interest Rate Parity (Paridad de Tasas de Inter\'es no Cubierta).\\
\\
\noindent
{\bf USD:} \indent
U.S. dollar (d\'olar estadounidense).\\
\\
\noindent
{\bf VAR:} \indent
vector autoregressive model (modelo vectorial autorregresivo).\\
\\
\noindent
{\bf VWAP:} \indent
volume-weighted average price (precio promedio ponderado por volumen).\\
\\
\noindent
{\bf YoY:} \indent
year-on-year (a\~{n}o a a\~{n}o).

\phantomsection
\cleardoublepage
\addcontentsline{toc}{section}{Algunas Notaciones Matem\'aticas}
\section*{Algunas Notaciones Matem\'aticas}

\noindent
iff \indent
si y solo si.\\
\\
\noindent
max (min) \indent
m\'aximo (m\'inimo).\\
\\
\noindent
$\mbox{floor}(x)$ \indent
el mayor entero menor o igual a $x$.\\
\\
\noindent
$\mbox{ceiling}(x)$ \indent
el menor entero mayor o igual a $x$.\\
\\
\noindent
$(x)^+$ \indent
$\mbox{max}(x, 0)$.\\
\\
\noindent
$\mbox{sign}(x)$ \indent
el signo de $x$, definido como: $+1$ si $x > 0$; $-1$ si $x < 0$; $0$ si $x = 0$.\\
\\
\noindent
$|x|$ \indent
el valor absoluto de $x$ si $x$ es un n\'umero real.\\
\\
\noindent
$\mbox{rank}(x_i)$ \indent
el rango de $x_i$ cuando $N$ valores $x_i$ ($i=1,\dots,N$) se ordenan de forma ascendente.\\
\\
\noindent
$\exp(x)$ o $e^x$ \indent
el exponente natural de $x$.\\
\\
\noindent
$\mbox{ln}(x)$ \indent
el logaritmo natural de $x$.\\
\\
\noindent
$\sum_{i=1}^N x_i$ \indent
la suma de $N$ valores $x_i$ ($i=1,\dots, N$).\\
\\
\noindent
$\prod_{i=1}^N x_i$ \indent
el producto de $N$ valores $x_i$ ($i=1,\dots, N$).\\
\\
\noindent
$\left.A\right|_{B = b}$ (or $\left.A\right|_b$)  \indent
el valor de $A$ cuando alguna cantidad $B$ de la que $A$ depende impl\'icitamente (generalmente se evidencia en el contexto) toma el valor $b$.\\
\\
\noindent
$f(x) \rightarrow \mbox{min (max)}$ \indent
minimizar (maximizar) $f(x)$ con respecto a $x$ (en donde $x$ puede, por ejemplo, ser un $N$-vector $x_i$, $i=1,\dots, N$).\\
\\
\noindent
$\mbox{argmax}_{\, z}~f(z)$ \indent
el valor de $z$ para cual $f(z)$ se maximiza.\\
\\
\noindent
$\partial f/\partial x$ \indent
la primera derivada parcial de la funci\'on $f$ (que puede depender de variables distintas de $x$) con respecto a $x$.\\
\\
\noindent
$\partial^2 f/\partial x^2$ \indent
la segunda derivada parcial de la funci\'on $f$ (que puede depender de variables distintas de $x$) con respecto a $x$.\\
\\
\noindent
$G: A \mapsto B$ \indent
$G$ es un mapa desde el conjunto $A$ al conjunto $B$.\\
\\
\noindent
$A \subset B$ \indent
el conjunto $A$ es un subconjunto del conjunto $B$.\\
\\
\noindent
$\{i|f(i) = a\}$ \indent
el conjunto de valores de $i$ tales que la condici\'on $f(i) = a$ se satisface.\\
\\
\noindent
$\mbox{min}(i: f(i) > a)$ \indent
el valor m\'inimo de $i$ tal que la condici\'on $f(i) > a$ se satisface.\\
\\
\noindent
$i \in J$ \indent
$i$ es un elemento del conjunto $J$.\\
\\
\noindent
$|J|$ \indent
el n\'umero de elementos de $J$ si $J$ es un conjunto finito.\\
\\
\noindent
$\delta_{AB}$ (o $\delta_{A,B})$ \indent
1 si $A = B$; de otra manera, 0 (delta de Kronecker).\\
\\
\noindent
$\mbox{diag}(x_i)$ \indent
la matriz diagonal con las dimensiones $N\times N$ con $x_i$ ($i=1,\dots,N$) en su diagonal.\\
\\
\noindent
$A^T$ \indent
la transposici\'on de la matriz $A$.\\
\\
\noindent
$A^{-1}$ \indent
la inversa de la matriz $A$.\\
\\
\noindent
$E_t(A)$ \indent
el valor esperado de $A$ al momento $t$.\\
\\
\noindent
$dX(t)$ \indent
un incremento infinitesimal de un proceso continuo $X(t)$.\\
\\
\noindent
$dt$ \indent
un incremento infinitesimal del tiempo $t$.\\
\\
\noindent
$P(A|B)$ \indent
la probabilidad condicional de $A$ que ocurre asumiendo que $B$ es verdad.\\
\\

\phantomsection
\cleardoublepage
\addcontentsline{toc}{section}{Referencias en Ingl\'es}
\section*{Referencias en Ingl\'es}

{}Notas explicativas: Algunas palabras/expresiones en el texto principal y en otras partes de este libro est\'an en ingl\'es, que es como se usan habitualmente en la jerga financiera/comercial en espa\~{n}ol. Aqu\'i brindamos (de forma aproximada) traducciones al espa\~{n}ol de tales palabras, aunque en muchos casos estas traducciones en espa\~{n}ol no se usan en el contexto de finanzas.\\

\noindent
{\bf ask:} \indent
compra (precio de compra).\\
\\
\noindent
{\bf backtest (o backtesting):} \indent
prueba retrospectiva.\\
\\
\noindent
{\bf barbell:} \indent
barra con pesas.\\
\\
\noindent
{\bf bid:} \indent
venta (precio de venta).\\
\\
\noindent
{\bf blockchain:} \indent
cadena de bloques.\\
\\
\noindent
{\bf book-to-market:} \indent
libro al mercado (valor de libros sobre valor de mercado).\\
\\
\noindent
{\bf book-to-price:} \indent
libro al precio (valor de libros por acciones en circulaci\'on sobre precio de mercado).\\
\\
\noindent
{\bf brownfield:} \indent
campo marr\'on.\\
\\
\noindent
{\bf bullet:} \indent
bala.\\
\\
\noindent
{\bf call:} \indent
opci\'on de compra.\\
\\
\noindent
{\bf carry:} \indent
acarreo.\\
\\
\noindent
{\bf Cash \& Carry:} \indent
efectivo y acarreo.\\
\\
\noindent
{\bf claim:} \indent
reclamo.\\
\\
\noindent
{\bf cluster:} \indent
grupo, racimo.\\
\\
\noindent
{\bf clustering:} \indent
agrupamiento.\\
\\
\noindent
{\bf commodity:} \indent
mercanc\'ia (o producto b\'asico, o materia prima).\\
\\
\noindent
{\bf distress:} \indent
dificultad (en el contexto de activos).\\
\\
\noindent
{\bf drawdown:} \indent
retroceso.\\
\\
\noindent
{\bf duration-targeting:} \indent
apuntando duraci\'on.\\
\\
\noindent
{\bf flattener:} \indent
aplanador.\\
\\
\noindent
{\bf forward-looking:} \indent
mirando hacia adelante.\\
\\
\noindent
{\bf Gamma scalping:} \indent
especulaci\'on con Gamma.\\
\\
\noindent
{\bf goes home flat:} \indent
se va a casa plana.\\
\\
\noindent
{\bf greenfield:} \indent
campo verde.\\
\\
\noindent
{\bf hard-to-borrow:} \indent
dif\'icil de pedir prestado.\\
\\
\noindent
{\bf Hard-to-Borrow List:} \indent
lista de activos dif\'iciles de pedir prestado.\\
\\
\noindent
{\bf healthy-minus-distressed:} \indent
saludable menos en dificultad.\\
\\
\noindent
{\bf hedger:} \indent
trader cuyo objetivo es lograr cobertura.\\
\\
\noindent
{\bf ladder:} \indent
escalera.\\
\\
\noindent
{\bf loan-to-own:} \indent
prestar a poseer.\\
\\
\noindent
{\bf lookback:} \indent
retroactivo.\\
\\
\noindent
{\bf market-making:} \indent
creaci\'on de mercado.\\
\\
\noindent
{\bf momentum:} \indent
impulso.\\
\\
\noindent
{\bf passthrough:} \indent
pasar por.\\
\\
\noindent
{\bf payoff:} \indent
valor a vencimiento (por ejemplo, de una opci\'on).\\
\\
\noindent
{\bf proxy:} \indent
indicador, aproximador.\\
\\
\noindent
{\bf put:} \indent
opci\'on de venta.\\
\\
\noindent
{\bf R Package for Statistical Computing:} \indent
Paquete R para Inform\'atica Estad\'istica.\\
\\
\noindent
{\bf rally:} \indent
mercado muy alcista.\\
\\
\noindent
{\bf ranking:} \indent
clasificaci\'on.\\
\\
\noindent
{\bf risk reversal:} \indent
reversi\'on de riesgo.\\
\\
\noindent
{\bf risk-on/risk-off:} \indent
tomar/evitar riesgo.\\
\\
\noindent
{\bf roll:} \indent
rollo.\\
\\
\noindent
{\bf roll yield:} \indent
rendimiento del rollo.\\
\\
\noindent
{\bf roll-down:} \indent
rodar hacia abajo.\\
\\
\noindent
{\bf slippage:} \indent
deslizamiento.\\
\\
\noindent
{\bf Small minus Big:} \indent
peque\~{n}o menos grande.\\
\\
\noindent
{\bf snap-back:} \indent
volver bruscamente.\\
\\
\noindent
{\bf spark:} \indent
chispa.\\
\\
\noindent
{\bf split:} \indent
divisi\'on.\\
\\
\noindent
{\bf spot:} \indent
contado.\\
\\
\noindent
{\bf steepener:} \indent
empinador.\\
\\
\noindent
{\bf stemming:} \indent
lematizaci\'on, reducir a la ra\'iz.\\
\\
\noindent
{\bf stop-loss:} \indent
detener la p\'erdida.\\
\\
\noindent
{\bf swap:} \indent
intercambio.\\
\\
\noindent
{\bf targeting:} \indent
apuntando.\\
\\
\noindent
{\bf ticker:} \indent
tablero de cotizaciones, s\'imbolo.\\
\\
\noindent
{\bf trader:} \indent
comerciante.\\
\\
\noindent
{\bf trading:} \indent
comercio (o compraventa).\\
\\
\noindent
{\bf tree boosting:} \indent
crecimiento de \'arboles.\\
\\
\noindent
{\bf value:} \indent
valor.\\
\\
\noindent
{\bf variable dummy:} \indent
dummy variable en ingl\'es, variable binaria o variable ficticia en espa\~{n}ol.\\
\\

\newpage

\phantomsection
\addcontentsline{toc}{section}{Notas Explicativas para el \'Indice}
\section*{Notas Explicativas para el \'Indice}

{}En las entradas del \'Indice, el plural en muchos casos (pero no en todos) se reduce a singular (por lo tanto, por ejemplo, ``commodity'' tambi\'en incluye ``commodities''). Los par\'entesis contienen acr\'onimos o definiciones, y en algunos casos (pero no en todos) ambas versiones est\'an presentes en el texto principal. La mayor\'ia (pero no todas) de las entradas del \'Indice con comas, por ejemplo, ``{\em sustantivo, adjetivo}'', corresponden a entradas de texto de modo que la cadena de texto precisa ``{\em adjetivo sustantivo}'' no est\'a directamente presente en el texto, pero est\'a presente indirectamente (por ejemplo, ``{\em adjetivo (...) sustantivo}'') o contextualmente.

\phantomsection
\cleardoublepage
\addcontentsline{toc}{section}{\indexname}
\raggedright \printindex
\phantomsection
\cleardoublepage



\addcontentsline{toc}{section}{Bibliograf\'ia}
\begin{thebibliography}{999}

\makeatletter
\def\@biblabel#1{}
\makeatother

\bibitem[Abken, 1989]{Abken1989} Abken, P.A. (1989)
An analysis of intra-market spreads in heating oil futures.
{\em Journal of Futures Markets} 9(1): 77-86.

\bibitem[Abken and Nandi, 1996]{Abken1996} Abken, P.A. and Nandi, S. (1996)
Options and Volatility.
{\em Federal Reserve Bank of Atlanta, Economic Review} 81(3): 21-35.

\bibitem[Abraham and Hendershott, 1993]{Abraham1993} Abraham, J.M. and Hendershott, P.H. (1993)
Patterns and Determinants of Metropolitan House Prices, 1977 to 1991.
In: Browne, L.E. and Rosengren, E.S. (eds.) {\em Real Estate and the Credit Crunch.} Boston, MA: Federal Reserve Bank of Boston, pp. 18-42.

\bibitem[Abraham and Hendershott, 1996]{Abraham1996} Abraham, J.M. and Hendershott, P.H. (1996)
Bubbles in Metropolitan Housing Markets.
{\em Journal of Housing Research} 7(2): 191-207.

\bibitem[Abreu and Brunnermeier, 2002]{Abreu2002} Abreu, D. and Brunnermeier, M.K. (2002)
Synchronization risk and delayed arbitrage.
{\em Journal of Financial Economics} 66(2-3): 341-360.

\bibitem[Accominotti and Chambers, 2014]{Accominotti2014} Accominotti, O. and Chambers, D. (2014)
Out-of-Sample Evidence on the Returns to Currency Trading.
{\em Working Paper.} Available online: \url{https://ssrn.com/abstract=2293684}.

\bibitem[Acharya, Almeida and Campello, 2007]{Acharya2007} Acharya, V.V., Almeida, H. and Campello, M. (2007)
Is cash negative debt? A hedging perspective on corporate financial policies.
{\em Journal of Financial Intermediation} 16(4): 515-554.

\bibitem[Ackert and Tian, 2000]{Ackert2000} Ackert, L.F. and Tian, Y.S. (2000)
Arbitrage and valuation in the market for Standard and Poor's Depositary Receipts.
{\em Financial Management} 29(3): 71-87.
	
\bibitem[Adam and Lin, 2001]{Adam2001} Adam, F. and Lin, L.H. (2001)
An Analysis of the Applications of Neural Networks in Finance.
{\em Interfaces} 31(4): 112-122.

\bibitem[Adams and Gl\"{u}ck, 2015]{Adams2015} Adams, Z. and Gl\"{u}ck, T. (2015)
Financialization in commodity markets: A passing trend or the new normal?
{\em Journal of Banking \& Finance} 60: 93-111.

\bibitem[Adrangi {\em et al}, 2006]{Adrangi2006} Adrangi, B., Chatrath, A., Song, F. and Szidarovszky, F. (2006)
Petroleum spreads and the term structure of futures prices.
{\em Applied Economics} 38(16): 1917-1929.

\bibitem[Adrian {\em et al}, 2013]{Adrian2013} Adrian, T., Begalle, B., Copeland, A. and Martin, A. (2013)
Repo and Securities Lending.
{\em Federal Reserve Bank of New York Staff Reports}, No. 529. Available online:\\
\url{https://www.newyorkfed.org/medialibrary/media/research/staff_reports/sr529.pdf}.

\bibitem[Adrian and Wu, 2010]{Adrian2010} Adrian, T.  and Wu, H. (2010)
The Term Structure of Inflation Expectations.
{\em Federal Reserve Bank of New York Staff Reports}, No. 362. Available online:\\ \url{https://www.newyorkfed.org/medialibrary/media/research/staff_reports/sr362.pdf}.

\bibitem[Agapova, 2011a]{Agapova2011a} Agapova, A. (2011a)
Conventional mutual funds versus exchange-traded funds.
{\em Journal of Financial Markets} 14(2): 323-343.

\bibitem[Agapova, 2011b]{Agapova2011b} Agapova, A. (2011b)
The Role of Money Market Mutual Funds in Mutual Fund Families.
{\em Journal of Applied Finance} 21(1): 87-102.

\bibitem[Agarwal {\em et al}, 2011]{Agarwal2011} Agarwal, V., Fung, W.H., Loon, Y.C. and Naik, N.Y. (2011)
Risk and return in convertible arbitrage: Evidence from the convertible bond market.
{\em Journal of Empirical Finance} 18(2): 175-194.

\bibitem[Ahmadi, Sharp and Walther, 1986]{Ahmadi1986} Ahmadi, H.Z., Sharp, P.A. and Walther, C.H. (1986)
The effectiveness of futures and options in hedging currency risk.
In: Fabozzi, F. (ed.) {\em Advances in Futures and Options Research}, Vol. 1, Part B. Greenwich, CT: JAI Press, Inc., pp. 171-191.

\bibitem[Ahmerkamp and Grant, 2013]{Ahmerkamp2013} Ahmerkamp, J.D. and Grant, J. (2013)
The Returns to Carry and Momentum Strategies.
{\em Working Paper.} Available online: \url{https://ssrn.com/abstract=2227387}.

\bibitem[Ahn {\em et al}, 2002]{Ahn2002} Ahn, D.-H., Boudoukh, J., Richardson, M. and Whitelaw, R.F. (2002)
Partial adjustment or stale prices? Implications from stock index and futures return autocorrelations.
{\em Review of Financial Studies} 15(2): 655-689.

\bibitem[Ahn, Conrad and Dittmar, 2003]{Ahn2003} Ahn, D.-H., Conrad, J. and Dittmar, R. (2003)
Risk Adjustment and Trading Strategies.
{\em Review of Financial Studies} 16(2): 459-485.

\bibitem[Ai and Bansal, 2016]{Ai2016} Ai, H. and Bansal, R. (2016)
Risk Preferences and the Macro Announcement Premium.
{\em Working Paper.} Available online: \url{https://ssrn.com/abstract=2827445}.

\bibitem[Aiba and Hatano, 2006]{Aiba2006} Aiba, Y. and Hatano, N. (2006)
A microscopic model of triangular arbitrage.
{\em Physica A: Statistical Mechanics and its Applications} 371(2): 572-584.

\bibitem[Aiba {\em et al}, 2002]{Aiba2002} Aiba, Y., Hatano, N., Takayasu, H., Marumo, K. and Shimizu, T. (2002)
Triangular arbitrage as an interaction among foreign exchange rates.
{\em Physica A: Statistical Mechanics and its Applications} 310(3-4): 467-479.

\bibitem[Aiba {\em et al}, 2003]{Aiba2003} Aiba, Y., Hatano, N., Takayasu, H., Marumo, K. and Shimizu, T. (2003)
Triangular arbitrage and negative auto-correlation of foreign exchange rates.
{\em Physica A: Statistical Mechanics and its Applications} 324(1-2): 253-257.

\bibitem[\"{A}ij\"{o}, 2008]{Aijo2008} \"{A}ij\"{o}, J. (2008)
Implied volatility term structure linkages between VDAX, VSMI and VSTOXX volatility indices.
{\em Global Finance Journal} 18(3): 290-302.

\bibitem[A\"{\i}t-Sahalia and Duarte, 2003]{Ait-Sahalia2003} A\"{\i}t-Sahalia, Y. and Duarte, J. (2003)
Nonparametric option pricing under shape restrictions.
{\em Journal of Econometrics} 116(1-2): 9-47.

\bibitem[A\"{\i}t-Sahalia, Karaman and Mancini, 2015]{Ait-Sahalia2015} A\"{\i}t-Sahalia, Y., Karaman, M. and Mancini, L. (2015)
The Term Structure of Variance Swaps and Risk Premia.
{\em Working Paper.} Available online: \url{https://ssrn.com/abstract=2136820}.

\bibitem[Akram, Rime and Sarno, 2008]{Akram2008} Akram, Q.F., Rime, D. and Sarno, L. (2008)
Arbitrage in the foreign exchange market: Turning on the microscope.
{\em Journal of International Economics} 76(2): 237-253.

\bibitem[Alaminos, del Castillo and Fern\'{a}ndez, 2016]{Alaminos2016} Alaminos, D., del Castillo, A. and Fern\'{a}ndez, M.\'{A}. (2016)
A Global Model for Bankruptcy Prediction.
{\em PLoS ONE} 11(11): e0166693.

\bibitem[Alaton, Djehiche and Stillberger, 2010]{Alaton2010} Alaton, P., Djehiche, B. and Stillberger, D. (2010)
On modelling and pricing weather derivatives.
{\em Applied Mathematical Finance} 9(1): 1-20.

\bibitem[Albeverio, Steblovskaya and Wallbaum, 2013]{Albeverio2013} Albeverio, S., Steblovskaya, V. and Wallbaum, K. (2013)
Investment instruments with volatility target mechanism.
{\em Quantitative Finance} 13(10): 1519-1528.

\bibitem[Albrecht, 1985]{Albrecht1985} Albrecht, P. (1985)
A note on immunization under a general stochastic equilibrium model of the term structure.
{\em Insurance: Mathematics and Economics} 4(4): 239-244.

\bibitem[Aldohni, 2013]{Aldohni2013} Aldohni, A.K. (2013)
Loan Sharks v. Short-term Lenders: How Do the Law and Regulators Draw the Line?
{\em Journal of Law and Society} 40(3): 420-449.

\bibitem[Aldridge, 2013]{Aldridge2013} Aldridge, I. (2013)
{\em High-Frequency Trading: A Practical Guide to Algorithmic Strategies and Trading Systems.} (2nd ed.)
Hoboken, NJ: John Wiley \& Sons, Inc.

\bibitem[Aldridge, 2016]{Aldridge2016} Aldridge, I. (2016)
ETFs, High-Frequency Trading, and Flash Crashes.
{\em Journal of Portfolio Management} 43(1): 17-28.

\bibitem[Alessandretti {\em et al}, 2018]{Alessandretti2018} Alessandretti, L., ElBahrawy, A., Aiello, L.M. and Baronchelli, A. (2018)
Machine Learning the Cryptocurrency Market.
{\em Working Paper.} Available online: \url{https://arxiv.org/pdf/1805.08550.pdf}.

\bibitem[Alexander and Korovilas, 2012]{Alexander2012} Alexander, C. and Korovilas, D. (2012)
Understanding ETNs on VIX Futures.
{\em Working Paper.} Available online: \url{https://ssrn.com/abstract=2043061}.

\bibitem[Alexander and Resnick, 1985]{Alexander1985} Alexander, G.J. and Resnick, B.G. (1985)
Using linear and goal programming to immunize bond portfolios.
{\em Journal of Banking \& Finance} 9(1): 35-54.

\bibitem[Allen and Michaely, 1995]{Allen1995} Allen, F. and Michaely, R. (1995)
Dividend Policy.
In: Jarrow, R.A., Maksimovic, V. and Ziemba, W.T. (eds.)
{\em Handbooks in Operations Research and Management Science}, Vol 9.
Amsterdam, The Netherlands: Elsevier, Chapter 25, pp. 793-837.

\bibitem[Almeida, Campello and Weisbach, 2005]{Almeida2005} Almeida, H., Campello, M. and Weisbach, M.S. (2005)
The Cash Flow Sensitivity of Cash.
{\em Journal of Finance} 59(4): 1777-1804.

\bibitem[Almgren {\em et al}, 2005]{Almgren2005} Almgren, R., Thum, C., Hauptmann, E. and Li, H. (2005)
Equity market impact. {\em Risk Magazine} 18(7): 57-62.

\bibitem[Altman, 1968]{Altman1968} Altman, E.I. (1968)
Financial Ratios, Discriminant Analysis and the Prediction of Corporate Bankruptcy.
{\em Journal of Finance} 23(4): 589-609.

\bibitem[Altman, 1993]{Altman1993} Altman, E. (1993)
{\em Corporate financial distress and bankruptcy.} (2nd ed.) Hoboken, NJ: John Wiley \& Sons, Inc.

\bibitem[Altman, 1998]{Altman1998} Altman, E.I. (1998)
Market Dynamics and Investment Performance of Distressed and Defaulted Debt Securities.
{\em Working Paper.} Available online: \url{https://ssrn.com/abstract=164502}.

\bibitem[Altman, 1992]{Altman1992} Altman, N.S. (1992)
An introduction to kernel and nearest-neighbor nonparametric regression.
{\em American Statistician} 46(3): 175-185.

\bibitem[Altman {\em et al}, 2005]{Altman2005} Altman, E.I., Brady, B., Resti, A. and Sironi, A. (2005)
The link between default and recovery rates: theory, empirical evidence and implications.
{\em Journal of Business} 78(6): 2203-2228.

\bibitem[Altman and Hotchkiss, 2006]{Altman2006} Altman, E.I. and Hotchkiss, E. (2006)
{\em Corporate Financial Distress and Bankruptcy: Predict and Avoid Bankruptcy, Analyze and Invest in Distressed Debt.}
Hoboken, NJ: John Wiley \& Sons, Inc.

\bibitem[Amaitiek, B\'{a}lint and Re\v{s}ovsk\'{y}, 2010]{Amaitiek2010}
Amaitiek, O.F.S., B\'{a}lint, T. and Re\v{s}ovsk\'{y}, M. (2010)
The Short Call Ladder strategy and its application in trading and hedging.
{\em Acta Montanistica Slovaca} 15(3): 171-182.

\bibitem[Amato and Gyntelberg, 2005]{Amato2005} Amato, J.D. and Gyntelberg, J. (2005)
CDS Index Tranches and the Pricing of Credit Risk Correlations.
{\em BIS Quarterly Review}, December 2005, pp. 73-87. Available online: \url{https://www.bis.org/publ/qtrpdf/r_qt0503g.pdf}.

\bibitem[Amato and Remolona, 2003]{Amato2003} Amato, J.D. and Remolona, E.M. (2003)
The credit spread puzzle. {\em BIS Quarterly Review}, December 2003, pp. 51-63. Available online: \url{https://www.bis.org/publ/qtrpdf/r_qt0312e.pdf}.

\bibitem[Ambrose, LaCour-Little and Sanders, 2004]{Ambrose2004} Ambrose, B., LaCour-Little, M. and Sanders, A. (2004)
The Effect of Conforming Loan Status on Mortgage Yield Spreads: A Loan Level Analysis.
{\em Real Estate Economics} 32(4): 541-569.

\bibitem[Amenc {\em et al}, 2016]{Amenc2016} Amenc, N., Ducoulombier, F., Goltz, F. and Ulahel, J. (2016)
Ten Misconceptions about Smart Beta.
{\em Working Paper.} Available online: \url{https://www.edhec.edu/sites/www.edhec-portail.pprod.net/files/publications/pdf/edhec-position-paper-ten-misconceptions-about-smart-beta\%5F1468395239135-pdfjpg}.

\bibitem[Amenc {\em et al}, 2015]{Amenc2015} Amenc, N., Goltz, F., Sivasubramanian, S. and Lodh, A. (2015)
Robustness of Smart Beta Strategies.
{\em Journal of Index Investing} 6(1): 17-38.

\bibitem[Amenc, Martellini and Ziemann, 2009]{Amenc2009} Amenc, N., Martellini, L. and Ziemann, V. (2009)
Inflation-Hedging Properties of Real Assets and Implications for Asset-Liability Management Decisions.
{\em Journal of Portfolio Management} 35(4): 94-110.

\bibitem[Amihud, 2002]{Amihud2002} Amihud, Y. (2002)
Illiquidity and stock returns: cross-section and time-series effects.
{\em Journal of Financial Markets} 5(1): 31-56.

\bibitem[Amihud and Goyenko, 2013]{Amihud2013} Amihud, Y. and Goyenko, R. (2013)
Mutual Fund's $R^2$ as Predictor of Performance.
{\em Review of Financial Studies} 26(3): 667-694.

\bibitem[Amihud and Murgia, 1997]{Amihud1997} Amihud, Y. and Murgia, M. (1997)
Dividends, Taxes, and Signaling: Evidence from Germany.
{\em Journal of Finance} 52(1): 397-408.

\bibitem[Amin and Kat, 2003]{Amin2003} Amin, G.S. and Kat, H.M. (2003)
Welcome to the Dark Side: Hedge Fund Attrition and Survivorship Bias over the Period 1994-2001.
{\em Journal of Alternative Investments} 6(1): 57-73.

\bibitem[Amiri {\em et al}, 2010]{Amiri2010} Amiri, M., Zandieh, M., Vahdani, B., Soltani, R. and Roshanaei, V. (2010)
An integrated eigenvector-DEA-TOPSIS methodology for portfolio risk evaluation in the FOREX spot market.
{\em Expert Systems with Applications} 37(1): 509-516.

\bibitem[Amjad and Shah, 2017]{Amjad2017} Amjad, M.J. and Shah, D. (2017)
Trading Bitcoin and Online Time Series Prediction.
{\em Working Paper.} Available online: \url{http://proceedings.mlr.press/v55/amjad16.pdf}.

\bibitem[Ammann, Kind and Seiz, 2010]{Ammann2010} Ammann, M., Kind, A. and Seiz, R. (2010)
What drives the performance of convertible-bond funds?
{\em Journal of Banking \& Finance} 34(11): 2600-2613.

\bibitem[Ammann, Kind and Wilde, 2003]{Ammann2003} Ammann, M., Kind, A. and Wilde, C. (2003)
Are convertible bonds underpriced? An analysis of the French market.
{\em Journal of Banking \& Finance} 27(4): 635-653.

\bibitem[An {\em et al}, 2014]{An2014} An, B.-J., Ang, A., Bali, T.G. and Cakici, N. (2014)
The Joint Cross Section of Stocks and Options.
{\em Journal of Finance} 69(5): 2279-2337.

\bibitem[Anacker, 2009]{Anacker2009} Anacker, K.B. (2009)
Big flipping schemes in small cities? The case of Mansfield, Ohio.
{\em Housing and Society} 36(1): 5-28.

\bibitem[Anacker and Schintler, 2015]{Anacker2015} Anacker, K.B. and Schintler, L.A. (2015)
Flip that house: visualising and analysing potential real estate property flipping transactions in a cold local housing market in the United States.
{\em International Journal of Housing Policy} 15(3): 285-303.

\bibitem[Anand and Venkataraman, 2016]{Anand2016} Anand, A. and Venkataraman, K. (2016)
Market Conditions, Fragility, and the Economics of Market Making.
{\em Journal of Financial Economics} 121(2): 327-349.

\bibitem[Andersen, 1999]{Andersen1999} Andersen, L. (1999)
A Simple Approach to the Pricing of Bermudan Swaptions in the Multi-factor Libor Market Model.
{\em Journal of Computational Finance} 3(2): 5-32.

\bibitem[Andersen, 2010]{Andersen2010} Andersen, L.B.G. (2010)
Markov models for commodity futures: theory and practice.
{\em Quantitative Finance} 10(8): 831-854.

\bibitem[Andersen and Sidenius, 2005]{Andersen2005} Andersen, L. and Sidenius, J. (2005)
Extensions to the Gaussian Copula: Random Recovery and Random Factor Loadings.
{\em Journal of Credit Risk} 1(1): 29-70.

\bibitem[Andersen, Sidenius and Basu, 2003]{Andersen2003} Andersen, L., Sidenius, J. and Basu, S. (2003)
All your hedges in one basket.
{\em Risk}, November 2003, pp. 67-72.

\bibitem[Anderson, Bianchi and Goldberg, 2014]{Anderson2014} Anderson, R.M., Bianchi, S.W. and Goldberg, L.R. (2014)	
Determinants of Levered Portfolio Performance.
{\em Financial Analysts Journal} 70(5): 53-72.

\bibitem[Anderson and Danthine, 1981]{Anderson1981} Anderson, R.W. and Danthine, J.P. (1981)
Cross Hedging.
{\em Journal of Political Economy} 89(6): 1182-1196.

\bibitem[Andrade, Mitchell and Stafford, 2001]{Andrade2001} Andrade, G., Mitchell, M. and Stafford, E. (2001)
New evidence and perspectives on mergers.
{\em Journal of Economic Perspectives} 15(2): 103-120.

\bibitem[Andrie\c{s} and V\^{\i}rlan, 2017]{Andries2017} Andrie\c{s}, A.M. and V\^{\i}rlan, C.A. (2017)
Risk arbitrage in emerging Europe: are cross-border mergers and acquisition deals more risky?
{\em Economic Research -- Ekonomska Istra\v{z}ivanja} 30(1): 1367-1389.

\bibitem[An\'{e} and Labidi, 2001]{Ane2001} An\'{e}, T. and Labidi, C. (2001)
Implied volatility surfaces and market activity over time.
{\em Journal of Economics and Finance} 25(3): 259-275.

\bibitem[Ang, Alles and Allen, 1998]{Ang1998} Ang, S., Alles, L. and Allen, D. (1998)
Riding the Yield Curve: An Analysis of International Evidence.
{\em Journal of Fixed Income} 8(3): 57-74.

\bibitem[Ang, Bekaert and Wei, 2008]{AngBW2008} Ang, A., Bekaert, G. and Wei, M. (2008)
The Term Structure of Real Rates and Expected Inflation.
{\em Journal of Finance} 63(2): 797-849.

\bibitem[Ang {\em et al}, 2017]{Ang2017} Ang, A., Green, R.C., Longstaff, F.A. and Xing, Y. (2017)
Advance Refundings of Municipal Bonds.
{\em Journal of Finance} 72(4): 1645-1682.

\bibitem[Ang {\em et al}, 2006]{Ang2006} Ang, A., Hodrick, R., Xing, Y. and Zhang, X. (2006)
The Cross-Section of Volatility and Expected Returns.
{\em Journal of Finance} 61(1): 259-299.

\bibitem[Ang {\em et al}, 2009]{Ang2009} Ang, A., Hodrick, R., Xing, Y. and Zhang, X. (2009)
High Idiosyncratic Volatility and Low Returns: International and Further U.S. Evidence.
{\em Journal of Financial Economics} 91(1): 1-23.

\bibitem[Ang and Quek, 2006]{AngQuek2006} Ang, K.K. and Quek, C. (2006)
Stock trading using RSPOP: A novel rough set-based neuro-fuzzy approach.
{\em IEEE Transactions on Neural Networks} 17(5): 1301-1315.

\bibitem[Anglin, Rutherford and Springer, 2003]{Anglin2003} Anglin, P.M., Rutherford, R. and Springer, T. (2003)
The Trade-off Between the Selling Price of Residential Properties and Time-on-the-Market: The Impact of Price Setting.
{\em Journal of Real Estate Finance and Economics} 26(1): 95-111.

\bibitem[Anker, 1999]{Anker1999} Anker, P. (1999)
Uncovered interest parity, monetary policy and time-varying risk premia.
{\em Journal of International Money and Finance} 18(6): 835-851.

\bibitem[Ankirchner {\em et al}, 2012]{Ankirchner2012} Ankirchner, S., Dimitroff, G., Heyne, G. and Pigorsch, C. (2012)
Futures Cross-Hedging with a Stationary Basis.
{\em Journal of Financial and Quantitative Analysis} 47(6): 1361-1395.

\bibitem[Ankirchner and Heyne, 2012]{AnkirchnerHeyne2012} Ankirchner, S. and Heyne, G. (2012) Cross Hedging with Stochastic Correlation.
{\em Finance and Stochastics} 16(1): 17-43.

\bibitem[Ansar {\em et al}, 2016]{Ansar2016} Ansar, A., Flyvbjerg, B., Budzier, A. and  Lunn, D. (2016)
Does infrastructure investment lead to economic growth or economic fragility? Evidence from China.
{\em Oxford Review of Economic Policy} 32(3): 360-390.

\bibitem[Anson, 1998]{Anson1998} Anson, M.J.P (1998)
Spot Returns, Roll Yield, and Diversification with Commodity Futures.
{\em Journal of Alternative Investments} 1(3): 16-32.

\bibitem[Anson, 2013]{Anson2013} Anson, M. (2013)
Performance Measurement in Private Equity: The Impact of FAS 157 on the Lagged Beta Effect.
{\em Journal of Private Equity} 17(1): 29-44.

\bibitem[Antonacci, 2014]{Antonacci2014} Antonacci, G. (2014)
{\em Dual Momentum Investing: An Innovative Strategy for Higher Returns with Lower Risk.}
New York, NY: McGraw-Hill, Inc.

\bibitem[Antonacci, 2017]{Antonacci2017} Antonacci, G. (2017)
Risk Premia Harvesting Through Dual Momentum.
{\em Journal of Management \& Entrepreneurship} 11(1): 27-55.

\bibitem[Antoniou and Holmes, 1995]{Antoniou1995} Antoniou, A. and Holmes, P. (1995)
Futures Trading, Information and Spot Price Volatility: Evidence from the FTSE 100 Stock Index Futures Contract using GARCH.
{\em Journal of Banking \& Finance} 19(1): 117-129.

\bibitem[Aragon {\em et al}, 2017]{Aragon2017} Aragon, G.O., Ergun, A.T., Getmansky, M. and Girardi, G. (2017)
Hedge Fund Liquidity Management.
{\em Working Paper.} Available online: \url{https://ssrn.com/abstract=3033930}.

\bibitem[Ardizzi {\em et al}, 2014]{Ardizzi2014} Ardizzi, G., Petraglia, C., Piacenza, M., Schneider, F. and Turati, G. (2014)
Money Laundering as a Crime in the Financial Sector: A New Approach to Quantitative Assessment, with an Application to Italy.
{\em Journal of Money, Credit and Banking} 46(8): 1555-1590.

\bibitem[Aretz and Pope, 2013]{Aretz2013} Aretz, K. and Pope, P.F. (2013)
Common factors in default risk across countries and industries.
{\em European Financial Management} 19(1): 108-152.

\bibitem[Arezki and Sy, 2016]{Arezki2016} Arezki, R. and Sy, A. (2016)
Financing Africa's Infrastructure Deficit: From Development Banking to Long-term Investing.
{\em Journal of African Economies} 25(S2): 59-73.

\bibitem[Armann and Weisdorf, 2008]{Armann2008} Armann, V. and Weisdorf, M. (2008)
Hedging Inflation with Infrastructure Assets.
In: Benaben, B. and Goldenberg, S. (eds.)
{\em Inflation Risk and Products: The Complete Guide.} London, UK: Risk Books, pp. 111-126.

\bibitem[Arnott {\em et al}, 2014]{Arnott2014} Arnott, R., Chaves, D., Gunzberg, J., Hsu, J. and Tsui, P. (2014)
Getting Smarter about Commodities: An index to counter the possible pitfalls.
{\em  Journal of Indexes}, November/December 2014, pp. 52-60.

\bibitem[Arnott {\em et al}, 2013]{Arnott2013} Arnott, R.D., Hsu, J., Kalesnik, V. and Tindall, P. (2013)
The Surprising Alpha from Malkiel's Monkey and Upside-Down Strategies.
{\em Journal of Portfolio Management} 39(4): 91-105.

\bibitem[Arnsdorf and Halperin, 2007]{Arnsdorf2007} Arnsdorf, M. and Halperin, I. (2007)
BSLP: Markovian bivariate spread-loss model for portfolio credit derivatives.
{\em Working Paper.} Available online: \url{https://arxiv.org/pdf/0901.3398}.

\bibitem[Asem and Tian, 2010]{Asem2010} Asem, E. and Tian, G. (2010)
Market Dynamics and Momentum Profits.
{\em Journal of Financial and Quantitative Analysis} 45(6): 1549-1562.

\bibitem[Asgharian {\em et al}, 2004]{Asgharian2004} Asgharian, M., Diz, F., Gregoriou, G.N. and Rouah, F. (2004)
The Global Macro Hedge Fund Cemetery.
{\em Journal of Derivatives Accounting} 1(2): 187-194.

\bibitem[Asgharian and Karlsson, 2008]{Asgharian2008} Asgharian, H. and Karlsson, S. (2008)
An Empirical Analysis of Factors Driving the Swap Spread.
{\em Journal of Fixed Income} 18(2): 41-56.

\bibitem[Asness, 1994]{Asness1994} Asness, C.S. (1994)
{\em Variables that Explain Stock Returns} (Ph.D. Thesis). Chicago, IL: University of Chicago.

\bibitem[Asness, 1995]{Asness1995} Asness, C.S. (1995)
The Power of Past Stock Returns to Explain Future Stock Returns.
{\em Working Paper} (unpublished). New York, NY: Goldman Sachs Asset Management.

\bibitem[Asness, 1997]{Asness1997} Asness, C. (1997)
The Interaction of Value and Momentum Strategies.
{\em Financial Analysts Journal} 53(2): 29-36.

\bibitem[Asness {\em et al}, 2014]{Asness2014} Asness, C., Frazzini, A.,  Israel, R. and Moskowitz, T. (2014)
Fact, Fiction, and Momentum Investing.
{\em Journal of Portfolio Management } 40(5): 75-92.

\bibitem[Asness {\em et al}, 2001]{Asness2001} Asness, C., Krail, R.J. and Liew, J.M. (2001)
Do Hedge Funds Hedge?
{\em Journal of Portfolio Management} 28(1): 6-19.

\bibitem[Asness, Moskowitz and Pedersen, 2013]{Asness2013} Asness, C., Moskowitz, T. and Pedersen, L.H. (2013)
Value and Momentum Everywhere.
{\em Journal of Finance} 68(3): 929-985.

\bibitem[Asness, Porter and Stevens, 2000]{Asness2000} Asness, C.S., Porter, R.B. and Stevens, R.L. (2000)
Predicting Stock Returns Using Industry-Relative Firm Characteristics.
{\em Working Paper.} Available online: \url{https://ssrn.com/abstract=213872}.

\bibitem[Augustin, Brenner and Subrahmanyam, 2015]{Augustin2015} Augustin, P., Brenner, B. and Subrahmanyam, M.G. (2015)
Informed Options Trading prior to M\&A Announcements: Insider Trading?
{\em Working Paper}. Available online: \url{https://ssrn.com/abstract=2441606}.

\bibitem[Aussenegg, G\"{o}tz and Jelic, 2014]{Aussenegg2014} Aussenegg, W., G\"{o}tz, L. and Jelic, R. (2014)
European asset swap spreads and the credit crisis.
{\em European Journal of Finance} 22(7): 572-600.

\bibitem[Avdjiev {\em et al}, 2016]{Avdjiev2016} Avdjiev, S., Du, W., Koch, C. and Shin, H.S. (2016)
The Dollar, Bank Leverage and the Deviation from Covered Interest Parity.
{\em Working Paper.} Available online: \url{https://ssrn.com/abstract=2870057}.

\bibitem[Avellaneda and Lee, 2010]{Avellaneda2010} Avellaneda, M. and Lee, J.H. (2010)
Statistical arbitrage in the U.S. equity market.
{\em Quantitative Finance} 10(7): 761-782.

\bibitem[Avellaneda and Papanicolaou, 2018]{Avellaneda2018} Avellaneda, M. and Papanicolaou, A. (2018)
Statistics of VIX Futures and Applications to Trading Volatility Exchange-Traded Products.
{\em Journal of Investment Strategies} 7(2): 1-33.

\bibitem[Avellaneda and Stoikov, 2008]{Avellaneda2008} Avellaneda, M. and Stoikov, S. (2008)
High frequency trading in a limit order book.
{\em Quantitative Finance} 8(3): 217-224.

\bibitem[Avellaneda and Zhang, 2010]{AvellanedaZhang2010} Avellaneda, M. and Zhang, S. (2010)
Path-Dependence of Leveraged ETF Returns.
{\em Journal on Financial Mathematics} 1(1): 586-603.

\bibitem[Ayache, Forsyth and Vetzal, 2003]{Ayache2013} Ayache, E., Forsyth, P.A. and Vetzal, K.R. (2003)
Valuation of Convertible Bonds With Credit Risk.
{\em Journal of Derivatives} 11(1): 9-29.

\bibitem[Ayuso and Restoy, 1996]{Ayuso1996} Ayuso, J. and Restoy, F. (1996)
Interest Rate Parity and Foreign Exchange Risk Premia in the ERM.
{\em Journal of International Money and Finance} 15(3): 369-382.

\bibitem[Azmat and Iqbal, 2017]{Azmat2017} Azmat, Q. and Iqbal, A.M. (2017)
The role of financial constraints on precautionary cash holdings: evidence from Pakistan.
{\em Economic Research -- Ekonomska Istra\v{z}ivanja} 30(1): 596-610.

\bibitem[Baba and Packer, 2009]{Baba2009} Baba, N. and Packer, F. (2009)
Interpreting deviations from covered interest parity during the financial market turmoil of 2007-08.
{\em Journal of Banking \& Finance} 33(11): 1953-1962.

\bibitem[Babbs and Nowman, 1999]{Babbs1999} Babbs, S.H. and Nowman, B.K. (1999)
Kalman filtering of generalized Vasicek term structure models.
{\em Journal of Financial and Quantitative Analysis} 34(1): 115-130.

\bibitem[Bacchetta and van Wincoop, 2006]{Bacchetta2006} Bacchetta, P. and van Wincoop, E. (2006)
Incomplete Information Processing: A Solution to the Forward Discount Puzzle.
{\em American Economic Review} 96(3): 552-576.

\bibitem[Bacchetta and van Wincoop, 2010]{Bacchetta2010} Bacchetta, P. and van Wincoop, E. (2010)
Infrequent Portfolio Decisions: A Solution to the Forward Discount Puzzle.
{\em American Economic Review} 100(3): 870-904.

\bibitem[Baek and Elbeck, 2014]{Baek2014} Baek, C. and Elbeck, M. (2014)
Bitcoins as an Investment or Speculative Vehicle? A First Look.
{\em Applied Economics Letters} 22(1): 30-34.

\bibitem[Bai, Bond and Hatch, 2015]{Bai2015} Bai, Q., Bond, S.A. and Hatch, B.C. (2015)
The Impact of Leveraged and Inverse ETFs on Underlying Real Estate Returns.
{\em Real Estate Economics} 43(1): 37-66.

\bibitem[Bai and Collin-Dufresne, 2013]{Bai2013} Bai, J. and Collin-Dufresne, P. (2013)
The CDS-Bond Basis. {\em Working Paper.} Available online: \url{https://ssrn.com/abstract=2024531}.

\bibitem[Baillie and Myers, 1991]{Baillie1991} Baillie, R.T. and Myers, R.J. (1991)
Bivariate GARCH estimation of the optimal commodity futures hedge.
{\em Journal of Applied Econometrics} 6(2): 109-124.

\bibitem[Baillie and Osterberg, 2000]{Baillie2000} Baillie, R.T. and Osterberg, W.P. (2000)
Deviations from daily uncovered interest rate parity and the role of intervention.
{\em Journal of International Financial Markets, Institutions and Money} 10(4): 363-379.

\bibitem[Baker, Bradley and Wurgler, 2011]{Baker2011} Baker, M., Bradley, B. and Wurgler, J. (2011)
Benchmarks as Limits to Arbitrage: Understanding the Low-Volatility Anomaly.
{\em Financial Analysts Journal} 67(1): 40-54.

\bibitem[Baker, Pan and Wurgler, 2012]{Baker2012} Baker, M., Pan, A. and Wurgler, J. (2012)
The effect of reference point prices on mergers and acquisitions.
{\em Journal of Financial Economics} 106(1): 49-71.

\bibitem[Baker and Sava\c{s}oglu, 2002]{Baker2002} Baker, M. and Sava\c{s}oglu, S. (2002)
Limited arbitrage in mergers and acquisitions.
{\em Journal of Financial Economics} 64(1): 91-115.

\bibitem[Bakshi and Kapadia, 2003a]{Bakshi2003a} Bakshi, G. and Kapadia, N. (2003a)
Delta-Hedged Gains and the Negative Market Volatility Risk Premium.
{\em Review of Financial Studies} 16(2): 527-566.

\bibitem[Bakshi and Kapadia, 2003b]{Bakshi2003b} Bakshi, G. and Kapadia, N. (2003b)
Volatility Risk Premiums Embedded in Individual Equity Options.
{\em Journal of Derivatives} 11(1): 45-54.

\bibitem[Bakshi, Kapadia and Madan, 2003]{BakshiKM2003} Bakshi, G., Kapadia, N. and Madan, D. (2003)
Stock Return Characteristics, Skew Laws, and the Differential Pricing of Individual Equity Options.
{\em Review of Financial Studies} 16(1): 101-143.

\bibitem[Bakshi and Panayotov, 2013]{Bakshi2013} Bakshi, G. and Panayotov, G. (2013)
Predictability of currency carry trades and asset pricing implications.
{\em Journal of Financial Economics} 110(1): 139-163.

\bibitem[Balb\'{a}s, Longarela and Lucia, 1999]{Balbas1999} Balb\'{a}s, A., Longarela, I.R. and Lucia, J.J. (1999)
How Financial Theory Applies to Catastrophe-Linked Derivatives -- An Empirical Test of Several Pricing Models.
{\em Journal of Risk and Insurance} 66(4): 551-582.

\bibitem[Bali and Demirtas, 2008]{Bali2008} Bali, T.G. and Demirtas, K.O. (2008)
Testing mean reversion in financial market volatility: Evidence from S\&P 500 index futures.
{\em Journal of Futures Markets} 28(1): 1-33.

\bibitem[Bali and Hovakimian, 2009]{Bali2009} Bali, T.G. and Hovakimian, A. (2009)
Volatility Spreads and Expected Stock Returns.
{\em Management Science} 55(11): 1797-1812.

\bibitem[Ballings {\em et al}, 2015]{Ballings2015} Ballings, M., Van den Poel, D., Hespeels, N. and Gryp, R. (2015)
Evaluating multiple classifiers for stock price direction prediction.
{\em Expert Systems with Applications} 42(20): 7046-7056.

\bibitem[Balta and Kosowski, 2013]{Balta2013} Balta, A.-N. and Kosowki, R. (2013)
Momentum Strategies in Futures Markets and Trend-Following Funds.
{\em Working Paper.} Available online: \url{https://www.edhec.edu/sites/www.edhec-portail.pprod.net/files/publications/pdf/edhec-working-paper-momentum-strategies-in-futures_1410350911195-pdfjpg}.

\bibitem[Bandarchuk and Hilscher, 2013]{Bandarchuk2013} Bandarchuk, P. and Hilscher, J. (2013)
Sources of Momentum Profits: Evidence on the Irrelevance of Characteristics.
{\em Review of Finance} 17(2): 809-845.

\bibitem[Banz, 1981]{Banz1981} Banz, R. (1981)
The relationship between return and market value of common stocks.
{\em Journal of Financial Economics} 9(1): 3-18.

\bibitem[Barber, Bennett and Gvozdeva, 2015]{Barber2015} Barber, J., Bennett, S. and Gvozdeva, E. (2015)
How to Choose a Strategic Multifactor Equity Portfolio?
{\em Journal of Index Investing} 6(2): 34-45.

\bibitem[Barberis, 2000]{Barberis2000} Barberis, N. (2000)
Investing for the Long Run when Returns Are Predictable.
{\em Journal of Finance} 55(1): 225-264.

\bibitem[Barberis and Huang, 2008]{Barberis2008} Barberis, N. and Huang, M. (2008)
Stocks as Lotteries: The Implications of Probability Weighting for Security Prices.
{\em American Economic Review} 98(5): 2066-2100.

\bibitem[Bardong and Lehnert, 2004]{Bardong2004} Bardong, F. and Lehnert, T. (2004)
TIPS, Break-Even Inflation, and Inflation Forecasts.
{\em Journal of Fixed Income} 14(3): 15-35.

\bibitem[Bariviera {\em et al}, 2017]{Bariviera2017} Bariviera, A.F., Basgall, M.J., Hasperu\'e, W. and Naiouf, M. (2017)
Some stylized facts of the Bitcoin market.
{\em Physica A: Statistical Mechanics and its Applications} 484: 82-90.

\bibitem[Barnes {\em et al}, 2010]{Barnes2010} Barnes, M.L., Bodie, Z., Triest, R.K. and Wang, J.C. (2010)
A TIPS Scorecard: Are They Accomplishing Their Objectives?
{\em Financial Analysts Journal} 66(5): 68-84.

\bibitem[Baron {\em et al}, 2014]{Baron2014} Baron, M., Brogaard, J., Hagstr\"{o}mer, B. and Kirilenko, A. (2014)
Risk and Return in High-Frequency Trading.
{\em Journal of Financial and Quantitative Analysis} (forthcoming). Available online: \url{https://ssrn.com/abstract=2433118}.

\bibitem[Barr and Campbell, 1997]{Barr1997} Barr, D.G. and Campbell, J.Y. (1997)
Inflation, real interest rates, and the bond market: A study of UK nominal and index-linked government bond prices.
{\em Journal of Monetary Economics} 39(3): 361-383.

\bibitem[Barrett and Kolb, 1995]{Barrett1995} Barrett, W.B. and Kolb, R.W. (1995)
Analysis of spreads in agricultural futures.
{\em Journal of Futures Markets} 15(1): 69-86.

\bibitem[Barrieu and El Karoui, 2002]{Barrieu2002} Barrieu, P. and El Karoui, N. (2002)
Optimal design of weather derivatives.
{\em ALGO Research} 5(1): 79-92.

\bibitem[Barrieu and Scaillet, 2010]{Barrieu2010} Barrieu, P. and Scaillet, O. (2010)
A Primer on Weather Derivatives.
In: Filar, J.A. and Haurie, A. (eds.) {\em Uncertainty and Environmental Decision Making: A Handbook of Research and Best Practice.} International Series in Operations Research \& Management Science, Vol. 138. New York, NY: Springer U.S.

\bibitem[Barroso and Santa-Clara, 2014]{Barroso2014} Barroso, P. and Santa-Clara, P. (2014)
Momentum Has Its Moments.
{\em Journal of Financial Economics} 116(1): 111-120.

\bibitem[Bartonov\'{a}, 2012]{Bartonova2012} Bartonov\'{a}, M. (2012)
Hedging of Sales by Zero-cost Collar and its Financial Impact.
{\em Journal of Competitiveness} 4(2): 111-127.

\bibitem[Bartov, Radhakrishnan and Krinsky, 2005]{Bartov2005} Bartov, E., Radhakrishnan, S. and Krinsky, I. (2005)
Investor Sophistication and Patterns in Stock Returns after Earnings Announcements.
{\em Accounting Review} 75(1): 289-319.

\bibitem[Basu, 1977]{Basu1977} Basu, S. (1977)
The investment performance of common stocks in relation to their price to earnings ratios: A test of the efficient market hypothesis.
{\em Journal of Finance} 32(3): 663-682.

\bibitem[Basu and Miffre, 2013]{Basu2013} Basu, D. and Miffre, J. (2013)
Capturing the risk premium of commodity futures: The role of hedging pressure.
{\em Journal of Banking \& Finance} 37(7): 2652-2664.

\bibitem[Batta, Chacko and Dharan, 2010]{Batta2010} Batta, G., Chacko, G. and Dharan, B. (2010)
A Liquidity-Based Explanation of Convertible Arbitrage Alphas.
{\em Journal of Fixed Income} 20(1): 28-43.

\bibitem[Battalio and Mendenhall, 2007]{Battalio2007} Battalio, R. and Mendenhall, R. (2007)
Post-Earnings Announcement Drift: Intra-Day Timing and Liquidity Costs.
{\em Working Paper.} Available online: \url{https://ssrn.com/abstract=937257}.

\bibitem[Batten and Ellis, 1996]{Batten1996} Batten, J. and Ellis, C. (1996)
Technical trading system performance in the Australian share market: Some empirical evidence.
{\em Asia Pacific Journal of Management} 13(1): 87-99.

\bibitem[Batten, Khaw and Young, 2014]{Batten2014} Batten, J.A., Khaw, K. and Young, M.R. (2014)
Convertible Bond Pricing Models.
{\em Journal of Economic Surveys} 28(5): 775-803.

\bibitem[Baxter and King, 1999]{Baxter1999} Baxter, M. and King, R. (1999)
Measuring business cycles: Approximate band-pass filters for economic time-series.
{\em Review of Economics and Statistics} 81(4): 575-593.

\bibitem[Baxter and Rennie, 1996]{Baxter1996} Baxter, M. and Rennie, A. (1996)
{\em Financial Calculus: An Introduction to Derivative Pricing.}
Cambridge, UK: Cambridge University Press.

\bibitem[Bayer {\em et al}, 2015]{Bayer2015} Bayer, P.J., Geissler, C., Mangum, K. and Roberts, J.W. (2015)
Speculators and Middlemen: The Strategy and Performance of Investors in the Housing Market.
{\em Working Paper.} Available online: \url{https://ssrn.com/abstract=1754003}.

\bibitem[Beaver, 1966]{Beaver1966} Beaver, W.H. (1966)
Financial ratios as predictors of failure.
{\em Journal of Accounting Research} 4: 71-111.

\bibitem[Beaver, McNichols and Rhie, 2005]{Beaver2005} Beaver, W.H., McNichols, M.F. and Rhie, J.-W. (2005)
Have financial statements become less informative? Evidence from the ability of financial ratios to predict bankruptcy.
{\em Review of Accounting Studies} 10(1): 93-122.

\bibitem[Bedendo, Cathcart and El-Jahel, 2007]{Bedendo2007} Bedendo, M., Cathcart, L. and El-Jahel, L. (2007)
The Slope of the Term Structure of Credit Spreads: An Empirical Investigation.
{\em Journal of Financial Research} 30(2): 237-257.

\bibitem[Beekhuizen {\em et al}, 2016]{Beekhuizen2016} Beekhuizen, P., Duyvesteyn, J., Martens, M. and Zomerdijk, C. (2016)
Carry Investing on the Yield Curve.
{\em Working Paper.} Available online: \url{http://ssrn.com/abstract=2808327}.

\bibitem[Bekaert and Wang, 2010]{Bekaert2010} Bekaert, G. and Wang, X. (2010)
Inflation Risk and the Inflation Risk Premium.
{\em Economic Policy} 25(64): 755-806.

\bibitem[Bekaert, Wei and Xing, 2007]{Bekaert2007} Bekaert, G., Wei, M. and Xing, Y. (2007)
Uncovered interest rate parity and the term structure.
{\em Journal of International Money and Finance} 26(6): 1038-1069.

\bibitem[Bekkers, Doeswijk and Lam, 2009]{Bekkers2009} Bekkers, N., Doeswijk, R.Q. and Lam, T.W. (2009)
Strategic Asset Allocation: Determining the Optimal Portfolio with Ten Asset Classes.
{\em Journal of Wealth Management} 12(3): 61-77.

\bibitem[Belgrade and Benhamou, 2004]{Belgrade2004} Belgrade, N. and Benhamou, E. (2004)
Reconciling Year on Year and Zero Coupon Inflation Swap: A Market Model Approach.
{\em Working Paper.} Available online: \url{https://ssrn.com/abstract=583641}.

\bibitem[Belgrade, Benhamou and Koehler, 2004]{BelgradeBK2004} Belgrade, N., Benhamou, E. and Koehler, E. (2004)
A Market Model for Inflation.
{\em Working Paper.} Available online: \url{https://ssrn.com/abstract=576081}.

\bibitem[Belkin, Suchover and Forest, 1998]{Belkin1998} Belkin, B., Suchover, S. and Forest, L. (1998)
A one-parameter representation of credit risk and transition matrices.
{\em Credit Metrics Monitor} 1(3): 46-56.

\bibitem[Bellamy, 1994]{Bellamy1994} Bellamy, D.E. (1994)
Evidence of imputation clienteles in the Australian equity market.
{\em Asia Pacific Journal of Management} 11(2): 275-287.

\bibitem[Bellovary, Giacomino and Akers, 2007]{Bellovary2007} Bellovary, J.L., Giacomino, D.E. and Akers, M.D. (2007)
A review of bankruptcy prediction studies: 1930 to present.
{\em Journal of Financial Education} 33(4): 3-41.

\bibitem[Benavides, 2009]{Benavides2009} Benavides, G. (2009)
Predictive Accuracy of Futures Options Implied Volatility: The Case of the Exchange Rate Futures Mexican Peso-US Dollar.
{\em Panorama Econ\'{o}mico} 5(9): 55-95.

\bibitem[Ben-David, Franzoni and Moussawi, 2012]{Ben-David2012} Ben-David, I., Franzoni, F.A. and Moussawi, R. (2012)
ETFs, Arbitrage, and Contagion.
{\em Working Paper.} Available online: \url{http://www.nccr-finrisk.uzh.ch/media/pdf/wp/WP793_B1.pdf}.

\bibitem[Ben-David, Franzoni and Moussawi, 2017]{Ben-David2017} Ben-David, I., Franzoni, F.A. and Moussawi, R. (2017)
Do ETFs Increase Volatility?
{\em Journal of Finance} (forthcoming). Available online: \url{https://ssrn.com/abstract=1967599}.

\bibitem[Beneish and Whaley, 1996]{Beneish1996} Beneish, M.D. and Whaley, R.E. (1996)
An Anatomy of the ``S\&P Game": The Effects of Changing the Rules.
{\em Journal of Finance} 51(5): 1909-1930.

\bibitem[Benet, 1990]{Benet1990} Benet, B.A. (1990)
Commodity futures cross hedging of foreign exchange exposure.
{\em Journal of Futures Markets} 10(3): 287-306.

\bibitem[Bengio, 2009]{Bengio2009} Bengio, Y. (2009)
Learning Deep Architectures for AI.
{\em Foundations and Trends in Machine Learning} 2(1): 1-127.

\bibitem[Benhamou, 2016]{Benhamou2016} Benhamou, E. (2016)
Trend Without Hiccups -- A Kalman Filter Approach.
{\em Working Paper.} Available online: \url{https://ssrn.com/abstract=2747102}.

\bibitem[Benos {\em et al}, 2017]{Benos2017} Benos, E., Brugler, J., Hjalmarsson, E. and Zikes, F. (2017)
Interactions among High-Frequency Traders.
{\em Journal of Financial and Quantitative Analysis} 52(4): 1375-1402.

\bibitem[Benos and Sagade, 2016]{Benos2016} Benos, E. and Sagade, S. (2016) Price Discovery and the Cross-Section of High-Frequency Trading.
{\em Journal of Financial Markets} 30: 54-77.

\bibitem[Benth, 2003]{Benth2003} Benth, F. (2003)
On arbitrage-free pricing of weather derivatives based on fractional Brownian motion.
{\em Applied Mathematical Finance} 10(4): 303-324.

\bibitem[Benth and Kettler, 2010]{Benth2011} Benth, F.E. and Kettler, P.C. (2010)
Dynamic copula models for the spark spread.
{\em Quantitative Finance} 11(3): 407-421.

\bibitem[Benth, Kholodnyi and Laurence, 2014]{Benth2014} Benth, F.E., Kholodnyi, V.A. and Laurence, P. (eds.) (2014)
{\em Quantitative Energy Finance: Modeling, Pricing, and Hedging in Energy and Commodity Markets.}
New York, NY: Springer-Verlag.

\bibitem[Benth and Saltyte-Benth, 2005]{Benth2005} Benth, F.E. and Saltyte-Benth, J. (2005)
Stochastic modelling of temperature variations with a view towards weather derivatives.
{\em Applied Mathematical Finance} 12(1): 53-85.

\bibitem[Benth and Saltyte-Benth, 2007]{Benth2007} Benth, F.E. and Saltyte-Benth, J. (2007)
The volatility of temperature and pricing of weather derivatives.
{\em Quantitative Finance} 7(5): 553-561.

\bibitem[Benth, Saltyte-Benth and Koekebakker, 2007]{BenthSK2007} Benth, F., Saltyte-Benth, J. and Koekebakker, S. (2007)
Putting a price on temperature.
{\em Scandinavian Journal of Statistics} 34(4): 746-767.

\bibitem[BenZion, Anan and Yagil, 2005]{BenZion2005} BenZion, U., Anan, S.D. and Yagil, J. (2005)
Box Spread Strategies and Arbitrage Opportunities.
{\em Journal of Derivatives} 12(3): 47-62.

\bibitem[BenZion {\em et al}, 2003]{BenZion2003} BenZion, U., Klein, P., Shachmurove, Y. and Yagil, J. (2003)
Efficiency differences between the S\&P 500 and the Tel-Aviv 25 indices: a moving average comparison.
{\em International Journal of Business} 8(3): 267-284.

\bibitem[Beracha and Downs, 2015]{Beracha2015} Beracha, E. and Downs, D.H. (2015)
Value and Momentum in Commercial Real Estate: A Market-Level Analysis.
{\em Journal of Portfolio Management} 41(6): 48-61.

\bibitem[Beracha and Skiba, 2011]{Beracha2011} Beracha, E. and Skiba, H. (2011)
Momentum in Residential Real Estate.
{\em Journal of Real Estate Finance and Economics} 43(3): 229-320.

\bibitem[Berk, Green and Naik, 1999]{Berk1999} Berk, J., Green, R. and Naik, V. (1999)
Optimal Investment, Growth Options and Security Returns.
{\em Journal of Finance} 54(5): 1153-1608.

\bibitem[Bernadell, Coche and Nyholm, 2005]{Bernadell2005} Bernadell, C., Coche, J. and Nyholm, K. (2005)
Yield curve prediction for the strategic investor.
{\em Working Paper Series}, No. 472. Frankfurt am Main, Germany: European Central Bank. Available online: \url{https://www.ecb.europa.eu/pub/pdf/scpwps/ecbwp472.pdf?1dc8846d9df4642959c54aa73cee81ad}.

\bibitem[Bernanke and Kuttner, 2005]{Bernanke2005} Bernanke, B.S. and Kuttner, K.N. (2005)
What Explains the Stock Market's Reaction to Federal Reserve Policy?
{\em Journal of Finance} 60(3): 1221-1257.

\bibitem[Bernard, Cui and Mcleish, 2014]{Bernard2014} Bernard, C., Cui, Z. and Mcleish, D. (2014)
Convergence of the discrete variance swap in time-homogeneous diffusion models.
{\em Quantitative Finance Letters} 2(1): 1-6.

\bibitem[Bernard and Thomas, 1989]{Bernard1989} Bernard, V.L. and Thomas, J.K. (1989)
Post-Earnings-Announcement Drift: Delayed Price Response or Risk Premium?
{\em Journal of Accounting Research} 27: 1-36.

\bibitem[Bernard and Thomas, 1990]{Bernard1990} Bernard, V.L. and Thomas, J.K. (1990)
Evidence That Stock Prices Do Not Fully Reflect the Implications of Current Earnings for Future Earnings.
{\em Journal of Accounting and Economics} 13(4): 305-340.

\bibitem[Bernardi, Leippold and Lohre, 2018]{Bernardi2018} Bernardi, S., Leippold, M. and Lohre, H. (2018)
Maximum Diversification Strategies along Commodity Risk Factors.
{\em European Financial Management} 24(1): 53-78.

\bibitem[Bernstein, 1990]{Bernstein1990} Bernstein, J. (1990)
{\em Jake Bernstein's seasonal futures spreads: high-probability seasonal spreads for futures traders.}
Hoboken, NJ: John Wiley \& Sons, Inc.

\bibitem[Bessembinder, 1992]{Bessembinder1992} Bessembinder, H. (1992)
Systematic risk, hedging pressure, and risk premiums in futures markets.
{\em Review of Financial Studies} 5(4): 637-667.

\bibitem[Bessembinder, 1993]{Bessembinder1993} Bessembinder, H. (1993)
An empirical analysis of risk premia in futures markets.
{\em Journal of Futures Markets} 13(6): 611-630.

\bibitem[Bessembinder and Chan, 1992]{BessembinderChan1992} Bessembinder, H. and Chan, K.	 (1992)
Time-varying risk premia and forecastable returns in futures markets.
{\em Journal of Financial Economics} 32(2): 169-193.

\bibitem[Bessembinder {\em et al}, 1995]{Bessembinder1995} Bessembinder, H., Coughenour, J.F., Seguin, P.J. and Smoller, M.M. (1995)
Mean reversion in equilibrium asset prices: evidence from the futures term structure.
{\em Journal of Finance} 50(1): 361-375.

\bibitem[Bessembinder and Maxwell, 2008]{Bessembinder2008} Bessembinder, H. and Maxwell, W. (2008)
Markets: Transparency and the Corporate Bond Market.
{\em Journal of Economic Perspectives} 22(2): 217-234.

\bibitem[Bessembinder and Seguin, 1993]{BessembinderS1993} Bessembinder, H. and Seguin, P.J. (1993)
Price volatility, trading volume, and market depth: Evidence from futures markets.
{\em Journal of Financial and Quantitative Analysis} 28(1): 21-39.

\bibitem[Bester, Martinez and Rosu, 2017]{Bester2017} Bester, A., Martinez, V.H. and Rosu, I. (2017)
Cash Mergers and the Volatility Smile.
{\em Working Paper.} Available online: \url{https://ssrn.com/abstract=1364491}.

\bibitem[Beyaert, Garc\'{\i}a-Solanes, and P\'{e}rez-Castej\'{o}n, 2007]{Beyaert2007}
Beyaert, A., Garc\'{\i}a-Solanes, J. and P\'{e}rez-Castej\'{o}n, J.J. (2007)
Uncovered interest parity with switching regimes.
{\em Economic Modelling} 24(2): 189-202.

\bibitem[Bharadwaj and Wiggins, 2001]{Bharadwaj2001} Bharadwaj, A. and Wiggins, J.B. (2001)
Box Spread and Put-Call Parity Tests for the S\&P 500 Index LEAPS Market.
{\em Journal of Derivatives} 8(4): 62-71.

\bibitem[Bhattacharya {\em et al}, 2017]{Bhattacharya2017} Bhattacharya, U., Loos, B., Meyer, S. and Hackethal, A. (2017)
Abusing ETFs.
{\em Review of Finance} 21(3): 1217-1250.

\bibitem[Bhojraj and Swaminathan, 2006]{Bhojraj2006} Bhojraj, S. and Swaminathan, B. (2006)
Macromomentum: Returns Predictability in International Equity Indices.
{\em Journal of Business} 79(1): 429-451.

\bibitem[Bhushan, 1994]{Bhushan1994} Bhushan, R. (1994)
An Informational Efficiency Perspective on the Post-Earnings Announcement Drift.
{\em Journal of Accounting and Economics} 18(1): 45-65.

\bibitem[Biais and Foucault, 2014]{Biais2014} Biais, B. and Foucault, T. (2014)
HFT and market quality.
{\em Bankers, Markets \& Investors} 128: 5-19.

\bibitem[Biais, Foucault and Moinas, 2014]{BiaisMoinas2014} Biais, B., Foucault, T. and Moinas, S. (2014)
Equilibrium Fast Trading.
{\em Working Paper.}
Available online: \url{https://ssrn.com/abstract=2024360}.

\bibitem[Bianchi, Drew and Fan, 2015]{Bianchi2015} Bianchi, R.J., Drew, M. and Fan, J. (2015)
Combining momentum with reversal in commodity futures.
{\em Journal of Banking \& Finance} 59: 423-444.

\bibitem[Biby, Modukuri and Hargrave, 2001]{Biby2001} Biby, J.D., Modukuri, S. and Hargrave, B. (2001)
Collateralized Borrowing via Dollar Rolls.
In: Fabozzi, F.J. (ed.) {\em The Handbook of Mortgage-Backed Securities.} (5th ed.) New York, NY: McGraw-Hill, Inc.

\bibitem[Bielecki, Brigo and Patras, 2011]{Bielecki2011} Bielecki, T.R., Brigo, D. and Patras, F.  (2011)
{\em Credit Risk Frontiers: Subprime Crisis, Pricing and Hedging, CVA, MBS, Ratings, and Liquidity.}
Hoboken, NJ: John Wiley \& Sons, Inc.

\bibitem[Bielecki, Jeanblanc and Rutkowski, 2007]{Bielecki2007} Bielecki, T., Jeanblanc, M. and Rutkowski, M. (2007)
Hedging of basket credit derivatives in the Credit Default Swap market.
{\em Journal of Credit Risk} 3(1): 91-132.

\bibitem[Bielecki, Vidozzi and Vidozzi, 2008]{Bielecki2008} Bielecki, T., Vidozzi, A. and Vidozzi, L. (2008)
A Markov copulae approach to pricing and hedging of credit index derivatives and ratings triggered step-up bonds.
{\em Journal of Credit Risk} 4(1): 47-76.

\bibitem[Bieri and Chincarini, 2004]{Bieri2004} Bieri, D.S. and Chincarini, L.B. (2004)
Riding the Yield Curve: Diversification of Strategies.
{\em Working Paper.} Available online: \url{https://ssrn.com/abstract=547682}.

\bibitem[Bieri and Chincarini, 2005]{Bieri2005} Bieri, D.S. and Chincarini, L.B. (2005)
Riding the Yield Curve: A Variety of Strategies.
{\em Journal of Fixed Income} 15(2): 6-35.

\bibitem[Bierwag, 1979]{Bierwag1979} Bierwag, G.O. (1979)
Dynamic portfolio immunization policies.
{\em Journal of Banking \& Finance} 3(1): 23-41.

\bibitem[Bierwag and Kaufman, 1978]{Bierwag1978} Bierwag, G.O. and Kaufman, G. (1978)
Bond Portfolio Strategy Simulations: A Critique.
{\em Journal of Financial and Quantitative Analysis} 13(3): 519-525.

\bibitem[Billingsley and Chance, 1985]{Billingsley1985} Billingsley, R.S. and Chance, D.M. (1985)
Options Market Efficiency and the Box Spread Strategy.
{\em Financial Review} 20(4): 287-301.

\bibitem[Billingsley and Chance, 1988]{Billingsley1988} Billingsley, R.S. and Chance, D.M. (1988)
The pricing and performance of stock index futures spreads.
{\em Journal of Futures Markets} 8(3): 303-318.

\bibitem[Bilson, 1981]{Bilson1981} Bilson, J.F.O. (1981)
The ``Speculative Efficiency" Hypothesis.
{\em Journal of Business} 54(3): 435-451.

\bibitem[Birari and Rode, 2014]{Birari2014} Birari, A. and Rode, M. (2014)
Edge Ratio of Nifty for Last 15 Years on Donchian Channel.
{\em SIJ Transactions on Industrial, Financial \& Business Management (IFBM)} 2(5): 247-254.

\bibitem[Bird, Liem and Thorp, 2014]{Bird2014} Bird, R., Liem, H. and Thorp, S. (2014)
Infrastructure: Real Assets and Real Returns.
{\em European Financial Management} 20(4): 802-824.

\bibitem[Bitsch, Buchner and Kaserer, 2010]{Bitsch2010} Bitsch, F., Buchner, A. and Kaserer, C. (2010)
Risk, Return and Cash Flow Characteristics of Infrastructure Fund Investments.
{\em EIB Papers} 15(1): 106-136.

\bibitem[Bjornson and Carter, 1997]{Bjornson1997} Bjornson, B. and Carter, C.A. (1997)
New Evidence on Agricultural Commodity Return Performance under Time-Varying Risk.
{\em American Journal of Agricultural Economics} 79(3): 918-930.

\bibitem[Black, 1972]{Black1972} Black, F. (1972)
Capital Market Equilibrium with Restricted Borrowing.
{\em Journal of Business} 45(3): 444-455.

\bibitem[Black and Litterman, 1991]{Black1991} Black, F. and Litterman, R. (1991)
Asset allocation: Combining investors' views with market equilibrium.
{\em Journal of Fixed Income} 1(2): 7-18.

\bibitem[Black and Litterman, 1992]{Black1992} Black, F. and Litterman, R. (1992)
Global portfolio optimization.
{\em Financial Analysts Journal} 48(5): 28-43.

\bibitem[Black and Scholes, 1973]{Black1973} Black, F. and Scholes, M. (1973)
The pricing of options and corporate liabilities.
{\em Journal of Political Economy} 81(3): 637-659.

\bibitem[Blake and Catlett, 1984]{Blake1984} Blake, M.L. and Catlett, L. (1984)
Cross Hedging Hay Using Corn Futures: An Empirical Test.
{\em Western Journal of Agricultural Economics} 9(1): 127-134.

\bibitem[Blanc-Brude, Hasan and Whittaker, 2016]{Blanc-Brude2016} Blanc-Brude, F., Hasan, M. and Whittaker, T. (2016)
Benchmarking infrastructure project finance: Objectives, roadmap, and recent progress.
{\em Journal of Alternative Investments} 19(2): 7-18.

\bibitem[Blanc-Brude, Whittaker and Wilde, 2017]{Blanc-Brude2017} Blanc-Brude, F., Whittaker, T. and Wilde, S. (2017)
Searching for a listed infrastructure asset class using mean-variance spanning.
{\em Financial Markets and Portfolio Management} 31(2): 137-179.

\bibitem[Blanchard and Gali, 2007]{Blanchard2007} Blanchard, O.J. and Gali, J. (2007)
The Macroeconomic Effects of Oil Shocks: Why are the 2000s So Different from the 1970s?
{\em Working Paper.} Available online: \url{http://www.nber.org/papers/w13368.pdf}.

\bibitem[Blanchard and Riggi, 2013]{Blanchard2013} Blanchard, O.J. and Riggi, M. (2013)
Why are the 2000s so different from the 1970s? A structural interpretation of changes in the macroeconomic effects of oil prices.
{\em Journal of the European Economic Association} 11(5): 1032-1052.

\bibitem[Blank, 1984]{Blank1984} Blank, S.C. (1984)
Cross Hedging Australian Cattle.
{\em Australian Journal of Agricultural Economics} 28(2-3): 153-162.

\bibitem[Blitz {\em et al}, 2013]{Blitz2013} Blitz, D., Huij, J., Lansdorp, S. and Verbeek, M. (2013)
Short-term residual reversal.
{\em Journal of Financial Markets} 16(3): 477-504.

\bibitem[Blitz, Huij and Martens, 2011]{Blitz2011} Blitz, D., Huij, J. and Martens, M. (2011)
Residual Momentum.
{\em Journal of Empirical Finance} 18(3): 506-521.

\bibitem[Blitz and van Vliet, 2007]{Blitz2007} Blitz, D.C. and van Vliet, P. (2007)
The Volatility Effect: Lower Risk without Lower Return.
{\em Journal of Portfolio Management} 34(1): 102-113.

\bibitem[Blitz and Van Vliet, 2008]{Blitz2008} Blitz, D. and Van Vliet, P. (2008)
Global Tactical Cross Asset Allocation: Applying Value and Momentum Across Asset Classes.
{\em Journal of Portfolio Management} 35(1): 23-28.

\bibitem[Block, 2011]{Block2011} Block, R.L. (2011)
{\em Investing in REITs: Real Estate Investment Trusts.}
New York, NY: Bloomberg Press.

\bibitem[Bloesch and Gourio, 2015]{Bloesch2015} Bloesch, J. and Gourio, F. (2015)
The effect of winter weather on U.S. economic activity.
{\em Federal Reserve Bank of Chicago, Economic Perspectives} 39(1): 1-20.

\bibitem[Bloom, Easley and O'Hara, 1994]{Bloom1994} Bloom, L., Easley, D. and O'Hara, M. (1994)
Market Statistics and Technical Analysis: The Role of Volume.
{\em Journal of Finance} 49(1): 153-181.

\bibitem[Blundell, 2006]{Blundell2006} Blundell, L. (2006)
Infrastructure investment: On the up.
{\em Property Australia} 20(9): 20-22.

\bibitem[Bobey, 2010]{Bobey2010} Bobey, B. (2010)
The Effects of Default Correlation on Corporate Bond Credit Spreads.
{\em Working Paper.} Available online: \url{https://ssrn.com/abstract=1510170}.

\bibitem[Bodie, 1983]{Bodie1983} Bodie, Z. (1983)
Commodity Futures as a Hedge against Inflation.
{\em Journal of Portfolio Management} 9(3): 12-17.

\bibitem[Bodie, Kane and Marcus, 1996]{Bodie1996} Bodie, Z., Kane, A. and Marcus, A.J. (1996)
{\em Investments.}
New York, NY: McGraw-Hill, Inc.

\bibitem[Bodie and Rosansky, 1980]{Bodie1980} Bodie, Z. and Rosansky, V.I. (1980)
Risk and Return in Commodity Futures.
{\em Financial Analysts Journal} 36(3): 27-39.

\bibitem[Bogomolov, 2013]{Bogomolov2013} Bogomolov, T. (2013)
Pairs trading based on statistical variability of the spread process.
{\em Quantitative Finance} 13(9): 1411-1430.

\bibitem[Bohlin and Strickland, 2004]{Bohlin2004} Bohlin, S. and Strickland, G. (2004)
Climbing the Ladder: How to Manage Risk in Your Bond Portfolio.
{\em American Association of Individual Investors Journal}, July 2004, pp. 5-8.

\bibitem[Bol, Rachev and W\"urth, 2009]{Bol2009} Bol, G., Rachev, S.T. and W\"urth, R. (eds.) (2009)
{\em Risk Assessment: Decisions in Banking and Finance.}
Heidelberg, Germany: Physica-Verlag.

\bibitem[Bollen and Busse, 2005]{Bollen2005} Bollen, N.P.B. and Busse, J.A. (2005)
Short-Term Persistence in Mutual Fund Performance.
{\em Review of Financial Studies} 18(2): 569-597.

\bibitem[Bollen and Mao, 2011]{Bollen2011} Bollen, J. and Mao, H. (2011)
Twitter mood as a stock market predictor.
{\em Computer} 44(10): 91-94.

\bibitem[Bollen, Mao and Zeng, 2011]{BollenMZ2011} Bollen, J., Mao, H. and Zeng, X. (2011)
Twitter mood predicts the stock market.
{\em Journal of Computational Science} 2(1): 1-8.

\bibitem[Bollen and Whaley, 2004]{Bollen2004} Bollen, N.P.B. and Whaley, R. (2004)
Does Net Buying Pressure Affect the Shape of Implied Volatility Functions?
{\em Journal of Finance} 59(2): 711-754.

\bibitem[Bollerslev, Gibson and Zhou, 2011]{Bollerslev2011} Bollerslev, T., Gibson, M. and Zhou, H. (2011)
Dynamic estimation of volatility risk premia and investor risk aversion from option-implied and realized volatilities.
{\em Journal of Econometrics} 160(1): 235-245.

\bibitem[Bologna and Cavallo, 2002]{Bologna2002} Bologna, P. and Cavallo, L. (2002)
Does the Introduction of Index Futures Effectively Reduce Stock Market Volatility? Is the Futures Effect Immediate? Evidence from the Italian Stock Exchange Using GARCH.
{\em Applied Financial Economics} 12(3): 183-192.

\bibitem[Bond and Seiler, 1998]{Bond1998} Bond, M.T. and Seiler, M.J. (1998)
Real Estate Returns and Inflation: An Added Variable Approach.
{\em Journal of Real Estate Research} 15(3): 327-338.

\bibitem[Bondarenko, 2014]{Bondarenko2014} Bondarenko, O. (2014)
Why Are Put Options So Expensive?
{\em Quarterly Journal of Finance} 4(3): 1450015.

\bibitem[Booth, 1987]{Booth1987} Booth, L.D. (1987)
The dividend tax credit and Canadian ownership objectives.
{\em Canadian Journal of Economics} 20(2): 321-339.

\bibitem[Booth and Johnston, 1984]{BoothJ1984} Booth, L.D. and Johnston, D.J. (1984)
The ex-dividend day behavior of Canadian stock prices: Tax changes and clientele effects.
{\em Journal of Finance} 39(2): 457-476.

\bibitem[Booth, Smith and Stolz, 1984]{Booth1984} Booth, J.R., Smith, R.L. and Stolz, R.W. (1984)
The Use of Interest Rate Futures by Financial Institutions.
{\em Journal of Bank Research} 15(1): 15-20.

\bibitem[Borovkova and Geman, 2006]{Borovkova2006} Borovkova, S. and Geman, H. (2006)
Seasonal and stochastic effects in commodity forward curves.
{\em Review of Derivatives Research} 9(2): 167-186.

\bibitem[Bos, 2000]{Bos2000} Bos, R. (2000)
{\em Index Calculation Primer.}
New York, NY: Standard and Poor's Quantitative Services.

\bibitem[Bos, Carter and Skiba, 2012]{Bos2012} Bos, M., Carter, S. and Skiba, P.M. (2012)
The Pawn Industry and Its Customers: The United States and Europe.
{\em Working Paper.} Available online: \url{https://ssrn.com/abstract=2149575}.

\bibitem[Boscher and Ward, 2002]{Boscher2002} Boscher, H. and Ward, I. (2002)
Long or short in CDOs.
{\em Risk}, June 2002, pp. 125-129.

\bibitem[Bossu, 2006]{Bossu2006} Bossu, S. (2006)
Introduction to Variance Swaps.
{\em Wilmott Magazine}, March 2006, pp. 50-55.

\bibitem[Boudoukh, Richardson and Whitelaw, 1994]{Boudoukh1994} Boudoukh, J., Richardson, M. and Whitelaw, R.F. (1994)
Industry Returns and the Fisher Effect.
{\em Journal of Finance} 49(5): 1595-1615.

\bibitem[Boudoukh {\em et al}, 1997]{Boudoukh1997} Boudoukh, J., Whitelaw, R., Richardson, M. and Stanton, R. (1997)
Pricing Mortgage-Backed Securities in a Multifactor Interest Rate Environment: A Multivariate Density Estimation Approach.
{\em Review of Financial Studies} 10(2): 405-446.

\bibitem[Boulos and Swanson, 1994]{Boulos1994} Boulos, N. and Swanson, P.E. (1994)
Interest Rate Parity in Times of Turbulence: The Issue Revisited.
{\em Journal of Financial and Strategic Decisions} 7(2): 43-52.

\bibitem[Bouman and Houtman, 1988]{Bouman1988} Bouman, F.J.A. and Houtman, R. (1988)
Pawnbroking as an Instrument of Rural Banking in the Third World.
{\em Economic Development and Cultural Change} 37(1): 69-89.

\bibitem[Bouoiyour, Selmi and Tiwari, 2015]{Bouoiyour2015} Bouoiyour, J., Selmi, R. and Tiwari, A.K. (2015)
Is Bitcoin business income or speculative foolery? New ideas through an improved frequency domain analysis.
{\em Annals of Financial Economics} 10(1): 1-23.

\bibitem[Bouoiyour {\em et al}, 2016]{Bouoiyour2016} Bouoiyour, J., Selmi, R., Tiwari, A.K. and Olayeni, O.R. (2016)
What drives Bitcoin price?
{\em Economics Bulletin} 36(2): 843-850.

\bibitem[Bouri {\em et al}, 2017a]{BouriGupta2017} Bouri, E., Gupta, R., Tiwari, A.K. and Roubaud, D. (2017a)
Does Bitcoin hedge global uncertainty? Evidence from wavelet-based quantile-in-quantile regressions.
{\em Finance Research Letters} 23: 87-95.

\bibitem[Bouri {\em et al}, 2017b]{Bouri2017} Bouri, E., Moln\'ar, P., Azzi, G., Roubaud, D. and Hagfors, L.I. (2017b)
On the hedge and safe haven properties of Bitcoin: Is it really more than a diversifier?
{\em Finance Research Letters} 20: 192-198.

\bibitem[Bouzoubaa and Osseiran, 2010]{Bouzoubaa2010} Bouzoubaa, M. and Osseiran, A. (2010)
{\em Exotic options and hybrids: a Guide to Structuring, Pricing and Trading.}
Chichester, UK: John Wiley \& Sons, Ltd.

\bibitem[Bowen and Hutchinson, 2016]{Bowen2016} Bowen, D.A. and Hutchinson, M.C. (2016)
Pairs trading in the UK equity market: Risk and return.
{\em European Journal of Finance} 22(14): 1363-1387.

\bibitem[Bowen, Hutchinson and O'Sullivan, 2010]{Bowen2010} Bowen, D., Hutchinson, M.C. and O'Sullivan, N. (2010)
High frequency equity pairs trading: Transaction costs, speed of execution and patterns in returns.
{\em Journal of Trading} 5(3): 31-38.

\bibitem[Bowsher, 1979]{Bowsher1979} Bowsher, N. (1979)
Repurchase Agreements.
{\em Federal Reserve Bank of St. Louis Review} 61(9): 17-22.

\bibitem[Boyarchenko, Fuster and Lucca, 2014]{Boyarchenko2014} Boyarchenko, N., Fuster, A. and Lucca, D.O. (2014)
Understanding Mortgage Spreads.
{\em Federal Reserve Bank of New York Staff Reports}, No. 674. Available online:\\ \url{https://www.newyorkfed.org/medialibrary/media/research/staff_reports/sr674.pdf}.

\bibitem[Boyd, Hu and Jagannathan, 2005]{Boyd2005} Boyd, J.H., Hu, J. and Jagannathan, R. (2005)
The Stock Market's Reaction to Unemployment News: Why Bad News Is Usually Good for Stocks.
{\em Journal of Finance} 60(2): 649-672.

\bibitem[Boyd and Mercer, 2010]{Boyd2010} Boyd, N.E. and Mercer, J.M. (2010)
Gains from Active Bond Portfolio Management Strategies.
{\em Journal of Fixed Income} 19(4): 73-83.

\bibitem[Boyle, 1978]{Boyle1978} Boyle, P.P. (1978) Immunization under stochastic models of the term structure.
{\em Journal of the Institute of Actuaries} 105(2): 177-187.

\bibitem[Bozdog {\em et al}, 2011]{Bozdog2011} Bozdog, D., Florescu, I., Khashanah, K. and Wang, J. (2011)
Rare Events Analysis of High-Frequency Equity Data.
{\em Wilmott Magazine} 2011(54): 74-81.

\bibitem[Bozic and Fortenbery, 2012]{Bozic2012} Bozic, M. and Fortenbery, T.R. (2012)
Creating Synthetic Cheese Futures: A Method for Matching Cash and Futures Prices in Dairy.
{\em Journal of Agribusiness} 30(2): 87-102.

\bibitem[Brandvold {\em et al}, 2015]{Brandvold2015} Brandvold, M., Moln\'{a}r, P., Vagstad, K. and Valstad, O.C.A. (2015)
Price discovery on Bitcoin exchanges.
{\em Journal of International Financial Markets, Institutions and Money} 36: 18-35.

\bibitem[Branger and Schlag, 2004]{Branger2004} Branger, N. and Schlag, C. (2004)
Why is the Index Smile So Steep?
{\em Review of Finance} 8(1): 109-127.

\bibitem[Brazil, 1988]{Brazil1988} Brazil, A.J. (1988)
Citicorp's mortgage valuation model: Option-adjusted spreads and option-based durations.
{\em Journal of Real Estate Finance and Economics} 1(2): 151-162.

\bibitem[Breeden and Litzenberger, 1978]{Breeden1978} Breeden, D.T. and Litzenberger, R.H. (1978)
Prices of state-contingent claims implicit in option prices.
{\em Journal of Business} 51(4): 621-651.

\bibitem[Brennan and Schwartz, 1977]{Brennan1977} Brennan, M.J. and Schwartz, E.S. (1977)
Convertible Bonds: Valuation and Optimal Strategies for Call and Conversion.
{\em Journal of Finance} 32(5): 1699-1715.

\bibitem[Brennan and Schwartz, 1985]{Brennan1985} Brennan, M.J. and Schwartz, E.S. (1985)
Determinants of GNMA Mortgage Prices.
{\em Real Estate Economics} 13(3): 209-228.

\bibitem[Brennan and Schwartz, 1988]{Brennan1988} Brennan, M.J. and Schwartz, E.S. (1988)
The case for convertibles.
{\em Journal of Applied Corporate Finance} 1(2): 55-64.

\bibitem[Brenner, Subrahmanyam and Uno, 1989]{Brenner1989} Brenner, M., Subrahmanyam, M.G. and Uno, J. (1989)
Stock index futures arbitrage in the Japanese markets.
{\em Japan and the World Economy} 1(3): 303-330.

\bibitem[Brezigar-Masten and Masten, 2012]{Brezigar2012} Brezigar-Masten, A. and Masten, P. (2012)
CART-based selection of bankruptcy predictors for the logit model.
{\em Expert Systems with Applications} 39(11): 10153-10159.

\bibitem[Bri\`ere, Oosterlinck and Szafarz, 2015]{Briere2015} Bri\`ere, M., Oosterlinck, K. and Szafarz, A. (2015)
Virtual currency, tangible return: Portfolio diversification with bitcoin.
{\em Journal of Asset Management} 16(6): 365-373.

\bibitem[Briys and Solnik, 1992]{Briys1992} Briys, E. and Solnik, B. (1992)
Optimal currency hedge ratios and interest rate risk.
{\em Journal of International Money and Finance} 11(5): 431-445.

\bibitem[Broadie and Jain, 2008]{Broadie2008} Broadie, M. and Jain, A. (2008)
The effect of jumps and discrete sampling on volatility and variance swaps.
{\em International Journal of Theoretical and Applied Finance} 11(8): 761-797.

\bibitem[Brock, Lakonishock and LeBaron, 1992]{Brock1992} Brock, W., Lakonishock, J. and LeBaron, B. (1992)
Simple technical trading rules and the stochastic properties of stock returns.
{\em Journal of Finance} 47(5): 1731-1764.

\bibitem[Brockett {\em et al}, 2010]{Brocket2010} Brockett, P., Golden, L.L., Wen, M. and Yang, C. (2010)
Pricing weather derivatives using the indifference pricing approach.
{\em North American Actuarial Journal} 13(3): 303-315.

\bibitem[Brockett, Wang and Yang, 2005]{Brockett2005} Brockett, P.L., Wang, M. and Yang, C. (2005)
Weather Derivatives And Weather Risk Management.
{\em Risk Management and Insurance Review} 8(1): 127-140.

\bibitem[Brody, Syroka and Zervos, 2002]{Brody2002} Brody, D., Syroka, J. and Zervos, M. (2002)
Dynamical pricing of weather derivatives.
{\em Quantitative Finance} 2(3): 189-198.

\bibitem[Brogaard and Garriott, 2018]{Brogaard2018} Brogaard, J. and Garriott, C. (2018)
High-Frequency Trading Competition.
{\em Working Paper.} Available online: \url{https://ssrn.com/abstract=2435999}.

\bibitem[Brogaard {\em et al}, 2015]{Brogaard2015}Brogaard, J., Hagstr\"{o}mer, B., Nord\'{e}n, L. and Riordan, R. (2015)
Trading Fast and Slow: Colocation and Liquidity.
{\em Review of Financial Studies} 28(12): 3407-3443.

\bibitem[Brogaard, Hendershott and Riordan, 2014]{Brogaard2014} Brogaard, J., Hendershott, T. and Riordan, R. (2014)
High-Frequency Trading and Price Discovery.
{\em Review of Financial Studies} 27(8): 2267-2306.

\bibitem[Brooks, 2017]{BrooksAQR2017} Brooks, J. (2017)
A Half Century of Macro Momentum.
{\em Working Paper.} Available online:\\
\url{https://www.aqr.com/-/media/AQR/Documents/Insights/White-Papers/A-Half-Century-of-Macro-Momentum.pdf}.

\bibitem[Brooks and Chong, 2001]{Brooks2001} Brooks, C. and Chong, J. (2001)
The Cross-Currency Hedging Performance of Implied Versus Statistical Forecasting Models.
{\em Journal of Futures Markets} 21(11): 1043-1069.

\bibitem[Brooks, Davies and Kim, 2007]{BrooksDK2007} Brooks, C., Davies, R.J. and Kim, S.S. (2007)
Cross Hedging with Single Stock Futures.
{\em Assurances et Gestion des Risques} 74(4): 473-504.

\bibitem[Brooks, Henry and Persand, 2002]{Brooks2002} Brooks, C., Henry, O.T. and Persand, G. (2002)
The Effect of Asymmetries on Optimal Hedge Ratios.
{\em Journal of Business} 75(2): 333-352.

\bibitem[Brooks and Moskowitz, 2017]{Brooks2017} Brooks, J. and Moskowitz, T.J. (2017)
Yield Curve Premia.
{\em Working Paper.} Available online: \url{https://ssrn.com/abstract=2956411}.

\bibitem[Brown, 1999]{Brown1999} Brown, D. (1999)
The Determinants of Expected Returns on Mortgage-backed Securities: An Empirical Analysis of Option-adjusted Spreads.
{\em Journal of Fixed Income} 9(2): 8-18.

\bibitem[Brown and Clarke, 1993]{Brown1993} Brown, P. and Clarke, A. (1993)
The Ex-Dividend Day Behaviour of Australian Share Prices Before and After Dividend Imputation.
{\em Australian Journal of Management} 18(1): 1-40.

\bibitem[Brown, Davies and Ringgenberg, 2018]{Brown2018} Brown, D.C., Davies, S. and Ringgenberg, M. (2018)
ETF Arbitrage and Return Predictability.
{\em Working Paper.} Available online: \url{https://ssrn.com/abstract=2872414}.

\bibitem[Brown {\em et al}, 1992]{Brown1992} Brown, S.J., Goetzmann, W., Ibbotson, R.G. and Ross, S.A. (1992)
Survivorship Bias in Performance Studies.
{\em Review of Financial Studies} 5(4): 553-580.

\bibitem[Brown {\em et al}, 2012]{Brown2012} Brown, S.J., Grundy, B.D., Lewis, C.M. and Verwijmeren, P. (2012)
Convertibles and hedge funds as distributors of equity exposure.
{\em Review of Financial Studies} 25(10): 3077-3112.

\bibitem[Brown and Raymond, 1986]{Brown1986} Brown, K.C. and Raymond, M.V. (1986)
Risk arbitrage and the prediction of successful corporate takeovers.
{\em Financial Management} 15(3): 54-63.

\bibitem[Browne, 2000]{Browne2000} Browne, S. (2000)
Risk-constrained dynamic active portfolio management.
{\em Management Science} 46(9): 1188-1199.

\bibitem[Br\"{u}ck and Fan, 2017]{Fan2017} Br\"{u}ck, E. and Fan, Y. (2017)
Smart Beta In Global Government Bonds And Its Risk Exposure.
{\em Working Paper.} Available online:\\ \url{https://www.cfasociety.org/France/Documents/QuantAwards2017_Etienne\%20BRUECK\%20and\%20Yuanting\%20FAN_EDHEC.pdf}.

\bibitem[Bruder {\em et al}, 2013]{Bruder2013} Bruder, B., Dao, T.-L., Richard, R.-J. and Roncalli, T. (2013)
Trend Filtering Methods for Momentum Strategies.
{\em Working Paper.} Available online: \url{https://ssrn.com/abstract=2289097}.

\bibitem[Brunnermeier, Nagel and Pedersen, 2008]{Brunnermeier2008} Brunnermeier, M.K., Nagel, S. and Pedersen, L.H. (2008)
Carry Trades and Currency Crashes.
{\em NBER Macroeconomics Annual} 23(1): 313-347.

\bibitem[Bu and Lacey, 2007]{Bu2007} Bu, Q. and Lacey, N. (2007)
Exposing Survivorship Bias in Mutual Fund Data.
{\em Journal of Business and Economics Studies} 13(1): 22-37.

\bibitem[Budish, Cramton and Shim, 2015]{Budish2015} Budish, E., Cramton, P. and Shim, J. (2015)
The High-Frequency Trading Arms Race: Frequent Batch Auctions as a Market Design Response.
{\em Quarterly Journal of Economics} 130(4): 1547-1621.

\bibitem[Buetow and Henderson, 2012]{Buetow2012} Buetow, G.W. and Henderson, B.J. (2012)
An empirical analysis of exchange-traded funds.
{\em Journal of Portfolio Management} 38(4): 112-127.

\bibitem[Buetow and Henderson, 2016]{Buetow2016} Buetow, G.W. and Henderson, B.J. (2016)
The VIX Futures Basis: Determinants and Implications.
{\em Journal of Portfolio Management} 42(2): 119-130.

\bibitem[B\"{u}hler and Kempf, 1995]{Buhler1995} B\"{u}hler, W. and Kempf, A. (1995)
DAX index futures: Mispricing and arbitrage in German markets.
{\em Journal of Futures Markets} 15(7): 833-859.

\bibitem[Bundgaard, 2013]{Bundgaard2013} Bundgaard, J. (2013)
Coordination Rules as a Weapon in the War against Cross-Border Tax Arbitrage -- The Case of Hybrid Entities and Hybrid Financial Instruments.
{\em Bulletin for International Taxation}, April/May 2013, pp. 200-204.

\bibitem[Buraschi and Jiltsov, 2005]{Buraschi2005} Buraschi, A. and Jiltsov, A. (2005)
Inflation Risk Premia and the Expectations Hypothesis.
{\em Journal of Financial Economics} 75(2): 429-490.

\bibitem[Burnside {\em et al}, 2011]{Burnside2011} Burnside, C., Eichenbaum, M., Kleshchelski, I. and Rebelo, S. (2011)
Do peso problems explain the returns to the carry trade?
{\em Review of Financial Studies} 24(3): 853-891.

\bibitem[Burnside, Eichenbaum and Rebelo, 2007]{Burnside2007} Burnside, C., Eichenbaum, M. and Rebelo, S. (2007)
The Returns to Currency Speculation in Emerging Markets.
{\em American Economic Review} 97(2): 333-338.

\bibitem[Burnside, Eichenbaum and Rebelo, 2008]{Burnside2008} Burnside, C., Eichenbaum, M. and Rebelo, S. (2008)
Carry Trade: The Gains of Diversification.
{\em Journal of the European Economic Association} 6(2/3): 581-588.

\bibitem[Burnside, Eichenbaum and Rebelo, 2011]{BurnsideEich2011} Burnside, C., Eichenbaum, M. and Rebelo, S. (2011)
Carry trade and momentum in currency markets.
{\em Annual Review of Financial Economics} 3: 511-535.

\bibitem[Burtshell, Gregory and Laurent, 2009)]{Burtshell2009} Burtshell, X., Gregory, J. and Laurent, J.-P. (2009)
A comparative analysis of CDO pricing models under the factor copula framework.
{\em Journal of Derivatives} 16(4): 9-37.

\bibitem[Busch, Christensen and Nielsen, 2011]{Busch2011} Busch, T., Christensen, B.J. and Nielsen, M.{\O}. (2011)
The Role of Implied Volatility in Forecasting Future Realized Volatility and Jumps in Foreign Exchange, Stock, and Bond Markets.
{\em Journal of Econometrics} 160(1): 48-57.

\bibitem[Buser and Hess, 1986]{Buser1986} Buser, S.A. and Hess, P.J. (1986)
Empirical determinants of the relative yields on taxable and tax-exempt securities.
{\em Journal of Financial Economics} 17(2): 335-355.

\bibitem[Butterworth and Holmes, 2010]{Butterworth2010} Butterworth, D. and Holmes, P. (2010)
Mispricing in stock index futures contracts: evidence for the FTSE 100 and FTSE mid 250 contracts.
{\em Applied Economics Letters} 7(12): 795-801.

\bibitem[Buttimer, Hyland and Sanders, 2005]{Buttimer2005} Buttimer, R.J., Hyland, D.C. and Sanders, A.B. (2005)
REITs, IPO Waves, and Long-Run Performance.
{\em Real Estate Economics} 33(1): 51-87.

\bibitem[Caginalp, DeSantis and Sayrak, 2014]{Caginalp2014} Caginalp, G., DeSantis, M. and Sayrak, A. (2014) The nonlinear price dynamics of US equity ETFs.
{\em Journal of Econometrics} 183(2): 193-201.

\bibitem[Calamos, 2003]{Calamos2003} Calamos, N.P. (2003)
{\em Convertible Arbitrage: Insights and Techniques for Successful Hedging.}
Hoboken, NJ: John Wiley \& Sons, Inc.

\bibitem[Caldeira and Moura, 2013]{Caldeira2013} Caldeira, J. and Moura, G.V. (2013)
Selection of a portfolio of pairs based on cointegration: A statistical arbitrage strategy.
{\em Working Paper.} Available online: \url{https://ssrn.com/abstract=2196391}.

\bibitem[Callaghan and Barry, 2003]{Callaghan2003} Callaghan, S.R. and Barry, C.B. (2003)
Tax-induced trading of equity securities: Evidence from the ADR market.
{\em Journal of Finance} 58(4): 1583-1611.

\bibitem[Callej\'{o}n {\em et al}, 2013]{Callejon2013} Callej\'{o}n, A.M., Casado, A.M., Fern\'{a}ndez, M.A. and Pel\'{a}ez, J.I. (2013)
A System of Insolvency Prediction for industrial companies using a financial alternative model with neural networks.
{\em International Journal of Computational Intelligence Systems} 6(1): 29-37.

\bibitem[Campbell, 1991]{Campbell1991} Campbell, J.Y. (1991)
A Variance Decomposition for Stock Returns.
{\em Economic Journal} 101(405): 157-179.

\bibitem[Campbell, Chan and Viceira, 2003]{Campbell2003} Campbell, J.Y., Chan, Y.L. and Viceira, L.M. (2003)
A multivariate model of strategic asset allocation.
{\em Journal of Financial Economics} 67(1): 41-80.

\bibitem[Campbell and Diebold, 2005]{Campbell2005} Campbell, S.D. and Diebold, F.X. (2005)
Weather forecasting for weather derivatives.
{\em Journal of the American Statistical Association} 100(469): 6-16.

\bibitem[Campbell, Hilscher and Sziglayi, 2008]{Campbell2008} Campbell, J.Y., Hilscher, J. and Sziglayi, J. (2008)
In Search of Distress Risk.
{\em Journal of Finance} 63(6): 2899-2939.

\bibitem[Campbell, Shiller and Viceira, 2009]{Campbell2009} Campbell, J.Y., Shiller, R.J. and Viceira, L.M. (2009)
Understanding Inflation-Indexed Bond Markets.
In: Romer, D. and Wolfers, J. (eds.)
{\em Brookings Papers on Economic Activity.} Washington, DC: Brookings Institution Press, pp. 79-120.

\bibitem[Campbell, Sunderam and Viceira, 2017]{Campbell2017} Campbell, J.Y., Sunderam, A. and Viceira, L.M. (2017)
Inflation Bets or Deflation Hedges? The Changing Risks of Nominal Bonds.
{\em Critical Finance Review} 6(2): 263-301.

\bibitem[Campbell and Viceira, 2004]{CampbellViceira2004} Campbell, J.Y. and Viceira, L.M. (2004)
Long-Horizon Mean-Variance Analysis: A User Guide.
{\em Working Paper.} Available online: \url{http://www.people.hbs.edu/lviceira/faj_cv_userguide.pdf}.

\bibitem[Campbell and Viceira, 2005]{CampbellViceira2005} Campbell, J.Y. and Viceira, L.M. (2005)
The Term Structure of the Risk: Return Trade-Off.
{\em Financial Analysts Journal} 61(1): 34-44.

\bibitem[Canina and Figlewski, 1993]{Canina1993} Canina, L. and Figlewski, S. (1993)
The Informational Content of Implied Volatility.
{\em Review of Financial Studies} 6(3): 659-681.

\bibitem[Cao {\em et al}, 2013]{Cao2013} Cao, C., Chen, Y., Liang, B. and Lo, A.W. (2013)
Can hedge funds time market liquidity?
{\em Journal of Financial Economics} 109(2): 493-516.

\bibitem[Cao {\em et al}, 2016]{Cao2016} Cao, C., Goldie, B., Liang, B. and Petrasek, L. (2016)
What Is the Nature of Hedge Fund Manager Skills? Evidence from the Risk-Arbitrage Strategy.
{\em Journal of Financial and Quantitative Analysis} 51(3): 929-957.

\bibitem[Cao and Wei, 2000]{Cao2000} Cao, M. and Wei, J. (2000)
Pricing the weather.
{\em Risk}, May 2000, pp. 67-70.

\bibitem[Cao and Wei, 2004]{Cao2004} Cao, M. and Wei, J. (2004)
Weather derivatives valuation and market price of weather risk.
{\em Journal of Futures Markets} 24(11): 1065-1089.

\bibitem[Caplin and Leahy, 2011]{Caplin2011} Caplin, A. and Leahy, J. (2011)
Trading Frictions and House Price Dynamics.
{\em Journal of Money, Credit and Banking} 43(7): 283-303.

\bibitem[Capozza, Hendershott and Mack, 2004]{Capozza2004} Capozza, D.R., Hendershott, P.H. and Mack, C. (2004)
An Anatomy of Price Dynamics in Illiquid Markets: Analysis and Evidence from Local Housing Markets.
{\em Real Estate Economics} 32(1): 1-32.

\bibitem[Carhart, 1997]{Carhart1997} Carhart, M.M. (1997)
Persistence in mutual fund performance.
{\em Journal of Finance} 52(1): 57-82.

\bibitem[Carhart {\em et al}, 2002]{Carhart2002} Carhart, M.M., Carpenter, J.N., Lynch, A.W. and Musto, D.K. (2002)
Mutual Fund Survivorship.
{\em Review of Financial Studies} 15(5): 1439-1463.

\bibitem[Carmona and Cr\'{e}pey, 2010]{Carmona2010} Carmona, R. and Cr\'{e}pey, S. (2010)
Particle methods for the estimation of credit portfolio loss distributions.
{\em International Journal of Theoretical and Applied Finance} 13(4): 577-602.

\bibitem[Carmona and Durrleman, 2003]{Carmona2003} Carmona, R. and Durrleman, V. (2003)
Pricing and Hedging Spread Options.
{\em SIAM Review} 45(4): 627-685.

\bibitem[Carr and Javaheri, 2005]{Carr2005} Carr, P. and Javaheri, A. (2005)
The forward PDE for European options on stocks with fixed fractional jumps.
{\em International Journal of Theoretical and Applied Finance} 8(2): 239-253.

\bibitem[Carr and Lee, 2007]{Carr2007} Carr, P. and Lee, R. (2007)
Realized volatility and variance: Options via swaps.
{\em Risk} 20(5): 76-83.

\bibitem[Carr and Lee, 2009]{CarrLee2009} Carr, P. and Lee, R. (2009)
Volatility Derivatives.
{\em Annual Review of Financial Economics} 1: 319-339.

\bibitem[Carr, Lee and Wu, 2012]{Carr2012} Carr, P., Lee, R. and Wu, L. (2012)
Variance swaps on time-changed L\'{e}vy processes.
{\em Finance and Stochastics} 16(2): 335-355.

\bibitem[Carr and Wu, 2009]{Carr2009} Carr, P. and Wu, L. (2009)
Variance risk premiums.
{\em Review of Financial Studies} 22(3): 1311-1341.

\bibitem[Carr and Wu, 2016]{Carr2016} Carr, P. and Wu, L. (2016)
Analyzing volatility risk and risk premium in option contracts: A new theory.
{\em Journal of Financial Economics} 120(1): 1-20.

\bibitem[Carrasco, 2007]{Carrasco2007} Carrasco, C.G. (2007)
Studying the properties of the correlation trades.
{\em Working Paper.} Available online: \url{https://mpra.ub.uni-muenchen.de/22318/1/MPRA_paper_22318.pdf}.

\bibitem[Carrion, 2013]{Carrion2013} Carrion, A. (2013)
Very fast money: High-frequency trading on the NASDAQ.
{\em Journal of Financial Markets} 16(4): 680-711.

\bibitem[Carrion and Kolay, 2017]{Carrion2017} Carrion, A. and Kolay, M. (2017)
Trade Signing in Fast Markets.
{\em Working Paper.}
Available online: \url{https://ssrn.com/abstract=2489868}.

\bibitem[Carron and Hogan, 1988]{Carron1988} Carron, A.S. and Hogan, M. (1988)
The option valuation approach to mortgage pricing.
{\em Journal of Real Estate Finance and Economics} 1(2): 131-149.

\bibitem[Cartea and Figueroa, 2005]{Cartea2005} Cartea, A. and Figueroa, M. (2005)
Pricing in electricity markets: a mean reverting jump diffusion model with seasonality.
{\em Applied Mathematical Finance} 12(4): 313-335.

\bibitem[Cartea and Pedraz, 2012]{Cartea2012} Cartea, A. and Pedraz, C.G. (2012)
How Much Should We Pay for Interconnecting Electricity Markets? A Real Options Approach.
{\em Energy Economics} 34(1): 14-30.

\bibitem[Carter, Rausser and Schmitz, 1983]{Carter1983} Carter, C., Rausser, G. and Schmitz, A. (1983)
Efficient asset portfolios and the theory of normal backwardation.
{\em Journal of Political Economy} 91(2): 319-331.

\bibitem[Casassus and Collin-Dufresne, 2005]{Casassus2005} Casassus, J. and Collin-Dufresne, P. (2005)
Stochastic convenience yield implied from commodity futures and interest rates.
{\em Journal of Finance} 60(5): 2283-2331.

\bibitem[Case and Shiller, 1987]{Case1987} Case, K.E. and Shiller, R.J. (1987)
Prices of Single Family Homes since 1970: New Indexes for Four Cities.
{\em Federal Reserve Bank of Boston, New England Economic Review}, September-October 1987, pp. 45-56.

\bibitem[Case and Shiller, 1989]{Case1989} Case, K.E. and Shiller, R.J. (1989)
The Efficiency of the Market for Single-Family Homes.
{\em American Economic Review} 79(1): 125-137.

\bibitem[Case and Shiller, 1990]{Case1990} Case, K.E. and Shiller, R.J. (1990)
Forecasting Prices and Excess Returns in the Housing Market.	
{\em Real Estate Economics} 18(3): 253-273.

\bibitem[Caskey, 1991]{Caskey1991} Caskey, J.P. (1991)
Pawnbroking in America: the Economics of a Forgotten Credit Market.
{\em Journal of Money, Credit and Banking} 23(1): 85-99.

\bibitem[Cassano and Sick, 2013]{Cassano2013} Cassano, M. and Sick, G. (2013)
Valuation of a spark spread: an LM6000 power plant.
{\em European Journal of Finance} 18(7-8): 689-714.

\bibitem[Castelino and Vora, 1984]{Castelino1984} Castelino, M.G and Vora, A. (1984)
Spread volatility in commodity futures: The length effect.
{\em Journal of Futures Markets} 4(1): 39-46.

\bibitem[Cavaglia and Vadim, 2002]{Cavaglia2002} Cavaglia, S. and Vadim, M. (2002)
Cross-Industry, Cross Country Allocation.
{\em Financial Analysts Journal} 58(6): 78-97.

\bibitem[Cecchetti, Cumby and Figlewski, 1988]{Cecchetti1988} Cecchetti, S.G., Cumby, R.E. and Figlewski, S. (1988)
Estimation of the Optimal Futures Hedge.
{\em Review of Economics and Statistics} 70(4): 623-630.

\bibitem[\v{C}erovi\'c and Pepi\'c, 2011]{Cerovic2011}\v{C}erovi\'c, S. and Pepi\'c, M. (2011)
Interest rate derivatives in developing countries in Europe.
{\em Perspectives of Innovation in Economics and Business} 9(3): 38-42.

\bibitem[\v{C}erovi\'c {\em et al}, 2014]{Cerovic2014}
\v{C}erovi\'c, S., Pepi\'c, M., \v{C}erovi\'c, S. and \v{C}erovi\'c, N. (2014)
Duration and convexity of bonds.
{\em Singidunum Journal of Applied Sciences} 11(1): 53-66.

\bibitem[Cerrato and Djennad, 2008]{Cerrato2008} Cerrato, M. and Djennad, A. (2008)
Dynamic Option Adjusted Spread and the Value of Mortgage Backed Securities.
{\em Working Paper.} Available online: \url{https://www.gla.ac.uk/media/media_71226_en.pdf}.

\bibitem[Chaboud and Wright, 2005]{Chaboud2005} Chaboud, A.P. and Wright, J.H. (2005)
Uncovered interest parity: it works, but not for long.
{\em Journal of International Economics} 66(2): 349-362.

\bibitem[Chaiyapo and Phewchean, 2017]{Chaiyapo2017} Chaiyapo, N. and Phewchean, N. (2017)
An application of Ornstein-Uhlenbeck process to commodity pricing in Thailand.
{\em Advances in Difference Equations} 2017: 179.

\bibitem[Chakravarty, Gulen and Mayhew, 2004]{Chakravarty2004} Chakravarty, S., Gulen, H. and Mayhew, S. (2004)
Informed Trading in Stock and Option Markets.
{\em Journal of Finance} 59(3): 1235-1257.

\bibitem[Chalmers, 1998]{Chalmers1998} Chalmers, J.M.R. (1998)
Default Risk Cannot Explain the Muni Puzzle: Evidence from Municipal Bonds that are Secured by U.S. Treasury Obligations.
{\em Review of Financial Studies} 11(2): 281-308.

\bibitem[Chambers {\em et al}, 2014]{Chambers2014} Chambers, D.R., Foy, M., Liebner, J. and Lu, Q. (2014)
Index Option Returns: Still Puzzling.
{\em Review of Financial Studies} 27(6): 1915-1928.

\bibitem[Chan, 2013]{Chan2013} Chan, E.P. (2013)
{\em Algorithmic Trading: Winning Strategies and Their Rationale.}
Hoboken, NJ: John Wiley \& Sons, Inc.

\bibitem[Chan and Chen, 2007]{Chan2007} Chan, A.W.H. and Chen, N.-F. (2007)
Convertible bond underpricing: Renegotiable covenants, seasoning, and convergence.
{\em Management Science} 53(11): 1793-1814.

\bibitem[Chan and Chen, 1991]{Chan1991} Chan, K.C. and Chen, N.-F. (1991)
Structural and Return Characteristics of Small and Large Firms.
{\em Journal of Finance} 46(4): 1467-1484.

\bibitem[Chan and Chung, 1993]{Chan1993} Chan, K. and Chung, Y.P. (1993)
Intraday relationships among index arbitrage, spot and futures price volatility, and spot market volume: A transactions data test.
{\em Journal of Banking \& Finance} 17(4): 663-687.

\bibitem[Chan, Hendershott and Sanders, 1990]{ChanHS1990} Chan, K.C., Hendershott, P.H. and Sanders, A.B. (1990)
Risk and Return on Real Estate: Evidence from Equity REITs.
{\em AREUEA Journal} 18(4): 431-452.

\bibitem[Chan, Jegadeesh and Lakonishok, 1996]{Chan1996} Chan, K.C., Jegadeesh, N. and Lakonishok, J. (1996)
Momentum Strategies.
{\em Journal of Finance} 51(5): 1681-1713.

\bibitem[Chan, Leung and Wang, 1998]{Chan1998} Chan, S.H., Leung, W.K. and Wang, K. (1998)
Institutional Investment in REITs: Evidence and Implications.
{\em Journal of Real Estate Research} 16(3): 357-374.

\bibitem[Chan {\em et al}, 2011]{Chan2011} Chan, K.F., Treepongkaruna, S., Brooks, R. and Gray, S. (2011)
Asset market linkages: Evidence from financial, commodity and real estate assets.
{\em Journal of Banking \& Finance} 35(6): 1415-1426.

\bibitem[Chance, 1994]{Chance1994} Chance, D. (1994)
{\em Managed Futures and Their Role in Investment Portfolios.}
Charlottesville, VA: The Research Foundation of the Institute of Chartered Financial Analysts.

\bibitem[Chance and Jordan, 1996]{Chance1996} Chance, D.M. and Jordan, J.V. (1996)
Duration, Convexity, and Time as Components of Bond Returns.
{\em Journal of Fixed Income} 6(2): 88-96.

\bibitem[Chandra, 2003]{Chandra2003} Chandra, P. (2003)
Sigmoidal Function Classes for Feedforward Artificial Neural Networks.
{\em Neural Processing Letters} 18(3): 205-215.

\bibitem[Chang, Cheng and Pinegar, 1999]{Chang1999} Chang, E.C., Cheng, J.W. and Pinegar, J.M. (1999)
Does Futures Trading Increase Stock Market Volatility? The Case of the Nikkei Stock Index Futures Exchange.
{\em Journal of Banking \& Finance} 23(5): 727-753.

\bibitem[Chang and Fang, 1990]{Chang1990} Chang, J.S. and Fang, H. (1990)
An intertemporal measure of hedging effectiveness.
{\em Journal of Futures Markets} 10(3): 307-321.

\bibitem[Chang {\em et al}, 2016]{Chang2016} Chang, R.P., Ko, K.-C., Nakano, S. and Rhee, S.G. (2016)
Residual Momentum and Investor Underreaction in Japan.
{\em Working Paper.} Available online: \url{http://sfm.finance.nsysu.edu.tw/php/Papers/CompletePaper/134-1136665035.pdf}.

\bibitem[Chaput and Ederington, 2003]{Chaput2003} Chaput, J.S. and Ederington, L.H. (2003)
Option Spread and Combination Trading.
{\em Journal of Derivatives} 10(4): 70-88.

\bibitem[Chaput and Ederington, 2005]{Chaput2005} Chaput, J.S. and Ederington, L.H. (2005)
Vertical Spread Design.
{\em Journal of Derivatives} 12(3): 28-46.

\bibitem[Chaput and Ederington, 2008]{Chaput2008} Chaput, J.S. and Ederington, L.H. (2008)
Ratio Spreads.
{\em Journal of Derivatives} 15(3): 41-57.

\bibitem[Charupat and Miu, 2011]{Charupat2011} Charupat, N. and Miu, P. (2011)
 The Pricing and Performance of Leveraged Exchange-Traded Funds.
{\em Journal of Banking \& Finance} 35(4): 966-977.

\bibitem[Chatterjee, Dhillon and Ram\'irez, 1996]{Chatterjee1996} Chatterjee, S., Dhillon, U.S. and Ram\'irez, G.G. (1996)
Resolution of Financial Distress: Debt Restructurings via Chapter 11, Prepackaged Bankruptcies, and Workouts.
{\em Financial Management} 25(1): 5-18.

\bibitem[Chaudhuri and De, 2011]{Chaudhuri2011} Chaudhuri, A. and De, K. (2011)
Fuzzy support vector machine for bankruptcy prediction.
{\em Applied Soft Computing} 11(2): 2472-2486.

\bibitem[Chaumont, Imkeller and M\"{u}ller, 2006]{Chaumont2006} Chaumont, S., Imkeller, P. and M\"{u}ller, M. (2006)
Equilibrium Trading of Climate and Weather Risk and Numerical Simulation in a Markovian Framework.
{\em Stochastic Environment Research and Risk Assessment} 20(3): 184-205.

\bibitem[Chava and Jarrow, 2004]{Chava2004} Chava, S. and Jarrow, R.A. (2004)
Bankruptcy Prediction with Industry Effects.	
{\em Review of Finance} 8(4): 537-569.

\bibitem[Chaves, 2012]{Chaves2012} Chaves, D.B. (2012)
Eureka! A Momentum Strategy that Also Works in Japan.
{\em Working Paper.} Available online:\\ \url{https://ssrn.com/abstract=1982100}.

\bibitem[Chaves and Viswanathan, 2016]{Chaves2016} Chaves, D.B. and Viswanathan, V. (2016)
Momentum and mean-reversion in commodity spot and futures markets.
{\em Journal of Commodity Markets} 3(1): 39-53.

\bibitem[Che, 2016]{Che2016} Che, Y.S. (2016)
{\em A study on the risk and return of option writing strategies} (Ph.D. Thesis).
{\em HKBU Institutional Repository. Open Access Theses and Dissertations.} 187. Hong Kong, China: Hong Kong Baptist University. Available online:
\url{https://repository.hkbu.edu.hk/etd_oa/187/}.

\bibitem[Che and Fung, 2011]{Che2011} Che, S.Y.S. and Fung, J.K.W. (2011)
The performance of alternative futures buy-write strategies.
{\em Journal of Futures Markets} 31(12): 1202-1227.

\bibitem[Cheah and Fry, 2015]{Cheah2015} Cheah, E.T. and Fry, J. (2015)
Speculative Bubbles in Bitcoin markets? An Empirical Investigation into the Fundamental Value of Bitcoin.
{\em Economics Letters} 130: 32-36.

\bibitem[Chen, 2014]{Chen2014} Chen, M.Y. (2014)
A high-order fuzzy time series forecasting model for internet stock trading.
{\em Future Generation Computer Systems} 37: 461-467.

\bibitem[Chen {\em et al}, 2017]{Chen2017} Chen, H.J., Chen, S.J., Chen, Z. and Li, F. (2017)
Empirical Investigation of an Equity Pairs Trading Strategy.
{\em Management Science} (forthcoming). DOI: \url{https://doi.org/10.1287/mnsc.2017.2825}.

\bibitem[Chen, Chen and Howell, 1999]{Chen1999} Chen, A.H.Y., Chen, K.C. and Howell, S. (1999)
An analysis of dividend enhanced convertible stocks.
{\em International Review of Economics and Finance} 8(3): 327-338.

\bibitem[Chen, Chung and Tsai, 2016]{Chen2016} Chen, T.F., Chung, S.L. and Tsai, W.C. (2016)
Option-Implied Equity Risk and the Cross-Section of Stock Returns.
{\em Financial Analysts Journal} 72(6): 42-55.

\bibitem[Chen {\em et al}, 1998]{ChenHVC1998} Chen, S.-J., Hsieh, C., Vines, T.W. and Chiou, S. (1998)
Macroeconomic Variables, Firm-Specific Variables and Returns to REITs.
{\em Journal of Real Estate Research} 16(3): 269-278.

\bibitem[Chen, Kang and Yang, 2005]{Chen2005} Chen, A.H., Kang, J. and Yang, B. (2005)
A Model for Convexity-Based Cross-Hedges with Treasury Futures.
{\em Journal of Fixed Income} 15(3): 68-79.

\bibitem[Chen, Lesmond and Wei, 2007]{Chen2007} Chen, L., Lesmond, D.A. and Wei, J. (2007)
Corporate Yield Spreads and Bond Liquidity.
{\em Journal of Finance} 62(1): 119-149.

\bibitem[Chen, Leung and Daouk, 2003]{Chen2003} Chen, A.S., Leung, M.T. and Daouk, H. (2003)
Application of neural networks to an emerging financial market: Forecasting and trading the Taiwan Stock Index.
{\em Computers \& Operations Research} 30(6): 901-923.

\bibitem[Chen, Liu and Cheng, 2010]{ChenLC2010} Chen, R.-R., Liu, B. and Cheng, X. (2010)
Pricing the Term Structure of Inflation Risk Premia: Theory and Evidence from TIPS.
{\em Journal of Empirical Finance} 17(4): 702-721.

\bibitem[Chen, Mao and Wang, 2010]{Chen2010} Chen, Z., Mao, C.X. and Wang, Y. (2010)
Why firms issue callable bonds: Hedging investment uncertainty.
{\em Journal of Corporate Finance} 16(4): 588-607.

\bibitem[Chen, Roberts and Thraen, 2006]{ChenRT2006} Chen, G., Roberts, M.C. and Thraen, C.S. (2006)
Managing dairy profit risk using weather derivatives.
{\em Journal of Agricultural and Resource Economics} 31(3): 653-666.

\bibitem[Chen and Selender, 1994]{Chen1994} Chen, A.H. and Selender, A.K. (1994)
Determination of Swap Spreads: An Empirical Analysis.
{\em Cox School of Business Historical Working Papers}, No. 170. Dallas, TX: Southern Methodist University.
Available online: \url{http://scholar.smu.edu/business_workingpapers/170}.

\bibitem[Chen and Sutcliffe, 2007]{ChenS2007} Chen, F. and Sutcliffe, C. (2007)
Better Cross Hedges With Composite Hedging? Hedging Equity Portfolios Using Financial and Commodity Futures.
{\em European Journal of Finance} 18(6): 575-595.

\bibitem[Chen {\em et al}, 2011]{ChenHL2011} Chen, H.-L., Yang, B., Wang, G., Liu, J., Xu, X., Wang, S.-J. and Liu D.-Y. (2011)
A novel bankruptcy prediction model based on an adaptive fuzzy k-nearest neighbor method.
{\em Knowledge-Based Systems} 24(8): 1348-1359.

\bibitem[Cheng and Madhavan, 2010]{Cheng2010} Cheng, M. and Madhavan, A. (2010)
The Dynamics of Leveraged and Inverse Exchange-Traded Funds.
{\em Journal of Investment Management} 7(4): 43-62.

\bibitem[Cheng and Xiong, 2013]{ChengXiong2013} Cheng, I.-H. and Xiong, W. (2013)
Why Do Hedgers Trade so Much?
{\em Working Paper.} Available online: \url{https://ssrn.com/abstract=2358762}.

\bibitem[Chernenko and Sunderam, 2016]{Chernenko2016} Chernenko, S. and Sunderam, A. (2016)
Liquidity Transformation in Asset Management: Evidence from the Cash Holdings of Mutual Funds.
{\em Working Paper.} Available online: \url{http://www.nber.org/papers/w22391}.

\bibitem[Chernov and Mueller, 2012]{Chernov2012} Chernov, M. and Mueller, P. (2012)
The Term Structure of Inflation Expectations.
{\em Journal of Financial Economics} 106(2): 367-394.

\bibitem[Cherry, 2004]{Cherry2004} Cherry, J. (2004)
The Limits of Arbitrage: Evidence from Exchange Traded Funds.
{\em Working Paper.} Available online: \url{https://ssrn.com/abstract=628061}.

\bibitem[Cheung, 2010]{Cheung2010} Cheung, W. (2010)
The Black-Litterman model explained.
{\em Journal of Asset Management} 11(4): 229-243.

\bibitem[Cheung, Kwan and Sarkar, 2010]{CheungKwan2010} Cheung, C.S., Kwan, C.C.Y. and Sarkar, S. (2010)
Bond Portfolio Laddering: A Mean-Variance Perspective.
{\em Journal of Applied Finance} 20(1): 103-109.

\bibitem[Cheung, Kwan and Yip, 1990]{Cheung1990} Cheung, C.W., Kwan, C.C. and Yip, P.C. (1990)
The hedging effectiveness of options and futures: a mean-gini approach.
{\em Journal of Futures Markets} 10(1): 61-73.

\bibitem[Cheung, Roca and Su, 2015]{CheungRS2015} Cheung, A., Roca, E. and Su, J.-J. (2015)
Crypto-currency Bubbles: an Application of the Phillips-Shi-Yu (2013) Methodology on Mt. Gox Bitcoin Prices.
{\em Applied Economics} 47(23): 2348-2358.

\bibitem[Chiang and Jiang, 1995]{Chiang1995} Chiang, T.C. and Jiang, C.X. (1995)
Foreign exchange returns over short and long horizons.
{\em International Review of Economics \& Finance} 4(3): 267-282.

\bibitem[Chiang and Wang, 2002]{Chiang2002} Chiang, M.H. and Wang, C.Y. (2002)
The Impact of Futures Trading on Spot Index Volatility: Evidence from Taiwan Index Futures.
{\em Applied Economics Letters} 9(6): 381-385.

\bibitem[Chidambaran, Fernando and Spindt, 2001]{Chidambaran2001} Chidambaran, N.K., Fernando, C.S. and Spindt, P.A. (2001)
Credit enhancement through financial engineering: Freeport McMoRan's gold-denominated depositary shares.
{\em Journal of Financial Economics} 60(2-3): 487-528.

\bibitem[Chin, Prevost and Gottesman, 2002]{Chin2002} Chin, J.Y.F., Prevost, A.K. and Gottesman, A.A. (2002)
Contrarian investing in a small capitalization market: Evidence from New Zealand.
{\em Financial Review} 37(3): 421-446.

\bibitem[Chinco and Mayer, 2012]{Chinco2012} Chinco, A. and Mayer, C. (2012)
Distant speculators and asset bubbles in the housing market.
{\em Working Paper.} Available online: \url{http://www.econ.yale.edu/~shiller/behfin/2012-04-11/Chinco_Mayer.pdf}.

\bibitem[Chinloy, 1989]{Chinloy1989} Chinloy, P. (1989)
The Probability of Prepayment.
{\em Journal of Real Estate Finance and Economics} 2(4): 267-283.

\bibitem[Cho, 1996]{Cho1996} Cho, M. (1996)
House Price Dynamics: A Survey of Theoretical and Empirical Issues.
{\em Journal of Housing Research} 7(2): 145-172.

\bibitem[Choi, 2011]{Choi2011} Choi, M.S. (2011)
Momentary exchange rate locked in a triangular mechanism of international currency.
{\em Applied Economics} 43(16): 2079-2087.

\bibitem[Choi {\em et al}, 2010]{Choi2010} Choi, D., Getmansky, M., Henderson, B. and Tookes, H. (2010)
Convertible bond arbitrageurs as suppliers of capital.
{\em Review of Financial Studies} 23(6): 2492-2522.

\bibitem[Choi, Getmansky and Tookes, 2009]{Choi2009} Choi, D., Getmansky, M. and Tookes, H. (2009)
Convertible bond arbitrage, liquidity externalities, and stock prices.
{\em Journal of Financial Economics} 91(2): 227-251.

\bibitem[Choi {\em et al}, 2014]{Choi2014} Choi, H.I., Kwon, S.-H., Kim, J.Y. and Jung, D.-S. (2014)
Commodity Futures Term Structure Model.
{\em Bulletin of the Korean Mathematical Society} 51(6): 1791-1804.

\bibitem[Chong, Han and Park, 2017]{Chong2017} Chong, E., Han, C. and Park, F.C. (2017)
Deep learning networks for stock market analysis and prediction: Methodology, data representations, and case studies.
{\em Expert Systems with Applications} 83: 187-205.

\bibitem[Chong and Miffre, 2010]{Chong2010} Chong, J. and Miffre, J. (2010)
Conditional Correlation and Volatility in Commodity Futures and Traditional Asset Markets.
{\em Journal of Alternative Investments} 12(3): 61-75.

\bibitem[Chordia {\em et al}, 2009]{Chordia2009} Chordia, T., Goyal, A., Sadka, G., Sadka, R. and Shivakumar, L. (2009)
Liquidity and the Post-Earnings-Announcement Drift.
{\em Financial Analysts Journal} 65(4): 18-32.

\bibitem[Chordia and Shivakumar, 2002]{Chordia2002} Chordia, T. and Shivakumar, L. (2002)
Momentum, Business Cycle, and Time-Varying Expected Returns.
{\em Journal of Finance} 57(2): 985-1019.

\bibitem[Chordia and Shivakumar, 2006]{Chordia2006} Chordia, T. and Shivakumar, L. (2006)
Earnings and price momentum.
{\em Journal of Financial Economics} 80(3): 627-656.

\bibitem[Choro\'{s}-Tomczyk, H\"ardle and Okhrin, 2016]{Choros-Tomczyk2016} Choro\'{s}-Tomczyk, B., H\"ardle, W.K. and Okhrin, O. (2016)
A semiparametric factor model for CDO surfaces dynamics.
{\em Journal of Multivariate Analysis} 146: 151-163.

\bibitem[Choudhry, 2004]{Choudhry2004} Choudhry, M. (2004)
The credit default swap basis: analysing the relationship between cash and synthetic credit markets.
{\em Journal of Derivatives Use, Trading and Regulation} 10(1): 8-26.

\bibitem[Choudhry, 2006]{Choudhry2006} Choudhry, M. (2006)
Revisiting the Credit Default Swap Basis: Further Analysis of the Cash and Synthetic Credit Market Differential.
{\em Journal of Structured Finance} 11(4): 21-32.

\bibitem[Choudhry, 2007]{Choudhry2007} Choudhry, M. (2007)
Trading the CDS Basis: Illustrating Positive and Negative Basis Arbitrage Trades.
{\em Journal of Trading} 2(1): 79-94.

\bibitem[Christensen, 1999]{Christensen1999} Christensen, M. (1999)
Duration and Convexity for Bond Portfolios.
{\em Finanzmarkt und Portfolio Management} 13(1): 66-72.

\bibitem[Christensen and Fabozzi, 1985]{Christensen1985} Christensen, P.E. and Fabozzi, F.J. (1985)
Bond Immunization: An Asset Liability Optimization Strategy.
In: Fabozzi, F.J. and Pollack, I.M. (eds.)
{\em The Handbook of Fixed Income Securities.} (2nd ed.) Homewood, IL: Dow Jones-Irwin, pp. 676-703.

\bibitem[Christensen and Gillan, 2012]{Christensen2012} Christensen, J.H.E. and Gillan, J.M. (2012)
Could the U.S. Treasury Benefit from Issuing More TIPS?
{\em Federal Reserve Bank of San Francisco, Working Papers Series}, No. 2011-16.
Available online: \url{https://www.frbsf.org/economic-research/files/wp11-16bk.pdf}.

\bibitem[Christensen, Lopez and Rudebusch, 2010]{Christensen2010} Christensen, J.H.E., Lopez, J.A. and Rudebusch, G.D. (2010)
Inflation Expectations and Risk Premiums in an Arbitrage-Free Model of Nominal and Real Bond Yields.
{\em Journal of Money, Credit, and Banking} 42(6): 143-178.

\bibitem[Christensen and Prabhala, 1998]{Christensen1998} Christensen, B.J. and Prabhala, N.R. (1998)
The relation between implied and realized volatility.
{\em Journal of Financial Economics} 50(2): 125-150.

\bibitem[Christiansen and Lund, 2005]{Christiansen2005} Christiansen, C. and Lund, J.  (2005)
Revisiting the Shape of the Yield Curve: The Effect of Interest Rate Volatility.
{\em Working Paper.} Available online: \url{https://ssrn.com/abstract=264139}.

\bibitem[Christie-David and Chaudry, 2001]{ChristieDavid2001} Christie-David, R. and Chaudry, M. (2001)
Coskewness and cokurtosis in futures markets.
{\em Journal of Empirical Finance} 8(1): 55-81.

\bibitem[Christoffersen {\em et al}, 2005]{Christoffersen2005} Christoffersen, S.E.K., G\'{e}czy, C.C., Musto, D.K. and Reed, A.V. (2005)
Crossborder Dividend Taxation and the Preferences of Taxable and Nontaxable Investors: Evidence From Canada.
{\em Journal of Financial Economics} 78(1): 121-144.

\bibitem[Christoffersen {\em et al}, 2003]{Christoffersen2003} Christoffersen, S.E.K., Reed, A.V., G\'{e}czy, C.C. and Musto, D.K. (2003)
The Limits to Dividend Arbitrage: Implications for Cross Border Investment.
{\em Working Paper.} Available online: \url{https://ssrn.com/abstract=413867}.

\bibitem[Chua, Koh and Ramaswamy, 2006]{Chua2006} Chua, C.T., Koh, W.T.H. and Ramaswamy, K. (2006)
Profiting from Mean-Reverting Yield Curve Trading Strategies.
{\em Journal of Fixed Income} 15(4): 20-33.

\bibitem[Chuang, 2015]{Chuang2015} Chuang, H. (2015)
Time Series Residual Momentum.
{\em Working Paper.} Available online: \url{http://www.econ.tohoku.ac.jp/econ/datascience/DDSR-DP/no38.pdf}.

\bibitem[Chuang and Ho, 2014]{Chuang2014} Chuang, H. and Ho, H.-C. (2014)
Implied Price Risk and Momentum Strategy.
{\em Review of Finance} 18(2): 591-622.

\bibitem[Chui, Titman and Wei, 2003a]{Chui2003a} Chui, A.C.W., Titman, S. and Wei, K.C.J. (2003a)
The Cross-Section of Expected REIT Returns.
{\em Real Estate Economics} 31(3): 451-479.

\bibitem[Chui, Titman and Wei, 2003b]{Chui2003b} Chui, A.C.W., Titman, S. and Wei, K.C.J. (2003b)
Intra-industry momentum: the case of REITs.
{\em Journal of Financial Markets} 6(3): 363-387.

\bibitem[Chung, 2000]{Chung2000} Chung, S.Y. (2000)
Review of Macro Trading and Investment Strategies: Macroeconomic Arbitrage in Global Markets.
{\em Journal of Alternative Investments} 3(1): 84-85.

\bibitem[Ciaian, Rajcaniova and Kancs, 2015]{Ciaian2015} Ciaian, P., Rajcaniova, M. and Kancs, D. (2015)
The economics of BitCoin price formation.
{\em Applied Economics} 48(19): 1799-1815.

\bibitem[Cirelli {\em et al}, 2017]{Cirelli2017} Cirelli, S., Vitali, S., Ortobelli Lozza, S. and Moriggia, V. (2017)
A conservative discontinuous target volatility strategy.
{\em Investment Management and Financial Innovations} 14(2-1): 176-190.

\bibitem[Clare, Ioannides and Skinner, 2000]{Clare2000} Clare, A.D., Ioannides, M. and Skinner, F.S. (2000)
Hedging Corporate Bonds with Stock Index Futures: A Word of Caution.
{\em Journal of Fixed Income} 10(2): 25-34.

\bibitem[Clarida, Davis and Pedersen, 2009]{Clarida2009} Clarida, R.H., Davis, J.M. and Pedersen, N. (2009)
Currency carry trade regimes: Beyond the Fama regression.
{\em Journal of International Money and Finance} 28(8): 1375-1389.

\bibitem[Clarida and Waldman, 2007]{Clarida2007} Clarida, R. and Waldman, D. (2007)
Is Bad News About Inflation Good News for the Exchange Rate?
{\em Working Paper.} Available online: \url{http://www.nber.org/papers/w13010.pdf}.

\bibitem[Clark, 2017]{Clark2017} Clark, G.L. (2017)
Financial intermediation, infrastructure investment and regional growth.
{\em Area Development and Policy} 2(3): 217-236.

\bibitem[Clark {\em et al}, 2012]{Clark2012} Clark, G.L., Monk, A.H.B., Orr, R. and Scott, W. (2012)
The new Era of infrastructure investing.
{\em Pensions: An International Journal} 17(2): 103-111.

\bibitem[Clark and Terry, 2010]{Clark2010} Clark, T.E. and Terry, S.J. (2010)
Time Variation in the Inflation Passthrough of Energy Prices.
{\em Journal of Finance} 42(7): 1419-1433.

\bibitem[Clark and Weinstein, 1983]{Clark1983} Clark, T.A. and Weinstein, M.I. (1983)
The behavior of the common stock of bankrupt firms.
{\em Journal of Finance} 38(2): 489-504.

\bibitem[Clarke, de Silva and Thorley, 2006]{Clarke2006} Clarke, R.G., de Silva, H. and Thorley, S. (2006)
Minimum-Variance Portfolios in the U.S. Equity Market.
{\em Journal of Portfolio Management} 33(1): 10-24.

\bibitem[Clarke, de Silva and Thorley, 2010]{Clarke2010} Clarke, R.G., de Silva, H. and Thorley, S. (2010)
Know Your VMS Exposure.
{\em Journal of Portfolio Management} 36(2): 52-59.

\bibitem[Clarke, de Silva and Thorley, 2013]{Clarke2013} Clarke, R.G., de Silva, H. and Thorley, S. (2013)
{\em Fundamentals of Futures and Options.} New York, NY: The Research Foundation of CFA Institute.

\bibitem[Clifford, Fulkerson and Jordan, 2014]{Clifford2014} Clifford, C.P., Fulkerson, J.A. and Jordan, B.D. (2014)
What Drives ETF Flows?
{\em Financial Review} 49(3): 619-642.

\bibitem[Clinton, 1988]{Clinton1988} Clinton, K. (1988)
Transactions costs and covered interest arbitrage: Theory and evidence.
{\em Journal of Political Economy} 96(2): 358-370.

\bibitem[Cochrane, 1999]{Cochrane1999} Cochrane, J.H. (1999)
Portfolio Advice for a Multifactor World.
{\em Federal Reserve Bank of Chicago, Economic Perspectives} 23(3): 59-78.

\bibitem[Cochrane and Piazzesi, 2005]{Cochrane2005} Cochrane, J.H. and Piazzesi, M. (2005)
Bond Risk Premia.
{\em American Economic Review} 95(1): 138-160.

\bibitem[Coffey, Hrung and Sarkar, 2009]{Coffey2009} Coffey, N., Hrung, W.B. and Sarkar, A. (2009)
Capital constraints, counterparty risk, and deviations from covered interest rate parity.
{\em Federal Reserve Bank of New York Staff Reports}, No. 393.
Available online:\\
\url{https://www.newyorkfed.org/medialibrary/media/research/staff_reports/sr393.pdf}.

\bibitem[Cohen, 2005]{Cohen2005} Cohen, G. (2005)
{\em The bible of options strategies: the definitive guide for practical trading strategies.} Upper Saddle River, NJ: Financial Times Prentice Hall.

\bibitem[Cole {\em et al}, 1999]{Cole1999} Cole, C.A., Kastens, T.L., Hampel, F.A. and Gow, L.R. (1999)
A calendar spread trading simulation of seasonal processing spreads.
In: {\em Proceedings of the NCCC-134 Conference on Applied Commodity Price Analysis, Forecasting, and Market Risk Management.}
Available online: \url{http://www.farmdoc.illinois.edu/nccc134/conf_1999/pdf/confp14-99.pdf}.

\bibitem[Cole and Young, 1995]{Cole1995} Cole, C.S. and Young, P.J. (1995)
Modified duration and convexity with semiannual compounding.
{\em Journal of Economics and Finance} 19(1): 1-15.

\bibitem[Colianni, Rosales and Signorotti, 2015]{Colianni2015} Colianni, S., Rosales, S. and Signorotti, M. (2015)
Algorithmic Trading of Cryptocurrency Based on Twitter Sentiment Analysis.
{\em Working Paper.} Available online: \url{http://cs229.stanford.edu/proj2015/029_report.pdf}.

\bibitem[Collin-Dufresne and Solnik, 2001]{Collin-Dufresne2001} Collin-Dufresne, P. and Solnik, B. (2001)
On the Term Structure of Default Premia in the Swap and LIBOR Markets.
{\em Journal of Finance}  56(3): 1095-1115.

\bibitem[Cong, Tan and Weng, 2013]{Cong2013} Cong, J., Tan, K.S. and Weng, C. (2013)
VAR-Based Optimal Partial Hedging.
{\em ASTIN Bulletin: The Journal of the IAA} 43(3): 271-299.

\bibitem[Cong, Tan and Weng, 2014]{Cong2014} Cong, J., Tan, K.S. and Weng, C. (2014)
CVaR-Based Optimal Partial Hedging.
{\em Journal of Risk} 16(3): 49-83.

\bibitem[Connor and Leland, 1995]{Connor1995} Connor, G. and Leland, H. (1995)
Cash Management for Index Tracking.
{\em Financial Analysts Journal} 51(6): 75-80.

\bibitem[Connor and Woo, 2004]{Connor2004} Connor, G. and Woo, M. (2004)
An Introduction to hedge funds.
{\em Working Paper.} Available online: \url{http://eprints.lse.ac.uk/24675/1/dp477.pdf}.

\bibitem[Conover {\em et al}, 2008]{Conover2008} Conover, C.M., Jensen, G., Johnson, R. and Mercer, M. (2008)
Sector Rotation and Monetary Conditions.
{\em Journal of Investing} 28(1): 34-46.

\bibitem[Conover {\em et al}, 2010]{Conover2010} Conover, C.M., Jensen, G.R., Johnson, R.R. and Mercer, J.M. (2010)
Is Now the Time to Add Commodities to Your Portfolio?
{\em Journal of Investing} 19(3): 10-19.

\bibitem[Conrad, Dittmar and Ghysels, 2013]{Conrad2013} Conrad, J., Dittmar, R.F. and Ghysels, E. (2013)
Ex Ante Skewness and Expected Stock Returns.
{\em Journal of Finance} 68(1): 85-124.

\bibitem[Conrad, Hameed and Niden, 2013]{Conrad1994} Conrad, J.S., Hameed, A. and Niden, C. (1994)
 Volume and autocovariances in short-horizon individual security returns.
{\em Journal of Finance} 49(4): 1305-1329.

\bibitem[Conrad and Kaul, 1998]{Conrad1998} Conrad, J. and Kaul, G. (1998)
An Anatomy of Trading Strategies.
{\em Review of Financial Studies} 11(3): 489-519.

\bibitem[Cont and Minca, 2013]{Cont2013} Cont, R. and Minca, A. (2013)	
Recovering Portfolio Default Intensities Implied by CDO Quotes.
{\em Mathematical Finance} 23(1): 94-121.

\bibitem[Cook and LaRoche, 1993]{Cook1993} Cook, T.Q. and LaRoche, R.B. (eds.) (1993)
{\em Instruments of the money market.} (7th ed.)
Richmond, Virginia: Federal Reserve Bank of Richmond.

\bibitem[Cook and Rowe, 1986]{Cook1986} Cook, T.Q. and Rowe, T.D. (eds.) (1986)
{\em Instruments of the money market.} (6th ed.)
Richmond, Virginia: Federal Reserve Bank of Richmond.

\bibitem[Cooper, 2010]{Cooper2010} Cooper, T. (2010)
Alpha Generation and Risk Smoothing Using Managed Volatility.
{\em Working Paper.} Available online: \url{https://ssrn.com/abstract=1664823}.

\bibitem[Cooper, Downs and Patterson, 1999]{Cooper1999} Cooper, M., Downs, D.H. and Patterson, G.A. (1999)
Real Estate Securities and a Filter-based, Short-term Trading Strategy.
{\em Journal of Real Estate Research} 18(2): 313-334.

\bibitem[Cooper, Gutierrez and Hameed, 2004]{Cooper2004} Cooper, M.J., Gutierrez, R.C., Jr. and Hameed, A. (2004)
Market States and Momentum.
{\em Journal of Finance} 59(3): 1345-1365.

\bibitem[Cooper and Priestley, 2008]{Cooper2008} Cooper, I. and Priestley, R. (2008)
Time-Varying Risk Premiums and the Output Gap.
{\em Review of Financial Studies} 22(7): 2801-2833.

\bibitem[Copeland and Galai, 1983]{Copeland1983} Copeland, T.E. and Galai, D. (1983)
Information Effects on the bid-ask spread.
{\em Journal of Finance} 38(5): 1457-1469.

\bibitem[Corbally and Dang, 2002]{Corbally2002} Corbally, M. and Dang, P. (2002)
 Underlying Markets and Indexes.
In: Banks, E. (ed.) {\em Weather Risk Management: Market, Products and Applications.} London, UK: Palgrave Macmillan.

\bibitem[Corbett, 2006] {Corbett2006} Corbett, M. (2006)
{\em Find it, fix it, flip it! Make millions in real estate -- one house at a time.}
New York, NY: Plume.

\bibitem[Cornell and French, 1983]{Cornell1983} Cornell, B. and French, K.R. (1983)
The pricing of stock index futures.
{\em Journal of Futures Markets} 3(1): 1-14.

\bibitem[Cornelli and Li, 2002]{Cornelli2002} Cornelli, F. and Li, D.D. (2002)
Risk Arbitrage in Takeovers.
{\em Review of Financial Studies} 15(3): 837-868.

\bibitem[Corrado and Miller, 2005]{Corrado2005} Corrado, C.J. and Miller, T.W., Jr. (2005)
The forecast quality of CBOE implied volatility indexes.
{\em Journal of Futures Markets} 25(4): 339-373.

\bibitem[Corrado and Su, 1997]{Corrado1997} Corrado, C.J. and Su, T. (1997)
Implied volatility skews and stock return skewness and kurtosis implied by stock option prices.
{\em European Journal of Finance} 3(1): 73-85.

\bibitem[Correia, Richardson and Tuna, 2012]{Correia2012} Correia, M.M., Richardson, S.A. and Tuna, A.I. (2012)
Value Investing in Credit Markets.
{\em Review of Accounting Studies} 17(3): 572-609.

\bibitem[Cosandier and Lang, 1981]{Cosandier1981} Cosandier, P.-A. and Lang, B.R. (1981)
Interest rate parity tests: Switzerland and some major western countries.
{\em Journal of Banking \& Finance} 5(2): 187-200.

\bibitem[Cousin and Laurent, 2012]{Cousin2012} Cousin, A. and Laurent, J. (2012)
Dynamic Hedging of Synthetic CDO Tranches: Bridging the Gap between Theory and Practice.
In: Bielecki, T.R., Brigo, D. and Patras, F. (eds.)
{\em Credit Risk Frontiers}. Hoboken, NJ: John Wiley \& Sons, Inc., Chapter 6.

\bibitem[Coval and Shumway, 2001]{Coval2001} Coval, J.D. and Shumway, T. (2001)
Expected options returns.
{\em Journal of Finance} 56(3): 983-1009.

\bibitem[Cover, 1984]{Cover1984} Cover, T.M. (1984)
An algorithm for maximizing expected log investment return.
{\em IEEE Transactions on Information Theory} 30(2): 369-373.

\bibitem[Cox, 2015]{Cox2015} Cox, D. (2015)
{\em Handbook of Anti Money Laundering.}
Chichester, UK: John Wiley \& Sons, Ltd.

\bibitem[Crabbe and Fabozzi, 2002]{Crabbe2002} Crabbe, L.E. and Fabozzi, F.J. (2002)
{\em Corporate Bond Portfolio Management.}
Hoboken, NJ: John Wiley \& Sons, Inc.

\bibitem[Creamer and Freund, 2007]{Creamer2007} Creamer, G.G. and Freund, Y. (2007)
A Boosting Approach for Automated Trading.
{\em Journal of Trading} 2(3): 84-96.

\bibitem[Creamer and Freund, 2010]{Creamer2010a} Creamer, G.G. and Freund, Y. (2010)
Automated Trading with Boosting and Expert Weighting.
{\em Quantitative Finance} 10(4): 401-420.

\bibitem[Cremers and Weinbaum, 2010]{Cremers2010} Cremers, M. and Weinbaum, D. (2010)
Deviations from Put-Call Parity and Stock Return Predictability.
{\em Journal of Financial and Quantitative Analysis} 45(2): 335-367.

\bibitem[Creti, Jo\"{e}ts and Mignon, 2013]{Creti2013} Creti, A., Jo\"{e}ts, M. and Mignon, V. (2013)
On the links between stock and commodity markets' volatility.
{\em Energy Economics} 37: 16-28.

\bibitem[Cross and Kozyakin, 2015]{Cross2015} Cross, R. and Kozyakin, V. (2015)
Fact and fictions in FX arbitrage processes.
{\em Journal of Physics: Conference Series} 585: 012015.

\bibitem[Cultrera and Br\'{e}dart, 2015]{Cultrera2015} Cultrera, L. and Br\'{e}dart, X. (2015)
Bankruptcy prediction: the case of Belgian SMEs.
{\em Review of Accounting and Finance} 15(1): 101-119.

\bibitem[Czaja, Kaufmann and Scholz, 2013]{Czaja2013} Czaja, M.-G., Kaufmann, P. and Scholz, H. (2013)
Enhancing the profitability of earnings momentum strategies: The role of price momentum, information diffusion and earnings uncertainty.
{\em Journal of Investment Strategies} 2(4): 3-57.

\bibitem[Dahlgran, 2000]{Dahlgran2000} Dahlgran, R.A. (2000)
Cross-hedging the cottonseed crush: A case study.
{\em Agribusiness} 16(2): 141-158.

\bibitem[Daigler, 2007]{Daigler2007} Daigler, R.T. (2007)
Spread volume for currency futures.
{\em Journal of Economics and Finance} 31(1): 12-19.

\bibitem[Daigler and Copper, 1998]{Daigler1998} Daigler, R.T. and Copper, M. (1998)
A Futures Duration-Convexity Hedging Method.
{\em Financial Review} 33(4): 61-80.

\bibitem[Dale and Currie, 2015]{Dale2015} Dale, A. and Currie, E. (2015)
An alternative funding model for agribusiness research in Canada.
{\em Agricultural Sciences} 6(9): 961-969.

\bibitem[Damghani and Kos, 2013]{Damghani2013} Damghani, B.M. and Kos, A. (2013)
De-arbitraging With a Weak Smile: Application to Skew Risk.
{\em Wilmott Magazine} 2013(64): 40-49.

\bibitem[Damiani, 2012]{Damiani2012} Damiani, D. (2012)
The Case for Cash.
{\em CFA Institute Magazine} 23(4): 8-9.

\bibitem[D'Amico, Kim and Wei, 2018]{DAmico2018} D'Amico, S., Kim, D. and Wei, M. (2018)
Tips from TIPS: The Informational Content of Treasury Inflation-Protected Security Prices.
{\em Journal of Financial and Quantitative Analysis} 53(1): 395-436.

\bibitem[Daniel, 2001]{Daniel2001} Daniel, K. (2001)
The Power and Size of Mean Reversion Tests.
{\em Journal of Empirical Finance} 8(5): 493-535.

\bibitem[Daniel and Moskowitz, 2016]{Daniel2016} Daniel, K. and Moskowitz, T.J. (2016)
Momentum crashes.
{\em Journal of Financial Economics} 122(2): 221-247.

\bibitem[D'Antonio, 2008]{DAntonio2008} D'Antonio, L. (2008)
Equity Collars as Alternative to Asset Allocation.
{\em Journal of Financial Service Professionals} 62(1): 67-76.

\bibitem[Dao, 2014]{Dao2014} Dao, T.-L. (2014)
Momentum strategies with the L1 filter.
{\em Journal of Investment Strategies} 3(4): 57-82.

\bibitem[Das, 2005]{Das2005} Das, S. (2005)
{\em Credit Derivatives: Trading \& Management of Credit \& Default Risk.} (3rd ed.)
Hoboken, NJ: John Wiley \& Sons, Inc.

\bibitem[da S. Gomes, Ludermir and Lima, 2011]{daSGomes2011} da S. Gomes, G.S., Ludermir, T.B. and Lima, L.M.M.R. (2011)
Comparison of new activation functions in neural network for forecasting financial time series.
{\em Neural Computing and Applications} 20(3): 417-439.

\bibitem[Dash and Dash, 2016]{Dash2016} Dash, R. and Dash, P.K. (2016)
A hybrid stock trading framework integrating technical analysis with machine learning techniques.
{\em Journal of Finance and Data Science} 2(1): 42-57.

\bibitem[Da Silva, Lee and Pornrojnangkool, 2009]{DaSilva2009} Da Silva, A.S., Lee, W. and Pornrojnangkool, B. (2009)
The Black-Litterman model for active portfolio management.
{\em Journal of Portfolio Management} 35(2): 61-70.

\bibitem[Daumas, 2017]{Daumas2017} Daumas, L.D. (2017)
Hedging stocks through commodity indexes: a DCC-GARCH approach.
{\em Working Paper.} Available online: \url{https://impa.br/wp-content/uploads/2017/11/RiO2017-PP_FAiube.pdf}.

\bibitem[Davidson, Herskovitz and Van Drunen, 1988]{Davidson1988} Davidson, A.S., Herskovitz, M.D. and Van Drunen, L.D. (1988)
The refinancing threshold pricing model: An economic approach to valuing MBS.
{\em Journal of Real Estate Finance and Economics} 1(2): 117-130.

\bibitem[Davis, 1996]{Davis1996} Davis, J.L. (1996)
The cross-section of stock returns and survivorship bias: Evidence from delisted stocks.
{\em Quarterly Review of Economics and Finance} 36(3): 365-375.

\bibitem[Davis, 2001]{DavisM2001} Davis, M. (2001)
Pricing Weather Derivatives by Marginal Value.
{\em Quantitative Finance} 1(3): 305-308.

\bibitem[Davis, 2006]{Davis2006} Davis, M.H.A. (2006)
Optimal Hedging with Basis Risk.
In: Kabanov, Y., Liptser, R. and Stoyanov, J. (eds.) {\em From Stochastic Calculus to Mathematical Finance.} Berlin, Germany: Springer.

\bibitem[Davis and Lleo, 2012]{Davis2012} Davis, M. and Lleo, S. (2012)
Fractional Kelly strategies in continuous time: Recent developments.
In: MacLean, L.C. and Ziemba, W. (eds.)
{\em Handbook of the Fundamentals of Financial Decision Making.} Singapore: World Scientific Publishing.

\bibitem[Davis and Lo, 2001]{Davis2001} Davis, M. and Lo, V. (2001)
Infectious defaults.
{\em Quantitative Finance} 1(4): 382-387.

\bibitem[Deacon, Derry and Mirfendereski, 2004]{Deacon2004} Deacon, M., Derry, A. and Mirfendereski, D. (2004)
{\em Inflation-indexed Securities: Bonds, Swaps and other Derivatives.}
Chichester, UK: John Wiley \& Sons, Ltd.

\bibitem[Deardorff, 1979]{Deardorff1979} Deardorff, A.V. (1979)
One-Way Arbitrage and Its Implications for the Foreign Exchange Markets.
{\em Journal of Political Economy} 87(2): 351-364.

\bibitem[de Boer {\em et al}, 2005]{deBoer2005} de Boer, P.-T., Kroese, D.P., Mannor, S. and Rubinstein, R.Y. (2005)
A Tutorial on the Cross-Entropy Method.
{\em Annals of Operations Research} 134(1): 19-67.

\bibitem[DeBondt and Thaler, 1985]{DeBondt1985} DeBondt, W.F.M. and Thaler, R.H. (1985)
Does stock market overreact?
{\em Journal of Finance} 40(3): 793-807.

\bibitem[De Carvalho {\em et al}, 2014]{DeCarvalho2014} De Carvalho, R.L., Dugnolle, P., Lu, X. and Moulin, P. (2014)
Low-Risk Anomalies in Global Fixed Income: Evidence from Major Broad Markets.
{\em Journal of Fixed Income} 23(4): 51-70.

\bibitem[Dechant and Finkenzeller, 2013]{Dechant2013} Dechant, T. and Finkenzeller, K. (2013)
How much into infrastructure? Evidence from dynamic asset allocation.
{\em Journal of Property Research} 30(2): 103-127.

\bibitem[Dechario {\em et al}, 2010]{Dechario2010} Dechario, T., Mosser, P., Tracy, J., Vickery, J. and Wright, J. (2010)
A Private Lender Cooperative Model for Residential Mortgage Finance.
{\em Federal Reserve Bank of New York Staff Reports}, No. 466.
Available online:\\ \url{https://www.newyorkfed.org/medialibrary/media/research/staff_reports/sr466.pdf}.

\bibitem[De Jong, Dutordoir and Verwijmeren, 2011]{DeJong2011} De Jong, A., Dutordoir, M. and Verwijmeren, P. (2011)
Why do convertible issuers simultaneously repurchase stock? An arbitrage-based explanation.
{\em Journal of Financial Economics} 100(1): 113-129.

\bibitem[De La Pe\~{n}a, Garayeta and Iturricastillo, 2017]{DeLaPena2017} De La Pe\~{n}a, J.I., Garayeta, A. and Iturricastillo, I. (2017)
Dynamic immunisation does not imply cash flow matching: a hard application to Spain.
{\em Economic Research -- Ekonomska Istra\v{z}ivanja} 30(1): 238-255.

\bibitem[DeLisle, Doran and Krieger, 2014]{DeLisle2014} DeLisle, J., Doran, J. and Krieger, K. (2014)
Volatility as an Asset Class: Holding VIX in a Portfolio.
{\em Working Paper.} Available online: \url{https://ssrn.com/abstract=2534081}.

\bibitem[DeMaskey, 1995]{DeMaskey1995} DeMaskey, A.L. (1995)
A Comparison of the Effectiveness of Currency Futures and Currency Options in the Context of Foreign Exchange Risk Management.
{\em Managerial Finance} 21(4): 40-51.

\bibitem[DeMaskey, 1997]{DeMaskey1997} DeMaskey, A.L. (1997)
Single and Multiple Portfolio Cross-Hedging with Currency Futures.
{\em Multinational Finance Journal} 1(1): 23-46.

\bibitem[DeMaskey and Pearce, 1998]{DeMaskey1998} DeMaskey, A.L. and Pearce, J.A. (1998)
Commodity and Currency Futures Cross-Hedging of ASEAN Currency Exposures.
{\em Journal of Transnational Management Development} 4(1): 5-24.

\bibitem[Demeterfi {\em et al},  1999]{Demeterfi1999} Demeterfi, K., Derman, E., Kamal, M. and Zou, J.  (1999)
A guide to volatility and variance swaps.
{\em Journal of Derivatives} 6(4): 9-32.

\bibitem[DeMiguel {\em et al}, 2013]{DeMiguel2013} DeMiguel, V., Plyakha, Y., Uppal, R. and Vilkov, G. (2013)
Improving Portfolio Selection Using Option-Implied Volatility and Skewness.
{\em Journal of Financial and Quantitative Analysis} 48(6): 1813-1845.

\bibitem[DeMoura, Pizzinga and Zubelli, 2016]{deMoura2016} DeMoura, C.E., Pizzinga, A. and Zubelli, J. (2016)
A pairs trading strategy based on linear state space models and the Kalman filter.
{\em Quantitative Finance} 16(10): 1559-1573.

\bibitem[Dempster and Jones, 2002]{Dempster2002} Dempster, M.A.H. and Jones, C.M. (2002)
Can channel pattern trading be profitably automated?
{\em European Journal of Finance} 8(3): 275-301.

\bibitem[Deng, 2008]{Deng2008} Deng, Q. (2008)
Volatility Dispersion Trading.
{\em Working Paper.} Available online: \url{https://ssrn.com/abstract=1156620}.

\bibitem[Deng, Johnson and Sogomonian, 2001]{Deng2001} Deng, S.-J., Johnson, B. and Sogomonian, A. (2001)
Exotic electricity options and the valuation of electricity generation and transmission assets.
{\em Decision Support Systems} 30(3): 383-392.

\bibitem[Deng, McCann and Wang, 2012]{Deng2012} Deng, G., McCann, C. and Wang, O. (2012)
Are VIX Futures ETPs Effective Hedges?
{\em Journal of Index Investing} 3(3): 35-48.

\bibitem[Dennis and Mayhew, 2002]{Dennis2002} Dennis, P. and Mayhew, S. (2002)
Risk-Neutral Skewness: Evidence from Stock Options.
{\em Journal of Financial and Quantitative Analysis} 37(3): 471-493.

\bibitem[Dennis, Mayhew and Stivers, 2006]{Dennis2006} Dennis, P., Mayhew, S. and Stivers, C.  (2006)
Stock Returns, Implied Volatility Innovations, and the Asymmetric Volatility Phenomenon.
{\em Journal of Financial and Quantitative Analysis} 41(2): 381-406.

\bibitem[Denton and Hung, 1996]{Denton1996} Denton, J.W. and Hung, M.S. (1996)
A comparison of nonlinear optimization methods for supervised learning in multilayer feedforward neural networks.
{\em European Journal of Operational Research} 93(2): 358-368.

\bibitem[de Oliveira, Nobre and Z\'arate, 2013]{Oliveira2013} de Oliveira, F.A., Nobre, C.N. and Z\'arate, L.E. (2013)
Applying Artificial Neural Networks to prediction of stock price and improvement of the directional prediction index -- Case study of PETR4, Petrobras, Brazil.
{\em Expert Systems with Applications} 40(18): 7596-7606.

\bibitem[Depken, Hollans and Swidler, 2009]{Depken2009} Depken, C.A., Hollans, H. and Swidler, S. (2009)
An empirical analysis of residential property flipping.
{\em Journal of Real Estate Finance and Economics} 39(3): 248-263.

\bibitem[Depken, Hollans and Swidler, 2011]{Depken2011} Depken, C.A., Hollans, H. and Swidler, S. (2011)
Flips, flops and foreclosures: Anatomy of a real estate bubble.
{\em Journal of Financial Economic Policy} 3(1): 49-65.

\bibitem[Derman and Kani, 1994]{Derman1994} Derman, E. and Kani, I. (1994)
Riding on a Smile.
{\em Risk} 7(2): 139-145.

\bibitem[de Roon, Nijman and Veld, 1998]{DeRoon1998} de Roon, F.A., Nijman, T.E. and Veld, C. (1998)
Pricing Term Structure Risk in Futures Markets.
{\em Journal of Financial and Quantitative Analysis} 33(1): 139-157.

\bibitem[de Roon, Nijman and Veld, 2000]{DeRoon2000} de Roon, F.A., Nijman, T.E. and Veld, C. (2000)
Hedging pressure effects in futures markets.
{\em Journal of Finance} 55(3): 1437-1456.

\bibitem[Derwall {\em et al}, 2009]{DerwallHBM2009} Derwall, J., Huij, J., Brounen, D. and Marquering, W. (2009)
REIT Momentum and the Performance of Real Estate Mutual Funds.
{\em Financial Analysts Journal} 65(5): 24-34.

\bibitem[Derwall, Huij and De Zwart, 2009]{Derwall2009} Derwall, J., Huij, J. and De Zwart, G.B. (2009)
The Short-Term Corporate Bond Anomaly.
{\em Working Paper.} Available online: \url{https://ssrn.com/abstract=1101070}.

\bibitem[D'Este, 2014]{DEste2014} D'Este, R. (2014)
The Effect of Stolen Goods Markets on Crime: Evidence from a Quasi-Natural Experiment.
{\em Working Paper.} Available online: \url{https://warwick.ac.uk/fac/soc/economics/research/workingpapers/2014/twerp_1040b_deste.pdf}.

\bibitem[Detemple and Rindisbacher, 2010]{Detemple2010} Detemple, J. and Rindisbacher, M. (2010)
Dynamic Asset Allocation: Portfolio Decomposition Formula and Applications.
{\em Review of Financial Studies} 23(1): 25-100.

\bibitem[Detlefsen and H\"{a}rdle, 2013]{Detlefsen2013} Detlefsen, K. and H\"{a}rdle, W.K. (2013)
Variance swap dynamics.
{\em Quantitative Finance} 13(5): 675-685.

\bibitem[Dewally, Ederington and Fernando, 2013]{Dewally2013} Dewally, M., Ederington, L.H. and Fernando, C.S. (2013)
Determinants of Trader Profits in Commodity Futures Markets.
{\em Review of Financial Studies} 26(10): 2648-2683.

\bibitem[De Wit, 2010]{DeWit2010} De Wit, I. (2010)
International Diversification Strategies for Direct Real Estate.
{\em Journal of Real Estate Finance and Economics} 41(4): 433-457.

\bibitem[De Wit, 2006]{DeWit2006} De Wit, J. (2006)
Exploring the CDS-Bond Basis.
{\em Working Paper.} Available online: \url{https://ssrn.com/abstract=1687659}.

\bibitem[de Wit and van der Klaauw, 2013]{deWit2013} de Wit, E.R. and van der Klaauw, B. (2013)
Asymmetric Information and List-Price Reductions in the Housing Market.
{\em Regional Science and Urban Economics}  43(3): 507-520.

\bibitem[De Zwart {\em et al}, 2009]{DeZwart2009} De Zwart, G., Markwat, T., Swinkels, L. and van Dijk, D. (2009)
The economic value of fundamental and technical information in emerging currency markets.
{\em Journal of International Money and Finance} 28(4): 581-604.

\bibitem[Dichev, 1998]{Dichev1998} Dichev, I. (1998)
Is the risk of bankruptcy a systematic risk?
{\em Journal of Finance} 53(3): 1131-1147.

\bibitem[Diebold and Li, 2002]{Diebold2002} Diebold, F.X. and Li, C. (2002)
Forecasting the term structure of government bond yields.
{\em  Journal of Econometrics} 130(2): 337-364.

\bibitem[Diebold, Rudebusch and Aruoba, 2006]{Diebold2006} Diebold, F.X., Rudebusch, G.D. and Aruoba, S.B. (2006)
The macroeconomy and the yield curve: a dynamic latent factor approach.
{\em  Journal of Econometrics} 131(1-2): 309-338.

\bibitem[Ding and Sherris, 2011]{Ding2011} Ding, J.J. and Sherris, M. (2011)
Comparison of market models for measuring and hedging synthetic CDO tranche spread risks.
{\em European Actuarial Journal} 1(S2): 261-281.

\bibitem[Disatnik, Duchin and Schmidt, 2014]{Disatnik2014} Disatnik, D., Duchin, R. and Schmidt, B. (2014)
Cash Flow Hedging and Liquidity Choices.
{\em Review of Finance} 18(2): 715-748.

\bibitem[Dischel, 1998a]{Dischel1998a} Dischel, B. (1998a)
At last: A model for weather risk.
{\em Energy and Power Risk Management} 11(3): 20-21.

\bibitem[Dischel, 1998b]{Dischel1998b} Dischel, B. (1998b)
Black-Scholes won't do.
{\em Energy and Power Risk Management} 11(10): 8-9.

\bibitem[Dischel, 1999]{Dischel1999} Dischel, B. (1999)
Shaping history for weather risk management.
{\em Energy and Power Risk Management} 12(8): 13-15.

\bibitem[Do and Faff, 2010]{Do2010} Do, B. and Faff, R. (2010)
Does simple pairs trading still work?
{\em Financial Analysts Journal} 66(4): 83-95.

\bibitem[Do and Faff, 2012]{Do2012} Do, B. and Faff, R. (2012)
Are pairs trading profits robust to trading costs?
{\em Journal of Financial Research} 35(2): 261-287.

\bibitem[Doan, Alexeev and Brooks, 2014]{Doan2014} Doan, M.P., Alexeev, V. and Brooks, R. (2014)
Concurrent momentum and contrarian strategies in the Australian stock market.
{\em Australian Journal of Management} 41(1): 77-106.

\bibitem[Dobson, 1984]{Dobson1984} Dobson, M.W.R. (1984)
Global Investment Portfolios: The United Kingdom and Scandinavia.
{\em ICFA Continuing Education Series} 1984(4): 56-60.

\bibitem[Doeswijk, Lam and Swinkels, 2014]{Doeswijk2014} Doeswijk, R., Lam, T. and Swinkels, L. (2014)
The Global Multi-Asset Market Portfolio, 1959-2012.
{\em Financial Analysts Journal} 70(2): 26-41.

\bibitem[Doeswijk and Vliet, 2011]{Doeswijk2011} Doeswijk, R. and van Vliet, P. (2011)
Global tactical sector allocation: a quantitative approach.
{\em Journal of Portfolio Management} 28(1): 29-47.

\bibitem[Dolan, 1999]{Dolan1999} Dolan, C.P. (1999)
Forecasting the Yield Curve Shape.
{\em Journal of Fixed Income} 9(1): 92-99.

\bibitem[Dolvin, 2009]{Dolvin2009} Dolvin, S.D. (2009)
ETFs: Arbitrage opportunities and market forecasting.
{\em Journal of Index Investing} 1(1): 107-116.

\bibitem[Dolvin and Kirby, 2011]{Dolvin2011} Dolvin, S. and Kirby, J. (2011)
Momentum Trading in Sector ETFs.
{\em Journal of Index Investing} 2(3): 50-57.

\bibitem[Donchian, 1960]{Donchian1960} Donchian, R.D. (1960)
High finance in copper.
{\em Financial Analysts Journal} 16(6): 133-142.

\bibitem[Dong {\em et al}, 2009]{Dong2009} Dong, J.-C., Liu, J.-X., Wang, C.-H., Yuan, H. and Wang, W.-J. (2009)
Pricing Mortgage-Backed Security: An Empirical Analysis.
{\em Systems Engineering -- Theory \& Practice} 29(12): 46-52.

\bibitem[Dong and Zhou, 2008]{Dong2008} Dong, Z. and Zhou, D.-X. (2008)
Learning gradients by a gradient descent algorithm.
{\em Journal of Mathematical Analysis and Applications} 341(2): 1018-1027.

\bibitem[Donier and Bouchaud, 2015]{Donier2015} Donier, J. and Bouchaud, J.-P. (2015)
Why Do Markets Crash? Bitcoin Data Offers Unprecedented Insights.
{\em PLoS ONE} 10(10): e0139356.

\bibitem[Donninger, 2014]{Donninger2014} Donninger, C. (2014) VIX Futures Basis Trading: The Calvados-Strategy 2.0.
{\em Working Paper.} Available online: \url{https://ssrn.com/abstract=2379985}.

\bibitem[Donninger, 2015]{Donninger2015} Donninger, C. (2015)
Trading the Patience of Mrs. Yellen. A Short Vix-Futures Strategy for FOMC Announcement Days.
{\em Working Paper.} Available online: \url{https://ssrn.com/abstract=2544445}.

\bibitem[Doran and Krieger, 2010]{Doran2010} Doran, J.S. and Krieger, K. (2010)
Implications for Asset Returns in the Implied Volatility Skew.
{\em Financial Analysts Journal} 66(1): 65-76.

\bibitem[Doran, Peterson and Tarrant, 2007]{Doran2007} Doran, J.S., Peterson, D.R. and Tarrant, B.C. (2007)
Is there information in the volatility skew?
{\em Journal of Futures Markets} 27(10): 921-959.

\bibitem[Dorfleitner and Wimmer, 2010]{Dorfleitner2010} Dorfleitner, G. and Wimmer, M. (2010)
The pricing of temperature futures at the Chicago Mercantile Exchange.
{\em Journal of Banking \& Finance} 34(6): 1360-1370.

\bibitem[Dornier and Queruel, 2000]{Dornier2000} Dornier, F. and Queruel, M. (2000)
Caution to the wind.
{\em Energy and Power Risk Management} 13(8): 30-32.

\bibitem[Doskov and Swinkels, 2015]{Doskov2015} Doskov, N. and Swinkels, L. (2015)
Empirical evidence on the currency carry trade, 1900-2012.
{\em Journal of International Money and Finance} 51: 370-389.

\bibitem[Douglas, 2007]{Douglas2007} Douglas, R. (ed.) (2007)
{\em Credit Derivative Strategies: New Thinking on Managing Risk and Return.}
New York, NY: Bloomberg Press.

\bibitem[Dowd and Hutchinson, 2015]{Dowd2015} Dowd, K. and Hutchinson, M. (2015)
Bitcoin Will Bite the Dust.
{\em Cato Journal} 35(2): 357-382.

\bibitem[Downing, Jaffee and Wallace, 2009]{Downing2009} Downing, C., Jaffee, D. and Wallace, N. (2009)
Is the Market for Mortgage-Backed Securities a Market for Lemons?
{\em Review of Financial Studies} 22(7): 2457-2494.

\bibitem[Doyle, Lundholm and Soliman, 2006]{Doyle2006} Doyle, J.T., Lundholm, R.J. and Soliman, M.T. (2006)
The extreme future stock returns following I/B/E/S earnings surprises.
{\em Journal of Accounting Research} 44(5): 849-887.

\bibitem[Draper, Faff and Hillier, 2006]{Draper2006} Draper, P., Faff, R.W. and Hillier, D. (2006)
Do Precious Metals Shine? An Investment Perspective.
{\em Financial Analysts Journal} 62(2): 98-106.

\bibitem[Dreyfus, 1990]{Dreyfus1990} Dreyfus, S.E. (1990)
Artificial neural networks, back propagation, and the Kelley-Bryson gradient procedure.
{\em Journal of Guidance, Control, and Dynamics} 13(5): 926-928.

\bibitem[Driessen, Maenhout and Vilkov, 2009]{Driessen2009} Driessen, J., Maenhout, P.J. and Vilkov, G. (2009)
The Price of Correlation Risk: Evidence from Equity Options.
{\em Journal of Finance} 64(3): 1377-1406.

\bibitem[Driessen, Nijman and Simon, 2017]{Driessen2017} Driessen, J., Nijman, T. and Simon, Z. (2017)
The Missing Piece of the Puzzle: Liquidity Premiums in Inflation-Indexed Markets.
{\em Working Paper.} Available online: \url{https://ssrn.com/abstract=3042506}.

\bibitem[Drobetz, 2001]{Drobetz2001} Drobetz, W. (2001)
How to Avoid the Pitfalls in Portfolio Optimization? Putting the Black-Litterman Approach at Work.
{\em Financial Markets and Portfolio Management} 15(1): 59-75.

\bibitem[Drobny, 2006]{Drobny2006} Drobny, S. (2006)
{\em Inside the House of Money: Top Hedge Fund Traders on Profiting in the Global Markets.}
Hoboken, NJ: John Wiley \& Sons, Inc.

\bibitem[Droms and Walker, 2001]{Droms2001} Droms, W.G. and Walker, D.A. (2001)
Performance persistence of international mutual funds.
{\em Global Finance Journal} 12(2): 237-248.

\bibitem[Du, Tepper and Verdelhan, 2018]{Du2018} Du, W., Tepper, A. and Verdelhan, A. (2018)
Deviations from Covered Interest Rate Parity.
{\em Journal of Finance} (forthcoming). DOI: \url{https://doi.org/10.1111/jofi.12620}. Available online: \url{https://ssrn.com/abstract=2768207}.

\bibitem[Duarte, Longstaff and Yu, 2006]{Duarte2006} Duarte, J., Longstaff, F.A. and Yu, F. (2006)
Risk and Return in Fixed-Income Arbitrage: Nickels in Front of a Steamroller?
{\em Review of Financial Studies} 20(3): 769-811.

\bibitem[Dubil, 2011]{Dubil2011} Dubil, R. (2011)
Hedge Funds: Alpha, Beta and Replication Strategies.
{\em Journal of Financial Planning} 24(10): 68-77.

\bibitem[Duca {\em et al}, 2012]{Duca2012} Duca, E., Dutordoir, M., Veld, C. and Verwijmeren, P. (2012)
Why are convertible bond announcements associated with increasingly negative issuer stock returns? An arbitrage based explanation.
{\em Journal of Banking \& Finance} 36(11): 2884-2899.

\bibitem[Duchin, 2010]{Duchin2010} Duchin, R. (2010)
Cash Holdings and Corporate Diversification.
{\em Journal of Finance} 65(3): 955-992.

\bibitem[Dudley, Roush and Steinberg, 2009]{Dudley2009} Dudley, W., Roush, J.E. and Steinberg, M. (2009)
The Case for Tips: An Examination of the Costs and Benefits.
{\em Federal Reserve Bank of New York, Economic Policy Review} 15(1): 1-17.

\bibitem[Duffie, 1996]{Duffie1996} Duffie, D. (1996)
Special repo rates.
{\em Journal of Finance} 51(2): 493-526.

\bibitem[Duffie, 2004]{Duffie2004} Duffie, D. (2004)
Time to adapt copula methods for modelling credit risk correlation.
{\em Risk}, April 2004, p. 77.

\bibitem[Duffie, 2017]{Duffie2017} Duffie, D. (2017)
The covered interest parity conundrum.
{\em Risk}, May 2017.
Available online:
\url{https://www.risk.net/4353726}.

\bibitem[Duffie and G\^{a}rleanu, 2001]{Duffie2001} Duffie, D. and G\^{a}rleanu, N. (2001)
Risk and Valuation of Collateralized Debt Obligations.
{\em Financial Analysts Journal} 57(1): 41-59.

\bibitem[Duffie and Huang, 1996]{DuffieHuang1996} Duffie, D. and Huang, M. (1996)
Swap Rates and Credit Quality.
{\em Journal of Finance} 51(2): 921-949.

\bibitem[Duffie, Saita and Wang, 2007]{Duffie2007} Duffie, D., Saita, L. and Wang, K. (2007)
Multi-period corporate default prediction with stochastic covariates.
{\em Journal of Financial Economics} 83(3): 635-665.

\bibitem[Duffie and Singleton, 1997a]{Duffie1997a} Duffie, D. and Singleton, K.J. (1997a)
Modeling term structures of defaultable bonds.
{\em Review of Financial Studies} 12(4): 687-720.

\bibitem[Duffie and Singleton, 1997b]{Duffie1997b} Duffie, D. and Singleton, K.J. (1997b)
An Econometric Model of the Term Structure of Interest Rate Swap Yields.
{\em Journal of Finance} 52(4): 1287-1321.

\bibitem[DuJardin, 2015]{DuJardin2015} DuJardin, P. (2015)
Bankruptcy prediction using terminal failure processes.
{\em European Journal of Operational Research} 242(1): 286-303.

\bibitem[Dukes, Frolich and Ma, 1992]{Dukes1992} Dukes, W.P., Frolich, C.J. and Ma, C.K. (1992)
Risk arbitrage in tender offers.
{\em Journal of Portfolio Management} 18(4): 47-55.

\bibitem[Dumas, Fleming and Whaley, 1998]{Dumas1998} Dumas, B., Fleming, J. and Whaley, R. (1998)
Implied Volatility Functions: Empirical Tests.
{\em Journal of Finance} 53(6): 2059-2106.

\bibitem[Dunis, Laws and Evans, 2006]{Dunis2006} Dunis, C., Laws, J. and Evans, B. (2006)
Trading futures spreads.
{\em Applied Financial Economics} 16(12): 903-914.

\bibitem[Dunis, Laws and Evans, 2010]{Dunis2010} Dunis, C., Laws, J. and Evans, B. (2010)
Trading and filtering futures spread portfolios.
{\em Journal of Derivatives \& Hedge Funds} 15(4): 274-287.

\bibitem[Dunis, Laws and Rudy, 2013]{Dunis2013} Dunis, C., Laws, J. and Rudy, J. (2013)
Mean Reversion Based on Autocorrelation: A Comparison Using the S\&P 100 Constituent Stocks and the 100 Most Liquid ETFs.
{\em ETF Risk}, October 2013, pp. 36-41.

\bibitem[Dunn and McConnell, 1981a]{Dunn1981a} Dunn, K.B. and McConnell, J.J. (1981a)
A Comparison of Alternative Models for Pricing GNMA Mortgage-Backed Securities.
{\em Journal of Finance} 36(2): 471-484.

\bibitem[Dunn and McConnell, 1981b]{Dunn1981b} Dunn, K.B. and McConnell, J.J. (1981b)
Valuation of GNMA Mortgage-Backed Securities.
{\em Journal of Finance} 36(3): 599-616.

\bibitem[Dupire, 1994]{Dupire1994} Dupire, B. (1994)
Pricing with a smile.
{\em Risk} 7(1): 18-20.

\bibitem[Dusak, 1973]{Dusak1973} Dusak, K. (1973)
Futures Trading and Investor Returns: An Investigation of Commodity Market Risk Premiums.
{\em Journal of Political Economy} 81(6): 1387-1406.

\bibitem[Dutordoir {\em et al}, 2014]{Dutordoir2014} Dutordoir, M., Lewis, C.M., Seward, J. and Veld, C. (2014)
What we do and do not know about convertible bond financing.
{\em Journal of Corporate Finance} 24: 3-20.

\bibitem[Dutt {\em et al}, 1997]{Dutt1997} Dutt, H.R., Fenton, J., Smith, J.D. and Wang, G.H.K. (1997)
Crop year influences and variability of the agricultural futures spreads.
{\em Journal of Futures Markets} 17(3): 341-367.

\bibitem[Dwyer, Locke and Yu, 1996]{Dwyer1996} Dwyer, G.P., Jr., Locke, P. and Yu, W. (1996)
Index Arbitrage and Nonlinear dynamics Between the S\&P 500 Futures and Cash.
{\em Review of Financial Studies} 9(1): 301-332.

\bibitem[Dyhrberg, 2015]{Dyhrberg2015}
Dyhrberg, A.H. (2015)
Bitcoin, gold and the dollar -- a GARCH volatility analysis.
{\em Finance Research Letters} 16: 85-92.

\bibitem[Dyhrberg, 2016]{Dyhrberg2016} Dyhrberg, A.H. (2016)
Hedging capabilities of bitcoin. Is it the virtual gold?
{\em Finance Research Letters} 16: 139-144.

\bibitem[Dyl and Joehnk, 1981]{Dyl1981} Dyl, E.A. and Joehnk, M.D. (1981)
Riding the Yield Curve: Does it Work?
{\em Journal of Portfolio Management} 7(3): 13-17.

\bibitem[Dyl and Martin, 1986]{Dyl1986} Dyl, E.A. and Martin, S.A. (1986)
Another Look at Barbells Versus Ladders.
{\em Journal of Portfolio Management} 12(3): 54-59.

\bibitem[Dynkin {\em et al}, 2001]{Dynkin2001} Dynkin, L., Hyman, J., Konstantinovsky, V. and Roth, N. (2001)
Building an MBS Index: Conventions and Calculations.
In: Fabozzi, F.J. (ed.) {\em The Handbook of Mortgage-Backed Securities.} (5th ed.) New York, NY: McGraw-Hill, Inc.

\bibitem[Dzikevi\v{c}ius and \v{S}aranda, 2010]{Dzikevicius2010} Dzikevi\v{c}ius, A. and \v{S}anranda, S. (2010)
EMA versus SMA: Usage to forecast Stock Markets: The Case of S\&P 500 and OMX Baltic Benchmark.
{\em Verslas: teorija ir praktika -- Business: theory and practice} 11(3): 248-255.

\bibitem[Easley, L\'{o}pez de Prado and O'Hara, 2011]{Easley2011} Easley, D., L\'{o}pez de Prado, M.M. and O'Hara, M. (2011)
The microstructure of the `flash crash': flow toxicity, liquidity crashes and the probability of informed trading.
{\em Journal of Portfolio Management} 37(2): 118-128.

\bibitem[Easley, L\'{o}pez de Prado and O'Hara, 2012]{Easley2012} Easley, D., L\'{o}pez de Prado, M.M. and O'Hara, M. (2012)
The volume clock: Insights into the high frequency paradigm.
{\em Journal of Portfolio Management} 39(1): 19-29.

\bibitem[Eastman and Lucey, 2008]{Eastman2008} Eastman, A.M. and Lucey, B.M. (2008)
Skewness and asymmetry in futures returns and volumes.
{\em Applied Financial Economics} 18(10): 777-800.

\bibitem[Eberhart, Altman and Aggarwal, 1999]{Eberhart1999} Eberhart, A., Altman, E. and Aggarwal, R. (1999)
The Equity Performance of Firms Emerging from Bankruptcy.
{\em Journal of Finance} 54(5): 1855-1868.

\bibitem[Eberhart and Sweeney, 1992]{Eberhart1992} Eberhart, A.C. and Sweeney, R.J. (1992)
Does the Bond Market Predict Bankruptcy Settlements?
{\em Journal of Finance} 47(3): 943-980.

\bibitem[Ebrahim and Rahman, 2005]{Ebrahim2005} Ebrahim, S. and Rahman, S. (2005)
On the pareto-optimality of futures contracts over Islamic forward contracts: Implications for the emerging Muslim economies.
{\em Journal of Economic Behavior \& Organization} 56(2): 273-295.

\bibitem[Ederington, 1979]{Ederington1979} Ederington, L.H. (1979)
The hedging performance of the new futures markets.
{\em Journal of Finance} 34(1): 157-170.

\bibitem[Edwards, 2009]{Edwards2009} Edwards, D.W. (2009)
{\em Energy Trading \& Investing: Trading, Risk Management and Structuring Deals in the Energy Market.}
New York, NY: McGraw-Hill, Inc.

\bibitem[Edwards, 1988]{Edwards1988} Edwards, F.R. (1988)
Futures Trading and Cash Market Volatility: Stock Index and Interest Rate Futures.
{\em Journal of Futures Markets} 8(4): 421-439.

\bibitem[Edwards and Magee, 1992]{Edwards1992} Edwards, R. and Magee, J. (1992)
{\em Technical Analysis of Stock Trends.}
New York, NY: New York Institute of Finance.

\bibitem[Edwards and Park, 1996]{Edwards1996} Edwards, F.R. and Park, J.M. (1996)
Do Managed Futures Make Good Investments?
{\em Journal of Futures Markets} 16(5): 475-517.

\bibitem[Edwards and Susmel, 2003]{Edwards2003} Edwards, S. and Susmel, R. (2003)
Interest-Rate Volatility in Emerging Markets.
{\em Review of Economics and Statistics} 85(2): 328-348.

\bibitem[Egginton, Van Ness and Van Ness, 2016]{Egginton2016} Egginton, J.F., Van Ness, B.F. and Van Ness, R.A. (2016)
Quote Stuffing.
{\em Financial Management} 45(3): 583-608.

\bibitem[Ehlgen, 1998]{Ehlgen1998} Ehlgen, J. (1998)
Distortionary effects of the optimal Hodrick-Prescott filter.
{\em Economics Letters} 61(3): 345-349.

\bibitem[Eichenbaum and Evans, 1995]{Eichenbaum1995} Eichenbaum, M. and Evans, C.L. (1995)
Some Empirical Evidence on the Effects of Shocks to Monetary Policy on Exchange Rates.
{\em Quarterly Journal of Economics} 110(4): 975-1009.

\bibitem[Eichholtz {\em et al}, 1995]{Eichholtz1995} Eichholtz, P.M.A., Hoesli, M., MacGregor, B.D. and Nanthakumaran, N. (1995)
Real estate portfolio diversification by property type and region.
{\em Journal of Property Finance} 6(3): 39-59.

\bibitem[Eisdorfer and Misirli, 2015]{Eisdorfer2015} Eisdorfer, A. and Misirli, E.  (2015)
Distressed Stocks in Distressed Times.
{\em Working Paper.} Available online: \url{https://ssrn.com/abstract=2697771}.

\bibitem[Eisl, Gasser and Weinmayer, 2015]{Eisl2015} Eisl, A., Gasser, S. and Weinmayer, K.  (2015)
Caveat Emptor: Does Bitcoin Improve Portfolio Diversification?
{\em Working Paper.} Available online: \url{https://ssrn.com/abstract=2408997}.

\bibitem[Elder, 2014]{Elder2014} Elder, A. (2014)
{\em The New Trading for a Living.}
Hoboken, NJ: John Wiley \& Sons, Inc.

\bibitem[Eldred, 2004]{Eldred2004} Eldred, G.W. (2004)
{\em The Beginner's Guide to Real Estate Investing.}
Hoboken, NJ: John Wiley \& Sons, Inc.

\bibitem[Elias, Wahab and Fang, 2016]{Elias2016} Elias, R.S., Wahab, M.I.M. and Fang, L. (2016)
The spark spread and clean spark spread option based valuation of a power plant with multiple turbines.
{\em Energy Economics} 59: 314-327.

\bibitem[El Kalak and Hudson, 2016]{ElKalak2016} El Kalak, I. and Hudson, R. (2016)
The effect of size on the failure probabilities of SMEs: An empirical study on the US market using discrete hazard model.
{\em International Review of Financial Analysis} 43: 135-145.

\bibitem[Elliott, Siu and Chan, 2007]{Elliott2007} Elliott, R., Siu, T. and Chan, L. (2007)
Pricing volatility swaps under Heston's stochastic volatility model with regime switching.
{\em Applied Mathematical Finance} 14(1): 41-62.

\bibitem[Elliott, Van Der Hoek and Malcolm, 2005]{Elliott2005} Elliott, R.J., van der Hoek, J. and Malcolm, W.P. (2005)
Pairs trading.
{\em Quantitative Finance} 5(3): 271-276.

\bibitem[Elton, Gruber and Blake, 1996a]{Elton1996} Elton, E.J., Gruber, M.J. and Blake, C.R. (1996a)
The Persistence of Risk-Adjusted Mutual Fund Performance.
{\em Journal of Business} 69(2): 133-157.

\bibitem[Elton, Gruber and Blake, 1996b]{Elton1996b} Elton, E.J., Gruber, M.J. and Blake, C.R. (1996b)
Survivor Bias and Mutual Fund Performance.
{\em Review of Financial Studies} 9(4): 1097-1120.

\bibitem[Elton, Gruber and Rentzler, 1987]{Elton1987} Elton, E.J., Gruber, M.J. and Rentzler, J.C. (1987)
Professionally Managed, Publicly Traded Commodity Funds.
{\em Journal of Business} 60(2): 175-199.

\bibitem[Emery and Liu, 2002]{Emery2002} Emery, G.W. and Liu, Q. (2002)
An analysis of the relationship between electricity and natural gas futures prices.
{\em Journal of Futures Markets} 22(2): 95-122.

\bibitem[Engel, 1996]{Engel1996} Engel, C. (1996)
The Forward Discount Anomaly and the Risk Premium: A Survey of Recent Evidence.
{\em Journal of Empirical Finance} 3(2): 123-192.

\bibitem[Engle and Granger, 1987]{Engle1987} Engle, R.F. and Granger, C.W.J. (1987)
Co-integration and error correction: Representation, estimation and testing.
{\em Econometrica} 55(2): 251-276.

\bibitem[Engle and Rosenberg, 2000]{Engle2000} Engle, R. and Rosenberg, J. (2000)
Testing the volatility term structure using option hedging criteria.
{\em Journal of Derivatives} 8(1): 10-28.

\bibitem[Engle and Watson, 1987]{EngleW1987} Engle, R.F. and Watson, M.W. (1987)
The Kalman Filter: applications to forecasting and rational-expectation models.
In: Bewley, T.F. (ed.) {\em Fifth World Conference: Advances in Econometrics}, Vol. 1. Cambridge, UK: Cambridge University Press.

\bibitem[Eraker, 2009]{Eraker2009} Eraker, B. (2009)
The Volatility Premium.
{\em Working Paper.} Available online: \url{http://www.nccr-finrisk.uzh.ch/media/pdf/Eraker_23-10.pdf}.

\bibitem[Eraker and Wu, 2014]{Eraker2014}Eraker, B. and Wu, Y. (2014)
Explaining the Negative Returns to VIX Futures and ETNs: An Equilibrium Approach.
 {\em Working Paper.} Available online: \url{https://ssrn.com/abstract=2340070}.

\bibitem[Erb and Harvey, 2006] {Erb2006} Erb, C. and Harvey, C. (2006)
The Strategic and Tactical Value of Commodity Futures.
{\em Financial Analysts Journal} 62(2): 69-97.

\bibitem[Erickson, Goolsbee and Maydew, 2003] {Erickson2003} Erickson, M., Goolsbee, A. and Maydew, E. (2003)
How Prevalent is Tax Arbitrage? Evidence from the Market for Municipal Bonds.
{\em National Tax Journal} 56(1): 259-270.

\bibitem[Ertugrul and Giambona, 2011]{Ertugrul2011} Ertugrul, M. and Giambona, E. (2011)
Property Segment and REIT Capital Structure.
{\em Journal of Real Estate Finance and Economics} 43(4): 505-526.

\bibitem[Espinoza and Luccioni, 2002]{Espinoza2002} Espinoza, R.D. and Luccioni, L. (2002)
Proper Risk Management: The Key To Successful Brownfield Development.
{\em WIT Transactions on Ecology and the Environment} 55: 297-306.

\bibitem[Eun and Sabherwal, 2003]{Eun2003} Eun, C.S. and Sabherwal, S. (2003)
Cross-border listings and price discovery: Evidence from U.S. listed Canadian stocks.
{\em Journal of Finance} 58(2): 549-575.

\bibitem[Evans, 1998]{Evans1998} Evans, M.D.D. (1998)
Real Rates, Expected Inflation, and Inflation Risk Premia.
{\em Journal of Finance} 53(1): 187-218.

\bibitem[Evans and Marshall, 2007] {Evans2007} Evans, C.L. and Marshall, D.A. (2007)
Economic determinants of the nominal treasury yield curve.
{\em Journal of Monetary Economics} 54(7): 1986-2003.

\bibitem[Faber, 2007]{Faber2007} Faber, M. (2007)
A Quantitative Approach to Tactical Asset Allocation.
{\em Journal of Wealth Management} 9(4): 69-79.

\bibitem[Faber, 2015]{Faber2015} Faber, M. (2015)
Learning to Play Offense and Defense: Combining Value and Momentum from the Bottom Up, and the Top Down.
{\em Working Paper.} Available online: \url{https://ssrn.com/abstract=2669202}.

\bibitem[Faber, 2016]{Faber2016} Faber, M. (2016)
The Trinity Portfolio: A Long-Term Investing Framework Engineered for Simplicity, Safety, and Outperformance.
{\em Working Paper.} Available online: \url{https://ssrn.com/abstract=2801856}.

\bibitem[Fabozzi, 2002]{Fabozzi2002} Fabozzi, F.J. (ed.) (2002)
{\em The Handbook of Financial Instruments.}
Hoboken, NJ: John Wiley \& Sons, Inc.

\bibitem[Fabozzi, 2006a]{Fabozzi2006a} Fabozzi, F.J. (2006a)
{\em Fixed Income Mathematics: Analytical \& Statistical Techniques.}
New York, NY: McGraw-Hill, Inc.

\bibitem[Fabozzi, 2006b]{Fabozzi2006b} Fabozzi, F.J. (ed.) (2006b)
{\em The Handbook of Mortgage-Backed Securities.}
New York, NY: McGraw-Hill, Inc.

\bibitem[Fabozzi, 2012a]{Fabozzi2012a} Fabozzi, F.J. (2012a)
{\em Bond markets, analysis, and strategies.}
Upper Saddle River, NJ: Prentice Hall.

\bibitem[Fabozzi, 2012b]{Fabozzi2012b} Fabozzi, F.J. (2012b)
{\em Institutional Investment Management: Equity and Bond Portfolio Strategies and Applications.}
Hoboken, NJ: John Wiley \& Sons, Inc.

\bibitem[Fabozzi, Focardi and Jonas, 2010]{FabozziFocardi2010} Fabozzi, F.J., Focardi, S.M. and Jonas, C. (2010)
{\em Investment Management after the Global Financial Crisis.}
Charlottesville, VA: The Research Foundation of CFA Institute.

\bibitem[Fabozzi and Mann, 2010]{Fabozzi2010} Fabozzi, F.J. and Mann, S.V. (2010)
{\em Introduction to Fixed Income Analytics: Relative Value Analysis, Risk Measures, and Valuation.}
Hoboken, NJ: John Wiley \& Sons, Inc.

\bibitem[Fabozzi, Martellini and Priaulet, 2006]{FabbozziMP2006} Fabozzi, F.J., Martellini, L. and Priaulet, P. (2006)
{\em Advanced Bond Portfolio Management. Best Practices in Modeling and Strategies.}
Hoboken, NJ: John Wiley \& Sons, Inc.

\bibitem[Falkenstein and Hanweck, 1996]{Falkenstein1996} Falkenstein, E. and Hanweck, J. (1996)
Minimizing Basis Risk from Non-Parallel Shifts in the Yield Curve.
{\em Journal of Fixed Income} 6(1): 60-68.

\bibitem[Fama, 1984]{Fama1984} Fama, E.F. (1984)
Forward and spot exchange rates.
{\em Journal of Monetary Economics} 14(3): 319-338.

\bibitem[Fama, 1996]{FamaEF1996} Fama, E.F. (1996)
Multifactor Portfolio Efficiency and Multifactor Asset Pricing.
{\em Journal of Financial and Quantitative Analysis} 31(4): 441-465.

\bibitem[Fama and French, 1987]{Fama1987} Fama, E.F. and French, K.R. (1987)
 Commodity futures prices: some evidence on forecast power, premiums, and the theory of storage.
{\em Journal of Business} 60(1): 55-73.

\bibitem[Fama and French, 1988]{Fama1988} Fama, E.F. and French, K.R. (1988)
Business Cycles and the Behavior of Metals Prices.
{\em Journal of Finance} 43(5): 1075-1093.

\bibitem[Fama and French, 1992]{Fama1992} Fama, E.F. and French, K.R. (1992)
The Cross-Section of Expected Stock Returns.
{\em Journal of Finance} 47(2): 427-465.

\bibitem[Fama and French, 1993]{Fama1993} Fama, E.F. and French, K.R. (1993)
Common Risk Factors in the Returns on Stocks and Bonds.
{\em Journal of Financial Economics} 33(1): 3-56.

\bibitem[Fama and French, 1996]{Fama1996} Fama, E.F. and French, K.R. (1996)
Multifactor Explanations of Asset Pricing Anomalies.
{\em Journal of Finance} 51(1): 55-84.

\bibitem[Fama and French, 1998]{Fama1998} Fama, E.F. and French, K.R. (1998)
Value versus Growth: The International Evidence.
{\em Journal of Finance} 53(6): 1975-1999.

\bibitem[Fama and French, 2008]{Fama2008} Fama, E.F. and French, K.R. (2008)
Dissecting Anomalies.
{\em Journal of Finance} 63(4): 1653-1678.

\bibitem[Fama and French, 2012]{Fama2012} Fama, E.F. and French, K.R. (2012)
Size, Value and Momentum in International Stock Returns.
{\em Journal of Financial Economics} 105(3): 457-472.

\bibitem[Fama and Schwert, 1977]{Fama1977} Fama, E.F. and Schwert, G.W. (1977)
Asset returns and inflation.
{\em Journal of Financial Economics} 5(2): 115-146.

\bibitem[Fass and Francis, 2004]{Fass2004} Fass, S.M. and Francis, J. (2004)
Where have all the hot goods gone? The role of pawnshops.
{\em Journal of Research in Crime and Delinquency} 41(2): 156-179.

\bibitem[Fassas, 2011]{Fassas2011} Fassas, A.P. (2011)
Mispricing in stock index futures markets -- the case of Greece.
{\em Investment Management and Financial Innovations} 8(2): 101-107.

\bibitem[Fedorova, Gilenko and Dovzhenko, 2013]{Fedorova2013} Fedorova, E., Gilenko, E. and Dovzhenko, S. (2013)
Bankruptcy prediction for Russian companies: Application of combined classifiers.
{\em Expert Systems with Applications} 40(18): 7285-7293.

\bibitem[Feldh\"utter and Lando, 2008]{Feldhutter2008} Feldh\"utter, P. and Lando, D. (2008)
Decomposing swap spreads.
{\em Journal of Financial Economics} 88(2): 375-405.

\bibitem[Feldman, 2003]{Feldman2003} Feldman, B.E. (2003)
Investment Policy for Securitized and Direct Real Estate.
{\em Journal of Portfolio Management} 29(5): 112-121.

\bibitem[Feldman and Roy, 2004]{Feldman2004} Feldman, B. and Roy, D. (2004)
Passive Options-based Investment Strategies: The Case of the CBOE S\&P 500 BuyWrite Index.
{\em ETF and Indexing} 38(1): 72-89.

\bibitem[Feldman and Till, 2006]{Feldman2006} Feldman, B. and Till, H. (2006)
Backwardation and Commodity Futures Performance: Evidence from Evolving Agricultural Markets.
{\em Journal of Alternative Investments} 9(3): 24-39.

\bibitem[F\'{e}lix and Rodr\'{\i}guez, 2008]{Felix2008} F\'{e}lix, J.A. and Rodr\'{\i}guez, F.F. (2008)
Improving moving average trading rules with boosting and statistical learning methods.
{\em Journal of Forecasting} 27(5): 433-449.

\bibitem[Fengler, Herwartz and Werner, 2012]{Fengler2012} Fengler, M.R., Herwartz, H. and Werner, C. (2012)
A Dynamic Copula Approach to Recovering the Index Implied Volatility Skew.
{\em Journal of Financial Econometrics} 10(3): 457-493.

\bibitem[Fenn {\em et al}, 2009]{Fenn2009} Fenn, D.J., Howison, S.D., Mcdonald, M., Williams, S. and Johnson, N.F. (2009)
The mirage of triangular arbitrage in the spot foreign exchange market.
{\em International Journal of Theoretical and Applied Finance} 12(8): 1105-1123.

\bibitem[Fernandez-Perez {\em et al}, 2018]{Fernandez-Perez2018} Fernandez-Perez, A., Frijns, B., Fuertes, A.M. and Miffre, J. (2018)
The skewness of commodity futures returns.
{\em Journal of Banking \& Finance} 86: 143-158.

\bibitem[Fernandez-Perez, Fuertes and Miffre, 2016]{Fernandez-Perez2016} Fernandez-Perez, A., Fuertes, A.M. and Miffre, J. (2016)
Is idiosyncratic volatility priced in commodity futures markets?
{\em International Review of Financial Analysis} 46: 219-226.

\bibitem[Ferreira, Grammatikos and Michala, 2016]{Ferreira2016} Ferreira, S., Grammatikos, T. and Michala, D. (2016)
Forecasting distress in Europe SME portfolios.
{\em Journal of Banking \& Finance} 64: 112-135.

\bibitem[Ferson and Mo, 2016]{Ferson2016} Ferson, W. and Mo, H. (2016)
Performance measurement with selectivity, market and volatility timing.
{\em Journal of Financial Economics} 121(1): 93-110.

\bibitem[Fifield, Power and Knipe, 2008]{Fifield2008} Fifield, S.G.M., Power, D.M. and Knipe, D.G.S. (2008)
The performance of moving average rules in emerging stock markets.
{\em Applied Financial Economics} 18(19): 1515-1532.

\bibitem[Figlewski, Chidambaran and Kaplan, 1993]{Figlewski1993} Figlewski, S., Chidambaran, N.K. and Kaplan, S. (1993)
Evaluating the Performance of the Protective Put Strategy.
{\em Financial Analysts Journal} 49(4): 46-56, 69.

\bibitem[Filipovi\'{c}, Gourier and Mancini, 2016]{Filipovic2016} Filipovi\'{c}, D., Gourier, E. and Mancini, L. (2016)
Quadratic variance swap models.
{\em Journal of Financial Economics} 119(1): 44-68.

\bibitem[Finger, 1999]{Finger1999} Finger, C.C. (1999)
Conditional approaches for credit metrics portfolio distributions.
{\em Credit Metrics Monitor} 2(1): 14-33.

\bibitem[Finkenzeller, Dechant and Sch\"{a}fers, 2010]{Finkenzeller2010} Finkenzeller, K., Dechant, T. and Sch\"{a}fers, W. (2010)
Infrastructure: a new dimension of real estate? An asset allocation analysis.
{\em Journal of Property Investment \& Finance} 28(4): 263-274.

\bibitem[Finnerty and Tu, 2017]{Finnerty2017} Finnerty, J.D. and Tu, M. (2017)
Valuing Convertible Bonds: A New Approach.
{\em Business Valuation Review} 36(3): 85-102.

\bibitem[Fiorenzani, 2006]{Fiorenzani2006} Fiorenzani, S. (2006)
{\em Quantitative Methods for Electricity Trading and Risk Management: Advanced Mathematical and Statistical Methods for Energy Finance.}
London, UK: Palgrave Macmillan.

\bibitem[Firstenberg, Ross and Zisler, 1988]{Firstenberg1988} Firstenberg, P.M., Ross, S.A. and Zisler, R.C. (1988)
Real estate: The whole story.
{\em Journal of Portfolio Management} 14(3): 22-34.

\bibitem[Fishe, Janzen and Smith, 2014]{Fishe2014} Fishe, R.P.H., Janzen, J.P. and Smith, A. (2014)
Hedging and Speculative Trading in Agricultural Futures Markets.
{\em American Journal of Agricultural Economics} 96(2): 542-556.

\bibitem[Fisher, 2002]{Fisher2002} Fisher, M. (2002)
Special Repo Rates: An Introduction.
{\em Federal Reserve Bank of Atlanta, Economic Review} 87(2): 27-43.

\bibitem[Fisher, Shah and Titman, 2016]{Fisher2016} Fisher, G., Shah, R. and Titman, S. (2016)
Combining Value and Momentum.
{\em Journal of Investment Management} 14(2): 33-48.

\bibitem[Fisher and Weil, 1971]{Fisher1971} Fisher, L. and Weil, R.L. (1971)
Coping with the Risk of Interest-Rate Fluctuations: Returns to Bondholders from Na\"{\i}ve and Optimal Strategies.
{\em Journal of Business} 44(4): 408-431.

\bibitem[Fleckenstein, 2012]{Fleckenstein2012} Fleckenstein, M. (2012)
The Inflation-Indexed Bond Puzzle.
{\em Working Paper.} Available online: \url{https://ssrn.com/abstract=2180251}.

\bibitem[Fleckenstein, Longstaff and Lustig, 2013]{Fleckenstein2013} Fleckenstein, M., Longstaff, F.A. and Lustig, H.N. (2013)
Why Does the Treasury Issue TIPS? The TIPS-Treasury Bond Puzzle.
{\em Journal of Finance} 69(5): 2151-2197.

\bibitem[Fleckenstein, Longstaff and Lustig, 2017]{Fleckenstein2017} Fleckenstein, M., Longstaff, F.A. and Lustig, H.N. (2017)
Deflation Risk.
{\em Review of Financial Studies} 30(8): 2719-2760.

\bibitem[Fleming and Krishnan, 2012]{Fleming2012} Fleming, M.J. and Krishnan, N. (2012)
The Microstructure of the TIPS Market.
{\em Federal Reserve Bank of New York, Economic Policy Review} 18(1): 27-45.

\bibitem[Fleming, Ostdiek and Whaley, 1995]{Fleming1995} Fleming, J., Ostdiek, B. and Whaley, R.E. (1995)
Predicting stock market volatility: A new measure.
{\em Journal of Futures Markets} 15(3): 265-302.

\bibitem[Fleming and Sporn, 2013]{Fleming2013} Fleming, M.J. and Sporn, J.R. (2013)
Trading Activity and Price Transparency in the Inflation Swap Market.
{\em Federal Reserve Bank of New York, Economic Policy Review} 19(1): 45-58.

\bibitem[Flint and Mar\'{e}, 2017]{Flint2017} Flint, E. and Mar\'{e}, E. (2017)
Fractional Black-Scholes option pricing, volatility calibration and implied Hurst exponents in South African context.
{\em South African Journal of Economic and Management Sciences} 20(1): a1532.

\bibitem[Fong and Vasicek, 1983]{Fong1983} Fong, H.G. and Vasicek, O.A. (1983).
The tradeoff between return and risk in immunized portfolios.
{\em Financial Analysts Journal} 39(5): 73-78.

\bibitem[Fong and Vasicek, 1984]{Fong1984} Fong, H.G. and Vasicek, O.A. (1984)
A Risk Minimizing Strategy for Portfolio Immunization.
{\em Journal of Finance} 39(5): 1541-1546.

\bibitem[Fong and Yong, 2005]{Fong2005} Fong, W.M. and Yong, L.H.M. (2005)
Chasing trends: recursive moving average trading rules and internet stocks.
{\em Journal of Empirical Finance} 12(1): 43-76.

\bibitem[Fontaine and Nolin, 2017]{Fontaine2017} Fontaine, J.-F. and Nolin, G. (2017)
Measuring Limits of Arbitrage in Fixed-Income Markets.
{\em Staff Working Paper}, No. 2017-44. Ottawa, Canada: Bank of Canada.

\bibitem[Fontana, 2010]{Fontana2010} Fontana, A. (2010)
The Persistent Negative CDS-Bond Basis during the 2007/08 Financial Crisis.
{\em Working Paper.} Available online:\\
 \url{http://www.unive.it/media/allegato/DIP/Economia/Working_papers/Working_papers_2010/WP_DSE_fontana_13_10.pdf}.

\bibitem[Fontana and Scheicher, 2016]{Fontana2016} Fontana, A. and Scheicher, M. (2016)
An analysis of euro area sovereign CDS and their relation with government bonds.
{\em Journal of Banking \& Finance} 62: 126-140.

\bibitem[Fortin and Khoury, 1984]{Fortin1984} Fortin, M. and Khoury, N. (1984)
Hedging Interest Rate Risks with Financial Futures.
{\em Canadian Journal of Administrative Sciences} 1(2): 367-382.

\bibitem[Foster, Olsen and Shevlin, 1984]{Foster1984} Foster, G., Olsen, C. and Shevlin, T. (1984)
Earnings releases, anomalies, and the behavior of security returns.
{\em Accounting Review} 59(4): 574-603.

\bibitem[Foster and Whiteman, 2002]{Foster2002} Foster, F.D. and Whiteman, C.H. (2002)
Bayesian Cross Hedging: An Example from the Soybean Market.
{\em Australian Journal of Management} 27(2): 95-122.

\bibitem[Frachot, 1996]{Frachot1996} Frachot, A. (1996)
A reexamination of the uncovered interest rate parity hypothesis.
{\em Journal of International Money and Finance} 15(3): 419-437.

\bibitem[Frankel, 2006]{Frankel2006} Frankel, J.A. (2006)
The Effect of Monetary Policy on Real Commodity Prices.
In: Campbell, J. (ed.) {\em Asset Prices and Monetary Policy.} Chicago, IL: University of Chicago Press, pp. 291-333.

\bibitem[Franken and Parcell, 2003]{Franken2003} Franken, J.R.V. and Parcell, J.L. (2003)
Cash Ethanol Cross-Hedging Opportunities.
{\em Journal of Agricultural and Applied Economics} 35(3): 509-516.

\bibitem[Frazzini and Pedersen, 2014]{Frazzini2014} Frazzini, A. and Pedersen, L.H. (2014)
Betting against Beta.
{\em Journal of Financial Economics} 111(1): 1-25.

\bibitem[Frenkel and Levich, 1975]{Frenkel1975} Frenkel, J.A. and Levich, R.M. (1975)
Covered interest arbitrage: Unexploited profits?
{\em Journal of Political Economy} 83(2): 325-338.

\bibitem[Frenkel and Levich, 1981]{Frenkel1981} Frenkel, J.A. and Levich, R.M. (1981)
Covered interest arbitrage in the 1970's.
{\em Economics Letters} 8(3): 267-274.

\bibitem[Frey and Backhaus, 2008]{Frey2008} Frey, R. and Backhaus, J. (2008)
Pricing and Hedging of Portfolio Credit Derivatives with Interacting Default Intensities.
{\em International Journal of Theoretical and Applied Finance} 11(6): 611-634.

\bibitem[Frey and Backhaus, 2010]{Frey2010} Frey, R. and Backhaus, J. (2010)
Dynamic hedging of synthetic CDO tranches with spread risk and default contagion.
{\em Journal of Economic Dynamics and Control} 34(4): 710-724.

\bibitem[Frey, McNeil and Nyfeler, 2001]{Frey2001} Frey, R., McNeil, A. and Nyfeler, N. (2001)
Copulas and Credit Models.
{\em Risk}, October 2001, pp. 111-114.

\bibitem[Fridson and Xu, 2014]{Fridson2014} Fridson, M.S. and Xu, X. (2014)
Duration Targeting: No Magic for High-Yield Investors.
{\em Financial Analysts Journal} 70(3): 28-33.

\bibitem[Friewald, Jankowitsch and Subrahmanyam, 2012]{Friewald2012} Friewald, N., Jankowitsch, R. and Subrahmanyam, M. (2012)
Illiquidity, or Credit Deterioration: A Study of Liquidity in the U.S. Bond Market during Financial Crises.
{\em Journal of Financial Economics} 105(1): 18-36.

\bibitem[Frino {\em et al}, 2004]{Frino2004} Frino, A., Gallagher, D.R., Neubert, A.S. and Oetomo, T.N. (2004)
Index Design and Implications for Index Tracking.
{\em Journal of Portfolio Management} 30(2): 89-95.

\bibitem[Frino and McKenzie, 2002]{Frino2002} Frino, A. and McKenzie, M. (2002)
The pricing of stock index futures spreads at contract expiration.
{\em Journal of Futures Markets} 22(5): 451-469.

\bibitem[Froot, Scharfstein and Stein, 1993]{Froot1993} Froot, K.A., Scharfstein, D.S. and Stein, J.C. (1993)
Risk Management: Coordinating Corporate Investment and Financing Policies.
{\em Journal of Finance} 48(5): 1629-1658.

\bibitem[Froot and Thaler, 1990]{Froot1990} Froot, K.A. and Thaler, R.H. (1990)
Anomalies: Foreign Exchange.
{\em Journal of Economic Perspectives} 4(3): 179-192.

\bibitem[Fry and Cheah, 2016]{Fry2016} Fry, J. and Cheah, E.T. (2016)
Negative bubbles and shocks in cryptocurrency markets.
{\em International Review of Financial Analysis} 47: 343-352.

\bibitem[Fu, 2009]{Fu2009} Fu, F. (2009)
Idiosyncratic Risk and the Cross-Section of Expected Stock Returns.
{\em Journal of Financial Economics} 91(1): 24-37.

\bibitem[Fu and Qian, 2014]{Fu2014} Fu, Y. and Qian, W. (2014)
Speculators and Price Overreaction in the Housing Market.
{\em Real Estate Economics} 42(4): 977-1007.

\bibitem[Fu, Sandri and Shackleton, 2016]{Fu2016} Fu, X., Sandri, M. and Shackleton, M.B. (2016)
Asymmetric Effects of Volatility Risk on Stock Returns: Evidence from VIX and VIX Futures.
{\em Journal of Futures Markets} 36(11): 1029-1056.

\bibitem[Fuertes, Miffre and Fernandez-Perez, 2015]{Fuertes2015} Fuertes, A., Miffre, J. and Fernandez-Perez, A. (2015)
Commodity Strategies Based on Momentum, Term Structure, and Idiosyncratic Volatility.
{\em Journal of Futures Markets} 35(3): 274-297.

\bibitem[Fuertes, Miffre and Rallis, 2010]{Fuertes2010} Fuertes, A., Miffre, J. and Rallis, G. (2010)
Tactical allocation in commodity futures markets: Combining momentum and term structure signals.
{\em Journal of Banking \& Finance} 34(10): 2530-2548.

\bibitem[Fugazza, Guidolin and Nicodano, 2007]{Fugazza2007} Fugazza, C., Guidolin, M. and Nicodano, G. (2007)
Investing for the Long-run in European Real Estate.
{\em Journal of Real Estate Finance and Economics} 34(1): 35-80.

\bibitem[Fulli-Lemaire, 2013]{Fulli-Lemaire2013} Fulli-Lemaire, N. (2013)
An Inflation Hedging Strategy with Commodities: A Core Driven Global Macro.
{\em Journal of Investment Strategies} 2(3): 23-50.

\bibitem[Fung and Hsieh, 1999]{Fung1999} Fung, W. and Hsieh, D.A. (1999)
A Primer on Hedge Funds.
{\em Journal of Empirical Finance} 6(3): 309-331.

\bibitem[Fung, Mok and Wong, 2004]{Fung2004} Fung, J.K.W., Mok, H.M.K. and Wong, K.C.K. (2004)
Pricing Efficiency in a Thin Market with Competitive Market Makers: Box Spread Strategies in the Hang Seng Index Options Market.
{\em Financial Review} 39(3): 435-454.

\bibitem[Fusaro and James, 2005]{Fusaro2005} Fusaro, P.C. and James, T. (2005)
{\em Energy Hedging in Asia: Market Structure and Trading Opportunities.}
London, UK: Palgrave Macmillan.

\bibitem[F\"uss and Nikitina, 2011]{Fuss2011} F\"uss, R. and Nikitina, O. (2011)
Explaining Yield Curve Dynamics.
{\em Journal of Fixed Income} 21(2): 68-87.

\bibitem[Gabaix, Krishnamurthy and Vigneron, 2007]{Gabaix2007} Gabaix, X., Krishnamurthy, A. and Vigneron, O. (2007)
Limits of arbitrage: theory and evidence from the mortgage-backed securities market.
{\em Journal of Finance} 62(2): 557-595.

\bibitem[Gajardo, Kristjanpoller and Minutolo, 2018]{Gajardo2018} Gajardo, G., Kristjanpoller, W.D. and Minutolo, M. (2018)
Does Bitcoin exhibit the same asymmetric multifractal cross-correlations with crude oil, gold and DJIA as the Euro, Great British Pound and Yen?
{\em Chaos, Solitons \& Fractals} 109: 195-205.

\bibitem[Gande, Altman and Saunders, 2010]{Gande2010} Gande, A., Altman, E. and Saunders, A. (2010)
Bank Debt vs. Bond Debt: Evidence from Secondary Market Prices.
{\em Journal of Money, Credit and Banking} 42(4): 755-767.

\bibitem[Gao and Ren, 2015]{Gao2015} Gao, B. and Ren, R.-E. (2015)
A New Sector Rotation Strategy and its Performance Evaluation: Based on a Principal Component Regression Model.
{\em Working Paper.} Available online: \url{https://ssrn.com/abstract=2628058}.

\bibitem[Gao, Xing and Zhang, 2017]{Gao2017} Gao, C., Xing, Y. and Zhang, X. (2017)
Anticipating Uncertainty: conos Around Earnings Announcements.
{\em Working Paper.} Available online: \url{https://ssrn.com/abstract=2204549}.

\bibitem[Garbade, 2004]{Garbade2004}Garbade, K.D. (2004)
Origins of the Federal Reserve Book-Entry System.
{\em Federal Reserve Bank of New York, Economic Policy Review} 10(3): 33-50.

\bibitem[Garcia and Gould, 1993]{Garcia1993} Garcia, C.B. and Gould, F.J. (1993)
Survivorship Bias.
{\em Journal of Portfolio Management} 19(3): 52-56.

\bibitem[Garcia and Schweitzer, 2015]{Garcia2015} Garcia, D. and Schweitzer, F. (2015)
Social signals and algorithmic trading of Bitcoin.
{\em Royal Society Open Science} 2(9): 150288.

\bibitem[Garcia {\em et al}, 2014]{Garcia2014}
Garcia, D., Tessone, C.J., Mavrodiev, P. and Perony, N. (2014)
The digital traces of bubbles: feedback cycles between socioeconomic signals in the Bitcoin economy.
{\em Journal of The Royal Society Interface} 11(99): 0623.

\bibitem[Garcia-Feij\'{o}o {\em et al}, 2015]{Garcia-Feijoo2015} Garcia-Feij\'{o}o, L., Kochard, L., Sullivan, R.N. and Wang, P. (2015)
Low-Volatility Cycles: The Influence of Valuation and Momentum on Low-Volatility Portfolios.
{\em Financial Analysts Journal} 71(3): 47-60.

\bibitem[Garlappi and Yan, 2011]{Garlappi2011} Garlappi, L. and Yan, H. (2011)
Financial Distress and the Cross-section of Equity Returns.
{\em Journal of Finance} 66(3): 789-822.

\bibitem[G\^{a}rleanu, Pedersen and Poteshman, 2009]{Garleanu2009} G\^{a}rleanu, N., Pedersen, L.H. and Poteshman, A.M. (2009)
Demand-Based Option Pricing.
{\em Review of Financial Studies} 22(10): 4259-4299.

\bibitem[Garvey and Wu, 2009]{Garvey2009} Garvey, R. and Wu, F. (2009)
Intraday time and order execution quality dimensions.
{\em Journal of Financial Markets} 12(2): 203-228.

\bibitem[Garyn-Tal, 2014a]{GarynTal2014a} Garyn-Tal, S. (2014a)
An Investment Strategy in Active ETFs.
{\em Journal of Index Investing} 4(1): 12-22.

\bibitem[Garyn-Tal, 2014b]{GarynTal2014b} Garyn-Tal, S. (2014b)
Explaining and Predicting ETFs Alphas: The $R^2$ Methodology.
{\em Journal of Index Investing} 4(4): 19-32.

\bibitem[Garzarelli {\em et al}, 2014]{Garzarelli2014} Garzarelli, F., Cristelli, M., Pompa, G., Zaccaria, A. and Pietronero, L. (2014)
Memory effects in stock price dynamics: evidences of technical trading.
{\em Scientific Reports} 4: 4487.

\bibitem[Gatev, Goetzmann and Rouwenhorst, 2006]{Gatev2006} Gatev, E., Goetzmann, W.N. and Rouwenhorst, K.G. (2006)
Pairs Trading: Performance of a Relative-Value Arbitrage Rule.
{\em Review of Financial Studies} 19(3): 797-827.

\bibitem[Gatheral and Jacquier, 2014]{Gatheral2014} Gatheral, J. and Jacquier, A. (2014)
Arbitrage-free SVI volatility surfaces.
{\em Quantitative Finance} 14(1): 59-71.

\bibitem[Gatzlaff and Tirtiroglu, 1995]{Gatzlaff1995} Gatzlaff, D.H. and Tirtiroglu, D. (1995)
Real Estate Market Efficiency: Issues and Evidence.
{\em Journal of Real Estate Literature} 3(2): 157-189.

\bibitem[Gay and Kolb, 1983]{GayKolb1983} Gay, G.D. and Kolb, R.W. (1983)
The Management of Interest Rate Risk.
{\em Journal of Portfolio Management} 9(2): 65-70.

\bibitem[Gay, Kolb and Chiang, 1983]{Gay1983} Gay, G.D., Kolb, R.W. and Chiang, R. (1983)
Interest Rate Hedging: An Empirical Test Of Alternative Strategies.
{\em Journal of Financial Research} 6(3): 187-197.

\bibitem[Ge, 2016]{Ge2016} Ge, W. (2016)
A Survey of Three Derivative-Based Methods to Harvest the Volatility Premium in Equity Markets.
{\em Journal of Investing} 25(3): 48-58.

\bibitem[G\'{e}czy, Minton and Schrand, 1997]{Geczy1997} G\'{e}czy, C., Minton, B.A. and Schrand, C. (1997)
Why Firms Use Currency Derivatives.
{\em Journal of Finance} 52(4): 1323-1354.

\bibitem[G\'{e}czy and Samonov, 2016]{Geczy2016} G\'{e}czy, C.C. and Samonov, M. (2016)
Two Centuries of Price-Return Momentum.
{\em Financial Analysts Journal} 72(5): 32-56.

\bibitem[Gehricke and Zhang, 2018]{Gehricke2018} Gehricke, S.A. and Zhang, J.E. (2018)
Modeling VXX.
{\em Journal of Futures Markets} 38(8): 958-976.

\bibitem[Geltner {\em et al}, 2006]{Geltner2006} Geltner, D.M., Miller, N.G., Clayton, J. and Eichholtz, P. (2006)
{\em Commercial Real Estate Analysis and Investments.} (2nd ed.)
Atlanta, GA: OnCourse Learning Publishing.

\bibitem[Geltner, Rodriguez and O'Connor, 1995]{Geltner1995} Geltner, D.M., Rodriguez, J.V. and O'Connor, D. (1995)
The Similar Genetics of Public and Private Real Estate and the Optimal Long-Horizon Portfolio Mix.
{\em Real Estate Finance} 12(3): 13-25.

\bibitem[Geman, 1998]{Geman1998} Geman, H. (1998)
{\em Insurance and Weather Derivatives: From Exotic Options to Exotic Underlyings.}
London, UK: Risk Books.

\bibitem[Geman and Leonardi, 2005]{Geman2005} Geman, H. and Leonardi, M.-P. (2005)
Alternative approaches to weather derivatives pricing.
{\em Managerial Finance} 31(6): 46-72.

\bibitem[Geman and Roncoroni, 2006]{Geman2006} Geman, H. and Roncoroni, A. (2006)
Understanding the fine structure of electricity prices.
{\em Journal of Business} 79(3): 1225-1261.

\bibitem[Gen\c{c}ay, 1996]{Gencay1996} Gen\c{c}ay, R. (1996)
Nonlinear prediction of security returns with moving average rules.
{\em Journal of Forecasting} 15(3): 165-174.

\bibitem[Gen\c{c}ay, 1998]{Gencay1998} Gen\c{c}ay, R. (1998)
The Predictability of securities returns with simple technical rules.
{\em Journal of Empirical Finance} 5(4): 347-359.

\bibitem[Gen\c{c}ay and Stengos, 1998]{GencayStengos1998} Gen\c{c}ay, R. and Stengos, T. (1998)
Moving average rules, volume and the predictability of security returns with feedforward networks.
{\em Journal of Forecasting} 17(5-6): 401-414.

\bibitem[Genesove and Han, 2012]{Genesove2012} Genesove, D. and Han, L. (2012)
Search and Matching in the Housing Market.
{\em Journal of Urban Economics} 72(1): 31-35.

\bibitem[Genesove and Mayer, 1997]{Genesove1997} Genesove, D. and Mayer, C. (1997)
Equity and Time to Sale in the Real Estate Market.
{\em American Economic Review} 87(3): 255-269.

\bibitem[Genesove and Mayer, 2001]{Genesove2001} Genesove, D. and Mayer, C. (2001)
Loss Aversion and Seller Behavior: Evidence From the Housing Market.
{\em Quarterly Journal of Economics} 116(4): 1233-1260.

\bibitem[George and Hwang, 2010]{George2010} George, T.J. and Hwang, C.-Y. (2010)
A resolution of the distress risk and leverage puzzles in the cross section of stock returns.
{\em Journal of Financial Economics} 96(1): 56-79.

\bibitem[Georgoula {\em et al}, 2015]{Georgoula2015} Georgoula, I., Pournarakis, D., Bilanakos, C., Sotiropoulos, D. and Giaglis, G.M. (2015)
Using Time-Series and Sentiment Analysis to Detect the Determinants of Bitcoin Prices.
{\em Working Paper.} Available online: \url{https://ssrn.com/abstract=2607167}.

\bibitem[Gerakos and Linnainmaa, 2012]{Gerakos2012} Gerakos, J. and Linnainmaa, J. (2012)
Decomposing Value.
{\em Working Paper.} Available online: \url{https://ssrn.com/abstract=2083166}.

\bibitem[Gervais and Odean, 2001]{Gervais2001} Gervais, S. and Odean, T. (2001)
Learning to Be Overconfident.
{\em Review of Financial Studies} 14(1): 1-27.

\bibitem[Geske and Pieptea, 1987]{Geske1987} Geske, R.L. and Pieptea, D.R. (1987)
Controlling Interest Rate Risk and Return with Futures.
{\em Review of Futures Markets} 6(1): 64-86.

\bibitem[Gestel {\em et al}, 2001]{Gestel2001}
Gestel, T., Suykens, J.A.K., Baestaend, D.E., Lambrechts, A., Lanckriet, G., Vandaele, B., Moor, B. and Vandewalle, J. (2001)
Financial time series prediction using least squares support vector machines within the evidence framework.
{\em IEEE Transactions on Neural Networks} 12(4): 809-821.

\bibitem[Ghiulnara and Viegas, 2010]{Ghiulnara2010} Ghiulnara, A. and Viegas, C. (2010)
Introduction of weather-derivative concepts: perspectives for Portugal.
{\em Journal of Risk Finance} 11(1): 9-19.

\bibitem[Ghosh, 1993]{Ghosh1993} Ghosh, A. (1993)
Hedging with stock index futures: Estimation and forecasting with error correction model.
{\em Journal of Futures Markets} 13(7): 743-752.

\bibitem[Ghosh, 2012]{Ghosh2012} Ghosh, A. (2012)
Comparative study of Financial Time Series Prediction by Artificial Neural Network with Gradient Descent Learning.
{\em International Journal of Scientific \& Engineering Research} 3(1): 41-49.

\bibitem[Gibson, 2004]{Gibson2004} Gibson, M.S. (2004)
Understanding the Risk of Synthetic CDOs.
{\em Finance and Economics Discussion Series (FEDS)}, Paper No. 2004-36. Washington, DC: Board of Governors of the Federal Reserve System.
Available online: \url{https://www.federalreserve.gov/pubs/feds/2004/200436/200436pap.pdf}.

\bibitem[Gibson and Pritsker, 2000]{Gibson2000} Gibson, M.S. and Pritsker, M. (2000)
Improving Grid-based Methods for Estimating Value at Risk of Fixed-Income Portfolios.
{\em Journal of Risk} 3(2): 65-89.

\bibitem[Gibson and Schwartz, 1990]{Gibson1990} Gibson, R. and Schwartz, E.S. (1990)
Stochastic convenience yield and the pricing of oil contingent claims.
{\em Journal of Finance} 15(3): 959-967.

\bibitem[Giese, 2012]{Giese2012} Giese, P. (2012)
Optimal design of volatility-driven algo-alpha trading strategies.
{\em Risk} 25(5): 68-73.

\bibitem[Giesecke and Weber, 2006]{Giesecke2006} Giesecke, K. and Weber, S. (2006)
Credit contagion and aggregate losses.
{\em Journal of Economic Dynamics and Control} 30(5): 741-767.

\bibitem[Gilbert, Jones and Morris 2006]{Gilbert2006} Gilbert, S., Jones, S.K. and Morris, G.H. (2006)
The impact of skewness in the hedging decision.
{\em Journal of Futures Markets} 26(5): 503-520.

\bibitem[Gilmour and Ridley, 2015]{Gilmour2015} Gilmour, N. and Ridley, N. (2015)
Everyday vulnerabilities -- money laundering through cash intensive businesses.
{\em Journal of Money Laundering Control} 18(3): 293-303.

\bibitem[Gilson, 1995]{Gilson1995} Gilson, S.C. (1995)
Investing in Distressed Situations: A Market Survey.
{\em Financial Analysts Journal} 51(6): 8-27.

\bibitem[Gilson, 2010]{Gilson2010} Gilson, S.C. (2010)
{\em Creating Value Through Corporate Restructuring: Case Studies in Bankruptcies, Buyouts, and Breakups.}
Hoboken, NJ: John Wiley \& Sons, Inc.

\bibitem[Gilson, 2012]{Gilson2012} Gilson, S. (2012)
Preserving Value by Restructuring Debt.
{\em Journal of Applied Corporate Finance} 24(4): 22-35.

\bibitem[Gilson, John and Lang, 1990]{Gilson1990} Gilson, S.C., John, K. and Lang, L.H.P. (1990)
Troubled debt restructurings: An empirical study of private reorganization of firms in default.
{\em Journal of Financial Economics} 27(2): 315-353.

\bibitem[Girma and Paulson, 1998]{Girma1998} Girma, P.B. and Paulson, A.S. (1998)
Seasonality in petroleum futures spreads.
{\em Journal of Futures Markets} 18(5): 581-598.

\bibitem[Glabadanidis, 2015]{Glabadanidis2015} Glabadanidis, P. (2015)
Market Timing With Moving Averages.
{\em International Review of Finance} 15(3): 387-425.

\bibitem[Glaeser and Kallal, 1997]{Glaeser1997} Glaeser, E.L. and Kallal, H.D. (1997)
Thin Markets, Asymmetric Information, and Mortgage-Backed Securities.
{\em Journal of Financial Intermediation} 6(1): 64-86.

\bibitem[Glasserman and Wu, 2010]{Glasserman2010} Glasserman, P. and Wu, Q. (2010)
Forward and future implied volatility.
{\em International Journal of Theoretical and Applied Finance} 14(3): 407-432.

\bibitem[Gliner, 2014]{Gliner2014} Gliner, G. (2014)
{\em Global Macro Trading: Profiting in a New World Economy.}
Hoboken, NJ: John Wiley \& Sons, Inc.

\bibitem[Glorot, Bordes and Bengio, 2011]{Glorot2011} Glorot, X., Bordes, A. and Bengio, Y. (2011)
Deep Sparse Rectifier Neural Networks.
{\em Proceedings of Machine Learning Research} 15: 315-323.

\bibitem[Godfrey and Brooks, 2015]{Godfrey2015} Godfrey, C. and Brooks, C. (2015)
The Negative Credit Risk Premium Puzzle: A Limits to Arbitrage Story.
{\em Working Paper.} Available online: \url{https://ssrn.com/abstract=2661232}.

\bibitem[Goebel {\em et al}, 2013]{Goebel2013} Goebel, P.R., Harrison, D.M., Mercer, J.M. and Whitby, R.J. (2013)
REIT Momentum and Characteristic-Related REIT Returns.
{\em Journal of Real Estate Finance and Economics} 47(3): 564-581.

\bibitem[Goetzmann and Ibbotson, 1990]{Goetzmann1990} Goetzmann, W.N. and Ibbotson, R.G. (1990)
The Performance of Real Estate as an Asset Class.
{\em Journal of Applied Corporate Finance} 3(1): 65-76.

\bibitem[Goetzmann and Ibbotson, 1994]{Goetzmann1994} Goetzmann, W.N. and Ibbotson, R.G. (1994)
Do Winners Repeat?
{\em Journal of Portfolio Management} 20(2): 9-18.

\bibitem[Golden, Wang and Yang, 2007]{Golden2007} Golden, L.L., Wang, M. and Yang, C. (2007)
Handling Weather Related Risks through the Financial Markets: Considerations of Credit Risk, Basis Risk, and Hedging.
{\em Journal of Risk and Insurance} 74(2): 319-346.

\bibitem[Goldstein, 1964]{Goldstein1964} Goldstein, H.N. (1964)
The Implications of Triangular Arbitrage for Forward Exchange Policy.
{\em Journal of Finance} 19(3): 544-551.

\bibitem[Goltz and Lai, 2009]{Goltz2009} Goltz, F. and Lai, W.N. (2009)
Empirical properties of straddle returns.
{\em Journal of Derivatives} 17(1): 38-48.

\bibitem[G\"{o}nc\"{u}, 2012]{Goncu2012} G\"{o}nc\"{u}, A. (2012)
Pricing temperature-based weather derivatives in China.
{\em Journal of Risk Finance} 13(1): 32-44.

\bibitem[Goodfellow {\em et al}, 2013]{Goodfellow2013} Goodfellow, I., Warde-Farley, D., Mirza, M., Courville, A. and Bengio, Y. (2013)
Maxout Networks.
{\em Proceedings of Machine Learning Research} 28(3): 1319-1327.

\bibitem[Goodfriend, 2011]{Goodfriend2011} Goodfriend, M. (2011)
Money Markets.
{\em Annual Review of Financial Economics} 3: 119-1137.

\bibitem[Goodman, 2002]{Goodman2002} Goodman, L.S. (2002)
Synthetic CDOs: An Introduction.
{\em Journal of Derivatives} 9(3): 60-72.

\bibitem[Goodman and Lucas, 2002]{GoodmanLucas2002} Goodman, L.S. and Lucas, D.J. (2002)
And When CDOs PIK?
{\em Journal of Fixed Income} 12(1): 96-102.

\bibitem[Gordini, 2014]{Gordini2014} Gordini, N. (2014)
A genetic algorithm approach for SMEs bankruptcy prediction: Empirical evidence from Italy.
{\em Expert Systems with Applications} 41(14): 6433-6455.

\bibitem[Gorton, Hayashi and Rouwenhorst, 2013]{Gorton2013} Gorton, G.B., Hayashi, F. and Rouwenhorst, K.G. (2013)
The Fundamentals of Commodity Futures Returns.
{\em Review of Finance} 17(1): 35-105.

\bibitem[Gorton and Metrick, 2012]{Gorton2012} Gorton, G. and Metrick, A. (2012)
Securitized banking and the run on repo.
{\em Journal of Financial Economics} 104(3): 425-451.

\bibitem[Gorton and Rouwenhorst, 2006]{Gorton2006} Gorton, G.B. and Rouwenhorst, K.G. (2006)
Facts and Fantasies about Commodity Futures.
{\em Financial Analysts Journal} 62(2): 47-68.

\bibitem[Gradojevic, Gen\c{c}ay and Erdemlioglu, 2017]{Gradojevic2017} Gradojevic, N., Gen\c{c}ay, R. and Erdemlioglu, D. (2017)
Robust Prediction of Triangular Currency Arbitrage with Liquidity and Realized Risk Measures: A New Wavelet-Based Ultra-High-Frequency Analysis.
{\em Working Paper.} Available online: \url{https://ssrn.com/abstract=3018815}.

\bibitem[Graff, Harrington and Young, 1999]{Graff1999} Graff, R., Harrington, A. and Young, M. (1999)
Serial Persistence in Disaggregated Australian Real Estate Returns.
{\em Journal of Real Estate Portfolio Management} 5(2): 113-128.

\bibitem[Graff and Young, 1997]{Graff1997} Graff, R.A. and Young, M.S. (1997)
Serial Persistence in Equity REIT Returns.
{\em Journal of Real Estate Research} 14(3): 183-214.

\bibitem[Graham, Nikkinen and Sahlstr\"{o}m, 2003]{Graham2003} Graham, M., Nikkinen, J. and Sahlstr\"{o}m, P. (2003)
Relative importance of scheduled macroeconomic news for stock market investors.
{\em Journal of Economics and Finance} 27(2): 153-165.

\bibitem[Graham and Pirie, 1994]{Graham1994} Graham, S. and Pirie, W. (1994)
Index Fund duraci\'on-d\'olar-neutral and Market Efficiency.
{\em Journal of Economics and Finance} 18(2): 219-229.

\bibitem[Grant, 2016]{Grant2016} Grant, J. (2016)
{\em Trading Strategies in Futures Markets} (Ph.D. Thesis). London, UK: Imperial College.
Available online: \url{https://spiral.imperial.ac.uk/bitstream/10044/1/32011/1/Grant-J-2016-PhD-Thesis.PDFA.pdf}.

\bibitem[Grantier, 1988]{Grantier1988} Grantier, B.J. (1988)
Convexity and Bond Performance: The Benter the Better.
{\em Financial Analysts Journal} 44(6): 79-81.

\bibitem[Grasselli and Wagalath, 2018]{Grasselli2018} Grasselli, M. and Wagalath, L. (2018)
VIX vs VXX: A Joint Analytical Framework.
{\em Working Paper.} Available online: \url{https://ssrn.com/abstract=3144526}.

\bibitem[Green and Rydqvist, 1999]{Green1999} Green, R.C. and Rydqvist, K. (1999)
Ex-day behavior with dividend preference and limitations to short-term arbitrage: the case of Swedish lottery bonds.
{\em Journal of Financial Economics} 53(2): 145-187.

\bibitem[Greenhaus, 1991]{Greenhaus1991} Greenhaus, S.F. (1991)
Approaches to Investing in Distressed Securities: Passive Approaches.
In: Bowman, T.A. (ed.)
{\em Analyzing Investment Opportunities in Distressed and Bankrupt Companies.} (AIMR Conference Proceedings, Vol. 1991, Iss. 1.) Chicago, IL: AIMR, pp. 47-52.

\bibitem[Greer, 1978]{Greer1978} Greer, R.J. (1978)
Conservative Commodities: A Key Inflation Hedge.
{\em Journal of Portfolio Management} 4(4): 26-29.

\bibitem[Greer, 2000]{Greer2000} Greer, R.J. (2000)
The Nature of Commodity Index Returns.
{\em Journal of Alternative Investments} 3(1): 45-52.

\bibitem[Greer, 2007]{Greer2007} Greer, R.J. (2007)
The Role of Commodities in Investment Portfolios.
{\em CFA Institute Conference Proceedings Quarterly} 24(4): 35-44.

\bibitem[Grieves, 1999]{Grieves1999} Grieves, R. (1999)
Butterfly Trades.
{\em Journal of Portfolio Management} 26(1): 87-95.

\bibitem[Grieves and Mann, 2004]{Grieves2004}Grieves, R. and Mann, S.V. (2004)
An Overlooked Coupon Effect in Treasury Futures Contracts.
{\em Journal of Derivatives} 12(2): 56-61.

\bibitem[Grieves {\em et al}, 1999]{GrievesMann1999} Grieves, R., Mann, S.V., Marcus, A.J. and Ramanlal, P. (1999)
Riding the Bill Curve.
{\em Journal of Portfolio Management} 25(3): 74-82.

\bibitem[Grieves and Marcus, 1992]{Grieves1992} Grieves, R. and Marcus, A.J. (1992)
Riding the Yield Curve: Reprise.
{\em Journal of Portfolio Management} 18(4): 67-76.

\bibitem[Grieves and Marcus, 2005]{Grieves2005} Grieves, R. and Marcus, A.J. (2005)
Delivery Options and Treasury-Bond Futures Hedge Ratios.
{\em Journal of Derivatives} 13(2): 70-76.

\bibitem[Griffin, Ji and Martin, 2003]{Griffin2003} Griffin, J.M., Ji, X. and Martin, J.S. (2003)
Momentum Investing and Business Cycle Risks: Evidence from Pole to Pole.
{\em Journal of Finance} 58(6): 2515-2547.

\bibitem[Griffin and Lemmon, 2002]{Griffin2002} Griffin, J.M. and Lemmon, M.L. (2002)
Book-to-Market Equity, Distress Risk, and Stock Returns.
{\em Journal of Finance} 57(5): 2317-2336.

\bibitem[Grigg, 2010]{Grigg2010} Grigg, N.S. (2010)
{\em Infrastructure Finance: The Business of Infrastructure for a Sustainable Future.}
Hoboken, NJ: John Wiley \& Sons, Inc.

\bibitem[Grimsey and Lewis, 2002]{Grimsey2002} Grimsey, D. and Lewis, M.K. (2002)
Evaluating the risks of public private partnerships for infrastructure projects.
{\em International Journal of Project Management} 20(2): 107-118.

\bibitem[Grinblatt and Moskowitz, 2004]{Grinblatt2004} Grinblatt, M. and Moskowitz, T.J. (2004)
Predicting Stock Price Movements from Past Returns: The Role of Consistency and Tax-Loss Selling.
{\em Journal of Financial Economics} 71(3): 541-579.

\bibitem[Grinblatt and Titman, 1992]{Grinblatt1992} Grinblatt, M. and Titman, S. (1992)
The Persistence of Mutual Fund Performance.
{\em Journal of Finance} 47(5): 1977-1984.

\bibitem[Grinold and Kahn, 2000]{Grinold2000} Grinold, R.C. and Kahn, R.N. (2000)
{\em Active Portfolio Management.} New York, NY: McGraw-Hill, Inc.

\bibitem[Grishchenko and Huang, 2013]{Grishchenko2013} Grishchenko, O.V. and Huang, J.-Z. (2013)
Inflation Risk Premium: Evidence from the TIPS Market.
{\em Journal of Fixed Income} 22(4): 5-30.

\bibitem[Grishchenko, Vanden and Zhang, 2016]{Grishchenko2016} Grishchenko, O.V., Vanden, J.M. and Zhang, J. (2016)
The Informational Content of the Embedded Deflation Option in TIPS.
{\em Journal of Banking \& Finance} 65: 1-26.

\bibitem[Grissom, Kuhle and Walther, 1987]{Grissom1987} Grissom, T.V., Kuhle, J.L. and Walther, C.H. (1987)
Diversification works in real estate, too.
{\em Journal of Portfolio Management} 13(2): 66-71.

\bibitem[Grobys, Heinonen and Kolari, 2016]{Grobys2016} Grobys, K., Heinonen, J.-P. and Kolari, J.W. (2016)
Is Currency Momentum a Hedge for Global Economic Risk?
{\em Working Paper.} Available online: \url{https://ssrn.com/abstract=2619146}.

\bibitem[Grudnitski and Osborn, 1993]{Grudnitski1993} Grudnitski, G. and Osborn, L. (1993)
Forecasting S\&P and Gold Futures Prices: An Application of Neural Networks.
{\em Journal of Futures Markets} 13(6): 631-643.

\bibitem[Grundy and Martin, 2001]{Grundy2001} Grundy, B.D. and Martin, J.S. (2001)
Understanding the Nature of the Risks and the Source of the Rewards to Momentum Investing.
{\em Review of Financial Studies} 14(1): 29-78.

\bibitem[Grundy and Verwijmeren, 2016]{Grundy2016} Grundy, B.D. and Verwijmeren, P. (2016)
Disappearing call delay and dividend-protected convertible bonds.
{\em Journal of Finance} 71(1): 195-224.

\bibitem[Gunasekarage and Power, 2001]{Gunasekarage2001} Gunasekarage, A. and Power, D.M. (2001)
The profitability of moving average trading rules in South Asian stock markets.
{\em Emerging Markets Review} 2(1): 17-33.

\bibitem[Gunasekarage, Power and Zhou, 2008]{Gunasekarage2008} Gunasekarage, A., Power, D.M. and Ting Zhou, T.T. (2008)
The long-term inflation hedging effectiveness of real estate and financial assets: A New Zealand investigation.
{\em Studies in Economics and Finance} 25(4): 267-278.

\bibitem[Guo, 2000]{Guo2000} Guo, D. (2000)
Dynamic Volatility Trading Strategies in the Currency Option Market.
{\em Review of Derivatives Research} 4(2): 133-154.

\bibitem[Gupta and Miller, 2012]{Gupta2012} Gupta, R. and Miller, S.M. (2012)
``Ripple effects" and forecasting home prices in Los Angeles, Las Vegas, and Phoenix.
{\em Annals of Regional Science} 48(3): 763-782.

\bibitem[Guren, 2014]{Guren2014} Guren, A.M. (2014)
The Causes and Consequences of House Price Momentum.
{\em Working Paper.} Available online: \url{http://scholar.harvard.edu/files/guren/files/gurenjmp.pdf}.

\bibitem[G\"{u}rkaynak, Sack and Wright, 2010]{Gurkaynak2010} G\"{u}rkaynak, R.S., Sack, B. and Wright, J.H. (2010)
The TIPS Yield Curve and Inflation Compensation.
{\em American Economic Journal: Macroeconomics} 2(1): 70-92.

\bibitem[Gutierrez and Prinsky, 2007]{Gutierrez2007} Gutierrez, R.C. and Prinsky, C.A. (2007)
Momentum, Reversal, and the Trading Behaviors of Institutions.
{\em Journal of Financial Markets} 10(1): 48-75.

\bibitem[Hafner and Wallmeier, 2007]{Hafner2007} Hafner, R. and Wallmeier, M. (2007)
Volatility as an Asset Class: European Evidence.
{\em European Journal of Finance} 13(7): 621-644.

\bibitem[Hagenstein, Mertz and Seifert, 2004]{Hagenstein2004} Hagenstein, F., Mertz, A. and Seifert, J. (2004)
{\em Investing in Corporate Bonds and Credit Risk.}
London, UK: Palgrave Macmillan.

\bibitem[Hagopian, 1999]{Hagopian1999} Hagopian, G.C. (1999)
Property-flipping and fraudulent appraisals: The phenomenon and the crackdown.
{\em Assessment Journal} 6(6): 33-39.

\bibitem[Hagstr\"{o}mer and Nord\'{e}n, 2013]{Hagstromer2013} Hagstr\"{o}mer, B. and Nord\'{e}n, L. (2013)
The Diversity of High-Frequency Traders.
{\em Journal of Financial Markets} 16(4): 741-770.

\bibitem[Hagstr\"{o}mer, Nord\'{e}n and Zhang, 2014]{Hagstromer2014} Hagstr\"{o}mer, B., Nord\'{e}n, L. and Zhang, D. (2014)
The Aggressiveness of High-Frequency Traders.
{\em Financial Review} 49(2): 395-419.

\bibitem[Hall, Park and Samworth, 2008]{Hall2008} Hall, P., Park, B.U. and Samworth, R.J. (2008)
Choice of neighbor order in nearest-neighbor classification.
{\em Annals of Statistics} 36(5): 2135-2152.

\bibitem[Hall, Pinnuck and Thorne, 2013]{Hall2013} Hall, J., Pinnuck, M. and Thorne, M. (2013)
Market risk exposure of merger arbitrage in Australia.
{\em Accounting \& Finance} 53(1): 185-215.

\bibitem[Hamelink and Hoesli, 1996]{Hamelink1996} Hamelink, F. and Hoesli, M. (1996)
Swiss real estate as a hedge against inflation: New evidence using hedonic and autoregressive models.
{\em Journal of Property Finance} 7(1): 33-49.

\bibitem[Hamerle, Igl and Plank, 2012]{Hamerle2012} Hamerle, A., Igl, A. and Plank, K. (2012)
Correlation smile, volatility skew, and systematic risk sensitivity of tranches.
{\em Journal of Derivatives} 19(3): 8-27.

\bibitem[Hamilton, 2003]{Hamilton2003} Hamilton, J. (2003)
What is an oil shock?
{\em Journal of Econometrics} 113(2): 363-398.

\bibitem[Hamisultane, 2009]{Hamisultane2009} Hamisultane, H. (2009)
Utility-based pricing of weather derivatives.
{\em European Journal of Finance} 16(6): 503-525.

\bibitem[Han and Qiu, 2007]{Han2007} Han, S. and Qiu, J. (2007)
Corporate precautionary cash holdings.
{\em Journal of Corporate Finance} 13(1): 43-57.

\bibitem[Hancock, 2013]{Hancock2013} Hancock, G.D. (2013)
VIX Futures ETNs: Three Dimensional Losers.
{\em Accounting and Finance Research} 2(3): 53-64.

\bibitem[Hanley, 1999]{Hanley1999} Hanley, M. (1999)
Hedging the Force of Nature.
{\em Risk Professional} 5(4): 21-25.

\bibitem[Hanly, Morales and Cassells, 2018]{Hanly2018} Hanly, J., Morales, L. and Cassells, D. (2018)
The efficacy of financial futures as a hedging tool in electricity markets.
{\em International Journal of Financial Economics} 23(1): 29-40.

\bibitem[Hansch, Naik and Viswanathan, 1998]{Hansch1998} Hansch, O., Naik, N.Y. and Viswanathan, S. (1998)
Do inventories matter in dealership markets? Evidence from the London stock exchange.
{\em Journal of Finance} 53(5): 1623-1656.

\bibitem[Hansen and Hodrick, 1980]{Hansen1980} Hansen, L.P. and Hodrick, R.J. (1980)
Forward Exchange Rates as Optimal Predictors of Future Spot Rates: An Econometric Analysis.
{\em Journal of Political Economy} 88(5): 829-853.

\bibitem[Happ, 1986]{Happ1986} Happ, S. (1986)
The Behavior of Rates on Federal Funds and Repurchase Agreements.
{\em American Economist} 30(2): 22-32.

\bibitem[Haran {\em et al}, 2011]{Haran2011} Haran, M., Newell, G., Adair, A., McGreal, S. and Berry, J. (2011)
The performance of UK regeneration property within a mixed asset portfolio.
{\em Journal of Property Research} 28(1): 75-95.

\bibitem[Har\v{c}arikov\'{a} and \v{S}olt\'{e}s, 2016]{Harcarikova2016} Har\v{c}arikov\'{a}, M. and \v{S}olt\'{e}s, M. (2016)
Risk Management in Energy Sector Using Short Call Ladder Strategy.
{\em Montenegrin Journal of Economics} 12(3): 39-54.

\bibitem[H\"{a}rdle and L\'{o}pez Cabrera, 2011]{Hardle2011} H\"{a}rdle, W.K. and L\'{o}pez Cabrera, B. (2011)
The Implied Market Price of Weather Risk.
{\em Applied Mathematical Finance} 19(1): 59-95.

\bibitem[H\"{a}rdle and Silyakova, 2010]{Hardle2010} H\"{a}rdle, W. and Silyakova, E. (2010)
Volatility Investing with Variance Swaps.
{\em Working Paper.} Available online: \url{https://ssrn.com/abstract=2894245}.

\bibitem[Hardy, 1978]{Hardy1978} Hardy, C.C. (1978)
{\em The Investor's Guide to Technical Analysis.}
New York, NY: McGraw-Hill, Inc.

\bibitem[Harford, 2005]{Harford2005} Harford, J. (2005)
What drives merger waves?
{\em Journal of Financial Economics} 77(3): 529-560.

\bibitem[Harner, 2008]{Harner2008} Harner, M.M. (2008)
The Corporate Governance and Public Policy Implications of Activist Distressed Debt Investing.
{\em Fordham Law Review} 77(2): 703-773.

\bibitem[Harner, 2011]{Harner2011} Harner, M.M. (2011)
Activist Distressed Debtholders: The New Barbarians at the Gate?
{\em Washington University Law Review} 89(1): 155-206.

\bibitem[Harris, Hubbard and Kemsley, 2001]{Harris2001} Harris, T.S., Hubbard, R.G. and Kemsley, D. (2001)
The Share Price Effects Of Dividend Taxes And Tax Imputation Credits.
{\em Journal of Public Economics} 79(3): 569-596.

\bibitem[Harris and Namvar, 2016]{Harris2016} Harris, L.E. and Namvar, E. (2016)
The Economics of Flash Orders and Trading.
{\em Journal of Investment Management} 14(4): 74-86.

\bibitem[Harris and Yilmaz, 2009]{Harris2009} Harris, R.D.F. and Yilmaz, F. (2009)
A momentum trading strategy based on the low frequency component of the exchange rate.
{\em Journal of Banking \& Finance} 33(9): 1575-1585.

\bibitem[Harrison and Pliska, 1981]{Harrison1981} Harrison, J.M. and Pliska, S.R. (1981)
Martingales and stochastic integrals in the theory of continuous trading.
{\em Stochastic Processes and Their Applications} 11(3): 215-260.

\bibitem[Hartigan, Prasad and De Francesco, 2011]{Hartigan2011} Hartigan, L.R., Prasad, R. and De Francesco, A.J. (2011)
Constructing an investment return series for the UK unlisted infrastructure market: estimation and application.
{\em Journal of Property Research} 28(1): 35-58.

\bibitem[Hartzell, Eichholtz and Selender, 2007]{Hartzell2007} Hartzell, D.J., Eichholtz, P. and Selender, A. (2007)
Economic diversification in European real estate portfolios.
{\em Journal of Property Research} 10(1): 5-25.

\bibitem[Hartzell, Hekman and Miles, 1986]{Hartzell1986} Hartzell, D., Hekman, J. and Miles, M. (1986)
Diversification Categories in Investment Real Estate.
{\em Real Estate Economics} 14(2): 230-254.

\bibitem[Hartzell, Hekman and Miles, 1987]{HartzellHekman1987} Hartzell, D., Hekman, J.S. and Miles, M.E. (1987)
Real Estate Returns and Inflation.
{\em Real Estate Economics} 15(1): 617-637.

\bibitem[Hartzell, Shulman and Wurtzebach, 1987]{Hartzell1987} Hartzell, D.J., Shulman, D.G. and Wurtzebach, C.H. (1987)
Refining the Analysis of Regional Diversification for Income-Producing Real Estate.
{\em Journal of Real Estate Research} 2(2): 85-95.

\bibitem[Hartzog, 1982]{Hartzog1982} Hartzog, J. (1982)
Controlling Profit Volatility: Hedging with GNMA Options.
{\em Federal Home Loan Bank Board Journal} 15(2): 10-14.

\bibitem[Harvey, 1984]{Harvey1984} Harvey, A.C. (1984)
A unified view of statistical forecasting procedures.
{\em Journal of Forecasting} 3(3): 245-275.

\bibitem[Harvey, 1990]{Harvey1990} Harvey, A.C. (1990)
{\em Forecasting, Structural Time Series Models and the Kalman Filter.}
Cambridge, UK: Cambridge University Press.

\bibitem[Harvery, 2014]{Harvey2014} Harvey, C.R. (2014)
Bitcoin Myths and Facts.
{\em Working Paper.} Available online: \url{https://ssrn.com/abstract=2479670}.

\bibitem[Harvey, 2016]{Harvey2016} Harvey, C.R. (2016)
Cryptofinance.
{\em Working Paper.} Available online: \url{https://ssrn.com/abstract=2438299}.

\bibitem[Harvey, 2015]{Harvey2015} Harvey, J.T. (2015)
Deviations from uncovered interest rate parity: a Post Keynesian explanation.
{\em Journal of Post Keynesian Economics} 27(1): 19-35.

\bibitem[Harvey and Trimbur, 2008]{Harvey2008} Harvey, A. and Trimbur, T. (2008)
Trend Estimation and the Hodrick-Prescott Filter.
{\em Journal of the Japan Statistical Society} 38(1): 41-49.

\bibitem[Hasbrouck and Saar, 2013]{Hasbrouck2013} Hasbrouck, J. and Saar, G. (2013)
Low-latency Trading.
{\em Journal of Financial Markets} 16(4): 646-679.

\bibitem[Hastings and Nordby, 2007]{Hastings2007} Hastings, A. and Nordby, H. (2007)
Benefits of Global Diversification on a Real Estate Portfolio.
{\em Journal of Portfolio Management} 33(5): 53-62.

\bibitem[Hatemi-J, and Roca, 2006]{HatemiJ2006} Hatemi-J, A. and Roca, E. (2006)
Calculating the optimal hedge ratio: constant, time varying and the Kalman Filter approach.
{\em Applied Economics Letters} 13(5): 293-299.

\bibitem[Hau, 2014]{Hau2014} Hau, H. (2014)
The exchange rate effect of multi-currency risk arbitrage.
{\em Journal of International Money and Finance} 47: 304-331.

\bibitem[Haubrich, Pennacchi and Ritchken, 2012]{Haubrich2012} Haubrich, J., Pennacchi, G. and Ritchken, P. (2012)
Inflation Expectations, Real Rates, and Risk Premia: Evidence from Inflation Swaps.
{\em Review of Financial Studies} 25(5): 1588-1629.

\bibitem[Haug, 2001]{Haug2001} Haug, E.G. (2001)
Closed form Valuation of American Barrier Options.
{\em International Journal of Theoretical and Applied Finance} 4(2): 355-359.

\bibitem[Haugen, 1995]{Haugen1995} Haugen, R.A. (1995)
{\em The New Finance: The Case Against Efficient Markets.} Upper Saddle River, NJ: Prentice Hall.

\bibitem[Haurin and Gill, 2002]{Haurin2002} Haurin, D.R. and Gill, H.L. (2002)
The Impact of Transaction Costs and the Expected Length of Stay on Homeownership.
{\em Journal of Urban Economics} 51(3): 563-584.

\bibitem[Haurin {\em et al}, 2010]{Haurin2010} Haurin, D.R., Haurin, J.L., Nadauld, T. and Sanders, A. (2010)
List Prices, Sale Prices and Marketing Time: An Application to U.S. Housing Markets.
{\em Real Estate Economics} 38(4): 659-685.

\bibitem[Hautcoeur, 2006]{Hautcoeur2006} Hautcoeur, P.C. (2006)
Why and how to measure stock market fluctuations? The early history of stock market indices, with special reference to the French case.
{\em Working Paper.} Available online: \url{https://halshs.archives-ouvertes.fr/halshs-00590522/PDF/wp200610.pdf}.

\bibitem[Hayes, 2011]{Hayes2011} Hayes, B. (2011)
Multiple time scale attribution for commodity trading advisor (CTA) funds.
{\em Journal of Investment Management} 9(2): 35-72.

\bibitem[Hayre, 1990]{Hayre1990} Hayre, L.S. (1990)
Understanding option-adjusted spreads and their use.
{\em Journal of Portfolio Management} 16(4): 68-69.

\bibitem[He, Hsu and Rue, 2015]{He2015} He, D.X., Hsu, J.C. and Rue, N. (2015)
Option-Writing Strategies in a Low-Volatility Framework.
{\em Journal of Investing} 24(3): 116-128.

\bibitem[He, Tang and Zhang, 2016]{He2016} He, J., Tang, Q. and Zhang, H. (2016)
Risk reducers in convex order.
{\em Insurance: Mathematics and Economics} 70: 80-88.

\bibitem[Head, Lloyd-Ellis and Sun, 2014]{Head2014} Head, A., Lloyd-Ellis, H. and Sun, H. (2014)
Search, Liquidity, and the Dynamics of House Prices and Construction.
{\em American Economic Review} 104(4): 1172-1210.

\bibitem[Heaton, 1988]{Heaton1988} Heaton, H. (1988)
On the possible tax-driven arbitrage opportunities in the new municipal bond futures contract.
{\em Journal of Futures Markets} 8(3): 291-302.

\bibitem[Hegde, 1982]{Hegde1982} Hegde, S.P. (1982)
The Impact of Interest Rate Level and Volatility on the Performance of Interest Rate Hedges.
{\em Journal of Futures Markets} 2(4): 341-356.

\bibitem[Heidari and Wu, 2003]{Heidari2003} Heidari, M. and Wu, L. (2003)
Are Interest Rate Derivatives Spanned by the Term Structure of Interest Rates?
{\em Journal of Fixed Income} 13(1): 75-86.

\bibitem[Helm, 2009]{Helm2009} Helm, D. (2009)
Infrastructure Investment, the Cost of Capital, and Regulation: an Assessment.
{\em Oxford Review of Economic Policy} 25(3): 307-326.

\bibitem[Helm and Tindall, 2009]{HelmTindall2009} Helm, D. and Tindall, T. (2009)
The Evolution of Infrastructure and Utility Ownership and its Implications.
{\em Oxford Review of Economic Policy} 25(3): 411-434.

\bibitem[Hemler and Miller, 1997]{Hemler1997} Hemler, M.L. and Miller, T.W., Jr. (1997)
Box spread arbitrage profits following the 1987 market crash.
{\em Journal of Financial and Quantitative Analysis} 32(1): 71-90.

\bibitem[Hemler and Miller, 2015]{Hemler2015} Hemler, M.L. and Miller, T.W., Jr. (2015)
The Performance of Options-Based Investment Strategies: Evidence for Individual Stocks During 2003-2013.
{\em Working Paper.} Available online:\\
\url{http://www.optionseducation.org/content/dam/oic/documents/literature/files/perf-options-strategies.pdf}.

\bibitem[Hendershott, Jones and Menkveld, 2011]{Hendershott2011} Hendershott, T., Jones, C. and Menkveld, A. (2011)
Does Algorithmic Trading Improve Liquidity?
{\em Journal of Finance} 66(1): 1-33.

\bibitem[Hendershott, Jones and Menkveld, 2013]{Hendershott2013} Hendershott, T., Jones, C. and Menkveld, A. (2013)
Implementation Shortfall with Transitory Price Effects.
In: Easley, D., L\'{o}pez de Prado, M. and O'Hara, M. (eds.)
{\em High Frequency Trading: New Realities for Traders, Markets and Regulators.}
London, UK: Risk Books, Chapter 9.

\bibitem[Hendershott and Moulton, 2011]{HendershottMoulton2011} Hendershott, T. and Moulton, P.C. (2011)
Automation, speed, and stock market quality: The NYSE's Hybrid.
{\em Journal of Financial Markets} 14(4): 568-604.

\bibitem[Hendershott and Riordan, 2013]{HendershottRiordan2013} Hendershott, T. and Riordan, R. (2013)
Algorithmic Trading and the Market for Liquidity.
{\em Journal of Financial and Quantitative Analysis} 48(4): 1001-1024.

\bibitem[Henderson, 2005]{Henderson2005} Henderson, B. (2005)
{\em Convertible Bonds: New Issue Performance and Arbitrage Opportunities} (Ph.D. Thesis). Urbana-Champaign IL: University of Illinois.

\bibitem[Henderson, 1924]{Henderson1924} Henderson, R. (1924)
A new method of graduation.
{\em Transactions of the Actuarial Society of America} 25: 29-40.

\bibitem[Henderson, 1925]{Henderson1925} Henderson, R. (1925)
Further remarks on graduation.
{\em Transactions of the Actuarial Society of America} 26: 52-57.

\bibitem[Henderson, 1938]{Henderson1938} Henderson, R. (1938)
{\em Mathematical Theory of Graduation.}
New York, NY: Actuarial Society of America.

\bibitem[Henderson, 2003]{Henderson2003} Henderson, T.M. (2003)
{\em Fixed Income Strategy: The Practitioner's Guide to Riding the Curve.}
Chichester, UK: John Wiley \& Sons, Ltd.

\bibitem[Henderson and Tookes, 2012]{Henderson2012} Henderson, B.J. and Tookes, H. (2012)
Do investment banks' relationships with investors impact pricing? The case of convertible bond issues.
{\em Management Science} 58(2): 2272-2291.

\bibitem[Henrard, 2006]{Henrard2006} Henrard, M.P.A. (2006)
A Semi-Explicit Approach to Canary Swaptions in HJM One-Factor Model.
{\em Applied Mathematical Finance} 13(1): 1-18.

\bibitem[Hensher and Jones, 2007]{Hensher2007} Hensher, D. and Jones, S. (2007)
Forecasting corporate bankruptcy: Optimizing the performance of the mixed logit model.
{\em Abacus} 43(3): 241-364.

\bibitem[Herbertsson, 2008]{Herbertsson2008} Herbertsson, A. (2008)
Pricing synthetic CDO tranches in a model with default contagion using the matrix-analytic approach.
{\em Journal of Credit Risk} 4(4): 3-35.

\bibitem[Herranz-Lonc\'an, 2007]{Herranz-Loncan2007} Herranz-Lonc\'an, A. (2007)
Infrastructure investment and Spanish economic growth, 1850-1935.
{\em Explorations in Economic History} 44(3): 452-468.

\bibitem[Hess, Huang and Niessen, 2008]{Hess2008} Hess, D., Huang, H. and Niessen, A. (2008)
How Do Commodity Futures Respond to Macroeconomic News?
{\em Financial Markets and Portfolio Management} 22(2): 127-146.

\bibitem[Hew {\em et al}, 1996]{Hew1996} Hew, D., Skerratt, L., Strong, N. and Walker, M. (1996)
Post-earnings-announcement drift: Some preliminary evidence for the UK.
{\em Accounting \& Business Research} 26(4): 283-293.

\bibitem[Hill {\em et al}, 2006]{Hill2006} Hill, J.M., Balasubramanian, V., Gregory, K. and Tierens, I. (2006)
Finding Alpha via Covered Call Writing.
{\em Financial Analysts Journal} 62(5): 29-46.

\bibitem[Hill, Nadig and Hougan, 2015]{Hill2015} Hill, J.M., Nadig, D. and Hougan, M. (2015)
A Comprehensive Guide to Exchange-Traded Funds (ETFs).
{\em Research Foundation Publications} 2015(3): 1-181.

\bibitem[Hillegeist {\em et al}, 2004]{Hillegeist2004} Hillegeist, S.A., Keating, E., Cram, D.P. and Lunstedt, K.G. (2004)
Assessing the probability of bankruptcy.
{\em Review of Accounting Studies} 9(1): 5-34.

\bibitem[Hilliard, 1984]{Hilliard1984} Hilliard, J.E. (1984)
Hedging Interest Rate Risk with Futures Portfolios under Term Structure Effects.
{\em Journal of Finance} 39(5): 1547-1569.

\bibitem[Hilliard and  Jordan, 1989]{Hilliard1989} Hilliard, J. and  Jordan, S. (1989)
Hedging Interest Rate Risk with Futures Portfolios under Full-Rank Assumptions.
{\em Journal of Financial and Quantitative Analysis} 24(2): 217-240.

\bibitem[Hilliard and Reis, 1998]{Hilliard1998} Hilliard, J. and Reis, J. (1998)
Valuation of commodity futures and options under stochastic convenience yields, interest rates, and jump diffusions on the spot.
{\em Journal of Financial and Quantitative Analysis} 33(1): 61-86.

\bibitem[Hinnerich, 2008]{Hinnerich2008} Hinnerich, M. (2008)
Inflation-indexed swaps and swaptions.
{\em Journal of Banking \& Finance} 32(11): 2293-2306.

\bibitem[Hirschey, 2018]{Hirschey2018} Hirschey, N. (2018)
Do High-Frequency Traders Anticipate Buying and Selling Pressure?
{\em Working Paper.} Available online: \url{https://ssrn.com/abstract=2238516}.

\bibitem[Hirshleifer, 1990]{Hirshleifer1990} Hirshleifer, D. (1990)
Hedging Pressure and Futures Price Movements in a General Equilibrium Model.
{\em Econometrica} 58(2): 411-428.

\bibitem[Hirshleifer, Lim and Teoh, 2009]{Hirshleifer2009} Hirshleifer, D., Lim, S.S. and Teoh, S.H. (2009)
Driven to distraction: Extraneous events and underreaction to earnings news.
{\em Journal of Finance} 64(5): 2289-2325.

\bibitem[Ho and Saunders, 1983]{Ho1983} Ho, T. and Saunders, A. (1983)
Fixed Rate Loan Commitments, Take-Down Risk, and the Dynamics of Hedging with Futures.
{\em Journal of Financial and Quantitative Analysis} 18(4): 499-516.

\bibitem[Hodges and Carverhill, 1993]{Hodges1993} Hodges, S. and Carverhill, A. (1993)
Quasi mean reversion in an efficient stock market: the characterization of economic equilibria which support Black-Scholes option pricing.
{\em Economic Journal} 103(417): 395-405.

\bibitem[Hodrick, 1987]{Hodrick1987} Hodrick, R.J. (1987)
{\em The Empirical Evidence on the Efficiency of Forward and Futures Foreign Exchange Markets.}
New York, NY: Harwood Academic.

\bibitem[Hodrick and Prescott, 1997]{Hodrick1997} Hodrick, R.J. and Prescott, E.C. (1997)
Postwar U.S. Business Cycles: An Empirical Investigation.
{\em Journal of Money, Credit and Banking} 29(1): 1-16.

\bibitem[Hoesli and Lekander, 2008]{Hoesli2008} Hoesli, M. and Lekander, J. (2008)
Real estate portfolio strategy and product innovation in Europe.
{\em Journal of Property Investment \& Finance} 26(2): 162-176.

\bibitem[Hoevenaars {\em et al}, 2008]{Hoevenaars2008} Hoevenaars, R.P.M.M., Molenaar, R.D.J., Schotman, P.C. and Steenkamp, T.B.M. (2008)
Strategic asset allocation with liabilities: Beyond stocks and bonds.
{\em Journal of Economic Dynamics and Control} 32(9): 2939-2970.

\bibitem[Holden and Jacobsen, 2014]{Holden2014} Holden, C.W. and Jacobsen, S. (2014)
Liquidity Measurement Problems in Fast Competitive Markets: Expensive and Cheap Solutions.
{\em Journal of Finance} 69(4): 1747-1885.

\bibitem[Holmes, 1996]{Holmes1996} Holmes, P. (1996)
Stock index futures hedging: hedge ratio estimation, duration effects, expiration effects and hedge ratio stability.
{\em Journal of Business Finance \& Accounting} 23(1): 63-77.

\bibitem[Hong, Torous and Valkanov, 2007]{Hong2007} Hong, H., Torous, W. and Valkanov, R. (2007)
Do Industries Lead Stock Markets?
{\em Journal of Financial Economics} 83(2): 367-396.

\bibitem[Hopton, 1999]{Hopton1999} Hopton, D. (1999)
Prevention of Money Laundering: The Practical Day-to-Day Problems and Some Solutions.
{\em Journal of Money Laundering Control} 2(3): 249-252.

\bibitem[H\"{o}rdahl and Tristani, 2012]{Hordahl2012} H\"{o}rdahl, P. and Tristani, O. (2012)
Inflation Risk Premia in the Term Structure of Interest Rates.
{\em Journal of the European Economic Association} 10(3): 634-657.

\bibitem[H\"{o}rdahl and Tristani, 2014]{Hordahl2014} H\"{o}rdahl, P. and Tristani, O. (2014)
Inflation Risk Premia in the Euro Area and the United States.
{\em International Journal of Central Banking} 10(3): 1-47.

\bibitem[Horvath, 1998]{Horvath1998} Horvath, P.A. (1998)
A Measurement of the Errors in Intra-Period Compounding and Bond Valuation: A Short Extension.
{\em Financial Review} 23(3): 359-363.

\bibitem[Hotchkiss and Mooradian, 1997]{Hotchkiss1997} Hotchkiss, E.S. and Mooradian, R.M. (1997)
Vulture Investors and the Market for Control of Distressed Firms.
{\em Journal of Financial Economics} 43(3): 401-432.

\bibitem[Hotchkiss and Ronen, 2002]{Hotchkiss2002} Hotchkiss, E.S. and Ronen, R. (2002)
The Informational Efficiency of the Corporate Bond Market: An Intraday Analysis.
{\em Review of Financial Studies} 15(5): 1325-1354.

\bibitem[Hou and Nord\'{e}n, 2018]{Hou2017} Hou, A.J. and Nord\'{e}n, L.L. (2018)
VIX futures calendar spreads.
{\em Journal of Futures Markets} 38(7): 822-838.

\bibitem[Houdain and Guegan, 2006]{Houdain2006} Houdain, J.P. and Guegan, D. (2006)
Hedging tranches index products: illustration of model dependency.
{\em ICFAI Journal of Derivatives Markets} 4: 39-61.

\bibitem[Houweling and van Vundert, 2017]{Houweling2017} Houweling, P. and van Vundert, J. (2017)
Factor Investing in the Corporate Bond Market.
{\em Financial Analysts Journal} 73(2): 100-115.

\bibitem[Howison, Reisinger and Witte, 2013]{Howison2013} Howison, S.D., Reisinger, C. and Witte, J.H. (2013)
The Effect of Nonsmooth Payoffs on the Penalty Approximation of American Options.
{\em SIAM Journal on Financial Mathematics} 4(1): 539-574.

\bibitem[Hsieh and Barmish, 2015]{Hsieh2015} Hsieh, C.H. and Barmish, B.R. (2015)
On Kelly betting: Some limitations.
In: {\em Proceeding of the 53rd Annual Allerton Conference on Communication, Control, and Computing.} Washington, DC: IEEE, pp. 165-172.

\bibitem[Hsieh, Barmish and Gubner, 2016]{Hsieh2016} Hsieh, C.H., Barmish, B.R. and Gubner, J.A. (2016)
Kelly betting can be too conservative.
In: {\em Proceedings of the 2016 Conference on Decision and Control (CDC).} Washington, DC: IEEE, pp. 3695-3701.

\bibitem[Hsieh and Walkling, 2005]{Hsieh2005}  Hsieh, J. and Walkling, R.A. (2005)
Determinants and Implications of Arbitrage Holdings in Acquisitions.
{\em Journal of Financial Economics} 77(3): 605-648.

\bibitem[Hsu, 1998]{Hsu1998}  Hsu, M. (1998)
Spark Spread Options Are Hot!
{\em Electricity Journal} 11(2): 28-39.

\bibitem[Hsu, Lin and Vincent, 2018]{Hsu2018} Hsu, Y.-C., Lin, H.-W. and Vincent, K. (2018)
Analyzing the Performance of Multi-Factor Investment Strategies under Multiple Testing Framework.
{\em Working Paper.} Available online:\\ \url{http://www.econ.sinica.edu.tw/UpFiles/2013092817175327692/Seminar_PDF2013093010102890633/17-A0001(all).pdf}.

\bibitem[Hu, 2001]{Hu2001} Hu, J. (2001)
{\em Basics of Mortgage-Backed Securities.} (2nd ed.)
Hoboken, NJ: John Wiley \& Sons, Inc.

\bibitem[Huang and Kong, 2003]{Huang2003} Huang, J-.Z. and Kong, W. (2003)
Explaining Credit Spread Changes: New Evidence From Option-Adjusted Bond Indexes.
{\em Journal of Derivatives} 11(1): 30-44.

\bibitem[Huang, Nakamori and Wang, 2005]{Huang2005} Huang, W., Nakamori, Y. and Wang, S.-Y. (2005)
Forecasting stock market movement direction with support vector machine.
{\em Computers \& Operation Research} 32(10): 2513-2522.

\bibitem[Huang, Shiu and Lin, 2008]{HuangSL2008} Huang, H., Shiu, Y. and Lin, P. (2008)
HDD and CDD option pricing with market price of weather risk for Taiwan.
{\em Journal of Futures Markets} 28(8): 790-814.

\bibitem[Huang and Tsai, 2009]{Huang2009} Huang, C.L. and Tsai, C.Y. (2009)
A hybrid SOFM-SVR with a filter-based feature selection for stock market forecasting.
{\em Expert Systems with Applications} 36(2): 1529-1539.

\bibitem[Huault and Rainelli-Weis, 2011]{Huault2011} Huault, I. and Rainelli-Weis, H. (2011)
A Market for Weather Risk? Conflicting Metrics, Attempts at compromise, and Limits to Commensuration.
{\em Organization Studies} 32(10): 1395-1419.

\bibitem[Huck, 2009]{Huck2009} Huck, N. (2009)
Pairs selection and outranking: An application to the S\&P 100 index.
{\em European Journal of Operational Research} 196(2): 819-825.

\bibitem[Huck, 2015]{Huck2015} Huck, N. (2015)
Pairs trading: Does volatility timing matter?
{\em Applied Economics} 47(57): 6239-6256.

\bibitem[Huck and Afawubo, 2014]{Huck2014} Huck, N. and Afawubo, K. (2014)
Pairs trading and selection methods: is cointegration superior?
{\em Applied Economics} 47(6): 599-613.

\bibitem[Hudson-Wilson, 1990]{Hudson-Wilson1990} Hudson-Wilson, S. (1990)
New Trends in Portfolio Theory.
{\em Journal of Property Management} 55(3): 57-58.

\bibitem[Hudson-Wilson {\em et al}, 2005]{Hudson-Wilson2005} Hudson-Wilson, S., Gordon, J.N., Fabozzi, F.J., Anson, M.J.P. and Giliberto, M. (2005)
Why Real Estate?
{\em Journal of Portfolio Management} 31(5): 12-21.

\bibitem[Huerta, Elkan and Corbacho, 2013]{Huerta2013} Huerta, R., Elkan, C. and Corbacho, F. (2013)
Nonlinear Support Vector Machines Can Systematically Identify Stocks with High and Low Future Returns.
{\em Algorithmic Finance} 2(1): 45-58.

\bibitem[H\"{u}hn and Scholz, 2017]{Huhn2017} H\"{u}hn, H. and Scholz, H. (2017)
Alpha Momentum and Price Momentum.
{\em Working Paper.} Available online: \url{https://ssrn.com/abstract=2287848}.

\bibitem[Huij and Lansdorp, 2017]{Huij2017} Huij, J. and Lansdorp, S. (2017)
Residual Momentum and Reversal Strategies Revisited.
{\em Working Paper.} Available online: \url{https://ssrn.com/abstract=2929306}.

\bibitem[Hull, 1996]{Hull1996} Hull, D.A. (1996)
Stemming algorithms: A case study for detailed evaluation.
{\em Journal of the American Society for Information Science and Technology} 47(1): 70-84.

\bibitem[Hull, 2012]{Hull2012} Hull, J.C. (2012)
{\em Options, Futures and Other Derivatives.}
Upper Saddle River, NJ: Prentice Hall.

\bibitem[Hull and White, 2004]{Hull2004} Hull, J.C. and White, A.D. (2004)
Valuation of a CDO and an $n^{th}$ to Default CDS without Monte Carlo Simulation.
{\em Journal of Derivatives} 12(2): 8-23.

\bibitem[Hull and White, 2006]{Hull2006} Hull, J.C. and White, A.D. (2006)
Valuing Credit Derivatives Using an Implied Copula Approach.
{\em Journal of Derivatives} 14(2): 8-28.

\bibitem[Hull and White, 2010]{Hull2010} Hull, J.C. and White, A.D. (2010)
An Improved Implied Copula Model and its Application to the Valuation of Bespoke CDO Tranches.
{\em Journal of Investment Management} 8(3): 11-31.

\bibitem[Hull, Predescu and White, 2005]{Hull2005} Hull, J., Predescu, M. and White, A. (2005)
Bond Prices, Default Probabilities and Risk Premiums.
{\em Journal of Credit Risk} 1(2): 53-60.

\bibitem[Hung, 2016]{Hung2016} Hung, N.H. (2016)
Various moving average convergence divergence trading strategies: a comparison.
{\em Investment Management and Financial Innovations} 13(2): 363-369.

\bibitem[Hunter, 1999]{Hunter1999} Hunter, R. (1999)
Managing Mother Nature.
{\em Derivatives Strategy} 4(2): 15-19.

\bibitem[Hunter and Simon, 2005]{Hunter2005} Hunter, D.M. and Simon, D.P. (2005)
Are TIPS the ``real" deal?: A conditional assessment of their role in a nominal portfolio.
{\em Journal of Banking \& Finance} 29(2): 347-368.

\bibitem[H\"{u}rlimann, 2002]{Hurlimann2002} H\"{u}rlimann, W. (2002)
On immunization, stop-loss order and the maximum Shiu measure.
{\em Insurance: Mathematics and Economics} 31(3): 315-325.

\bibitem[H\"{u}rlimann, 2012]{Hurlimann2012} H\"{u}rlimann, W. (2012)
On directional immunization and exact matching.
{\em Communications in Mathematical Finance} 1(1): 1-12

\bibitem[Hurst, Ooi and Pedersen, 2017]{Hurst2017} Hurst, B., Ooi, Y.H. and Pedersen, L.H. (2017)
A Century of Evidence on Trend-Following Investing.
{\em Journal of Portfolio Management} 44(1): 15-29.

\bibitem[Husson and McCann, 2011]{Husson2011} Husson, T. and McCann, C.J.  (2011)
The VXX ETN and Volatility Exposure.
{\em PIABA Bar Journal} 18(2): 235-252.

\bibitem[Huston, 2000]{Huston2000} Hutson, E. (2000)
Takeover targets and the probability of bid success: Evidence from the Australian market.
{\em International Review of Financial Analysis} 9(1): 45-65.

\bibitem[Hwang and George, 2004]{Hwang2004} Hwang, C.-Y. and George, T.J. (2004)
The 52-Week High and Momentum Investing.
{\em Journal of Finance} 59(5): 2145-2176.

\bibitem[Idzorek, 2007]{Idzorek2007} Idzorek, T. (2007)
A Step-by-Step Guide to the Black-Litterman Model.
In: Satchell, S. (ed.) {\em Forecasting Expected Returns in the Financial Markets.}
Waltham, MA: Academic Press.

\bibitem[Illueca and Lafuente, 2003]{Illueca2003} Illueca, M. and Lafuente, J.A. (2003)
The Effect of Spot and Futures Trading on Stock Index Volatility: A Non-parametric Approach.
{\em Journal of Futures Markets} 23(9): 841-858.

\bibitem[Ilmanen, 2011]{Ilmanen2011} Ilmanen, A. (2011)
{\em Expected Returns: An Investor's Guide to Harvesting Market Rewards.}
Hoboken, NJ: John Wiley \& Sons, Inc.

\bibitem[Ilmanen {\em et al}, 2004]{Ilmanen2004} Ilmanen, A., Byrne, R., Gunasekera, H. and Minikin, R. (2004)
Which Risks Have Been Best Rewarded?
{\em Journal of Portfolio Management} 30(2): 53-57.

\bibitem[Ilut, 2012]{Ilut2012} Ilut, C. (2012)
Ambiguity Aversion: Implications for the Uncovered Interest Rate Parity Puzzle.
{\em American Economic Journal: Macroeconomics} 4(3): 33-65.

\bibitem[Inderst, 2010a]{Inderst2010} Inderst, G. (2010a)
Infrastructure as an Asset Class.
{\em EIB Papers} 15(1): 70-105.

\bibitem[Inderst, 2010b]{Inderst2010b} Inderst, G. (2010b)
Pension fund investment in infrastructure: What have we learnt?
{\em Pensions: An International Journal} 15(2): 89-99.

\bibitem[Ingersoll, 1977]{Ingersoll1977} Ingersoll, J. (1977)
A contingent-claims valuation of convertible securities.
{\em Journal of Financial Economics} 4(3): 289-322.

\bibitem[Irwin, Zulauf and Jackson, 1996]{Irwin1996} Irwin, S.H., Zulauf, C.R. and Jackson, T.E. (1996)
Monte Carlo analysis of mean reversion in commodity futures prices.
{\em American Journal of Agricultural Economics} 78(2): 387-399.

\bibitem[Israelov, 2017]{Israelov2017} Israelov, R. (2017)
Pathetic Protection: The Elusive Benefits of Protective Puts.
{\em Working Paper}. Available online: \url{https://ssrn.com/abstract=2934538}.

\bibitem[Israelov and Klein, 2016]{Israelov2016} Israelov, R. and Klein, M. (2016)
Risk and Return of Equity Index Collar Strategies.
{\em Journal of Alternative Investments} 19(1): 41-54.

\bibitem[Israelov and Nielsen, 2014]{Israelov2014} Israelov, R. and Nielsen, L.N. (2014)
Covered Call Strategies: One Fact and Eight Myths.
{\em Financial Analysts Journal} 70(6): 23-31.

\bibitem[Israelov and Nielsen, 2015a]{Israelov2015a} Israelov, R. and Nielsen, L.N. (2015a)
Covered Calls Uncovered.
{\em Financial Analysts Journal}  71(6): 44-57.

\bibitem[Israelov and Nielsen, 2015b]{Israelov2015b} Israelov, R. and Nielsen, L.N. (2015b)
Still Not Cheap: Portfolio Protection in Calm Markets.
{\em Journal of Portfolio Management} 41(4): 108-120.

\bibitem[Israelov, Nielsen and Villalon, 2017]{Israelov2017.1}
Israelov, R., Nielsen L.N. and Villalon, D. (2017) Embracing Downside Risk.
{\em Journal of Alternative Investments} 19(3): 59-67.

\bibitem[Ito {\em et al}, 2012]{Ito2012} Ito, T., Yamada, K., Takayasu, M. and Takayasu, H.  (2012)
Free Lunch! Arbitrage Opportunities in the Foreign Exchange Markets.
{\em Working Paper.} Available online: \url{http://www.nber.org/papers/w18541}.

\bibitem[Iturricastillo and De La Pe\~{n}a, 2010]{Iturricastillo2010} Iturricastillo, I. and De La Pe\~{n}a, J.I. (2010)
Absolute Immunization Risk as general measure of immunization risk.
{\em An\'{a}lisis Financiero} 114(3): 42-59.

\bibitem[Ivanov and Lenkey, 2014]{Ivanov2014} Ivanov, I.T. and Lenkey, S.L. (2014)
Are Concerns About Leveraged ETFs Overblown?
{\em Finance and Economics Discussion Series (FEDS)}, Paper No. 2014-106.
Washington, DC: Board of Governors of the Federal Reserve System.
Available online: \url{https://www.federalreserve.gov/econresdata/feds/2014/files/2014106pap.pdf}.

\bibitem[Jabbour and Budwick, 2010]{Jabbour2010} Jabbour, G. and Budwick, P. (2010)
{\em The option trader handbook: strategies and trade adjustments.}
(2nd ed.) Hoboken, NJ: John Wiley \& Sons, Inc.

\bibitem[Jackwerth, 2000]{Jackwerth2000} Jackwerth, J.C. (2000)
Recovering Risk Aversion from Option Prices and Realized Returns.
{\em Review of Financial Studies} 13(2): 433-451.

\bibitem[Jacobs and Weber, 2015]{Jacobs2015} Jacobs, H. and Weber, M. (2015)
On the determinants of pairs trading profitability.
{\em Journal of Financial Markets} 23: 75-97.

\bibitem[Jacoby and Shiller, 2008]{Jacoby2008} Jacoby, G. and Shiller, I. (2008)
Duration and Pricing of TIPS.
{\em Journal of Fixed Income} 18(2): 71-84.

\bibitem[Jain and Baile, 2000]{Jain2000} Jain, G. and Baile, C. (2000)
Managing weather risks.
{\em Strategic Risk}, September 2000, pp. 28-31.

\bibitem[James, 1968]{James1968} James, F.E., Jr. (1968)
 Monthly moving averages -- An effective investment tool?
{\em Journal of Financial and Quantitative Analysis} 3(3): 315-326.

\bibitem[James, 2003]{James2003} James, T. (2003)
{\em Energy Price Risk: Trading and Price Risk Management.}
London, UK: Palgrave Macmillan.

\bibitem[Jan and Hung, 2004]{Jan2004} Jan, T.C. and Hung, M.W. (2004)
Short-Run and Long-Run Persistence in Mutual Funds.
{\em Journal of Investing} 13(1): 67-71.

\bibitem[Jankowitsch and Nettekoven, 2008]{Jankowitsch2008} Jankowitsch, R. and Nettekoven, M. (2008)
Trading strategies based on term structure model residuals.
{\em European Journal of Finance} 14(4): 281-298.

\bibitem[Jansen and Nikiforov, 2016]{Jansen2016} Jansen, I.P. and Nikiforov, A.L. (2016)
Fear and Greed: A Returns-Based Trading Strategy around Earnings Announcements.
{\em Journal of Portfolio Management} 42(4): 88-95.

\bibitem[Jarrow, 2010]{Jarrow2010} Jarrow, R.A. (2010)
Understanding the risk of leveraged ETFs.
{\em Finance Research Letters} 7(3): 135-139.

\bibitem[Jarrow {\em et al}, 2013]{Jarrow2013} Jarrow, R., Kchia, Y., Larsson, M. and Protter, P. (2013)
Discretely sampled variance and volatility swaps versus their continuous approximations.
{\em Finance and Stochastics} 17(2): 305-324.

\bibitem[Jarrow, Lando and Turnbull, 1997]{Jarrow1997} Jarrow, R., Lando, D. and Turnbull, S. (1997)
A Markov model for the term structure of credit spreads.
{\em Review of Financial Studies} 10(2): 481-523.

\bibitem[Jarrow and Protter, 2012]{Jarrow2012} Jarrow, R.A. and Protter, P. (2012)
A Dysfunctional Role of High Frequency Trading in Electronic Markets.
{\em International Journal of Theoretical and Applied Finance} 15(3): 1250022.

\bibitem[Jarrow and Turnbull, 1995]{Jarrow1995} Jarrow, R.A. and Turnbull, S.M. (1995)
Pricing Derivatives on Financial Securities Subject to Credit Risk.
{\em Journal of Finance} 50(1): 53-85.

\bibitem[Jarrow and Yildirim, 2003]{Jarrow2003} Jarrow, R. and Yildirim, Y. (2003)
Pricing treasury inflation protected securities and related derivatives using an HJM model.
{\em Journal of Financial and Quantitative Analysis} 38(2): 409-430.

\bibitem[Jasemi and Kimiagari, 2012]{Jasemi2012} Jasemi, M. and Kimiagari, A.M. (2012)
An investigation of model selection criteria for technical analysis of moving average.
{\em Journal of Industrial Engineering International} 8: 5.

\bibitem[Jegadeesh, 1990]{Jegadeesh1990} Jegadeesh, N. (1990)
Evidence of Predictable Behavior of Security Returns.
{\em Journal of Finance} 45(3): 881-898.

\bibitem[Jegadeesh and Titman, 1993]{Jegadeesh1993} Jegadeesh, N. and Titman, S. (1993)
Returns to Buying Winners and Selling Losers: Implications for Stock Market Efficiency.
{\em Journal of Finance} 48(1): 65-91.

\bibitem[Jegadeesh and Titman, 1995]{Jegadeesh1995} Jegadeesh, N. and Titman, S. (1995)
Overreaction, delayed reaction, and contrarian profits.
{\em Review of Financial Studies} 8(4): 973-993.

\bibitem[Jegadeesh and Titman, 2001]{Jegadeesh2001} Jegadeesh, N. and Titman, S. (2001)
Profitability of Momentum Strategies: An Evaluation of Alternative Explanations.
{\em Journal of Finance} 56(2): 699-720.

\bibitem[Jensen, 1968]{Jensen1968} Jensen, M.C. (1968)
The Performance of Mutual Funds in the Period 1945-1964.
{\em Journal of Finance} 23(2): 389-416.

\bibitem[Jensen, Johnson and Mercer, 2000]{Jensen2000} Jensen, G.R., Johnson, R.R. and Mercer, J.M. (2000)
Efficient use of commodity futures in diversified portfolios.
{\em Journal of Futures Markets} 20(5): 489-506.

\bibitem[Jensen, Johnson and Mercer, 2002]{Jensen2002} Jensen, G.R., Johnson, R.R. and Mercer, J.M. (2002)
Tactical Asset Allocation and Commodity Futures.
{\em Journal of Portfolio Management} 28(4): 100-111.

\bibitem[Jermann, 2016]{Jermann2016} Jermann, U.J. (2016)
Negative Swap Spreads and Limited Arbitrage.
{\em Working Paper.} Available online: \url{https://ssrn.com/abstract=2737408}.

\bibitem[Jetley and Ji, 2010]{Jetley2010} Jetley, G. and Ji, X. (2010)
The shrinking merger arbitrage spread: Reasons and implications.
{\em Financial Analysts Journal} 66(2): 54-68.

\bibitem[Jewson, 2004a]{Jewson2004a} Jewson, S. (2004a)
Weather Derivative Pricing and the Distributions of Standard Weather Indices on US Temperatures.
{\em Working Paper.} Available online: \url{https://ssrn.com/abstract=535982}.

\bibitem[Jewson, 2004b]{Jewson2004b} Jewson, S. (2004b)
Introduction to Weather Derivative Pricing.
{\em Working Paper.} Available online: \url{https://ssrn.com/abstract=557831}.

\bibitem[Jewson, Brix and Ziehmann, 2005]{Jewson2005} Jewson, S., Brix, A. and Ziehmann, C. (2005)
{\em Weather Derivative Valuation: The Meteorological, Statistical, Financial and Mathematical Foundations.}
Cambridge, UK: Cambridge University Press.

\bibitem[Jewson and Caballero, 2003]{Jewson2003} Jewson, S. and Caballero, R. (2003)
Seasonality in the statistics of surface air temperature and the pricing of weather derivatives.
{\em Meteorological Applications} 10(4): 367-376.

\bibitem[Jha and Kalimipal, 2010]{Jha2010} Jha, R. and Kalimipal, M. (2010)
The economic significance of conditional skewness in index option markets.
{\em Journal of Futures Markets} 30(4): 378-406.

\bibitem[Jiang, Li and Wang, 2017]{JiangLi2017} Jiang, H., Li, D. and Wang, A. (2017)
Dynamic Liquidity Management by Corporate Bond Mutual Funds.
{\em Working Paper.} Available online: \url{https://ssrn.com/abstract=2776829}.

\bibitem[Jiang, Li and Wang, 2012]{Jiang2012} Jiang, W., Li, K. and Wang, W. (2012)
Hedge Funds and Chapter 11.
{\em Journal of Finance} 67(2): 513-560.

\bibitem[Jiang and Liang, 2017]{JiangLiang2017} Jiang, Z. and Liang, J. (2017)
Cryptocurrency Portfolio Management with Deep Reinforcement Learning.
{\em Working Paper.} Available online: \url{https://arxiv.org/pdf/1612.01277.pdf}.

\bibitem[Jiang and Peterburgsky, 2017]{Jiang2017} Jiang, X. and Peterburgsky, S. (2017)
Investment performance of shorted leveraged ETF pairs.
{\em Applied Economics} 49(44): 4410-4427.

\bibitem[Jo, Han and Lee, 1997]{Jo1997} Jo, H., Han, I. and Lee, H. (1997)
Bankruptcy prediction using case-based reasoning, neural networks, and discriminant analysis.
{\em Expert Systems with Applications} 13(2): 97-108.

\bibitem[Jobst, 2005]{Jobst2005} Jobst, A. (2005)
Tranche Pricing in Subordinated Loan Securitization.
{\em Journal of Structured Finance} 11(2): 64-96.

\bibitem[Jobst, 2006a]{Jobst2006a} Jobst, A. (2006a)
European Securitization: A GARCH Model of Secondary Market Spreads.
{\em Journal of Structured Finance} 12(1): 55-80.

\bibitem[Jobst, 2006b]{Jobst2006b} Jobst, A. (2006b)
Sovereign Securitization in Emerging Markets.
{\em Journal of Structured Finance} 12(3): 2-13.

\bibitem[Jobst, 2006c]{Jobst2006c} Jobst, A. (2006c)
Correlation, Price Discovery and Co-movement of ABS and Equity.
{\em Derivatives Use, Trading \& Regulation} 12(1-2): 60-101.

\bibitem[Jobst, 2007]{Jobst2007} Jobst, A. (2007)
A Primer on Structured Finance.
{\em Journal of Derivatives \& Hedge Funds} 13(3): 199-213.

\bibitem[John and Brigitte, 2009]{John2009} John, W. and Brigitte, U. (2009)
Measuring Global Money Laundering: ``The Walker Gravity Model".
{\em Review of Law \& Economics} 5(2): 821-853.

\bibitem[Johnson, 1979]{Johnson1979} Johnson, H.F. (1979)
Is It Better to Go Naked on the Street? A Primer on the Options Market.
{\em Notre Dame Lawyer (Notre Dame Law Review)} 55(1): 7-32.

\bibitem[Johnson, 2002]{Johnson2002} Johnson, T.C. (2002)
Rational Momentum Effects.
{\em Journal of Finance} 57(2): 585-608.

\bibitem[Johnson, 2008]{Johnson2008} Johnson, T.C. (2008)
Volume, liquidity, and liquidity risk.
{\em Journal of Financial Economics} 87(2): 388-417.

\bibitem[Jones, 1991]{Jones1991} Jones, F.J. (1991)
Yield Curve Strategies.
{\em Journal of Fixed Income} 1(2): 43-48.

\bibitem[Jones, Lamont and Lumsdaine, 1998]{Jones1998} Jones, C.M., Lamont, O. and Lumsdaine, R.L. (1998)
Macroeconomic news and bond market volatility.
{\em Journal of Financial Economics} 47(3): 315-337.

\bibitem[Jongadsayakul, 2016]{Jongadsayakul2016} Jongadsayakul, W. (2016)
A Box Spread Test of the SET50 Index Options Market Efficiency: Evidence from the Thailand Futures Exchange.
{\em International Journal of Economics and Financial Issues} 6(4): 1744-1749.

\bibitem[Jongadsayakul, 2017]{Jongadsayakul2017} Jongadsayakul, W. (2017)
Arbitrage Opportunity In Thailand Futures Exchange: An Empirical Study of SET50 Index Options.
In: {\em 2017 IACB, ICE \& ISEC Proceedings}, Paper No. 381. Littleton, CO: Clute Institute.

\bibitem[Jonsson and Fridson, 1996]{Jonsson1996} Jonsson, J. and Fridson, M. (1996)
Forecasting Default Rates on High Yield Bonds.
{\em Journal of Fixed Income} 6(1): 69-77.

\bibitem[Jordan and Jordan, 1997]{Jordan1997} Jordan, B.D. and Jordan, S. (1997)
Special repo rates: An empirical analysis.
{\em Journal of Finance} 52(5): 2051-2072.

\bibitem[Joseph, 1952]{Joseph1952} Joseph, A. (1952)
The Whittaker-Henderson Method of Graduation.
{\em Journal of the Institute of Actuaries} 78(1): 99-114.

\bibitem[Joshi and Lambert, 2011]{Joshi2011} Joshi, N.N. and Lambert, J.H. (2011)
Diversification of infrastructure projects for emergent and unknown non-systematic risks.
{\em Journal of Risk Research} 14(6): 717-733.

\bibitem[Joslin and Konchitchki, 2018]{Joslin2017} Joslin, S. and Konchitchki, Y. (2018)
Interest rate volatility, the yield curve, and the macroeconomy.
{\em Journal of Financial Economics} 128(2): 344-362.

\bibitem[Joslin, Priebsch and Singleton, 2014]{Joslin2014} Joslin, S., Priebsch, M. and Singleton, K.J. (2014)
Risk Premiums in Dynamic Term Structure Models with Unspanned Macro Risks.
{\em Journal of Finance} 69(3): 1197-1233.

\bibitem[Jostarndt and Sautner, 2010]{Jostarndt2010} Jostarndt, P. and Sautner, Z. (2010)
Out-of-Court Restructuring versus Formal Bankruptcy in a Non-Interventionist Bankruptcy Setting.
{\em Review of Finance} 14(4): 623-668.

\bibitem[Jostova {\em et al}, 2013]{Jostova2013} Jostova, G., Nikolova, S., Philipov, A. and Stahel, C.W. (2013)
Momentum in Corporate Bond Returns.
{\em Review of Financial Studies} 26(7): 1649-1693.

\bibitem[Joyce, Lildholdt and Sorensen, 2010]{Joyce2010} Joyce, M., Lildholdt, P. and Sorensen, S. (2010)
Extracting Inflation Expectations and Inflation Risk Premia from the Term Structure: A Joint Model of the UK Nominal and Real Yield Curves.
{\em Journal of Banking \& Finance} 34(2): 281-294.

\bibitem[Judd, Kubler and Schmedders, 2011]{Judd2011} Judd, K.L., Kubler, F. and Schmedders, K. (2011)
Bond Ladders and Optimal Portfolios.
{\em Review of Financial Studies} 24(12): 4123-4166.

\bibitem[Julio, Hassan and Ngene, 2013]{Julio2013} Julio, I.F., Hassan, M.K. and Ngene, G.M. (2013)
Trading Strategies in Futures Markets.
{\em Global Journal of Finance and Economics} 10(1): 1-12.

\bibitem[Junkus, 1991]{Junkus1991} Junkus, J.C. (1991)
Systematic skewness in futures contracts.
{\em Journal of Futures Markets} 11(1): 9-24.

\bibitem[Jurek, 2014]{Jurek2014} Jurek, J.W. (2014)
Crash-neutral currency carry trades.
{\em Journal of Financial Economics} 113(3): 325-347.

\bibitem[Kablan, 2009]{Kablan2009} Kablan, A. (2009)
Adaptive Neuro-Fuzzy Inference System for Financial Trading using Intraday Seasonality Observation Model.
{\em International Journal of Economics and Management Engineering} 3(10): 1909-1918.

\bibitem[Kahn and Lemmon, 2015]{Kahn2015} Kahn, R.N. and Lemmon, M. (2015)
Smart Beta: The Owner's Manual.
{\em Journal of Portfolio Management} 41(2): 76-83.

\bibitem[Kahn and Lemmon, 2016]{Kahn2016} Kahn, R.N. and Lemmon, M. (2016)
The Asset Manager's Dilemma: How Smart Beta Is Disrupting the Investment Management Industry.
{\em Financial Analysts Journal} 72(1): 15-20.

\bibitem[Kahneman and Tversky, 1979]{Kahneman1979} Kahneman, D. and Tversky, A. (1979)
Prospect theory: an analysis of decision under risk.
{\em Econometrica} 47(2): 263-292.

\bibitem[Kakodkar {\em et al}, 2006]{Kakodkar2006} Kakodkar, A., Galiani, S., J\'{o}nsson, J.G. and Gallo, A. (2006)
{\em Credit Derivatives Handbook 2006 -- Vol. 2: A Guide to the Exotics Credit Derivatives Market.} New York, NY: Credit Derivatives Strategy, Merrill Lynch.

\bibitem[Kakushadze, 2015a]{Kakushadze2015a} Kakushadze, Z. (2015a)
Phynance.
{\em Universal Journal of Physics and Application} 9(2): 64-133.
Available online: \url{https://ssrn.com/abstract=2433826}.

\bibitem[Kakushadze, 2015b]{Kakushadze2015b} Kakushadze, Z. (2015b)
Mean-Reversion and Optimization.
{\em Journal of Asset Management} 16(1): 14-40.
Available online: \url{https://ssrn.com/abstract=2478345}.

\bibitem[Kakushadze, 2015c]{Kakushadze2015c} Kakushadze, Z. (2015c)
4-Factor Model for Overnight Returns.
{\em Wilmott Magazine} 2015(79): 56-62. Available online: \url{https://ssrn.com/abstract=2511874}.

\bibitem[Kakushadze, 2015d]{Kakushadze2015d} Kakushadze, Z. (2015d)
On Origins of Alpha.
{\em Hedge Fund Journal} 108: 47-50. Available online: \url{https://ssrn.com/abstract=2575007}.

\bibitem[Kakushadze, 2015e]{Kakushadze2015e} Kakushadze, Z. (2015e)
Heterotic Risk Models.
{\em Wilmott Magazine} 2015(80): 40-55.
Available online: \url{https://ssrn.com/abstract=2600798}.

\bibitem[Kakushadze, 2016]{Kakushadze2016} Kakushadze, Z. (2016)
101 Formulaic Alphas.
{\em Wilmott Magazine} 2016(84): 72-80.
Available online: \url{https://ssrn.com/abstract=2701346}.

\bibitem[Kakushadze and Tulchinsky, 2016]{KakushadzeTulchinsky2016} Kakushadze, Z. and Tulchinsky, I. (2016)
Performance v. Turnover: A Story by 4,000 Alphas.
{\em Journal of Investment Strategies} 5(2): 75-89.
Available online: \url{http://ssrn.com/abstract=2657603}.

\bibitem[Kakushadze and Yu, 2016a]{KakushadzeYu2016a} Kakushadze, Z. and Yu, W. (2016a)
Multifactor Risk Models and Heterotic CAPM.
{\em Journal of Investment Strategies} 5(4): 1-49.
Available online: \url{https://ssrn.com/abstract=2722093}.

\bibitem[Kakushadze and Yu, 2016b]{KakushadzeYu2016b} Kakushadze, Z. and Yu, W. (2016b)	
Statistical Industry Classification.
{\em Journal of Risk \& Control} 3(1): 17-65.
Available online: \url{https://ssrn.com/abstract=2802753}.

\bibitem[Kakushadze and Yu, 2017a]{KakushadzeYu2017a} Kakushadze, Z. and Yu, W. (2017a)
Statistical Risk Models.
{\em Journal of Investment Strategies} 6(2): 1-40.
Available online: \url{https://ssrn.com/abstract=2732453}.

\bibitem[Kakushadze and Yu, 2017b]{KakushadzeYu2017b} Kakushadze, Z. and Yu, W. (2017b)
How to Combine a Billion Alphas.
{\em Journal of Asset Management} 18(1): 64-80.
Available online: \url{https://ssrn.com/abstract=2739219}.

\bibitem[Kakushadze and Yu, 2017c]{KakushadzeYu2017c} Kakushadze, Z. and Yu, W. (2017c)	
*K-Means and Cluster Models for Cancer Signatures.
{\em Biomolecular Detection and Quantification} 13: 7-31.
Available online: \url{https://ssrn.com/abstract=2908286}.

\bibitem[Kakushadze and Yu, 2018a]{KakushadzeYu2018} Kakushadze, Z. and Yu, W. (2018a)
Decoding Stock Market with Quant Alphas.
{\em Journal of Asset Management} 19(1): 38-48. Available online: \url{https://ssrn.com/abstract=2965224}.

\bibitem[Kakushadze and Yu, 2018b]{KakushadzeYu2018b} Kakushadze, Z. and Yu, W. (2018b)
Notes on Fano Ratio and Portfolio Optimization.
{\em Journal of Risk \& Control} 5(1): 1-33. Available online: \url{https://ssrn.com/abstract=3050140}.

\bibitem[Kalev and Inder, 2006]{Kalev2006} Kalev, P.S. and Inder, B.A. (2006)
The information content of the term structure of interest rates.
{\em Applied Economics} 38(1): 33-45.

\bibitem[Kallberg, Liu and Trzcinka, 2000]{Kallberg2000} Kallberg, J.G., Liu, C.L. and Trzcinka, C. (2000)
The Value Added from Investment Managers: An Examination of Funds of REITs.
{\em Journal of Financial and Quantitative Analysis} 35(3): 387-408.

\bibitem[Kalman, 1960]{Kalman1960} Kalman, P.E. (1960)
A New Approach to Linear Filtering and Prediction Problems.
{\em Journal of Basic Engineering} 82(1): 35-45.

\bibitem[Kambhu, 2006]{Kambhu2006} Kambhu, J. (2006)
Trading Risk, Market Liquidity, and Convergence Trading in the Interest Rate Swap Spread.
{\em Federal Reserve Bank of New York, Economic Policy Review} 12(1): 1-13.

\bibitem[Kaminski, 2004]{Kaminski2004} Kaminski, V. (2004)
{\em Managing Energy Price Risk: The New Challenges and Solutions.}
London, UK: Risk Books.

\bibitem[Kandel, Ofer and Sarig, 1996]{Kandel1996} Kandel, S., Ofer, A.R. and Sarig, O. (1996)
Real Interest Rates and Inflation: An Ex-Ante Empirical Analysis.
{\em Journal of Finance} 51(1): 205-225.

\bibitem[Kandel and Stambaugh, 1987]{Kandel1987} Kandel, S. and Stambaugh, R.F. (1987)
Long-horizon Returns and Short-horizon Models.
{\em CRSP Working Paper No. 222.} Chicago, IL: University of Chicago.

\bibitem[Kang and Gardner, 1989]{Kang1989} Kang, H.B. and Gardner, J. (1989)
Selling Price and Marketing Time in the Residential Real Estate Market.
{\em Journal of Real Estate Research} 4(1): 21-35.

\bibitem[Kang and Lee, 1996]{Kang1996} Kang, J.K. and Lee, Y.W. (1996)
The pricing of convertible debt offerings.
{\em Journal of Financial Economics} 41(2): 231-248.

\bibitem[Kang, Liu and Ni, 2002]{Kang2002} Kang, J., Liu, M.H. and Ni, S.X. (2002)
Contrarian and momentum strategies in the China stock market: 1993-2000.
{\em Pacific-Basin Finance Journal} 10(3): 243-265.

\bibitem[Kapadia and Szado, 2007]{Kapadia2007} Kapadia, N. and Szado, E. (2007)
The Risk Return Characteristics of the Buy-Write Strategy on the Russell 2000 Index.
{\em Journal of Alternative Investments} 9(4): 39-56.

\bibitem[Kaplan and Lummer, 1998]{Kaplan1998} Kaplan, P. and Lummer, S.L. (1998)
Update: GSCI Collateralized Futures as a Hedging Diversification Tool for Institutional Portfolios.
{\em Journal of Investing} 7(4): 11-18.

\bibitem[Kara, Boyacioglu and Baykan, 2011]{Kara2011} Kara, Y., Boyacioglu, M.A. and Baykan, O.K. (2011)
Predicting direction of stock price index movement using artificial neural networks and support vector machines: The sample of the Istanbul Stock Exchange.
{\em Expert Systems with Applications} 38(5): 5311-5319.

\bibitem[Karlik and Vehbi, 2011]{Karlik2011} Karlik, B. and Vehbi, A. (2011)
Performance Analysis of Various Activation Functions in Generalized MLP Architectures of Neural Networks.
{\em International Journal of Artificial Intelligence and Expert Systems} 1(4): 111-122.

\bibitem[Karolyi and Kho, 2004]{Karolyi2004} Karolyi, G.A. and Kho, B.C. (2004)
Momentum strategies: Some bootstrap tests.
{\em Journal of Empirical Finance} 11(4): 509-536.

\bibitem[Karolyi and Sanders, 1998]{Karolyi1998} Karolyi, G.A. and Sanders, A.B. (1998)
The Variation of Economic Risk Premiums in Real Estate Returns.
{\em Journal of Real Estate Finance and Economics} 17(3): 245-262.

\bibitem[Karolyi and Shannon, 1999]{Karolyi1999} Karolyi, G.A. and Shannon, J. (1999)
Where's the Risk in Risk Arbitrage?
{\em Canadian Investment Review} 12(2): 12-18.

\bibitem[Kau {\em et al}, 1995]{Kau1995} Kau, J.B., Keenan, D.C., Muller, W.J., III and Epperson, J.F. (1995)
The valuation at origination of fixed-rate mortgages with default and prepayment.
{\em Journal of Real Estate Finance and Economics} 11(1): 5-36.

\bibitem[Kawaller, Koch and Ludan, 2002]{Kawaller2002}
Kawaller, I.G., Koch, P.D. and Ludan, L. (2002)
Calendar spreads, outright futures positions and risk.
{\em Journal of Alternative Investments} 5(3): 59-74.

\bibitem[Kazemi and Li, 2009]{Kazemi2009} Kazemi, H. and Li, Y. (2009)
Market timing of CTAs: An examination of systematic CTAs vs. discretionary CTAs.
{\em Journal of Futures Markets} 29(11): 1067-1099.

\bibitem[Keane, 1996]{Keane1996} Keane, F. (1996)
Repo rate patterns for new Treasury notes.
{\em Federal Reserve Bank of New York, Current Issues in Economics and Finance} 2(10): 1-6.

\bibitem[Kelly, 1956]{Kelly1956} Kelly, J.L. (1956)
A New Interpretation of Information Rate.
{\em Bell System Technical Journal} 35(4): 917-926.

\bibitem[Kemp, 2007]{Kemp2007} Kemp, K. (2007)
{\em Flipping confidential: The secrets of renovating property for profit in any market.}
Hoboken, NJ: John Wiley \& Sons, Inc.

\bibitem[Kenett {\em et al}, 2013]{Kenett2013} Kenett, D.Y., Ben-Jacob, E., Stanley, H.E. and gur-Gershgoren, G. (2013)
How High Frequency Trading Affects a Market Index.
{\em Scientific Reports} 3: 2110.

\bibitem[Kenyon, 2008]{Kenyon2008} Kenyon, C. (2008)
Inflation is normal.
{\em Risk}, July 2008, pp. 76-82.

\bibitem[Khan, 2002]{Khan2002} Khan, S.A. (2002)
Merger Arbitrage: A Long-Term Investment Strategy.
{\em Journal of Wealth Management} 4(4): 76-81.

\bibitem[Khandani and Lo, 2011]{Khandani2011}
Khandani, A. and Lo, A.W. (2011)
What Happened to the Quants in August 2007? Evidence from Factors and Transactions Data.
{\em Journal of Financial Markets} 14(1): 1-46.

\bibitem[Khang, 1983]{Khang1983} Khang, C.H. (1983)
A dynamic global portfolio immunization strategy in the world of multiple interest rate changes: A dynamic immunization and minimax theorem.
{\em Journal of Financial and Quantitative Analysis} 18(3): 355-363.

\bibitem[Khuzwayo and Mar\'{e}, 2014]{Khuzwayo2014} Khuzwayo, B. and Mar\'{e}, E. (2014)
Aspects of volatility targeting for South African equity investors.
{\em South African Journal of Economic and Management Sciences} 17(5): 691-699.

\bibitem[Kidd, 2014]{Kidd2014} Kidd, D. (2014)
Global Tactical Asset Allocation: One Strategy Fits All?
In: {\em Investment Risk and Performance.} Charlottesville, VA: CFA Institute.

\bibitem[Kilgallen, 2012]{Kilgallen2012} Kilgallen, T. (2012)
Testing the Simple Moving Average across Commodities, Global Stock Indices, and Currencies.
{\em Journal of Wealth Management} 15(1): 82-100.

\bibitem[Kim, 1990]{Kim1990} Kim, I.J. (1990)
The analytic valuation of American options.
{\em Review of Financial Studies} 3(4): 547-572.

\bibitem[Kim, 2011]{Kim2011} Kim, K. (2011)
Performance Analysis of Pairs Trading Strategy Utilizing High Frequency Data with an Application to KOSPI 100 Equities.
{\em Working Paper.} Available online: \url{https://ssrn.com/abstract=1913707}.

\bibitem[Kim, 2003]{Kim2003} Kim, K.J. (2003)
Financial time series forecasting using support vector machines.
{\em Neurocomputing} 55(1-2): 307-319.

\bibitem[Kim, 2006]{Kim2006} Kim, K.J. (2006)
Artificial neural networks with evolutionary instance selection for financial forecasting.
{\em Expert Systems with Applications} 30(3): 519-526.

\bibitem[Kim and Enke, 2016]{KimEnke2016} Kim, Y. and Enke, D. (2016)
Using neural networks to forecast volatility for an asset allocation strategy based on the target volatility.
{\em Procedia Computer Science} 95: 281-286.

\bibitem[Kim and Han, 2000]{Kim2000} Kim, K. and Han, I. (2000)
Genetic algorithms approach to feature discretization in artificial neural networks for the prediction of stock price index.
{\em Expert Systems with Applications} 19(2): 125-132.

\bibitem[Kim and Leuthold, 1997]{Kim1997} Kim, M.-K. and Leuthold, R.M. (1997)
The Distributional Behavior of Futures Price Spread Changes: Parametric and Nonparametric Tests for Gold, T-Bonds, Corn, and Live Cattle.
{\em Working Paper.} Available online: \url{https://ageconsearch.umn.edu/bitstream/14767/1/aceo9703.pdf}.

\bibitem[Kim, Li and Zhang, 2016]{Kim2016} Kim, G.H., Li, H. and Zhang, W. (2016)
CDS-Bond Basis and Bond Return Predictability.
{\em Journal of Empirical Finance} 38: 307-337.

\bibitem[Kim, Li and Zhang, 2017]{Kim2017} Kim, G.H., Li, H. and Zhang, W. (2017)
The CDS-Bond Basis Arbitrage and the Cross Section of Corporate Bond Returns.
{\em Journal of Futures Markets} 37(8): 836-861.

\bibitem[Kim {\em et al}, 2016]{KimKim2016} Kim, Y.B., Kim, J.G., Kim, W., Im, J.H., Kim, T.H., Kang, S.J. and Kim, C.H. (2016)
Predicting Fluctuations in Cryptocurrency Transactions Based on User Comments and Replies.
{\em PLoS ONE} 11(8): e0161197.

\bibitem[King, 1986]{King1986} King, R. (1986)
Convertible Bond Valuation: An Empirical Test.
{\em Journal of Financial Research} 9(1): 53-69.

\bibitem[King and Mauer, 2014]{King2014} King, T.H.D. and Mauer, D.C. (2014)
Determinants of corporate call policy for convertible bonds.
{\em Journal of Corporate Finance} 24: 112-134.

\bibitem[Kingma and Ba, 2014]{Kingma2014} Kingma, D.P. and Ba, J. (2014)
Adam: A Method for Stochastic Optimization.
{\em Working Paper.} Available online: \url{https://arxiv.org/pdf/1412.6980}.

\bibitem[Kirby and Ostdiek, 2012]{Kirby2012} Kirby, C. and Ostdiek, B. (2012)
It's All in the Timing: Simple Active Portfolio Strategies that Outperform Na\"{i}ve Diversification.
{\em Journal of Financial and Quantitative Analysis} 47(2): 437-467.

\bibitem[Kirilenko {\em et al}, 2017]{Kirilenko2017} Kirilenko, A., Kyle, A., Samadi, M. and Tuzun, T. (2017)
The Flash Crash: High-Frequency Trading in an Electronic Market.
{\em Journal of Finance} 72(3): 967-998.

\bibitem[Kishore, 2012]{Kishore2012} Kishore, V. (2012)
Optimizing Pairs Trading of US Equities in a High Frequency Setting.
{\em Working Paper.} Available online:\\
\url{https://repository.upenn.edu/cgi/viewcontent.cgi?article=1095&context=wharton_research_scholars}.

\bibitem[Kitsul and Wright, 2013]{Kitsul2013} Kitsul, Y. and Wright, J.H. (2013)
The Economics of Options-Implied Inflation Probability Density Functions.
{\em Journal of Financial Economics} 110(3): 696-711.

\bibitem[Klingler and Sundaresan, 2016]{Klinger2016} Klingler, S. and Sundaresan, S.M. (2016)
An Explanation of Negative Swap Spreads: Demand for Duration from Underfunded Pension Plans.
{\em Working Paper.} Available online: \url{https://ssrn.com/abstract=2814975}.

\bibitem[Knight, 2002]{Knight2002} Knight, J.R. (2002)
Listing Price, Time on Market, and Ultimate Selling Price: Causes and Effects of Listing Price Changes.
{\em Real Estate Economics} 30(2): 213-237.

\bibitem[Kobor, Shi and Zelenko, 2005]{Kobor2005} Kobor, A., Shi, L. and Zelenko, I. (2005)
What Determines U.S. Swap Spreads?
{\em World Bank Working Paper No. 62.} Washington, DC: World Bank.

\bibitem[Kocherlakota, Rosenbloom and Shiu, 1988]{Kocherlakota1988} Kocherlakota, R., Rosenbloom, E. and Shiu, E. (1988)
Algorithms for cash-flow matching.
{\em Transactions of Society of Actuaries} 40: 477-484.

\bibitem[Kocherlakota, Rosenbloom and Shiu, 1990]{Kocherlakota1990} Kocherlakota, R., Rosenbloom, E. and Shiu, E. (1990)
Cash-flow matching and linear programming duality.
{\em Transactions of Society of Actuaries} 42: 281-293.

\bibitem[Kochin and Parks, 1988]{Kochin1988} Kochin, L. and Parks, R. (1988)
Was the tax-exempt bond market inefficient or were future expected tax rates negative?
{\em Journal of Finance} 43(4): 913-931.

\bibitem[Koijen, Moskowitz, Pedersen and Vrugt, 2018]{Koijen2018} Koijen, R.S.J., Moskowitz, T.J., Pedersen, L.H. and Vrugt, E.B. (2018)
Carry.
{\em Journal of Financial Economics} 127(2): 197-225.

\bibitem[Kolb and Chiang, 1981]{Kolb1981} Kolb, R.W. and Chiang, R. (1981)
Improving Hedging Performance Using Interest Rate Futures.
{\em Financial Management} 10(3): 72-79.

\bibitem[Kolb and Chiang, 1982]{Kolb1982} Kolb, R.W. and Chiang, R. (1982)
Duration, Immunization, and Hedging with Interest Rate Futures.
{\em Journal of Financial Research} 5(2): 161-170.

\bibitem[Konstantinidi and Skiadopoulos, 2016]{Konstantinidi2016} Konstantinidi, E. and Skiadopoulos, G. (2016)
How does the market variance risk premium vary over time? Evidence from S\&P 500 variance swap investment returns.
{\em Journal of Banking \& Finance} 62: 62-75.

\bibitem[Koopman, Lucas and Schwaab, 2012]{Koopman2012} Koopman, S.J., Lucas, A. and Schwaab, B. (2012)
Dynamic factor models with macro, frailty, and industry effects for U.S. default counts: The credit crisis of 2008.
{\em Econometric Reviews} 30(4): 521-532.

\bibitem[Korajczyk and Murphy, 2017]{Korajczyk2017} Korajczyk, R.A. and Murphy, D. (2017)
High Frequency Market Making to Large Institutional Trades.
{\em Working Paper.} Available online: \url{https://ssrn.com/abstract=2567016}.

\bibitem[Korajczyk and Sadka, 2004]{Korajczyk2004} Korajczyk, R.A. and Sadka, R. (2004)
Are momentum profits robust to trading costs?
{\em Journal of Finance} 59(3): 1039-1082.

\bibitem[Kordonis, Symeonidis and Arampatzis, 2016]{Kordonis2016} Kordonis, J., Symeonidis, A. and Arampatzis, A. (2016)
Stock Price Forecasting via Sentiment Analysis on Twitter.
In: {\em Proceedings of the 20th Pan-Hellenic Conference on Informatics (PCI'16).} New York, NY: ACM, Article No. 36.

\bibitem[Kordos and Cwiok, 2011]{Kordos2011} Kordos, M. and Cwiok, A. (2011)
A new approach to neural network based stock trading strategy.
In: Yin, H., Wang, W. and Rayward-Smith, V. (eds.) {\em Intelligent Data Engineering and Automated Learning-IDEAL.}
Berlin, Germany: Springer, pp. 429-436.

\bibitem[Korkeamaki and Michael, 2013]{Korkeamaki2013} Korkeamaki, T. and Michael, T.B. (2013)
Where are they now? An analysis of the life cycle of convertible bonds.
{\em Financial Review} 48(3): 489-509.

\bibitem[Korol, 2013]{Korol2013} Korol, T. (2013)
Early warning models against bankruptcy risk for Central European and Latin American enterprises.
{\em Economic Modelling} 31: 22-30.

\bibitem[Kozhan, Neuberger and Schneider, 2013]{Kozhan2013} Kozhan, R., Neuberger, A. and Schneider, P. (2013)
The Skew Risk Premium in the Equity Index Market.
{\em Review of Financial Studies} 26(9): 2174-2203.

\bibitem[Kozhan and Tham, 2012]{Kozhan2012} Kozhan, R. and Tham, W.W. (2012)
Execution Risk in High-Frequency Arbitrage.
{\em Management Science} 58(11): 2131-2149.

\bibitem[Kozhemiakin, 2007]{Kozhemiakin2007} Kozhemiakin, A.V. (2007)
The Risk Premium of Corporate Bonds.
{\em Journal of Portfolio Management} 33(2): 101-109.

\bibitem[Kozicki and Tinsley, 2012]{Kozicki2012}Kozicki, S. and Tinsley, P.A. (2012)
Effective Use of Survey Information in Estimating the Evolution of Expected Inflation.
{\em Journal of Money, Credit and Banking} 44(1): 145-169.

\bibitem[Kozlov and Petajisto, 2013]{Kozlov2013} Kozlov, M. and Petajisto, A. (2013)
Global Return Premiums on Earnings Quality, Value, and Size.
{\em Working Paper.} Available online: \url{https://ssrn.com/abstract=2179247}.

\bibitem[Kraenzlin, 2007]{Kraenzlin2007} Kraenzlin, S. (2007)
The characteristics and development of the Swiss franc repurchase agreement market.
{\em Financial Markets and Portfolio Management} 21(2): 241-261.

\bibitem[Krainer, 2001]{Krainer2001} Krainer, J. (2001)
A Theory of Liquidity in Residential Real Estate Markets.
{\em Journal of Urban Economics} 49(1): 32-53.

\bibitem[Krause, Ehsani and Lien, 2014]{Krause2014} Krause, T., Ehsani, S. and Lien, D. (2014)
Exchange-traded funds, liquidity and volatility.
{\em Applied Financial Economics} 24(24): 1617-1630.

\bibitem[Krauss, 2017]{Krauss2017} Krauss, C. (2017)
Statistical arbitrage pairs trading strategies: Review and outlook.
{\em Journal of Economic Surveys} 31(2): 513-545.

\bibitem[Krauss and St\"{u}binger, 2017]{KraussStubinger2017} Krauss, C. and St\"{u}binger, J. (2017)
Non-linear dependence modelling with bivariate copulas: Statistical arbitrage pairs trading on the S\&P 100.
{\em Applied Economics} 23(1): 1-18.

\bibitem[Krishnamurthy, 2002]{Krishnamurthy2002} Krishnamurthy, A. (2002)
The Bond/Old-Bond Spread.
{\em Journal of Financial Economics} 66(2): 463-506.

\bibitem[Kristoufek, 2015]{Kristoufek2015} Kristoufek, L. (2015)
What Are the Main Drivers of the Bitcoin Price? Evidence from Wavelet Coherence Analysis.
{\em PLoS ONE} 10(4): e0123923.

\bibitem[Kroner and Sultan, 1993]{Kroner1993} Kroner, K.F. and Sultan, J. (1993)
Time-Varying Distributions and Dynamic Hedging with Foreign Currency Futures.
{\em Journal of Financial and Quantitative Analysis} 28(4): 535-551.

\bibitem[Kruttli, Monin and Watugala, 2018]{Kruttli2018} Kruttli, M., Monin, P. and Watugala, S.W. (2018)
Investor Concentration, Flows, and Cash Holdings: Evidence from Hedge Funds.
{\em Working Paper.} Available online: \url{https://ssrn.com/abstract=3031663}.

\bibitem[Kryzanowski, Galler and Wright, 1993]{Kryzanowski1993} Kryzanowski, L., Galler, M. and Wright, D. (1993)
Using Artificial Neural Networks to Pick Stocks.
{\em Financial Analysts Journal} 49(4): 21-27.

\bibitem[Kuberek and Pefley, 1983]{Kuberek1983} Kuberek, R.C. and Pefley, N.G. (1983)
Hedging Corporate Debt with U.S. Treasury Bond Futures.
{\em Journal of Futures Markets} 3(4): 345-353.

\bibitem[Kudryavtsev, 2012]{Kudryavtsev2012} Kudryavtsev, A. (2012)
Overnight stock price reversals.
{\em Journal of Advanced Studies in Finance} 3(2): 162-170.

\bibitem[Kuhle and Alvayay, 2000]{Kuhle2000} Kuhle, J. and Alvayay, J. (2000)
The Efficiency of Equity REIT Prices.
{\em Journal of Real Estate Portfolio Management} 6(4): 349-354.

\bibitem[Kumar, 2009]{Kumar2009} Kumar, A. (2009)
Who Gambles in the Stock Market?
{\em Journal of Finance} 64(4): 1889-1933.

\bibitem[Kumar, 2012]{Kumar2012} Kumar, V.A. (2012)
Money Laundering: Concept, Significance and its Impact.
{\em European Journal of Business and Management} 4(2): 113-119.

\bibitem[Kumar and Thenmozhi, 2001]{Kumar2001} Kumar, M. and Thenmozhi, M. (2001)
Forecasting Stock Index Movement: A Comparison of Support Vector Machines and Random Forest.
{\em Working Paper.} Available online: \url{https://ssrn.com/abstract=876544}.

\bibitem[Kwok, 2014]{Kwok2014} Kwok, Y.K. (2014)
Game option models of convertible bonds: Determinants of call policies.
{\em Journal of Financial Engineering} 1(4): 1450029.

\bibitem[Lafuente, 2013]{Lafuente2013} Lafuente, J.A. (2013)
Optimal cross-hedging under futures mispricing: A note.
{\em Journal of Derivatives \& Hedge Funds} 19(3): 181-188.

\bibitem[Lahmiri, 2014]{Lahmiri2014} Lahmiri, S. (2014)
Wavelet low- and high-frequency components as features for predicting stock prices with backpropagation neural networks.
{\em Journal of King Saud University -- Computer and Information Sciences} 26(2): 218-227.

\bibitem[Lai, Tseng and Huang, 2016]{Lai2016} Lai, H.-C., Tseng, T.-C. and Huang, S.-C. (2016)
Combining value averaging and Bollinger Band for an ETF trading strategy.
{\em Applied Economics} 48(37): 3550-3557.

\bibitem[Laitinen and Laitinen, 2000]{Laitinen2000} Laitinen, E.K. and Laitinen, T. (2000)
Bankruptcy prediction application of the Taylor's expansion in logistic regression.
{\em International Review of Financial Analysis} 9(4): 327-349.

\bibitem[Lakonishok {\em et al}, 2007]{Lakonishok2007} Lakonishok, J., Lee, I., Pearson, N.D. and Poteshman, A.M. (2007)
Option market activity.
{\em Review of Financial Studies} 20(3): 813-857.

\bibitem[Lakonishok, Shleifer and Vishny, 1994]{Lakonishok1994} Lakonishok, J., Shleifer, A. and Vishny, R.W. (1994)
Contrarian investment, extrapolation, and risk.
{\em Journal of Finance} 49(5): 1541-1578.

\bibitem[Lakonishok and Vermaelen, 1986]{Lakonishok1986} Lakonishok, J. and Vermaelen, T. (1986)
Tax-Induced Trading Around the Ex-Day.
{\em Journal of Financial Economics} 16(3): 287-319.

\bibitem[Lambert, Papageorgiou and Platania, 2006]{Lambert2006} Lambert, M., Papageorgiou, N. and Platania, F. (2006)
Market Efficiency and Hedge Fund Trading Strategies.
{\em Working Paper.} Available online: \url{https://www.edhec.edu/sites/www.edhec-portail.pprod.net/files/edhec_working_paper_market_efficiency_and_hedge_fund_trading_strategies_f.compressed.pdf}.

\bibitem[Lamoureux and Lastrapes, 1993]{Lamoureux1993} Lamoureux, C.G. and Lastrapes, W. (1993)
Forecasting stock return variance: towards understanding stochastic implied volatility.
{\em Review of Financial Studies} 6(2): 293-326.

\bibitem[Lamoureux and Wansley, 1987]{Lamoureux1987} Lamoureux, C. and Wansley, J. (1987)
Market Effects of Changes in the S\&P 500 Index.
{\em Financial Review} 22(1): 53-69.

\bibitem[Landes, Stoffels and Seifert, 1985]{Landes1985} Landes, W.J., Stoffels, J.D. and Seifert, J.A. (1985)
An Empirical Test of a Duration-Based Hedge: The Case of Corporate Bonds.
{\em Journal of Futures Markets} 5(2): 173-182.

\bibitem[Lang, Litzenberger and Liu, 1998]{Lang1998} Lang, L.H.P., Litzenberger, R.H. and Liu, A.L. (1998)
Determinants of Interest Rate Swap Spreads.
{\em Journal of Banking \& Finance} 22(12): 1507-1532.

\bibitem[Langetieg, Leibowitz and Kogelman, 1990]{Langetieg1990} Langetieg, T.C., Leibowitz, L. and Kogelman, S. (1990)
Duration Targeting and the Management of Multiperiod Returns.
{\em Financial Analysts Journal} 46(5): 35-45.

\bibitem[Larker and Lys, 1987]{Larker1987} Larker, D. and Lys, T. (1987)
An empirical analysis of the incentives to engage in costly information acquisition: The case of risk arbitrage.
{\em Journal of Financial Economics} 18(1): 111-126.

\bibitem[Larkin, Babin and Rose, 2004]{Larkin2004} Larkin, D.E., Babin, M.L. and Rose, C.A. (2004)
Structuring European real estate private equity funds.
{\em Briefings in Real Estate Finance} 3(3): 229-235.

\bibitem[Larsen and Resnick, 1998]{Larsen1998} Larsen, G. and Resnick, B. (1998)
Empirical Insights on Indexing.
{\em Journal of Portfolio Management} 25(1): 51-60.

\bibitem[Larsson and Flohr, 2011]{Larsson2011} Larsson, P. and Flohr, L. (2011) Optimal proxy-hedging of options on illiquid baskets.
{\em Working Paper.} Available online: \url{https://www.math.kth.se/matstat/seminarier/reports/M-exjobb11/110131a.pdf}.

\bibitem[Lasfer, 1995]{Lasfer1995} Lasfer, M.A. (1995) Ex-Day Behavior: Tax or Short-Term Trading Effects.
{\em Journal of Finance} 50(3): 875-897.

\bibitem[Laurent and Gregory, 2005]{Laurent2005} Laurent, J.-P. and Gregory, J. (2005)
Basket Default Swaps, CDOs and Factor Copulas.
{\em Journal of Risk} 7(4): 8-23.

\bibitem[Laurent, Cousin and Fermanian, 2011]{Laurent2011} Laurent, J.-P., Cousin, A. and Fermanian, J.D. (2011)
Hedging default risks of CDOs in Markovian contagion models.
{\em Quantitative Finance} 11(12): 1773-1791.

\bibitem[Laureti, Medo and Zhang, 2010]{Laureti2010} Laureti, P., Medo, M. and Zhang, Y.-C. (2010)
Analysis of Kelly-optimal portfolios.
{\em Quantitative Finance} 10(7): 689-697.

\bibitem[Lautier and Galli, 2004]{Lautier2004} Lautier, D. and Galli, A. (2004)
Simple and extended Kalman filters: an application to term structures of commodity prices.
{\em Applied Financial Economics} 14(13): 963-973.

\bibitem[Lazo {\em et al}, 2011]{Lazo2011} Lazo, J.K., Lawson, M., Larsen, P.H. and Waldman, D.M. (2011)
U.S. Economic Sensitivity to Weather Variability.
{\em Bulletin of the American Meteorological Society} 92(6): 709-720.

\bibitem[Lebeck, 1978]{Lebeck1978} Lebeck, W.W. (1978)
Futures trading and hedging.
{\em Food Policy} 3(1): 29-35.

\bibitem[Lee, 2010]{Lee2010} Lee, S. (2010)
The Changing Benefit of REITs to the Multi-Asset Portfolio.
{\em Journal of Real Estate Portfolio Management} 16(3): 201-215.

\bibitem[Lee, Guo and Wang, 2018]{Chuen2018} Lee, D.K.C., Guo, L. and Wang, Y.  (2018)
Cryptocurrency: A New Investment Opportunity?
{\em Journal of Alternative Investments} 20(3): 16-40.

\bibitem[Lee, Liao and Tung, 2017]{Lee2017} Lee, H., Liao, T. and Tung, P. (2017)
Investors' Heterogeneity in Beliefs, the VIX Futures Basis, and S\&P 500 Index Futures Returns.
{\em Journal of Futures Markets} 37(9): 939-960.

\bibitem[Lee and Oh, 1993]{Lee1993} Lee, S.B. and Oh, S.H. (1993)
Managing non-parallel shift risk of yield curve with interest rate futures.
{\em Journal of Futures Markets} 13(5): 515-526.

\bibitem[Lee and Oren, 2009]{Lee2009} Lee, Y. and Oren, S. (2009)
An equilibrium pricing model for weather derivatives in a multi-commodity setting.
{\em Energy Economics} 31(5): 702-713.

\bibitem[Lee and Stevenson, 2005]{Lee2005} Lee, S. and Stevenson, S. (2005)
The Case for REITs in the Mixed-Asset Portfolio in the Short and Long Run.
{\em Journal of Real Estate Portfolio Management} 11(1): 55-80.

\bibitem[Leggio and Lien, 2002]{Leggio2002} Leggio, K. and Lien, D. (2002)
Hedging gas bills with weather derivatives.
{\em Journal of Economics and Finance} 26(1): 88-100.

\bibitem[Lehecka, 2013]{Lehecka2013} Lehecka, G.V. (2013)
Hedging and Speculative Pressures: An Investigation of the Relationships among Trading Positions and Prices in Commodity Futures Markets.
In: {\em Proceedings of the NCCC-134 Conference on Applied Commodity Price Analysis, Forecasting, and Market Risk Management.}
Available online: \url{http://www.farmdoc.illinois.edu/nccc134/conf_2013/pdf/Lehecka_NCCC-134_2013.pdf}.

\bibitem[Lehmann, 1990]{Lehmann1990} Lehmann, B.N. (1990)
Fads, Martingales, and Market Efficiency.
{\em Quarterly Journal of Economics} 105(1): 1-28.

\bibitem[Leibowitz and Bova, 2013]{Leibowitz2013} Leibowitz, M.L. and Bova, A. (2013)
Duration Targeting and Index Convergence.
{\em Morgan Stanley Investment Management Journal} 3(1): 73-80.

\bibitem[Leibowitz, Bova and Kogelman, 2014]{Leibowitz2014} Leibowitz, M.L., Bova, A. and Kogelman, S. (2014)
Long-Term Bond Returns under Duration Targeting.
{\em Financial Analysts Journal} 70(1): 31-51.

\bibitem[Leibowitz, Bova and Kogelman, 2015]{Leibowitz2015} Leibowitz, M.L., Bova, A. and Kogelman, S. (2015)
Bond Ladders and Rolling Yield Convergence.
{\em Financial Analysts Journal} 71(2): 32-46.

\bibitem[Leigland, 2018]{Leigland2018} Leigland, J. (2018)
Changing Perceptions of PPP Risk and Return: The Case of Brownfield Concessions.
{\em Journal of Structured Finance} 23(4): 47-56.

\bibitem[Leland and Connor, 1995]{Leland1995} Leland, H. and Connor, G. (1995)
Optimal Cash Management for Investment Funds.
{\em Research Program in Finance Working Papers}, No. RPF-244. Berkeley, CA: University of California at Berkeley.

\bibitem[Leland and Panos, 1997]{Leland1997} Leland, E.C. and Panos, N. (1997)
The Puttable Bond Market: Structure, Historical Experience, and Strategies.
{\em Journal of Fixed Income} 7(3): 47-60.

\bibitem[Le Moigne and Viveiros, 2008]{LeMoigne2008} Le Moigne, C. and Viveiros, \'E. (2008)
Private Real Estate as an Inflation Hedge: An Updated Look with a Global Perspective.
{\em Journal of Real Estate Portfolio Management} 14(4): 263-286.

\bibitem[Leontsinis and Alexander, 2016]{Leontsinis2016} Leontsinis, S. and Alexander, C. (2016) Arithmetic variance swaps.
{\em Quantitative Finance} 17(4): 551-569.

\bibitem[Lessambo, 2016]{Lessambo2016} Lessambo, F.I. (2016)
{\em International Aspects of the US Taxation System.}
New York, NY: Palgrave Macmillan.

\bibitem[Leung {\em et al}, 2016]{Leung2016} Leung, T., Li, J., Li, X. and Wang, Z. (2016)
Speculative Futures Trading under Mean Reversion.
{\em Asia-Pacific Financial Markets} 23(4): 281-304.

\bibitem[Leung and Tse, 2013]{Leung2013} Leung, C.K.Y. and Tse, C.-Y. (2013)
Flippers in housing market search.
{\em Working Paper.} Available online: \url{https://hub.hku.hk/bitstream/10722/190689/1/Content.pdf}.

\bibitem[Levi and Reuter, 2006]{Levi2006} Levi, M. and Reuter, P. (2006)
Money Laundering.
{\em Crime and Justice} 34(1): 289-375.

\bibitem[Levin and Davidson, 2005]{Levin2005} Levin, A. and Davidson, A. (2005)
Prepayment Risk-and Option-Adjusted Valuation of MBS.
{\em Journal of Portfolio Management} 31(4): 73-85.

\bibitem[Levine and Pedersen, 2016]{Levine2016} Levine, A. and Pedersen, L.H. (2016)
Which Trend Is Your Friend?
{\em Financial Analysts Journal} 72(3): 51-66.

\bibitem[Levis and Liodakis, 1999]{Levis1999} Levis, M. and Liodakis, M. (1999)
The Profitability of Style Rotation Strategies in the United Kingdom.
{\em Journal of Portfolio Management} 26(1): 73-86.

\bibitem[Levitt and Syverson, 2008]{Levitt2008} Levitt, S.D. and Syverson, C. (2008)
Market Distortions When Agents Are Better Informed: The Value of Information in Real Estate Transactions.
{\em Review of Economics and Statistics} 90(4): 599-611.

\bibitem[Levy, 1991]{Levy1991}Levy, P.S. (1991)
Approaches to Investing in Distressed Securities: Active Approaches.
In: Bowman, T.A. (ed.)
{\em Analyzing Investment Opportunities in Distressed and Bankrupt Companies.} (AIMR Conference Proceedings, Vol. 1991, Iss. 1.) Chicago, IL: AIMR, pp. 44-46.

\bibitem[Levy and Lieberman, 2013]{Levy2013}Levy, A. and Lieberman, O. (2013)
Overreaction of country ETFs to US market returns: Intraday vs. daily horizons and the role of synchronized trading.
{\em Journal of Banking \& Finance} 37(5): 1412-1421.

\bibitem[Lewis, 1995]{Lewis1995} Lewis, K. (1995)
Puzzles in International Financial Markets.
In: Grossman G.M. and Rogoff, K. (eds.) {\em Handbook of International Economics}, Vol. 3.
Amsterdam, The Netherlands: North-Holland, Chapter 37.

\bibitem[Lewis, 2014]{Lewis2014} Lewis, M. (2014)
{\em Flash Boys: A Wall Street Revolt.}
New York, NY: W.W. Norton \& Company, Inc.

\bibitem[Lewis, Rogalski and Seward, 1999]{Lewis1999} Lewis, C.M., Rogalski, R.J. and Seward, J.K. (1999)
Is convertible debt a substitute for straight debt or for common equity?
{\em Financial Management} 28(3): 5-27.

\bibitem[Lewis and Verwijmeren, 2011]{Lewis2011} Lewis, C.M. and Verwijmeren, P. (2011)
Convertible security design and contract innovation.
{\em Journal of Corporate Finance} 17(4): 809-831.

\bibitem[Lhabitant, 2002]{Lhabitant2002} Lhabitant, F.-S. (2002)
{\em Hedge Funds: Myths and Limits.}
Chichester, UK: John Wiley \& Sons, Ltd.

\bibitem[L'Hoir and Boulhabel, 2010]{LHoir2010}  L'Hoir, M. and Boulhabel, M. (2010)
A Bond-Picking Model for Corporate Bond Allocation.
{\em Journal of Portfolio Management} 36(3): 131-139.

\bibitem[Li, 2000]{Li2000} Li, D.X. (2000)
On default correlation: a copula function approach.
{\em Journal of Fixed Income} 9(4): 43-54.

\bibitem[Li {\em et al}, 2018]{Li2018} Li, T.R., Chamrajnagar, A.S., Fong, X.R., Rizik, N.R. and Fu, F. (2018)
Sentiment-Based Prediction of Alternative Cryptocurrency Price Fluctuations Using Gradient Boosting Tree Model.
{\em Working Paper.} Available online: \url{https://arxiv.org/pdf/1805.00558.pdf}.

\bibitem[Li {\em et al}, 2014]{LiDeng2014} Li, X., Deng, X., Zhu, S., Wang, F. and Xie, H. (2014)
An intelligent market making strategy in algorithmic trading.
{\em Frontiers of Computer Science} 8(4): 596-608.

\bibitem[Li {\em et al}, 2015]{Li2015} Li, B., Hoi, S.C.H., Sahoo, D. and Liu, Z.-Y. (2015)
Moving average reversion strategy for on-line portfolio selection.
{\em Artificial Intelligence} 222: 104-123.

\bibitem[Li and Kleindorfer, 2009]{Li2009} Li, L. and Kleindorfer, P.R. (2009)
On hedging spark spread options in electricity markets.
{\em Risk and Decision Analysis} 1(4): 211-220.

\bibitem[Li, Sullivan and Garcia-Feij\'{o}o, 2014]{Li2014} Li, X., Sullivan, R.N. and Garcia-Feij\'{o}o, L. (2014)
The Limits to Arbitrage and the Low-Volatility Anomaly.
{\em Financial Analysts Journal} 70(1): 52-63.

\bibitem[Li, Sullivan and Garcia-Feij\'{o}o, 2016]{Li2016} Li, X., Sullivan, R.N. and Garcia-Feij\'{o}o, L. (2016)
The Low-Volatility Anomaly: Market Evidence on Systematic Risk vs. Mispricing.
{\em Financial Analysts Journal} 72(1): 36-47.

\bibitem[Li and Wang, 1995]{LiWang1995} Li, Y. and Wang, K. (1995)
The Predictability of REIT Returns and Market Segmentation.
{\em Journal of Real Estate Research} 10(5): 471-482.

\bibitem[Li and Yang, 2017]{Li2017} Li, P. and Yang, J. (2017)
Pricing Collar Options with Stochastic Volatility.
{\em Discrete Dynamics in Nature and Society} 2017: 9673630.

\bibitem[Li {\em et al}, 2012]{Li2012} Li, B., Zhao, P., Hoi, S.C.H. and Gopalkrishnan, V. (2012)
PAMR: Passive aggressive mean reversion strategy for portfolio selection.
{\em Machine Learning} 87(2): 221-258.

\bibitem[Liao, 2016]{Liao2016} Liao, G.Y. (2016)
Credit migration and covered interest rate parity.
{\em Working Paper.} Available online: \url{http://scholar.harvard.edu/files/gliao/files/creditcip.pdf}.

\bibitem[Lien, 1992]{Lien1992} Lien, D. (1992) Optimal Hedging and Spreading in Cointegrated Markets.
{\em Economics Letters} 40(1): 91-95.

\bibitem[Lien, 2004]{Lien2004} Lien, D. (2004) Cointegration and the Optimal Hedge Ratio: The General Case.
{\em Quarterly Review of Economics and Finance} 44(5): 654-658.

\bibitem[Lien, 2010]{Lien2010} Lien, D. (2010)
The effects of skewness on optimal production and hedging decisions: An application of the skew-normal distribution.
{\em Journal of Futures Markets} 30(3): 278-289.

\bibitem[Lien and Luo, 1993]{Lien1993} Lien, D. and Luo, X. (1993)
Estimating Multiperiod Hedge Ratios in Cointegrated Markets.
{\em Journal of Futures Markets} 13(8): 909-920.

\bibitem[Lien and Tse, 2000]{LienTse2000} Lien, D. and Tse, Y.K. (2000)
Hedging downside risk with futures contracts.
{\em Applied Financial Economics} 10(2): 163-170.

\bibitem[Lien and Wang, 2015]{Lien2015}
Lien, D. and Wang, Y. (2015)
Effects of skewness and kurtosis on production and hedging decisions: A skewed $t$ distribution approach.
{\em European Journal of Finance} 21(13-14): 1132-1143.

\bibitem[Liew and Budav\'{a}ri, 2016] {LiewB2016}Liew, J.K.-S. and Budav\'{a}ri, T. (2016)
Do Tweet Sentiments Still Predict the Stock Market?
{\em Working Paper.} Available online: \url{https://ssrn.com/abstract=2820269}.

\bibitem[Liew, Li and Budav\'{a}ri, 2018] {LiewLB2018}Liew, J.K.-S., Li, R.Z. and Budav\'{a}ri, T. (2018)
Crypto-Currency Investing Examined.
{\em Working Paper.} Available online: \url{https://ssrn.com/abstract=3157926}.

\bibitem[Liew and Mayster, 2018]{Liew2018} Liew, J.K.-S. and Mayster, B. (2018)
Forecasting ETFs with Machine Learning Algorithms.
{\em Journal of Alternative Investments} 20(3): 58-78.

\bibitem[Liew and Roberts, 2013]{LiewRoberts2013} Liew, J. and Roberts, R. (2013)
U.S. Equity Mean-Reversion Examined.
{\em Risks} 1(3): 162-175.

\bibitem[Liew and Vassalou, 2000]{Liew2000} Liew, J. and Vassalou, M. (2000)
Can Book-to-Market, Size and Momentum be Risk Factors that Predict Economic Growth?
{\em Journal of Financial Economics} 57(2): 221-245.

\bibitem[Liew and Wu, 2013]{Liew2013} Liew, R. and Wu, Y. (2013)
Pairs trading: A copula approach.
{\em Journal of Derivatives \& Hedge Funds} 19(1): 12-30.

\bibitem[Lin, Lan and Chuang, 2013]{Lin2013} Lin, L., Lan, L.-H. and Chuang, S.-s. (2013)
An Option-Based Approach to Risk Arbitrage in Emerging Markets: Evidence from Taiwan Takeover Attempts.
{\em Journal of Forecasting} 32(6): 512-521.

\bibitem[Lin, McCrae and Gulati, 2006]{Lin2006} Lin, Y.-X., McCrae, M. and Gulati, C. (2006)
Loss protection in pairs trading through minimum profit bounds: A cointegration approach.
{\em Journal of Applied Mathematics and Decision Sciences} 2006(4): 1-14.

\bibitem[Lin and Shyy, 2008]{Lin2008} Lin, S.-Y. and Shyy, G. (2008)
Credit Spreads, Default Correlations and CDO Tranching: New Evidence from CDS Quotes.
{\em Working Paper.} Available online: \url{https://ssrn.com/abstract=496225}.

\bibitem[Lin and Yung, 2004]{LinY2004} Lin, C.Y. and Yung, K. (2004)
Real Estate Mutual Funds: Performance and Persistence.
{\em Journal of Real Estate Research} 26(1): 69-93.

\bibitem[Lindahl, 1992]{Lindahl1992} Lindahl, M. (1992)
Minimum variance hedge ratios for stock index futures: duration and expiration effects.
{\em Journal of Futures Markets} 12(1): 33-53.

\bibitem[Lioui and Poncet, 2005]{Lioui2005} Lioui, A. and Poncet, P. (2005)
General equilibrium pricing of CPI derivatives.
{\em Journal of Banking \& Finance} 29(5): 1265-1294.

\bibitem[Litterman and Scheinkman, 1991]{Litterman1991} Litterman, R.B. and Scheinkman, J. (1991)
Common Factors Affecting Bond Returns.
{\em Journal of Fixed Income} 1(1): 54-61.

\bibitem[Litzenberger and Rabinowitz, 1995]{Litzenberger1995} Litzenberger, R.H. and Rabinowitz, N. (1995)
Backwardation in oil futures markets: Theory and empirical evidence.
{\em Journal of Finance} 50(3): 1517-1545.

\bibitem[Liu, Chang and Geman, 2017]{Liu2017} Liu, B., Chang, L.B. and Geman, H. (2017)
Intraday pairs trading strategies on high frequency data: The case of oil companies.
{\em Quantitative Finance} 17(1): 87-100.

\bibitem[Liu and Dash, 2012]{Liu2012} Liu, B. and Dash, S. (2012)
Volatility ETFs and ETNs.
{\em Journal of Trading} 7(1): 43-48.

\bibitem[Liu, Longstaff and Mandell, 2006)]{Liu2006} Liu, J., Longstaff, F.A. and Mandell, R.E. (2006)
The Market Price of Risk in Interest Rate Swaps: The Roles of Default and Liquidity Risks.
{\em Journal of Business} 79(5): 2337-2360.

\bibitem[Liu and Mei, 1992]{LiuMei1992} Liu, C.H. and Mei, J. (1992)
The Predictability of Returns on Equity REITs and their Co-Movement with Other Assets.
{\em Journal of Real Estate Finance and Economics} 5(4): 401-418.

\bibitem[Liu, Pantelous and von Mettenheim, 2018]{Liu2018} Liu, F., Pantelous, A.A. and von Mettenheim, H.-J. (2018)
Forecasting and trading high frequency volatility on large indices.
{\em Quantitative Finance} 18(5): 737-748.

\bibitem[Liu and Tang, 2010]{Liu2010} Liu, P. and Tang, K. (2010)
No-arbitrage conditions for storable commodities and the models of futures term structures.
{\em Journal of Banking \& Finance} 34(7): 1675-1687.

\bibitem[Liu and Tang, 2011]{Liu2011} Liu, P. and Tang, K. (2011)
The stochastic behavior of commodity prices with heteroscedasticity in the convenience yield.
{\em Journal of Empirical Finance} 18(2): 211-224.

\bibitem[Liu and van der Heijden, 2016]{Liu2016} Liu, Z.F. and van der Heijden, T. (2016)
Model-Free Risk-Neutral Moments and Proxies.
{\em Working Paper.} Available online: \url{https://ssrn.com/abstract=2641559}.

\bibitem[Liu and Xu, 1998]{LiuXu1998} Liu, J.-G. and Xu, E. (1998)
Pricing of mortgage-backed securities with option-adjusted spread.
{\em Managerial Finance} 24(9-10): 94-109.

\bibitem[Liu and Zhang, 2008]{Liu2008} Liu, L.X. and Zhang, L. (2008)
Momentum Profits, Factor Pricing, and Macroeconomic Risk.
{\em Review of Financial Studies} 21(6): 2417-2448.

\bibitem[Liverance, 2010]{Liverance2010} Liverance, E. (2010) Variance Swap.
In: Cont, R. (ed.) {\em Encyclopedia of Quantitative Finance.} Hoboken, NJ: John Wiley \& Sons, Inc.

\bibitem[Livnat and Mendenhall, 2006]{Livnat2006} Livnat, J. and Mendenhall, R.R. (2006)
Comparing the post-earnings announcement drift for surprises calculated from analyst and time series forecasts.
{\em Journal of Accounting Research} 44(1): 177-205.

\bibitem[Lo, 2008]{Lo2008} Lo, A. (2008)
Where Do Alphas Come From?: A New Measure of the Value of Active Investment Management.
{\em Journal of Investment Management} 6(2): 1-29.

\bibitem[Lo, 2016]{Lo2016} Lo, A. (2016)
What Is an Index?
{\em Journal of Portfolio Management} 42(2): 21-36.

\bibitem[Lo and MacKinlay, 1990]{Lo1990} Lo, A.W. and MacKinlay, A.C. (1990)
When Are Contrarian Profits Due to Stock Market Overreaction?
{\em Review of Financial Studies} 3(3): 175-205.

\bibitem[Lo, Mamaysky and Wang, 2000]{Lo2000} Lo, A., Mamaysky, H. and Wang, J. (2000)
Foundations of Technical Analysis: Computational Algorithms, Statistical Inference, and Empirical Implementation.
{\em Journal of Finance} 55(4): 1705-1765.

\bibitem[Lo, Orr and Zhang, 2017]{LoOZ2017} Lo, A.W., Orr, A. and Zhang, R. (2017)
The Growth of Relative Wealth and the Kelly Criterion.
{\em Working Paper.} Available online: \url{https://ssrn.com/abstract=2900509}.

\bibitem[Loh and Warachka, 2012]{Loh2012} Loh, R.K. and Warachka, M. (2012)
Streaks in earnings surprises and the cross-section of stock returns.
{\em Management Science} 58(7): 1305-1321.

\bibitem[Loncarski, ter Horst and Veld, 2006]{Loncarski2006} Loncarski, I., ter Horst, J.R. and Veld, C.H. (2006)
The Convertible Arbitrage Strategy Analyzed.
{\em Working Paper.} Available online: \url{https://pure.uvt.nl/ws/files/779871/98.pdf}.

\bibitem[Loncarski, ter Horst and Veld, 2009]{Loncarski2009} Loncarski, I., ter Horst, J. and Veld, C. (2009)
The Rise and Demise of the Convertible Arbitrage Strategy.
{\em Financial Analysts Journal} 65(5): 35-50.

\bibitem[Longstaff, 2005]{Longstaff2005} Longstaff, F. (2005)
Borrower Credit and the Valuation of Mortgage-Backed Securities.
{\em Real Estate Economics} 33(4): 619-661.

\bibitem[Longstaff, 2011]{Longstaff2011} Longstaff, F.A. (2011)
Municipal Debt and Marginal Tax Rates: Is There a Tax Premium in Asset Prices?
{\em Journal of Finance} 66(3): 721-751.

\bibitem[Low {\em et al}, 2002]{Low2002} Low, A., Muthuswamy, J., Sakar, S. and Terry, E. (2002)
Multiperiod hedging with futures contracts.
{\em Journal of Futures Markets} 22(12): 1179-1203.

\bibitem[Lozovaia and Hizhniakova, 2005]{Lozovaia2005} Lozovaia, T. and Hizhniakova, H. (2005)
How to Extend Modern Portfolio Theory to Make Money from Trading Equity Options.
{\em Working Paper.} Available online:\\ \url{http://www.ivolatility.com/doc/Dispersion_Article.pdf}.

\bibitem[Lu, Lee and Chiu, 2009]{Lu2009} Lu, C.J., Lee, T.S. and Chiu, C. (2009)
Financial time series forecasting using independent component analysis and support vector regression.
{\em Decision Support Systems} 47(2): 115-125.

\bibitem[Lu, Wang and Zhang, 2012]{Lu2012} Lu, L., Wang, J. and Zhang, G. (2012)
Long term performance of leveraged ETFs.
{\em Financial Services Review} 21(1): 63-80.

\bibitem[Lucas, Goodman and Fabozzi, 2006]{Lucas2006} Lucas, D.J., Goodman, L.S. and Fabozzi, F.J. (eds.) (2006)
{\em Collateralized Debt Obligations: Structures and Analysis.}
Hoboken, NJ: John Wiley \& Sons, Inc.

\bibitem[Lucca and Moench, 2012]{Lucca2012} Lucca, D.O. and Moench, E. (2012)
The Pre-FOMC Announcement Drift.
{\em Journal of Finance} 70(1): 329-371.

\bibitem[Lummer and Siegel, 1993]{Lummer1993} Lummer, S.L. and Siegel, L.B. (1993)
GSCI Collateralized Futures: A Hedging and Diversification Tool for Institutional Portfolio.
{\em Journal of Investing} 2(2): 75-82.

\bibitem[Lumpkin, 1987]{Lumpkin1987} Lumpkin, S.A. (1987)
Repurchase and Reverse Repurchase Agreements.
{\em Federal Reserve Bank of Richmond, Economic Review} 73(1): 15-23.

\bibitem[Lustig, Roussanov and Verdelhan, 2011]{Lustig2011} Lustig, H., Roussanov, N. and Verdelhan, A. (2011)
Common Risk Factors in Currency Markets.
{\em Review of Financial Studies} 24(11): 3731-3777.

\bibitem[Lustig, Roussanov and Verdelhan, 2014]{Lustig2014} Lustig, H., Roussanov, N. and Verdelhan, A. (2014)
Countercyclical currency risk premia.	
{\em Journal of Financial Economics} 111(3): 527-553.

\bibitem[Lustig and Verdelhan, 2007]{Lustig2007} Lustig, H. and Verdelhan, A. (2007)
The Cross-Section of Foreign Currency Risk Premia and US Consumption Growth Risk.
{\em American Economic Review} 97(1): 89-117.

\bibitem[Ma, Mercer and Walker, 1992]{Ma1992} Ma, K., Mercer, M. and Walker, M. (1992)
Rolling over futures contracts: A note.
{\em Journal of Futures Markets} 12(2): 203-217.

\bibitem[Maaravi and Levy, 2017]{Maaravi2017} Maaravi, Y. and Levy, A. (2017)
When your anchor sinks your boat: Information asymmetry in distributive negotiations and the disadvantage of making the first offer.
{\em Judgment and Decision Making} 12(5): 420-429.

\bibitem[Macaulay, 1938]{Macaulay1938} Macaulay, F.R. (1938)
{\em Some theoretical problems suggested by the movements of interest rates, bond yields and stock prices in the United States since 1856.}
New York, NY: NBER, Inc.

\bibitem[MacKinnon and Al Zaman, 2009]{MacKinnon2009} MacKinnon, G.H. and Al Zaman, A. (2009)
Real estate for the long term: the effect of return predictability on long-horizon allocations.
{\em Real Estate Economics} 37(1): 117-153.

\bibitem[Mackintosh, 2017]{Mackintosh2017} Mackintosh, P. (2017)
It's all about active ETFs.
{\em Journal of Index Investing} 7(4): 6-15.

\bibitem[Madhavan, 2012]{Madhavan2012} Madhavan, A. (2012)
Exchange-Traded Funds, Market Structure, and the Flash Crash.
{\em Financial Analysts Journal} 68(4): 20-35.

\bibitem[Madhavan, 2016]{Madhavan2016} Madhavan, A.N. (2016)
{\em Exchange-Traded Funds and the New Dynamics of Investing.}
Oxford, UK: Oxford University Press.

\bibitem[Madura and Ngo, 2008]{Madura2008} Madura, J. and Ngo, T. (2008)
Impact of ETF inception on the valuation and trading of component stocks.
{\em Applied Financial Economics} 18(12): 995-1007.

\bibitem[Maghrebi, Kim and Nishina, 2007]{Maghrebi2007} Maghrebi, N., Kim, M. and Nishina, K. (2007)
The KOSPI200 Implied Volatility Index: Evidence of Regime Shifts in Expected Volatility.
{\em Asia-Pacific Journal of Financial Studies} 36(2): 163-187.

\bibitem[Maheswaran and Yeoh, 2005]{Maheswaran2005} Maheswaran, K. and Yeoh, S.C. (2005)
The Profitability of Merger Arbitrage: Some Australian Evidence.
{\em Australian Journal of Management} 30(1): 111-126.

\bibitem[Malizia and Simons, 1991]{Malizia1991} Malizia, E.E. and Simons, R.A. (1991)
Comparing Regional Classifications for Real Estate Portfolio Diversification.
{\em Journal of Real Estate Research} 6(1): 53-77.

\bibitem[Malkiel, 2014]{Malkiel2014} Malkiel, B.G. (2014)
Is Smart Beta Really Smart?
{\em Journal of Portfolio Management} 40(5): 127-134.

\bibitem[Malpezzi, 1999]{Malpezzi1999} Malpezzi, S. (1999)
A Simple Error Correction Model of House Prices.
{\em Journal of Housing Economics} 8(1): 27-62.

\bibitem[Maluf and Albuquerque, 2013]{Maluf2013} Maluf, Y.S. and Albuquerque, P.H.M. (2013)
Empirical evidence: arbitrage with Exchange-traded Funds (ETFs) on the Brazilian market.
{\em Revista Contabilidade \& Finan\c{c}as} 24(61): 64-74.

\bibitem[Mancini-Griffoli and Ranaldo, 2011]{Mancini-Griffoli2011} Mancini-Griffoli, T. and Ranaldo, A. (2011)
Limits to Arbitrage During the Crisis: Funding Liquidity Constraints and Covered Interest Parity.
{\em Working Paper.} Available online: \url{https://ssrn.com/abstract=1549668}.

\bibitem[Mankiw and Summers, 1984]{Mankiw1984} Mankiw, N.G. and Summers, L.H. (1984)
Do Long-Term Interest Rates Overreact to Short-Term Interest Rates?
{\em Brookings Papers on Economic Activity}, No. 1, pp. 223-242.

\bibitem[Mann and Ramanlal, 1997]{Mann1997} Mann, S.V. and Ramanlal, P. (1997)
The relative performance of yield curve strategies.
{\em Journal of Portfolio Management} 23(4): 64-70.

\bibitem[Manoliu, 2004]{Manoliu2004} Manoliu, M. (2004)
Storage options valuation using multilevel trees and calendar spreads.
{\em International Journal of Theoretical and Applied Finance} 7(4): 425-464.

\bibitem[Maribu, Galli and Armstrong, 2007]{Maribu2007} Maribu, K.M., Galli, A. and Armstrong, M. (2007)
Valuation of spark-spread options with mean reversion and stochastic volatility.
{\em International Journal of Electronic Business Management} 5(3): 173-181.

\bibitem[Mark and Wu, 2001]{Mark2001} Mark, N.C. and Wu, Y. (2001)
Rethinking Deviations From Uncovered Interest Parity: the Role of Covariance Risk and Noise.
{\em Economic Journal} 108(451): 1686-1706.

\bibitem[Markowitz, 1952]{Markowitz1952} Markowitz, H. (1952)
Portfolio Selection.
{\em Journal of Finance} 7(1): 77-91.

\bibitem[Markwardt, Lopez and DeVol, 2016]{Markwardt2016} Markwardt, D., Lopez, C. and DeVol, R. (2016)
The Economic Impact of Chapter 11 Bankruptcy versus Out-of-Court Restructuring.
{\em Journal of Applied Corporate Finance} 28(4): 124-128.

\bibitem[Marques, Neves and Sarmento, 2003]{Marques2003} Marques, C.R., Neves, P.D. and Sarmento, L.M. (2003)
Evaluating core inflation indicators.
{\em Economic Modelling} 20(4): 765-775.

\bibitem[Marshall, 2008]{Marshall2008} Marshall, C.M. (2008)
{\em Volatility trading: Hedge funds and the search for alpha (new challenges to the efficient markets hypothesis)} (Ph.D. Thesis).
New York, NY: Fordham University. Available online: \url{https://fordham.bepress.com/dissertations/AAI3353774/}.

\bibitem[Marshall, 2009]{Marshall2009} Marshall, C.M. (2009)
Dispersion trading: Empirical evidence from U.S. options markets.
{\em Global Finance Journal} 20(3): 289-301.

\bibitem[Marshall, Cahan and Cahan, 2008]{MarshallCahan2008} Marshall, B.R., Cahan, R.H. and Cahan, J.M. (2008)
Can commodity futures be profitably traded with quantitative market timing strategies?
{\em Journal of Banking \& Finance} 32(9): 1810-1819.

\bibitem[Marshall, Nguyen and Visaltanachoti]{Marshall2013} Marshall, B.R., Nguyen, N.H. and Visaltanachoti, N. (2013)
ETF arbitrage: Intraday evidence.
{\em Journal of Banking \& Finance} 37(9): 3486-3498.

\bibitem[Martellini, Milhau and Tarelli, 2015]{Martellini2015} Martellini, L., Milhau, V. and Tarelli, A. (2015)
Hedging Inflation-Linked Liabilities without Inflation-Linked Instruments through Long/Short Investments in Nominal Bonds.
{\em Journal of Fixed Income} 24(3): 5-29.

\bibitem[Martellini, Priaulet and Priaulet, 2002]{Martellini2002} Martellini, L., Priaulet, P. and Priaulet, S. (2002)
Understanding the butterfly strategy.
{\em Journal of Bond Trading and Management} 1(1): 9-19.

\bibitem[Martellini, Priaulet and Priaulet, 2003]{Martellini2003} Martellini, L., Priaulet, P. and Priaulet, S. (2003)
{\em Fixed Income Securities: Valuation, Risk Management and Portfolio Strategies.}
Hoboken, NJ: John Wiley \& Sons, Inc.

\bibitem[Martin, 2010]{Martin2010} Martin, G. (2010)
The Long-Horizon Benefits of Traditional and New Real Assets in the Institutional Portfolio.
{\em Journal of Alternative Investments} 13(1): 6-29.

\bibitem[Martin, 2011]{Martin2011} Martin, I. (2011)
Simple Variance Swaps.
{\em Working Paper.} Available online: \url{http://www.nber.org/papers/w16884}.

\bibitem[Martinelli and Rhoads, 2010]{Martinelli2010} Martinelli, R. and Rhoads, N. (2010)
Predicting Market Data Using The Kalman Filter, Part 1 and Part 2.
{\em Technical Analysis of Stocks \& Commodities} 28(1): 44-47; {\em ibid.} 28(2): 46-51.

\bibitem[Mart\'inez and Torr\'o, 2018]{Martinez2018} Mart\'inez, B. and Torr\'o, H. (2018)
Hedging spark spread risk with futures.
{\em Energy Policy} 113: 731-746.

\bibitem[Maslov and Zhang, 1998]{Maslov1998} Maslov, S. and Zhang, Y.-C. (1998)
Optimal investment strategy for risky assets.
{\em International Journal of Theoretical and Applied Finance} 1(3): 377-387.

\bibitem[Matsypura and Timkovsky, 2010]{Matsypura2010} Matsypura, D. and Timkovsky, V.G. (2010)
Combinatorics of Option Spreads: The Margining Aspect.
{\em Working Paper.} Available online:\\
\url{https://ses.library.usyd.edu.au/bitstream/2123/8172/1/OMWP_2010_04.pdf}.

\bibitem[Mauer and Sebastian, 2002]{Mauer2002} Mauer, R. and Sebastian, S. (2002)
Inflation Risk Analysis of European Real Estate Securities.
{\em Journal of Real Estate Research} 24(1): 47-78.

\bibitem[Mayers, 1998]{Mayers1998} Mayers, D. (1998)
Why firms issue convertible bonds: The matching of financial and real investment options.
{\em Journal of Financial Economics} 47(1): 83-102.

\bibitem[Mayhew, 1995]{Mayhew1995} Mayhew, S. (1995)
Implied Volatility.
{\em Financial Analysts Journal} 51(4): 8-20.

\bibitem[Maze, 2012]{Maze2012} Maze, S. (2012)
Dispersion Trading in South Africa: An Analysis of Profitability and a Strategy Comparison.
{\em Working Paper.} Available online: \url{https://ssrn.com/abstract=2398223}.

\bibitem[Mazurczak, 2011]{Mazurczak2011} Mazurczak, A. (2011)
Development of Real Estate Investment Trust (REIT) regimes in Europe.
{\em Journal of International Studies} 4(1): 115-123.

\bibitem[McCants, 2007]{McCants2007} McCants, A. (2007)
Goods at Pawn: The Overlapping Worlds of Material Possessions and Family Finance in Early Modern Amsterdam.
{\em Social Science History} 31(2): 213-238.

\bibitem[McComas, 2003]{McComas2003} McComas, A. (2003)
Getting technical with spreads.
{\em Futures Magazine}, July 2013, pp. 52-55.

\bibitem[McConnell and Buser, 2011]{McConnell2011} McConnell, J.J. and Buser, S.A. (2011)
The Origins and Evolution of the Market for Mortgage-Backed Securities.
{\em Annual Review of Financial Economics} 3: 173-192.

\bibitem[McConnell and Schwartz, 1986]{McConnell1986} McConnell, J.J. and Schwartz, E.S. (1986)
LYON Taming.
{\em Journal of Finance} 41(3): 561-577.

\bibitem[McDevitt and Kirwan, 2008]{McDevitt2008} McDevitt, D. and Kirwan, J. (2008)
Corporate and Infrastructure-backed Inflation-linked Bonds.
In: Benaben, B. and Goldenberg, S. (eds.)
{\em Inflation Risk and Products: The Complete Guide.} London, UK: Risk Books, pp. 621-641.

\bibitem[McDonald, 2001]{McDonald2001} McDonald, R.L.  (2001)
Cross-Border Investing with Tax Arbitrage: The Case of German Dividend Tax Credits.
{\em Review of Financial Studies} 14(3): 617-657.

\bibitem[Mcelroy, 2008]{Mcelroy2008} Mcelroy, T.  (2008)
Exact formulas for the Hodrick-Prescott Filter.
{\em Econometrics Journal} 11(1): 208-217.

\bibitem[McEnally and Rice, 1979]{McEnally1979} McEnally, R.W. and Rice, M.L. (1979)
Hedging Possibilities in the Flotation of Debt Securities.
{\em Financial Management} 8(4): 12-18.

\bibitem[McKee and Lensberg, 2002]{McKey2002} McKee, T.E. and Lensberg, T. (2002)
Genetic programming and rough sets: A hybrid approach to bankruptcy classification.
{\em European Journal of Operational Research} 138(2): 436-451.

\bibitem[McKenzie, 2002]{McKenzie2002} McKenzie, J.A. (2002)
A Reconsideration of the Jumbo/Non-Jumbo Mortgage Rate Differential.
{\em Journal of Real Estate Finance and Economics} 25(2-3): 197-213.

\bibitem[McMillan, 2002]{McMillan2002} McMillan, L.G. (2002)
{\em Options as a Strategic Investment.} (4th ed.) New York, NY: New York Institute of Finance.

\bibitem[Meen, 2002]{Meen2002} Meen, G. (2002)
The Time-Series Behavior of House Prices: A Transatlantic Divide?
{\em Journal of Housing Economics} 11(1): 1-23.

\bibitem[Mehra, 2002]{Mehra2002} Mehra, Y.P. (2002)
Survey Measures of Expected Inflation: Revisiting the Issues of Predictive Content and Rationality.
{\em Federal Reserve Bank of Richmond, Economic Quarterly} 88(3): 17-36.

\bibitem[Mei and Gao, 1995]{Mei1995} Mei, J. and Gao, B. (1995)
Price Reversals, Transaction costs and Arbitrage Profits in the Real Estate Securities Market.
{\em Journal of Real Estate Finance and Economics} 11(2): 153-165.

\bibitem[Mei and Liao, 1998]{Mei1998} Mei, J. and Liao, H.H. (1998)
Risk Characteristics of Real Estate Related Securities: An Extension of Liu and Mei (1992).
{\em Journal of Real Estate Research} 16(3): 279-290.

\bibitem[Meissner, 2008]{Meissner2008} Meissner, G. (ed.) (2008)
{\em The Definitive Guide to CDOs.}
London, UK: Incisive Media.

\bibitem[Meissner, 2016]{Meissner2016} Meissner, G. (2016)
Correlation Trading Strategies: Opportunities and Limitations.
{\em Journal of Trading} 11(4): 14-32.

\bibitem[Mendenhall, 2004]{Mendenhall2004} Mendenhall, R. (2004)
Arbitrage Risk and the Post-Earnings-Announcement Drift.
{\em Journal of Business} 77(6): 875-894.

\bibitem[Menkhoff {\em et al}, 2012]{Menkhoff2012} Menkhoff, L., Sarno, L., Schmeling, M. and Schrimpf, A. (2012)
 Currency momentum strategies.
{\em Journal of Financial Economics} 106(3): 660-684.

\bibitem[Menkveld, 2013]{Menkveld2013} Menkveld, A.J. (2013)
High Frequency Trading and the New Market Makers.
{\em Journal of Financial Markets} 16(4): 712-740.

\bibitem[Menkveld, 2016]{Menkveld2016} Menkveld, A.J. (2016)
The Economics of High-Frequency Trading: Taking Stock.
{\em Annual Review of Financial Economics} 8: 1-24.

\bibitem[Mercurio, 2005]{Mercurio2005} Mercurio, F. (2005)
Pricing inflation-indexed derivatives.
{\em Quantitative Finance} 5(3): 289-302.

\bibitem[Mercurio and Moreni, 2006]{Mercurio2006} Mercurio, F. and Moreni, N. (2006)
Inflation with a smile.
{\em Risk} 19(3): 70-75.

\bibitem[Mercurio and Moreni, 2009]{Mercurio2009} Mercurio, F. and Moreni, N. (2009)
Inflation modelling with SABR dynamics.
{\em Risk}, June 2009, pp. 106-111.

\bibitem[Mercurio and Yildirim, 2008]{Mercurio2008} Mercurio, F. and Yildirim, Y. (2008)
Modelling Inflation.
In: Benaben, B. and Goldenberg, S. (eds.) {\em Inflation Risks and Products: The Complete Guide.} London, UK: Risk Books.

\bibitem[Merton, 1987]{Merton1987} Merton, R.C. (1987)
A Simple Model of Capital Market Equilibrium with Incomplete Information.
{\em Journal of Finance} 42(3): 483-510.

\bibitem[Metghalchi, Marcucci and Chang, 2012]{Metghalchi2012} Metghalchi, M., Marcucci, J. and Chang, Y.-H. (2012)
Are moving average trading rules profitable? Evidence from the European stock markets.
{\em Applied Economics} 44(12): 1539-1559.

\bibitem[Meziani, 2015]{Meziani2015} Meziani, A.S. (2015)
Active exchange-traded funds: Are we there yet?
{\em Journal of Index Investing} 6(2): 86-98.

\bibitem[Mhaskar and Micchelli, 1993]{Mhaskar1993} Mhaskar, H.N. and Micchelli, C.A. (1993)
How to choose an activation function.
In: {\em Proceedings of the 6th International Conference on Neural Information Processing Systems (NIPS'93).}
San Francisco, CA: Morgan Kaufmann Publishers, Inc., pp. 319-326.

\bibitem[Miao, 2014]{Miao2014} Miao, G.J. (2014)
High frequency and dynamic pairs trading based on statistical arbitrage using a two-stage correlation and cointegration approach.
{\em International Journal of Economics and Finance} 6(3): 96-110.

\bibitem[Miao, Wei and Zhou 2012]{Miao2012} Miao, G.J., Wei, B. and Zhou, H. (2012)
Ambiguity Aversion and Variance Premium.
{\em Working Paper.} Available online: \url{https://ssrn.com/abstract=2023765}.

\bibitem[Miffre, 2012]{Miffre2012} Miffre, J. (2012)
Hedging pressure-based long/short commodity strategy used for third generation commodity index.
{\em Risk}, January 2012.
Available online: \url{https://www.risk.net/2247251}.

\bibitem[Miffre and Rallis, 2007]{Miffre2007} Miffre, J. and Rallis, G. (2007)
Momentum strategies in commodity futures markets.
{\em Journal of Banking \& Finance} 31(6): 1863-1886.

\bibitem[Milanov {\em et al}, 2013]{Milanov2013} Milanov, K., Kounchev, O., Fabozzi, F.J., Kim, Y.S. and Rachev, S.T. (2013)
A Binomial-Tree Model for Convertible Bond Pricing.
{\em Journal of Fixed Income} 22(3): 79-94.

\bibitem[Miles and Mahoney, 1997]{Miles1997} Miles, M. and Mahoney, J. (1997)
Is commercial real estate an inflation hedge?
{\em Real Estate Finance} 13(4): 31-45.

\bibitem[Miles and McCue, 1984]{Miles1984} Miles, M. and McCue, T. (1984)
Commercial Real Estate Returns.
{\em Real Estate Economics} 12(3): 355-377.

\bibitem[Miller, 1977]{Miller1977} Miller, M.H. (1977)
Debt and taxes.
{\em Journal of Finance} 32(2): 261-275.

\bibitem[Milonas, 1991]{Milonas1991} Milonas, N.T. (1991)
Measuring seasonalities in commodity markets and the half-month effect.
{\em Journal of Futures Markets} 11(3): 331-346.

\bibitem[Milosevic, 2016]{Milosevic2016} Milosevic, N. (2016)
Equity Forecast: Predicting Long Term Stock Price Movement using Machine Learning.
{\em Journal of Economics Library} 3(2): 288-294.

\bibitem[Miltersen and Schwartz, 1998]{Miltersen1998} Miltersen, K.R. and Schwartz, E.S. (1998)
Pricing of options on commodity futures with stochastic term structures of convenience yield and interest rates.
{\em Journal of Financial and Quantitative Analysis} 33(1): 33-59.

\bibitem[Min, Lee and Han, 2006]{Min2006} Min, S., Lee, J. and Han, I. (2006)
Hybrid genetic algorithms and support vector machines for bankruptcy prediction.
{\em Expert Systems with Applications} 31(3): 652-660.

\bibitem[Minton, 1997]{Minton1997} Minton, B.A. (1997)
An empirical examination of basic valuation models for plain vanilla U.S. interest rate swaps.
{\em Journal of Financial Economics} 44(2): 251-277.

\bibitem[Mitchell and Pulvino, 2001]{Mitchell2001} Mitchell, M. and Pulvino, T. (2001)
Characteristics of Risk and Return in Risk Arbitrage.
{\em Journal of Finance} 56(6): 2135-2175.

\bibitem[Mittal and Goel, 2012]{Mittal2012} Mittal, A. and Goel, A. (2012)
Stock Prediction Using Twitter Sentiment Analysis.
{\em Working Paper.} Palo Alto, CA: Stanford University.

\bibitem[Mitton and Vorkink, 2007]{Mitton2007} Mitton, T. and Vorkink, K. (2007)
Equilibrium Underdiversification and the Preference for Skewness.
{\em Review of Financial Studies} 20(4): 1255-1288.

\bibitem[Mixon, 2007]{Mixon2007} Mixon, S. (2007)
The implied volatility term structure of stock index options.
{\em Journal of Empirical Finance} 14(3): 333-354.

\bibitem[Mixon, 2011]{Mixon2011} Mixon, S. (2011)
What Does Implied Volatility Skew Measure?
{\em Journal of Derivatives} 18(4): 9-25.

\bibitem[Mladina, 2014]{Mladina2014} Mladina, P. (2014)
Dynamic Asset Allocation with Horizon Risk: Revisiting Glide Path Construction.
{\em Journal of Wealth Management} 16(4): 18-26.

\bibitem[Monkhouse, 1993]{Monkhouse1993} Monkhouse, P.H.L. (1993)
The Cost of Equity Under the Australian Dividend Imputation Tax System.
{\em Accounting and Finance} 33(2): 1-18.

\bibitem[Monoyios, 2004]{Monoyios2004} Monoyios, M. (2004)
Performance of Utility-Based Strategies for Hedging Basis Risk.
{\em Quantitative Finance} 4(3): 245-255.

\bibitem[Monoyios and Sarno, 2002]{Monoyios2002} Monoyios, M. and Sarno, L. (2002)
Mean reversion in stock index futures markets: a nonlinear analysis.
{\em Journal of Futures Markets} 22(4): 285-314.

\bibitem[Montelongo and Chang, 2008]{Montelongo2008} Montelongo, A. and Chang, H.K. (2008)
{\em Flip and grow rich: The heart and mind of real estate investing.}
San Antonio, TX: Armondo Montelongo Worldwide, Inc.

\bibitem[Montrucchio and Peccati, 1991]{Montrucchio1991} Montrucchio, L. and Peccati, L. (1991)
A note on Shiu-Fisher-Weil immunization theorem.
{\em Insurance: Mathematics and Economics} 10(2): 125-131.

\bibitem[Moore, Toepke and Colley, 2006]{Moore2006} Moore, S., Toepke, J. and Colley, N. (2006)
{\em The encyclopedia of commodity and financial spreads.}
Hoboken, NJ: John Wiley \& Sons, Inc.

\bibitem[Moosa, 2001]{Moosa2001} Moosa, I. (2001)
Triangular Arbitrage in the Spot and Forward Foreign Exchange Markets.
{\em Quantitative Finance} 1(4): 387-390.

\bibitem[Moosa, 2003a]{Moosa2003} Moosa, I.A. (2003a)
Two-Currency, Three-Currency and Multi-Currency Arbitrage.
In: {\em International Financial Operations: Arbitrage, Hedging, Speculation, Financing and Investment.}
Finance and Capital Markets Series. London, UK: Palgrave Macmillan, Chapter 1, pp. 1-18.

\bibitem[Moosa, 2003b]{Moosa2003b} Moosa, I.A. (2003b)
The sensitivity of the optimal hedge ratio to model specification.
{\em Finance Letters} 1(1): 15-20.

\bibitem[Moran and Dash, 2007]{Moran2007} Moran, M.T. and Dash, S. (2007)
VIX Futures and Options: Pricing and Using Volatility Products to Manage Downside Risk and Improve Efficiency in Equity Portfolios.
{\em Journal of Trading} 2(3): 96-105.

\bibitem[Morisawa, 2009]{Morisawa2009} Morisawa, Y. (2009)
Toward a Geometric Formulation of Triangular Arbitrage: An Introduction to Gauge Theory of Arbitrage.
{\em Progress of Theoretical Physics Supplement} 179: 209-215.

\bibitem[Morse and Shaw, 1988]{Morse1988} Morse, D. and Shaw, W. (1988)
Investing in Bankrupt Firms.
{\em Journal of Finance} 43(5): 1193-1206.

\bibitem[Moskowitz and Grinblatt, 1999]{Moskowitz1999} Moskowitz, T.J. and Grinblatt, M. (1999)
Do Industries Explain Momentum?
{\em Journal of Finance} 54(4): 1249-1290.

\bibitem[Moskowitz, Ooi and Pedersen, 2012]{Moskowitz2012} Moskowitz, T.J., Ooi, Y.H. and Pedersen, L.H. (2012)
Time Series Momentum.
{\em Journal of Financial Economics} 104(2): 228-250.

\bibitem[Moss {\em et al}, 2015]{Moss2015} Moss, A., Clare, A., Thomas, S. and Seaton, J. (2015)
Trend Following and Momentum Strategies for Global REITs.
{\em Journal of Real Estate Portfolio Management} 21(1): 21-31.

\bibitem[Mossman {\em et al}, 1998]{Mossman1998} Mossman, C.E., Bell, G.G., Swartz, L.M. and Turtle, H. (1998)
An empirical comparison of bankruptcy models.
{\em Financial Review} 33(2): 35-54.

\bibitem[Mou, 2010]{Mou2010} Mou, Y. (2010)
Limits to Arbitrage and Commodity Index Investment: Front-Running the Goldman Roll.
{\em Working Paper.} Available online: \url{https://ssrn.com/abstract=1716841}.

\bibitem[Mouakhar and Roberge, 2010]{Mouakhar2010} Mouakhar, T. and Roberge, M. (2010)
The Optimal Approach to Futures Contract Roll in Commodity Portfolios.
{\em Journal of Alternative Investments} 12(3): 51-60.

\bibitem[Moyer, Martin and Martin, 2012]{Moyer2012} Moyer, S.G., Martin, D. and Martin, J. (2012)
A Primer on Distressed Investing: Buying Companies by Acquiring Their Debt.
{\em Journal of Applied Corporate Finance} 24(4): 59-76.

\bibitem[Mraoua, 2007]{Mraoua2007} Mraoua, M. (2007)
Temperature stochastic modelling and weather derivatives pricing: empirical study with Moroccan data.
{\em Afrika Statistika} 2(1): 22-43.

\bibitem[Mueller, 1993]{Mueller1993} Mueller, G.R. (1993)
Refining Economic Diversification Strategies for Real Estate Portfolios.
{\em Journal of Real Estate Research} 8(1): 55-68.

\bibitem[Mueller and Laposa, 1995]{Mueller1995} Mueller, G.R. and Laposa, S.P. (1995)
Property-Type Diversification in Real Estate Portfolios: A Size and Return Perspective.
{\em Journal of Real Estate Portfolio Management} 1(1): 39-50.

\bibitem[Mueller and Mueller, 2003]{Mueller2003} Mueller, A. and Mueller, G. (2003)
Public and Private Real Estate in a Mixed-Asset Portfolio.
{\em Journal of Real Estate Portfolio Management} 9(3): 193-203.

\bibitem[Mugwagwa {\em et al}, 2012]{Mugwagwa2012} Mugwagwa, T., Ramiah, V., Naughton, T. and Moosa, I. (2012)
The efficiency of the buy-write strategy: Evidence from Australia.
{\em Journal of International Financial Markets, Institutions and Money} 22(2): 305-328.

\bibitem[M\"{u}ller and Grandi, 2000]{Muller2000} M\"{u}ller, A. and Grandi, M. (2000)
Weather Derivatives: A Risk Management Tool for Weather-sensitive Industries.
{\em Geneva Papers on Risk and Insurance} 25(2): 273-287.

\bibitem[Mun, 2016]{Mun2016} Mun, K.-C. (2016)
Hedging bank market risk with futures and forwards.
{\em Quarterly Review of Economics and Finance} 61: 112-125.

\bibitem[Mun and Morgan, 1997]{Mun1997} Mun, K.-C. and Morgan, G.E. (1997)
Cross-hedging foreign exchange rate risks: The case of deposit money banks in emerging Asian countries.
{\em Pacific-Basin Finance Journal} 5(2): 215-230.

\bibitem[Mun, Vasconcellos and Kish, 2000]{Mun2000} Mun, J.C., Vasconcellos, G.M. and Kish, R. (2000)
The contrarian overreaction hypothesis: An analysis of the US and Canadian stock markets.
{\em Global Finance Journal} 11(1-2): 53-72.

\bibitem[Murphy, 1986]{Murphy1986} Murphy, J.J. (1986)
{\em Technical analysis of the futures markets: A comprehensive guide to trading methods and applications.}
New York, NY: New York Institute of Finance.

\bibitem[Muthuswamy {\em et al}, 2011]{Muthuswamy2011} Muthuswamy, J., Palmer, J., Richie, N. and Webb, R. (2011)
High-Frequency Trading: Implications for Markets, Regulators, and Efficiency.
{\em Journal of Trading} 6(1): 87-97.

\bibitem[Mwangi and Duncan, 2012]{Mwangi2012} Mwangi, C.I. and Duncan, M.O. (2012)
An investigation into the existence of exchange rate arbitrage in the Mombasa spot market.
{\em International Journal of Humanities and Social Science} 2(21): 182-196.

\bibitem[Myers, 1991]{Myers1991} Myers, R.J. (1991)
Estimating time-varying optimal hedge ratios on futures markets.
{\em Journal of Futures Markets} 11(1): 39-53.

\bibitem[Nakamoto, 2008]{Nakamoto2008} Nakamoto, S. (2008)
Bitcoin: A Peer-to-Peer Electronic Cash System.
{\em Working Paper.} Available online: \url{https://bitcoin.org/bitcoin.pdf}.

\bibitem[Nakano, Takahashi and Takahashi, 2018]{Nakano2018} Nakano, M., Takahashi, A. and Takahashi, S. (2018)
Bitcoin Technical Trading With Artificial Neural Network.
{\em Working Paper.} Available online: \url{https://ssrn.com/abstract=3128726}.

\bibitem[Nandy and Chattopadhyay, 2016]{Nandy2016} Nandy (Pal), S. and Chattopadhyay, A.Kr. (2016)
Impact of Individual Stock Derivatives Introduction in India on Its Underlying Spot Market Volatility.	
{\em Asia-Pacific Journal of Management Research and Innovation} 12(2): 109-133.

\bibitem[Nartea and Eves, 2010]{Nartea2010} Nartea, G. and Eves, C. (2010)
Role of farm real estate in a globally diversified asset portfolio.
{\em Journal of Property Investment \& Finance} 28(3): 198-220.

\bibitem[Nashikkar, Mahanti, 2011]{Nashikkar2011} Nashikkar, A., Subrahmanyam, M.G. and Mahanti, S. (2011)
Liquidity and Arbitrage in the Market for Credit Risk.
{\em Journal of Financial and Quantitative Analysis} 46(3): 627-656.

\bibitem[Nawalkha and Chambers, 1996]{Nawalkha1996} Nawalkha, S.K. and Chambers, D.R. (1996)
An improved immunization strategy: M-absolute.
{\em Financial Analysts Journal} 52(5): 69-76.

\bibitem[Nekrasov, 2014]{Nekrasov2014} Nekrasov, V. (2014)
Kelly Criterion for Multivariate Portfolios: A Model-Free Approach.
{\em Working Paper.} Available online: \url{https://ssrn.com/abstract=2259133}.

\bibitem[Nelken, 2006]{Nelken2006} Nelken, I. (2006)
Variance swap volatility dispersion.
{\em Derivatives Use, Trading \& Regulation} 11(4): 334-344.

\bibitem[Nelling and Gyourko, 1998]{Nelling1998} Nelling, E. and Gyourko, J. (1998)
The Predictability of Equity REIT Returns.
{\em Journal of Real Estate Research} 16(3): 251-268.

\bibitem[Newell, 1996]{Newell1996} Newell, G. (1996)
The inflation-hedging characteristics of Australian commercial property: 1984-1995.
{\em Journal of Property Finance} 7(1): 6-20.

\bibitem[Newell, Chau and Wong, 2009]{Newell2009} Newell, G., Chau, K.W. and Wong, S.K. (2009)
The significance and performance of infrastructure in China.
{\em Journal of Property Investment \& Finance} 27(2): 180-202.

\bibitem[Newell and Peng, 2008]{Newell2008} Newell, G. and Peng, H.W. (2008)
The role of US infrastructure in investment portfolios.
{\em Journal of Real Estate Portfolio Management} 14(1): 21-34.

\bibitem[Newell, Peng and De Francesco, 2011]{Newell2011} Newell, G., Peng, H.W. and De Francesco, A. (2011)
The performance of unlisted infrastructure in investment portfolios.
{\em Journal of Property Research} 28(1): 59-74.

\bibitem[Ng and Phelps, 2015]{Ng2015} Ng, K.Y. and Phelps, B.D. (2015)
The Hunt for a Low-Risk Anomaly in the USD Corporate Bond Market.
{\em Journal of Portfolio Management} 42(1): 63-84.

\bibitem[Ng and Pirrong, 1994]{Ng1994} Ng, V.K. and Pirrong, S.C. (1994)
Fundamentals and volatility: storage, spreads, and the dynamics of metals prices.
{\em Journal of Business} 67(2): 203-230.

\bibitem[Ng, Rusticus and Verdi, 2008]{Ng2008} Ng, J., Rusticus, T. and Verdi, R. (2008)
Implications of Transaction Costs for the Post-Earnings Announcement Drift.
{\em Journal of Accounting Research} 46(3): 661-696.

\bibitem[Nguyen and Sercu, 2010]{Nguyen2010} Nguyen, V.T.T. and Sercu, P. (2010)
Tactical Asset Allocation with Commodity Futures: Implications of Business Cycle and Monetary Policy.
{\em Working Paper.} Available online: \url{https://ssrn.com/abstract=1695889}.

\bibitem[Niblock, 2017]{Niblock2017} Niblock, S.J. (2017)
Flight of the c\'ondors: Evidence on the Performance of c\'ondor Option Spreads in Australia.
{\em Applied Finance Letters} 6(1): 38-53.

\bibitem[Nielsen and Schwartz, 2004]{Nielsen2044} Nielsen, M.J. and Schwartz, E.S. (2004)
Theory of storage and the pricing of commodity claims.
{\em Review of Derivatives Research} 7(1): 5-24.

\bibitem[Nisar and Yeung, 2018]{Nisar2018} Nisar, T.M. and Yeung, M. (2018)
Twitter as a tool for forecasting stock market movements: A short-window event study.
{\em Journal of Finance and Data Science} 4(2): 101-119.

\bibitem[Noh, Engle and Kane, 1994]{Noh1994} Noh, J., Engle, R.F. and Kane, A. (1994)
Forecasting volatility and option prices of the S\&P500 Index.
{\em Journal of Derivatives} 2(1): 17-30.

\bibitem[Nossman and Wilhelmsson, 2009]{Nossman2009} Nossman, M. and Wilhelmsson, A. (2009)
Is the VIX Futures Market Able to Predict the VIX Index? A Test of the Expectation Hypothesis.
{\em Journal of Alternative Investments} 12(2): 54-67.

\bibitem[Nothaft, Lekkas and Wang, 1995]{Nothaft1995} Nothaft, F.E., Lekkas, V. and Wang, G.H.K. (1995)
The Failure of the Mortgage-Backed Futures Contract.
{\em Journal of Futures Markets} 15(5): 585-603.

\bibitem[Novak and Velu\v{s}\c{c}ek, 2016]{Novak2016} Novak, M.G. and Velu\v{s}\c{c}ek, D. (2016)
Prediction of stock price movement based on daily high prices.
{\em Quantitative Finance} 16(5): 793-826.

\bibitem[Novy-Marx, 2009]{NovyMarx2009} Novy-Marx, R. (2009)
Hot and Cold Markets.
{\em Real Estate Economics} 37(1): 1-22.

\bibitem[Novy-Marx, 2013]{Novy-Marx2013} Novy-Marx, R. (2013)
The other side of value: The gross profitability premium.
{\em Journal of Financial Economics} 108(1): 1-28.

\bibitem[Nyaradi, 2010]{Nyaradi2010} Nyaradi, J. (2010)
{\em Super Sectors: How to Outsmart the Market Using Sector Rotation and ETFs.}
Hoboken, NJ: John Wiley \& Sons, Inc.

\bibitem[Odean, 2002]{Odean2002} Odean, T. (2002)
Volume, Volatility, Price, and Profit When All Traders Are Above Average.
{\em Journal of Finance} 53(6): 1887-1934.

\bibitem[O'Doherty, 2012]{ODoherty2012} O'Doherty, M.S. (2012)
On the Conditional Risk and Performance of Financially Distressed Stocks.
{\em Management Science} 58(8): 1502-1520.

\bibitem[Odom and Sharda, 1990]{Odom1990} Odom, M.D. and Sharda, R. (1990)
A neural network model for bankruptcy prediction.
In: {\em Proceedings of the International Joint Conference on Neural Networks}, Vol. 2.  Washington, DC: IEEE, pp. 163-168.

\bibitem[Oetomo and Stevenson, 2005]{Oetomo2005} Oetomo, T. and Stevenson, M. (2005)
Hot or cold? A comparison of different approaches to the pricing of weather derivatives.
{\em Journal of Emerging Market Finance} 4(2): 101-133.

\bibitem[Officer, 2004]{Officer2004} Officer, M.S. (2004)
Collars and renegotiation in mergers and acquisitions.
{\em Journal of Finance} 59(6): 2719-2743.

\bibitem[Officer, 2006]{Officer2006} Officer, M.S. (2006)
The market pricing of implicit options in merger collars.
{\em Journal of Business} 79(1): 115-136.

\bibitem[O'Hara, 2015]{OHara2015} O'Hara, M. (2015)
High frequency market microstructure.
{\em Journal of Financial Economics} 116(2): 257-270.

\bibitem[Ohlson, 1980]{Ohlson1980} Ohlson, J.A. (1980)
Financial Ratios and the Probabilistic Prediction of Bankruptcy.
{\em Journal of Accounting Research} 18(1): 109-131.

\bibitem[Okunev and White, 2003]{Okunev2003} Okunev, J. and White, D. (2003)
Do Momentum-Based Strategies Still Work in Foreign Currency Markets?
{\em Journal of Financial and Quantitative Analysis} 38(2): 425-447.

\bibitem[Olmo and Pilbeam, 2009]{Olmo2009} Olmo, J. and Pilbeam, K. (2009)
The profitability of carry trades.
{\em Annals of Finance} 5(2): 231-241.

\bibitem[Olszweski and Zhou, 2013]{Olszweski2013} Olszweski, F. and Zhou, G. (2013)
Strategy diversification: Combining momentum and carry strategies within a foreign exchange portfolio.
{\em Journal of Derivatives \& Hedge Funds} 19(4): 311-320.

\bibitem[O'Neal, 2000]{ONeal2000} O'Neal, E.S. (2000)
Industry Momentum and Sector Mutual Funds.
{\em Financial Analysts Journal} 56(4): 37-49.

\bibitem[Opler {\em et al}, 1999]{Opler1999} Opler, T., Pinkowitz, L., Stulz, R. and Williamson, R. (1999)
The determinants and implications of corporate cash holdings.
{\em Journal of Financial Economics} 52(1): 3-46.

\bibitem[Opp, 2017]{Opp2017} Opp, C.C. (2017)
Learning, Optimal Default, and the Pricing of Distress Risk.
{\em Working Paper.} Available online: \url{https://ssrn.com/abstract=2181441}.

\bibitem[Ortalo-Magn\'{e} and Rady, 2006]{OrtaloMagne2006} Ortalo-Magn\'{e}, F. and Rady, S. (2006)
Housing Market Dynamics: On the Contribution of Income Shocks and Credit Constraints.
{\em Review of Economic Studies} 73(2): 459-485.

\bibitem[Ortisi, 2016]{Ortisi2016} Ortisi, M. (2016)
Bitcoin Market Volatility Analysis Using Grand Canonical Minority Game.
{\em Ledger} 1: 111-118.

\bibitem[Osborne, 2005]{Osborne2005} Osborne, M.J. (2005)
On the computation of a formula for the duration of a bond that yields precise results.
{\em Quarterly Review of Economics and Finance} 45(1): 161-183.

\bibitem[Osler, 2000]{Osler2000} Osler, C.L. (2000)
Support for Resistance: Technical Analysis and Intraday Exchange Rates.
{\em Federal Reserve Bank of New York, Economic Policy Review} 6(2): 53-68.

\bibitem[Osler, 2003]{Osler2003} Osler, C.L. (2003)
Currency Orders and Exchange Rate Dynamics: An explanation for the predictive success of Technical Analysis.
{\em Journal of Finance} 58(5): 1791-1819.

\bibitem[Osteryoung, McCarty and Roberts, 1981]{Osteryoung1981} Osteryoung, J.S., McCarty, D.E. and Roberts, G.S. (1981)
Riding the Yield Curve with Treasury Bills.
{\em Financial Review} 16(3): 57-66.

\bibitem[Osu, 2010]{Osu2010} Osu, B.O. (2010)
Currency Cross Rate and Triangular Arbitrage in Nigerian Exchange Market.
{\em International Journal of Trade, Economics and Finance} 1(4): 345-348.

\bibitem[O'Tool, 2013]{OTool2013} O'Tool, R. (2013)
The Black-Litterman model: A risk budgeting perspective.
{\em Journal of Asset Management} 14(1): 2-13.

\bibitem[Ou and Wang, 2009]{Ou2009} Ou, P. and Wang, H. (2009)
Prediction of stock market index movement by ten data mining techniques.
{\em Modern Applied Science} 3(12): 28-42.

\bibitem[Oyedele, Adair and McGreal, 2014]{Oyedele2014} Oyedele, J.B., Adair, A. and McGreal, S. (2014)
Performance of global listed infrastructure investment in a mixed asset portfolio.
{\em Journal of Property Research} 31(1): 1-25.

\bibitem[Ozdagli, 2010]{Ozdagli2010} Ozdagli, A.K. (2010)
The Distress Premium Puzzle.
{\em Working Paper.} Available online: \url{https://ssrn.com/abstract=1713449}.

\bibitem[Oztekin {\em et al}, 2017]{Oztekin2017} Oztekin, A.S., Mishra, S., Jain, P.K., Daigler, R.T., Strobl, S. and Holowczak, R.D. (2017)
Price Discovery and Liquidity Characteristics for U.S. Electronic Futures and ETF Markets.
{\em Journal of Trading} 12(2): 59-72.

\bibitem[Packer and Zhu, 2005]{Packer2005} Packer, F. and Zhu, H. (2005)
Contractual terms and CDS pricing.
{\em BIS Quarterly Review}, March 2005, pp. 89-100. Available online: \url{https://www.bis.org/publ/qtrpdf/r_qt0503h.pdf}.

\bibitem[Pagnotta and Philippon, 2012]{Pagnotta2012} Pagnotta, E. and Philippon, T. (2012)
Competing on Speed.
{\em Working Paper.} Available online: \url{https://ssrn.com/abstract=1972807}.

\bibitem[Pagolu {\em et al}, 2016]{Pagolu2016} Pagolu, V.S., Reddy, K.N., Panda, G. and Majhi, B. (2016)
Sentiment analysis of Twitter data for predicting stock market movements.
In: {\em Proceedings of the 2016 International Conference on Signal Processing, Communication, Power and Embedded System (SCOPES).} Washington, DC: IEEE, pp. 1345-1350.

\bibitem[Pagonidis, 2014]{Pagonidis2014} Pagonidis, A.S. (2014)
The IBS Effect: Mean Reversion in Equity ETFs.
{\em Working Paper.} Available online:\\ \url{http://www.naaim.org/wp-content/uploads/2014/04/00V_Alexander_Pagonidis_The-IBS-Effect-Mean-Reversion-in-Equity-ETFs-1.pdf}.

\bibitem[Pan and Poteshman, 2006]{Pan2006} Pan, J. and Poteshman, A.M. (2006)
The Information in Option Volume for Future Stock Prices.
{\em Review of Financial Studies} 19(3): 871-908.

\bibitem[Panayiotou and Medda, 2014]{Panayiotou2014} Panayiotou, A. and Medda, F.R. (2014)
Attracting Private Sector Participation in Transport Investment.
{\em Procedia -- Social and Behavioral Sciences} 111: 424-431.

\bibitem[Panayiotou and Medda, 2016]{Panayiotou2016} Panayiotou, A. and Medda, F. (2016)
Portfolio of Infrastructure Investments: Analysis of European Infrastructure.
{\em Journal of Infrastructure Systems} 22(3): 04016011.

\bibitem[Pantalone and Platt, 1984]{Pantalone1984} Pantalone, C. and Platt, H. (1984)
Riding the Yield Curve.
{\em Journal of Financial Education}, No. 13, pp. 5-9.

\bibitem[Papageorgiou, Reeves and Sherris, 2017]{Papageorgiou2017} Papageorgiou, N.A., Reeves, J.J. and Sherris, M. (2017)
Equity investing with targeted constant volatility exposure.
{\em Working Paper.} Available online: \url{https://ssrn.com/abstract=2614828}.

\bibitem[Park, Jung and Lee, 2018]{Park2017} Park, K., Jung, M. and Lee, S. (2018)
Credit ratings and convertible bond prices: a simulation-based valuation.
{\em European Journal of Finance} 24(12): 1001-1025.

\bibitem[Parnaudeau and Bertrand, 2018]{Parnaudeau2018} Parnaudeau, M. and Bertrand, J.-L. (2018)
The contribution of weather variability to economic sectors.
{\em Applied Economics} 50(43): 4632-4649.

\bibitem[Pascalau and Poirier, 2015]{Pascalau2015} Pascalau, R. and Poirier, R. (2015)
Bootstrapping the Relative Performance of Yield Curve Strategies.
{\em Journal of Investment Strategies} 4(2): 55-81.

\bibitem[Paschke and Prokopczuk, 2012]{Paschke2012} Paschke, R. and Prokopczuk, M. (2012)
Investing in commodity futures markets: can pricing models help?
{\em European Journal of Finance} 18(1): 59-87.

\bibitem[Passmore, Sherlund and Burgess, 2005]{Passmore2005} Passmore, W., Sherlund, S.M. and Burgess, G. (2005)
The Effect of Housing Government-Sponsored Enterprises on Mortgage Rates.
{\em Real Estate Economics} 33(3): 427-463.

\bibitem[P\'{a}stor and Stambaugh, 2003]{Pastor2003} P\'{a}stor, L'. and Stambaugh, R.F. (2003)
Liquidity Risk and Expected Stock Returns.
{\em Journal of Political Economy} 111(3): 642-685.

\bibitem[P\"{a}t\"{a}ri and Vilska, 2014]{Patari2014} P\"{a}t\"{a}ri, E. and Vilska, M. (2014)
Performance of moving average trading strategies over varying stock market conditions: the Finnish evidence.
{\em Applied Economics} 46(24): 2851-2872.

\bibitem[Pelaez, 1997]{Pelaez1997} Pelaez, R.F. (1997)
Riding the yield curve: Term premiums and excess returns.
{\em Review of Financial Economics} 6(1): 113-119.

\bibitem[Peng and Newell, 2007]{Peng2007} Peng, H.W. and Newell, G. (2007)
The Significance of Infrastructure in Australian Investment Portfolios.
{\em Pacific Rim Property Research Journal} 13(4): 423-450.

\bibitem[Pennacchi, 1991]{Pennacchi1991} Pennacchi, G.G. (1991)
Identifying the Dynamics of Real Interest Rates and Inflation: Evidence Using Survey Data.
{\em Review of Financial Studies} 4(1): 53-86.

\bibitem[Pepi\'{c}, 2014]{Pepic2014} Pepi\'{c}, M. (2014)
Managing interest rate risk with interest rate futures.
{\em Ekonomika preduze\'{c}a} 62(3-4): 201-209.

\bibitem[Perchanok, 2012]{Pechanok2012} Perchanok, K. (2012)
{\em Futures spreads: theory and praxis} (Ph.D. Thesis).
Northampton, UK: The University of Northampton. Available online: \url{http://nectar.northampton.ac.uk/4963/1/Perchanok20124963.pdf}.

\bibitem[Perchanok and Kakabadse, 2013]{Perchanok2013} Perchanok, K. and Kakabadse, N. (2013)
Causes of Market Anomalies of Crude Oil Calendar Spreads: Does Theory of Storage Address the Issue?
{\em Problems and Perspectives in Management} 11(2): 35-47.

\bibitem[Perchet, de Carvalho and Moulin, 2014]{Perchet2014} Perchet, R., de Carvalho, R.L. and Moulin, P. (2014)
Intertemporal risk parity: a constant volatility framework for factor investing.
{\em Journal of Investment Strategies} 4(1): 19-41.

\bibitem[Perez-Gonzalez and Yun, 2010]{PerezGonzalez2010} Perez-Gonzalez, F. and Yun, H. (2010)
Risk Management and Firm Value: Evidence from Weather Derivatives.
{\em Working Paper.} Available online: \url{https://ssrn.com/abstract=1357385}.

\bibitem[Peri\'{c}, 2015]{Peric2015} Peri\'{c}, M.R. (2015)
{\em Ekonomski aspekti korporativnih bankrotstava i ste\v{c}ajnih procesa.}
Belgrade, Serbia: Modern Business School.

\bibitem[Perlin, 2009]{Perlin2009} Perlin, M.S. (2009)
Evaluation of pairs-trading strategy at the Brazilian financial market.
{\em Journal of Derivatives \& Hedge Funds} 15(2): 122-136.

\bibitem[Person, 2007]{Person2007} Person, J.L. (2007)
{\em Candlestick and Pivot Point Trading Triggers.}
Hoboken, NJ: John Wiley \& Sons, Inc.

\bibitem[Peterson and Hsieh, 1997]{Peterson1997} Peterson, J.D. and Hsieh, C.-H. (1997)
Do Common Risk Factors in the Returns on Stocks and Bonds Explain Returns on REITs?
{\em Real Estate Economics} 25(2): 321-345.

\bibitem[Petre, 2015]{Petre2015} Petre, G. (2015)
A Case for Dynamic Asset Allocation for Long Term Investors.
{\em Procedia Economics and Finance} 29: 41-55.

\bibitem[Pflueger and Viceira, 2011]{Pflueger2011} Pflueger, C.E. and Viceira, L.M. (2011)
Inflation-Indexed Bonds and the Expectations Hypothesis.
{\em Annual Review of Financial Economics} 3: 139-158.

\bibitem[Philosophov and Philosophov, 2005]{Philosophov2005} Philosophov, L.V. and Philosophov, V.L. (2005)
Optimization of a firm's capital structure: A quantitative approach based on a probabilistic prognosis of risk and time of bankruptcy.
{\em International Review of Financial Analysis} 14(2): 191-209.

\bibitem[Piazzesi and Schneider, 2009]{Piazzesi2009} Piazzesi, M. and Schneider, M. (2009)
Momentum Traders in the Housing Market: Survey Evidence and a Search Model.
{\em American Economic Review} 99(2): 406-411.

\bibitem[Picou, 1981]{Picou1981} Picou, G. (1981)
Managing Interest Rate Risk with Interest Rate Futures.
{\em Bankers Magazine}, Vol. 164, May-June 1981, pp. 76-81.

\bibitem[Pindado, Rodrigues and de la Torre, 2008]{Pindado2008} Pindado, J., Rodrigues, L. and de la Torre, C. (2008)
Estimating financial distress likelihood.
{\em Journal of Business Research} 61(9): 995-1003.

\bibitem[Pindyck, 2001]{Pindyck2001} Pindyck, R.S. (2001)
The dynamics of commodity spot and futures markets: a primer.
{\em Energy Journal} 22(3): 1-30.

\bibitem[Piotroski, 2000]{Piotroski2000} Piotroski, J.D. (2000)
Value investing: The use of historical financial statement information to separate winners from losers.
{\em Journal of Accounting Research} 38: 1-41.

\bibitem[Piotroski and So, 2012]{Piotroski2012} Piotroski, J.D. and So, E.C. (2012)
Identifying Expectation Errors in Value/ Glamour Strategies: A Fundamental Analysis Approach.
{\em Review of Financial Studies} 25(9): 2841-2875.

\bibitem[Pirrong, 2005]{Pirrong2005} Pirrong, C. (2005)
Momentum in Futures Markets.
{\em Working Paper.} Available online: \url{https://ssrn.com/abstract=671841}.

\bibitem[Pirrong, 2017]{Pirrong2017} Pirrong, C. (2017)
The economics of commodity market manipulation: A survey.
{\em Journal of Commodity Markets} 5: 1-17.

\bibitem[Pisani, 2010]{Pisani2010} Pisani, B. (2010)
Man Vs. Machine: How Stock Trading Got So Complex.
{\em CNBC} (September 13, 2010). Available online: \url{https://www.cnbc.com/id/38978686}.

\bibitem[Pitts, 1985]{Pitts1985} Pitts, M. (1985)
The Management of Interest Rate Risk: Comment.
{\em Journal of Portfolio Management} 11(4): 67-69.

\bibitem[Pivar, 2003]{Pivar2003} Pivar, W. (2003)
{\em Real Estate Investing From A to Z: The Most Comprehensive, Practical, and Readable Guide to Investing Profitably in Real Estate.}
New York, NY: McGraw-Hill, Inc.

\bibitem[Pizzutilo, 2013]{Pizzutilo2013} Pizzutilo, F. (2013)
A note on the effectiveness of pairs trading for individual investors.
{\em International Journal of Economics and Financial Issues} 3(3): 763-771.

\bibitem[Podobnik {\em et al}, 2010]{Podobnik2010} Podobnik, B., Horvatic, D., Petersen, A.M., Uro\v{s}evi\'{c}, B. and Stanley, H.E. (2010)
Bankruptcy risk model and empirical tests.
{\em Proceedings of the National Academy of
Sciences} 107(43): 18325-18330.

\bibitem[Poitras, 1990]{Poitras1990} Poitras, G. (1990)
The distribution of gold futures spreads.
{\em Journal of Futures Markets} 10(6): 643-659.

\bibitem[Pole, 2007]{Pole2007} Pole, A. (2007)
{\em Statistical arbitrage: algorithmic trading insights and techniques.}
Hoboken, NJ: John Wiley \& Sons, Inc.

\bibitem[Popper, 1993]{Popper1993} Popper, H. (1993)
Long-term covered interest parity: evidence from currency swaps.
{\em Journal of International Money and Finance} 12(4): 439-448.

\bibitem[Porter, 1980]{Porter1980} Porter, M.F. (1980)
An Algorithm for Suffix Stripping.
{\em Program} 14(3): 130-137.

\bibitem[Poterba, 1986]{Poterba1986} Poterba, J. (1986)
Explaining the yield spread between taxable and tax exempt bonds.
In: Rosen, H. (ed.) {\em Studies in State and Local Public Finance.} Chicago, IL: University of Chicago Press, pp. 5-48.

\bibitem[Poterba, 1989]{Poterba1989} Poterba, J. (1989)
Tax reform and the market for tax-exempt debt.
{\em Regional Science and Urban Economics} 19(3): 537-562.

\bibitem[Poterba and Sinai, 2008]{PoterbaS2008} Poterba, J. and Sinai, T. (2008)
Tax Expenditures for Owner-Occupied Housing: Deductions for Property Taxes and Mortgage Interest and the Exclusion of Imputed Rental Income.
{\em American Economic Review} 98(2): 84-89.

\bibitem[Poterba and Summers, 1988]{Poterba1988} Poterba, J.M. and Summers, L.H. (1988)
Mean reversion in stock prices: evidence and implications.
{\em Journal of Financial Economics} 22(1): 27-59.

\bibitem[Potjer and Gould, 2007]{Potjer2007} Potjer, D. and Gould, C. (2007)
{\em Global Tactical Asset Allocation: Exploiting the opportunity of relative movements across asset classes and financial markets.}
London, UK: Risk Books.

\bibitem[Pounds, 1978]{Pounds1978} Pounds, H. (1978)
Covered Call Option Writing Strategies and Results.
{\em Journal of Portfolio Management} 4(2): 31-42.

\bibitem[Prince, 2005]{Prince2005} Prince, J.T. (2005)
Investing in Collateralized Debt Obligations.
{\em CFA Institute Conference Proceedings} 2005(1): 52-61.

\bibitem[Pring, 1985]{Pring1985} Pring, M.J. (1985)
{\em Technical analysis explained: The successful investor's guide to spotting investment trends and turning points.} (3rd ed.)
New York, NY: McGraw-Hill, Inc.

\bibitem[Prokopczuk and Simen, 2014]{Prokopczuk2014} Prokopczuk, M. and Simen, C.W. (2014)
The importance of the volatility risk premium for volatility forecasting.
{\em Journal of Banking \& Finance} 40: 303-320.

\bibitem[Putnam, 1991]{Putnam1991} Putnam, G., III (1991)
Investment Opportunities in Distressed Equities.
In: Levine, S. (ed.) {\em Handbook of Turnaround and Bankruptcy Investing.} New York, NY: HarperCollins, pp. 196-207.

\bibitem[Puttonen, 1993]{Puttonen1993} Puttonen, V. (1993)
The ex ante profitability of index arbitrage in the new Finnish markets.
{\em Scandinavian Journal of Management} 9(S1): 117-127.

\bibitem[Quintero, 1989]{Quintero1989} Quintero, R.G. (1989)
Acquiring the Turnaround Candidate.
In: Levine, S. (ed.) {\em The Acquisitions Manual.} New York, NY: New York Institute of Finance, pp. 379-441.

\bibitem[Rad, Low and Faff, 2016]{Rad2016} Rad, H., Low, R.K.Y. and Faff, R. (2016)
The profitability of pairs trading strategies: distance, cointegration and copula methods.
{\em Quantitative Finance} 16(10): 1541-1558.

\bibitem[Rajan, McDermott and Roy, 2007]{Rajan2007} Rajan, A., McDermott, G. and Roy, R. (eds.) (2007)
{\em The Structured Credit Handbook.}
Hoboken, NJ: John Wiley \& Sons, Inc.

\bibitem[Ramamurti and Doh, 2004]{Ramamurti2004} Ramamurti, R. and Doh, J. (2004)
Rethinking Foreign Infrastructure Investment in Developing Countries.
{\em Journal of World Business} 39(2): 151-167.

\bibitem[Rao, 2011]{Rao2011} Rao, V.K. (2011)
Multiperiod Hedging using Futures: Mean Reversion and the Optimal Hedging Path.
{\em Journal of Risk and Financial Management} 4(1): 133-161.

\bibitem[Rao and Srivastava, 2012]{Rao2012} Rao, T. and Srivastava, S. (2012)
Analyzing stock market movements using twitter sentiment analysis.
In: {\em Proceedings of the 2012 International Conference on Advances in Social Networks Analysis and Mining (ASONAM 2012).}
Washington, DC: IEEE, pp. 119-123.

\bibitem[Raulji and Saini, 2016]{Raulji2016} Raulji, J.K. and Saini, J.R. (2016)
Stop-Word Removal Algorithm and its Implementation for Sanskrit Language.
{\em International Journal of Computer Applications} 150(2): 15-17.

\bibitem[Ready, Roussanov and Ward, 2017]{Ready2017} Ready, R., Roussanov, N. and Ward, C. (2017)
Commodity Trade and the Carry Trade: A Tale of Two Countries.
{\em Journal of Finance} 72(6): 2629-2684.

\bibitem[Reddington, 1952]{Reddington1952} Reddington, F.M. (1952)
Review of the Principles of Life Insurance Valuations.
{\em Journal of the Institute of Actuaries} 78(3): 286-340.

\bibitem[Refenes, Zapranis and Francis, 1994]{Refenes1994} Refenes, A.N., Zapranis, A.S. and Francis, G. (1994)
Stock Performance Modeling Using Neural Networks: Comparative Study with Regressive Models.
{\em Neural Networks} 7(2): 375-388.

\bibitem[Rehring, 2012]{Rehring2012} Rehring, C. (2012)
Real Estate in a Mixed-Asset Portfolio: The Role of the Investment Horizon.
{\em Real Estate Economics} 40(1): 65-95.

\bibitem[Reiss and Phelps, 1991]{Reiss1991} Reiss, M.F. and Phelps, T.G. (1991)
Identifying a Troubled Company.
In: Dinapoli, D., Sigoloff, S.C. and Cushman, R.F. (eds.)
{\em Workouts and Turnarounds: The Handbook of Restructuring and Investing in Distressed Companies.} Homewood, IL: Business One-Irwin, pp. 7-43.

\bibitem[Reitano, 1996]{Reitano1996} Reitano, R. (1996)
Non-parallel yield curve shifts and stochastic immunization.
{\em Journal of Portfolio Management} 22(2): 71-78.

\bibitem[Remolona, Wickens and Gong, 1998]{Remolona1998} Remolona, E.M., Wickens, M.R. and Gong, F.F. (1998)
What Was the Market's View of UK Monetary Policy? Estimating Inflation Risk and Expected Inflation with Indexed Bonds.
{\em Federal Reserve Bank of New York Staff Reports}, No. 57. Available online: \url{https://ssrn.com/abstract=937350}.

\bibitem[Rendleman, 1999]{Rendleman1999} Rendleman, R.J. (1999)
Duration-Based Hedging with Treasury Bond Futures.
{\em Journal of Fixed Income} 9(1): 84-91.

\bibitem[Rendleman, Jones and Latan\'{e}, 1982]{Rendleman1982} Rendleman, R.J., Jones, C.P. and Latan\'{e}, H.A. (1982)
Empirical anomalies based on unexpected earnings and the importance of risk adjustments.
{\em Journal of Financial Economics} 10(3): 269-287.

\bibitem[Reynauld and Tessier, 1984]{Reynauld1984} Reynauld, J. and Tessier, J. (1984)
Risk Premiums in Futures Markets: An Empirical Investigation.
{\em Journal of Futures Markets} 4(2): 189-211.

\bibitem[Rhee and Chang, 1992]{Rhee1992} Rhee, S.G. and Chang, R.P. (1992)
Intra-Day Arbitrage Opportunities in Foreign Exchange and Eurocurrency Markets.
{\em Journal of Finance} 47(1): 363-379.

\bibitem[Ribeiro {\em et al}, 2012]{Ribeiro2012} Ribeiro, B., Silva, C., Chen, N., Vieira, A. and das Neves, J.C. (2012)
Enhanced default risk models with SVM+.
{\em Expert Systems with Applications} 39(11): 10140-10152.

\bibitem[Richard and Roll, 1989]{Richard1989} Richard, S.F. and Roll, R. (1989)
Prepayments on fixed-rate mortgage-backed securities.
{\em Journal of Portfolio Management} 15(3): 73-82.

\bibitem[Richards, Manfredo and Sanders, 2004]{Richards2004} Richards, T., Manfredo, M. and Sanders, D. (2004)
Pricing weather derivatives.
{\em American Journal of Agricultural Economics} 86(4): 1005-1017.

\bibitem[Richie, Daigler and Gleason, 2008]{Richie2008} Richie, N., Daigler, R.T. and Gleason, K.C. (2008)
The limits to stock index arbitrage: Examining S\&P 500 futures and SPDRS.
{\em Journal of Futures Markets} 28(12): 1182-1205.

\bibitem[Rickards, 2008]{Rickards2008} Rickards, D. (2008)
Global Infrastructure -- A Growth Story.
In: Davis, H. (ed.) {\em Infrastructure Finance: Trends and Techniques.} London, UK: Euromoney Books, pp. 1-47.

\bibitem[Rime, Schrimpf and Syrstad, 2017]{Rime2017} Rime, D., Schrimpf, A. and Syrstad, O. (2017)
Segmented Money Markets and Covered Interest Parity Arbitrage.
{\em Working Paper.} Available online: \url{https://ssrn.com/abstract=2879904}.

\bibitem[Riordan and Storkenmaier, 2012]{Riordan2012} Riordan, R. and Storkenmaier, A. (2012)
Latency, liquidity and price discovery.
{\em Journal of Financial Markets} 15(4): 416-437.

\bibitem[Rising and Wyner, 2012]{Rising2012} Rising, J.K. and Wyner, A.J. (2012)
Partial Kelly portfolios and shrinkage estimators.
In: {\em Proceedings of the 2012 International Symposium on Information Theory (ISIT).} Washington, DC: IEEE, pp. 1618-1622.

\bibitem[R\"{o}del and Rothballer, 2012]{Rodel2012} R\"{o}del, M. and Rothballer, C. (2012)
Infrastructure as Hedge against Inflation -- Fact or Fantasy?
{\em Journal of Alternative Investments} 15(1): 110-123.

\bibitem[Rodr\'{\i}guez-Gonz\'{a}lez {\em et al}, 2011]{RodriguezGonzalez2011}
Rodr\'{\i}guez-Gonz\'{a}lez, A., Garc\'{\i}a-Crespo, \'{A}., Colomo-Palacios, R., Iglesias, F.G. and G\'{o}mez-Berb\'{\i}s, J.M. (2011)
CAST: Using neural networks to improve trading systems based on technical analysis by means of the RSI financial indicator.
{\em Expert Systems with Applications} 38(9): 11489-11500.

\bibitem[Rogers and Shi, 1995]{Rogers1995} Rogers, L.C.G. and Shi, Z. (1995)
The Value of an Asian Option.
{\em Journal of Applied Probability} 32(4): 1077-1088.

\bibitem[Roll, 1996]{Roll1996} Roll, R. (1996)
U.S. Treasury Inflation-Indexed Bonds: The Design of a New Security.
{\em Journal of Fixed Income} 6(3): 9-28.

\bibitem[Roll, 2004]{Roll2004} Roll, R. (2004)
Empirical TIPS.
{\em Financial Analysts Journal} 60(1): 31-53.

\bibitem[Roll and Yan, 2008]{Roll2008} Roll, R. and Yan, S. (2008)
An explanation of the forward premium `puzzle'.
{\em European Financial Management} 6(2): 121-148.

\bibitem[Rompotis, 2011a]{Rompotis2011a} Rompotis, G.G. (2011a)
The performance of actively managed exchange traded funds.
{\em Journal of Index Investing} 1(4): 53-65.

\bibitem[Rompotis, 2011b]{Rompotis2011b} Rompotis, G.G. (2011b)
Active vs. passive management: New evidence from exchange traded funds.
{\em International Review of Applied Financial Issues and Economics} 3(1): 169-186.

\bibitem[Ronn and Ronn, 1989]{Ronn1989} Ronn, A.G. and Ronn, E.I. (1989)
The Box Spread Arbitrage Conditions: Theory, Tests, and Investment Strategies.
{\em Review of Financial Studies} 2(1): 91-108.

\bibitem[Rosales and McMillan, 2017]{Rosales2017} Rosales, E.B. and McMillan, D. (2017)
Time-series and cross-sectional momentum and contrarian strategies within the commodity futures markets.
{\em Cogent Economics \& Finance} 5(1): 1339772.

\bibitem[Rosenberg, 1992]{Rosenberg1992}Rosenberg, H. (1992)
{\em Vulture Investors.}
New York, NY: HarperCollins.

\bibitem[Rosenberg, Reid and Lanstein, 1985]{Rosenberg1985} Rosenberg, B., Reid, K. and Lanstein, R. (1985)
Persuasive evidence of market inefficiency.
{\em Journal of Portfolio Management} 11(3): 9-16.

\bibitem[Ross, 2006]{Ross2006} Ross, J. (2006) Exploiting spread trades.
{\em Futures Magazine}, December 2006, pp. 34-36.

\bibitem[Ross and Zisler, 1991]{Ross1991} Ross, S. and Zisler, R. (1991)
Risk and return in real estate.
{\em Journal of Real Estate Finance and Economics} 4(2): 175-190.

\bibitem[Rothballer and Kaserer, 2012]{Rothballer2012} Rothballer, C. and Kaserer, C. (2012)
The Risk Profile of Infrastructure Investments: Challenging Conventional Wisdom.
{\em Journal of Structured Finance} 18(2): 95-109.

\bibitem[Routledge, Seppi and Spatt, 2000]{Routledge2000} Routledge, B., Seppi, D.J. and Spatt, C. (2000)
Equilibrium forward curves for commodities.
{\em Journal of Finance} 55(3): 1297-1338.

\bibitem[Rouwenhorst, 1998]{Rouwenhorst1998} Rouwenhorst, K.G. (1998)
International Momentum Strategies.
{\em Journal of Finance} 53(1): 267-284.

\bibitem[Roy and Vetterli, 2007]{Roy2007} Roy, O. and Vetterli, M. (2007)
The effective rank: A measure of effective dimensionality.
In: {\em Proceedings -- EUSIPCO 2007, 15th European Signal Processing Conference.} Pozna\'n, Poland (September 3-7), pp. 606-610.

\bibitem[Ruan, Durresi and Alfantoukh, 2018]{Ruan2018} Ruan, Y., Durresi, A. and Alfantoukh, L. (2018)
Using Twitter trust network for stock market analysis.
{\em Knowledge-Based Systems} 145: 207-218.

\bibitem[Ruchin, 2011]{Ruchin2011} Ruchin, A. (2011)
Can Securities Lending Transactions Substitute for Repurchase Agreement Transactions?
{\em Banking Law Journal} 128(5): 450-480.

\bibitem[Ruder, 2017]{Ruder2017} Ruder, S. (2017)
An overview of gradient descent optimization algorithms.
{\em Working Paper.} Available online:\\
\url{https://arxiv.org/pdf/1609.04747.pdf}.

\bibitem[Rudy, Dunis and Laws, 2010]{Rudy2010} Rudy, J., Dunis, C. and Laws, J. (2010)
Profitable Pair Trading: A Comparison Using the S\&P 100 Constituent Stocks and the 100 Most Liquid ETFs.
{\em Working Paper.} Available online: \url{https://ssrn.com/abstract=2272791}.

\bibitem[Rujivan and Zhu, 2012]{Rujivan2012} Rujivan, S. and Zhu, S.P. (2012)
A simplified analytical approach for pricing discretely sampled variance swaps with stochastic volatility.
{\em Applied Mathematics Letters} 25(11): 1644-1650.

\bibitem[Rumelhart, Hinton and Williams, 1986]{Rumelhart1986} Rumelhart, D.E., Hinton, G.E. and Williams, R.J. (1986)
Learning representations by back-propagating errors.
{\em Nature} 323(6088): 533-536.

\bibitem[Rusn\'{a}kov\'{a} and \v{S}olt\'{e}s, 2012]{Rusnakova2012} Rusn\'{a}kov\'{a}, M. and \v{S}olt\'{e}s, V. (2012)
Long strangle strategy using barrier options and its application in hedging.
{\em Actual Problems of Economics} 134(8): 452-465.

\bibitem[Rusn\'{a}kov\'{a}, \v{S}olt\'{e}s and Szabo, 2015]{Rusnakova2015}
Rusn\'{a}kov\'{a}, M., \v{S}olt\'{e}s, V. and Szabo, Z.K. (2015)
Short Combo Strategy Using Barrier Options and its Application in Hedging.
{\em Procedia Economics and Finance} 32: 166-179.

\bibitem[Ryabkov, 2015]{Ryabkov2015} Ryabkov, N. (2015)
Hedge Fund Price Pressure in Convertible Bond Markets.
{\em Working Paper.} Available online: \url{https://ssrn.com/abstract=2539929}.

\bibitem[Saad, Prokhorov and Wunsch, 1998]{Saad1998} Saad, E.W., Prokhorov, D.V. and Wunsch, D.C. (1998)
Comparative study of stock trend prediction using time delay, recurrent and probabilistic neural networks.
{\em IEEE Transactions on Neural Networks} 9(6): 1456-1470.

\bibitem[Sack and Elsasser, 2004]{Sack2004} Sack, B. and Elsasser, R. (2004)
Treasury Inflation-Indexed Debt: A Review of the U.S. Experience.
{\em Federal Reserve Bank of New York, Economic Policy Review} 10(1): 47-63.

\bibitem[Sadka, 2002]{Sadka2002} Sadka, R. (2002)
The Seasonality of Momentum: Analysis of Tradability.
{\em Working Paper.} Available online: \url{https://ssrn.com/abstract=306371}.

\bibitem[Sagi and Seasholes, 2007]{Sagi2007} Sagi, J. and Seasholes, M. (2007)
Firm-specific Attributes and the Cross-section of Momentum.
{\em Journal of Financial Economics} 84(2): 389-434.

\bibitem[Salcedo, 2004]{Salcedo2004} Salcedo, Y. (2004)
Spreads for the fall.
{\em Futures Magazine}, September 2004, pp. 54-57.

\bibitem[Saltyte-Benth and Benth, 2012]{SaltyteBenth2012} Saltyte-Benth, J. and Benth, F.E. (2012)
A critical view on temperature modelling for application in weather derivatives markets.
{\em Energy Economics} 34(2): 592-602.

\bibitem[Samuelson, 1945]{Samuelson1945} Samuelson, P.A. (1945)
The effect of interest rate increases on the banking system.
{\em American Economic Review} 35(1): 16-27.

\bibitem[Samuelson, 1971]{Samuelson1971} Samuelson, P. (1971)
The ``fallacy" of maximizing the geometric mean in long sequences of investing or gambling.
{\em Proceedings of the National Academy of Sciences} 68(10): 2493-2496.

\bibitem[Samuelson and Rosenthal, 1986]{Samuelson1986} Samuelson, W. and Rosenthal, L. (1986)
Price Movements as Indicators of Tender Offer Success.
{\em Journal of Finance} 41(2): 481-499.

\bibitem[Samworth, 2012]{Samworth2012} Samworth, R.J. (2012)
Optimal weighted nearest neighbour classifiers.
{\em Annals of Statistics} 40(5): 2733-2763.

\bibitem[Sanchez-Robles, 1998]{Sanchez-Robles1998} Sanchez-Robles, B. (1998)
Infrastructure Investment and Growth: Some Empirical Evidence.
{\em Contemporary Economic Policy} 16(1): 98-108.

\bibitem[Saretto and Goyal, 2009]{Saretto2009} Saretto, A. and Goyal, A. (2009)
Cross-section of option returns and volatility.
{\em Journal of Financial Economics} 94(2): 310-326.

\bibitem[Sassetti and Tani, 2006]{Sassetti2006} Sassetti, P. and Tani, M. (2006)
Dynamic Asset Allocation Using Systematic Sector Rotation.
{\em Journal of Wealth Management} 8(4): 59-70.

\bibitem[Satchell and Scowcroft, 2000]{Satchell2000} Satchell, S. and Scowcroft, A. (2000)
A demystification of the Black-Litterman model: Managing quantitative and traditional portfolio construction.
{\em Journal of Asset Management} 1(2): 138-150.

\bibitem[Savor and Wilson, 2013]{Savor2013} Savor, P. and Wilson, M. (2013)
How Much Do Investors Care About Macroeconomic Risk? Evidence from Scheduled Economic Announcements.
{\em Journal of Financial and Quantitative Analysis} 48(2): 343-375.

\bibitem[Sawant, 2010a]{Sawant2010} Sawant, R.J. (2010a)
{\em Infrastructure Investing: Managing Risks \& Rewards for Pensions, Insurance Companies \& Endowments.}
Hoboken, NJ: John Wiley \& Sons, Inc.

\bibitem[Sawant, 2010b]{Sawant2010b} Sawant, R.J. (2010b)
Emerging Market Infrastructure Project Bonds: Their Risks and Returns.
{\em Journal of Structured Finance} 15(4): 75-83.

\bibitem[Schaede, 1990]{Schaede1990} Schaede, U. (1990)
The introduction of commercial paper -- a case study in the liberalisation of the Japanese financial markets.
{\em Japan Forum} 2(2): 215-234.

\bibitem[Schap, 2005]{Schap2005} Schap, K. (2005)
{\em The complete guide to spread trading.}
New York, NY: McGraw-Hill, Inc.

\bibitem[Schatz, 2012]{Schatz2012} Schatz, H.R. (2012)
The Characterization of Repurchase Agreements in the Context of the Federal Securities Laws.
{\em St. John's Law Review} 61(2): 290-310.

\bibitem[Schiereck, Bondt and Weber, 1999]{Schiereck1999} Schiereck, D., Bondt, W.D. and Weber, M. (1999)
Contrarian and momentum strategies in Germany.
{\em Financial Analysts Journal} 55(6): 104-116.

\bibitem[Schiller, Seidler and Wimmer, 2010]{Schiller2010} Schiller, F., Seidler, G. and Wimmer, M. (2010)
Temperature models for pricing weather derivatives.
{\em Quantitative Finance} 12(3): 489-500.

\bibitem[Schizas, 2014]{Schizas2014} Schizas, P. (2014)
Active ETFs and their performance vis-\`{a}-vis passive ETFs, mutual funds, and hedge funds.
{\em Journal of Wealth Management} 17(3): 84-98.

\bibitem[Schizas, Thomakos and Wang, 2011]{Schizas2011} Schizas, P., Thomakos, D.D. and Wang, T. (2011)
Pairs Trading on International ETFs.
{\em Working Paper.} Available online: \url{https://ssrn.com/abstract=1958546}.

\bibitem[Schmidhuber, 2015]{Schmidhuber2015} Schmidhuber, J. (2015)
Deep learning in neural networks: An overview.
{\em Neural Networks} 61: 85-117.

\bibitem[Schmidt and Ward, 2002]{Schmidt2002} Schmidt, W. and Ward, I. (2002)
Pricing default baskets.
{\em Risk}, January 2002, pp. 111-114.

\bibitem[Schneeweis and Gupta, 2006]{Schneeweis2006} Schneeweis, T. and Gupta, R. (2006)
Diversification benefits of managed futures.
{\em Journal of Investment Consulting} 8(1): 53-62.

\bibitem[Schneider and Windischbauer, 2008]{Schneider2008} Schneider, F. and Windischbauer, U. (2008)
Money laundering: some facts.
{\em European Journal of Law and Economics} 26(3): 387-404.

\bibitem[Scholes and Williams, 1977]{Scholes1977} Scholes, M. and Williams, J. (1977)
Estimating Betas from Nonsynchronous Data.
{\em Journal of Financial Economics} 5(3): 309-327.

\bibitem[Sch\"{o}nbucher, 2003]{Schonbucher2003} Sch\"{o}nbucher, P.J. (2003)
{\em Credit Derivatives Pricing Models.} Hoboken, NJ: John Wiley \& Sons, Inc.

\bibitem[Schoutens, 2005]{Schoutens2005} Schoutens, W. (2005)
Moment swaps.
{\em Quantitative Finance} 5(6): 525-530.

\bibitem[Schultz, 2016]{Schultz2016} Schultz, G.M. (2016)
{\em Investing in Mortgage-Backed and Asset-Backed Securities: Financial Modeling with R and Open Source Analytics + Website.}
Hoboken, NJ: John Wiley \& Sons, Inc.

\bibitem[Schumaker and Chen, 2010]{Schumaker2010} Schumaker, R.P. and Chen, H. (2010)
A Discrete Stock Price Prediction Engine Based on Financial News.
{\em Computer} 43(1): 51-56.

\bibitem[Schwartz, 1997]{Schwartz1997} Schwartz, E.S. (1997)
The Stochastic Behavior of Commodity Prices: Implications for Valuation and Hedging.
{\em Journal of Finance} 52(3): 923-973.

\bibitem[Schwartz, 1998]{Schwartz1998} Schwartz, E.S. (1998)
Valuing long-term commodity assets.
{\em Journal of Energy Finance \& Development} 3(2): 85-99.

\bibitem[Schwartz and Laatsch, 1991]{Schwartz1991} Schwartz, T.V. and Laatsch, F. (1991)
Price Discovery and Risk Transfer in Stock Index Cash and Futures Markets.
{\em Journal of Futures Markets} 11(6): 669-683.

\bibitem[Schwartz and Smith, 2000]{Schwartz2000} Schwartz, E.S. and Smith, J.E. (2000)
Short-term variations and long-term dynamics in commodity prices.
{\em Management Science} 46(7): 893-911.

\bibitem[Schwartz and Torous, 1989]{Schwartz1989} Schwartz, E.S. and Torous, W.N. (1989)
Prepayment and the Valuation of Mortgage-Backed Securities.
{\em Journal of Finance} 44(2): 375-392.

\bibitem[Schwartz and Torous, 1992]{Schwartz1992} Schwartz, E.S. and Torous, W.N. (1992)
Prepayment, Default, and the Valuation of Mortgage Pass-through Securities.
{\em Journal of Business} 65(2): 221-239.

\bibitem[Schwert, 2003]{Schwert2003} Schwert, G.W. (2003)
Anomalies and market efficiency.
In: Constantinides, G.M., Harris, M. and Stulz, R.M. (eds.)
{\em Handbook of the Economics of Finance, Vol 1B.} (1st ed.) Amsterdam, The Netherlands: Elsevier, Chapter 15, pp. 939-974.

\bibitem[Sefton and Scowcroft, 2005]{Sefton2005} Sefton, J.A. and Scowcroft, A. (2005)
Understanding Momentum.
{\em Financial Analysts Journal} 61(2): 64-82.

\bibitem[Seiler, Webb and Myer, 1999]{Seiler1999} Seiler, M.J., Webb, J.R. and Myer, F.C.N. (1999)
Diversification Issues in Real Estate Investment.
{\em Journal of Real Estate Literature} 7(2): 163-179.

\bibitem[Sepp\"{a}l\"{a}, 2004]{Seppala2004} Sepp\"{a}l\"{a}, J. (2004)
The term structure of real interest rates: theory and evidence from UK index-linked bonds.
{\em Journal of Monetary Economics} 51(7): 1509-1549.

\bibitem[Serban, 2010]{Serban2010} Serban, A.F. (2010)
Combining mean reversion and momentum trading strategies in foreign exchange markets.
{\em Journal of Banking \& Finance} 34(11): 2720-2727.

\bibitem[Seymour, 2008]{Seymour2008} Seymour, B. (2008)
Global Money Laundering.
{\em Journal of Applied Security Research} 3(3-4): 373-387.

\bibitem[Sezer, Ozbayoglu and Dogdu, 2017]{Sezer2017} Sezer, O.B., Ozbayoglu, M. and Dogdu, E. (2017)
A Deep Neural-Network Based Stock Trading System Based on Evolutionary Optimized Technical Analysis Parameters.
{\em Procedia Computer Science} 114: 473-480.

\bibitem[Shackman and Tenney, 2006]{Shackman2006} Shackman, J.D. and Tenney, G. (2006)
The Effects of Government Regulations on the Supply of Pawn Loans: Evidence from 51 Jurisdictions in the US.
{\em Journal of Financial Services Research} 30(1): 69-91.

\bibitem[Shah, 2017]{Shah2017} Shah, A. (2017)
Hedging of a Portfolio of Rainfall Insurances using Rainfall Bonds and European Call Options (Bull Spread).
{\em Working Paper.} Available online: \url{https://ssrn.com/abstract=2778647}.

\bibitem[Shah and Zhang, 2014]{Shah2014} Shah, D. and Zhang, K. (2014)
Bayesian regression and Bitcoin.
{\em Working Paper.} Available online: \url{https://arxiv.org/pdf/1410.1231.pdf}.

\bibitem[Shaikh and Padhi, 2015]{Shaikh2015} Shaikh, I. and Padhi, P. (2015)
The implied volatility index: Is `{\em investor fear gauge}' or `{\em forward-looking}'?
{\em Borsa Istanbul Review} 15(1): 44-52.

\bibitem[Shan, Garvin and Kumar, 2010]{Shan2010} Shan, L., Garvin, M.J. and Kumar, R. (2010)
Collar options to manage revenue risks in real toll public-private partnership transportation projects.
{\em Construction Management and Economics} 28(10): 1057-1069.

\bibitem[Sharpe, 1966]{Sharpe1966} Sharpe, W.F. (1966)
Mutual Fund Performance.
{\em Journal of Business} 39(1): 119-138.

\bibitem[Sharpe, 1994]{Sharpe1994} Sharpe, W.F. (1994)
The Sharpe Ratio.
{\em Journal of Portfolio Management} 21(1): 49-58.

\bibitem[Sharpe, 2009]{Sharpe2009} Sharpe, W.F. (2009)
Adaptive Asset Allocation Policies.
{\em Financial Analysts Journal} 66(3): 45-59.

\bibitem[Sharpe and Perold, 2009]{Sharpe1988} Sharpe, W.F. and Perold, A.F. (1988)
Dynamic Strategies for Asset Allocation.
{\em Financial Analysts Journal} 44(1): 16-27.

\bibitem[Shaviro, 2002]{Shaviro2002} Shaviro, D. (2002)
Dynamic Strategies for Asset Allocation.
{\em Chicago Journal of International Law} 3(2): 317-331.

\bibitem[Shen, 2006]{Shen2006} Shen, P. (2006)
Liquidity Risk Premia and Breakeven Inflation Rates.
{\em Federal Reserve Bank of Kansas City, Economic Review} 91(2): 29-54.

\bibitem[Shen and Corning, 2001]{Shen2001} Shen, P. and Corning, J. (2001)
Can TIPS Help Identify Long-Term Inflation Expectations?
{\em Federal Reserve Bank of Kansas City, Economic Review} 86(4): 61-87.

\bibitem[Sher, 2014]{Sher2014} Sher, G. (2014)
Cashing in for Growth: Corporate Cash Holdings as an Opportunity for Investment in Japan.
{\em Working Paper.} Available online: \url{https://ssrn.com/abstract=2561246}.

\bibitem[Sherrill and Upton, 2018]{Sherrill2018} Sherrill, D.E. and Upton, K. (2018)
Actively managed ETFs vs actively managed mutual funds.
{\em Managerial Finance} 44(3): 303-325.

\bibitem[Shi, Jiang and Zhou, 2015]{Shi2015} Shi, H.-L., Jiang, Z.-Q. and Zhou, W.-X. (2015)
Profitability of Contrarian Strategies in the Chinese Stock Market.
{\em PLoS ONE} 10(9): e0137892.

\bibitem[Shiller, 1979]{Shiller1979} Shiller, R.J. (1979)
The Volatility of Long-Term Interest Rates and Expectations Models of the Term Structure.
{\em Journal of Political Economy } 87(6): 1190-1219.

\bibitem[Shiller and Modigliani, 1979]{ShillerMod1979} Shiller, R.J. and Modigliani, F. (1979)
Coupon and tax effects on new and seasoned bond yields and the measurement of the cost of debt capital.
{\em Journal of Financial Economics} 7(3): 297-318.

\bibitem[Shimko, 1994]{Shimko1994} Shimko, D.C. (1994)
Options on futures spreads: Hedging, speculation, and valuation.
{\em Journal of Futures Markets} 14(2): 183-213.

\bibitem[Shin and Lee, 2002]{Shin2002} Shin, K. and Lee, Y. (2002)
A genetic algorithm application in bankruptcy prediction modeling.
{\em Expert Systems with Applications} 23(3): 321-328.

\bibitem[Shiu, 1987]{Shiu1987} Shiu, E.S.W. (1987)
On the Fisher-Weil immunization theorem.
{\em Insurance: Mathematics and Economics} 6(4): 259-266.

\bibitem[Shiu, 1988]{Shiu1988} Shiu, E.S.W. (1988)
Immunization of multiple liabilities.
{\em Insurance: Mathematics and Economics} 7(4): 219-224.

\bibitem[Shiu and Lu, 2011]{Shiu2011} Shiu, Y.-M. and Lu, T.-H. (2011)
Pinpoint and Synergistic Trading Strategies of Candlesticks.
{\em International Journal of Economics and Finance} 3(1): 234-244.

\bibitem[Shum {\em et al}, 2016]{Shum2016} Shum, P., Hejazi, W., Haryanto, E. and Rodier, A. (2016)
Intraday Share Price Volatility and Leveraged ETF Rebalancing.
{\em Review of Finance} 20(6): 2379-2409.

\bibitem[Shumway, 2001]{Shumway2001} Shumway, T. (2001)
Forecasting Bankruptcy More Accurately: A Simple Hazard Model.
{\em Journal of Business} 74(1): 101-104.

\bibitem[Siganos and Chelley-Steeley, 2006]{Siganos2006}
Siganos, A. and Chelley-Steeley, P. (2006)
Momentum Profits Following Bull and Bear Markets.
{\em Journal of Asset Management} 6(5): 381-388.

\bibitem[Sill, 1996]{Sill1996} Sill, K. (1996)
The Cyclical Volatility of Interest Rates.
{\em Business Review of the Federal Reserve Bank of Philadelphia}, January/February 1996, pp. 15-29.

\bibitem[Simmons, 1954]{Simmons1954} Simmons, E. (1954)
Sales of Government Securities to Federal Reserve Banks Under Repurchase Agreements.
{\em Journal of Finance} 9(1): 23-40.

\bibitem[Simon and Campasano, 2014]{Simon2014} Simon, D.P. and Campasano, J. (2014)
The VIX Futures Basis: Evidence and Trading Strategies.
{\em Journal of Derivatives} 21(3): 54-69.

\bibitem[Simpson and Grossman, 2016]{Simpson2016} Simpson, M.W. and Grossman, A. (2016)
The Role of Industry Effects in Simultaneous Reversal and Momentum Patterns in One-Month Stock Returns.
{\em Journal of Behavioral Finance} 17(4): 309-320.

\bibitem[Simutin, 2014]{Simutin2014} Simutin, M. (2014)
Cash Holdings and Mutual Fund Performance.
{\em Review of Finance} 18(4): 1425-1464.

\bibitem[Sing and Low, 2000]{Sing2000} Sing, T.-F. and Low, S.-H.Y. (2000)
The inflation-hedging characteristics of real estate and financial assets in Singapore.
{\em Journal of Real Estate Portfolio Management} 6(4): 373-386.

\bibitem[Singh and Chandra, 2003]{Singh2003} Singh, Y. and Chandra, P. (2003)
A class +1 sigmoidal activation functions for FFANNs.
{\em Journal of Economic Dynamics and Control} 28(1): 183-187.

\bibitem[Singhal, Newell and Nguyen, 2011]{Singhal2011} Singhal, S., Newell, G. and Nguyen, T.K. (2011)
The significance and performance of infrastructure in India.
{\em Journal of Property Research} 28(1): 15-34.

\bibitem[Siriopoulos and Fassas, 2009]{Siriopoulos2009} Siriopoulos, C. and Fassas, A. (2009)
Implied Volatility Indices -- A Review.
{\em Working Paper.} Available online: \url{https://ssrn.com/abstract=1421202}.

\bibitem[Skelton, 1983]{Skelton1983} Skelton, J.L. (1983)
Banks, firms and the relative pricing of tax-exempt and taxable bonds.
{\em Journal of Financial Economics} 12(3): 343-355.

\bibitem[Skiadopoulos, 2004]{Skiadopoulos2004} Skiadopoulos, G. (2004)
The Greek implied volatility index: construction and properties.
{\em Applied Financial Economics} 14(16): 1187-1196.

\bibitem[Skiadopoulos, Hodges and Clewlow, 1999]{Skiadopoulos1999} Skiadopoulos, G., Hodges, S. and Clewlow, L. (1999)
The Dynamics of the S\&P 500 Implied Volatility Surface.
{\em Review of Derivatives Research} 3(3): 263-282.

\bibitem[Slowinski and Zopounidis, 1995]{Slowinski1995} Slowinski, R. and Zopounidis, C. (1995)
Application of the rough set approach to evaluation of bankruptcy risk.
{\em Intelligent Systems in Accounting, Finance and Management} 4(1): 27-41.

\bibitem[Smit and Trigeorgis, 2009]{Smit2009} Smit, H.T.J. and Trigeorgis, L. (2009)
Valuing infrastructure investment: An option games approach.
{\em California Management Review} 51(2): 82-104.

\bibitem[Smith and Pantilei, 2015]{Smith2015} Smith, D.M. and Pantilei, V.S. (2015)
Do ``Dogs of the World" Bark or Bite? Evidence from Single-Country ETFs.
{\em Journal of Investing} 24(1): 7-15.

\bibitem[Smith and Shulman, 1976]{Smith1976} Smith, K.V. and Shulman, D. (1976)
Institutions Beware: The Performance of Equity Real Estate Investment Trusts.
{\em Financial Analysts Journal} 32(5): 61-66.

\bibitem[Sollinger, 1994]{Sollinger1994} Sollinger, A. (1994)
The Triparty Is Just Beginning.
{\em Institutional Investor} 28(1): 133-135.

\bibitem[\v{S}olt\'{e}s, 2010]{Soltes2010} \v{S}olt\'{e}s, M. (2010)
Relationship of speed certificates and inverse vertical ratio call back spread option strategy.
{\em E+M Ekonomie a Management} 13(2): 119-124.

\bibitem[\v{S}olt\'{e}s, 2011]{Soltes2011} \v{S}olt\'{e}s, V. (2011)
The application of the long and short combo option strategies in the building of structured products. In: Kocourek, A. (ed.)
{\em Proceedings of the 10th International Conference: Liberec Economic Forum 2011}. Liberec, Czech Republic: Technical University of Liberec, pp. 481-487.

\bibitem[\v{S}olt\'{e}s and Amaitiek, 2010a]{Soltes2010a} \v{S}olt\'{e}s, V. and Amaitiek, O.F.S. (2010a)
The Short Put Ladder Strategy and its Application in Trading and Hedging.
{\em Club of Economics in Miskolc: Theory, Methodology, Practice} 6(2): 77-85.

\bibitem[\v{S}olt\'{e}s and Amaitiek, 2010b]{Soltes2010b} \v{S}olt\'{e}s, V. and Amaitiek, O.F.S. (2010b)
Inverse Vertical Ratio Put Spread Strategy and its Application in Hedging against a Price Drop.
{\em Journal of Advanced Studies in Finance} 1(1): 100-107.

\bibitem[\v{S}olt\'{e}s and Rusn\'{a}kov\'{a}, 2012]{Soltes2012} \v{S}olt\'{e}s, V. and Rusn\'{a}kov\'{a}, M. (2012)
Long Combo strategy using barrier options and its application in hedging against a price drop.
{\em Acta Montanistica Slovaca} 17(1): 17-32.

\bibitem[\v{S}olt\'{e}s and Rusn\'{a}kov\'{a}, 2013]{Soltes2013} \v{S}olt\'{e}s, V. and Rusn\'{a}kov\'{a}, M. (2013)
Hedging Against a Price Drop Using the Inverse Vertical Ratio Put Spread Strategy Formed by Barrier Options.
{\em Engineering Economics} 24(1): 18-27.

\bibitem[S{\o}rensen, 1999]{Sorensen1999} S{\o}rensen, C. (1999)
Dynamic Asset Allocation and Fixed Income Management.
{\em Journal of Financial and Quantitative Analysis} 34(4): 513-531.

\bibitem[Sorensen and Burke, 1986]{Sorensen1986} Sorensen, E.H. and Burke, T. (1986)
Portfolio Returns from Active Industry Group Rotation.
{\em Financial Analysts Journal} 42(5): 43-50.

\bibitem[S{\o}rensen and Trolle, 2005]{Sorensen2005} S{\o}rensen, C. and Trolle, A.B. (2005)
A General Model of Dynamic Asset Allocation with Incomplete Information and Learning.
{\em Working paper.} Available online: \url{https://ssrn.com/abstract=675625}.

\bibitem[S\"{o}rensson, 1993]{Sorensson1993} S\"{o}rensson, T. (1993)
Two methods for valuing convertible bonds -- A comparison.
{\em Scandinavian Journal of Management} 9(S1): 129-139.

\bibitem[Soudijn, 2016]{Soudijn2016} Soudijn, M.R.J. (2016)
Rethinking money laundering and drug trafficking: Some implications for investigators, policy makers and researchers.
{\em Journal of Money Laundering Control} 19(3): 298-310.

\bibitem[Sprenger {\em et al}, 2014]{Sprenger2014} Sprenger, T.O., Tumasjan, A., Sandner, P.G. and Welpe, I.M. (2014)
Tweets and trades: The information content of stock microblogs.
{\em European Financial Management} 20(5): 926-957.

\bibitem[Spyrou, 2005]{Spyrou2005} Spyrou, S.I. (2005)
Index Futures Trading and Spot Price Volatility: Evidence from an Emerging Market.
{\em Journal of Emerging Market Finance} 4(2): 151-167.

\bibitem[Staal {\em et al}, 2015]{Staal2015} Staal, A., Corsi, M., Shores, S. and Woida, C. (2015)
A Factor Approach to Smart Beta Development in Fixed Income.
{\em Journal of Index Investing} 6(1): 98-110.

\bibitem[Stambaugh, 1988]{Stambaugh1988} Stambaugh, R.F. (1988)
The information in forward rates: Implications for models of the term structure.
{\em Journal of Financial Economics} 21(1): 41-70.

\bibitem[Stanton, 1995]{Stanton1995} Stanton, R. (1995)
Rational Prepayment and the Valuation of Mortgage-Backed Securities.
{\em Review of Financial Studies} 8(3): 677-708.

\bibitem[Statman, Thorley and Vorkink, 2006]{Statman2006} Statman, M., Thorley, S. and Vorkink, K. (2006)
Investor Overconfidence and Trading Volume.
{\em Review of Financial Studies} 19(4): 1531-1565.

\bibitem[Stattman, 1980]{Stattman1980} Stattman, D. (1980)
Book Values and Stock Returns.
{\em Chicago MBA: A Journal of Selected Papers} 1980(4): 25-45.

\bibitem[Stefanini, 2006]{Stefanini2006} Stefanini, F. (2006)
{\em Investment Strategies of Hedge Funds.}
Chichester, UK: John Wiley \& Sons, Ltd.

\bibitem[Stein, 1992]{Stein1992} Stein, J.C. (1992)
Convertible bonds as backdoor equity financing.
{\em Journal of Financial Economics} 32(1): 3-21.

\bibitem[Stein, 1995]{Stein1995} Stein, J.C. (1995)
Prices and Trading Volume in the Housing Market: A Model with Down-Payment Effects.
{\em Quarterly Journal of Economics} 110(2): 379-406.

\bibitem[Steinert and Crowe, 2001]{Steinert2001} Steinert, M. and Crowe, S.  (2001)
Global Real Estate Investment: Characteristics, Optimal Portfolio Allocation and Future Trends.
{\em Pacific Rim Property Research Journal} 7(4): 223-239.

\bibitem[Stevenson, 2001]{Stevenson2001} Stevenson, S. (2001)
Bayes-Stein Estimators and International Real Estate Asset Allocation.
{\em Journal of Real Estate Research} 21(1/2): 89-104.

\bibitem[Stevenson, 2002]{Stevenson2002} Stevenson, S. (2002)
Momentum Effects and Mean Reversion in Real Estate Securities.
{\em Journal of Real Estate Research} 23(1/2): 47-64.

\bibitem[Stickel, 1991]{Stickel1991} Stickel, S.E. (1991)
Common stock returns surrounding earnings forecast revisions: More puzzling evidence.
{\em Accounting Review} 66(2): 402-416.

\bibitem[Stivers and Sun, 2010]{Stivers2010} Stivers, C. and Sun, L. (2010)
Cross-Sectional Return Dispersion and Time Variation in Value and Momentum Premiums.
{\em Journal of Financial and Quantitative Analysis} 45(4): 987-1014.

\bibitem[Stoll, 1969]{Stoll1969} Stoll, H.R. (1969)
The Relationship Between Put and Call Option Prices.
{\em Journal of Finance} 24(5): 801-824.

\bibitem[Stotz, 2016]{Stotz2016} Stotz, O. (2016)
Investment strategies and macroeconomic news announcement days.
{\em Journal of Asset Management} 17(1): 45-56.

\bibitem[Stoval, 1996]{Stoval1996} Stovall, S. (1996)
{\em Sector Investing.} New York, NY: McGraw Hill, Inc.

\bibitem[Stroebel and Taylor 2012]{Stroebel2012} Stroebel, J. and Taylor, J.B. (2012)
Estimated Impact of the Federal Reserve's Mortgage-Backed Securities Purchase Program.
{\em International Journal of Central Banking} 8(2): 1-42.

\bibitem[St\"{u}binger and Bredthauer, 2017]{StuingerBredthauer2017} St\"{u}binger, J. and Bredthauer, J. (2017)
Statistical Arbitrage Pairs Trading with High-frequency Data.
{\em International Journal of Economics and Financial Issues} 7(4): 650-662.

\bibitem[St\"{u}binger and Endres, 2017]{Stubinger2017} St\"{u}binger, J. and Endres, S. (2017)
Pairs trading with a mean-reverting jump-diffusion model on high-frequency data.
{\em Quantitative Finance} (forthcoming). DOI: \url{https://doi.org/10.1080/14697688.2017.1417624}.

\bibitem[Stulz, 1996]{Stulz1996} Stulz, R.M. (1996)
Rethinking risk management.
{\em Journal of Applied Corporate Finance} 9(3): 8-25.

\bibitem[Stulz, 2010]{Stulz2010} Stulz, R.M. (2010)
Credit Default Swaps and the Credit Crisis.
{\em Journal of Economic Perspectives} 24(1): 73-92.

\bibitem[Su, 2006]{Su2006} Su, X. (2006)
Hedging basket options by using a subset of underlying assets.
{\em Working paper.} Available online: \url{https://www.econstor.eu/bitstream/10419/22959/1/bgse14_2006.pdf}.

\bibitem[Su and Knowles, 2010]{Su2010} Su, E. and Knowles, T.W. (2010)
Measuring Bond Portfolio Value at Risk and Expected Shortfall in US Treasury Market.
{\em Asia Pacific Management Review} 15(4): 477-501.

\bibitem[Subha and Nambi, 2012]{Subha2012} Subha, M. and Nambi, S. (2012)
Classification of stock index movement using k-Nearest Neighbours (k-NN) algorithm.
{\em WSEAS Transactions on Information Science and Applications} 9(9): 261-270.

\bibitem[Subramanian, 2004]{Subramanian2004} Subramanian, A. (2004)
Option pricing on stocks in mergers and acquisitions.
{\em Journal of Finance} 59(2): 795-829.

\bibitem[Suhonen, Lennkh and Perez, 2017]{Suhonen2017} Suhonen, A., Lennkh, M. and Perez, F. (2017)
Quantifying Backtest Overfitting in Alternative Beta Strategies.
{\em Journal of Portfolio Management} 43(2): 90-104.

\bibitem[Sul, Dennis and Yuan, 2017]{Sul2017} Sul, H.K., Dennis, A.R. and Yuan, L.(I). (2017)
Trading on Twitter: Using Social Media Sentiment to Predict Stock Returns.
{\em Decision Sciences} 48(3): 454-488.

\bibitem[Sullivan, Timmermann and White, 1999]{Sullivan1999} Sullivan, R., Timmermann, A. and White, H. (1999)
Data-snooping, technical trading rule performance, and the bootstrap.
{\em Journal of Finance} 54(5): 1647-1691.

\bibitem[Summers, 1980]{Summers1980} Summers, B.J. (1980)
Negotiable Certificates of Deposit.
{\em Federal Reserve Bank of Richmond, Economic Review} 66(4): 8-19.

\bibitem[Suresh, 2015]{Suresh2015} Suresh, A.S. (2015)
Analysis of Option Combination Strategies.
{\em Management Insight} 11(1): 31-40.

\bibitem[Svec and Stevenson, 2007]{Svec2007} Svec, J. and Stevenson, M. (2007)
Modelling and forecasting temperature based weather derivatives.
{\em Global Finance Journal} 18(2): 185-204.

\bibitem[Swank and Root, 1995]{Swank1995} Swank, T.A. and Root, T.H. (1995)
Bonds in Default: Is Patience a Virtue?
{\em Journal of Fixed Income} 5(1): 26-31.

\bibitem[Swinkels, 2002]{Swinkels2002} Swinkels, L. (2002)
International Industry Momentum.
{\em Journal of Asset Management} 3(2): 124-141.

\bibitem[Swishchuk and Cui, 2013]{Swishchuk2013} Swishchuk, A. and Cui, K. (2013)
Weather derivatives with applications to Canadian data.
{\em Journal of Mathematical Finance} 3(1): 81-95.

\bibitem[Switzer and Jiang, 2010]{Switzer2010} Switzer, L.N. and Jiang, H. (2010)
Market Efficiency and the Risks and Returns of Dynamic Trading Strategies with Commodity Futures.	
In: Stanley, H.E. (ed.) {\em Proceedings Of The First Interdisciplinary Chess Interactions Conference.} Singapore: World Scientific Publishing, pp. 127-156.

\bibitem[Symeonidis {\em et al}, 2012]{Symeonidis2012} Symeonidis, L., Prokopczuk, M., Brooks, C. and Lazar, E. (2012)
Futures basis, inventory and commodity price volatility: An empirical analysis.
{\em Economic Modelling} 29(6): 2651-2663.

\bibitem[Szado and Schneeweis, 2010]{Szado2010} Szado, E. and Schneeweis, T. (2010)
Loosening Your Collar: Alternative Implementations of QQQ Collars.
{\em Journal of Trading} 5(2): 35-56.

\bibitem[Szado and Schneeweis, 2011]{Szado2011} Szado, E. and Schneeweis, T. (2011)
An Update of `Loosening Your Collar: Alternative Implementations of QQQ Collars': Credit Crisis and Out-of-Sample Performance.
{\em Working Paper.} Available online: \url{http://ssrn.com/abstract=1507991}.

\bibitem[Szakmary, Shen and Sharma, 2010]{Szakmary2010} Szakmary, A.C., Shen, Q. and Sharma, S.C. (2010)
Trend-following trading strategies in commodity futures: A re-examination.
{\em Journal of Banking \& Finance} 34(2): 409-426.

\bibitem[Szakmary and Zhou, 2015]{Szakmary2015} Szakmary, A.C. and Zhou, X. (2015)
Industry momentum in an earlier time: Evidence from the Cowles data.
{\em Journal of Financial Research} 38(3): 319-347.

\bibitem[Tang and Jang, 2011]{Tang2011} Tang, C.H. and Jang, S.H. (2011)
Weather risk management in ski resorts: Financial hedging and Geographical diversification.
{\em International Journal of Hospitality Management} 30(2): 301-311

\bibitem[Tang and Xu, 2013]{Tang2013} Tang, H. and Xu, X.E. (2013)
Solving the Return Deviation Conundrum of Leveraged Exchange-Traded Funds.
{\em Journal of Financial and Quantitative Analysis} 48(1): 309-342.

\bibitem[Tavakoli, 1998]{Tavakoli1998} Tavakoli, J.M. (1998)
{\em Credit Derivatives \& Synthetic Structures: A Guide to Instruments and Applications.} (2nd ed.)
Hoboken, NJ: John Wiley \& Sons, Inc.

\bibitem[Tay and Cao, 2001]{Tay2001a} Tay, F.E.H. and Cao, L. (2001)
Application of support vector machines in financial time series forecasting.
{\em Omega} 29(4): 309-317.

\bibitem[Taylor, 1999]{Taylor1999} Taylor, C.R. (1999)
Time-on-the-Market as a Sign of Quality.
{\em Review of Economic Studies} 66(3): 555-578.

\bibitem[Taylor, 2004]{Taylor2004} Taylor, N. (2004)
Modeling discontinuous periodic conditional volatility: Evidence from the commodity futures market.
{\em Journal of Futures Markets} 24(9): 805-834.

\bibitem[Taylor, 2016]{Taylor2016} Taylor, N. (2016)
Roll strategy efficiency in commodity futures markets.
{\em Journal of Commodity Markets} 1(1): 14-34.

\bibitem[Taylor and Allen, 1992]{Taylor1992} Taylor, M.P. and Allen, H. (1992)
The use of technical analysis in the foreign exchange market.
{\em Journal of International Money and Finance} 11(3): 304-314.

\bibitem[Teixeira and de Oliveira, 2010]{Teixeira2010} Teixeira, L.A. and de Oliveira, A.L.I. (2010)
A method for automatic stock trading combining technical analysis and nearest neighbor classification.
{\em Expert Systems with Applications} 37(10): 6885-6890.

\bibitem[Telser, 1958]{Telser1958} Telser, L.G. (1958)
Futures Trading and the Storage of Cotton and Wheat.
{\em Journal of Political Economy} 66(3): 233-255.

\bibitem[The Options Institute, 1995]{OI1995} The Options Institute (1995)
{\em Options: Essential Concepts and Trading Strategies.} (2nd ed.) Chicago, IL: Richard D. Irwin, Inc.

\bibitem[Thibodeau and Giliberto, 1989]{Thibodeau1989} Thibodeau, T.G. and Giliberto, S.M. (1989)
Modeling Conventional Residential Mortgage Refinancing.
{\em Journal of Real Estate Finance and Economics} 2(4): 285-299.

\bibitem[Thomsett, 2003]{Thomsett2003} Thomsett, M.C. (2003)
{\em Support and Resistance Simplified.}
Columbia, MD: Marketplace Books.

\bibitem[Thornes, 2006]{Thornes2006} Thornes, J.E. (2006)
An introduction to weather and climate derivatives.
{\em Weather} 58(5): 193-196.

\bibitem[Thorp, 2006]{Thorp2006} Thorp, E.O. (2006)
The Kelly criterion in blackjack, sports betting, and the stock market.
In: Zenios, S.A. and Ziemba, W.T. (eds.)
{\em Handbook of Asset and Liability Management: Theory and Methodology (Vol. 1).} Amsterdam, The Netherlands: Elsevier, pp. 385-428.

\bibitem[Thorp and Kassouf, 1967]{Thorp1967} Thorp, E.O. and Kassouf, S.T. (1967)
{\em Beat the Market: A Scientific Stock Market System.}
New York, NY: Random House.

\bibitem[Till, 2008]{Till2008} Till, H. (2008)
Case Studies and Risk Management Lessons in Commodity Derivatives Trading.
In: Geman, H. (ed.) {\em Risk Management in Commodity Markets: From Shipping to Agriculturals and Energy.}
Chichester, UK: John Wiley \& Sons, Ltd., pp. 255-291.

\bibitem[Till and Eagleeye, 2017]{Till2017} Till, H. and Eagleeye, J. (2017)
Commodity Futures Trading Strategies: Trend-Following and Calendar Spreads.
{\em Working Paper.} Available online: \url{https://ssrn.com/abstract=2942340}.

\bibitem[Tille, Stoffels and Gorbachev, 2001]{Tille2001} Tille, C., Stoffels, N. and Gorbachev, O. (2001)
To What Extent Does Productivity Drive the Dollar?
{\em Current Issues in Economics and Finance} 7(8): 1-6.

\bibitem[Timmermans, Schumacher and Ponds, 2017]{Timmermans2017} Timmermans, S.H.J.T., Schumacher, J.M. and Ponds, E.H.M. (2017)
A multi-objective decision framework for lifecycle investment.
{\em Working Paper.} Available online: \url{http://ssrn.com/abstract=3038803}.

\bibitem[Tinoco and Wilson, 2013]{Tinoco2013} Tinoco, M.H. and Wilson, N. (2013)
Financial distress and bankruptcy prediction among listed companies using accounting, market and macroeconomic variables.
{\em International Review of Financial Analysis} 30: 394-419.

\bibitem[Titman and Warga, 1986]{Titman1986} Titman, S. and Warga, A. (1986)
Risk and the Performance of Real Estate Investment Trusts: A Multiple Index Approach.
{\em AREUEA Journal} 14(3): 414-431.

\bibitem[Todorov, 2010]{Todorov2010} Todorov, V. (2010)
Variance Risk-Premium Dynamics: The Role of Jumps.
{\em Review of Financial Studies} 23(1): 345-383.

\bibitem[Toevs and Jacob, 1986]{Toevs1986} Toevs, A. and Jacob, D. (1986)
Futures and Alternative Hedge Ratio Methodologies.
{\em Journal of Portfolio Management} 12(3): 60-70.

\bibitem[Tokic, 2013]{Tokic2013} Tokic, D. (2013)
Crude oil futures markets: Another look into traders' positions.
{\em Journal of Derivatives \& Hedge Funds} 19(4): 321-342.

\bibitem[Topaloglou, Vladimirou and Zenios, 2011]{Topaloglou2011} Topaloglou, N., Vladimirou, H. and Zenios, S.A. (2011)
Optimizing International Portfolios with Options and Forwards.
{\em Journal of Banking \& Finance} 35(12): 3188-3201.

\bibitem[Torrance, 2007]{Torrance2007} Torrance, M.I. (2007)
The Power of Governance in Financial Relationships: Governing Tensions in Exotic Infrastructure Territory.
{\em Growth and Change} 38(4): 671-695.

\bibitem[Torricelli, 2018]{Torricelli2018} Torricelli, L. (2018)
Volatility Targeting Using Delayed Diffusions.
{\em Working Paper.} Available online: \url{https://ssrn.com/abstract=2902063}.

\bibitem[Trainer, 1983]{Trainer1983} Trainer, F.H., Jr. (1983)
The Uses of Treasury Bond Futures in Fixed Income Portfolio Management.
{\em Financial Analysts Journal} 39(1): 27-34.

\bibitem[Trainor, 2010]{Trainor2010} Trainor, W.J., Jr. (2010)
Do Leveraged ETFs Increase Volatility?
{\em Technology and Investment} 1(3): 215-220.

\bibitem[Trehan, 2005]{Trehan2005} Trehan, B. (2005)
Oil price shocks and inflation.
{\em Federal Reserve Bank of San Francisco, Economic Letter}, No. 2005-28. Available online: \url{https://www.frbsf.org/economic-research/files/el2005-28.pdf}.

\bibitem[Trifonov {\em et al}, 2011]{Trifonov2011} Trifonov, Y., Yashin, S., Koshelev, E. and Podshibyakin, D. (2011)
Application of Synthetic conos for Equity Risk Management.
In: \v{C}ern\'{a}k, Z. (ed.) {\em Materi\'{a}ly VII mezin\'{a}rodn\'{\i} v\v{e}decko -- praktick\'{a} konference
``Zpr\'{a}vy v\v{e}deck\'{e} ideje -- 2011".} Prague, Czech Republic: Education and Science.

\bibitem[Trifonov {\em et al}, 2014]{Trifonov2014} Trifonov, Y., Yashin, S., Koshelev, E. and Podshibyakin, D. (2014)
Testing the Technology of Synthetic conos.
{\em Working Paper.} Available online: \url{https://ssrn.com/abstract=2429657}.

\bibitem[Tripathi and Garg, 2016]{Tripathi2016} Tripathi, V. and Garg, S. (2016)
A Cross-Country Analysis of Pricing Efficiency of Exchange Traded Funds.
{\em Journal of Applied Finance} 22(3): 41-63.

\bibitem[Trzcinka, 1982]{Trzcinka1982} Trzcinka, C. (1982)
The Pricing of Tax-Exempt Bonds and the Miller Hypothesis.
{\em Journal of Finance} 37(4): 907-923.

\bibitem[Tsai and Hsiao, 2010]{Tsai2010} Tsai, C.F. and Hsiao, Y.C. (2010)
Combining multiple feature selection methods for stock prediction: Union, intersection, and multi-intersection approaches.
{\em Decision Support Systems} 50(1): 258-269.

\bibitem[Tsai, Hsu and Yen, 2014]{Tsai2014} Tsai, C., Hsu, Y. and Yen, D.C. (2014)
A comparative study of classifier ensembles for bankruptcy prediction.
{\em Applied Soft Computing} 24: 977-984.

\bibitem[Tse, 2017]{Tse2017} Tse, Y. (2017)
Return predictability and contrarian profits of international index futures.
{\em Journal of Futures Markets} 38(7): 788-803.

\bibitem[Tsiveriotis and Fernandes, 1998]{Tsiveriotis1998} Tsiveriotis, K. and Fernandes, C. (1998)
Valuing convertible bonds with credit risk.
{\em Journal of Fixed Income} 8(2): 95-102.

\bibitem[Tuckman and Serrat, 2015]{Tuckman2011} Tuckman, B. and Serrat, A. (2012)
{\em Fixed Income Securities: Tools for Today's Markets.} (3rd ed.)
Hoboken, NJ: John Wiley \& Sons, Inc.

\bibitem[Tulchinsky {\em et al}, 2015]{Tulchinsky2015} Tulchinsky, I. {\em et al}. (2015)
{\em Finding Alphas: A Quantitative Approach to Building Trading Strategies.}
New York, NY: John Wiley \& Sons, Inc.

\bibitem[Turnovsky, 1989]{Turnovsky1989} Turnovsky, S.J. (1989)
The Term Structure of Interest Rates and the Effects of Macroeconomic Policy.
{\em Journal of Money, Credit and Banking} 21(3): 321-347.

\bibitem[Tuzun, 2013]{Tuzun2013} Tuzun, T. (2013)
Are Leveraged and Inverse ETFs the New Portfolio Insurers?
{\em Finance and Economics Discussion Series (FEDS)}, Paper No. 2013-48.
Washington, DC: Board of Governors of the Federal Reserve System.
Available online: \url{https://www.federalreserve.gov/pubs/feds/2013/201348/201348pap.pdf}.

\bibitem[Tversky and Kahneman, 1992]{Tversky1992} Tversky, A. and Kahneman, D. (1992)
Advances in prospect theory: Cumulative representation of uncertainty.
{\em Journal of Risk and Uncertainty} 5(4): 297-323.

\bibitem[Uhlenbeck and Ornstein, 1930]{Uhlenbeck1930} Uhlenbeck, G.E. and Ornstein, L.S. (1930)
On the Theory of the Brownian Motion.
{\em Physical Review} 36(5): 823-841.

\bibitem[Vaitonis and Masteika, 2016]{Vaitonis2016} Vaitonis, M. and Masteika, S. (2016)
Research in high frequency trading and pairs selection algorithm with Baltic region stocks.
In: Dregvaite, G. and Damasevicius, R. (eds.)
{\em Proceedings of the 22nd International Conference on Information and Software Technologies (ICIST 2016).}
Cham, Switzerland: Springer, pp. 208-217.

\bibitem[Van Alstyne, 2014]{VanAlstyne2014} Van Alstyne, M. (2014)
Why Bitcoin has value.
{\em Communications of the ACM} 57(5): 30-32.

\bibitem[van den Goorbergh, 2004]{vandenGoorbergh2004} van den Goorbergh, R.W.J. (2004)
{\em Essays on optimal hedging and investment strategies and on derivative pricing} (Ph.D. Thesis).
Tilburg, The Netherlands: Tilburg University.

\bibitem[van den Noord and Andr\'e, 2004]{Noord2007} van den Noord, P. and Andr\'e, C. (2007)
Why has Core Inflation Remained so Muted in the Face of the Oil Shock?
{\em Working Paper.} Available online: \url{http://dx.doi.org/10.1787/206408110285}.

\bibitem[Van Kervel and Menkveld, 2017]{VanKervel2017} Van Kervel, V. and Menkveld, A.J. (2017)
High-Frequency Trading around Large Institutional Orders.
{\em Journal of Finance} (forthcoming). Available online: \url{https://ssrn.com/abstract=2619686}.

\bibitem[van Marle and Verwijmeren, 2017]{Marle2017} van Marle, M. and Verwijmeren, P. (2017)
The long and the short of convertible arbitrage: An empirical examination of arbitrageurs' holding periods.
{\em Journal of Empirical Finance} 44: 237-249.

\bibitem[Van Oord, 2016]{VanOord2016} Van Oord, J.A. (2016)
{\em Essays on Momentum Strategies in Finance} (Ph.D. Thesis). Rotterdam, The Netherlands: Erasmus University.
Available online: \url{https://repub.eur.nl/pub/80036/EPS2016380F-A9789058924445.pdf}.

\bibitem[Vanstone and Finnie, 2009]{Vanstone2009} Vanstone, B. and Finnie, G. (2009)
An empirical methodology for developing stockmarket trading systems using artificial neural networks.
{\em Expert Systems with Applications} 36(3): 6668-6680.

\bibitem[Van Tassel, 2016]{VanTassel2016} Van Tassel, P. (2016)
Merger Options and Risk Arbitrage.
{\em Federal Reserve Bank of New York Staff Reports}, No. 761. Available online:\\ \url{https://www.newyorkfed.org/medialibrary/media/research/staff_reports/sr761.pdf?la=en}.

\bibitem[Vasicek, 2015]{Vasicek2015} Vasicek, O.A. (2015)
Probability of Loss on Loan Portfolio.
In: Vasicek, O.A. (ed.)
{\em Finance, Economics and Mathematics.} Hoboken, NJ: John Wiley \& Sons, Inc., Chapter 17.

\bibitem[Vassalou and Xing, 2004]{Vassalou2004} Vassalou, M. and Xing, Y. (2004)
Default Risk in Equity Returns.
{\em Journal of Finance} 59(2): 831-868.

\bibitem[Vedenov and Barnett, 2004]{Vedenov2004} Vedenov, D.V. and Barnett, B.J. (2004)
Efficiency of Weather Derivatives as Primary Crop Insurance Instruments.
{\em Journal of Agricultural Economics} 29(3): 387-403.

\bibitem[Vickery and Wright, 2010]{Vickery2010} Vickery, J. and Wright, J. (2010)
TBA Trading and Liquidity in the Agency MBS Market.
{\em Federal Reserve Bank of New York Staff Reports}, No. 468.
Available online: \url{https://www.newyorkfed.org/medialibrary/media/research/staff_reports/sr468.pdf}.

\bibitem[Vidyamurthy, 2004]{Vidyamurthy2004} Vidyamurthy, G. (2004)
{\em Pairs Trading: Quantitative Methods and Analysis.}
Hoboken, NJ: John Wiley \& Sons, Inc.

\bibitem[Viezer, 2000]{Viezer2000} Viezer, T.W. (2000)
Evaluating ``Within Real Estate" Diversification Strategies.
{\em Journal of Real Estate Portfolio Management} 6(1): 75-95.

\bibitem[Villani and Davis, 2006]{Villani2006} Villani, R. and Davis, C. (2006)
{\em FLIP: How to find, fix, and sell houses for profit.}
New York, NY: McGraw-Hill, Inc.

\bibitem[Vipul, 2009]{Vipul2009} Vipul (2009)
Box-spread arbitrage efficiency of Nifty index options: The Indian evidence.
{\em Journal of Futures Markets} 29(6): 544-562.

\bibitem[Viswanath, 1993]{Viswanath1993} Viswanath, P.V. (1993)
Efficient Use of Information, Convergence Adjustments, and Regression Estimates of Hedge Ratios.
{\em Journal of Futures Markets} 13(1): 43-53.

\bibitem[Vives, 1999]{Vives1999} Vives, A. (1999)
Pension Funds in Infrastructure Project Finance: Regulations and Instrument Design.
{\em Journal of Structured Finance} 5(2): 37-52.

\bibitem[Volpert, 1991]{Volpert1991} Volpert, B.S. (1991)
Opportunities for Investing in Troubled Companies.
In: Dinapoli, D., Sigoloff, S.C. and Cushman, R.F. (eds.)
{\em Workouts and Turnarounds: The Handbook of Restructuring and Investing in Distressed Companies.} Homewood, IL: Business One-Irwin, pp. 514-542.

\bibitem[Vrugt {\em et al}, 2007]{Vrugt2007} Vrugt, E.B., Bauer, R., Molenaar, R. and Steenkamp, T. (2007)
Dynamic commodity trading strategies.
In: Till, H. and Eagleeye, J. (eds.)
{\em Intelligent Commodity Investing: New Strategies and Practical Insights for Informed Decision Making.} London, UK: Risk Books, Chapter 16.

\bibitem[Walker, 1999]{Walker1999} Walker, J. (1999)
How Big is Global Money Laundering?
{\em Journal of Money Laundering Control} 3(1): 25-37.

\bibitem[Walker, 2008]{Walker2008} Walker, M.B. (2008)
The static hedging of CDO tranche correlation risk.
{\em International Journal of Computer Mathematics} 86(6): 940-954.

\bibitem[Walkling, 1985]{Walkling1985} Walkling, R.A. (1985)
Predicting tender offer success: A logistic analysis.
{\em Journal of Financial and Quantitative Analysis} 20(4): 461-478.

\bibitem[Wang, 2005]{Wang2005} Wang, K.Q. (2005)
Multifactor Evaluation of Style Rotation.
{\em Journal of Financial and Quantitative Analysis} 40(2): 349-372.

\bibitem[Wang, 2014]{Wang2014} Wang, L. (2014)
Margin-Based Asset Pricing and the Determinants of the CDS Basis.
{\em Journal of Fixed Income} 24(2): 61-78.

\bibitem[Wang {\em et al}, 2017]{Wang2017} Wang, J., Brooks, R., Lu, X. and Holzhauer, H.M. (2017)
Sector Momentum.
{\em Journal of Investing} 26(2): 48-60.

\bibitem[Wang and Min, 2013]{Wang2013} Wang, C.-H. and Min, K.J. (2013)
Electric Power Plant Valuation Based on Day-Ahead Spark Spreads.
{\em Engineering Economist} 58(3): 157-178.

\bibitem[Wang and Vergne, 2017]{WangVergne2017} Wang, S. and Vergne, J.-P. (2017)
Buzz factor or innovation potential: What explains cryptocurrencies returns?
{\em PLoS ONE} 12(1): e0169556.

\bibitem[Wang and Yu, 2004]{Wang2004} Wang, C. and Yu, M. (2004)
Trading activity and price reversals in futures markets.
{\em Journal of Banking \& Finance} 28(6): 1337-1361.

\bibitem[Ward and Griepentrog, 1993]{Ward1993}
Ward, D. and Griepentrog, G. (1993)
Risk and Return in Defaulted Bonds.
{\em Financial Analysts Journal} 49(3): 61-65.

\bibitem[Watts, 1978]{Watts1978} Watts, R.L. (1978)
Systematic `abnormal' returns after quarterly earnings announcements.
{\em Journal of Financial Economics} 6(2-3): 127-150.

\bibitem[Webb, Curcio and Rubens, 1988]{Webb1988} Webb, J.R., Curcio, R.J. and Rubens, J.H. (1988)
Diversification gains from including real estate in mixed-asset portfolios.
{\em Decision Sciences} 19(2): 434-452.

\bibitem[Weber, Adair and McGreal, 2008]{Weber2008} Weber, B.R., Adair, A. and McGreal, S. (2008)
Solutions to the five key brownfield valuation problems.
{\em Journal of Property Investment \& Finance} 26(1): 8-37.

\bibitem[Weber, Staub-Bisang and Alfen, 2016]{Weber2016} Weber, B., Staub-Bisang, M. and Alfen, H.W. (2016)
{\em Infrastructure as an Asset Class: Investment Strategy, Sustainability, Project Finance and PPP.}
Chichester, UK: John Wiley \& Sons, Ltd.

\bibitem[Weinert, 2007]{Weinert2007} Weinert, H.L. (2007)
Efficient computation for Whittaker-Henderson smoothing.
{\em Computational Statistics \& Data Analysis} 52(2): 959-974.

\bibitem[Weiser, 2003]{Weiser2003} Weiser, S. (2003)
The strategic case for commodities in portfolio diversification.
{\em Commodities Now}, September 2003, pp. 7-11.

\bibitem[Weller, Friesen and Dunham, 2009]{Weller2009} Weller, P.A., Friesen, G.C. and Dunham, L.M. (2009)
Price trends and patterns in technical analysis: a theoretical and empirical examination.
{\em Journal of Banking \& Finance} 6(33): 1089-1100.

\bibitem[Wells, 2016]{Wells2016} Wells, B. (2016) The Foreign Tax Credit War.
{\em Brigham Young University Law Review} 2016(6): 1895-1965.

\bibitem[Whaley, 2000]{Whaley2000} Whaley, R.E. (2000)
The Investor Fear Gauge.
{\em Journal of Portfolio Management} 26(3): 12-16.

\bibitem[Whaley, 2002]{Whaley2002} Whaley, R.E. (2002)
Return and Risk of CBOE Buy Write Monthly Index.
{\em Journal of Derivatives} 10(2): 35-42.

\bibitem[Whaley, 2009]{Whaley2009} Whaley, R.E. (2009)
Understanding the VIX.
{\em Journal of Portfolio Management} 35(3): 98-105.

\bibitem[Wheaton, 1990]{Wheaton1990} Wheaton, W.C. (1990)
Vacancy, Search, and Prices in a Housing Market Matching Model.
{\em Journal of Political Economy} 98(6): 1270-1292.

\bibitem[White, 2015]{White2015} White, L.H. (2015)
The Market for Cryptocurrencies.
{\em Cato Journal} 35(2): 383-402.

\bibitem[Whittaker, 1923]{Whittaker1923} Whittaker, E.T. (1923)
On a New Method of Graduations.
{\em Proceedings of the Edinburgh Mathematical Society} 41: 63-75.

\bibitem[Whittaker, 1924]{Whittaker1924} Whittaker, E.T. (1924)
On the theory of graduation.
{\em Proceedings of the Royal Society of Edinburgh} 44: 77-83.

\bibitem[Wilder, 1978]{Wilder1978} Wilder, J.W., Jr. (1978)
{\em New Concepts in Technical Trading Systems.}
Greensboro, NC: Trend Research.

\bibitem[Willett, 2006]{Willett2006} Willett, P. (2006)
The Porter stemming algorithm: then and now.
{\em Program: Electronic Library and Information Systems} 40(3): 219-223.

\bibitem[Wilner, 1996]{Wilner1996} Wilner, R. (1996)
 A new tool for portfolio managers: Level, slope, and curvature durations.
{\em Journal of Fixed Income} 6(1): 48-59.

\bibitem[Wilson, 2016]{Wilson2016} Wilson, D.J. (2016)
The Impact of Weather on Local Employment: Using Big Data on Small Places.
{\em Federal Reserve Bank of San Francisco Working Papers Series}, No. 2016-21. Available online: \url{https://www.frbsf.org/economic-research/files/wp2016-21.pdf}.

\bibitem[Wilson {\em et al}, 2018]{Wilson2018} Wilson, A.C., Roelofs, R., Stern, M., Stern, N. and Recht, B. (2018)
The Marginal Value of Adaptive Gradient Methods in Machine Learning.
{\em Working Paper.} Available online: \url{https://arxiv.org/pdf/1705.08292.pdf}.

\bibitem[Wilson and Sharda, 1994]{Wilson1994} Wilson, R.L. and Sharda, R. (1994)
Bankruptcy prediction using neural networks.
{\em Decision Support Systems} 11(5): 545-557.

\bibitem[Windas, 2007]{Windas2007} Windas, T. (2007)
{\em An Introduction to Option-Adjusted Spread Analysis.} (Miller, T. (ed.) Revised and Expanded Third Edition.)
Princeton, NJ: Bloomberg Press.

\bibitem[Wolf, 1987]{Wolf1987} Wolf, A. (1987)
Optimal hedging with futures options.
{\em Journal of Economics and Business} 39(2): 141-158.

\bibitem[Wolf, 2014]{Wolf2014} Wolf, V. (2014)
{\em Comparison of Markovian Price Processes and Optimality of Payoffs} (Ph.D. Thesis).
Freiburg im Breisgau, Germany: Albert-Ludwigs-Universit\"{a}t Freiburg.
Available online: \url{https://freidok.uni-freiburg.de/fedora/objects/freidok:9664/datastreams/FILE1/content}.

\bibitem[Wong, Thompson and Teh, 2011]{Wong2011} Wong, W.-K., Thompson, H.E. and Teh, K. (2011)
Was There Abnormal Trading in the S\&P 500 Index Options Prior to the September 11 Attacks?
{\em Multinational Finance Journal} 15(3/4): 1-46.

\bibitem[Wood, 1997]{Wood1997} Wood, J. (1997)
A Simple Model for Pricing Imputation Tax Credits Under Australia's Dividend Imputation Tax System.
{\em Pacific-Basin Finance Journal} 5(4): 465-480.

\bibitem[Woodard and Garcia, 2008]{Woodard2008} Woodard, J. and Garcia, P. (2008)
Weather Derivatives, Spatial Aggregation, and Systemic Risk: Implications for Reinsurance Hedging.
{\em Journal of Agricultural and Resource Economics} 33(1): 34-51.

\bibitem[Woodlock and Dangol, 2014]{Woodlock2014} Woodlock, P. and Dangol, R. (2014)
Managing bankruptcy and default risk.
{\em Journal of Corporate Accounting \& Finance} 26(1): 33-38.

\bibitem[Woodward, 1990]{Woodward1990} Woodward, G.T. (1990)
The Real Thing: A Dynamic Profile of the Term Structure of Real Interest Rates and Inflation.
{\em Journal of Business} 63(3): 373-398.

\bibitem[Working, 1953]{Working1953} Working, H. (1953)
Futures Trading and Hedging.
{\em American Economic Review} 43(3): 314-434.

\bibitem[Worzala and Newell, 1997]{Worzala1997} Worzala, E. and Newell, G. (1997)
International real estate: A review of strategic investment issues.
{\em Journal of Real Estate Portfolio Management} 3(2): 87-96.

\bibitem[Wright {\em et al}, 2017]{Wright2017} Wright, R., Tekin, E., Topalli, V., McClellan, C., Dickinson, T. and Rosenfeld, R. (2017)
Less Cash, Less Crime: Evidence from the Electronic Benefit Transfer Program.
{\em Journal of Law and Economics} 60(2): 361-383.

\bibitem[Wu, 2009]{Wu2009} Wu, H. (2009)
Global stability analysis of a general class of discontinuous neural networks with linear growth activation functions.
{\em Information Sciences} 179(19): 3432-3441.

\bibitem[Wu, 2003]{Wu2003} Wu, L. (2003)
Jumps and Dynamic Asset Allocation.
{\em Review of Quantitative Finance and Accounting} 20(3): 207-243.

\bibitem[Wurstbauer {\em et al}, 2016]{Wurstbauer2016} Wurstbauer, D., Lang, S., Rothballer, C. and Sch\"{a}fers, W. (2016)
Can common risk factors explain infrastructure equity returns? Evidence from European capital markets.
{\em Journal of Property Research} 33(2): 97-120.

\bibitem[Wurstbauer and Sch\"{a}fers, 2015]{Wurstbauer2015} Wurstbauer, D. and Sch\"{a}fers, W. (2015)
Inflation hedging and protection characteristics of infrastructure and real estate assets.
{\em Journal of Property Investment \& Finance} 33(1): 19-44.

\bibitem[Wurtzebach, Mueller and Machi, 1991]{Wurtzebach1991}Wurtzebach, C.H., Mueller, G.R. and Machi, D. (1991)
The Impact of Inflation and Vacancy on Real Estate Returns.
{\em Journal of Real Estate Research} 6(2): 153-168.

\bibitem[Wystup, 2017]{Wystup2017} Wystup, U. (2017)
{\em FX Options and Structured Products.} (2nd ed.) eBook: John Wiley \& Sons, Inc.

\bibitem[Wystup and  Zhou, 2014]{Wystup2014} Wystup, U. and  Zhou, Q. (2014)
Volatility as investment -- crash protection with calendar spreads of variance swaps.
{\em Journal of Applied Operational Research} 6(4): 243-254.

\bibitem[Xiao, 2013]{Xiao2013} Xiao, T. (2013)
A simple and precise method for pricing convertible bond with credit risk.
{\em Journal of Derivatives \& Hedge Funds} 19(4): 259-277.

\bibitem[Xie {\em et al}, 2014]{Xie2014} Xie, W., Liew, Q.R., Wu, Y. and Zou, X. (2014)
Pairs Trading with Copulas.
{\em Working Paper.} Available online: \url{https://ssrn.com/abstract=2383185}.

\bibitem[Xing, Zhang and Zhao, 2010]{Xing2010} Xing, Y., Zhang, X. and Zhao, R. (2010)
What Does Individual Option Volatility Smirk Tell Us About Future Equity Returns?
{\em Journal of Financial and Quantitative Analysis} 45(3): 641-662.

\bibitem[Yadav and Pope, 1990]{Yadav1990} Yadav, P.K. and Pope, P.F. (1990)
Stock index futures arbitrage: International evidence.
{\em Journal of Futures Markets} 10(6): 573-603.

\bibitem[Yadav and Pope, 1994]{Yadav1994} Yadav, P.K. and Pope, P.F. (1994)
Stock index futures mispricing: profit opportunities or risk premia?
{\em Journal of Banking \& Finance} 18(5): 921-953.

\bibitem[Yamada, 1999]{Yamada1999} Yamada, S. (1999)
Risk Premiums in the JGB Market and Application to Investment Strategies.
{\em Journal of Fixed Income} 9(2): 20-41.

\bibitem[Yan, 2006]{Yan2006} Yan, X. (2006)
The Determinants and Implications of Mutual Fund Cash Holdings: Theory and Evidence.
{\em Financial Management} 35(2): 67-91.

\bibitem[Yang, Brockett and Wen, 2009]{Yang2009} Yang, C.C., Brockett, P.L. and Wen, M.-M. (2009)
Basis risk and hedging efficiency of weather derivatives.
{\em Journal of Risk Finance} 10(5): 517-536.

\bibitem[Yang, You and Ji, 2011]{Yang2011} Yang, Z., You, W. and Ji, G. (2011)
Using partial least squares and support vector machines for bankruptcy prediction.
{\em Expert Systems with Applications} 38(7): 8336-8342.

\bibitem[Yao, 2012]{Yao2012} Yao, Y. (2012)
Momentum, contrarian, and the January seasonality.
{\em Journal of Banking \& Finance} 36(10): 2757-2769.

\bibitem[Yao and Tan, 2000]{Yao2000} Yao, J. and Tan, C.L. (2000)
A case study on using neural networks to perform technical forecasting of forex.
{\em Neurocomputing} 34(1-4): 79-98.

\bibitem[Yao, Tan and Poh, 1999]{Yao1999} Yao, J., Tan, C.L. and Poh, H.L. (1999)
Neural networks for technical analysis: a study on KLCI.
{\em International Journal of Theoretical and Applied Finance} 2(2): 221-241.

\bibitem[Yared and Veronesi, 1999]{Veronesi1999} Yared, F. and Veronesi, P. (1999)
Short and Long Horizon Term and Inflation Risk Premia in the US Term Structure: Evidence from an Integrated Model for Nominal and Real Bond Prices Under Regime Shifts.
{\em Working Paper.} Available online: \url{https://ssrn.com/abstract=199448}.

\bibitem[Yavas and Yang, 1995]{Yavas1995} Yavas, A. and Yang, S. (1995)
The Strategic Role of Listing Price in Marketing Real Estate: Theory and Evidence.
{\em Real Estate Economics} 23(3): 347-368.

\bibitem[Yawitz, Maloney and Ederington, 1985]{YawitzMaloney1985} Yawitz, J.B., Maloney, K.J. and Ederington, L.H. (1985)
Taxes, Default Risk, and Yield Spreads.
{\em Journal of Finance} 40(4): 1127-1140.

\bibitem[Yawitz and Marshall, 1985]{Yawitz1985} Yawitz, J.B. and Marshall, W.B. (1985)
The Use of Futures in Immunized Portfolios.
{\em Journal of Portfolio Management} 11(2): 51-55.

\bibitem[Yeutter and Dew, 1982]{Yeutter1982} Yeutter, C. and Dew, J.K. (1982)
The Use of Futures in Bank Loans.
In: Prochnow, H.V. (ed.) {\em Bank Credit.} New York, NY: Harper and Row.

\bibitem[Yim {\em et al}, 2011]{Yim2011} Yim, H.L., Lee, S.H., Yoo, S.K. and Kim, J.J. (2011)
A zero-cost collar option applied to materials procurement contracts to reduce price fluctuation risks in construction.
{\em International Journal of Social, Behavioral, Educational, Economic, Business and Industrial Engineering} 5(12): 1769-1774.

\bibitem[Yo, 2001]{Yo2001} Yo, S.W. (2001)
Index Futures Trading and Spot Price Volatility.
{\em Applied Economics Letters} 8(3): 183-186.

\bibitem[Yoshikawa, 2017]{Yoshikawa2017} Yoshikawa, D. (2017) An Entropic Approach for Pair Trading.
{\em Entropy} 19(7): 320.

\bibitem[Youbi, Pindza and Mar\'{e}, 2017]{Youbi2017} Youbi, F., Pindza, E. and Mar\'{e}, E. (2017)
A Comparative Study of Spectral Methods for Valuing Financial Options.
{\em Applied Mathematics \& Information Sciences} 11(3): 939-950.

\bibitem[Young and Graff, 1996]{Young1996} Young, M. and Graff, R.A. (1996)
Systematic Behavior in Real Estate Investment Risk: Performance Persistence in NCREIF Returns.
{\em Journal of Real Estate Research} 12(3): 369-381.

\bibitem[Ysmailov, 2017]{Ysmailov2017} Ysmailov, B. (2017)
Interest Rates, Cash and Short-Term Investments.
{\em Working Paper.} Available online: \url{http://www.fmaconferences.org/Boston/Interest_Rates_Cash_and_ShortTermInv.pdf}.

\bibitem[Yu, Wang and Lai, 2005]{Yu2005}
Yu, L., Wang, S. and Lai, K.K. (2005)
Mining Stock Market Tendency Using GA-Based Support Vector Machines.
In: Deng, X. and Ye, Y. (eds.) {\em Internet and Network Economics. WINE 2005. Lecture Notes in Computer Science, Vol. 3828.} Berlin, Germany: Springer, pp. 336-345.

\bibitem[Yu and Webb, 2014]{Yu2014} Yu, S. and Webb, G. (2014)
The Profitability of Pairs Trading Strategies Based on ETFs.
{\em Working Paper.} Available online: \url{http://swfa2015.uno.edu/B_Asset_Pricing_III/paper_196.pdf}.

\bibitem[Zabolotnyuk, Jones and Veld, 2010]{Zabolotnyuk2010} Zabolotnyuk, Y., Jones, R. and Veld, C. (2010)
An empirical comparison of convertible bond valuation models.
{\em Financial Management} 39(2): 675-706.

\bibitem[Zakamulin, 2014a]{Zakamulin2014a} Zakamulin, V. (2014a)
The Real-Life Performance of Market Timing with Moving Average and Time-Series Momentum Rules.
{\em Journal of Asset Management} 15(4): 261-278.

\bibitem[Zakamulin, 2014b]{Zakamulin2014b} Zakamulin, V. (2014b)
Dynamic Asset Allocation Strategies Based on Unexpected Volatility.
{\em Journal of Alternative Investments} 16(4): 37-50.

\bibitem[Zakamulin, 2015]{Zakamulin2015} Zakamulin, V. (2015)
A Comprehensive Look at the Empirical Performance of Moving Average Trading Strategies.
{\em Working Paper.} Available online: \url{https://ssrn.com/abstract=2677212}.

\bibitem[Zapranis and Alexandridis, 2008]{Zapranis2008} Zapranis, A. and Alexandridis, A. (2008)
Modelling the temperature time-dependent speed of mean reversion in the context of weather derivatives pricing.
{\em Applied Mathematical Finance} 15(3-4): 355-386.

\bibitem[Zapranis and Alexandridis, 2009]{Zapranis2009} Zapranis, A. and Alexandridis, A. (2009)
Weather derivatives pricing: modeling the seasonal residual variance of an Ornstein-Uhlenbeck temperature process with neural networks.
{\em Neurocomputing} 73(1-3): 37-48.

\bibitem[Zapranis and Tsinaslanidis, 2012]{Zapranis2012} Zapranis, A. and Tsinaslanidis, P.E. (2012)
Identifying and evaluating horizontal support and resistance levels: an empirical study on US stock markets.
{\em Applied Financial Economics} 22(19): 1571-1585.

\bibitem[Zaremba, 2014]{Zaremba2014} Zaremba, A. (2014)
A Performance Evaluation Model for Global Macro Funds.
{\em International Journal of Finance \& Banking Studies} 3(1): 161-171.

\bibitem[Zeng, 2000]{Zeng2000} Zeng, L. (2000)
Pricing weather derivatives.
{\em Journal of Risk Finance} 1(3): 72-78.

\bibitem[Zeng and Lee, 2014]{Zeng2014} Zeng, Z. and Lee, C.G. (2014)
Pairs trading: Optimal thresholds and profitability.
{\em Quantitative Finance} 14(11): 1881-1893.

\bibitem[Zhang, 2015]{Zhang2015} Zhang, C. (2015)
Using Excel's Data Table and Chart Tools Effectively in Finance Courses.
{\em Journal of Accounting and Finance} 15(7): 79-93.

\bibitem[Zhang, 2005]{Zhang2005} Zhang, L. (2005)
The Value Premium.
{\em Journal of Finance} 60(1): 67-103.

\bibitem[Zhang, 2014]{Zhang2014} Zhang, L. (2014)
A closed-form pricing formula for variance swaps with mean-reverting Gaussian volatility.
{\em ANZIAM Journal} 55(4): 362-382.

\bibitem[Zhang, 2006]{Zhang2006} Zhang, X.F. (2006)
Information Uncertainty and Stock Returns.
{\em Journal of Finance} 61(1): 105-136.

\bibitem[Zhang, Fargher and Hou, 2018]{Zhang2018}Zhang, J.Z., Fargher, N.L. and Hou, W. (2018)
Do Banks Audited by Specialists Engage in Less Real Activities Management? Evidence from Repurchase Agreements.
{\em AUDITING: A Journal of Practice \& Theory} (forthcoming). DOI: \url{https://doi.org/10.2308/ajpt-52017}.

\bibitem[Zhang, Fuehres and Gloor, 2011]{ZhangFG2011} Zhang, X., Fuehres, H. and Gloor, P.A. (2011)
Predicting Stock Market Indicators Through Twitter ``I hope it is not as bad as I fear".
{\em Procedia -- Social and Behavioral Sciences} 26: 55-62.

\bibitem[Zhang, Shu and Brenner, 2010]{Zhang2010} Zhang, J.E., Shu, J. and Brenner, M. (2010)
The new market for volatility trading.
{\em Journal of Futures Markets} 30(9): 809-833.

\bibitem[Zhang and Xiang, 2008]{Zhang2008} Zhang, J.E. and Xiang, Y. (2008)
The implied volatility smirk.
{\em Quantitative Finance} 8(3): 263-284.

\bibitem[Zhang and Zhu, 2006]{ZhangZhu2006} Zhang, J.E. and Zhu, Y. (2006)
VIX futures.
{\em Journal of Futures Markets} 26(6): 521-531.

\bibitem[Zheng and Kwok, 2014]{Zheng2014} Zheng, W. and Kwok, Y.K. (2014)
Closed form pricing formulas for discretely sampled generalized variance swaps.
{\em Mathematical Finance} 24(4): 855-881.

\bibitem[Zheng, Thomas and Allen, 2003]{Zheng2003} Zheng, H., Thomas, L.C. and Allen, D.E. (2003)
The Duration Derby: A Comparison of Duration Strategies in Asset Liability Management.
{\em Journal of Bond Trading and Management} 1(4): 371-380.

\bibitem[Zhou, 2013]{Zhou2013} Zhou, L. (2013)
Performance of corporate bankruptcy prediction models on imbalanced dataset: The effect of sampling methods.
{\em Knowledge-Based Systems} 41: 16-25.

\bibitem[Zhou {\em et al}, 2016]{Zhou2016} Zhou, L., Yao, S., Wang, J. and Ou, J. (2016)
Global financial crisis and China's pawnbroking industry.
{\em Journal of Chinese Economic and Business Studies} 14(2): 151-164.

\bibitem[Zhu, 2006]{Zhu2006} Zhu, H. (2006)
An Empirical Comparison of Credit Spreads between the Bond Market and the Credit Default Swap Market.
{\em Journal of Financial Services Research} 29(3): 211-235.

\bibitem[Zmijewski, 1984]{Zmijewski1984} Zmijewski, M.E. (1984)
Methodological Issues Related to the Estimation of Financial Distress Prediction Models.
{\em Journal of Accounting Research} 22: 59-82.

\end{thebibliography}
\end{document}